**Aix-Marseille Université**

**UFR Sciences**

**THÈSE**

Pour obtenir le grade de
Docteur de l'UNIVERSITE d'Aix-Marseille

**Specialité : Astrophysique et Cosmologie**

Laboratoires d'accueil : **Laboratoire d'Astrophysique de Marseille et Institut d'Astrophysique de Paris**

Ecole Doctorale : **Physique et Sciences de la Matière**

Présentée et soutenue publiquement par

**Cyrille Bazin**

Le 10 Octobre 2013

# L'interface photosphère solaire/chromosphère et couronne : apport des éclipses et des images EUV

Thèse dirigée par **Dr. Philippe Lamy et Dr. Serge Koutchmy**

**Jury :**

| | |
|---|---|
| **Prof. Magali Deleuil** | **Présidente du jury** |
| **Dr. Jean-Claude Vial** | **Rapporteur** |
| **Prof. Igor Veselovsky** | **Rapporteur** |
| **Dr. Philippe Lamy** | **Directeur de thèse** |
| **Dr. Serge Koutchmy** | **co-Directeur de thèse** |
| **Dr. Sergey Kuzin,** | **Examinateur** |
| **Prof. Claude Aime** | **Examinateur** |

# « L'interface photosphère solaire/chromosphère et couronne : apport des éclipses et des images EUV »

## Sommaire

## Remerciements

Je tiens à remercier Messieurs Serge Koutchmy et Philippe Lamy pour m'avoir dirigé durant mes années de thèse, de m'avoir guidé dans l'amélioration des résultats, et soutenu dans mes missions d'éclipses et publication des résultats.
Je remercie Mr Jean-Claude Vial et Mr Igor Veselovsky pour avoir accepté d'être rapporteur, Mr Serguey Kuzin, et Mr Claude Aime comme examinateurs et Madame Magali Deleuil pour avoir accepté de présider mon jury. Je remercie Jean-Yves Prado du CNES – Toulouse pour avoir financé mes missions d'éclipses.
Je tiens à remercier Mr François Sevre et Sylvain Weiller pour m'avoir aidé dans les traitements et redressements des spectres éclairs, Patrick Rocher et Xavier Jubier pour avoir fourni les éphémérides et calculs du profil lunaire pour les éclipses. Je tiens aussi à remercier Thierry Legault, Jean Mouette, Jean-Marie Munier et Jean-Marc Lecleire pour leurs images en lumière blanche aux éclipses de 2008, 2009, 2010 et 2012. Je remercie Christian Viladrich pour ses conseils sur les paramètres d'acquisition des caméras CCD et Monsieur Heiko Wilkens pour le logiciel d'acquisition d'images Astrofactum et Lucam Recorder. Je remercie Christian Buil pour le logiciel IRIS qui m'a permis de réaliser mes traitements de données. Je remercie Antoine Llebaria pour son aide avec le programme IDL, et Gilles Missonier pour avoir résolu les nombreux « bugs » informatiques. Je remercie Alice Pisani pour les échanges sur les méthodes d'inversion d'intégrales d'Abel et Ehsan Tavabi pour les analyses d'images Hinode avec l'algorithme Madmax.
Je remercie Ahmed Hady et Luc Damé pour m'avoir invité aux conférences IAGA et pour leur aide dans la mise en forme d'articles, Je remercie Anick Degroof, David Berghmans et l'équipe de SWAP/Proba2 pour m'avoir facilité l'accès aux données aidé dans la préparation d'un article et pour m'avoir invité lors d'un meeting. Je remercie Mr John Leibacher du NSO pour m'avoir aidé pour la publication de mon premier article dans Solar Physics et Léon Golub du CFA de Cambridge - USA pour ses nombreux conseils.
Je remercie Zadig Mouradian, Ralph Krikorian et Götz Stellmacher pour les discussions sur les protubérances, les raies H & K du calcium et sur les spectres éclair.
Je tiens à remercier Thierry Roudier, Jacques-Clair Nöens pour avoir facilité mes missions au Pic du Midi sur le spectro-coronographe MSCO, et Jean-François Roudier et Christophe Guillou pour m'avoir aidé sur place et Luc Dettwiller pour mes prochaines missions.
Je remercie Arturo Lopez, et Bernard Gely pour m'avoir accueilli lors de ma mission sur Thémis à Ténérife. Je remercie Xu Zhi et l'équipe du téléscope solaire au NVST de l'observatoire de Yunnan en Chine, pour m'avoir acceuilli et mis à notre disposition le télescope solaire, lors du dernier passage de Vénus du 6 Juin 2012. Je remercie aussi David Valls-Gabaud pour son aide lors des acquisitions des spectres sur place et le traitement des données. Je remercie Marianne Faurobert pour m'avoir invité lors du French-Chinese Meeting de Nice du 4 au 8 Novembre 2011.
Je remercie Madame Brigitte Schmieder et Monsieur Guillaume Aulanier pour m'avoir invité au symposium de l'Union Astronomique Internationale UAI, « Prominences and their role in Space Weather » qui a eu lieu du 10 au 14 Juin 2013 à Paris
Je remercie Mr Laurent Vigroux pour m'avoir autorisé à fréquenter l'IAP, Mr Juan Zorec, et Mr Arnaud Cassan pour m'avoir accueilli dans l'équipe Stella.
Je remercie Monsieur Talby et Madame Francia pour m'avoir accueilli et facilité mes démarches à l'Ecole Doctorale Physique et Sciences de la Matière ED 352 de l'Université de Marseille Provence. Je remercie Olga Koutchmy pour son accueil et pour m'avoir (très souvent) invité à diner durant ces trois années.
Je remercie Simone Thijs pour les discussions sur les étalonnages des spectres de 2012.
Je remercie toutes celles et tous ceux que j'aurai oubliés.

# Résumé

**Contexte:** Les régions d'interface du Soleil de la photosphère à la chromosphère et au delà de la basse couronne ont été étudiées depuis longtemps à partir des spectres éclairs obtenus durant les éclipses totales de Soleil. Les éclipses sont les plus adaptées à ce type d'observation, car l'occultation a lieu en dehors de l'atmosphère terrestre et sont exemptes de lumière parasite provenant du « disque occulteur » (c'est-à-dire la Lune), une propriété de grande importance lorsqu'on observe très près du limbe. La résolution temporelle des premiers spectres éclairs ne permettait pas de résoudre les basses couches de la région de transition et étaient dégradés par des effets non-linéaires  qui affectaient les films photographiques. Les images Extrême-UV des régions du limbe obtenues récemment dans l'espace sont analysées avec des modèles hydrostatiques à une dimension, comme les modèles VAL, mais cette méthode ne tient pas compte du phénomène d'émergence du champ magnétique, associé au réseau chromosphérique qui est responsable de: i) les spicules et le milieu interspiculaire, ii) les jets coronaux et macrospicules, et iii) l'ovalisation de la chromosphère. Les composants de la région d'interface sont dynamiques et différents types d'ondes et de reconnexions magnétiques sont supposées agir. Un saut de température de 0.01 à 1 MK est observé autour de 2 Mm d'altitude plus loin, et produit plus loin le flot du vent solaire permanent. Le processus de chauffage responsable du saut de température et la source du vent solaire ne sont pas encore compris. Dans cette thèse, nous traitons ces problèmes à partir de spectres éclairs récents réalisés avec les technologies actuelles de détecteurs CCD rapides, images d'éclipse en lumière blanche et des images EUV obtenues avec des instruments de missions spatiales. Nous illustrons les mécanismes des émissions des raies à faible potentiel de première ionisation (FIP) présents dans les basses couches de l'atmosphère solaire. Nous identifions plus précisément les raies associées aux éléments low FIP à la fois à l'intérieur et en dehors des protubérances. Nous caractérisons en détail les enveloppes d'hélium dans les interfaces.

**Méthodes:**
1) technique des spectres éclairs sans fente avec imagerie CCD rapide (éclipses 2006, 2008, 2009, 2010 et 2012).
2) Analyses des spectres du continu entre la myriade de raies d'émission au delà du limbe solaire et construction de courbes de lumière de quelques raies d'émission low FIP et high FIP.
3) Evaluations d'inversions d'intégrales d'Abel pour déduire des échelles de hauteurs et discussion de variations de température et de densité.
4) Analyse d'images EUV obtenues aux mêmes instants depuis AIA/SDO, SWAP, SOT/Hinode des missions spatiales, images en lumière blanche pour discuter des constituants de la couronne.

**Principaux résultats:**
i) Le bord du Soleil et la bifurcation de température : le vrai continu à partir du spectre observé aux altitudes de 400 à 600 km au dessus du limbe  dans le contexte de de mesures de diamètre solaire et processus d'émission. ii) raies d'émission visibles dans les régions d'interface comprenant les raies He I et surtout la raie He II P$\alpha$ visible à partir de 800 km au dessus du limbe, produite par photo-ionisation, montrant des enveloppes autour du Soleil et permettant le sondange de l'interface protubérance-couronne. iii) La contribution de structures de petite taille comme les spicules et macrospicules commençant à 1 Mm au dessus du limbe et montrant que les modèles hydrostatiques stratifiés 1D ne sont pas adaptés pour les couches supérieures.

**Conclusion:** Nous montrons que les raies low FIP sont sur-abondantes dans l'interface photosphère-chromosphère, que la couronne solaire est alimentée en permanence par ces éléments. Le titane est un élément abondant dans le milieu interspiculaire, et une analogie sur les gradients de température entre les interfaces photosphère-chromosphère et protubérance-couronne.

# Abstract

**Context:** The solar interface region from the photosphere to the chromosphere and beyond to the lower corona has been best studied for a long time on the basis of flash spectra obtained during solar total eclipses. Eclipses are very favourable for this type of observation as the occultation takes place outside the Earth atmosphere and are therefore free of parasitic scattered light coming from the "occulting disk" (i.e., the Moon) a property of paramount importance when observing near the solar limb. The resolution of past flash spectra did not allow resolving the low layers of the transition region and were further hampered by non-linear effects affecting photographic films. Independently, EUV filtergrams of the limb region obtained in space were analyzed using one dimensional hydrostatic models like the VAL models but this method ignores the ubiquitous magnetic field emergence phenomenon associated with the chromospheric network and responsible for: i) spicules and interspicular regions, ii) coronal jets and macrospicules, and iii) the chromospheric prolateness. The components of the solar interface region are dynamical and different type of waves and magnetic reconnections are suggested to be at work. A jump of temperature from 0.01 to 1 MK is observed near the 2 Mm heights and higher, further producing a permanent solar wind flow. The heating processes responsible for this temperature jump and for the flow are not yet fully understood. In this thesis, we reconsider these problems on the basis of original, superior flash spectra which benefit from present technology such as CCD detectors, white light (W-L) eclipse images and new EUV images obtained with space-borne instruments. We illustrate the mechanisms of low First Ionisation Potential (FIP) emission lines present in the low layers of the solar atmosphere. We identify more precisely low FIP lines both inside and nearby prominences. We characterize in detail the He shells and the solar interface region.

**Method:**

1) High cadence CCD slitless flash spectra made during the total solar eclipses of 2008, 2009, 2010 and slit spectra made in 2012.
2) Analysis of the continuum between the myriad of faint emission lines outside the solar limb and construction of the light curves of some low FIP line emissions.
3) Abell inversion evaluations to deduce the scales heights and discussion of the variations of temperature and density.
4) Analysis of simultaneously obtained EUV high resolution images from the Trace, AIA/SDO and SOT/Hinode space missions and W-L eclipse images to discuss the coronal components.

**Main results:**

i) The solar edge and the temperature bifurcation: structuration and discussion of the spectral behaviour of the analysed low FIP lines and the true continuum in the 400 to 600 km heights above the limb in the frame of the solar diameter measurements, emission mechanisms; ii) Visible mesospheric and chromospheric emission lines in the interface regions including the He I lines and especially the Pα HeII line starting at 800 km above the limb, produced by photo-ionisation showing shells around the Sun and probing the prominence-corona interface; iii) The contribution of small scale dynamical structures like spicules and macro-spicules starting at 1 Mm above the limb showing that the 1D hydrostatic stratified models are not adapted to the upper layers.

# Chapitre I)  Introduction

Cette thèse a pour but d'étudier et analyser les « structurations » des couches profondes de la région de transition photosphère solaire-chromosphère et couronne , en considérant l'interface photosphère-couronne  et la chromosphère, qui sont 2 régions différentes. Ces couches et interfaces sont examinées à l'aide des éclipses totales de Soleil en utilisant le mouvement naturel du bord de la Lune comme occulteur. Le mouvement de la Lune qui recouvre le disque solaire permet ainsi de réaliser des mesures à partir et au dessus du bord solaire pour bien fixer les altitudes des différentes couches juste au dessus du limbe solaire. Cette méthode d'occultation naturelle par la Lune a l'avantage de ne pas présenter de lumière diffractée. Des techniques d'occultation avec des instruments basés sur le principe de Lyot et d'apodisation ont été améliorées (Aime 2005 et 2007), mais il reste de la lumière parasite lorsque qu'on observe très proche du bord du disque salire, même en ayant oculté le disque solaire avec des coronographes au sol.
C'est pourquoi, les éclipses totales de Soleil ont cet avantage de ne pas présenter de lumière parasite, la Lune située à environ  400000 km oculte le disque solaire. L'importance des éclipses, les méthodes et précautions pour les observer, et leur intérêt pour l'étude de la chromosphère et la couronne solaire sont décrites dans l'ouvrage Guillermier Koutchmy 1999.
L'étude et l'accès à ces couches très proches du bord du Soleil sont très difficiles à cause du taux de lumière parasite très important provenant du disque solaire étendu et très brillant. C'est pourquoi les éclipses sont utilisées car le phénomène a lieu dans l'espace, la Lune recouvre exactement le disque du Soleil, et il n'y a plus de lumière parasite provenant du disque solaire alors occulté, et ces couches profondes de l'atmosphère solaire sur la ligne de visée deviennent accessibles durant les courts instants des contacts, de durée 10 secondes, au début de la totalité et à la fin de la totalité de l'éclipse totale.
Pour fixer les idées sur ce qui est actuellement publié sur ces régions profondes de l'atmosphère solaire, citons le terme « Comosphere » qui a été introduit par Ayres et al 2002, pour montrer un problème de physique solaire non résolu: à quoi ressemble la stratification de la température dans la chromosphère solaire ? La présence de monoxyde de carbone CO observée par Newkirk 1957 dans la haute photosphère,  indique des températures de 3700 K dans des régions chromosphériques, étudiées dans l'infra-rouge. Les raies moléculaires de CO ont aussi été observées sur le disque solaire dans l'UV entre 1600 et 2000Å Goldberg et al 1965, sur les spectres de Tousey 1964. Ces analyses de la molécule CO dans les couches profondes de l'atmosphère solaire, permettent d'introduire le concept du minimum de température, où cette molécule existe, et est observée sur les spectres UV (thèse de Denys Samain, soutenue en 1978).
Un autre aspect important est la bifurcation de la température dans la chromosphère, mise en évidence ou plutôt introduite plus récemment par Holzreuter et al 2006, avec une étude complémentaire sur raies du Ca II K, voir aussi Carlsson et al 2008-2010. Les figures N° I-1 illustrent la problématique de bifurcation de température sur le bord du Soleil. La bifurcation de température signifie un mélange sur la ligne de visée d'une atmosphère non magnétique encore hydrostatique qui est en contact plus haut avec la couronne, et une atmosphère dynamique où les modèles à une dimension ne sont plus valables, où la température remonte au dessus du minimum de température. Cette atmosphère dynamique, où émerge le champ magnétique, fournit la composante chromosphère (avec les couleurs associées aux raies $H\alpha$, $H\beta$, He I D3…) avec les pieds des spicules, dont l'origine n'est pas établie.

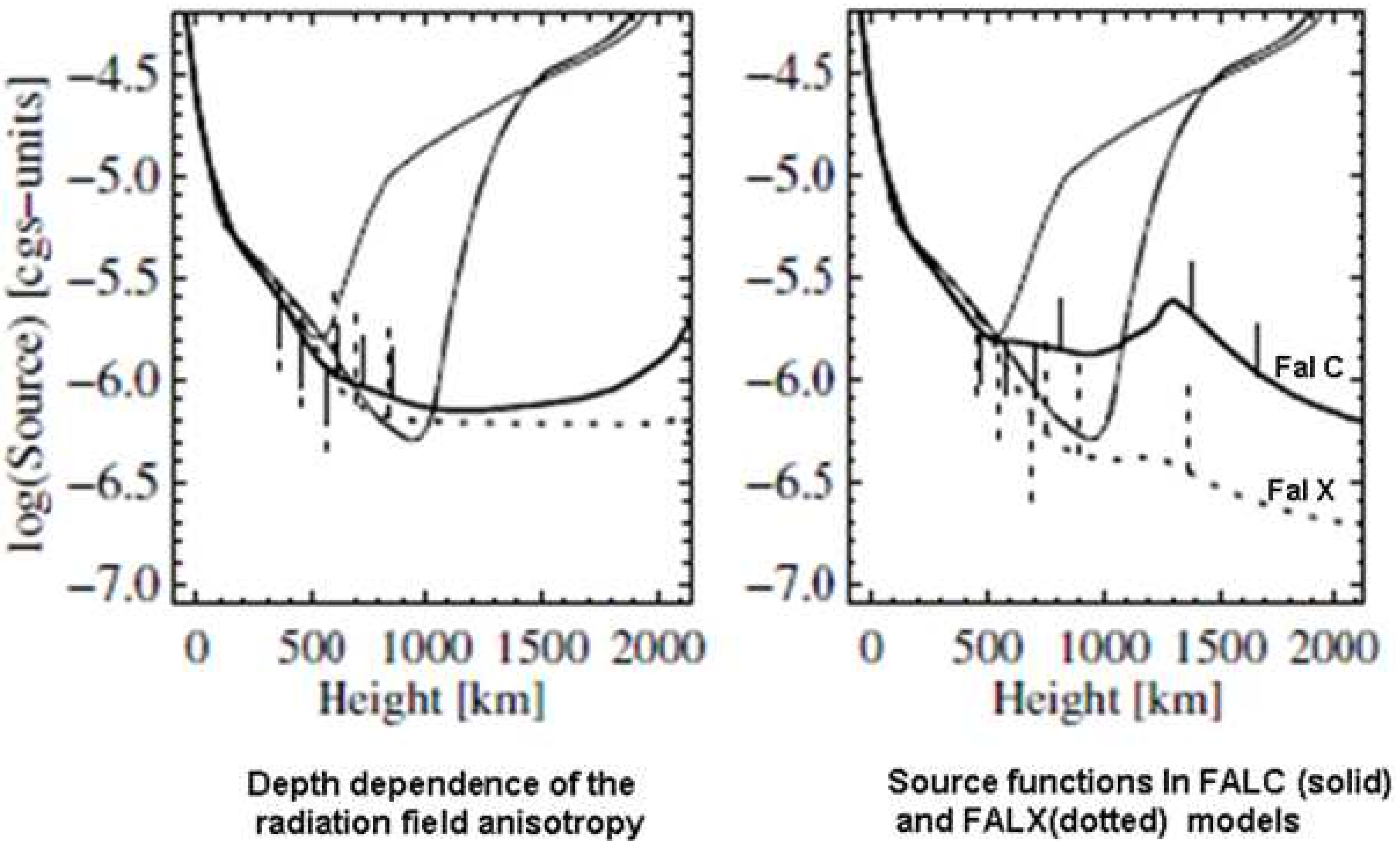

**Figure I-1:** *d'après Holzreuter R., Fluri D.M., and Stenflo, 2006.*
*Graphiques montrant la bifurcation de température dans la raie Ca II K à partir de 500 km au dessus de* $\tau_{5000}$ *=1. Le graphique de gauche est issu du modele FAL C et celui de droite du modèle FAL X. La fonction source de Planck est indiquée avec des traits fins, et le modèle à 2 composantes est représenté en traits épais, les observations avec les barres d'erreurs.*

La signification de $\tau_{5000}$ = 1 est liée à la définition d'une couche d'atmosphère solaire ayant une épaisseur optique de 1 à la longueur d'onde de 5000 Å. Plus de détails sont donnés en Annexe N°24.
Le minimum de température puis la remontée en température sur des échelles de hauteurs de 500, 1000 à 2000 km au dessus du bord solaire constituent un problème complexe de la physique solaire. Ces régions avec bifurcation apparaissent en introduisant une dissipation de l'onde de choc modélisée en 1 D par Judge Carlsson 2010, et une autre composante continue qui décroit en température. Aux altitudes plus élevées à partir de 1000 km au dessus du bord solaire, le comportement n'est pas seulement thermique, et les modèles 1 D ne sont plus valables.
Dans cette approche autour du minimum de température vers 600 km d'altitude, des modèles à une dimension sont considérés, et ils sont présentés pour introduire une description hydrostatique, homogène et stratifiée pour appréhender les couches profondes de l'interface qui sont étudiées dans cette thèse.
Nous apportons des observations nouvelles dans cette thèse pour analyser les structurations de la couche « renversante » qui au XX$^{ième}$ siècle correspondait à une couche de 500 km d'épaisseur non résolue, où lors des contacts d'éclipse, les raies d'absorption de Fraunöfer devenaient visibles en émission. Cette couche est étudiée plus en détails grâce aux spectres éclairs par imagerie CCD rapide, pour mieux définir son extension. Nous apportons une nouvelle définition du bord du Soleil, et apportons de nouvelles observations où la couche renversante, correspond à une « mésosphère », que nous définirons dans cette thèse.

La région de transition chromosphère couronne est analysée plus en détail avec les nouvelles données d'éclipse et d'observations spatiales, TRACE, SOT de Hinode, AIA/SDO.

# I-1) Description des couches extérieures de l'atmosphère solaire

## I-1-1) L'atmosphère solaire : photosphère, minimum de température, mésosphère et au-delà; le modèle VAL à une dimension pour décrire l'interface

Ces modèles consistent à décrire en fonction de l'altitude, la densité, la température et pression qui règnent dans l'atmosphère solaire. La structuration des couches des régions d'interface est simplifiée en considérant un modèle à une dimension, hydrostatique et stratifié. Un paramètre permettant de décrire la nature du plasma est introduit, le $\beta$ du plasma comme étant:

$$\beta = \text{pression du gaz / pression magnétique} = \frac{2nkT}{\frac{B^2}{8\pi}}$$

Ce paramètre est important pour caractériser l'influence du champ magnétique dans le processus de chauffage, et dans l'interface de transition photosphère-chromosphère/couronne solaire. D'autres paramètres comme l'abondance en éléments métalliques à « low FIP » à bas potentiel de première ionisation sont étudiés dans les interfaces, où des phénomènes de diffusion ambipolaire du plasma se produisent. Les éléctrons et les ions positifs sont dissociés sous l'effet du champ magnétique.

La figure I-1-1 décrit les variations de ce paramètre béta en fonction de l'altitude pour montrer où se situent les régions où $\beta < 1$, c'est-à-dire où le champ magnétique domine.

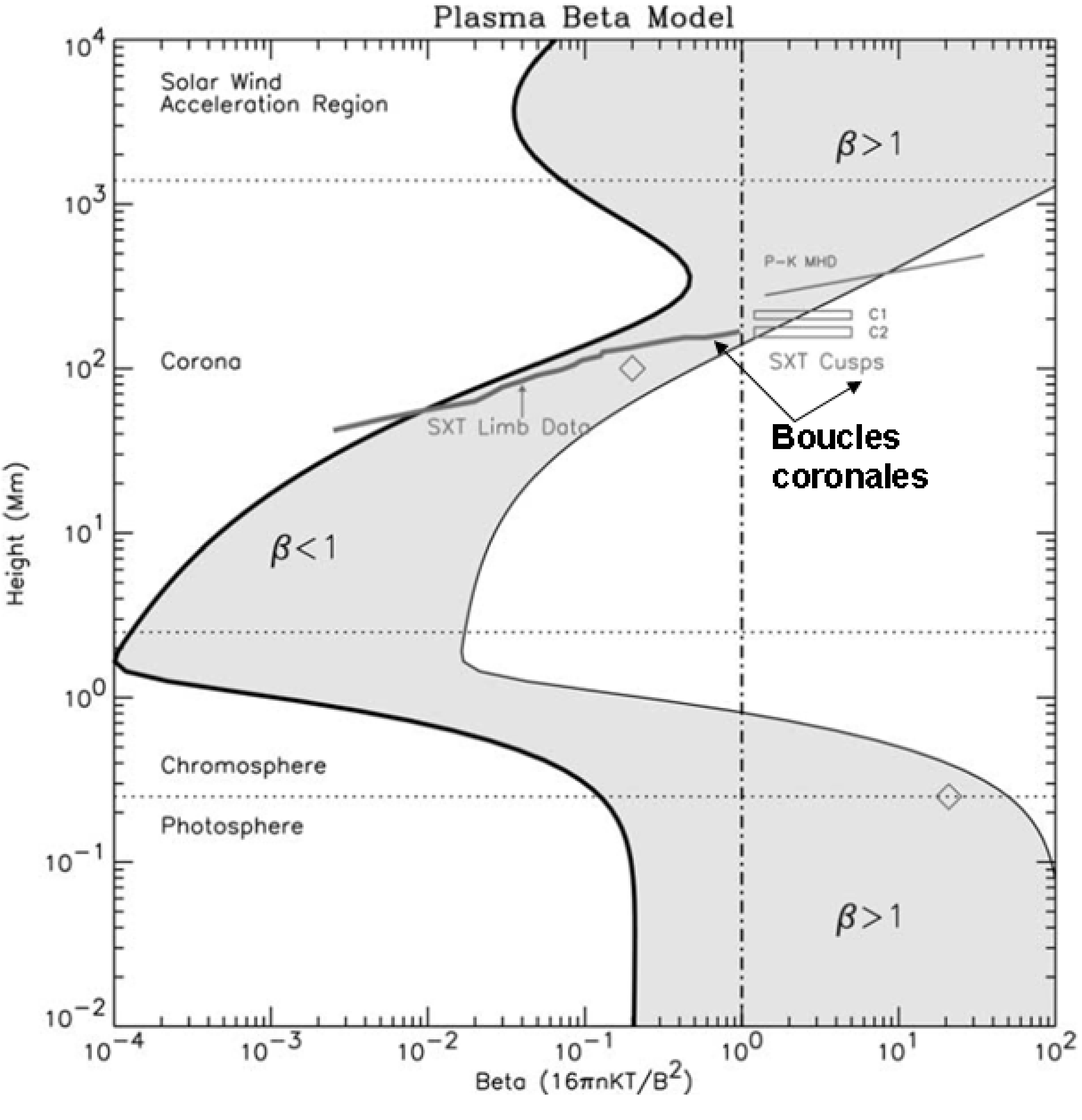

**Figure I-1-1:** *extrait du modèle du béta du plasma sur une région active de G.A. Gary, 2001. Le béta du plasma en fonction de la hauteur est représenté ombragé pour des lignes de champ ouvertes et fermées provenant d'une région entre une tache solaire de 2500 Gauss et une plage de 150 Gauss. Pour le Soleil calme,* $\beta$ *= 100. Les symboles en « diamants » marquent l'exemple photosphérique et coronal. Des données variées indiquent que* $\beta$ *est proche de 1 à des altitudes relativement basses.*

Le contexte d'études et d'analyses présentées dans cette thèse se situe essentiellement dans la gamme d'altitudes situées entre le bord solaire $h$ = 0 km et $h$ = 1000 km où $\beta < 1$. En effet, ces altitudes correspondent à la présence et formation d'une myriade de raies d'émission à faible potentiel de première ionisation (low FIP - First Ionisation Potential) et les altitudes inférieures à 1000 km correspondent au minimum de température.
La figure I-1-2 est extraite du modèle VAL pour décrire les variations de la température dans les couches d'interface, en faisant l'hypothèse d'un modèle hydrostatique et stratifié:

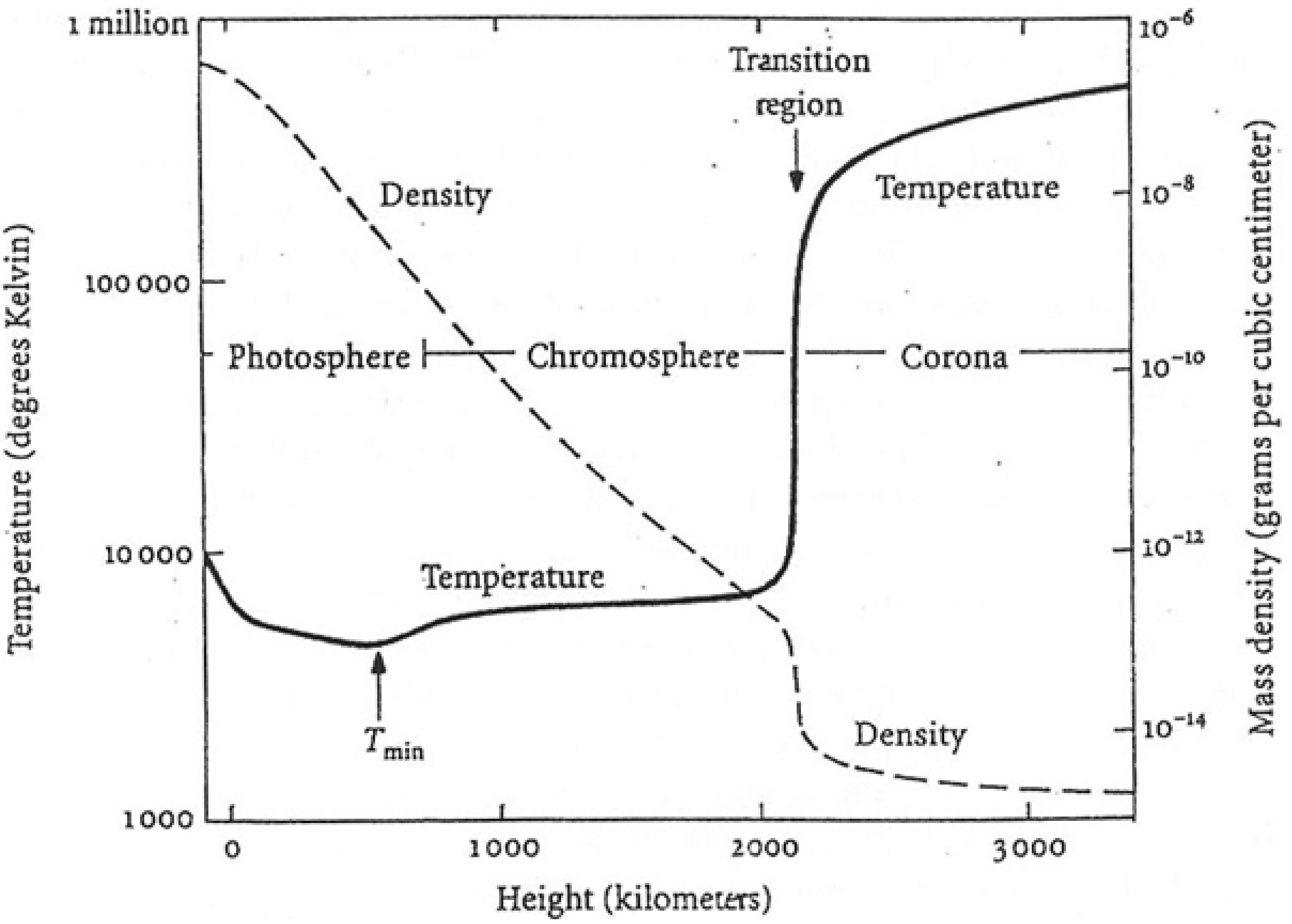

**Figure I-1-2:** *courbes extraites du modèle VAL à une dimension représentant les variations de température et de densité dans les régions d'interface photosphère chromosphère entre l'altitude 0 km et jusqu'à 1000 km au dessus du bord, avec un minimum de température, puis l'interface de transition entre la chromosphère et la couronne solaire au dessus de 3000 km*

Les différences entre les altitudes limitant les extensions des interfaces de transition photosphère-chromosphère présentées aux figures I-1-1 et I-1-2 sont incertaines, et peuvent varier de plusieurs centaines de kilomètres. Dans cette thèse, grâce à l'utilisation des spectres éclairs avec spectrographes sans fente et en utilisant le mouvement naturel de la Lune aux éclipses, nous allons donner plus de précisions sur les variations d'intensité, fluctuations, effets d'enveloppes, à partir d'analyses des courbes de lumière effectuées sur plusieurs raies d'émission et le continu, pour mieux décrire les interfaces photosphère-couronne d'une part et chromosphère d'autre part. Nous montrerons que le continu pris entre la myriade de raies d'émission, a permis de tracer des courbes de lumière pour contribuer à définir le « vrai » bord du Soleil non affecté par la lumière parasite provenant du reste du disque solaire et aussi sans la lumière résiduelle produite par les raies « low FIP » produites très proches du limbe. Ces analyses ont pour finalité de déduire la structure des couches avec une résolution de 20 à 25 km en altitude sur le limbe solaire, en utilisant des modèles atmosphériques.

## I-1-2) La couche renversante et la chromosphère

Durant une dizaine de secondes, avant et après la totalité d'une éclipse de Soleil, la Lune recouvre quasiment entièrement le disque solaire. Un très fin croissant de lumière de couleur rosée est visible; Il résulte de la superposition de la raie rouge de la série Balmer H alpha, de la raie orange D3 de l'hélium, et d'autres raies plus faibles, en réalité une myriade de petites raies en émission, qui sont les éléments les plus abondants par rapport à l'hydrogène dans les couches au dessus de la photosphère à la chromosphère. La lumière blanche résiduelle provenant du continu de l'extrême limbe solaire, grains de Baily est dominante puis décroit rapidement au début de la totalité, et croit à la fin de la totalité. Lorsqu'on réalise un spectre sans fente de ce fin croissant, les raies d'absorption sont encore visibles tant que l'extrême limbe photosphérique est visible. Lorsque la photosphère est totalement masquée par le bord de la Lune, les raies d'absorption sont vues en émission sur la ligne de visée, et constituent ce qui était autrefois appellé « couche renversante ». Ce spectre chromosphérique ou spectre éclair, a longtemps été observé, depuis la fin du XIX$^{ième}$ siècle avec des petits spectrographes à prismes en visuel. Ces spectres éclair observés autrefois comportaient plus de 3500 raies identifiées et correspondent essentiellement au spectre de Fraunhofer.
La couche renversante est la partie intermédiaire entre la photosphère et la chromosphère. Le concept de couche renversante a depuis ces époques été défini comme une fine couche de 500 km d'épaisseur, où les raies d'absorption de Fraunöfer sont vues en émission au moment des contacts avant ou après la totalité, lorsque le bord de la Lune couvre quasi-totalement le disque solaire. Dans les chapitres suivants, nous rappellons les résultats d'observations marquants, d'un point de vue historique, et comment les observations ont été améliorées au cours du temps. Nous montrerons que cette couche renversante au dessus de la haute photosphère peut être associée à une « mésosphère » où commencent à se former une myriade de petites raies d'émission et situées aux altitudes correspondant au minimum de température. Cette « mésosphère » est aussi associée aux cellules de mésogranulations, par analogie, sur les couches photosphériques, et au-delà.
La pression y est 1000 fois plus faible que sur Terre au niveau de la mer. Son étendue est au plus 1'' d'arc (Paul Couderc 1932).
La chromosphère est l'extension de la couche renversante. Son épaisseur est de 10'' d'arc au plus soit un peu moins de 1/100 du rayon solaire. Les modèles à une dimension ont été utilisés pour expliquer les observations solaires aux insterfaces chromosphère-couronne, avec les instruments à bord de la mission Skylab, et où les résultats ont été publiés dans l'ouvrage « a New Sun, the solar results from Skylab » J. Eddy et al 1979. Les modèles ont contribué à décrire des structures dynamiques, les spicules, qui présentent des inhomogénéités dans l'interface de transition chromosphère-couronne solaire, voir figure I-1-2-1.

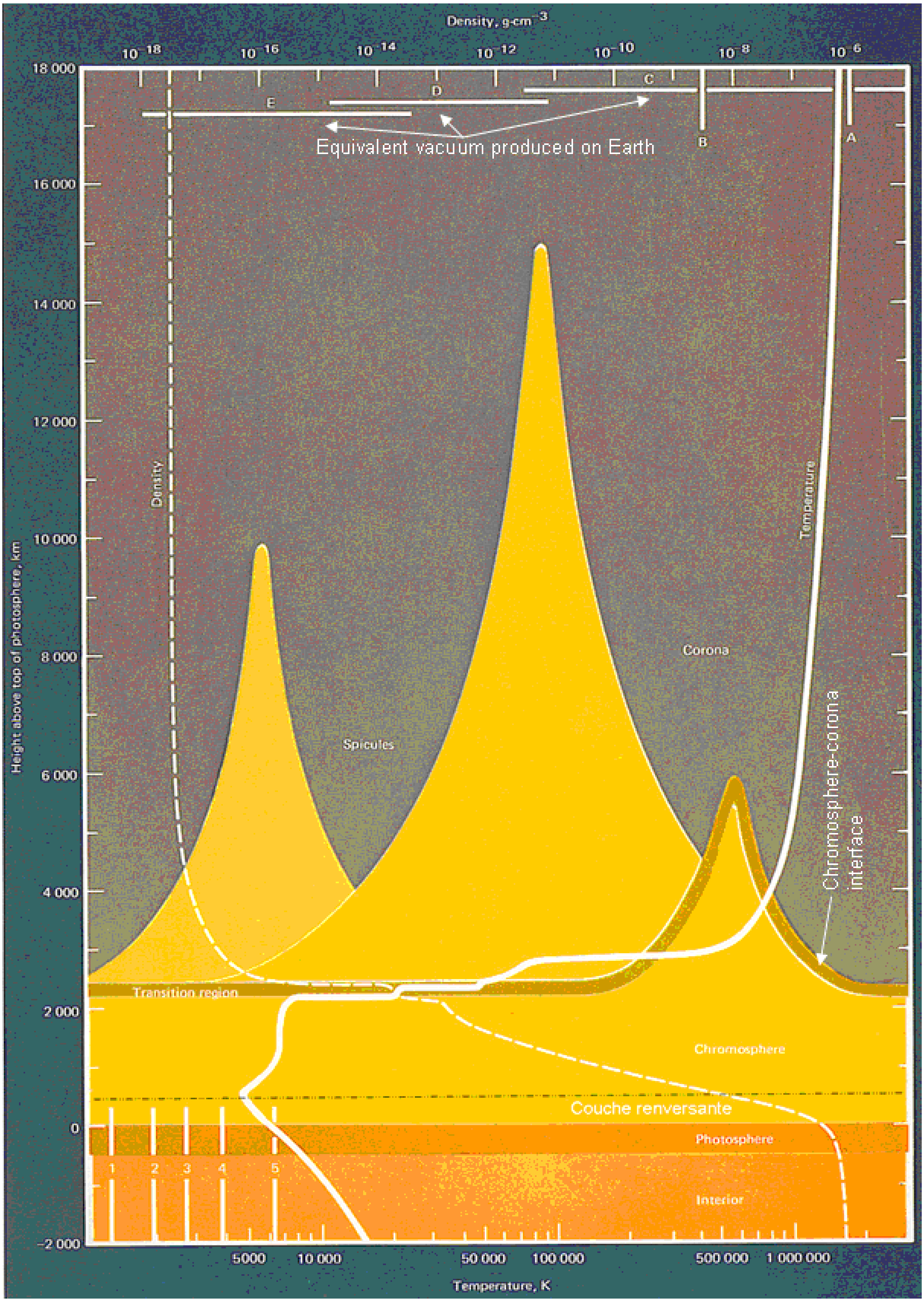

**Figure I-1-2-1:** *Courbes de températures et densités en fonction de l'altitude, d'après l'ouvrage « A new Sun, the solar results from Skylab » John A. Eddy 1979 page 2. Plus d'informations sur ce modèle issu du VAL 1D sont données ci-après.*

Les courbes de la figure I-1-2-1 montrent en jaune et orange les spicules qui à cette époque étaient supposés apparaître plus haut dans la chromosphère sous forme de sommets. Au chapitre II-4-1 il est indiqué que les spicules sont formés à des altitudes encore plus basses au dessus du limbe, à partir de 1000 km environ.

La région de transition entre la chromosphère (pieds) et la couronne est montrée en jaune foncé. La région de transition entre la chromosphère et la couronne est représentée avec une extension de seulement une centaine de kilomètres en suivant le contour des spicules. L'altitude zéro est définie au sommet, au dessus de la photosphère. La température est de l'ordre de 6000 K. Au dessus de la photosphère, là où se situe la couche renversante, se trouve le minimum de température, d'étendue plus étroite, environ 500 km, jusqu'où commencent les basses couches de la chromosphère. Ensuite, la température commence à remonter lentement dans la haute chromosphère. Ensuite au-delà d'une altitude de 5000 km environ où la couronne devient dominante, la température de 1 MK est atteinte.

La densité du plasma de l'atmosphère solaire décroit rapidement avec l'altitude au dessus de la photosphère. Entre la photosphère et la partie supérieure de la région de transition, sur une étendue de moins de 3000 km, la densité chute de 10 ordres de grandeurs, c'est-à-dire de $10^{-7}$ à $10^{-17}$ grammes/cm$^3$. Même dans la photosphère relativement plus dense, le gaz dans les couches profondes de l'atmosphère solaire, est assimilé à un vide équivalent très poussé sur Terre.

Les lettres au sommet de la figure indiquent le vide équivalent qui est obtenu sur Terre, correspondant à une altitude de 50 km dans l'atmosphère terrestre (A), à 90 km d'altitude dans l'atmosphère terrestre. Les lettres C, D et E indiquent les vides équivalents obtenus selon l'utilisation d'une pompe à vide mécanique, d'une pompe à diffusion, et d'une pompe ionique (10-7 Torr).

Les résolutions spatiales sur le disque solaire obtenues par Skylab étaient encore insuffisantes sur le limbe dans les raies « froides » pour résoudre les couches plus profondes de l'interface chromosphère-couronne solaire, et notamment pour l'analyse des embrillancements du limbe dans les raies UV.

Les embrillancements constatés dans les UV sont situés à des altitudes au-dessus de la région du minimum de température dans l'interface de transition entre la photosphère et la chromosphère. Cette interface pour les couches situées au dessus la la photosphère et avant la chromosphère permet d'apporter des informations pour l'interprétation des couches où se forment les raies low FIP (low First Ionisation Potential). En effet, ces résultats obtenus hors éclipse totale de Soleil, ont l'avantage d'être moins affectés par la lumière parasite du disque qui est encore présente dans les UV mais moins importante que dans le visible.

L'auteur utilise des modèles issus du VAL à une dimension et les résultats issus des observations qu'il obtient sont en très bon accord avec ces modèles. La méthode pour définir l'altitude $h = 0$ au dessus de la photosphère n'est pas explicitée en détail. Ceci montre que la définition du bord du Soleil proche du minimum de température est difficile, même dans la région UV. Dans cette thèse, nous avons tenté de définir l'altitude de référence $h = 0$ grâce aux spectres éclairs obtenus aux éclipses totales de Soleil, en l'absence de lumière parasite. Par ailleurs, des profils d'intensité ont été relevés sur les images de Hinode en H –Ca II, sur le bord solaire. Cette raie de température inférieure à 10000 K permet de déduire les altitudes des embrillancements dans des conditions hors éclipses, et en effectuant les dérivées de ces mêmes profils d'intensité. Nous montrerons que nos analyses à partir des résultats d'Hinode obtenus en H – Ca II concernent les couches à partir de 400 km au dessus de la photosphère mais présentent une incertitude de $\pm$ 100 km. Les images du disque solaire dans ces raies « froides », (par rapport aux températures des raies coronales à 1 MK), sont utilisées pour l'observation de phénomènes dynamiques comme les protubérances qui présentent aussi une interface avec la couronne ambiante.

## I-1-3) Les protubérances, l'interface protubérance-couronne, la couronne et le vent solaire

Il est établi par Hirayama 1978, que le centre des protubérances est un noyau plus « froid » à 4500-8500 K, et les régions externes, plus chaudes se trouvent dans le milieu coronal à 1 ou 2 MK. Les densités électroniques varient de $10^{10}$ à $10^{11.4}$ atomes/cm$^{3}$ (thèse de Jean-Claude Vial, 1981), et le degré d'ionisation $\frac{n_e}{n_1}$ varie de 1 à 3. Dans le cadre de cette thèse, les analyses de profils de raies sont peu traitées, sauf pour l'éclipse de 2012 où un spectrographe à fente a été utilisé. Aux éclipses de 2008, 2009 et 2010, nous avons utilisé un spectrographe sans fente de type réseau-objectif, et où nous observions en imagerie 3D (directions X, Y spatiales et $\lambda$ en longueur d'onde) dans les raies d'émission. Les spectres-images situées proches du bord du Soleil ont permis d'observer des protubérances monochromatiques formées dans l'hélium neutre He I 4713Å, He II 4686Å, Ti II. Ces mêmes spectres, permettent simultanément d'observer aussi le continu, correspondant aux couches le plus élevées de la photosphère (disparition des grains de Baily), jusqu'au continu coronal, et une myriade de raies d'émission correspondant chacune à un croissant de la basse atmosphère limité par le bord de la Lune. Quelques unes, parmi ces nombreuses raies ont été étudiées, comme le Ti II, Fe II, Mg I, Ba II, y compris les raies plus chaudes comme l'hélium neutre He I et une fois ionisé He II que nous avions observées comme des enveloppes et formées au dessus de ces petites raies d'émission. L'abondance de l'hélium dans les couches profondes de l'atmosphère solaire et dans les protubérances a suscité de nombreuses questions sur leur possible origine dans les couches inférieures de l'atmosphère solaire.

Les observations radio, UV et X (Schmahl 1978) ont montré la présence d'autres structures, les cavités coronales situées au dessus et autour des protubérances, où la pression du plasma dans l'interface protubérance-couronne est inférieure à la pression dans la transition chromosphère-couronne. Dans la région de transition chromosphère-couronne, Kanno et al 1980, Orrall et Schmahl 1980 déduisent que le degré d'ionisation de l'hélium est inférieur ou égal à 2. Ce résultat est important par rapport à nos observations et analyses dans les raies optiquement minces dans le visible, ainsi que les raies low FIP. Nous nous sommes aussi intéréssés à la comparaison des raies low FIP situées dans les interfaces photosphère-chromosphère/couronne et interface protubérance/couronne solaire, car ce travail a été effectué dans le domaine UV et Extrême UV (EUV) par Cirigliano, Vial, Rovira, 2004 avec les données SUMER.
Les spectres sans fente dans le domaine UV obtenus lors du vol fusée durant l'éclipse de 1970 (Gabriel 1971) montrent le gradient de température dans l'interface protubérance – couronne solaire, associé au degré d'ionisation d'éléments chimiques comme l'oxygène, le silicium, aluminium, carbone, et en fonction des largeurs en kilomètres des protubérances, voir figure I-1-3-1.

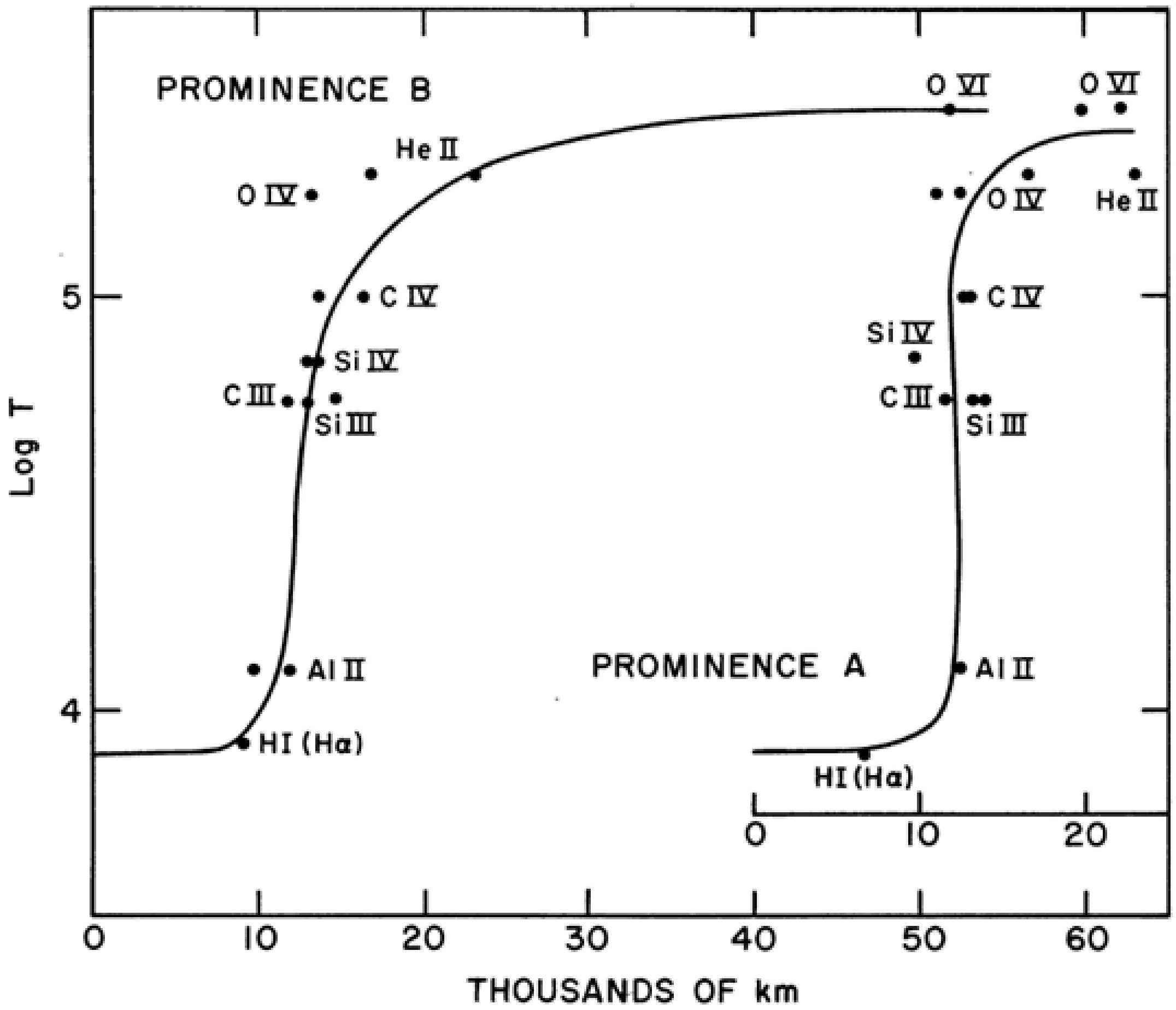

**Figure I-1-3-1:** *largeurs de protubérances en fonction de la température, obtenues avec un spectrographe sans fente, durant un vol fusée lors de l'éclipse totale du 7 Mars 1970 Ref : Orrall, F.Q. and Speer,1974*

Ces graphiques (Figure I-1-3-1) montrent que la protubérance A est moins large que la protubérance B.

Lors de l'éclipse totale du 11 Juillet 2010, nous avons analysé des profils d'intensité pris le long des protubérances dans les raies de l'hélium neutre 4713Å et hélium une fois ionisé 4686Å, en vue d'étudier des variations des rapports d'intensités dans ces raies à partir des images monochromatiques et la façon dont elles varient dans l'interface protubérance-couronne.

Les analyses de ces variations servent à évaluer des gradients de température pour tenter d'apporter quelques diagnostics pour traiter une partie du problème concernant le brusque saut de température, 10000 K à 1 MK, dans la région de transition photosphère-chromosphère/couronne, encore mal comprise à ce jour. Des importants travaux sur les mesures du champ magnétique dans les protubérances ont été réalisés par Leroy, 1981, et Labrosse et al 2010.

### I-1-4) Problématique contemporaine sur l’atmosphère solaire: origine du chauffage; source du vent solaire. Les couches profondes et la génération du champ magnétique à petite échelle.

Les degrés d’ionisation élevés observés dans la couronne, par exemple la raie du Fe XIV à 5303Å, qui a perdu 13 électrons, ou la raie du Fer X à 6374Å, indiquent que le plasma est chauffé à une température de plusieurs millions de degrés pour arracher la moitié et plus de la moitié des électrons aux atomes du gaz coronal. Ceci a été démontré par Gotrian 1939 et Edlen, 1936. Le plasma coronal est peu dense et les densités faibles $10^{11}$ particules/$m^3$, à comparer avec $10^{25}$ particules/ $m^3$ dans l’atmosphère terrestre. (Simon Mitton et Jean Audouze, 1980). Les mécanismes du chauffage du plasma coronal, de densité faible, restent encore à ce jour mal expliqués. Une source d’énergie apportée dans la couronne solaire et responsable de son chauffage, pourrait être sous forme d’ondes de choc et de turbulence qui prendraient naissance dans la photosphère, et les couches supérieures, où le champ magnétique est très concentré. Le champ magnétique présent à la surface du Soleil, généré par le plasma en mouvement, reste encore difficile à mesurer dans la haute photosphère. Les lignes du champ magnétique émergeant subissent des perturbations, qui peuvent se traduire par la génération d’ondes, de type Alfven, et qui en se propageant pourraient contribuer au chauffage coronal, mais les processus permettant d’apporter cette énergie ne sont pas encore totalement établis.
Les observations de la couronne ont révélé la présence de structures fines radiales, comme les plumes, et des mouvements rapides (jusqu’à 400 km/s) de jets dans les raies Extrême-UV, spicules et macrospicules dans la zone de transition. Certaines de ces structures (hors équilibre) sont décrites dans la Thèse de Guy Simon, « Dynamique de l’atmosphère solaire » 1987.
La plupart du rayonnement coronal se situe dans le domaine des rayons X, EUV, correspondant aux ions comme le Fer et Nickel ayant perdu la moitié de leurs électrons, car la couronne est très fortement ionisée et très chaude, au-delà de 2 MK.
La couronne, en équilibre dynamique, soumise à une force de pression qui s’oppose au champ gravitationnel du Soleil, est en expansion dans l’espace interplanétaire quasiment vide. Cette expansion de matière en dehors de la couronne est appelée vent solaire Parker 1965.
Les éjections sporadiques de plasma associées ou non aux éruptions, peuvent contribuer au vent solaire. Dans le cadre de cette thèse, nous avons comparé les interfaces protubérance -couronne solaire et photosphère - couronne, et montrerons des similitudes entre les processus d’ionisation du plasma pour ces 2 types d’interfaces en utilisant les observations des raies low FIP visibles dans chacune de ces interfaces. Nous avons aussi observé des cavités coronales autour des protubérances, qui pourraient être associées à un renforcement de chauffage coronal (voir article Bazin et al Annexe 1). La nature du plasma est encore mal comprise, aussi bien que les processus responsables du déficit de plasma. Cependant nous montrons que l’extension des cavités coronales dépend de la température du plasma. Ces cavités peuvent jouer un rôle dans le vent solaire, mais leur étude approfondie n’entre pas dans le cadre de cette thèse.
Dans la région de transition chromosphère-couronne, le flux conductif issu de la couronne est supérieur au flux radiatif dans un modèle homogène, $F_{conduction} = F_{radiation}$ pour $T \geq 10^5$ K. Un modèle à une dimension 1D hétérogène et statique a été introduit par Gabriel, 1976 dans lequel le flux conductif est canalisé et dirigé par les structures magnétiques. L’énergie dissipée est voisine du flux radiatif. Cependant, dans les couches de la partie supérieure de la région de transition chromosphère - couronne, les flux conductifs coronaux et les flux des ondes acoustiques sont trop faibles pour équilibrer les pertes radiatives. D'autres sources d'énergie peuvent être considérées comme la dissipation des ondes d'Alfven, les courants électriques dans les structures induites par le champ magnétique concentré. Jordan 1980 suggère qu'une partie de la chaleur peut être convertie par des mouvements turbulents, ce qui expliquerait l'élargissement non thermique des raies des éléments comme le fer, nickel, silicium, calcium, oxygène, présents dans les régions d'interface photosphère-chromosphère et couronne. Athay,1981 montre que le flux d'énergie nécessaire pour porter les spicules à la température de $10^6$ K est comparable au flux conductif donné par la plupart des modèles à une dimension.
Le tableau N° I-1-4-1 indique les pertes d'énergie dans la chromosphère et la couronne

| Pertes energétiques: Flux | Soleil calme (QS) | Trou coronal (CH) | Region Active (AR) |
|---|---|---|---|
| **Couronne** (ergs*cm $^{-2}$*s $^{-1}$) | | | |
| Flux conductif | $2*10^5$ | $6*10^4$ | $10^5$ à $10^7$ |
| Flux radiatif | $10^5$ | $10^4$ | $5*10^6$ |
| Flux du vent solaire | $< 5*10^4$ | $7*10^5$ | $< 10^5$ |
| Pertes radiatives de la **chromosphère (**ergs*cm $^{-2}$*s $^{-1}$) | | | |
| Basse chromosphère | $2*10^6$ | $2*10^6$ | $> 10^7$ |
| Chromosphère moyenne | $2*10^6$ | $2*10^6$ | $10^7$ |
| Haute chromosphère | $3*10^5$ | $3*10^5$ | $2*10^6$ |

**Tableau I-1-4-1 :** *Pertes d'énergie dans la chromosphère et la couronne solaire. D'après Withbroe et Noyes 1977 et utilisé la thèse de Guy Simon.*

Dans la chromosphère les pertes radiatives sont 10 à 100 fois plus importantes que dans la couronne. Cela signifie que cette interface requiert 10 à 100 fois plus d'énergie que dans la couronne. L'origine du chauffage de la chromosphère est encore plus compliquée que celle de la couronne.
Les modèles 1 D qui ont été perfectionnés ont des limites, et ne permettent pas de décrire les observations à la surface du Soleil plus modernes montrant des structures dynamiques complexes et inhomogènes, grâce aux images de meilleure résolution. D'autre part, Athay 1981 et 1982 a analysé la géométrie de l'atmosphère solaire et les mouvements dans les structures, mais n'ont pas permis d'expliquer le déficit en énergie dans la haute chromosphère et la région de transition pour $T \leq 2.5*10^5$ K. Il conclut que d'autres concepts liés à la dynamique de l'atmosphère sont nécéssaires, avec des gradients de température plus faibles et équilibre d'ionisation dépendant du temps.

## I-1-5) Désaccord des modèles à une dimension avec les observations après 1960

Ces modèles de type VAL à une dimension qui ont étés établis depuis Vernazza, Avrett Loeser 1976 et à partir d'autres observations publiées par Matsuno, K. et Hirayama 1988 montrent qu'à partir de spectres éclair d'éclipses totales passées, un total désaccord relatif aux températures, a été constaté avec ces modèles. La figure I-1-5-1 montre clairement ce désaccord.

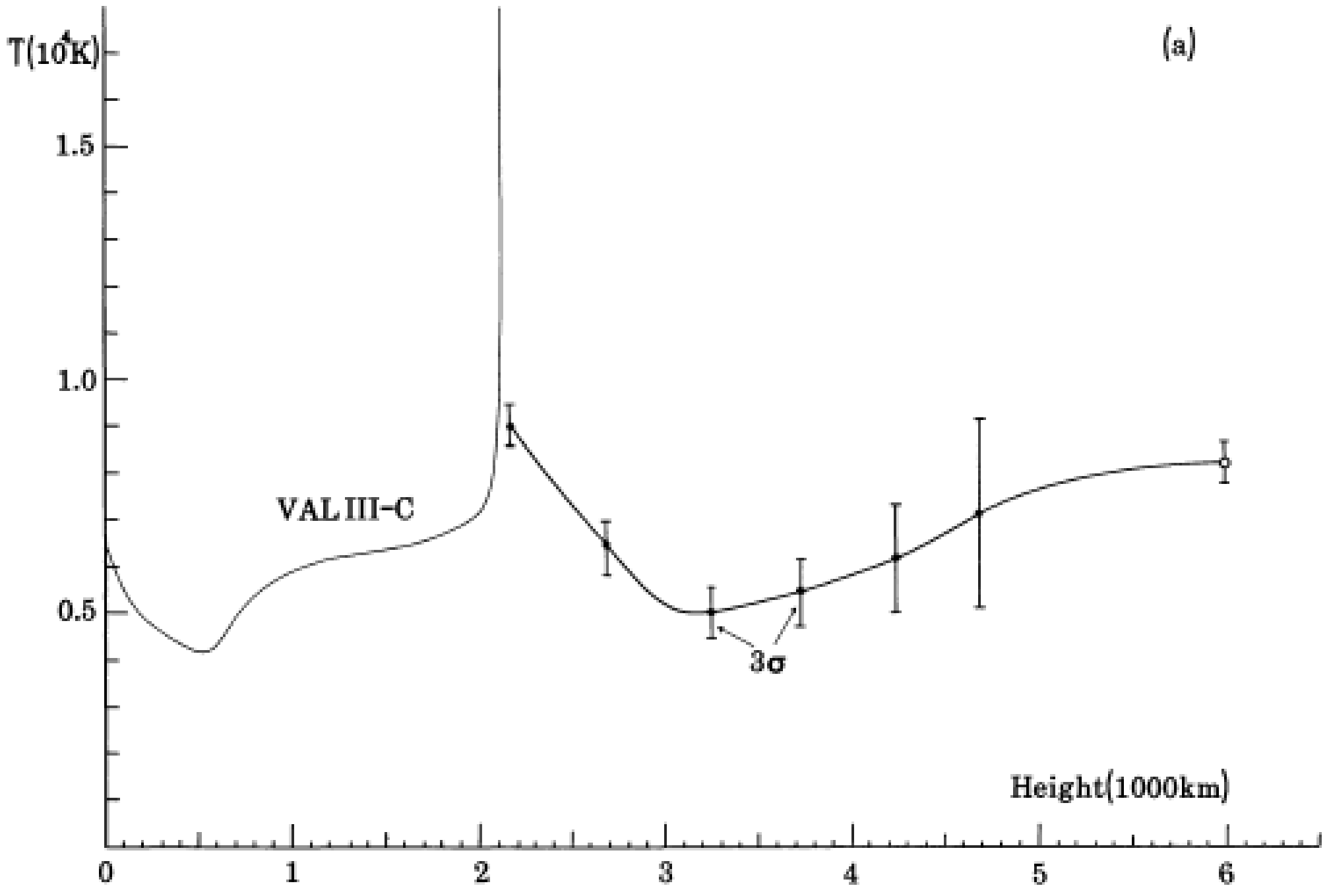

**Figure I-1-5-1:** *distributions des températures et des vitesses en fonction de l'altitude, d'après la figure 5-a de l'article Matsuno Hirayama 1988. L'altitude est prise pour une profondeur optique à 5000 Å,* $\tau_{5000}$ *= 1 dans le sens radial. Les barres d'erreurs, indiquées sur le graphique entre 2200 et 6000 km sont à* $3\sigma$ *de la valeur moyenne, et en traits fins est représentée la distribution des températures avec le modèle VAL III C, et pour les raies H et K du calcium ionisé Ca II.*

Ces résultats démontrent un désaccord entre les observations effectuées d'une part et le modèle VAL III C à une dimension où les 2 courbes de distributions des températures sont en désaccord.
Ces modèles ne correspondent pas non plus aux autres observations plus récentes montrant les spicules, macrospicules  spikes et des jets dans le domaine des rayons UV, EUV avec les nouvelles images de TRACE dans la raie du carbone C IV à 1548 Å ainsi que les nouvelles images du Solar Optical Telescope SOT de Hinode dans la raie H du Ca II à 3970 Å.
Ce désaccord se confirme avec les images figures I-1-5-2 et I-1-5-3 qui montrent que ces modèles VAL à une dimension sont inappropriés pour décrire les couches d'interfaces du bord solaire composé de spicules dynamiques, dont la visibilité a été rendue possible avec les nouveaux algorithmes de traitements d'images, preséntés en figures I-1-5-2 et I-1-5-3 utilisés par Tavabi Koutchmy 2012.

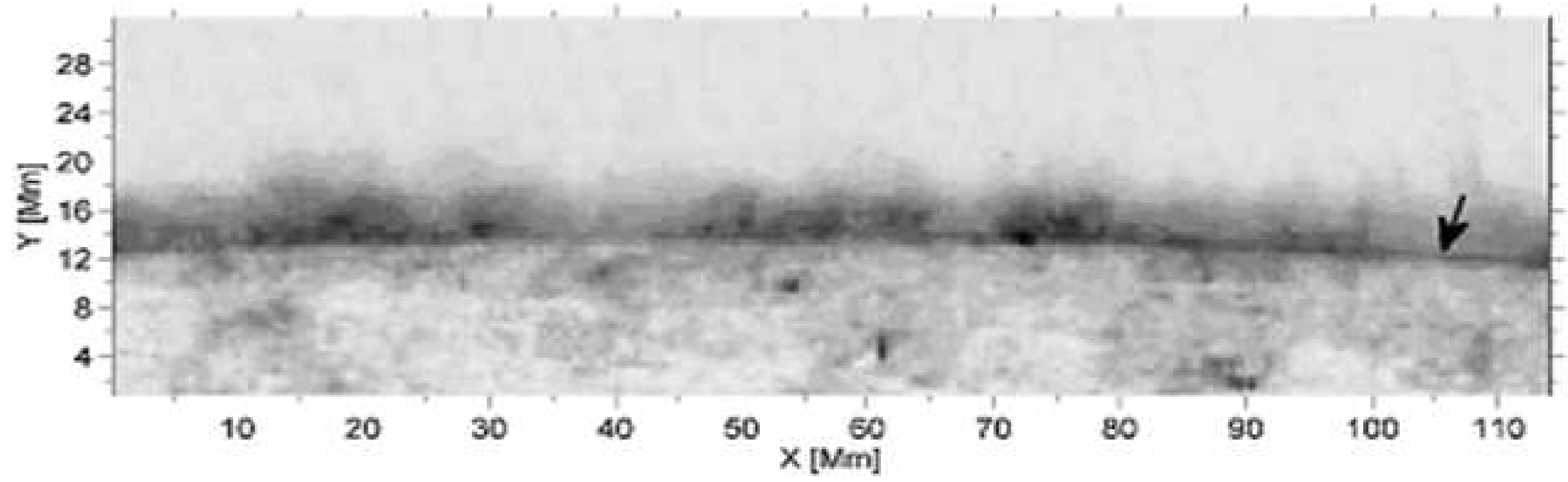

**Figure I-1-5-2:** *image en négatif du limbe solaire dans les régions polaires avec la mission spatiale TRACE, extraite d'une séquence, obtenue après soustraction de 2 images obtenues à 1550 Å et 1700 Å pour montrer les émissions du Carbone C IV dans la région de transition. Aucun traitement d'image n'a été effectué. La flèche montre une région limitée où des faibles émissions sont observées juste au dessus du limbe, d'après Tavabi E, Koutchmy, S 2012***.**

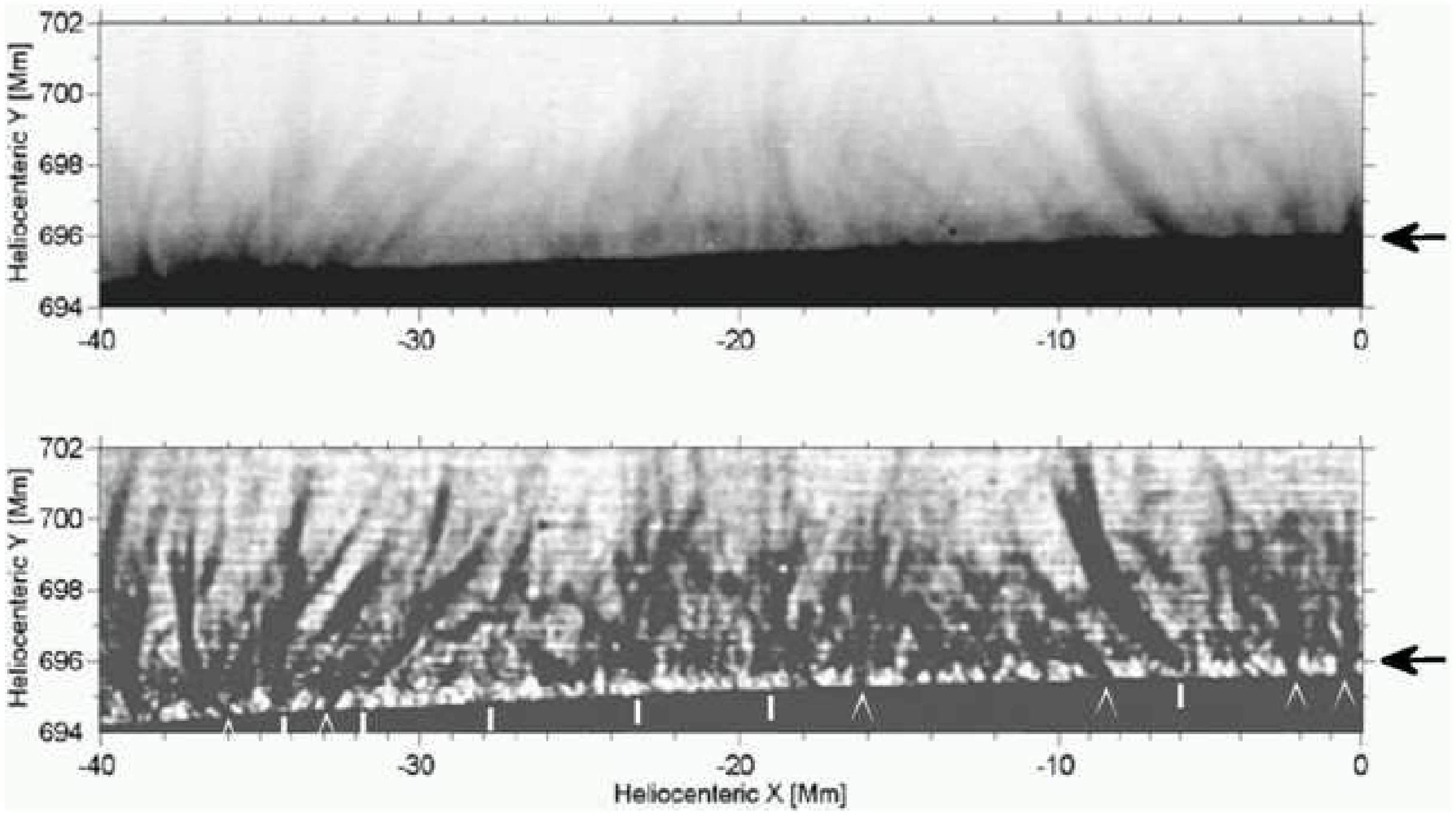

**Figure I-1-5-3:** *Images du limbe solaire avec le SOT de Hinode dans la raie H du Ca II intégré sur 5 minutes (partie supérieure) et l'image reconstruite ayant subi un traitement avec Madmax, voir Koutchmy, O 1988, par superposition d'images prises durant 5 minutes avec 8 secondes d'intervalle, pour renforcer les faibles variations d'intensité.*
*Noter à droite de chaque image, la flèche placée exactement à la même hauteur pour indiquer le travail effectué avec l'algorithme Madmax pour montrer les structures très proches du limbe. Sur l'image du bas ayant subi le traitement Madmax, sous le limbe, des marques sont indiquées pour aider à reconnaître les structures supposées provenir derrière ou au-delà (symbole ^) ou en avant du limbe photosphérique opaque. D'après Tavabi Koutchmy 2012.*

Ces images récentes ne montrent pas de structuration en forme d'étranglement, c'est à dire de “constriction” des lignes de force comme l'avaient prédit A. Gabriel et plus récement G. Athay, 1981 sur leurs modèles de flots pour expliquer l'apport d'énergie dans la couronne, comme le montre la figure I-1-5-4 extraite de l'article de G. Athay (1981).

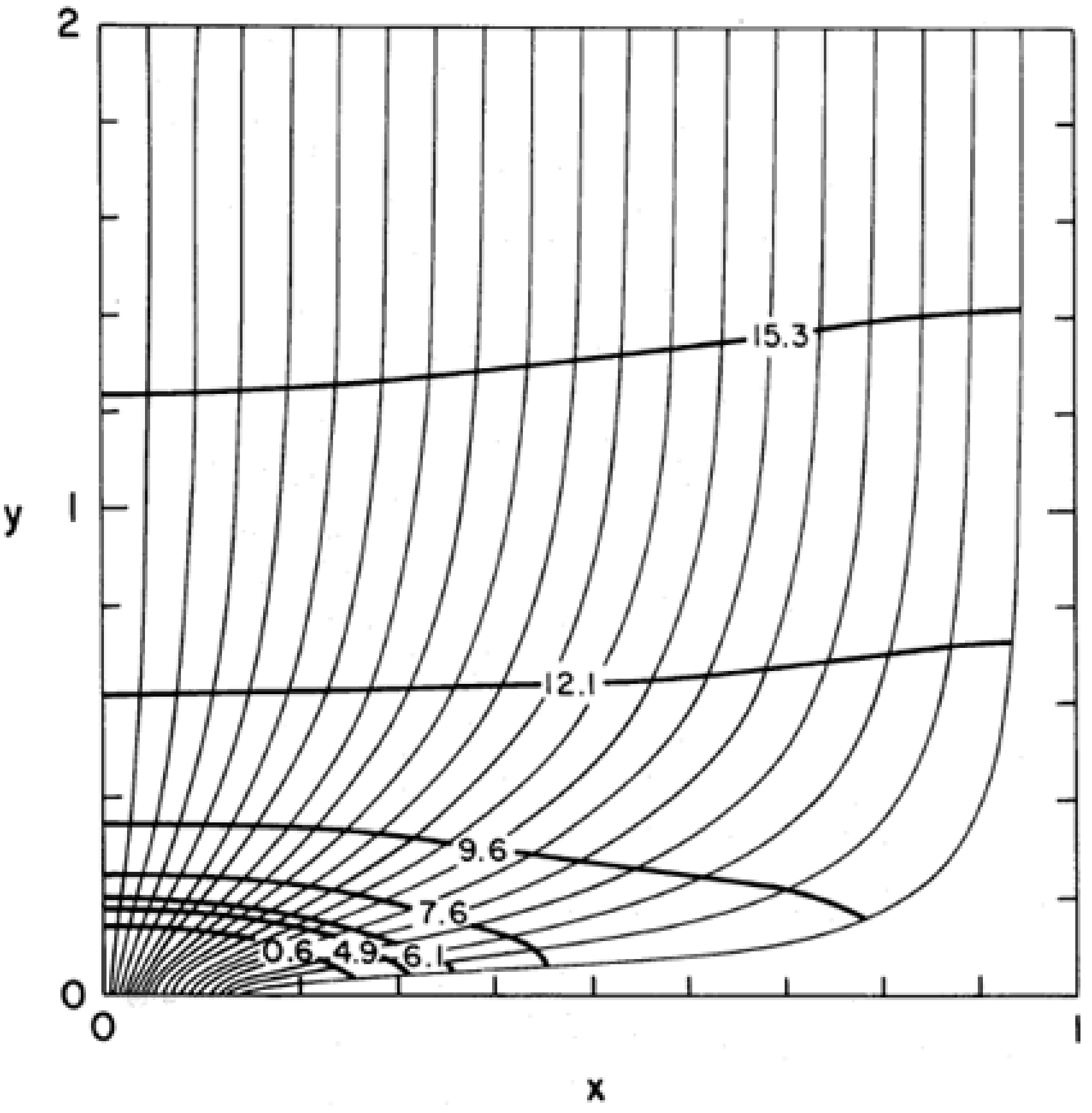

**Figure I-1-5-4:** *lignes de champ magnétique prenant leur source au centre d'un segment de réseau (x=0) jusqu'au centre d'une cellule de supergranule (x=1). Extrait de l'article de G. Athay 1981.*

Cette modélisation avec étranglement des lignes de force, proposée par Gabriel 1976, ne correspond pas aux observations récentes du limbe solaire avec spicules dans la raie H – Ca II avec le Solar Optical Telescope SOT de Hinode. Ces observations sont effectuées au bord. Ces modélisations sont différentes des observations effectuées par TRACE, le SOT de Hinode pour décrire la région de transition. Les phénomènes dynamiques sont importants, avec notamment les boucles, spicules et macrospicules, qui sont liés à l'émergence du champ magnétique en surface.
Ces constatations nous amènent à nous poser des questions pour comment expliquer le chauffage de la couronne, avec la brusque remontée de température dans l'interface de transition photosphère – chromosphère et couronne solaire.
La problématique essentielle repose sur la nature des phénomènes et les mécanismes sous-jacents responsables du chauffage de la couronne et chromosphère solaire, et qui sont associés à la remontée en température dans les basses couches de l'atmosphère solaire. Ces couches constituent une interface de transition entre la photosphère et la chromosphère, puis de la chromosphère à la couronne solaire. C'est pourquoi une majeure partie du travail de cette thèse est consacrée à l'observation et l'étude détaillée de ces couches dans les raies optiquement minces, situées situées dans l'interface de transition photosphère-chromosphère en utilisant les résultats de spectres éclairs uniques obtenus aux éclipses totales de Soleil. Cette région de transition correspond à un brusque saut de température passant de 0.01 MK à 1 MK sur 3000 km au dessus du bord solaire voir figure I-1-2. L'analyse des images dans les raies d'émission par la spectroscopie sans fente, avec le seul mouvement naturel de la Lune occultant le disque intense du Soleil, est un moyen d'analyser avec précision les infimes variations des flux de lumière émis dans les couches profondes de l'atmosphère solaire optiquement minces, accessibles seulement aux éclipses totales de Soleil en l'absence de lumière parasite. A la température de 60000K, les électrons libres ont suffisamment d'énergie pour ioniser l'hélium neutre He I, comme le décrit Dere 2000.
C'est dans ces régions de brusque transition et remontée de température qu'apparaissent des éjections de petite échelle, dynamiques, encore mal comprises à ce jour que sont les spicules apparaissant à partir de 800 à 1000 km. Aux altitudes plus élevées, 1500 à 2000km, apparaissent des macrospicules. Les analyses de ces strutctures sont développées au chapitre II-4 et surtout au chapitre IV-2-3
Les macrospicules sont environ 10 fois plus larges que les spicules et ces structures sont caractéristiques de la chromosphère. Ces phénomènes traduisent des inhomogénéités dans les couches d'atmosphère solaire et qui ne peuvent pas être représentées dans des modèles hydrostatiques stratifiés comme le VAL. Ces structures inhomogènes et dynamiques sont associées à l'émergence du champ magnétique concentré dans les couches au dessus de la photosphère solaire.
Dans cette thèse, des images CCD à cadence élevées (10 à 25 images/seconde) sont effectuées grâce aux nouvelles technologies de façon continue. Cela permet d'enregistrer les séquences des spectres éclairs durant les contacts, n'excédant pas 10 secondes, précédant et succédant la phase de totalité. Durant chaque contact, 100 à 150 spectres CCD sont réalisés, et où chacun est chronodaté par GPS. Un échantillonnage d'altitudes limitées par le seul mouvement naturel de la Lune, permet un sondage des couches d'atmosphère au dessus du limbe avec une résolution comprise entre 0.02'' à 0.04'' d'arc (intervalles d'altitudes inférieures à 30 km entre chaque spectre) selon la cadence utilisée. Le phénomène a lieu hors atmosphère terrestre mais est observé au sol, avec comme inconvénient principal, les aléas météorologiques.
Les éclipses totales de Soleil ont depuis longtemps été utilisées pour analyser l'atmosphère du Soleil, avec la méthode des spectres éclair. Il était difficile de déterminer précisément à quel moment déclencher les prises de vues avec les méthodes des plaques photographiques, pour obtenir des spectres éclairs réussis, car seulement un ou deux spectres éclair pouvaient être réalisés par contact au début du XX$^{ième}$ siècle.

## I-1-6) Problèmes pour définir le bord solaire aux éclipses en absence de lumière parasite

Bien que des progrès aient été réalisés sur les méthodes d'acquisition des spectres éclairs, leur interprétation reste difficile, de par la complexité des nombreuses structurations qu'ils révèlent: myriade de raies d'émission en forme de croissants, continu de la haute photosphère modulé par le relief du bord lunaire, protubérances monochromatiques dans le prolongement de certaines raies, modulations dans les continus chromosphérique et coronal, etc.
L'extrait figure I-1-6-1 d'un passage de l'article de B. Lites 1983 montre combien même il est difficile de définir avec exactitude une référence d'altitude *h* = 0 km sur des observations d'éclipse où des modèles sont calculés:

AN ESTIMATION OF THE FLUCTUATIONS IN THE EXTREME LIMB OF THE SUN 197

**3. The Extreme Limb Computed from an Average Solar Model**

Empirical models of the quiet solar atmosphere predict some properties of the extreme limb that will be of interest when very accurate limb-darkening measurements become available. This section is devoted to the discussion of limb curves computed from a model solar atmosphere that is homogeneous along the surface, yet including the effects of sphericity. Since the object of this study is to make a differential comparison of limb curves computed for various wavelengths from a homogeneous model of the atmosphere, but not to attempt to verify this average model, I make no attempt to compare the computed limb curves with observed limb curves. It is doubtful that either eclipse measurements (e.g. Kristenson, 1955) or direct observations of the solar limb (e.g. Dunn, 1959) attain the precision of measurement of the limb curve that would make a comparison with the observations meaningful. Such observations also suffer from the drawback of where to define a zero height for comparison to the computed curves.

**Figure I-1-6-1:** *Extrait du passage de l'article de B. W. Lites 1983 précisant les difficultés et inconvénients pour définir l'altitude zéro sur les courbes de lumière sur des modèles et sur des observations d'éclipses.*

Le problème de définition du bord du Soleil aux éclipses totales et soulevé par B. W. Lites a depuis longtemps été traité, analysé à partir d'observations, mais le relief lunaire rend très difficile l'évaluation des références exactes des altitudes; les observations anciennes tenaient peu compte du profil accidenté du bord de la Lune et des moyennes étaient effectuées, aussi bien au moyen des techniques de spectres sans fente (prisme-objectif), que le « mooving-plate » avec une fente, décrits au chapitre I-2.
A partir des analyses des spectres éclair CCD sans fente récemment obtenus aux éclipses totales de Soleil de 2008, 2009 et 2010, les références d'altitudes ont été déterminées grâce aux courbes de lumière dans les spectres des grains de Baily et décrits dans les chapitres III-3, III-4, III-5. Des analyses de spectres éclair ont été aussi effectuées avec un spectrographe à fente à l'éclipse de 2012, voir chapitre III-6, et les références des altitudes ont nécéssité une évaluation précise de la position de la fente sur le bord solaire, ce qui a permis d'exploiter ensuite ces spectres. Des moyennes dans le relief lunaire ont été considérées.
Pour des altitudes inférieures à 1000 km, les analyses des raies et structurations des basses couches de l'atmosphère solaire sont encore affectées par une quantité de lumière continue et diffuse issue des grains de Baily. Le rayonnement se mélange à l'atmosphère, et cela rend les analyses difficiles. Cette lumière continue correspond au bord photosphérique du Soleil modulé par le relief lunaire qui n'est pas totalement occulté par la Lune. L'extrait suivant de l'article Athay Menzel 1955 mentionne que de telles analyses à des altitudes inférieures à 1000 km sont difficiles. Plus précisément, les courbes de lumière des raies de l'hélium neutre 4713 Å et hélium une fois ionisé 4686 Å sont affectées par la lumière parasite pour les altitudes inférieures à 1000 km au dessus du limbe.

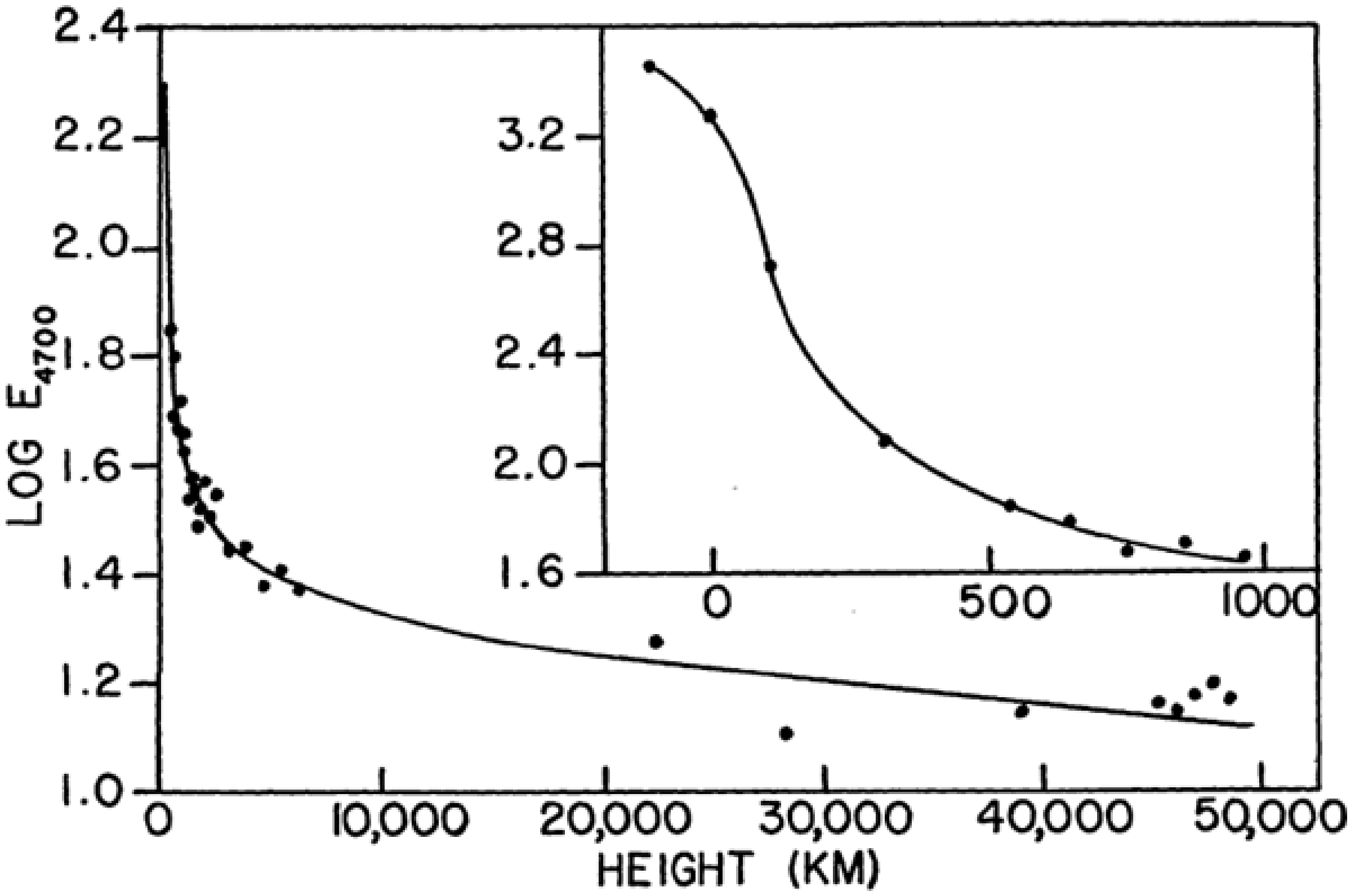

**Figure I-1-6-2 :** *extrait de l'article de Athay 1955, figure 3 pour montrer l'émission du continu à 4700 Å, et qu'aux altitudes inférieures à 1000 km, le continu devient plus intense, et rend difficile les mesures des raies d'hélium neutre 4713Å et ionisé 4686Å dont les intensités sont 10 à 100 fois plus faibles que le continu en dessous de 1000 km. La présence du point d'inflexion est due à des effets de saturation et de non-linéarité des films photométriques utilisés à cette époque. E est l'émission totale sur une tranche de 1 cm en ergs/s/frequence.*

Cet extrait montre et décrit de manière explicite que les analyses des courbes de lumière en dessous de 1000 km restent une tâche difficile car le spectre continu se superpose aux raies d'émission des éléments formés plus haut.

Des expériences de spectres éclairs ont été reproduites durant des décennies pour essayer de mieux comprendre la nature des couches situées très proches du limbe solaire, par des techniques variées qui chacune montraient une ingéniosité sans cesse améliorée pour les moyens techniques initiés au début du XX$^{ième}$ siècle avec la technique des plaques mobile (mooving plates) et l'utilisation du cinématographe. Dans cette thèse, nous avons amélioré ces techniques et analyses, grâce à l'obtention d'un nombre important de spectres éclairs pour chaque contact (avant et après la totalité) au moyen des nouvelles technologies d'imagerie CCD rapides CMos (cadences de 10 à 25 images/seconde et dynamique élevée). Les images des spectres éclair dans les différentes raies de l'atmosphère solaire aux altitudes inférieures à 1000 km au dessus du limbe ont permis une analyse détaillée et comparative des différentes couches profondes de l'atmosphère solaire, avec l'obtention et analyse des courbes de lumière déduites des spectres.

Dans les chapitres suivants, nous passons en revue les différentes observations marquantes des spectres éclair historiques, les évolutions, les limitations notamment des plaques photographiques.

## I-2) Observations historiques et observations modernes marquantes

Des expériences  historiques très importantes pour les moyens de l'époque ont été réalisées avec  des spectrographes utilisant des prismes qui se suivent ou « train de prismes », qui étaient répandus avant l'apparition des réseaux de diffraction, afin d'améliorer la dispersion. Les résultats sont en général très spectaculaires et méritent d'être examinés, mais ils sont plutôt qualitatifs. La figure I-2-1 décrit le montage du prisme-objectif, avec un train de 2 prismes qui se suivent :

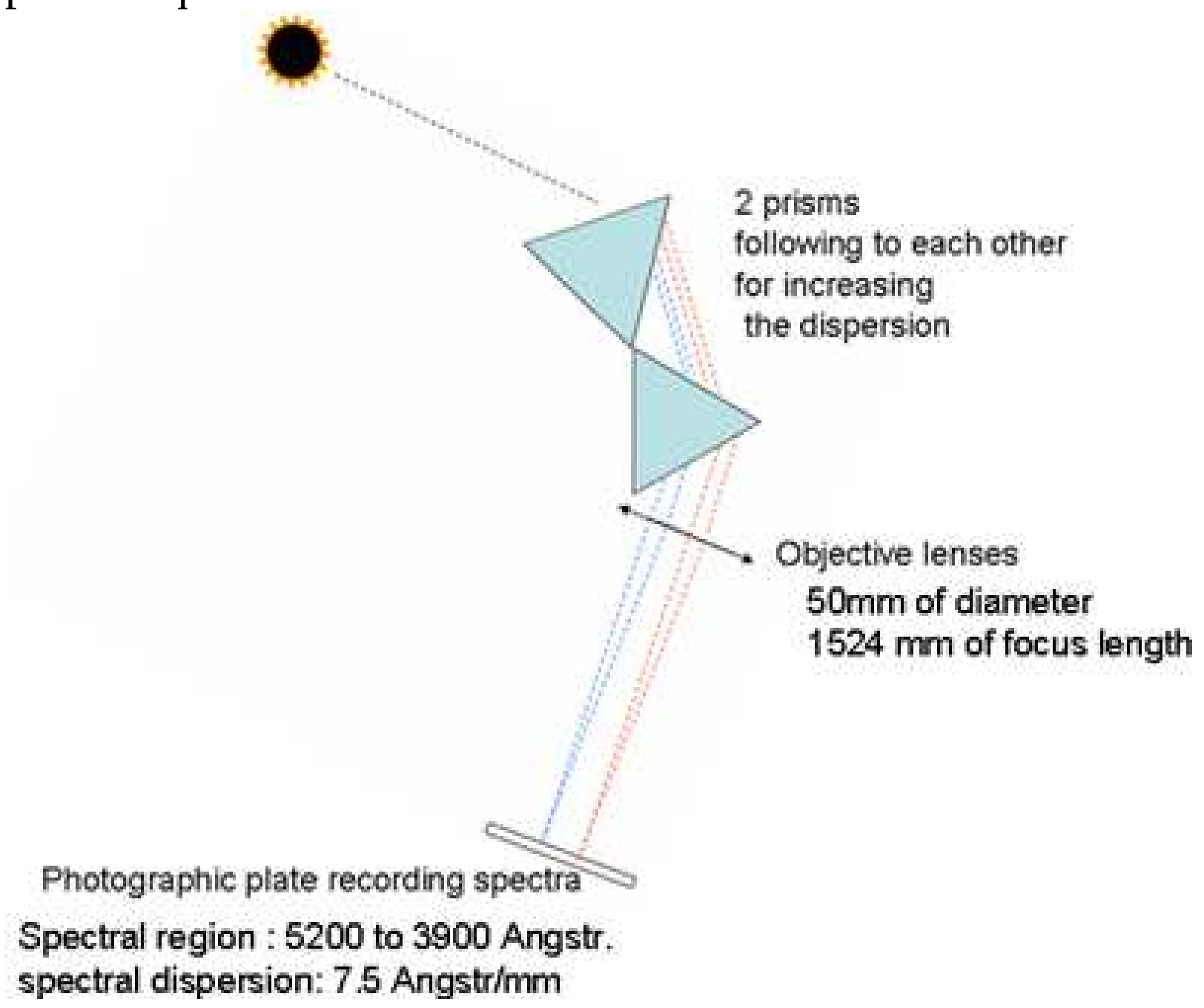

**Figure I-2-1:** *schéma du prisme objectif tel qu'utilisé à l'éclipse du 28 Mai 1900  par Lord Campbell, Thomaston, Georgia où les 2 prismes qui se suivent permettaient d'augmenter la dispersion,  plus importante en allant vers le bleu-violet. Campbell, 1900.*

Les premiers spectres utilisant  la technique du « moving plate » ont été réalisés en 1905 par Lord Campbell. Ces résultats sont décrits dans les publications de l'Observatoire de Lick 1931 par D. Menzel. Cette méthode d'enregistrement consiste à assurer le déplacement mécanique d'une plaque photographique, dont la vitesse de translation devant l'ouverture d'une fente étroite, permet d'enregistrer toute la séquence du spectre éclair, mais sur un domaine spectral et étendue spatiale limités.
Le schéma I-2-2 extrait de cet ouvrage décrit le fonctionnement du spectrographe a plaques mobiles:

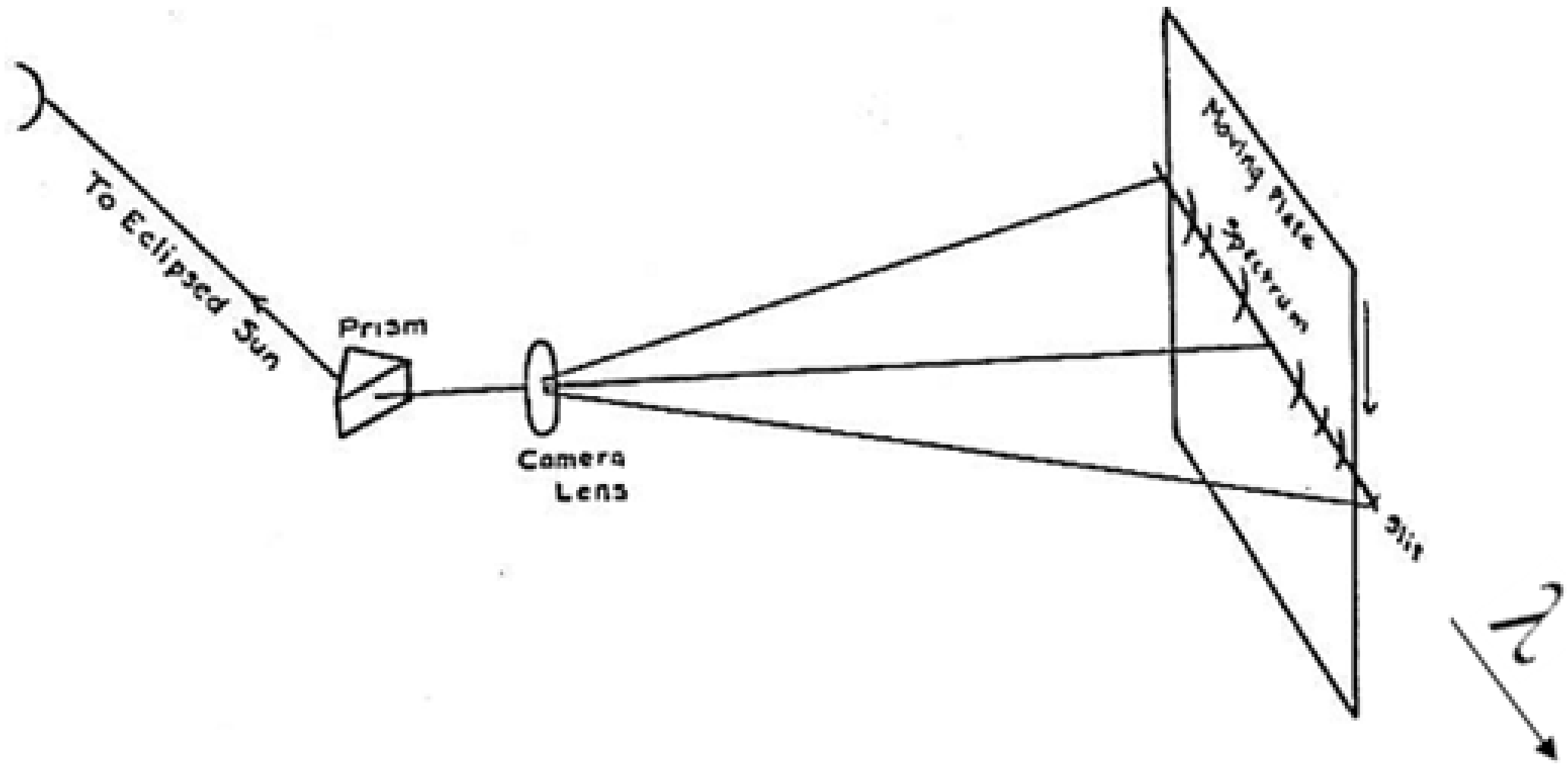

**Figure I-2-2:** *Schéma du montage du prisme-objectif à plaque mobile extrait de Publication of Lick Observatory 1931, WW Campbell and Menzel, D. Le spectre est maintenu fixe au foyer. Une fente sélectionne la lumière en provenance du spectre éclair, pendant que la plaque photographique est en mouvement dans la direction perpendiculaire au sens de dispersion spectral, permettant ainsi d'enregistrer les variations de lumière durant les contacts de l'éclipse.*

Le schéma I-2-3 indique plus précisément la position de la fente le long du sens de dispersion :

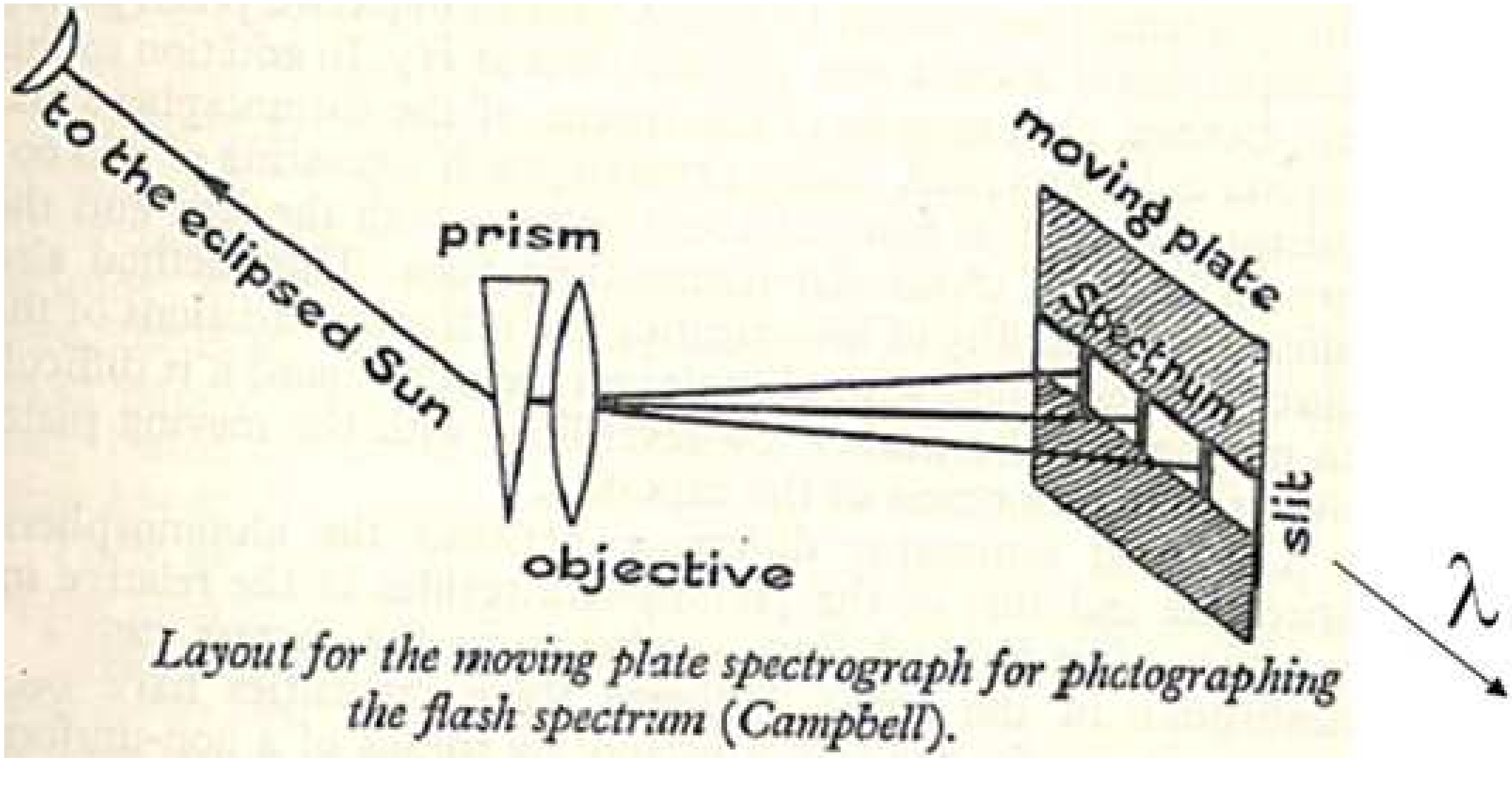

**Figure I-2-3:** *Description de l'expérience de spectrographe à plaque mobiles avec montage à prisme-objectif. Le spectre est maintenu fixe grâce à une monture assurant le guidage qui n'est pas représentée. Une fente de largeur limitée sélectionne une partie limitée de la lumière provenant des spectres flash lorsque la plaque photographique est en mouvement. Ceci permettait d'enregistrer les variations brusques de lumière pendant toute la durée des contacts lorsque la Lune couvre le bord du Soleil. D'après l'ouvrage de Giorgio Abetti, The Sun 1951.*

Le sens de dispersion spectral coïncide avec la longueur de la fente. La largeur de la fente permet de selectionner un domaine spatial limité sur la largeur du spectre éclair, pour ensuite effectuer un balayage linéaire sur les plaques mobiles, et obtenir les spectres voir figure I-2-4. Il fallait assurer un guidage sur le spectre éclair, de telle manière à ce que ce spectre reste centré et à la même position sur la fente, pendant que la plaque photographique d'enregistrement se déplaçait à vitesse constante sous la fente. Cette technique a été utilisée lors des éclipses totales de Soleil de 1898, 1900 et 1905 avec des plaques photographiques. Par exemple, la dispersion au voisinage de la raie H$\gamma$ 4340Å était de 5,7 Å/mm. Cette technique très originale pour l'époque, rappelle peut-être d'une certaine façon, une méthode ancienne de « scanner », où toute la séquence des spectres éclairs limitée par la largeur de la fente (comme échantillonnage) pouvait être enregistré sur une seule et même plaque photographique en la déplaçant. Les spectres obtenus avec cette technique illustrent le concept de couche renversante qui est aujourd'hui dépassé. Les raies d'absorption, et le spectre du limbe photosphérique visibles sur une étendue de quelques centaines de km, passent en émission et montrent une myriade de petites raies d'émission.

Un exemple des spectres historiques obtenus avec le « mooving plate » est illustré en figure I-2-4, pour un intervalle de longueurs d'onde situé dans la partie violette du spectre visible.

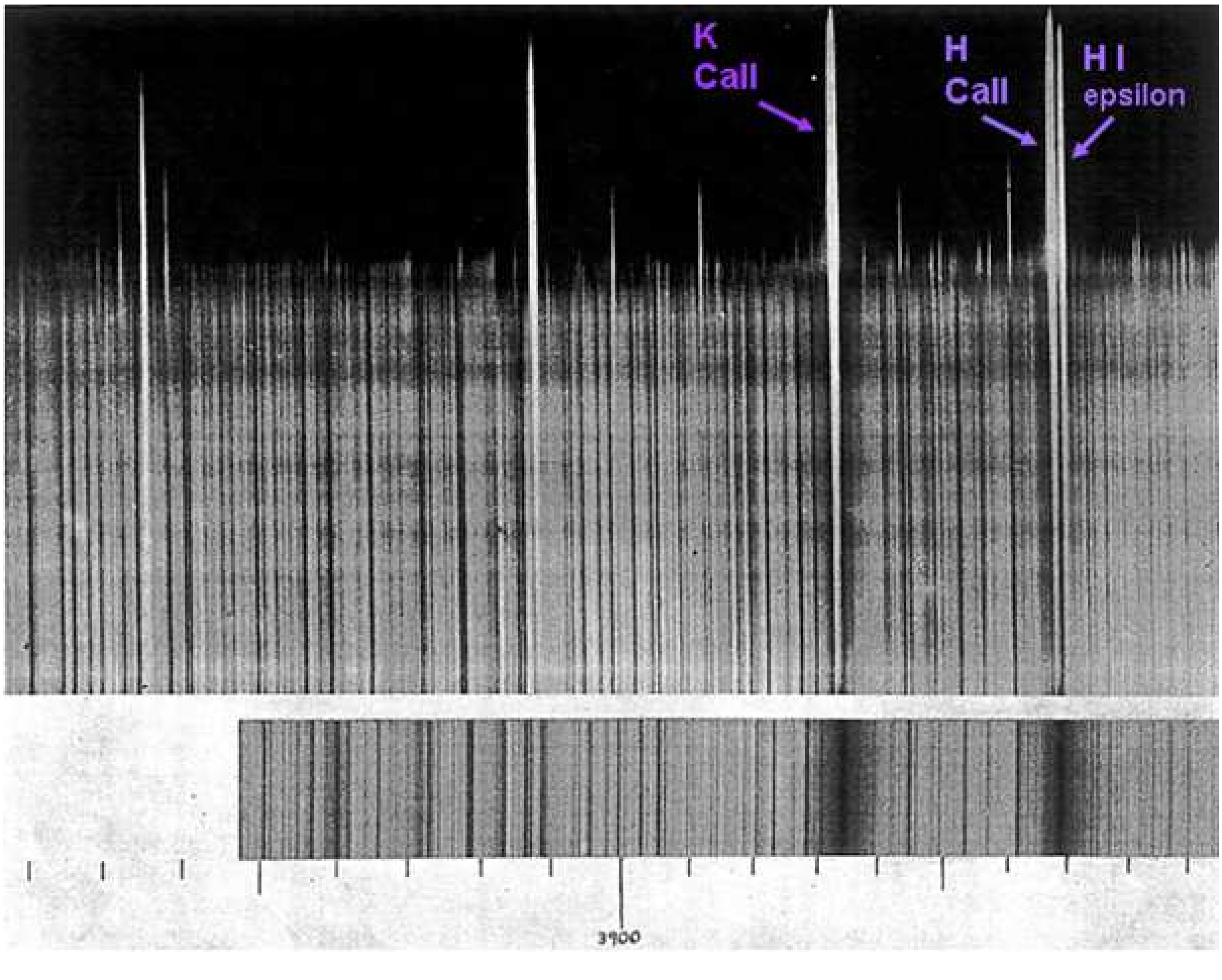

**Figure I-2-4:** *Spectres éclairs du 30 Août 1905 sur plaques mobiles «moving plate» défilant à la vitesse de 0.06 pouce/s (15 mm/seconde). La partie inférieure montre la comparaison avec un spectre normal dans les raies H&K. Association de 2 prismes à 60° placés devant un objectif de 2 pouces (51 mm) de diamètre et 60 pouces (1500 mm) de distance focale comme décrit sur la figure précédente. D'après la publication de la Partie I A "study of the solar chromosphere" by Donald H. Menzel (with an introduction by W.W. Campbell, 1931).*

Les limitations proviennent du fait que des prismes ont été utilisés, avec une dispersion non linéaire, mais devient meilleure dans le bleu-violet. La dispersion diminue en allant vers les longueurs d'onde correspondant au vert et diminue encore d'avantage vers le rouge; la dispersion est non linéaire avec les prismes et de la forme Cauchy, ci-après

$$D(\lambda) = A + \frac{B}{\lambda^2} + \frac{C}{\lambda^4}$$

A et B sont des constantes dépendant de l'angle et matériau du prisme et le terme en $\frac{C}{\lambda^4}$ est négligé devant les autres. Mitchell, S.A. a critiqué cette méthode des plaques movibles en expliquant que les croissants des spectres obtenus sur une plaque fixe sont beaucoup plus utiles et indispensables pour une analyse scientifique, c'est-à-dire pour résoudre spatialement le limbe solaire limité par le bord de la Lune, avec les images monochromatiques dans chaque raie d'émission, permettant l'analyse des basses couches de l'atmosphère solaire. Ainsi, aucune précision n'est atteinte sur les hauteurs analysées avec la technique des plaques mobiles, car le mouvement de la plaque est seulement destiné à compenser la grande variation des intensités radiales. Par ailleurs, la technique des plaques mobiles « moving plate » ne permet pas d'atteindre les niveaux d'intensité proches du continu coronal, ce qui a été aussi fortement argumenté par Mitchell, qui préconisait de réaliser plutôt des spectres éclairs sans fente. Le spectre figure I-2-5 a été réalisé dès 1905 avec un réseau objectif, et toujours sur plaque.

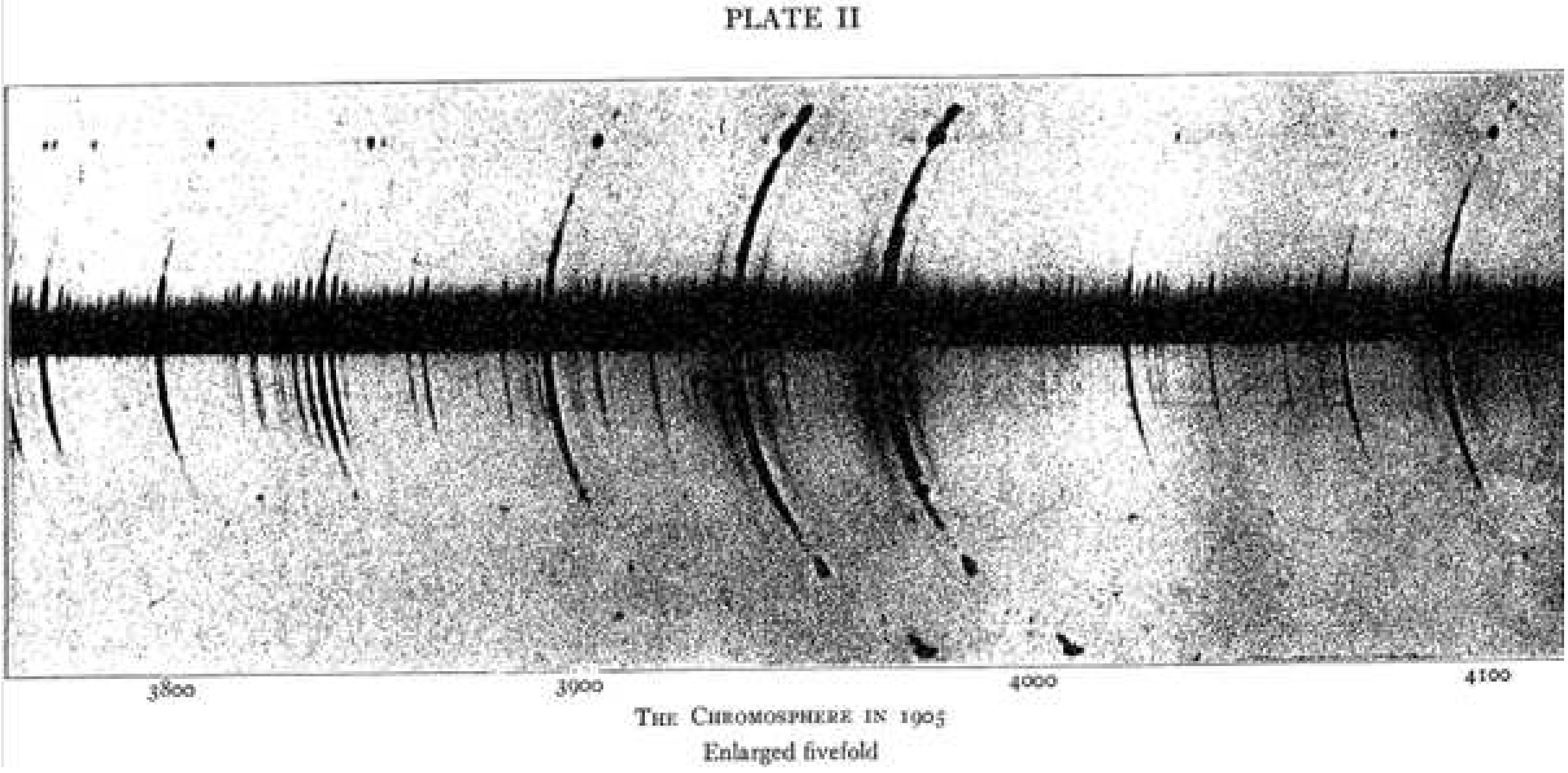

**Figure I-2-5:** *spectre éclair réalisé avec un réseau objectif, et exposition de 0.5 seconde. Ce spectre a été obtenu en 1905 dans des conditions de seeing excellentes d'après l'article de Mitchell, 1935, et les nombreuses petites raies visibles ont été identifiées et indiquées dans les tables, avec les potentiels d'ionisation.*

Les spectres éclairs obtenus par Mitchell consistent en un spectre sans fente et sans plaque mobile. Il a utilisé un réseau à la place d'un prisme pour obtenir une dispersion linéaire ce qui permet de réaliser plus facilement une identification des nombreuses raies d'émission, et atteindre une meilleure résolution spectrale. La figure I-3-6 montre un extrait d'une séquence de spectre flash obtenue par Mitchell et al en 1925 avec des indications et légendes des structures observées sur le spectre et telles que nous les étudierons avec les spectres flash récents obtenus avec des caméras CCD. Dans mes travaux de thèse j'ai utilisé la même méthode que Mitchell du spectrographe sans fente, mais avec une caméra CCD à cadence rapide d’images à la place des plaques photographiques, et aussi avec une expérience plus compacte et transportable dans un bagage pour le transport en avion sur les sites d’observation. L’avantage de cette méthode du spectrographe sans fente, est d’utiliser l’occultation naturelle de la Lune devant le Soleil, où les spectres éclair sont directement enregistrés et se présentent sous forme d’une multitude de croissants correspondant aux raies d’émission présentes dans les basses couches de l’atmosphère solaire. Un exemple de spectre éclair historique est donné en figure I-2-6.

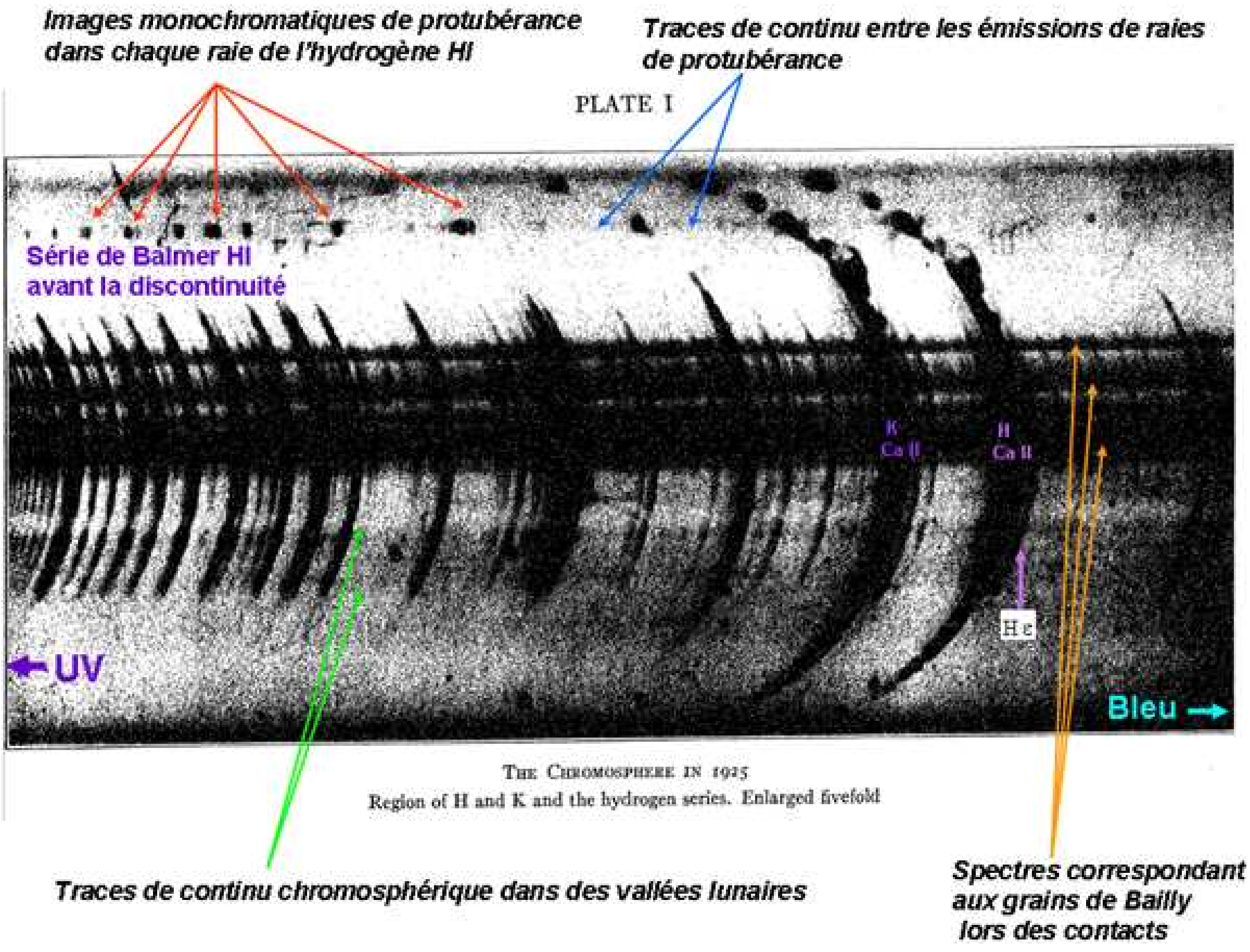

**Figure I-2-6:** *extrait du spectre éclair de l’éclipse totale de 1925 avec annotations et légendes dans les régions H&K et série de Balmer de l’hydrogène avant la discontinuité montrant les spectres du continu des grains de Baily, de nombreuses petites raies en émission avec les structures observées et l’effet des grains de Baily (vallées profondes du profil de la lune). Les raies de plus grande étendue montrent la structuration de l’enveloppe chromosphérique. Notamment dans l’hydrogène et du Calcium K à 393 nm, et dans la raie, on peut distinguer très nettement les images monochromatiques de protubérances. La durée d’exposition des spectres flash était de une demi-seconde avant la totalité et une demi-seconde après la totalité durant les contacts. D’après Mitchell, S.A. 1935.*

Cette structure de spectres avec les raies d’émission intenses sous forme de « croissants » s’explique par le mouvement de la Lune couvrant le disque solaire où il reste un fin croissant avant et après la totalité. La Lune occulte tout le disque solaire sauf les couches d’atmosphère sur le limbe, à la fin de la totalité de l’éclipse, comme le montre la figure I-2-6. Les plaques photographiques utilisées à cette époque étaient plus sensibles dans les parties bleu-violet du spectre visible et donc bien adaptées pour la région spectrale étudiée.

Le continu chromosphérique et la myriade de petites raies plus faibles ont été observés de façon satisfaisante durant cette éclipse même si dans l'article de Mitchell le « seeing » était moins bon qu'en 1905, compte-tenu de la durée d'exposition des plaques photographiques. Par comparaison entre ces 2 spectres, celui de 1905 ne montre qu'un continu beaucoup plus faible, entre des protubérances, mais ne montre pas de continu chromosphérique comme en 1925. Il est possible que le spectre de 1925 ait été pris quelques secondes plus tôt dans l'instant du contact où les derniers grains de Baily photosphériques étaient encore visibles, tandis que celui de 1905 pourrait avoir été enregistré quelques secondes plus tard au moment du tout dernier grain de Baily plus fin avant la totalité. Nous n'avons pas eu ce problème aux éclipses de 2008, 2009 et 2010, car les acquisitions des images de spectres CCD étaient effectués en continu, le début des enregistrements était engagé durant les phases partielles précédent le second contact, et il n'y avait pas à se préoccupper du problème de déterminer l'instant pour déclencher les prises de vues aux contacts avant et après la totalité.
Par ailleurs les nombreuses protubérances observées le long des raies en 1925 témoignent aussi d'un Soleil plus actif qu'en 1905. Cependant il était très difficile de conclure sur les mesures des intensités des raies en comparant ces 2 éclipses.
L'auteur cite la raie du Fe XIV 5303 Å et du Fe X 6374 Å qui ont été identifiées plus tard, et l'auteur prétend qu'elles étaient plus intenses en 1925 qu'en 1905, ce qui indiquerait une couronne plus active en 1925 qu'en 1905. La comparaison entre les spectres en 1905 et 1925 montre un « seeing » excellent en 1905 mais médiocre en 1925. L'identification des raies des spectres éclair est rapportée en Annexe 8.
Lors de l'éclipse totale de 1927, Pannekoek et Minnaert ont réalisé des photographies excellentes du spectre éclair et ont utilisé pour la première fois un microphotomètre enregistreur pour analyser avec précision les variations d'intensité dans les raies et continu des spectres aux contacts avant et après la totalité.
Chacune des phases de spectres flash sera beaucoup plus détaillée ultérieurement grâce à l'imagerie CCD rapide, à l'aide d'expériences beaucoup moins considérables en dimensions et beaucoup moins coûteuses.

L'avantage de la technique du spectrographe sans fente, est qu'elle permet d'obtenir des séries d'images monochromatiques 3D dans chaque raie d'éléments chimiques différentes, c'est-à-dire selon les directions spatiales X,Y des arcs, et spectrale $\lambda$, des couches de la chromosphère, et des images de protubérances. Les plaques photographiques de cette époque n'avaient pas une sensibilité suffisante pour bien observer et déceler les faibles niveaux du continu entre les protubérances, et les niveaux plus faibles du continu coronal. Cependant, sur l'extrait du spectre de Mitchell, certaines parties du continu entre les protubérances peut être vu, mais de manière uniquement qualitative.
La nature accidentée du relief lunaire ajoute des difficultés pour définir le bord solaire et déterminer l'origine des altitudes $h = 0$ km sur le limbe et aux instants des contacts, a été considéré pendant les phases partielles, précédant la totalité d'une éclipse. Menzel 1930 a tenté de définir le profil du bord de la Lune au moyen des phases partielles. Ce travail postérieur aux éclipses de 1905 et 1925 a permis de mieux analyser et établir les amplitudes des modulations sur le bord solaire partiellement occulté. A partir de ces résultats, des paramètres relatifs aux circonstances des éclipses ont pu être calculés.
En considérant la Lune et le Soleil comme des cercles parfaits (en négligeant le relief lunaire, montagne et vallées) sur la figure I-2-7, soit $a$ le rayon du Soleil et $c$ celui de la Lune. Pour une éclipse de Soleil, $c > a$.

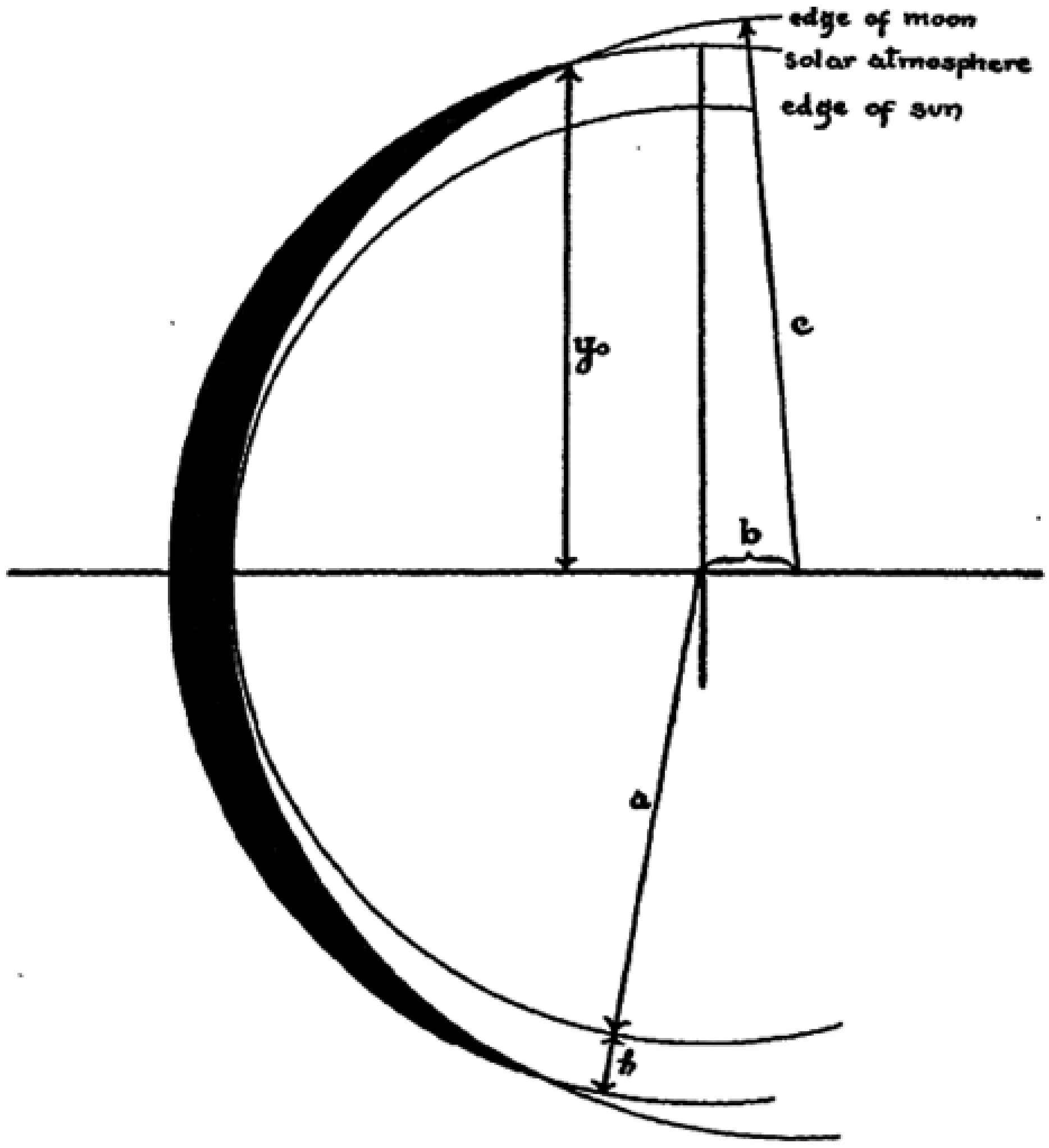

**Figure I-2-7:** *Relation géométrique des positions des centres de courbure de la Lune et du Soleil. D'après Menzel, D. H. 1930, dans le but d'expliquer la forme des images en "croissants" des spectres de la myriade de raies d'émission avec un spectrographe sans fente, utilisé lors des contacts précédant ou succédant la phase de totalité des éclipses.*

Les centres sont séparés par la distance b. En prenant l'origine au centre du Soleil, on peut déduire l'équation représentant l'atmosphère solaire à l'altitude *h* au dessus de la surface photosphérique :

$x^2 + y^2 = (a+h)^2$

et l'équation du bord de la Lune est :

$(x'+b)^2 + y'^2 = c^2$

En combinant les 2 équations précédentes, en posant $x = x'$ *et* $y = y'$, en négligeant les carrés des petites quantités très inférieures à 1, on obtient:

$$y_0/a = \sqrt{1 - \frac{(c-a)^2}{b^2} + 2*\frac{(c-a)}{b}*\frac{h}{b} - \frac{h^2}{b^2}}$$

en introduisant $c\text{-}a = b+f$, où *f* est la distance entre le bord de la Lune et le bord du Soleil.

Au second ou troisième contact (c'est-à-dire juste avant ou juste après la totalité de l'éclipse), lorsque les 2 astres sont tangents à l'intérieur, $f = 0$, on obtient:

$$y_0/a \sim \sqrt{-\frac{2f}{c-a}-\frac{f^2}{(c-a)^2}+\frac{2h}{c-a}+\frac{h^2}{(c-a)^2}}$$

Les termes en $f^2$ et $h^2$ sont fréquemment négligeables, généralement $f << b$
Les quantités *c* et *a* sont connues grâce aux circonstances de l'éclipse.
Par exemple, à l'éclipse de 1905, leurs valeurs respectives étaient 16.597' et 15.845'.
En prenant le rayon solaire à 695500 km, on pouvait vérifier facilement que $c\text{-}a$ = 33000 km
A partir des mesures des spectres d'éclipse, il est possible de déterminer la quantité $y_0/a$ pour n'importe quelle raie. De plus, si *f* est connu, il est possible de calculer l'altitude *h* pour laquelle cette quantité peut être tracée. La phase de l'éclipse est déterminée par *f*. Lors de l'éclipse de 1905, le bord de la Lune se déplaçait à 300 km/s à la distance du Soleil. Il était possible de calculer les valeurs de $y_0/a$ pour différentes phases. La figure I-2-8 présente les relations pour différentes phases partielles sur des intervalles de 1 seconde avec les altitudes en kilomètres et la manière dont les auteurs à l'époque ont essayé d'en tenir compte :

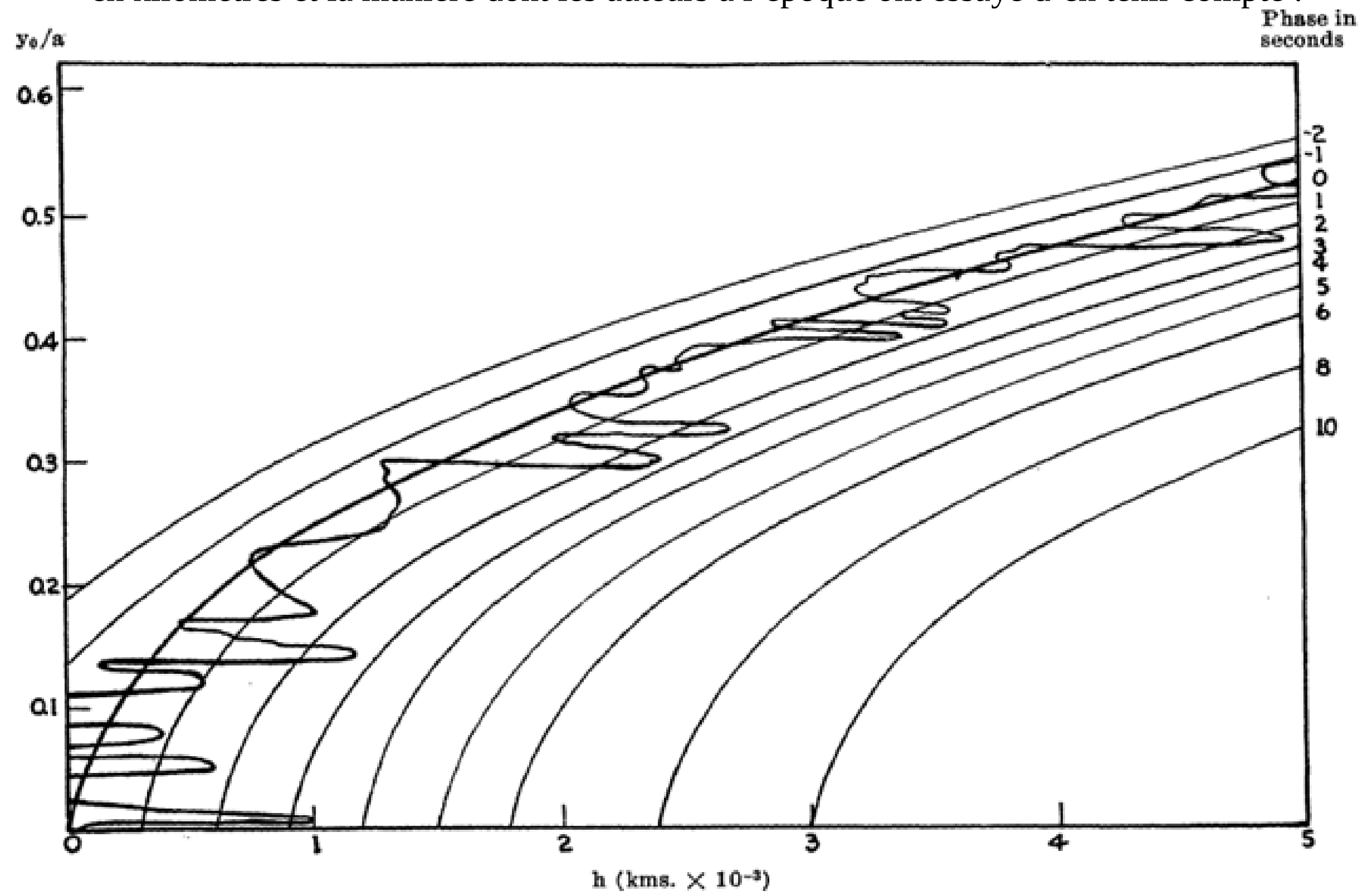

**Figure I-2-8:** *Effets de phase sur les hauteurs déterminées par les croissants au moment des phases partielles, estimés sur des images en direct. D'après Menzel, D. H. 1930.*

Les courbes indiquent les positions successives de la surface lunaire, considérée comme sphérique, sur des intervalles de 1 seconde. Mais en réalité la Lune n'est pas exactement sphérique à cause de faibles déformations et de son relief accidenté au bord. Les irrégularités de plusieurs kilomètres sont fréquentes. 1 km d'irrégularité sur la Lune correspond à environ 400 kilomètres au bord du Soleil. Les plus grandes vallées font 2.5 km et les plus petites font ¾ de km. Les vapeurs métalliques correspondant aux images dans les raies d'émission en croissants au dessus de 2000 km ont des chances de réapparaître dans la vallée suivante avec une incertitude de 700 ou 800 km d'après la figure I-2-8.
Ces profils ont permis une première approche des problèmes du relief lunaire.
Les spectres éclairs obtenus à cette époque ont permis d'analyser les caractéristiques des petites raies. Un travail considérable d'analyses et d'identification des éléments chimiques associés aux raies d'émission a été effectué, et les tables en Annexe 8 choisies sur des intervalles correspondent aux longueurs d'onde étudiées aux éclipses récentes de 2008, 2009 et 2010. La méthode d'imagerie CCD rapide permet d'analyser plus précisément, et avec une dynamique de $2^8$ à $2^{12}$ niveaux d'intensité les variations d'intensité dans ces petites raies au moment des instants des contacts précédant et succédant la totalité.
A cette époque une difficulté importante était de déclencher la photographie du spectre éclair au bon instant.
Lorsque c'était trop tôt avant la totalité, les spectres étaient surexposés et affectés des grains de Baily intenses, les images étaient saturées et les raies de plus faible émission étaient mélangées avec la lumière des spectres continus très intenses. Quand la photographie était déclenchée quelques secondes trop tard juste au début de la totalité, seule la frange chromosphérique était enregistrée, avec beaucoup moins de petites raies d'émission.
Comme les temps de pose sur les plaques photographiques étaient de l'ordre de la seconde, seulement une ou deux vues étaient possibles pour chaque contact avec les moyens de l'époque.
En résumé, les spectres sans fente obtenus à cette époque, montrent des arcs avec les protubérances, et il est possible d'interpréter ces données directement, tandis que les spectres enregistrés avec les plaques amovibles ne permettaient pas d'analyse approfondie sur les spectres dans l'atmosphère solaire par eux-mêmes, car on ne peut pas savoir si les spectres correspondent à une région perturbée ou si c'est une portion représentant les conditions solaires moyennes. (Mitchell et al 1935).

<u>Les conclusions scientifiques tirées de ces expériences Mitchell et al sont les suivantes:</u>

- à partir des spectres éclairs, chaque croissant en émission correspond aux raies d'un atome qui est formé à une altitude bien définie (voir figure I-3-1) et ces couches correspondent aux vapeurs métalliques. Ces couches ont des extensions au dessus de la photosphère.
- Le Fe I constitue 37.6 % des raies observées dans tout le spectre éclair visible
- Le Ti II comporte 226 raies d'émission et le Ti I 195 raies. Le titane est responsable de 15% des raies
- Beaucoup de différences apparaissent dans les arcs pour les différentes raies du spectre éclair. Les arcs les plus intenses correspondent aux gaz (vapeurs métalliques) dans les couches profondes, et les arcs de très faible intensité, comme He II 4686 (high FIP). Il est difficile d'expliquer comment cette raie ayant un potentiel d'excitation de 48.16 eV peut être trouvée à toutes les températures des couches de l'atmosphère solaire.

Des nouvelles expériences de spectres éclair ont été retentées dans les années 1940 et plus tard, dans le but d'étudier les profils du continu du bord du Soleil ainsi que les variations d'intensité des raies d'émission.

## I-3) Expériences des spectres éclairs 1940 publiés par R. Wildt en 1947

Cette expérience de spectres éclairs provient d'une observation d'une éclipse très peu connue ayant eu lieu le 1er Octobre 1940 et publiée par R.O. Redman 1942. Des profils de courbes de lumière ont été mesurés, par Ruppert Wildt 1947.
Les raies d'émission de spectres éclairs ont été également analysées par Wildt. où les altitudes des hauteurs d'émission des raies métalliques ont été déduites et l'auteur a défini une expression de gradients d'émission moyenne « Av. Emission Gradient $\Delta\log(I)$ per 100 km » voir figure I-3-1.

| Spectrum | Av. Emission Gradient $\Delta \log I$ per 100 km | Relevant Range in Height (Km) |
|---|---|---|
| *Sc* II. . . . . . . . . . . . . . . | 0.16 | 800–1500 |
| *Ti* II. . . . . . . . . . . . . . . | .19 | 1000–1500 |
| *V* II. . . . . . . . . . . . . . . | .18 | 500–1200 |
| *Cr* I. . . . . . . . . . . . . . . | .20 | 500–1000 |
| *Cr* II. . . . . . . . . . . . . . . | .12::* | 500–1200 |
| *Mn* II. . . . . . . . . . . . . . | .14 | 800–1500 |
| *Fe* I. . . . . . . . . . . . . . . | .16 | 500–1200 |
| *Fe* II. . . . . . . . . . . . . . . | .18 | 500–1200 |
| *Ni* I. . . . . . . . . . . . . . . | .18 | 500–1000 |
| *Y* II. . . . . . . . . . . . . . . | .18 | 500–1200 |
| *Zr* II. . . . . . . . . . . . . . . | .18 | 500–1200 |
| *Mg* I. . . . . . . . . . . . . . . | .05 | 2000–6000 |
| *Ca* I. . . . . . . . . . . . . . . | .12 | 500–1200 |
| *Ca* II. . . . . . . . . . . . . . . | .04 | 10,000–14,000 |
| *Sr* II. . . . . . . . . . . . . . . | .07 | 1000–6000 |
| *Bu* II. . . . . . . . . . . . . . | 0.17:: | 1000–2500 |

**Figure I-3-1:** *Gradients d'émission moyens, sur une plage de hauteurs limitée, pour les métaux alcalins et les terres rares, $\Delta$Log(I)/100 km qui prend la valeur 0.04 pour l'hydrogène. D'après l'éclipse totale de 1940.*

Ces calculs de gradients d'émission moyens sur une étendue de 100 km sont définis comme équivalents à des échelles de hauteurs, où les auteurs indiquent qu'il faut multiplier les valeurs de la colonne $\Delta$Log(I)/100 km par $2.3*10^{-7}$ cm$^{-1}$ pour déduire les valeurs du coefficient $\beta = \frac{1}{H}$ qui est l'inverse de l'échelle de hauteur, et selon les équations suivantes :

$n(h) = \int_{-\infty}^{\infty} c(h)dh = \sqrt{\ }(\frac{2\pi Rsun}{\beta})*c_0*e^{-\beta h}$, où *n(h)* est la densité d'atomes à l'altitude h au dessus du limbe, c(h) sont le nombre d'atomes émissifs, et $\beta$ est l'inverse de l'échelle de hauteur en cm$^{-1}$ pour laquelle la densité à diminué d'une valeur de *1/e.*
Le calcul de $\Delta$Log(I)/100 km, gradient logarithmique, permet de déduire $\beta$ comme étant l'inverse d'une échelle de hauteur. Les auteurs ont fait l'hypothèse de l'équilibre thermodynamique. Cela conduit à des expressions de la densité d'atomes comme étant une exponentielle décroissante proportionnelle à une densité spatiale :

$c(h) = c_0*e^{-\beta h}$

Le terme pré-exponentiel est multiplié par $\sqrt{\ }(\frac{2\pi Rsun}{\beta})$, pour tirer *n(h)*

Pour davantage de clarifications, nous avons reproduit le tableau précédent en donnant directement les valeurs des échelles de hauteurs d'après les mesures effectuées par les auteurs depuis cette éclipse de 1940. par exemple, pour $\Delta$Log(I)/100 = 0.16 sur la première ligne, on obtient $\beta = 0.16*2.3*10^{-7} = 3.68*10^{-8}$ cm$^{-1}$

| element | Av Emission Gradient $\Delta$logI per 100km | $\beta$ (cm$^{-1}$) | Scale of height H (km) ~ 1/$\beta$ | Relevant Range In Height (km) |
|---|---|---|---|---|
| Sc II | 0.16 | $3.68*10^{-8}$ | 271 $\pm$84 | 800 - 1500 |
| Ti II | 0.19 | $4.37*10^{-8}$ | 228 $\pm$60 | 1000 - 1500 |
| V II | 0.18 | $4.14*10^{-8}$ | 241 $\pm$84 | 500 - 1200 |
| Cr I | 0.20 | $4.60*10^{-8}$ | 217 $\pm$84 | 500 - 1000 |
| Cr II | 0.12 | $2.76*10^{-8}$ | 362 $\pm$84 | 500 - 1200 |
| Mn II | 0.14 | $3.22*10^{-8}$ | 310 $\pm$84 | 800 - 1500 |
| Fe I | 0.16 | $3.68*10^{-8}$ | 271 $\pm$56 | 500 - 1200 |
| Fe II | 0.18 | $4.14*10^{-8}$ | 241 $\pm$56 | 500 - 1200 |
| Ni I | 0.18 | $4.14*10^{-8}$ | 241 $\pm$84 | 500 - 1200 |
| Y II | 0.18 | $4.14*10^{-8}$ | 241 $\pm$84 | 500 - 1200 |
| Zr II | 0.18 | $4.14*10^{-8}$ | 241 $\pm$84 | 500 - 1200 |
| | | | | |
| Mg I | 0.05 | $1.15*10^{-8}$ | 241 $\pm$84 | 2000 - 6000 |
| Ca I | 0.12 | $2.76*10^{-8}$ | 362 $\pm$84 | 500 - 1200 |
| Ca II | 0.04 | $9.2*10^{-9}$ | 1086 $\pm$84 | 10000 - 14000 |
| Sr II | 0.07 | $1.61*10^{-8}$ | 621 $\pm$84 | 1000 - 6000 |
| Bu II | 0.17 | $3.91*10^{-8}$ | 255 $\pm$84 | 1000 - 2500 |

**Tableau I-3-2:** *Gradients d'émission moyens, complétés avec les calculs des échelles de hauteurs déduites du tableau I-3-1, sur une plage de hauteurs limitée, pour les métaux alcalins et les terres rares.* $\Delta$*Log(I)/100 km prend la valeur 0.04 pour l'hydrogène.*

Les raies à faible potentiel de première ionisation, dites low First Ionisation Potential « low FIP » sont situées entre 500 et 1200 km d'après ces relevés aux altitudes plus élevées. Plus haut ce sont les raies chromosphériques comme le Mg I, Ca II, H I, He I qui dominent.
La précision des relevés sur les échelles de hauteurs *H* est de $\pm$ 84 km pour le Ca I et $\pm$ 56 km pour les autres raies métalliques comme Fe II et $\pm$ 60 km pour le Ti II
La sensibilité des plaques photographiques utilisées par les auteurs était insuffisante pour observer et détecter d'autres raies d'hélium comme celle de He I 4713 Å ou He II 4686 Å.
Les gradients d'émission des métaux du premier groupe (c'est-à-dire la partie supérieure du tableau I-3-2) sont en accord avec ceux dérivés du Fer et du Titane, aux altitudes inférieures à 1500 km. Les gradients d'émission du Ba II et Ca I sont sensiblement identiques à ceux du premier groupe. Les gradients plus faibles du Mg I, Ca II, Sr II correspondent à la haute chromosphère. Au dessus de 2000 km les gradients d'émission des métaux sont plus faibles qu'en dessous de 1500 km d'altitude.
Dans cette thèse, nos observations et analyses ont été effectuées aux altitudes commençant à 400 km au dessus du limbe photosphérique.
Nous montrerons que l'utilisation des caméras CCD numériques avec une dynamique suffisante à partir de 8 bit et ensuite 12 bit, et à cadence d'images élevées a permis de sonder les altitudes plus basses. En effet, la réponse linéaire des nouvelles caméras CCD, les temps d'exposition très courts utilisés, c'est-à-dire entre 40 et 60 millisecondes, permettent d'accéder aux couches plus basses de l'atmosphère solaire et plus intenses, et le niveau de saturation se produit à environ 300 à 400 km au dessus de la haute photosphère. Ces altitudes étaient quasiment impossibles à enregistrer et inaccessibles avec les plaques photographiques, où les durées d'exposition étaient de l'ordre de 0.5 s à 1 seconde. La réponse en sensibilité des émulsions photographique n'était pas linéaire et les étalonnages difficiles, sans compter que le nombre de prises de vues était limité pour chaque contact avant et après la totalité.
Les analyses des spectres de l'éclipse du 1er Octobre 1940 ont contribué à définir des profils d'intensité en fonction des altitudes. Les variations des flux au limbe solaire enregistrés lors du second contact indiquent la présence d'un point d'inflexion sur lequel la référence des altitudes $h$ = 0 km a été choisie our le continu voisin de la raie H$\beta$ de H I 4861Å.

Le profil du bord solaire estimé à cette époque est décrit en Figure I-3-3:

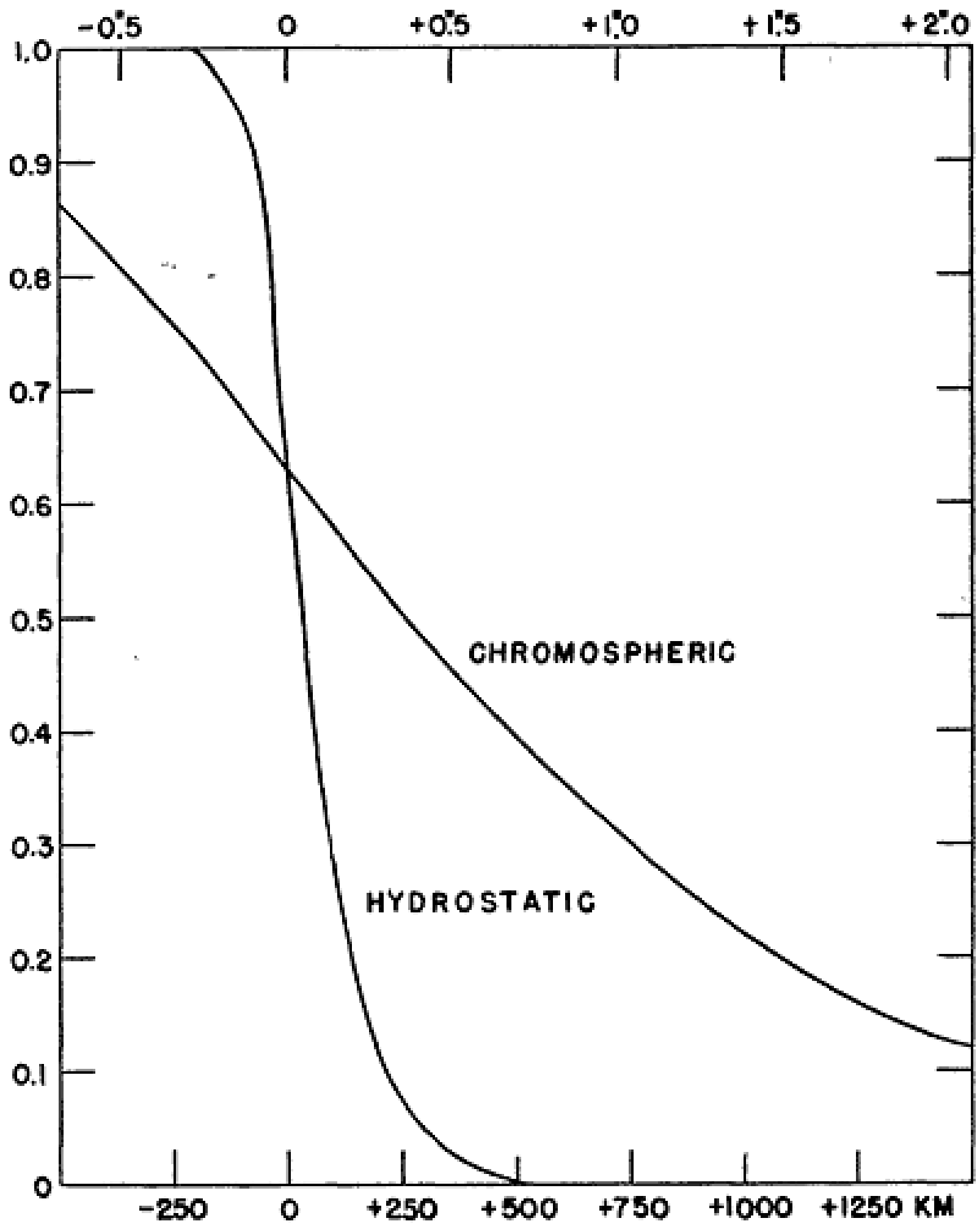

**Figure I-3-3:** *pente abrupte prédite du bord solaire. En abscisse, distance radiale à partir du limbe solaire, et en ordonnées, intensités en unités de brillance de surface du limbe solaire.*

Ces profils montrent un point d'inflexion, pour lequel l'altitude $h = 0$ est prise. L'existence de ce point d'inflexion prédite sur le bord solaire sera discutée à partir des courbes de lumière obtenues sur les spectres éclair, mais ces observations restent difficiles à réaliser, car très proches du limbe, les spectres des grains de Baily produisent encore beaucoup de lumière provenant de la photosphère, et saturent les films photographiques et les intensités plus élevées n'ont pas pu être mesurées.

En page 82 de l'article de R. Wildt cité plus haut, une note est donnée où le Professeur B. Lindblad a introduit un extrait de lettre dans laquelle il traite de la question de l'étroitesse du limbe solaire, et il communique des premiers résultats de mesures photométriques du continu au voisinage de la raie H$\beta$ 4861Å, sur le bord solaire à l'éclipse du 9 Juillet 1945. Ces enregistrements cinématographiques du spectre flash ont été effectués au second contact de cette éclipse et des relevés d'intensités en fonction des altitudes ont été mesurés. Ces relevés à partir des résultats de l'auteur, sont indiqués et reportés en figure I-3-4 pour une meilleure lecture. En ordonnées est donné le rapport $I/I_0$ où $I_0$ représente l'intensité enregistrée par l'émulsion du film la plus haute, et en abscisses l'altitude au dessus du limbe sans tenir compte de l'épaisseur optique:

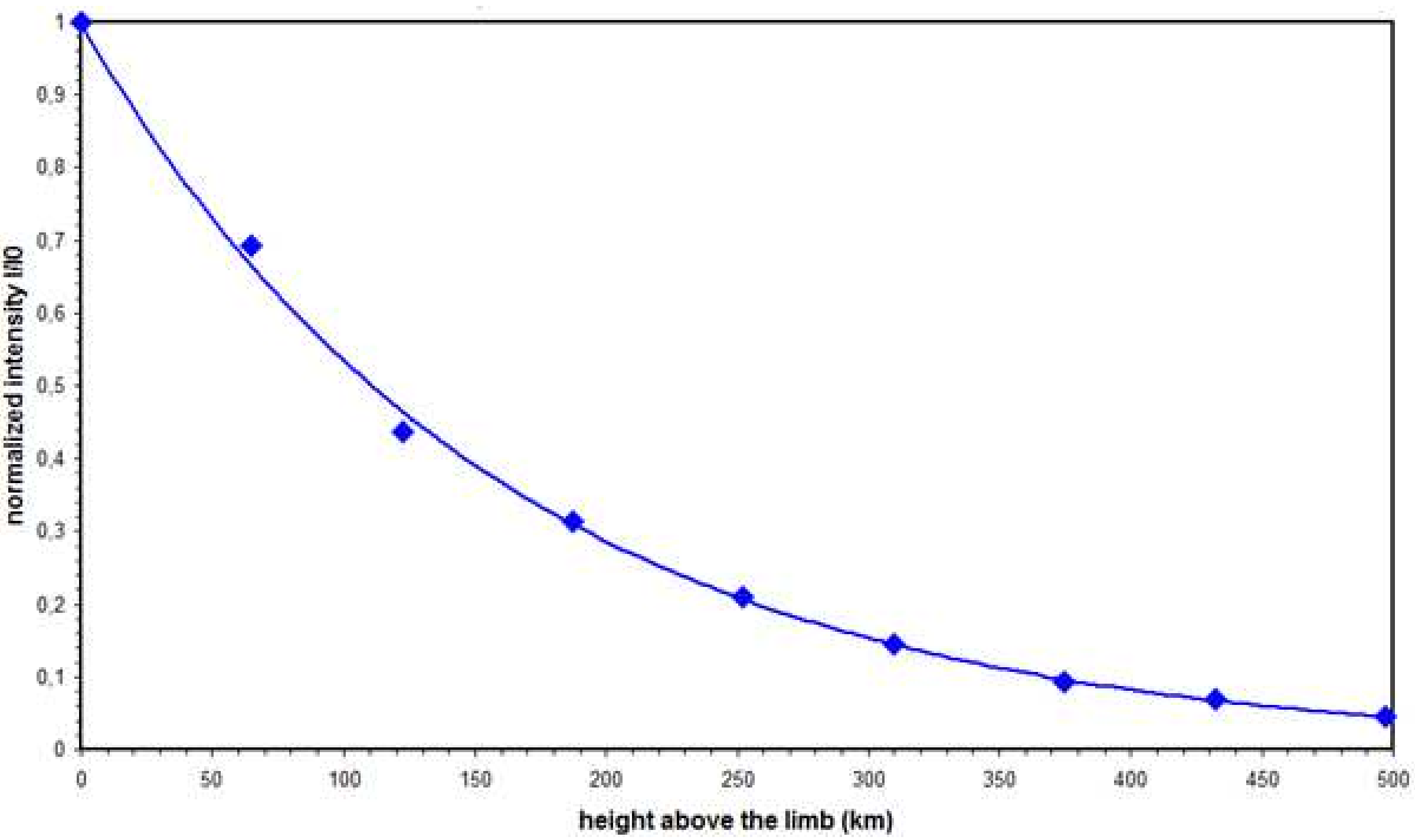

**Figure I-3-4:** *courbe à partir des relevés du Pr. B. Lindblad, du rapport d'intensités normalisées du continu en fonction de l'altitude à partir de séquences de spectres éclair au voisinage de la raie H$\beta$, au second contact de l'éclipse totale du 9 Juillet 1945.*

Le Professeur B. Lindblad précise qu'au le début de la courbe où $I/I_0$ est proche de 1 qu'en réalité l'intensité doit encore augmenter et être plus élevée, et pour les parties correspondant aux altitudes négatives sur le disque solaire. La valeur de $I_0$ est liée à une valeur d'intensité maximale de l'émulsion du film proche de la saturation. Malgré la réponse non linéaire du film, ces mesures sont remarquables.
A partir du graphique figure I-3-4, nous avons effectué un ajustement exponentiel décroissant, dont l'équation est :
$I(h) = 0.9968 * e^{-0.0062h} = 0.9968 * e^{-h/161}$
Ces résultats permettent la détermination d'une échelle de hauteur $H$ = 161 km pour le continu voisin de 4860 Å, et les résultats de cette époque et avec les moyens utilisés sont en très bon accord avec nos résultats obtenus récément avec les techniques CCD.
Ces résultats obtenus avec l'ajustement d'une exponentielle décroissante sont en bon accord avec les modèles hydrostatiques 1D qui ont sans douté déjà été utilisés depuis cette époque.
A noter que la gamme de longueur d'onde autour de 4700 Å que nous avons choisie récemment pour la réalisation des spectres éclair se justifie par les trois points suivants, et exposés dans le livre de Thomas- Athay 1961:

1) Il y a relativement peu de raies spectrales dans cet intervalle de 4700Å
2) L'importance relative de la contribution des électrons libres est maximale dans cette gamme de longueurs d'onde. Le continu à 4700Å provient de la diffusion d'électrons libres et de l'émission $H^-$ (Athay, Menzel 1955). Les contributions des continus Thomson et Paschen peuvent aussi être mesurées.
3) L'intensité du continu photosphérique varie peu avec la longueur d'onde dans ce domaine spectral limité à 120 Å d'étendue pour notre expérience.

4) Et par ailleurs ce choix se justifie aussi par la présence des 2 raies de l'hélium neutre à 4713 Å, et hélium ionisé He II 4686 Å, qui sont voisines, et espacées seulement de 27 Å dans cet intervalle. Avec un réseau objectif de 600 mm de focale muni d'un réseau de 600 tr/mm, ces 2 raies entrent en entier sur le même chip CCD avec une résolution suffisante de 0.12Å/pixel. L'intervalle spectral est de l'ordre de 120 Å autour de 4700Å, ce qui sera très pratique pour les comparer simultanément sur chaque séquence de spectres.

Les tableaux I-3-5 et I-3-6 donnent les caractéristiques des spectres éclair réalisés sur plusieurs années pour indiquer le nombre de raies identifiées, et comparer entre les résultats obtenus avec la technique du « moving plate » et des spectrographes à fentes étroites.

PUBLISHED LISTS OF FLASH-SPECTRUM LINES

| Eclipse | Method* | λ (A) | No. of Lines | Author | Year of Pub. |
|---|---|---|---|---|---|
| 1926 | *ns* | 3070–4200 | 960 | Davidson and Stratton | 1927 |
| 1927 | *ns* | 4150–4770 | 900 | Pannekoek and Minnaert | 1928 |
| 1901, 1905, 1925 | *ca* | 3070–7060 | 3250 | Mitchell | 1930 |
| 1900, 1905, 1908 | *mp*† | 3230–5320 | 3000 | Menzel | 1931 |
| 1930, 1937 | *ca*† | 3060–8860 | 3500 | Mitchell | 1947 |
| 1936 | *ns* | 5550–6570 | 364 | Fracastoro | 1939 |
| 1940 | *ca* | 3200–5400 | 1100 | Kiess | 1942 |
| 1941, 1945 | ....... | ............... | ........ | Melnikov | 1949 |
| 1945 | *ns* | 3120–5180 | 335 | Krat | 1951 |
| No eclipse | *ns* | 4800–6600 | 1200 | Adams and Burwell | 1915 |

* *ca* = chromospheric arc; *mp* = moving plate; *ns* = narrow slit.
† Most complete results. Most observers who employed other methods used the *ca* method besides.

**Tableau I-3-5:** *comparaison des techniques utilisées pour plusieurs eclipses totales. La mention « ca » signifie chromospheric arc, « mp » signifie moving plate, « ns » signifie narrow slit.« jf » signifie jumping film.Van de Hulst 1953.*

La méthode « ca » des arcs chromosphériques est une méthode brute pour déterminer les altitudes auxquelles les raies disparaissent, au moment des contacts. Elle est associée à la méthode de spectrographe sans fente. Un ou plusieurs temps d'exposition des images monochromatiques de chaque raie d'émission de la chromosphère encore visible, permet de montrer les arcs, tandis que les raies coronales sont sous forme d'anneaux complets.
La mention « jf » pour jumping film signifie une succession rapide d'arcs chromosphériques sur un film, donnant en enregistrement continu les spectres complets des arcs.
La méthode « mp » de plaques mobiles, permet d'obtenir une portion étroite au centre des arcs. Cette portion de spectre quasiment linéaire éclaire une plaque photographique se déplaçant de manière continue. Un enregistrement complet est obtenu lors du changement des spectres de cette portion de spectre avec raies d'absorption puis en émission en fonction du temps. Cette technique donne une portion de la chromosphère, des échelles de hauteurs correctes. L'enregistrement en continu montre aussi l'altitude $h = 0$ km où le spectre éclair avec raies en émission devient un spectre d'absorption sur le limbe du disque solaire.
Le procédé utilisant la fente étroite « ns » consiste en un spectrographe à fente, et devant laquelle un objectif produit l'image du Soleil éclipsé sur la fente. Celle-ci est placée tangentiellement au limbe lunaire. C'est la seule méthode qui permet de mesurer les longueurs d’ondes précisément, et les profils des raies, mais cette méthode est difficile à manipuler et à mettre en œuvre. Ce procédé est aussi utilisé hors contitions d’éclipses.

Le tableau I-3-6 donne une description de quelques raies et leurs extensions observées lors des spectres éclairs dans les années 1930.

INDIVIDUAL CHARACTERISTICS OF FLASH-SPECTRUM LINES

| Designation | | Occurs with | Characteristics |
|---|---|---|---|
| Menzel | Kiess | | |
| a........ | a | $He$, $He^+$ | Emission extends high in flash, sometimes reaching maximum intensity at great height; emission weak near limb; nothing seen on disk |
| c........ | ....... | $Eu^+$, $Ce^+$ | Emission drops sharply with height in flash but extends large distance into disk spectrum; two special lines of this type have been studied without an eclipse by Thackeray (1935) |
| b........ | a | $H$, $Sr^+$, $Fe^+$, $Ti^+$ | Emission may extend to great height in flash and also extends some distance into disk spectrum; no self-reversal seen |
| b........ | b | $Ti^+$, $Y^+$ | Emission of flash extends again into disk spectrum but exhibits signs of self-reversal; this may not be true self-reversal (see Sec. 5) |
| d, dw.... | c | $Fe$, $Ca$, $Mn$, $Cr$ | Emission in flash, absorption on disk, with fairly sudden transition |

**Tableau I-3-6:** *indications extraites des caractéristiques de quelques raies d’émission obtenues par Menzel et al à l’éclipse de 1930 d’après « The Sun », 1953.*

Ce tableau I-3-6 sert à décrire comment se présentaient les raies d’émission, leur extensions, comment pouvait varier leur intensité, avec les moyens des plaques photographiques utilisées à cette époque. Les raies low FIP comme le Titane, Magnésium sont indiquées ainsi que les raies high FIP comme l’hélium neutre et ionisé et hydrogène.
L’éclipse de 1932 a été réalisée avec un prisme-objectif équivalent à un spectrographe sans fente. La résolution spectrale a été ensuite améliorée, et l’équipe de R.B. Dunn de Sacramento Peak a réalisé des spectres de référence.
D’autres expériences d’éclipses ont été effectuées par Redman et Zanstra à l’éclipse de 1952, (Zanstra Redman, 1952) dans la région spectrale de 4340 Å à 3170 Å incluant la discontinuité de Balmer. L’originalité de cette expérience est d’avoir observé le spectre continu d’une protubérance intense en dehors des raies d’émission. Des mesures des densités électroniques ont été déduites d’après ces observations et les auteurs ont obtenu $0.7*10^{10}/cm^3$ en supposant une température éléctronique de 5000 K et $5*10^{10}/cm^3$ en supposant une température de 15000K. Cependant il n’est pas clair si les discontinuités observées autour du continu de cette protubérance pourraient être liées au relief lunaire, qui module le continu chromosphérique et coronal, et qui se superposent au continu de la protubérance. Il n’est pas mentionné de soustraction du continu coronal au continu de cette protubérance à partir des relevés sur les films au microdensitomètre. Toute fois les auteurs ont évalué ce continu dans une région en dehors des raies d’émission, mais le nombre d’images est limité et la résolution insuffisante. Nous avons réalisé ce type d’analyse du continu d’une protubérance à l’éclipse de 2010, avec une qualité photométrique, voir Annexe 1.

Les conclusions scientifiques tirées de ces expériences de Wildt et al sont les suivantes :

- La décroissance radiale de la densité de l'hydrogène H I dans la chromosphère est strictement exponentielle jusqu'à h = 15000 km et est proportionelle à $e^{-h/1086}$
- Les gradients de densité des métaux sont plus raides que l'hydrogène, mais entre 3000 et 6000 km, ils ont le même gradient que H I.
- Les températures d'excitation du Fe I et Fe II sont de l'ordre de 3000 à 4000 K et augmentent avec l'accroissement des potentiels d'ionisation.
- A l'altitude h = 500 km, $n_{HI} = 10^{15.63}/cm^3$ et $n_{FeI} = 10^{9.53}/cm^3$
- Pour 2000 < h < 6000 km, le gradient de densité de H I est quatre fois supérieur à celui de He I. Les observations ne permettent pas de confirmer que le max de densité de He I est en dessous de 2000 km.
- Les gradients de densité des métaux étant cinq fois plus élevés que ceux de H I, cela suggère l'existance de turbulence chromosphérique, indépendamment du poids atomique.
- La distribution radiale du rayonnement continu est dééterminée par le gradient de densité de H I. La pente du bord solaire à h = 500 km est sept fois plus grande que le gradient chromosphérique, avec l'hypothèse d'équilibre hydrostatique.

## I-4) Spectres de R.B. Dunn et al à l'éclipse de 1962

Une expérience de spectres éclairs sans fente de haute résolution spectrale a été réalisée à l'éclipse totale de Février 1962 en Nouvelle Guinée (voir R. B. Dunn et al 1968), dans le domaine spectral 3200 à 9100Å et 3500 raies d'émission ont été obtenues. Cette expérience comprenait un coelostat avec un miroir concave de 305 mm de diamètre et 5746 mm de focale. Un réseau de 1200 traits/mm par réflexion était utilisé et l'image du disque solaire faisait 54 mm au foyer à l'entrée des spectrographes. Des lentilles de quartz de 70 mm de diamètre et 710 mm de focale utilisées comme collimateur reprenaient l'image pour éclairer le réseau en faisceaux parallèles et produisaient les spectres après séparation, dans un domaine Infra-rouge et Ultra-Violet. Du film « Tri X » Kodak était utilisé pour les spectres UV et un film « HIR » Kodak était utilisé pour les spectres IR. Plus de détails de cette expérience remarquable et unique sont donnés dans l'article de R.B. Dunn, et al 1968.
Ces observations de spectres sont historiques et uniques de part leur résolution et qualité, et des tables détaillées (contribution des auteurs cités précédemment) donnant les intensités de chaque raie identifiée en fonction du numéro de plaque et de l'altitude ont été réalisées. Hélas, ces résultats très remarquables et sans égal dans la littérature ont été peu utilisés par la suite car ce sont des images instantanées prises aux courts instants des contacts, et l'exploitation des résultats réprésentait un travail considérable par rapport aux moyens d'analyses avec les (micro-densitomètres) de cette période, compte-tenu du grand nombre de raies (des milliers) obtenues dans l'UV et le proche infra-rouge.
Pour rendre les résultats de R.B. Dunn accessibles, nous les avons utilisés pour construire des graphiques montrant les intensités en fonction des hauteurs. Nous avons effectué les tracés des courbes de lumière $I = f(h)$ pour plusieurs raies d'émission d'après les tables de Dunn. Les courbes, figure I-4-1, ont été tracées d'après les plaques photographiques pour plusieurs éléments chimiques et où chaque plaque avait été étalonnée en flux et en altitude.

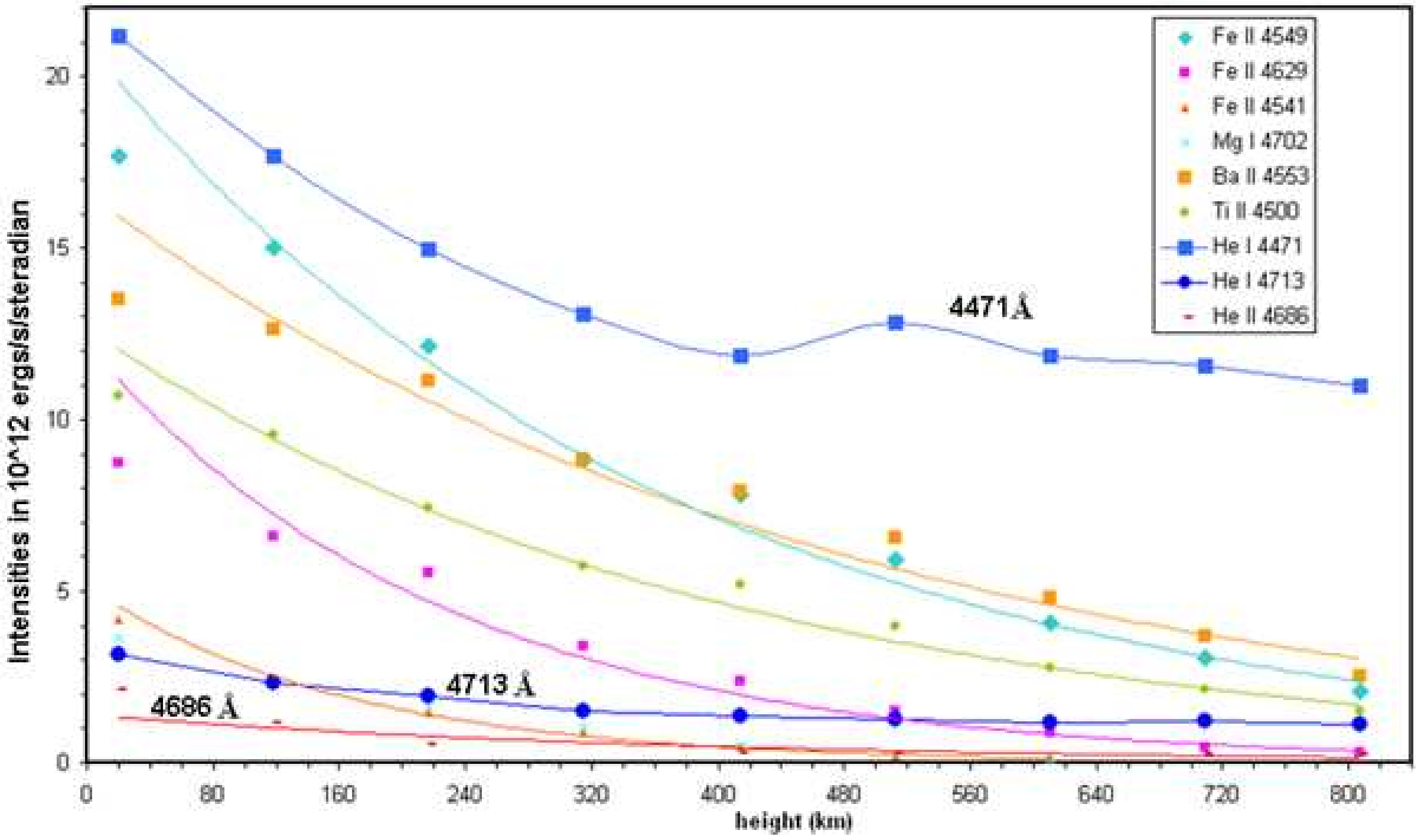

**Figure I-4-1:** *tracés des profils d'intensité I(h) en fonction de l'altitude après exploitation des plaques photographiques des spectres éclairs obtenus par R.B. Dunn à l'éclipse de 1962, sur une coupe en tranche de la chromosphère. 1 unité du disque solaire moyen équivaut à $3.77358*10^{14}$ ergs/cm$^2$ /s /sr/Å. Une division en ordonnées vaut $10^{12}$ ergs soit 0.00265 unités du disque solaire. La largeur de fente effective utilisée fait 10µm (0.001 cm). L'intervalle des points dans les spectres UV était de 0.053 Å de 0.1 Å dans les spectres IR. Les intensités sont exprimées en ergs/s/stéradian.*

La conversion en unités du disque solaire moyen et valeur en « ergs » a été effectuée à partir des données de Allen, 1973, où 1 unité du disque solaire moyen déduite à partir des données de Allen vaut $3.77358*10^{14}$ ergs/cm$^2$ /s/sr/Å. Par ailleurs, les intensités relatives sont exprimées en unités d'intensité du disque solaire moyen. Ce choix repose sur les unités de brillance du disque solaire adoptées, voir Annexes 28 et 29. Il faut prendre en compte le flux par unité de longueur d'onde, et prendre en compte la largeur de la fente pour évaluer les flux. Les graphiques figure I-4-2 représentent la conversion en échelle logarithme à partir de la figure I-4-1 précédente:

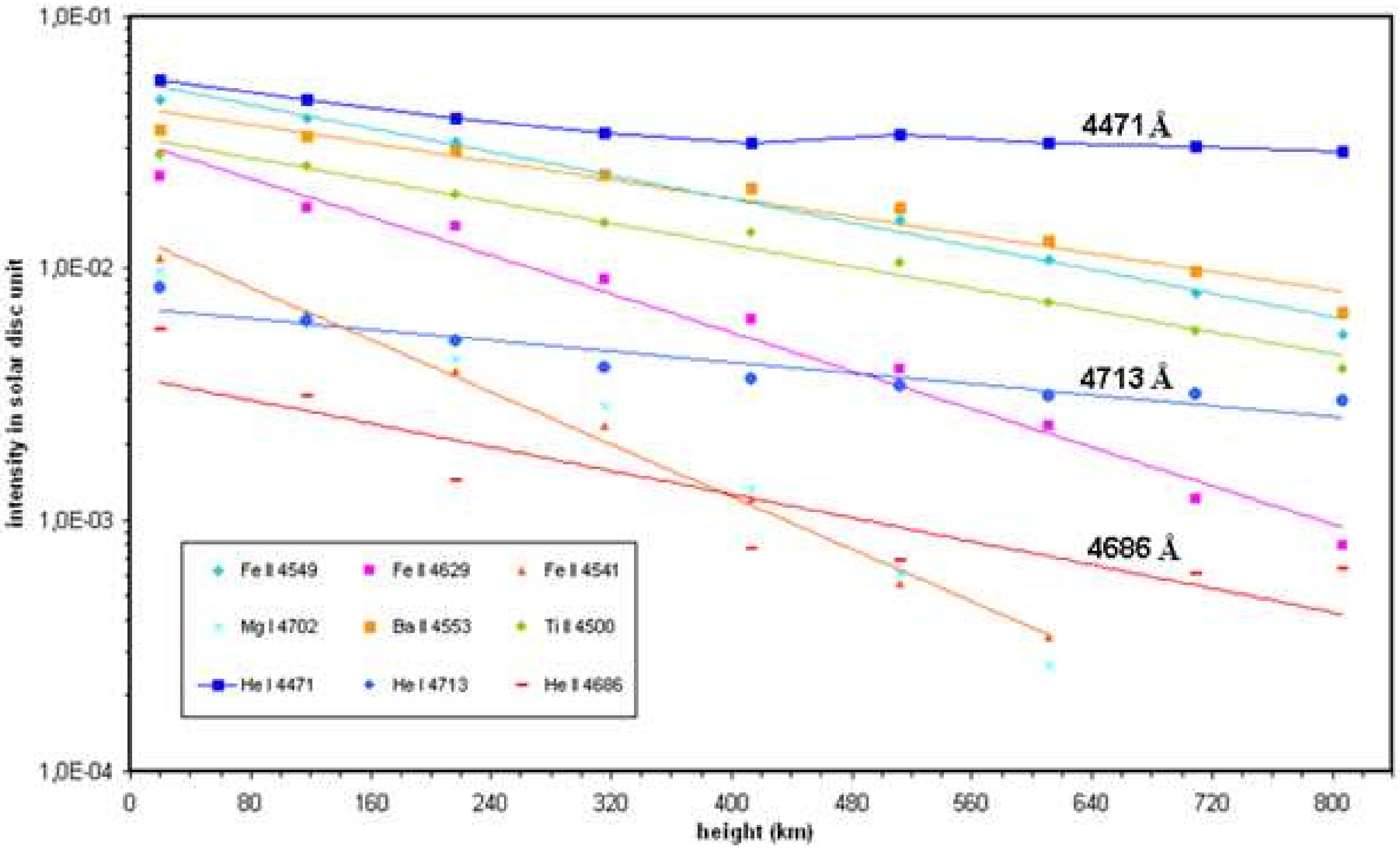

**Figure I-4-2:** *profils d'intensités I(h) pour différents ions en fonction des altitudes déduits à partir des analyses et étalonnages des spectres de D. Dunn à l'éclipse de 1962 publié en 1968. Spectrum of the chromosphere.*

Les choix des éléments chimiques relevés sur les spectres de R.B. Dunn et al correspondent à ceux que nous avons analysés à partir de nos résultats obtenus aux éclipses de 2008, 2009 et 2010.
L'altitude h = 0 km sur la figure I-4-2 déduits des résultats de Dunn et al correspond au point d'inflexion qui était utilisé pour définir le bord solaire.
Nos observations ont montré finalement qu'il n'y a pas de point d'inflexion dans cette région où les flux encore élevés des grains de Baily lors des contacts, provoquant la saturation des émulsions des plaques photographiques.
Le premier point de mesure pour la raie d'hélium He I 4471Å est donné pour l'altitude 20 km au dessus du limbe avec une valeur de $21*10^{12}$ ergs ce qui correspond à 0.05565 unités du disque solaire moyen. Cette altitude et cette valeur d'intensité de cette raie He I 4471Å n'a pas de sens aussi proche du bord, car les spectres du continu des grains de Baily sont très intenses, proches de la saturation et cette raie n'est pas visible aux altitudes indiquées à cause de la lumière encore très intense provenant de la haute photosphère.
De même pour les autres raies comme le Ti II 4500 Å à h = 600 km, l'intensité déduite des courbes de R.B. Dunn est voisine de $8*10^{-3}$ unités du disque solaire moyen, et nos résultats nous donnent une valeur de $1.4*10^{-4}$ unités du disque solaire moyen pour cette même altitude.
Les auteurs se réfèrent à l'ouvrage de Thomas et Athay 1961, qui ont effectué des mesures complémentaires de courbes de lumière du continu à 4700 Å et se servent des graphiques donnés en figure I-1-6-2 comme référence.

Dunn et ses collaborateurs déterminent en 1968 des altitudes dites “zéro” quelque part dans la basse chromosphère en se servant des profils du continu 4700 Å donnés par Athay 1955.
Cette définition sera clarifiée par la suite avec les analyses de spectres éclair à cadence élevée.

Les conclusions scientifiques tirées de ces expériences de Dunn et al sont les suivantes :

- Les identifications des raies ont été difficiles à réaliser dans la partie violette des spectres à cause des très nombreuses raies métalliques présentes, et des chevauchements (mélanges) des raies. Des spectres chromosphériques à très haute dispersion hors éclipses étaient nécéssaires (voir Keith Pierce, Worden).

## I-5) Spectres éclair réalisés par Suemoto, Z. & Hiei, E

Ces observations sont rapportées dans la publication « éclipse totale du 12 Octobre 1958 », Suemoto, Z. et Hiei, 1962. La figure I-5-1 est extraite de cet article, pour montrer la résolution et détails des spectres flash dans les raies de l'hélium et d'hydrogène autour de 3888Å obtenus par les auteurs, avec une myriade de raies fines en émission superposée au continu du spectre correspondant aux derniers grains de Baily.

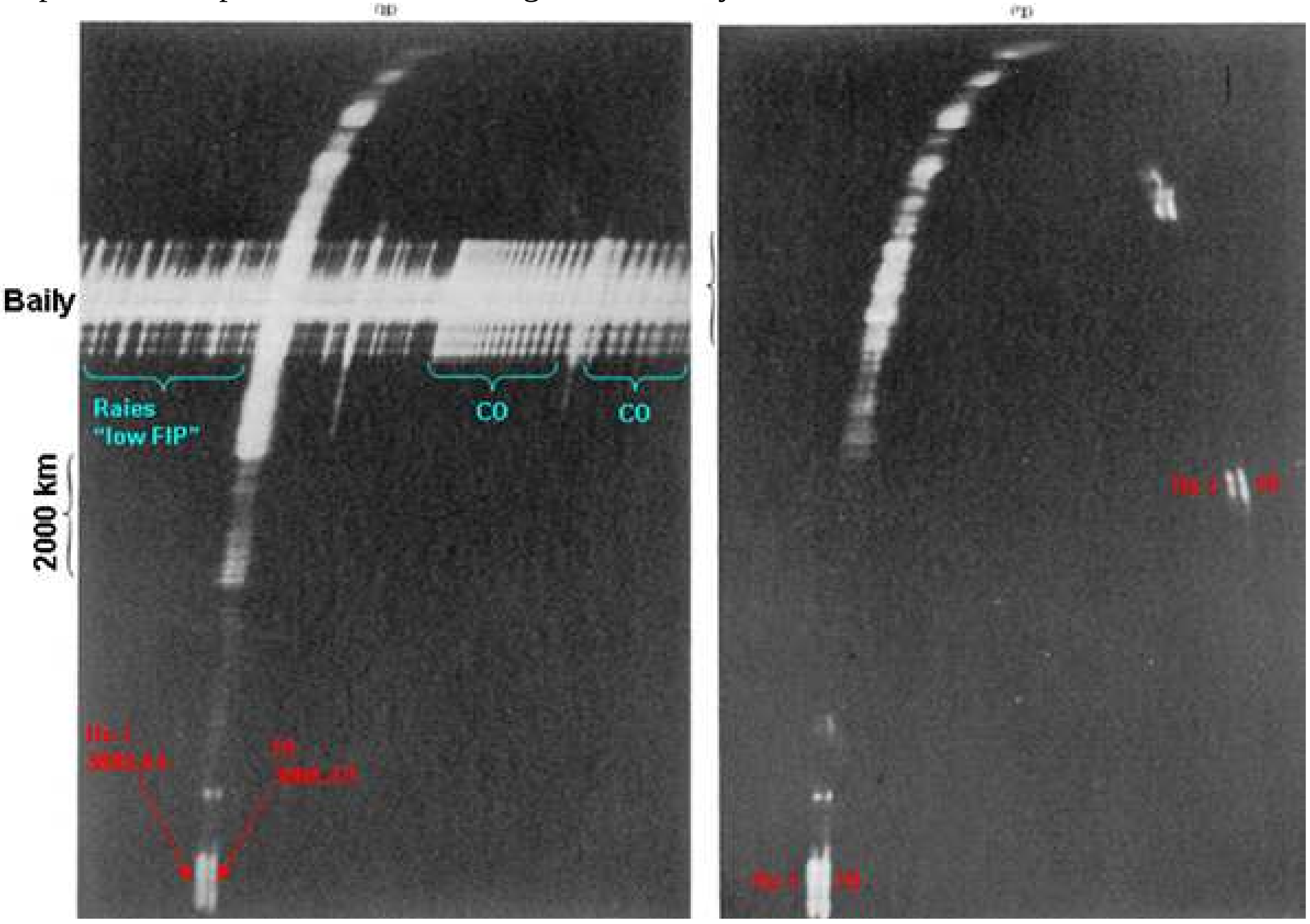

**Figure I-5-1:** *extrait de spectres flash, (page 41 de l'article Suemoto & Hiei 1962) montrant des structures de spicules, à droite dans la raie H8 de H I à 3888.49 Å, et à gauche dans la raie de l'hélium neutre He I 3888.64 Å.*
*Noter les similitudes des structures de ces 2 raies intenses dans les bas niveaux. Ces parties de croissants indiqués par des parenthèses au bord extérieur des images correspondent à des altitudes de l'ordre de 2000 km. La dispersion était de 0.23 Å /mm, et l'échelle verticale sur cette impression est de 7secondes d'arc/mm. La myriade de petites raies en émission est superposée au continu correspondant au spectre du dernier grain de Baily avant la totalité (disparition de la haute photosphère), avec les signatures de raies « low FIP » et les séries de raies correspondant à la molécule CO.*

Ces spectres montrent une structuration de macrospicules et spicules dans le profil des raies high FIP de He I et $H_8$ de H I qui sont très proches l'une de l'autre et ont pu être résolues. Ces structurations sont étudiées au chapitre IV-2-3 à partir des résultats de l'éclipse du 22 Juillet 2009 sur la raie He I 4471.
L'inconvénient sur les données de Suemoto et Hiei est le nombre très limité de ces spectres inédits, et les expériences étaient très coûteuses et volumineuses. Nous avons obtenu des spectres de résolution spectrale 100 fois moindre, de l'ordre de 24 Å /mm, c'est-à-dire 0.12 Å /pixel (taille du pixel 5 $\mu$m), mais cependant avec une cadence de 10 à 15 spectres par seconde avec les nouvelles caméras CCD rapides.
D'autres mesures sur les spectres ont été réalisées lors de l'éclipse totale de 1970, notamment sur le continu dans le but de mieux définir les variations d'intensité du continu jusque dans les altitudes atteignant la basse chromosphère.

<u>Des conclusions scientifiques très importantes sont tirées des expériences de Suemoto Hiei :</u>

- Hiei 1963 suggère que la chromosphère est composée seulement de spicules et que le milieu interspiculaire fait partie de la couronne. Les spicules, sous forme de gaines, apparaissant au dessus de 1000 km sont les mêmes pour n'importe quelle raie, et c'est vrai pour les raies d'hélium et hydrogène de températures différentes.
- L'émission du continu Paschen $\alpha$ dans le visible permet de déduire la densité de H I
- L'échelle de hauteur de l'ion $H^-$ est située autour de 50 km, et ce résultat est déduit à partir des mesures dans le continu de Balmer et discontinuite dans le domaine spectral 3600 à 3800 Å. Le continu de Balmer « libre- lié » donne la densité éléctronique.
- Aux altitudes inférieures à 700 km, la chromosphère pourrait être en équilibre hydrostatique en incluant une turbulence $\xi_0 < 10$ km/s (Hiei 1963)

## I-6) Expériences de Makita à l’éclipse de 1970 pour montrer les variations du continu chromosphérique

Les graphiques figure I-6-1 ont été obtenus par Makita et ses collaborateurs sur les variations du continu à partir du bord du Soleil, et où l’altitude *h* = 0 a été déterminée:

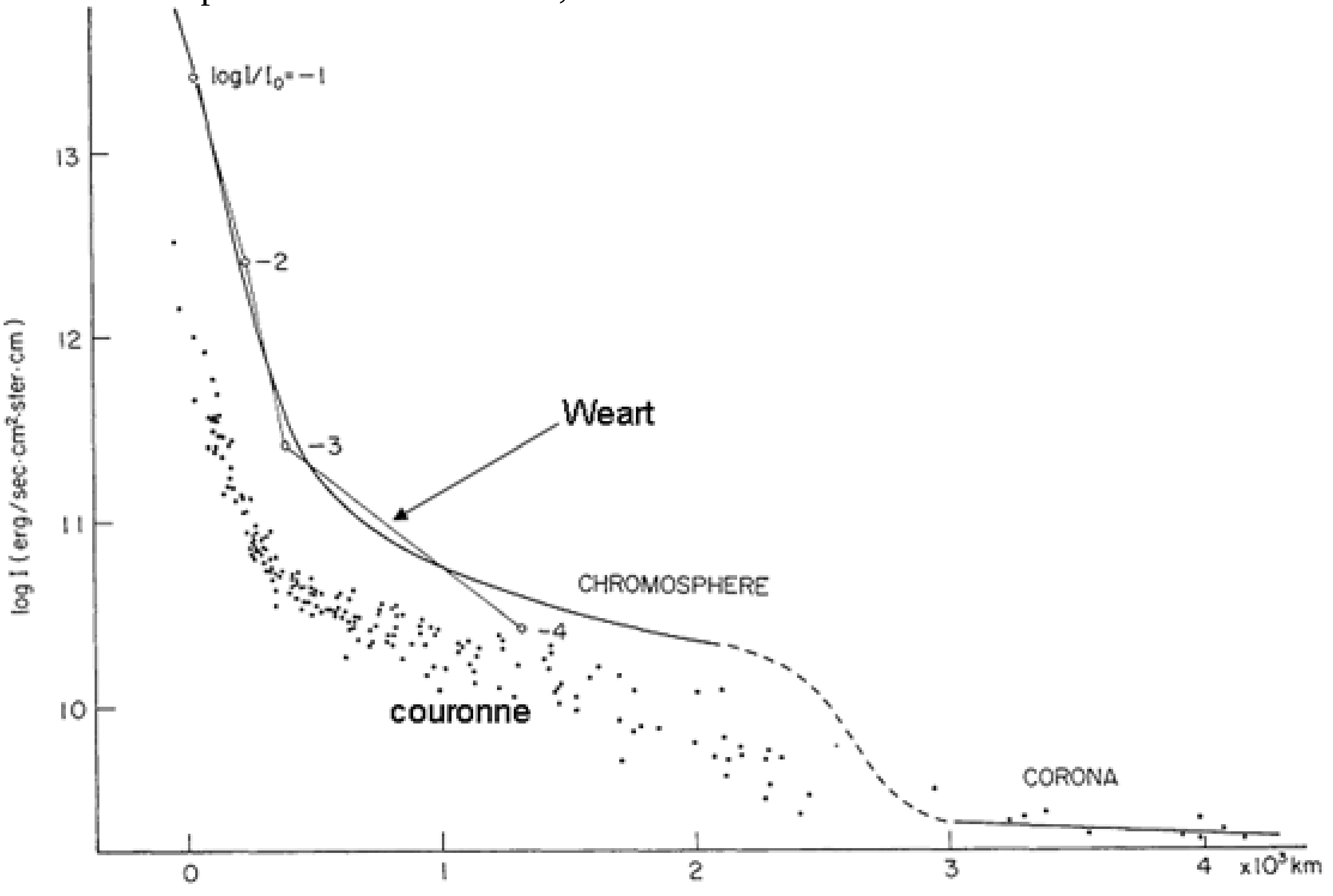

**Figure I-6-1:** *profils de lumière à 6900 Å pour montrer les variations du continu et distribution d’intensité des couches externes du Soleil. Les courbes en traits pleins reliées avec des courbes en pointillés montrent les variations dans les différentes couches de l’atmosphère solaire obtenus.Une chute d’intensité à 2500 km montre la limite supérieure de la chromosphère. Les cercles ouverts reliés avec des lignes fines sont les valeurs prises sur les résultats de Weart, normalisés à l’intensité au centre du disque solaire* $I_0$*. D’après Makita et al 1972.*

L’interprétation de la transition entre le continu chromosphérique et le continu coronal n’est pas claire. Cette transition est décrite dans le chapitre IV-2-3. La représentation avec les traits en pointillés et avec un point d’inflexion est discutable. Les reliefs et irrégularités du bord de la Lune ne sont toutefois pas suffisamment discutés par les auteurs.

<u>Les conclusions scientifiques tirées de ces expériences Makita et Tanaka sont les suivantes :</u>

- En dessous de 1000 km, l’émission $H^-$ est prédominante, et au-delà de cette altitude, la diffusion électronique et l’émission libre-liée joue un rôle principal.
- Aucune indication remarquable a été trouvée autour de h = 2500 km comme limite supérieure de la chromosphère.
- D’après le renvoi vers les travaux suivants de Tanaka Hiei 1972, la raie du Fe I serait formée à 50 km au dessus du limbe, la basse chromosphère jusqu’à 1000 km pourrait être une atmosphère homogène. Les régions brillantes qui sont concentrées à la limite de la supergranulation devraient être plus chaudes en s’élevant en altitude.

Les observations d'éclipses restent des expériences difficiles, et demandent beaucoup d'habilité des expérimentateurs pour les réussir, et pour obtenir de tels résultats de qualité photométrique.
Des observations des spectres du bord du Soleil ont été tentées dans des conditions hors éclipse avec les plus grands télescope solaires en 1972, afin de rechercher si des raies de faible intensité, hélium neutre He I 4713 Å et He II 4686 Å visibles en conditions d'éclipse, pourraient être visibles en dehors des conditions d'éclipse. Des résultats sont présentés par Worden B. et Hirayama 1972, T. à Sacramento Peak.

## I-7) Tentatives infructueuses d'observation des raies He I 4713 Å et He II 4686 Å par B. Worden et T. Hirayama dans des conditions hors éclipse avec le télescope sous vide de Sacramento Peak

Des spectres à haute résolution spectrale onté été tentés à l'observatoire solaire de Sacramento Peak aux Etats-Unis, pour rechercher la présence des raies de l'hélium neutre He I 4713Å et ionisé He II 4686Å très proches du limbe solaire, afin de savoir si ces raies peuvent être vraiment observées en dehors des éclipses. Les extraits figure I-7-1 montrent les spectres obtenus par les auteurs où des raies d'émission sont visibles:

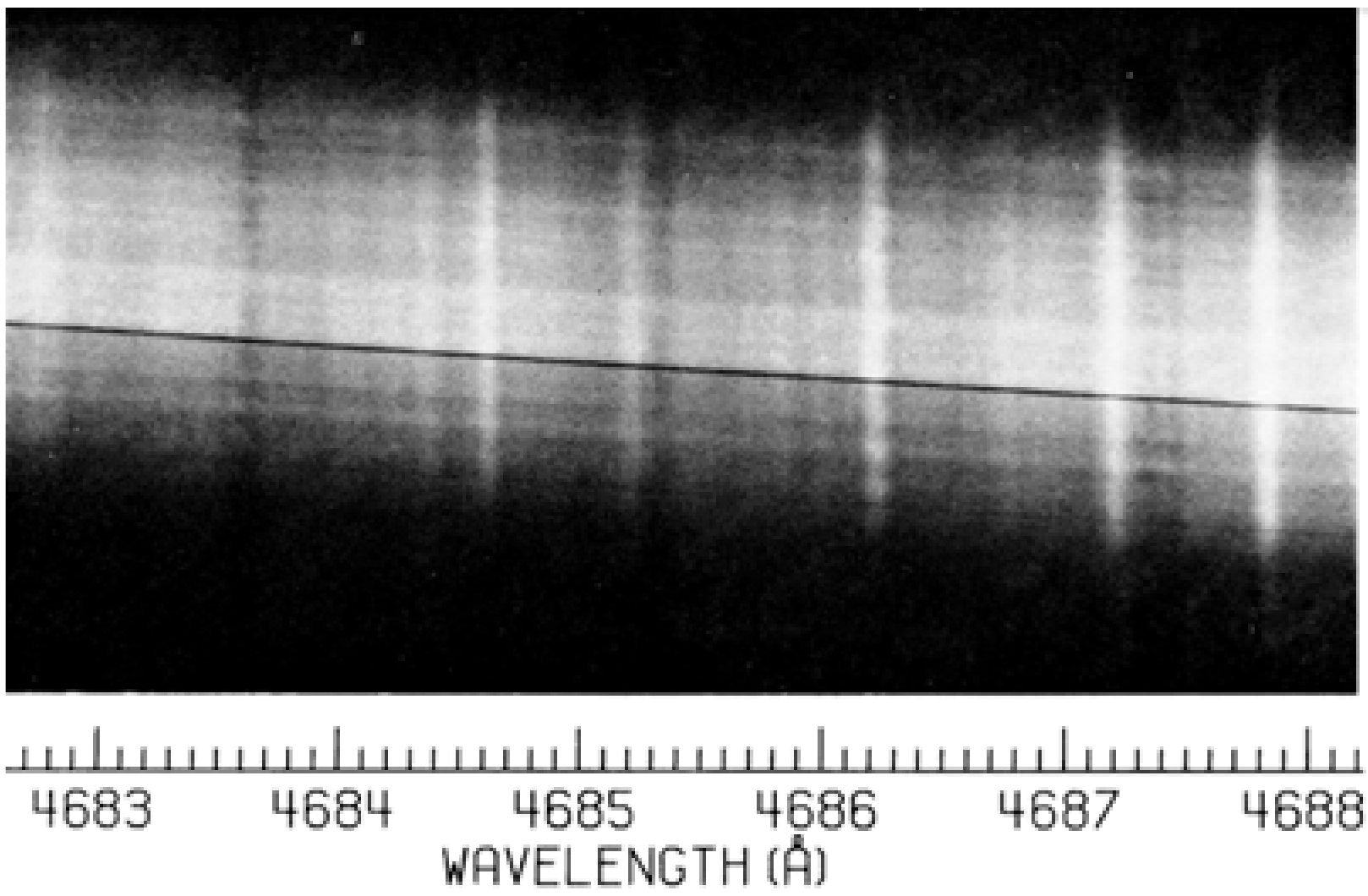

**Figure I-7-1:** *spectres juste au dessus du limbe dans la région à 4686 Å le 16 Janvier 1972. Dispersion 10.1 mm/ Å à 4686 Å fente de 200µm, 3 secondes de pose sur film Kodak 103a. D'après l'article Worden Beckers Hirayama 1972.*

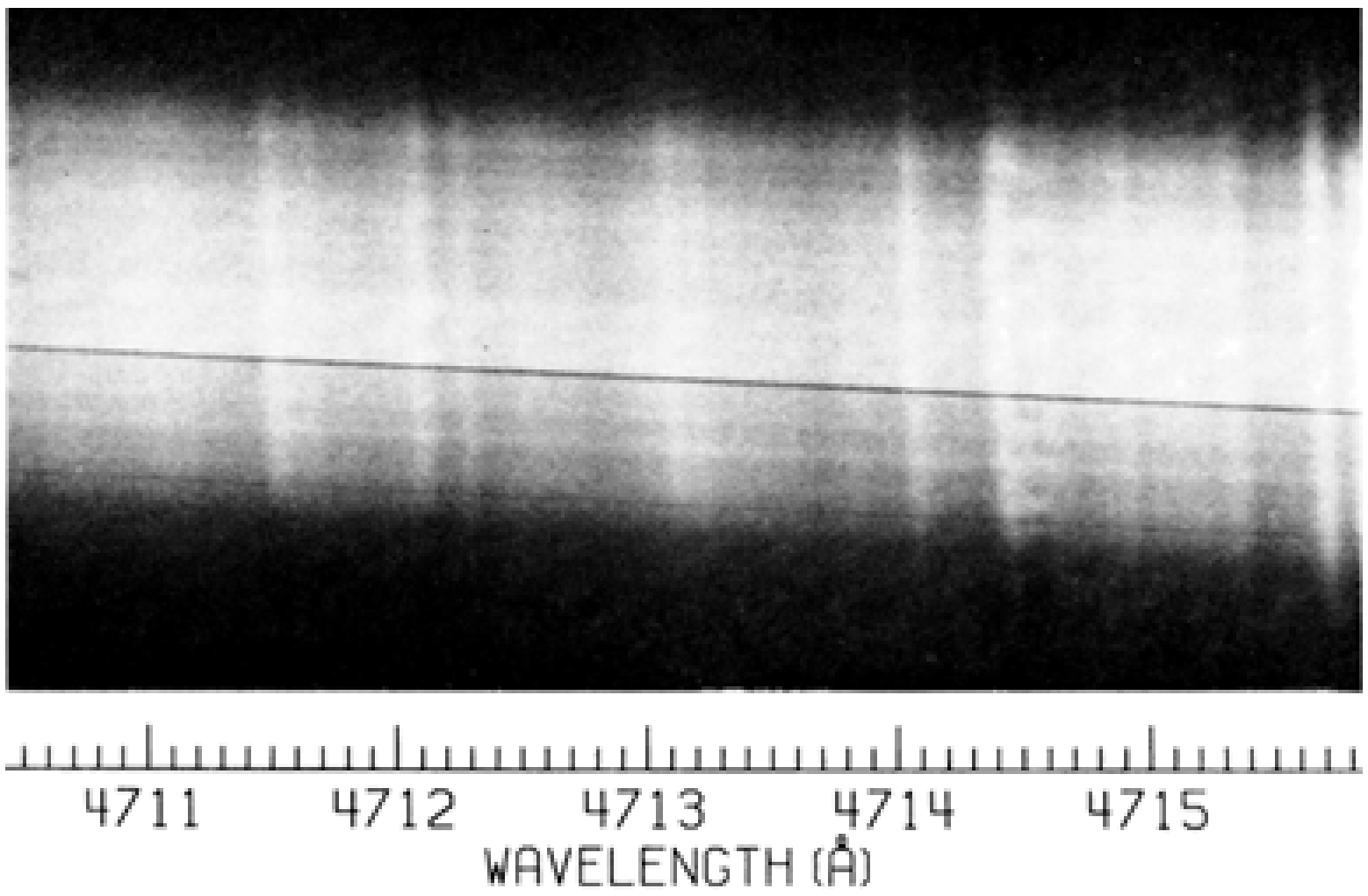

**Figure I-7-2:** *spectres juste au dessus du limbe dans la région à 4713 Å le 16 Janvier 1972. Dispersion 10.3 mm/ Å à 4713 Å fente de 200µm, 3 secondes de pose sur film Kodak 103a D'après l'article Worden, S. P. Beckers, J.M. Hirayama, 1973.*

La hauteur moyenne de la fente parallèle était située à environ 400 km au dessus du limbe, avec des variations de +/- 300 km. Bien que de nombreuses raies d'émission associées à des raies low FIP aient été enregistrées, les raies d'émission He I 4713 Å et He II 4686 Å n'ont pas été observées. Les auteurs ont montré qu'il était impossible d'observer les raies de l'hélium He I 4713 Å et ionisé He II 4686 Å en dehors des éclipses totales de Soleil et que les observations antérieures prétendant les avoir observées sur le limbe hors éclipse étaient erronées.
Cependant, de nombreuses raies d'émission chromosphériques comme le chrome, cérium, titane, fer, nickel, samarium, correspondant aux raies low FIP ont été observées. Keith Pierce et ses collaborateurs ont réalisé un atlas très précis de ces raies, voir Keith Pierce 1968. Ces résultats ont confirmé que les raies d'hélium He I 4713 Å et He II 4686 Å ne pouvaient pas être observées en dehors des éclipses totales de Soleil. La lumière parasite provenant du disque photosphérique est 20000 à 100000 fois plus intense que ces raies. Cette lumière parasite qui ne peut être réduite même avec les meilleurs coronographes, masque complétement ces raies d'hélium de faible intensité, formées très proches du limbe solaire. Ces résultats montrent de façon évidente que les conditions d'éclipses totales sont indispensables pour l'étude des raies « low FIP » et « high FIP » comme l'hélium He I 4713 Å et He II 4686Å qui sont optiquement minces.
Cependant il est toute fois possible d'observer la présence de ces raies d'hélium dans des protubérances intenses, celles-ci étant plus éloignées du limbe solaire,
Des nouvelles observations et mesures des raies d'hélium neutre et ionisé ont été renouvelées avec succès par Hirayama à partir de spectres d'éclipses totales pour mesurer les vitesses des ions dans la chromosphère.

<u>Les conclusions scientifiques tirées de ces expériences Worden et al sont les suivantes :</u>

- Les raies de l'hélium He II 4686Å et He I 4713 Å n'ont pas été détectées dans la basse chromosphère ni dans les spicules avec les résolutions spatiales et spectrales les plus élevées possibles à Sacramento Peak, en dehors des conditions d'éclipses. Les identifications de cette raie de par le passé, et hors éclipses étaient érronnées. Toutefois d'autres raies d'émission métalliques ont été détectées.

## I-8) Spectres éclairs sans fente de Houtgast et al, éclipses de Mars 1970 et du 30 Juin 1973

Des spectres éclairs plus récents ont été obtenus par Houtgast, 1979. Ils ont observé simultanément des raies « low FIP », et de terres rares, avec des spectres à cadence faible durant les contacts, montrant simultanément les raies d'émission (Terres rares) et raies d'absorption comme le montre l'extrait figure I-8-1 et I-8-2 :

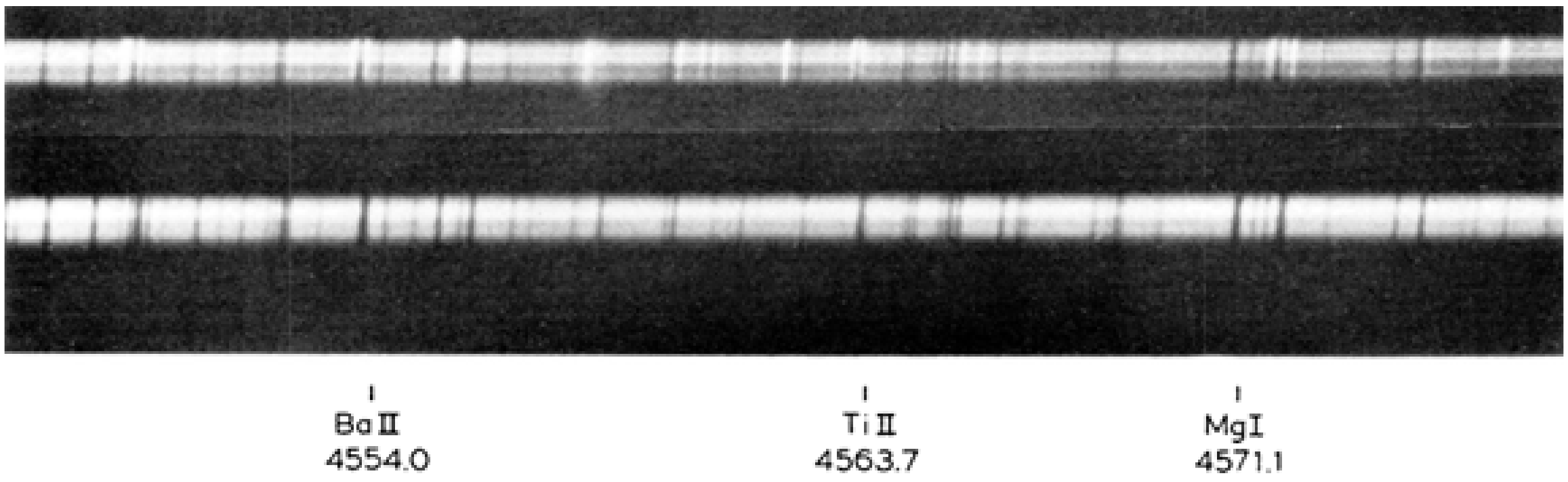

**Figure I-8-1:** *extrait de2 spectres éclairs Houtgast Namba de l'éclise de 1970: spectre du haut pose de 1.037s entre 768 et 456 km au dessus du limb. Spectre du bas, pose de 0.285 s entre 2430 et 2350 km au dessus du limbe. Spectrographe échelle.*

Ces spectres montrent le changement entre l'absorption et l'émission des raies low FIP, comme le Ba II 4554Å, Ti II 4564 Å, Ti II 4568 Å et Mg I 4571.1 Å.

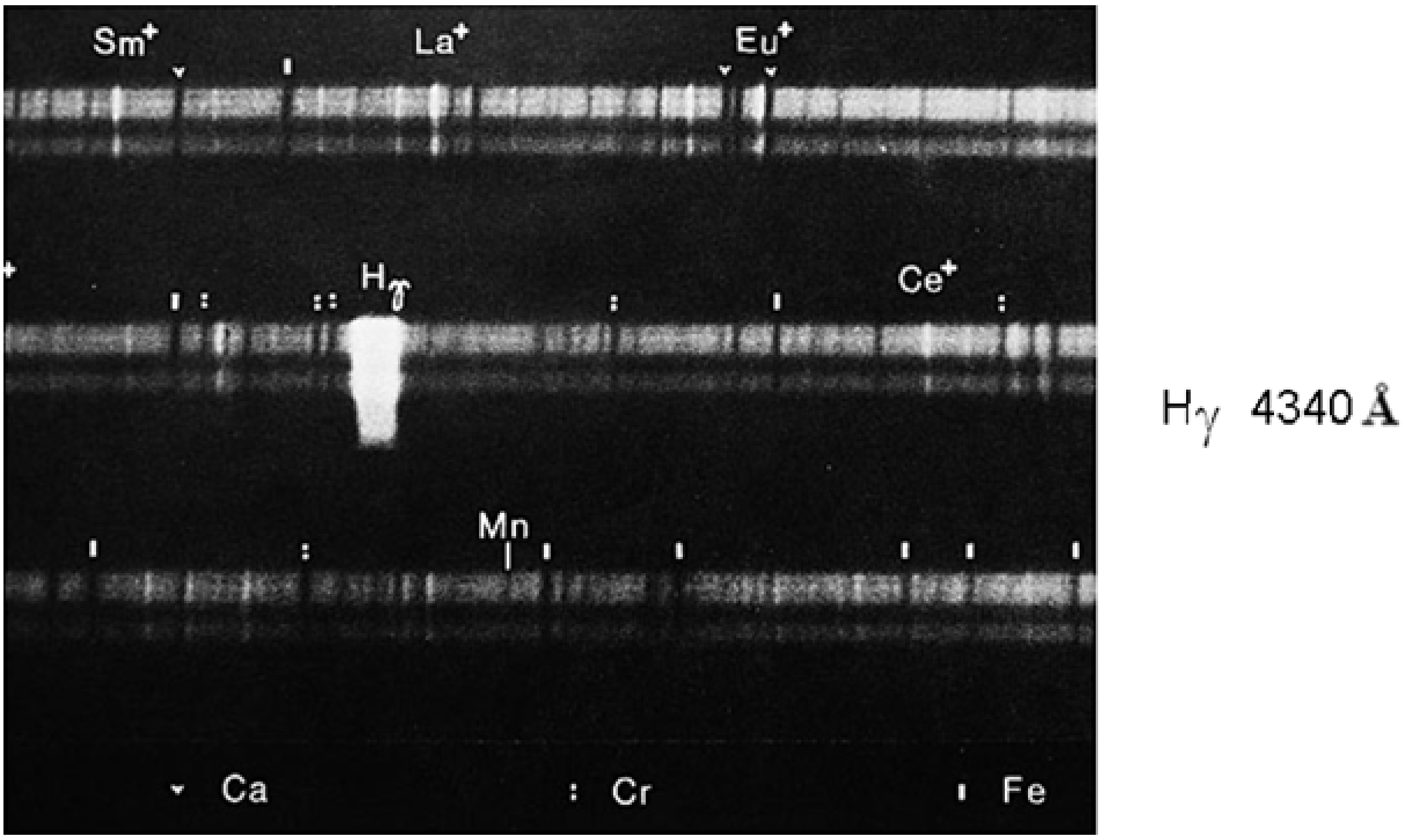

**Figure I-8-2:** *extrait de spectres éclairs de Houtgastet Numba pris juste avant le second contact de l'éclipse du 30 Juin 1973 en Mauritanie.*

Ces spectres remarquables démontrent qu'il est possible d'observer simultanément les raies d'émission et les raies d'absorption lors des séquences des spectres éclair. Ces spectres sont réalisés avec un spectrographe à fente. La région spectrale examinée se situe autour de la raie H$\gamma$ à 4340 Å. Des expériences utilisant un spectrographe à fente ont été renouvelées à l'éclipse de 1980 par T. Hirayama dans le domaine spectral correspondant aux raies de l'hélium neutre 4713 Å et ionisé 4686 Å.

Les conclusions scientifiques suivantes sont tirées des expériences de Houtgast et al :

- pour les raies une fois ionisées, plus la raie est intense, plus la transition absorption – émission a lieu profondément, comme le Ti II 4568 qui est très intense, cette transition a lieu à 25 km en dessous du limbe, tandis que pour le Ti II 4564 plus faible, la transition a lieu vers 1000 km au dessus du limbe.

## I-9) Mesures des largeurs des raies d'hélium neutre He I 4713 Å et He II 4686 Å par T. Hirayama à l'éclipse de 1980

Des nouvelles expériences de spectres éclairs ont été réalisées avec un réseau-objectif à incidence rasante, dans le but de mesurer avec la plus grande précision possible la largeur à mi-hauteur (FWHM) des raies d'émission des raies de l'hélium neutre He I 4713 Å et hélium ionisé He II 4686 Å, afin de déterminer les variations de vitesses Doppler des ions dans les couches de la chromosphère.
La procédure utilisée par les auteurs pour déterminer l'altitude zéro consistait à mesurer l'intensité du continu pour chaque image, et ensuite de mesurer le gradient le plus élevé de l'intensité en fonction de la hauteur pour chaque angle de position. Cela revient à effectuer la dérivée des intensités par rapport aux hauteurs, et tracer les variations des nombres dérivées, en fonction de l'altitude. Cette méthode a été utilisée au chapitre II-4 pour déterminer les positions des limbes à partir des images dans les raies H de Ca II et EUV. L'intervalle de temps entre 2 images successives était relativement long, une seconde, ce qui correspond à une différence de hauteur de 600 km entre 2 vues consécutives.
Les durées d'exposition des spectres étaient de une seconde avec du film photographique Kodac Tri-X relativement sensible dans le bleu.
Les auteurs ont effectué des moyennes sur le profil des spectres autour de l'instant de contact, et ont évalué la position du bord solaire donnée dans les graphiques figure I-9-1.

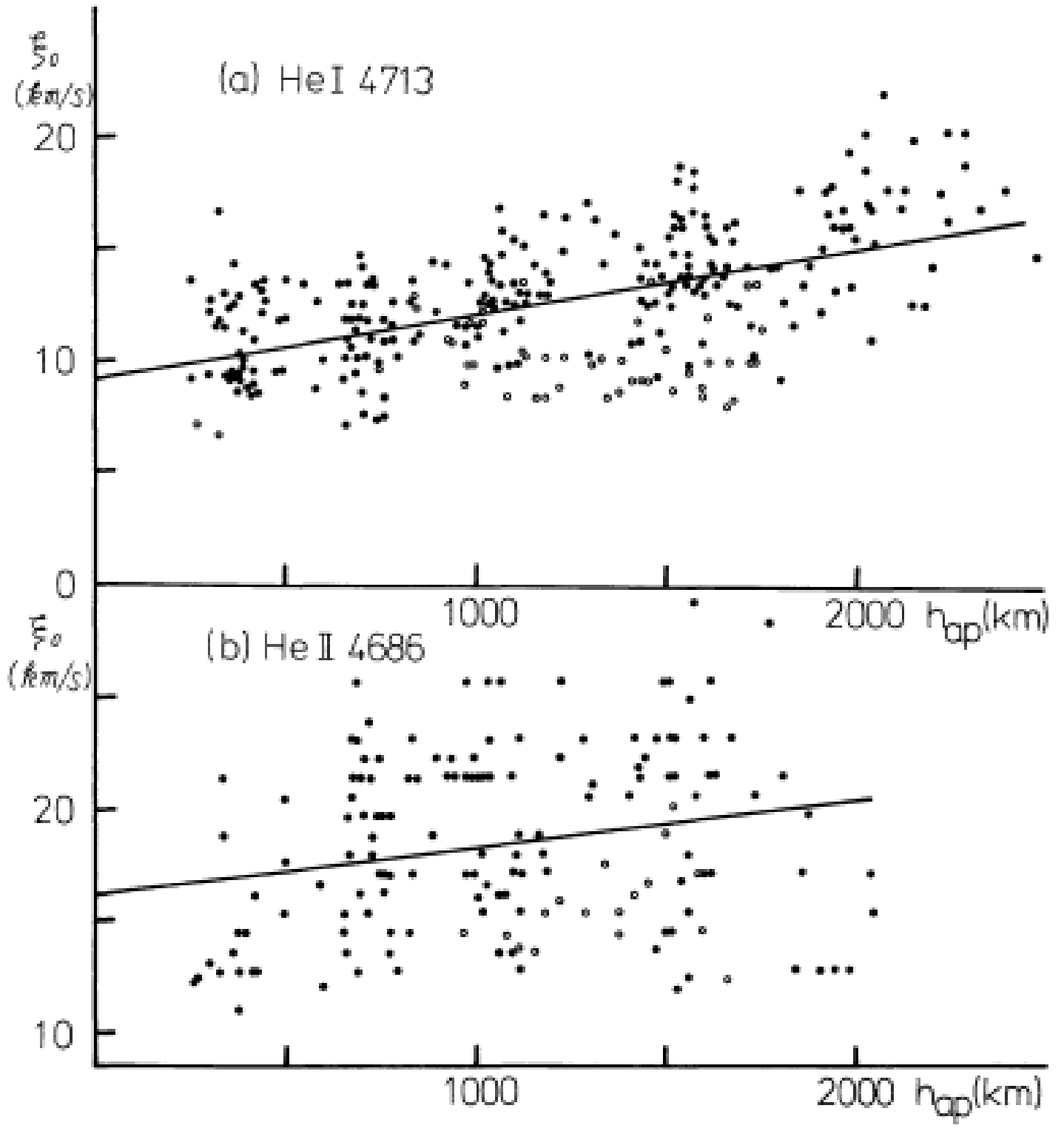

**Figure I-9-1:** *Largeurs Doppler observées pour l'hélium neutre He I 4713 Å et He II 4686 Å en fonction de l'altitude, au troisième contact de l'éclipse de 1962. D'après Hirayama et Irie 1984.*

La figure I-9-1 décrit les variations des largeurs Doppler mesurées pour He I 4713 Å et He II 4686 Å. La largeur Doppler est définie comme étant la largeur à 1/e du profil de la raie $\Delta\lambda_D$, et la conversion à $\xi_0$ est donnée par $\xi_0 = c\,\Delta\lambda_D / \lambda = (\xi_t^2 + 2RT/\mu)^{1/2}$. Les auteurs ont comparé la vitesse non thermique chromosphérique obtenue à partir des mesures des raies d'hélium neutre He I 4713 Å et ionisé He II 4686 Å. Il a été nécessaire de convertir la hauteur apparente $h_{ap}$ à la hauteur mesurée *h* radialement à partir de la profondeur optique unité à 5000 Å, $\tau_{5000} = 1$ définie en Annexe 24. La conversion de hauteur $h_{ap}$ à *h* est la suivante :

$h = 300 + h_{ap} + \frac{3}{2} H(km)$ avec *H* échelle de hauteur d'après Van de Hulst, 1953, page 229 et cité dans l'article Hirayama et Irie 1984.

Les graphiques en figure I-9-2 représentent les courbes de lumière relevées sur les raies d'hélium, en fonction de l'altitude, et les fluctuations autour de la courbe d'ajustement sont bien visibles.

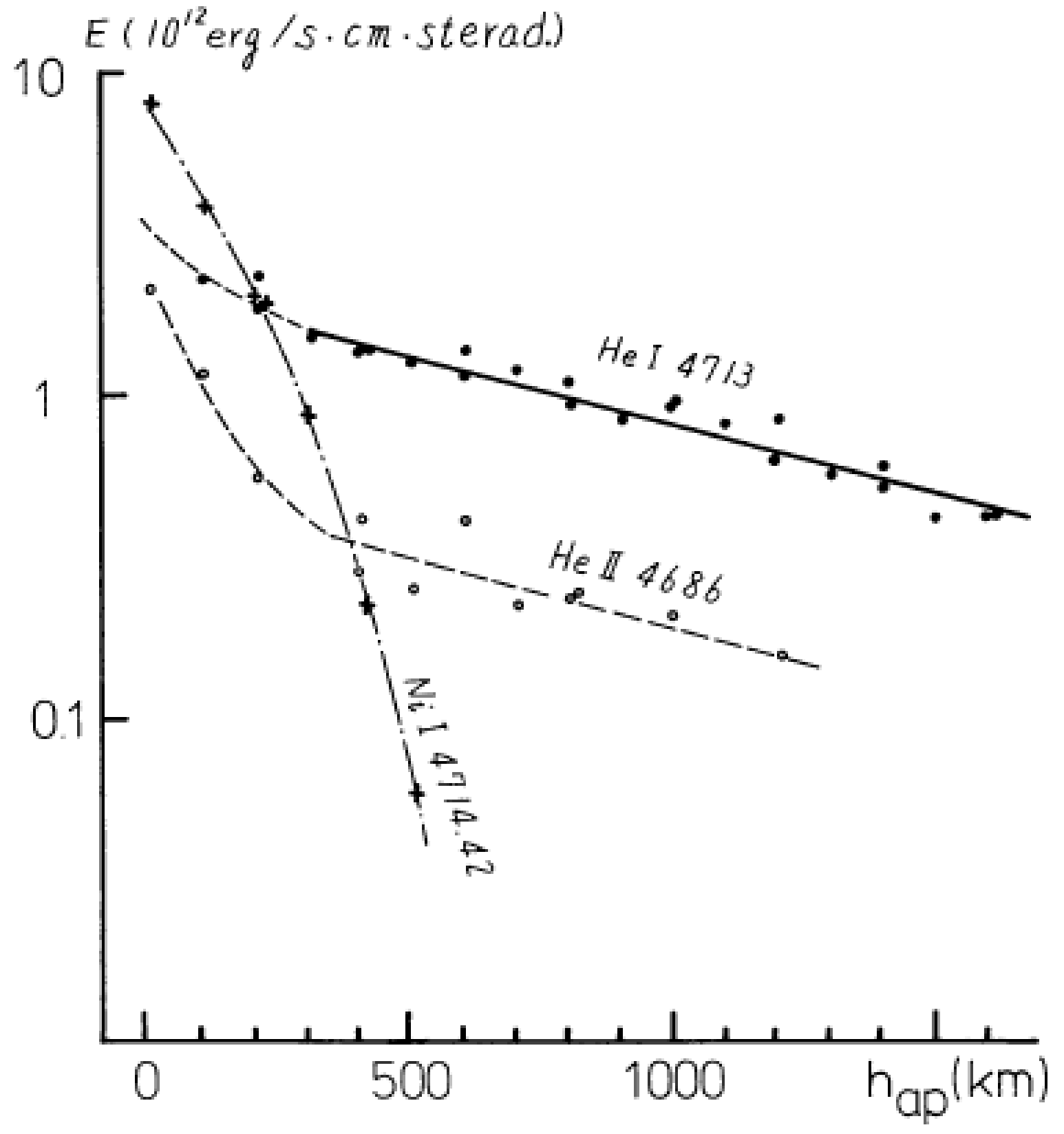

**Figure I-9-2:** *Profils d'intensités en fonction de la hauteur pour les raies He I 4713 Å et He II 4686 Å lors de l'éclipse de 1980. D'après Hirayama et Irie 1984.*

Les auteurs ont mesuré l'échelle de hauteur *H* qui vaut pour l'hélium neutre He I 4713 Å 4000 km. Elle est définie comme étant *H = -dh/dlnE*, comme étant la distance pour laquelle l'intensité totale d'une raie diminue d'un facteur *e*

Nous avons mesuré récemment ces échelles de hauteur à partir de nos spectres éclair CCD de 2008, 2009 et 2010 et avons trouvé des valeurs plus faibles, inférieures à 1000 km. D'après les résultats de Hirayama Irie 1984, les altitudes auxquelles les raies d'hélium apparaissent sont situées sur le limbe et ces résultats seront discutés dans cette thèse où nous les avons mesurées sur certaines courbes de lumière, à partir de 600 km au dessus du limbe.
Nous n'avons pas analysé de profils de largeurs de raies, car nous avons utilisé un spectrographe sans fente aux éclipses de 2008 à 2010, où nous avons obtenu des images dans ces mêmes raies, permettant une étude complémentaire des variations des émissivités grâce au mouvement naturel de la Lune.

<u>Les conclusions scientifiques tirées de ces expériences de Hirayama et al sont les suivantes :</u>

- La largeur Doppler du triplet He I 4173 Å augmente avec l'altitude et la largeur moyenne est compatible avec celle des raies métalliques et de l'hydrogène, suggérant que la température cinétique du triplet He I soit inférieur à 8000 K à l'altitude de 2000 km, dans cette région où elle est émise. Les vitesses non thermiques à 2000 km sont de 8 km/s et 14.8 km/s à 4000 km. Cela peut être expliqué par la recombinaison après photo-ionisation par les radiations UV coronales.
- La largeur Doppler de la raie Paschen $\alpha$ de He II 4686 Å est de 18.4 km/s, sans correction des sous-niveaux de cette raie et vitesses non – thermiques. Cette largeur de raie montre aussi une tendance d'augmentation avec l'altitude. Après comparaison des largeurs Doppler de He I 4713 Å et des raies EUV, et soustraction des vitesses non thermiques, la raie He II 4686 Å est émise dans une région (dans les spicules et chromosphère) où la température atteind 20000 K, et cette raie est émise aussi par un procédé de photo-ionisation à cause des raies coronales X – UV, en dessous de 228 Å.

- Toute fois la possibilité que la raie He II 4686 Å soit émise à une température inférieure à 10000 K n'est pas exclue. Le potentiel d'ionisation de He III avec les raies coronales X – UV est trop élevé pour une ionisation collisionnelle à cette température de 10000 K.

## I-10) Observation fusée Eclipse totale de 1970

Une expérience inédite de spectrographe sans fente a été effectuée dans le domaine UV du spectre à partir d'un vol fusée, où à haute altitude, l'atmosphère terrestre absorbe moins les rayonnements UV. Des observations des spectres éclairs dans l'UV ont été obtenus par A. H. Gabriel et al 1971.
A. H. Gabriel et ses collaborateurs, ont analysé les distributions d'intensités observées sur la myriade d'images monochromatiques dans les raies d'émission des spectres éclair UV obtenus lors du vol fusée historique à l'éclipse de 1970. Pour les ions O II de la région de transition, les élargissements Doppler sont évalués à 30 km/s à $T = 3*10^5$ K avec l'expérience fusée réalisée à l'éclipse du 7 Mars 1970. La figure suivante I-10-1 présente les résultats des spectres éclair UV obtenus par les auteurs:

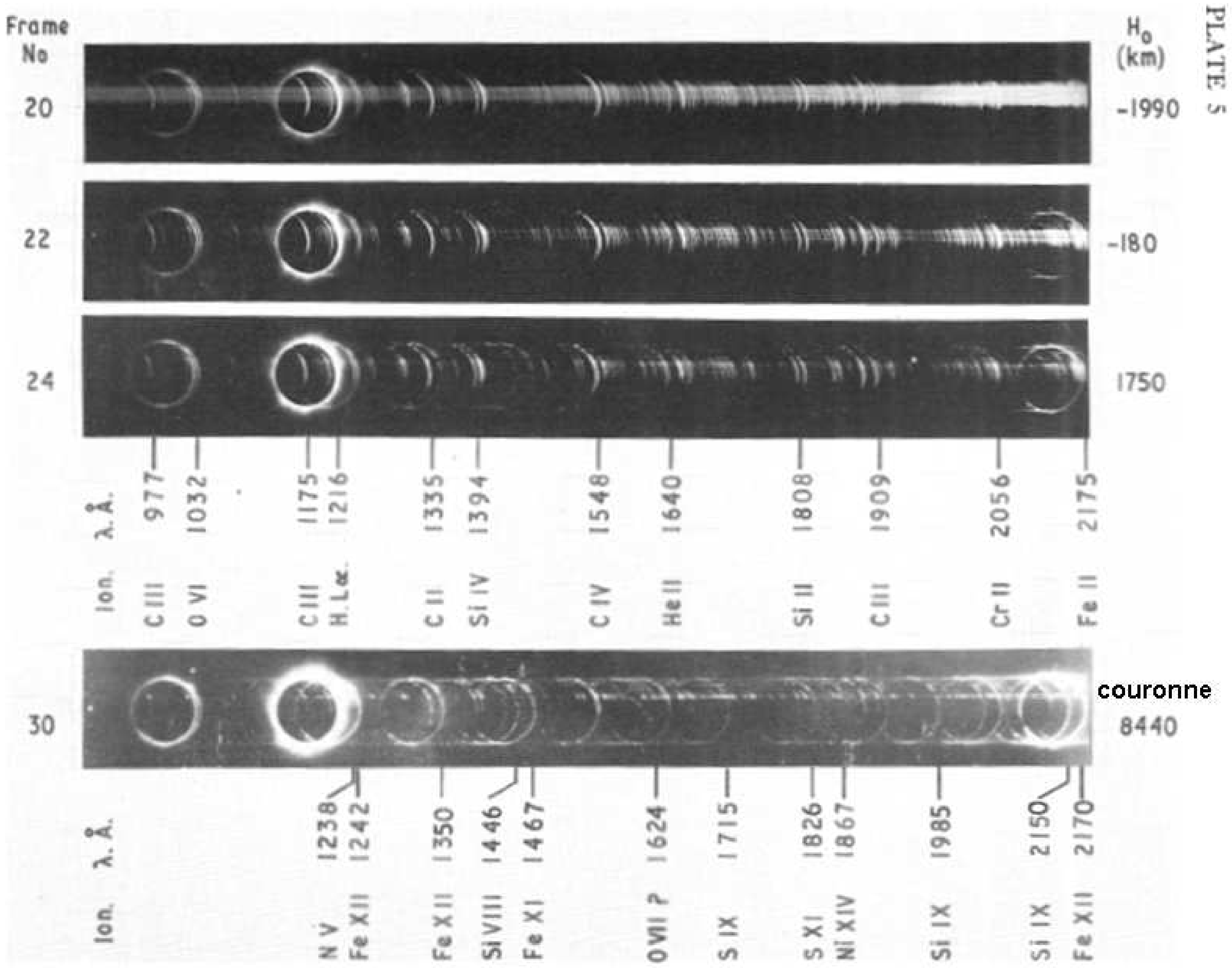

FIG. 4.—Four selected frames from the data. Frames 20, 22, and 24 are around second contact. Frame 30 is within totality and shows a predominantly coronal spectrum. The wavelengths marked refer to the right-hand limb of the Moon, as shown here.

**Figure I-10-1:** *extrait des spectres éclairs historiques obtenus dans le domaine UV depuis un vol fuse à l'éclipse totale du 7 mars 1970. D'après Gabriel et al 1971.*

L'expérience utilisée a consisté à réaliser des spectres sans fente dans les UV. Des raies du Mg, S, Si, Fe, Ni, etc.. qui étaient connues depuis Skylab et OSO ont été analysées. ont été . Elles ont des degrés d'ionisation variés, et indiquent plusieurs gammes de températures, allant de 10000 K à 100000 K. Les gammes de températures associées aux ions sont données dans le tableau Annexe N° 33. L'intérêt de ces spectres UV est d'identifier des raies low FIP comme le chrome une fois ionisé Cr II 2056Å, le fer une fois ionisé Fe II 2175Å et 2170 Å, pour retrouver et comparer ces mêmes éléments dans les raies des spectres éclair réalisés dans le domaine visible. Les spectres figure I-10-1 montrent simulatanément des raies low FIP de température inférieure à 10000 K et des raies plus chaudes comme le carbone C IV de high FIP, c'est-à-dire supérieur à 10 eV, dont les températures se situent autour de 100000 K. L'intensité relative des raies formées à différentes températures, en fonction de la position autour du limbe, est une fonction sensible à la température, et ce comportement a été important pour effectuer les identifications des raies. Les fluctuations d'intensités des profils du relief du limbe lunaire et du spectre éclair correspondant sont décrites sur la figure I-10-2.

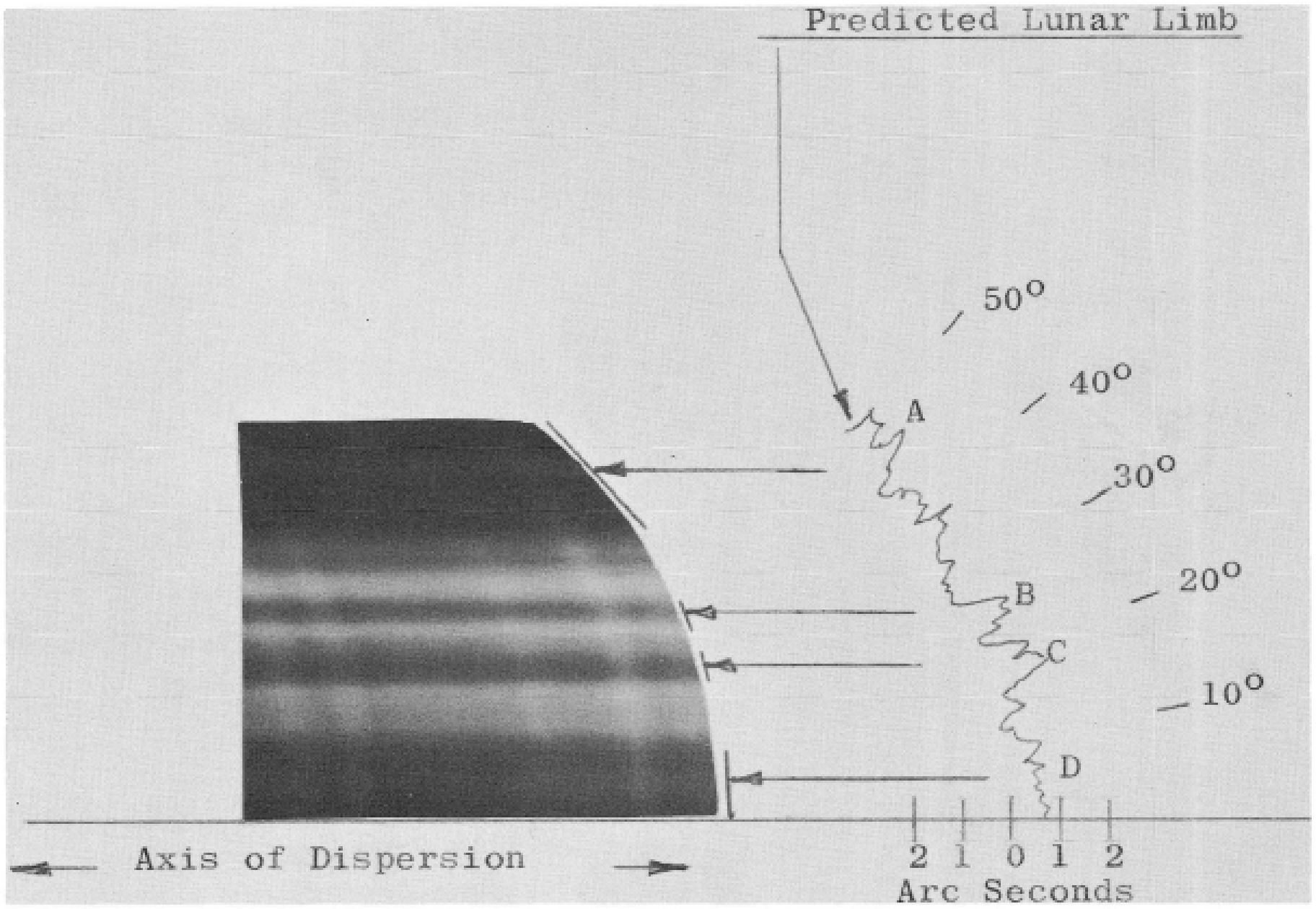

PLATE 4

FIG. 3.—Effect of lunar limb irregularities

GABRIEL *et al.* (*see* page 600)

**Figure I-10-2:** *extrait de la publication de Gabriel et al 1971, montrant les prédictions du limbe lunaire et par comparaison avec les spectres éclairs: la résolution spatiale est insuffisante pour tenir compte de tous les détails du profil du bord lunaire. Les montagnes du profil lunaire produisent l'assombrissement.*

La figure I-10-2 présente la correspondance entre les vallées du limbe lunaire et graduations en degrés d'une part et d'autre part les maxima d'intensité des spectres continus des grains de Baily, sur la partie de gauche avec l'axe de dispersion spectral. Ces spectres sont produits durant les quelques secondes où la haute photosphère n'est pas encore complétement occultée par la Lune. Des raies d'émission relativement intenses sont aussi visibles superposées au spectre continu.
La figure I-10-3 agrandie représente l'orientation du disque solaire et les principales structures (protubérances annotées) telles qu'elles sont observées dans chaque image monochromatique dans les spectres de raies UV de la figure I-10-1.
Ces figures servent à indiquer la présence de protubérances au dessus du limbe et dans l'interface de transition chromosphère-couronne.

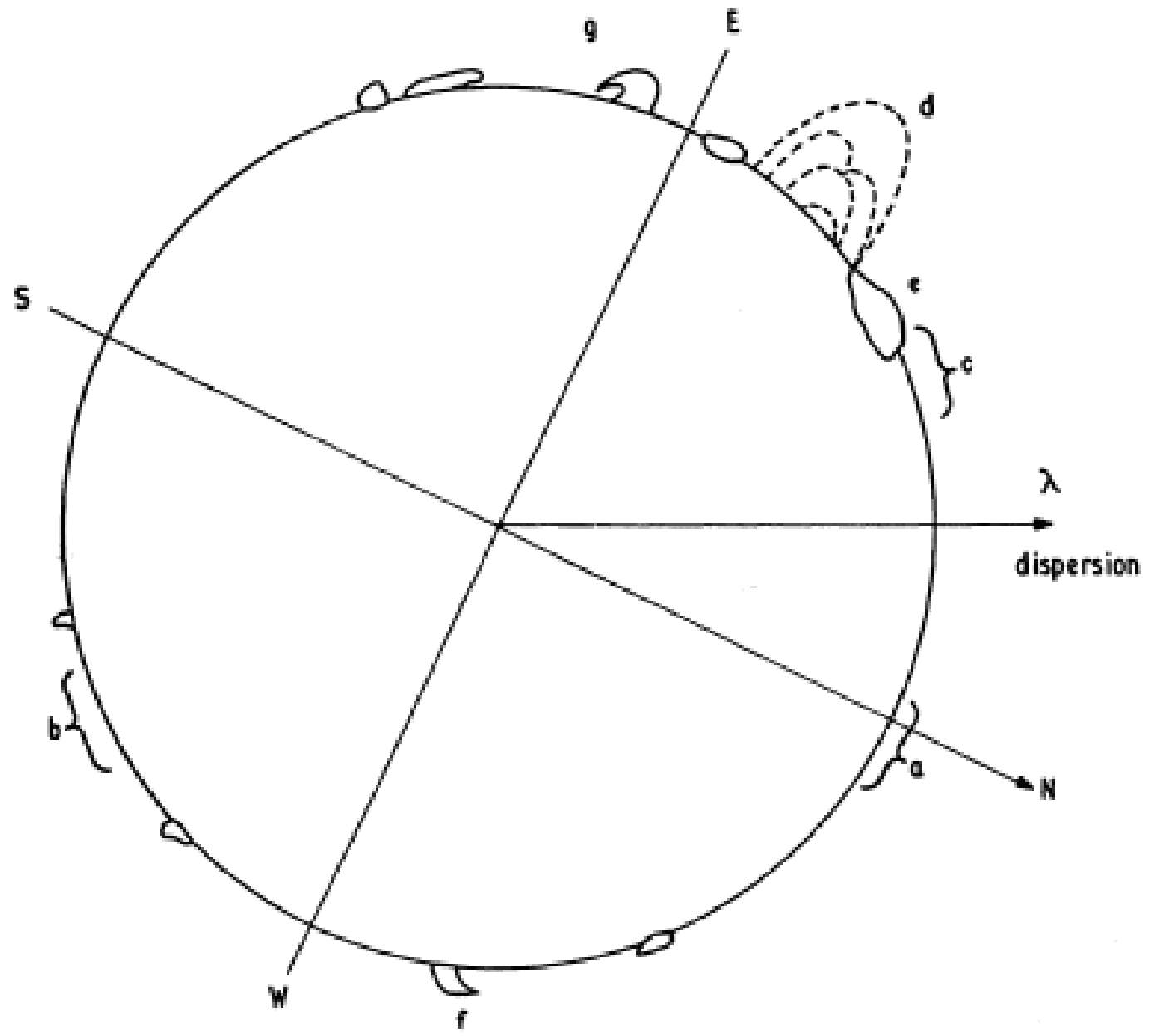

**Figure I-10-3:** *Diagramme du Soleil avec la même orientation qu'en figure I-10-1, montrant le pôle nord solaire, le sens de la dispersion spectrale, et les protubérances. D'après Gabriel et al 1971.*

Dans l'article Gabriel et al 1971, les auteurs ont par la suite déduit la brillance, par inversion d'intégrale d'Abel de la raie Lyman α de l'H I dans la couronne.
La figure I-10-4 indique les variations des brillances que les auteurs ont déduites pour Lyman α de H I, et pour des altitudes bien plus élevées, dans la couronne:

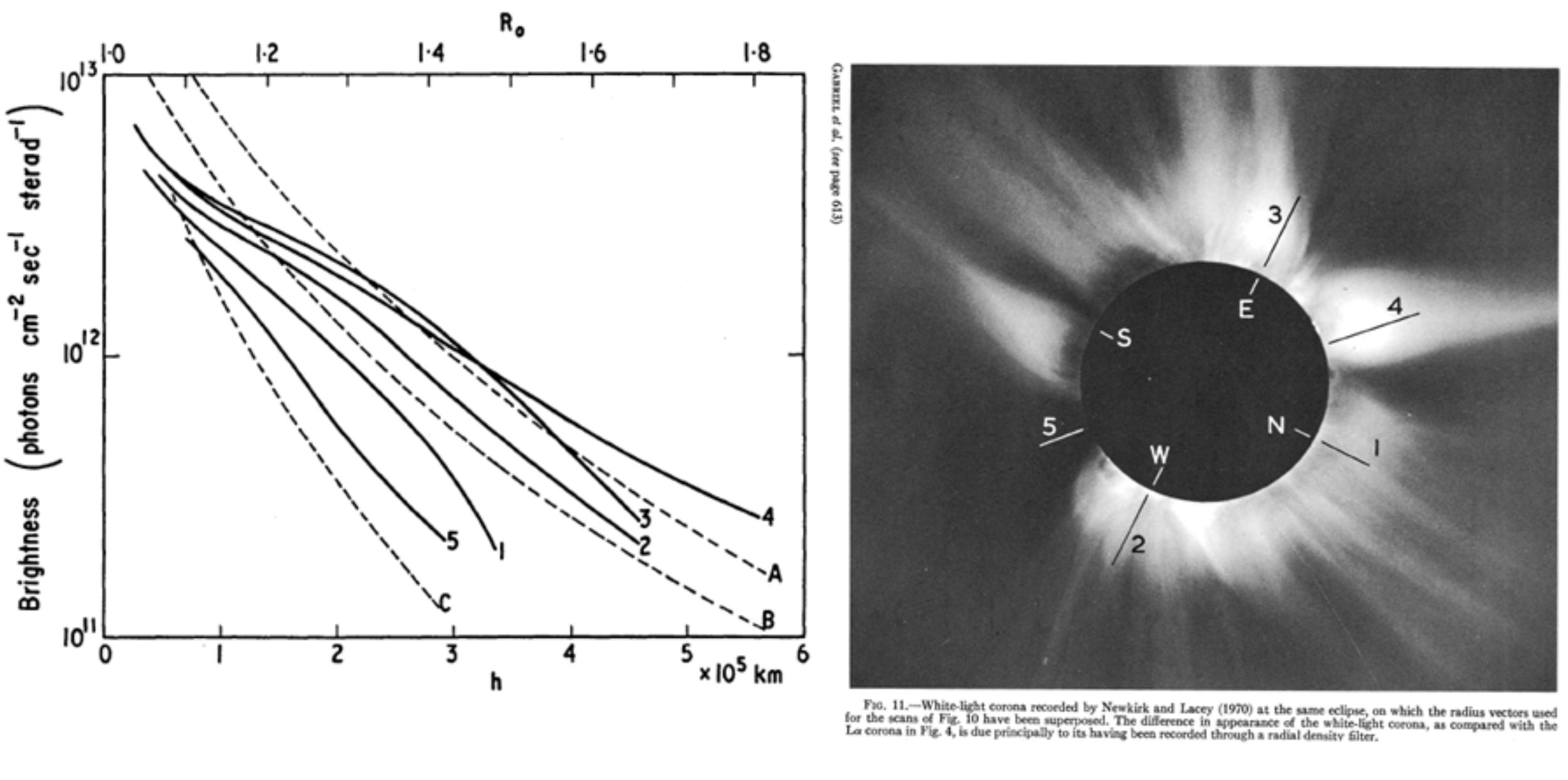

**Figure I-10-4:** *extrait de la publication de Gabriel et al 1971, montrant les variations de la brillance de la couronne en Lyman α. Les courbes 1, 2 et 3 montrent les observations le long d'un rayon vecteur au Nord, Ouest et Est solaire. La courbe 4 montre la brillance dans un jet intense en lumière blanche, et la courbe 5 dans une région calme. A, B et C sont des brillances prédites d'après Allen 1969.*

Les conclusions scientifiques suivantes depuis l'expérience de Gabriel et al sont tirées :

- 25 raies coronales dans les transitions interdites ont été étudiées pour les éléments abondants comme l'oxygène, silicium, néon, fer.
- La couronne Lyman $\alpha$ est interprétée comme dispersion résonnante du disque Ly$\alpha$ de H I coronal : cela entraîne des analyses futures des condensations coronales.

Ces premières images de la couronne dans les UV ont par la suite motivé de nouveaux programmes d'observation de la couronne solaire dans les rayonnements EUV et X correspondant à des températures de plusieurs millions de Kelvin. Des images de la couronne en rayons X avait été réalisés par Tousey 1962. Un vol fusée a été réalisé en 1991 ayant permis la superposition d'images de la couronne dans les rayons X et du disque en visible.

## I-11) Expérience marquante de Daw, Golub Deluca 1991

Cette expérience historique réalisée en 1991 a consisté à utiliser le Télescope à rayons X en Incidence Normale, NIXT, pour effectuer l'acquisition d'une image dans les rayons X à 63.5 Å simultanément superposée et alignée avec une image effectuée dans le visible à 6800 Å, comme le montre l'image figure I-11-1.

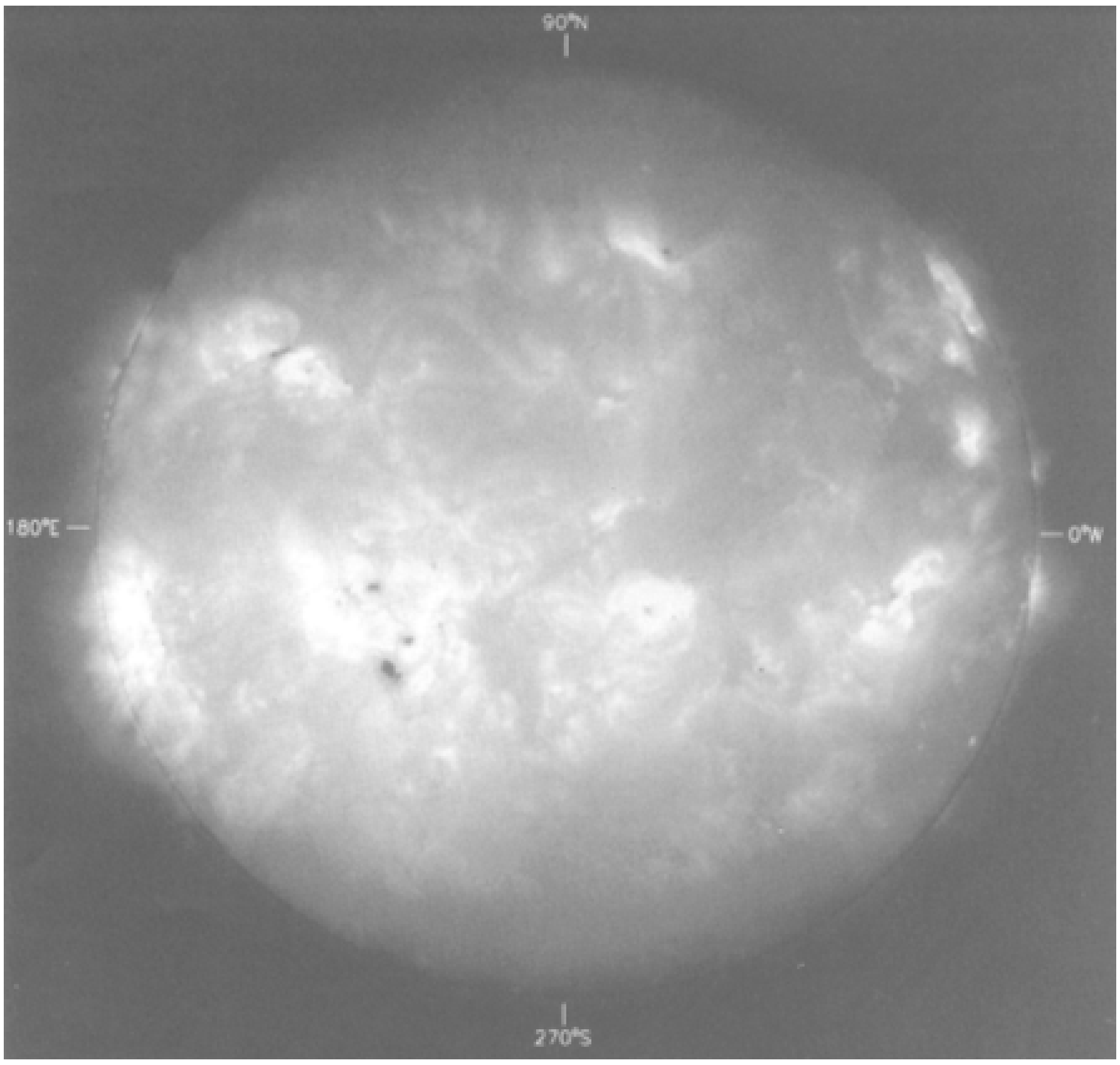

**Figure I-11-1:** *image de la couronne solaire réalisée le 22 Février 1991 à 19h50 TU avec NIXT dans les rayons X à 63.5 Å, avec la superposition du disque en lumière continue visible à 6800 Å. Les alignements ont été parfaitement ajustés, et révélant pour la première fois des images de la couronne dans les rayons X avec le bord intérieur à 6800 Å. Les coordonnées héliocentriques en degrés sont indiquées sur le limbe. D'après A. Daw, E.E. Deluca and L. Golub 1995.*

Les images en rayonnement X de la couronne avec NIXT ont rendu possible l'obtention de profils d'intensité en fonction de la distance radiale. L'origine a été prise sur le limbe. Deux types de régions de l'atmosphère solaire ont été analysés: régions calmes, et trous coronaux. Les graphiques présentés en figure I-11-3 ont été obtenus en utilisant un modèle à symétrie sphérique décrit en figure I-11-2 pour décrire les couches d'atmosphère solaire analysées.

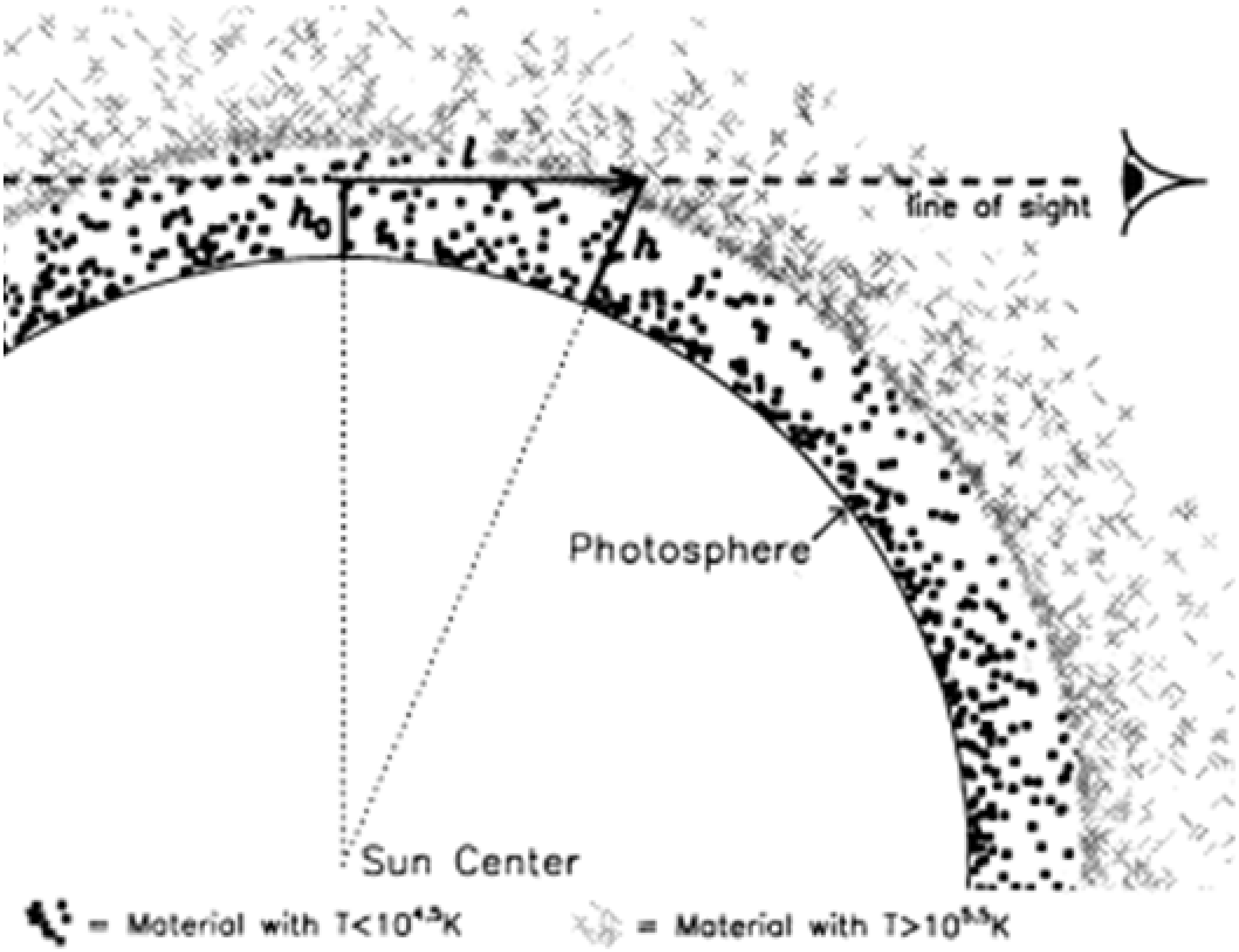

**Figure: I-11-2:** *modèle à symétrie sphérique permettant de calculer les intensités des rayons X observés par NIXT. L'émission de rayons X est intégrée le long de la ligne de visée. L'émission est définie comme étant le produit de l'émissivité par le carré de la densité électronique. Pour les lignes de visée traversant la chromosphère, comme représentée sur la figure ci-dessus, l'émission intégrée depuis l'arrière du limbe est atténuée par l'absorption chromosphérique avant d'arriver à l'observateur. La distance* $h_0$ *dans la partie supérieure de la figure, la ligne de visée l au dessus du limbe visible et l'altitude h sont indiquées. D'après Daw, Deluca et Golub 1995.*

Un modèle d'atmosphère à symétrie sphérique est considéré, pour lequel la densité et la température ne sont seulement fonction que de l'altitude.
Les profils d'intensités radiales, figure I-11-3, sont les résultats observés par NIXT dans plusieurs régions autour du disque solaire, en tenant compte de la résolution spatiale de 625 km pour les rayons X. Ces profils présentent un embrillancement du limbe aux altitudes comprises entre 4000 et 7000 km avec un maximum vers 5000 km.

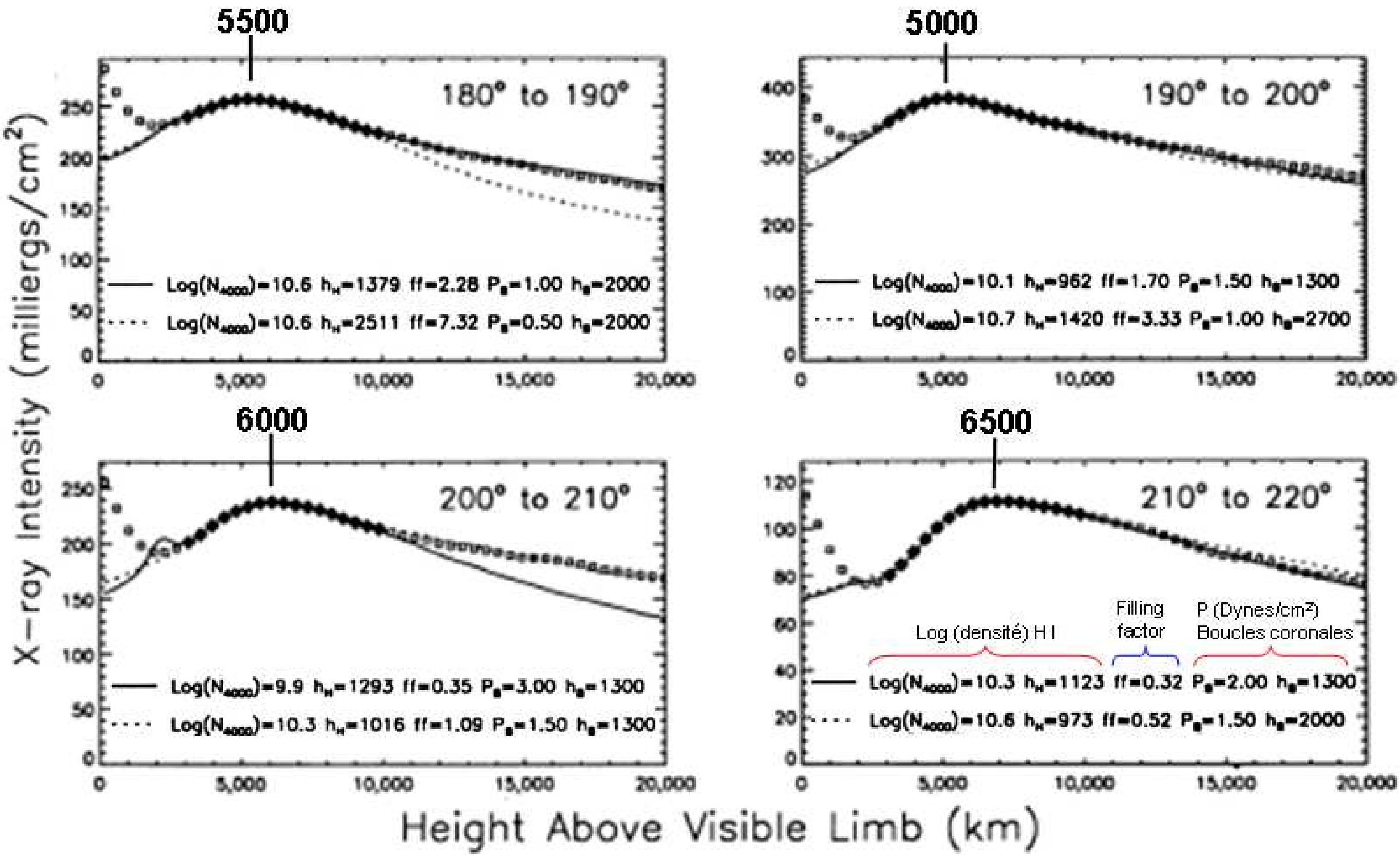

**Figure I-11-3:** *extrait de régions situées autour du disque solaire, observées par NIXT en rayons X entre 0 et 20000 km au dessus du limbe solaire visible. La localisation de chacune des régions sur le Soleil est indiquée sur la partie supérieure à droite de chaque tracé, par son angle en coordonnées héliocentriques. Sont représentées aussi sous les accolades:*
*1) le logarithme de la densité de l'hydrogène neutre en cm $^{-3}$ entre 960 et 2500 km.*
*2) l'échelle de hauteur de la densité de l'hydrogène neutre,*
*3) le facteur de remplissage (ff) en pourcents de boucle coronale,*
*4) la pression des boucles coronales en dynes/cm $^{2}$ pour plusieurs altitudes situées entre 960 et 2500km*
*5) l'altitude de la base des boucles coronales en kilomètres. D'après A. Daw, E.E. Deluca and L. Golub 1995.*

D'après ces profils (figure I-11-3), une absorption aux altitudes situées autour de 2000 km au dessus du limbe, suivie d'un embrillancement plus haut vers 5000 km et au delà sont observés à partir de ces profils radiaux à 63 Å. Ces quantités de rayonnement X absorbés entre 1000 et 3000 km au dessus du limbe peuvent s'expliquer par la présence de plasma « froid » composé d'hélium He II 228Å, de spicules, et macrospicules qui sont des structures (associées à du plasma inhomogène) dynamiques, associées au champ magnétique. Les températures de ces structures sont situées autour de 10000 K. Leur extension est comprise entre 1000 et 5000 km environ, et ce plasma plus froid qui absorbe ce rayonnement X à 63Å produit par la couronne chaude à plusieurs M K.
L'embrillancement si tué au dessus de l'absorption, entre 5000 et 6500 km peut être produit par du plasma coronal qui émet du rayonnement EUV et X émis par des ions ayant perdu plus de la moitié de leurs électrons, situés à des altitudes plus élevées au dessus de la région de transition chromosphère-couronne.

Des profils d'intensités comparables ont été également relevés dans le sens radial au chapitre V, figures V-5-9 et V-5-10, en vue de comparer les extensions des embrillancements à partir d'images de la mission spatiale SDO/AIA dans le Fe XII à 193Å, le Si XII à 131Å, le Fe IX/ fe X à 171Å et He II 304Å. Ces observatoires spatiaux ont permis d'atteindre une meilleure résolution spatiale et dynamique pour les analyses photométriques que les premiers vols fusée historiques, et Skylab. Dans le chapitre II suivant, les techniques d'observations modernes des spectres éclairs dans le domaine visible sont décrites, où les raies d'émission étudiées sont optiquement minces, et formées aux altitudes de 0 à 2000 km, et correspondent aussi aux observations des absorptions dans les profils (figure I-11-3) des images NIXT dans les rayons X.

Les conclusions scientifiques tirées de ces expériences de Daw et al sont les suivantes :

- Les interprétations des observations permettent de déduire que pour les boucles coronales, l'émission se produit au-delà des pieds des boucles coronales. Les pieds des boucles, lorsqu'ils sont observés sur le limbe, sont absorbés par le matériel chromosphérique sur la ligne de visée. Il est supposé une atmosphère dans laquelle les pieds des boucles coronales sont intercéptées le long de la ligne de visée par du matériel chromosphérique plus froid, et qui s'étend à des hauteurs au-delà des pieds des boucles.
- Le coefficient d'absorption des rayons X de NIXT par le matériel chromosphérique est proportionnel à la densité de l'hydrogène neutre H I entre 3000 et 10000 km. Pour h = 3000 km au dessus du limbe, la masse de H I atténue de 25 % les rayons X provenant loin du limbe. La masse chromosphérique au-delà de 3000 km n'est pas complétement confinée dans les spicules. L'absorption se produit pas seulement au dessus du limbe de la région de transition, mais au-delà jusqu'à la basse couronne.
- Le modèle « FAL » à 1 dimension hydrostatique n'explique pas l'absorption X de NIXT.

# Chapitre II) L'approche observationnelle moderne du limbe solaire- visible au sol

# II-1) La technique du spectre éclair sans fente

La technique des spectres éclairs a été depuis longtemps utilisée avec les spectrographes sans fente comme décrits au chapitre I-2.
Nous avons développé une nouvelle méthode d'acquisition des spectres éclairs sans fente lors des éclipses totales de Soleil, que j'ai améliorée depuis celle du 29 Mars 2006. Lors des éclipses de 2008, 2009 et 2010, le même instrument a été utilisé, et surtout une caméra CCD numérique qui enregistre en continu à une cadence de 15 images/seconde, et avec un logiciel qui régule le débit de données sans interrompre le processus d'acquisition, dont ensuite on n'a plus à s'occuper, ce qui est un avantage considérable. Nous sommes ainsi certains d'obtenir systématiquement toute la séquence des 10 secondes correspondant aux spectres éclairs.
Cette méthode consiste en une succession d'étapes préalables lors des phases partielles qui précédent les instants des contacts avant et après la totalité des éclipses de Soleil.
Dans un premier temps il faut obtenir l'image éblouissante du spectre au premier ordre du Soleil. Cela s'effectue par projection sur une feuille blanche, où le spectre coloré très intense apparaît. Le sens de dispersion spectral doit être confondu avec le sens du mouvement apparent du Soleil dans le ciel, ce qui nécéssite d'orienter le réseau de diffraction. Par ailleurs, un entraînement est nécéssaire afin de compenser le mouvement diurne. Une monture de type équatoriale est utilisée, voir figure II-1-3. Une fois cette opération effectuée, il faut ensuite régler au centre du tube de la lunette, le domaine bleu à 4700Å du spectre, pour qu'ensuite en insérant les filtres densité neutre 2 et filtre 4700Å, on puisse avoir exactement le segment spectral correspondant à la gamme de longueurs d'onde que l'on souhaite obtenir, centré sur l'axe optique de la Lunette de 50 mm de diamètre et 600 mm de focale. Il faut donc ensuite maintenir cette position durant toutes les phases précédant la totalité, en assurant le guidage de l'instrument.

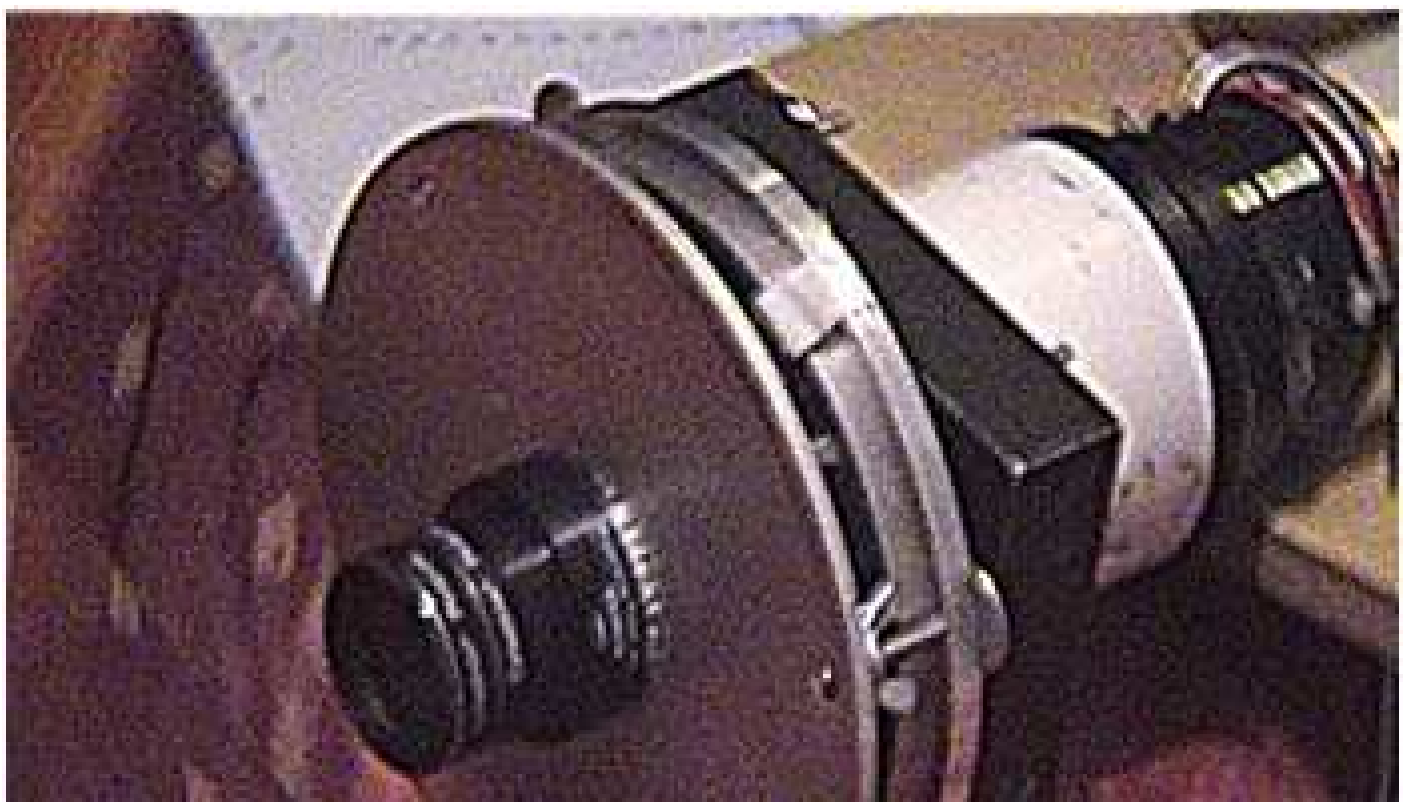

**Figure II-1-1:** *vue des filtres amovibles de 25 mm de diamètre, comprenant une densité neutre de densité 2 suivie d'un filtre de 4700Å et 120 Å de bande passate, placé devant le réseau qui peut tourner autour de l'axe optique de la Lunette.*

Plus précisément le domaine spectral dans la région bleue du spectre est obtenu grâce à un filtre centré sur 4700Å et de 100 Å de FWHM associé à un filtre densité neutre D= 2. Ces filtres sont placés devant le réseau par transmission. L'atténuation est suffisante pour ne pas s'éblouir lors du pointage car cette opération peut s'effectuer aussi visuellement et ne pas saturer la caméra CCD avant la totalité où les spectres éclair seront réalisés. En effet il faut manipuler la lunette de 50 mm de diamètre et 600 mm de focale installée sur la monture équatoriale, de telle manière à trouver et conserver le spectre à l'ordre 1, du côté du blaze, maximum de brillance.

Un avantage au niveau de la conception de l'instrument, est que le réseau par transmission peut pivoter autour de l'axe optique, ce qui facilité les opérations de centrage et orientation du spectre dans le sens du mouvement horaire, comme le montre la figure suivante:

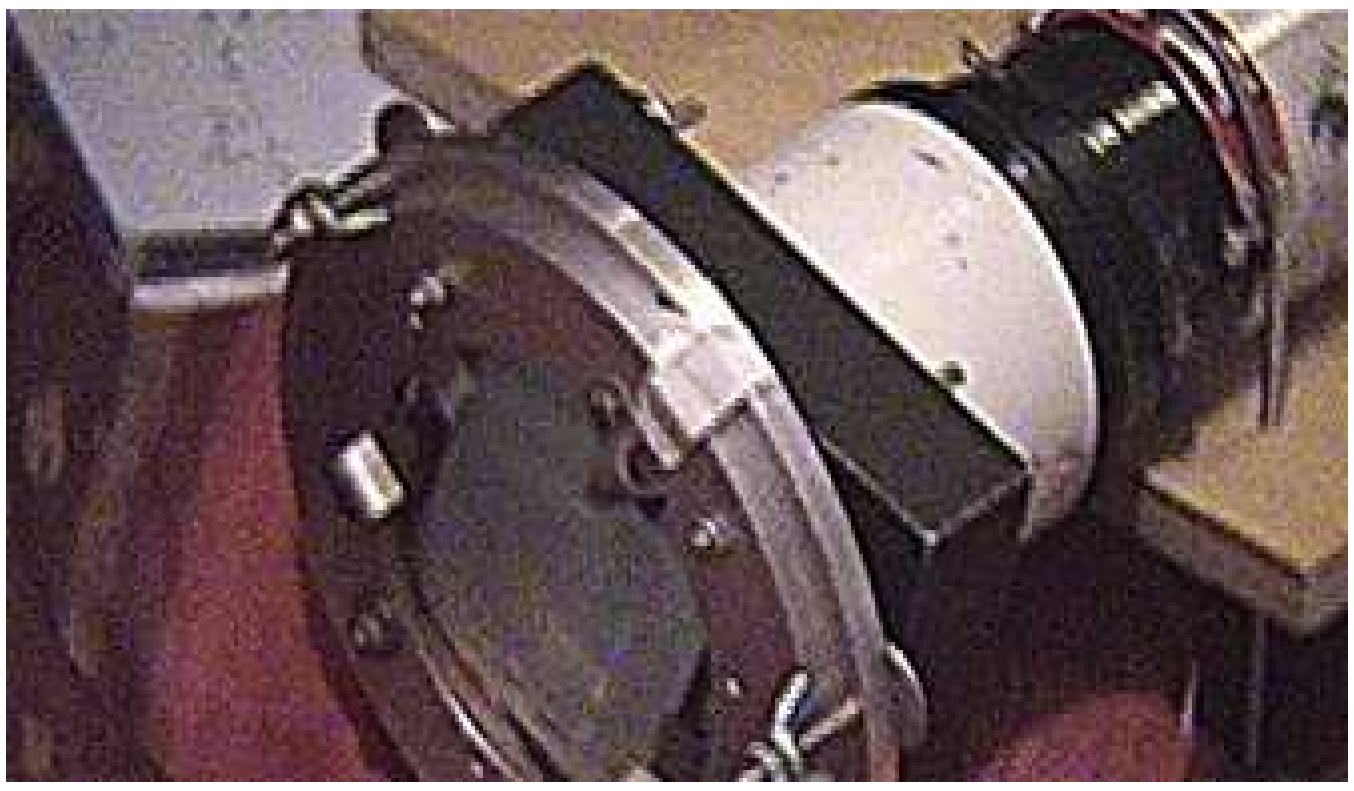

**Figure II-1-2:** *vue agrandie de l'objectif lunette + réseau tournant par transmission lorsque la densité neutre est retirée. Le réseau peut tourner autour de l'axe optique de la lunette de 50 mm de diamètre et 600 mm de focale ce qui facilite l'orientation du spectre à l'ordre 1.*

Le pointage est réussi, lorsqu'on voit à travers l'instrument, le halo bleuté correspondant à l'intervalle spectral de 100 Å à 4700 Å du spectre solaire. Il arrive qu'en cas de voiles nuageux trop épais, cette opération soit nécéssaire si le signal fourni par la caméra CCD est trop faible.
Une fois l'opération effectuée, on insère la caméra CCD, dont la mise au point à l'infini a été réalisée au préalable sur une étoile ou le croissant lunaire, à travers le réseau par transmission à l'ordre zéro. La butée de la bague correspond à la butée du support de la CCD. La caméra CCD peut ainsi être retirée ou remise, et la mise au point est conservée.
La figure II-1-3 présente l'instrument en service lors des phases du Soleil partiellement éclipsé, et le filtre de densité neutre associé à celui de 4700Å amovibles sont en place devant le trajet des faisceaux.

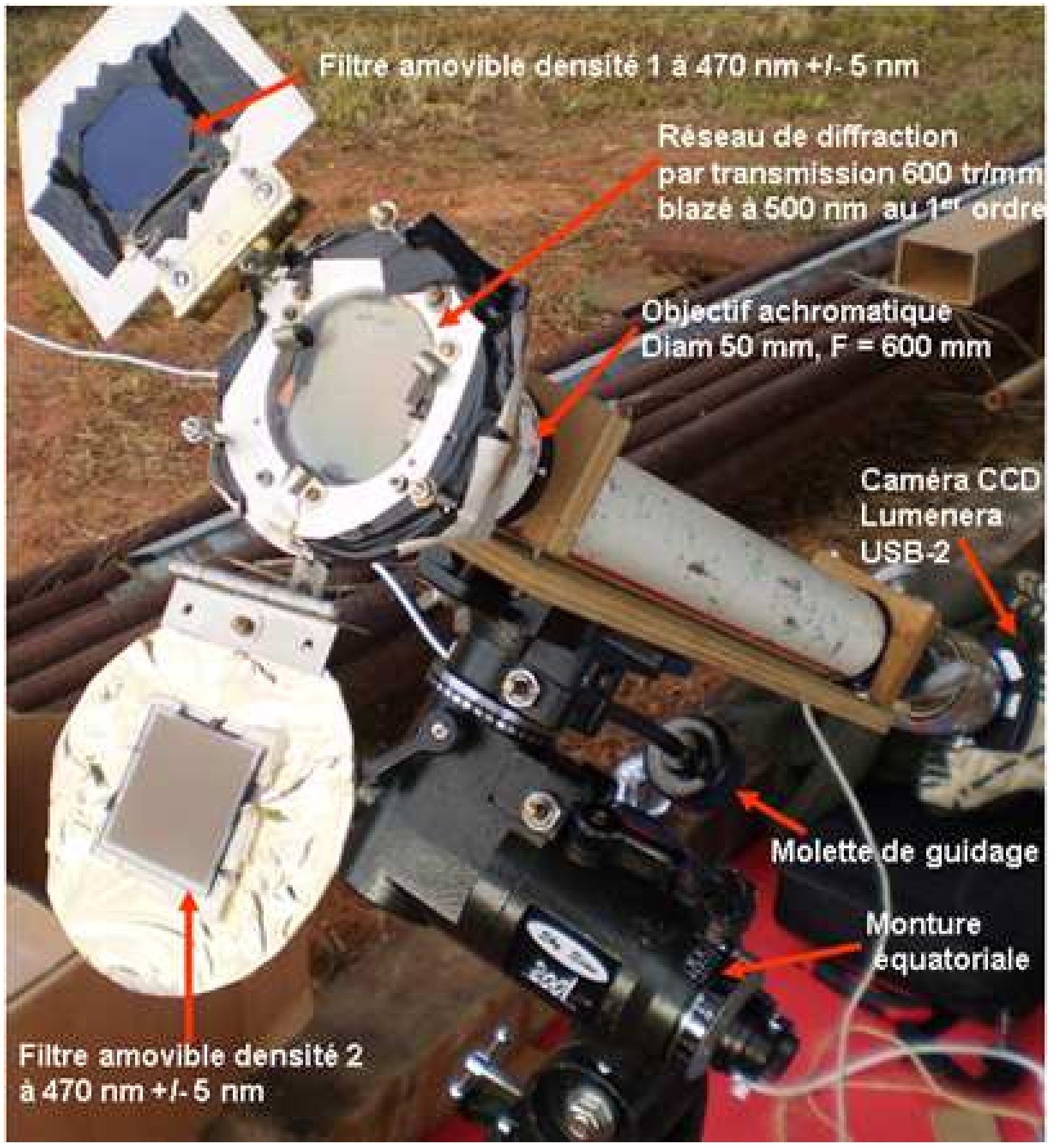

**Figure II-1-3:** *vue d'ensemble de l'expérience des spectres éclairs avec spectrographe sans fente, avant l'éclipse le matin du 13 Novembre 2012 en Australie. On peut distinguer la caméra CCD au foyer de l'instrument, le réseau de diffraction par transmission, les filtres amovibles dirigés vers le Soleil, à droite de l'image. L'expérience du réseau objectif est installée sur une monture équatoriale, et l'on peut voir les flexibles qui permettent d'assurer le guidage et entraînement manuel. L'instrument a été mis en station les jours précédant l'éclipse..*

L'ordinateur permettant les acquisitions a été protégé par des planches en carton afin de mieux isoler l'écran de visualisation de la lumière du jour.
Les filtres amovibles de densité neutre associés au filtre interférentiel centré sur 4700 Å, permettent de sélectionner facilement le segment spectral, correspondant à l'intervalle de longueur d'onde dans lequel se situent les raies d'hélium He I 4713 Å et He II 4686 Å.

Ces opérations demandent une certaine habilité et préparation de l'observateur.
En effet, les expériences d'éclipse nécéssitent de nombreuses répétitions sur le terrain, car le phénomène des phases précédant et succédant la totalité de l'éclipse de Soleil sont très rapides, la durée des contacts est de 10 secondes, et demandent une certaine maîtrise du maniement des instruments. L'expérience des spectres éclairs a été améliorée à l'éclipse totale de Soleil du 1er Août 2008 en Sibérie, figure II-1-4 où une caméra CCD Watec de sensibilité 10µ Lux a été utilisée avec une cadence de 25 images/seconde. La dimension des pixels fait 9µm de côté, et cette caméra a rendu possible l'enregistrement de la raie de faible intensité de l'hélium ionisée He II 4686 Å (entre $10^{-4}$ et $10^{-5}$ fois plus faible que l'intensité au bord du disque solaire) lors de cette éclipse totale du 1er Août 2008. En 2009 et 2010, la taille du pixel était de 4.6µ, comme indiqué sur la figure II-1-4. Le maniement et guidage manuels sont effectués plus facilement avec les molettes flexibles de la monture équatoriale de ce réseau-objectif, en se servant des phases partielles, c'est-à-dire quand la Lune ne couvre pas encore totalement le disque solaire. Ceci est réalisé en surveillant la progression de phases de l'éclipse avec un verre de soudeur de densité 5, avant le début de la totalité. Cela permet d'assurer les centrages du segment spectral, d'effectuer les recentrages, corrections des faibles dérives de mise en station (l'axe horaire de la monture équatoriale doit correspondre au pôle celeste). Avec l'expérience d'évaluation de l'assombrissement du ciel, le filtre de densité neutre amovible est retiré au bon instant précédant la totalité, pour obtenir les spectres flash dans la région spectrale autour de 4700 Å.

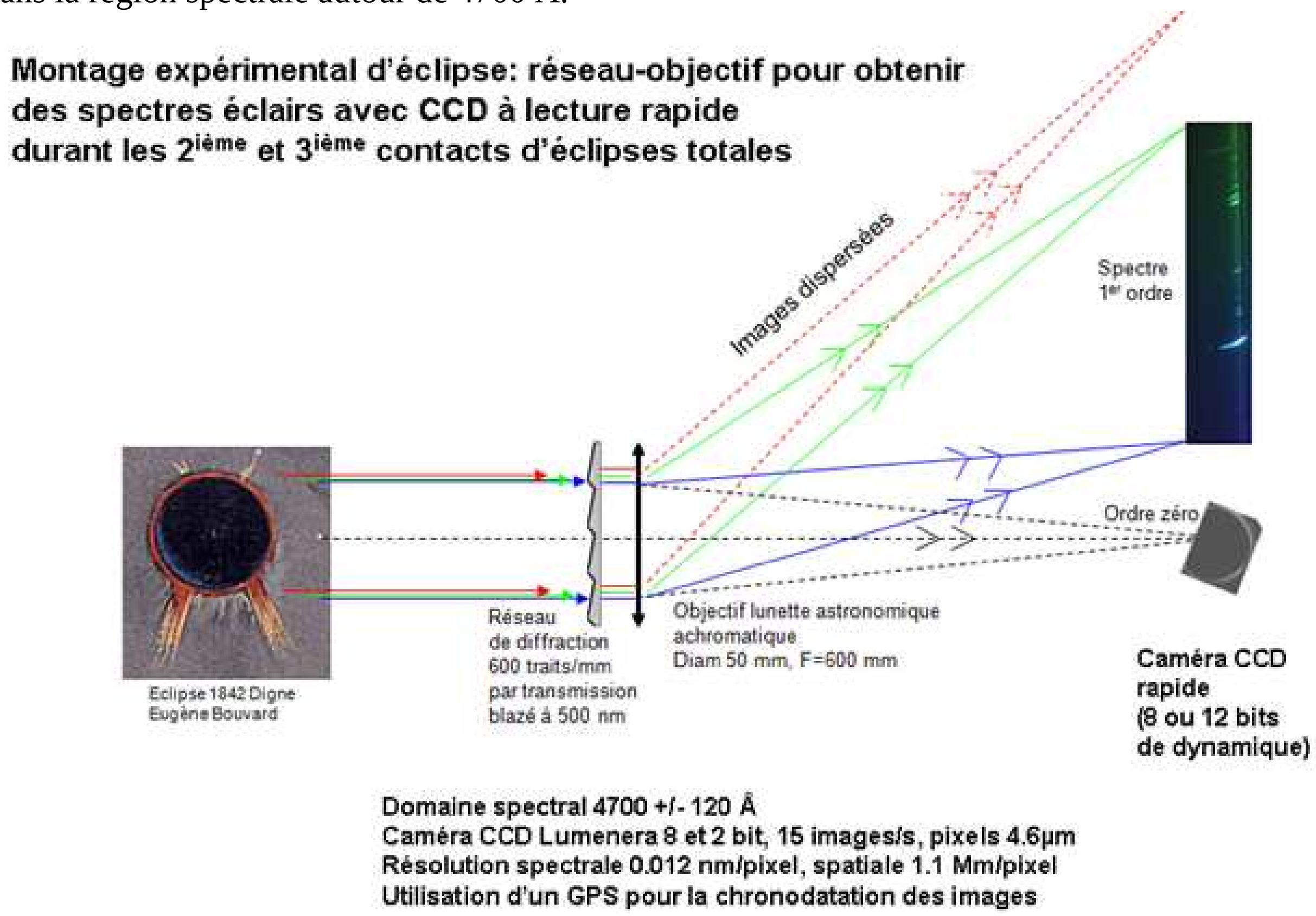

**Figure II-1-4:** *Schéma de l'expérience des spectres éclairs utilisé aux éclipses de 2008, 2009, 2010 et 2012. Le montage, consistant en un réseau de diffraction par transmission (utilisé au 1er ordre), placé devant un objectif réfracteur, est fixé sur une monture équatoriale pour le guidage. L'angle d'approximativement 25° indique la déviation du spectre solaire avec l'axe optique du réfracteur. La caméra CCD Watec 120N+ utilisée en 2008 a été remplacée par la caméra Lumenera Skynyx 2.1M en 2009 et 2010.*

La cadence d'acquisition a été de 25 images/seconde avec la caméra Watec 120 N+ à l'éclipse de 2008 mais cette caméra ne permettait pas d'obtenir une résolution spatiale meilleure que 1800 km/pixel. La dynamique était limitée à 8 bit. Ensuite lors de l'éclipse du 22 Juillet 2009, une caméra Lumenera Skynyx 2.1 M ayant des pixels de 4.6 µm a été utilisée, où la résolution spatiale a été de 1150 km/pixel, mais a été utilisée avec seulement 8 bit de dynamique et avec une cadence de 15 images/seconde pilotée et contrôlée avec le logiciel Iris. Ce logiciel gratuit ne permettait pas d'utiliser la caméra avec 12 bit de dynamique en mode d'acquisitions en continu. Ce logiciel est d'avantage dédié au traitement de données, et est par ailleurs conçu pour effectuer des acquisitions avec des appareils photographiques numériques et d'autres modèles de caméras CCD en longue pose pour des objets du ciel profond. Par la suite, à l'éclipse du 11 Juillet 2010 en Polynésie Française, cette caméra Lumenera Skynyx 2.1 M a été utilisée en 12 bit, à la cadence de 15 images/seconde en acquisitions continues grâce au Logiciel Lucam Recorder qui permet d'effectuer des acquisitions en continu sur une durée illimitée. La seule limitation est la capacité de stockage des données sur le disque dur de l'ordinateur portable de type PC.
Cependant, lors de l'éclipse du 11 Juillet 2010, en raison du débit élevé de données via la liaison USB-2, et la saturation au début des contacts, la cadence d'acquisition d'images 12 bit qui était de 15 images/seconde avant l'éjection des filtres a été ralentie, et est passée à environ 10 images/seconde. Cette gestion du débit de données est assurée par le logiciel Lucam Recorder, ce qui est un avantage considérable, car il a permis de ne pas interrompre le processus d'acquisition d'images en continu durant toute l'éclipse. D'autre part ce logiciel de prise en charge de la caméra CCD Lumenera, présente un autre avantage notable:
le début des acquisitions est lancé au moins 30 minutes avant le début de la totalité ce qui permet d'obtenir les instants des contacts avant et après la totalité où se produisent les spectres éclairs qui durent environ 10 secondes. L'utilisateur n'a pas plus à s'occuper du moment de déclenchement des acquisitions de la caméra. Le répertoire contenant les fichiers de données n'occupe pas tout l'espace disque de l'ordinateur d'acquisition. En effet, lors des éclipses passées, les opérateurs avaient beaucoup de difficultés, même avec les éphémérides et calculs des circonstances du lieu, pour déterminer à quel moment il fallait déclencher les prises de vues pour photographier les spectres éclairs.
Le logiciel Lucam Recorder permet d'intégrer les chronodatations, et celles-ci sont indiquées sur les en-têtes de chaque image de spectre enregistré. C'est une amélioration importante dans le procédé d'acquisition. La chronodatation est très importante pour déterminer les instants des contacts lors des analyses des spectres éclairs sans fente. D'autre-part, les nouveaux ordinateurs portables (2007 et après) ont des capacités mémoires 300 Giga octets, vitesse d'écriture de 5400 tours/minute, pour enregistrer et stocker les données durant plus de 40 minutes sans interruption. Typiquement la taille des séquences des fichiers des spectres éclairs si situe autour de 30 à 50 Giga octets.

La mise en œuvre de cette expérience requiert un guidage manuel qui s'est révélé le plus adapté car chaque contact a une durée de 10 secondes, et les corrections des dérives à effectuer sont plus faciles à effectuer à la main durant les minutes précédant l'évènement. L'inconvénient d'une motorisation est qu'il faut quasiment re-pointer manuellement avant le troisième contact pour déplacer l'instrument d'un angle de 32 minutes d'arc afin d'anticiper et bien obtenir dans le champ l'intervalle spectral de $4700 \pm 50$Å et cette manipulation de guidage motorisé est trop risquée vue la durée limitée des phases de totalité. En effet, le repointage du spectre au troisième contact nécécite d'être effectué très rapidement durant la totalité. Si la motorisation est assurée, l'entraînement s'effectue à vitesse constante. L'action de retirer l'engrenage du moteur pour ensuite pointer manuellement prend du temps, et est une opération difficile, car des erreurs de manipulations sont probables. C'est pourquoi, le guidage manuel reste une méthode plus simple de mise en œuvre, et des amplitudes de mouvements différents peuvent être assurées pour repointer, ce qui simplifie les opérations.
L'intervalle spectral est maintenu dans le champ du CCD au moyen des filtres amovibles, durant toutes les phases partielles précédant la totalité, en effectuant un guidage sur la bande de spectre correspondant au croissant solaire atténué par les filtres, et dont il faut maintenir l'image sur le CCD. L'avantage du Logiciel Lucam recorder, est qu'il permet d'avoir une visualisation des spectres en continu sur l'écran du PC en permanence, tout en effectuant simultanément les acquisitions des spectres en continu. C'est un avantage très important car l'opérateur n'a plus à se soucier de l'acquisition des images et la gestion de la caméra CCD, et il suffit de suivre l'image des spectres à l'écran.
La figure II-1-5 indique les coefficients de transmission des filtres de densité neutres utilisés aux éclipses précédentes permettant d'assurer simultanément l'étalonnage des intensités du disque solaire et de guider le spectre durant les phases partielles précédant le début de la totalité de l'éclipse. Les régions 4686 Å et 4713 Å sont situées dans la partie des plus courtes longueurs d'onde de cette courbe de transmission.

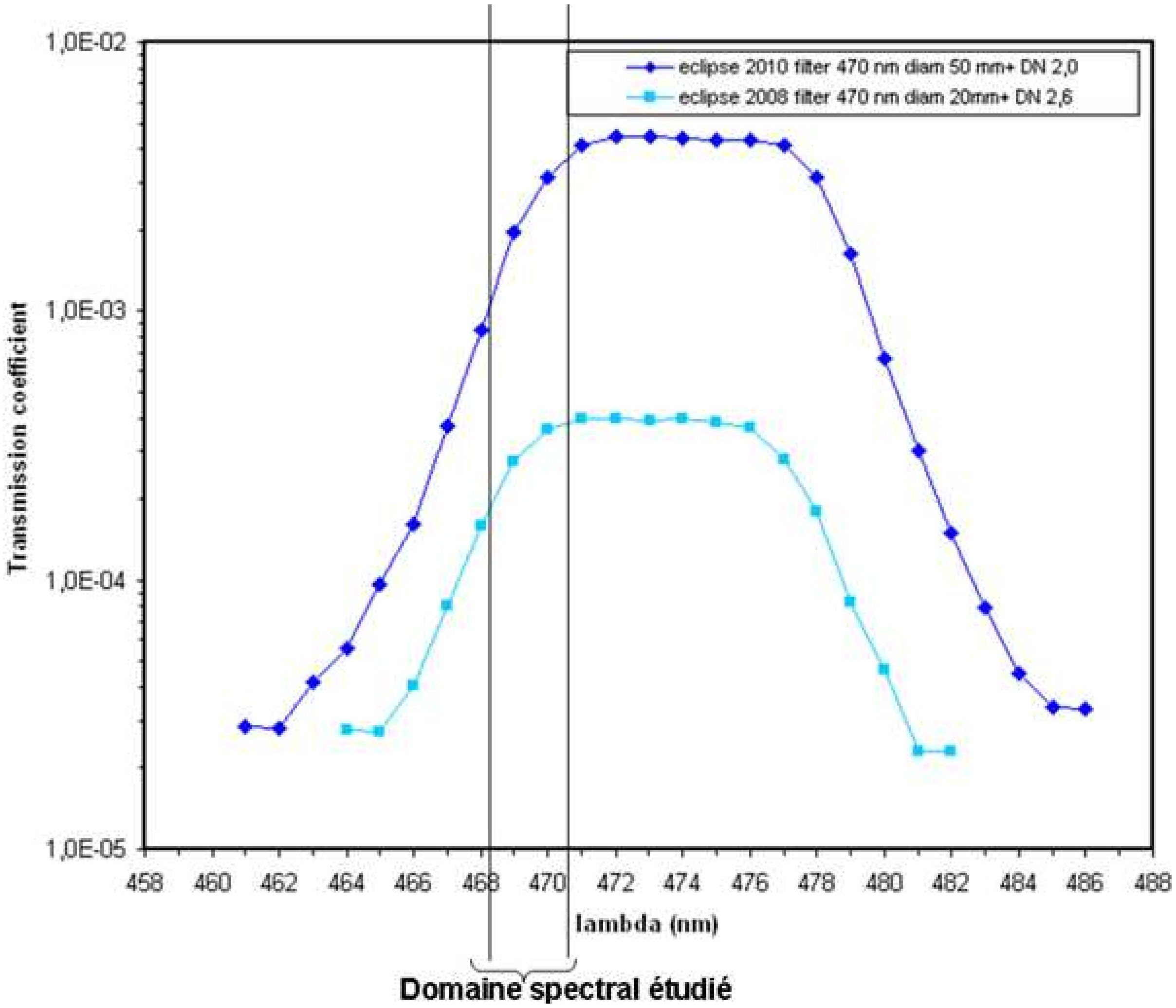

**Figure II-1-5:** *coefficients de transmission des filtres, en fonction de la longueur d'onde, utilisés aux éclipses de 2008 et 2010*

La gamme de longueurs d'onde 4680 à 4710Å était maintenue le plus possible en centre du champ de la caméra CCD, car en absence d'entraînement manuel, les spectres avaient une dérive des longueurs d'onde se déplaçant vers le rouge. En entraînant la molette, cela permet de recentrer le segment spectral en le déplaçant en direction de l'UV. Cette méthode permet d'ajuster le domaine spectral désiré quelques instants avant le second contact avant la totalité.
Une fois que le segment spectral est dans le champ de la caméra et affiché à l'écran, il faut effectuer un guidage manuel et des ajustements fins pour: i/ s'assurer que le sens de dispersion spectrale est bien parallèle à l'axe des pixels de la caméra CCD ii/ assurer les corrections des dérives du spectre au cours des phases partielles, en orientant le réseau par transmission, pour que le sens de dispersion reste perpendiculaire au croissant solaire quelques minutes avant et après la totalité.
Lorsque l'instant de contact approche, et selon l'épaisseur nuageuse (conditions météorologiques), il faut se préparer à l'éjection des filtres de densité neutre, tout en assurant le guidage du segment spectral dont l'intensité a fortement diminué à mesure que l'on s'approche du second contact.
L'appréciation de l'instant où les filtres sont éjectés, est liée à l'expérience de l'observateur, selon les circonstances du moment, compte-tenu de l'assombrissement ambiant, car l'ombre arrive très vite avant le début du second contact précédant la totalité. Les filtres sont ejectés quelques secondes avant lorsque tout s'assombrit, les images alors apparaissent totalement saturées à l'écran durant quelques secondes précédant l'instant du contact, puis quelques secondes ensuite les spectres éclair apparaissent et ensuite l'écran devient noir au début de la totalité où les raies d'émission les plus intenses ont disparu.
Ensuite durant la totalité, il faut continuer à assurer le guidage manuel de l'instrument, pour préparer et anticiper ensuite le troisième contact et se trouver aux bonnes longueurs d'onde. Cette opération est effectuée en connaissant la vitesse où il faut tourner la molette du flexible, (c'est-à-dire un quart de tour en une minute), et ce maniement a été préparé et répété à l'avance en comptant les secondes mentalement ou en écoutant la trotteuse d'un réveil contre l'oreille. Ensuite en ayant effectué ces corrections et guidages manuels nécéssaires, les spectres éclairs réapparaissent à nouveau, et s'intensifient. Quand les images deviennent trop saturées, les filtres de densité neutre associés aux filtres 4700 Å sont remis en place devant le réseau-objectif. Ces filtres amovibles ont été étalonnés au préalable, et permettent aussi d'exprimer les intensités des raies du spectre éclair en unités du disque solaire moyen, comme indiqué en Annexes 28 et 29. Grâce aux phases partielles, où les spectres de la photosphère sont atténués, les raies d'absorption telluriques et de Fraunhofer sont visibles, et permettent ces étalonnages.
Ensuite la caméra CCD est retirée de la bague située avant le foyer du réseau-objectif. Cette opération est effectuée en masquant le CCD de la lumière ambiante, et on insère devant le CCD un objectif de courte focale, 35 mm, pour filmer l'écran d'un GPS. Cela permet de voir les images de l'afficheur du GPS: heure, minutes et secondes dans une séquence, et sur chacune des images, la chronodatation est indiquée dans les en-têtes. Cela permet ensuite de convertir les datations des images en heure GPS.

L'expérience a bien fonctionné lors de l'éclipse de 2008, et les résultats obtenus ont permis d'obtenir des résultats quantitatifs, et effectuer des analyses photométriques. La linéarité de la réponse des caméras CCD utilisées a été mesurée, voir Annexe N°17.
Le procédé a été reproduit aux éclipses de 2009, 2010 mais n'a pas pu être reproduit à l'éclipse du 13 Novembre 2012 sur un site à Port Douglas en Australie (Queen's Land) à cause de nuages trop épais qui ont complétement masqué l'éclipse et aucun spectre éclair sans fente n'a pu être enregistré. Par ailleurs, j'avais ajouté un autre filtre de densité neutre intermédiaire de densité 1, afin de réduire la durée de saturation après l'éjection du filtre de densité neutre D = 2 durant les secondes précédant les spectres éclair. Cette manipulation n'a pas pu être effectuée à l'éclipse du 13 Novembre 2012, car des nuages trop épais et étendus ont complètement masqué l'éclipse, et aucun spectre éclair n'a pu être enregistré.
D'autres expériences effectuée lors de cette éclipse du 13 Novembre 2012 ont donné des résultats avec le spectrographe à fente, installé sur un autre site à environ 50 kilomètres plus au Sud-Ouest de Port Douglas, qui a bénéficié d'un ciel dégagé durant les contacts et la totalité.

## II-2) Spectrographe à fente utilisé en 2012

Un spectrographe miniature à fente a été construit pour l'éclipse du 13 Novembre 2012, afin d'étudier à résolution suffisante le profil de la raie verte du Fer XIV à 5302.86 Å dans la région de transition dans la basse couronne et la couronne moyenne.
Un intervalle spectral de 4650 Å à 5400 Å a été choisi pour étudier à la fois la raie verte, mais aussi une autre raie coronale, celle du Nickel XIII à 5140 Å, ainsi que certaines petites raies d'émission low FIP de l'interface.
Le schéma II-1-6 décrit le spectrographe utilisé par Serge Koutchmy et Jonathan Cirtain qui étaient installés sur un autre site « Howe Plantations » sur la bande de totalité de l'éclipse.

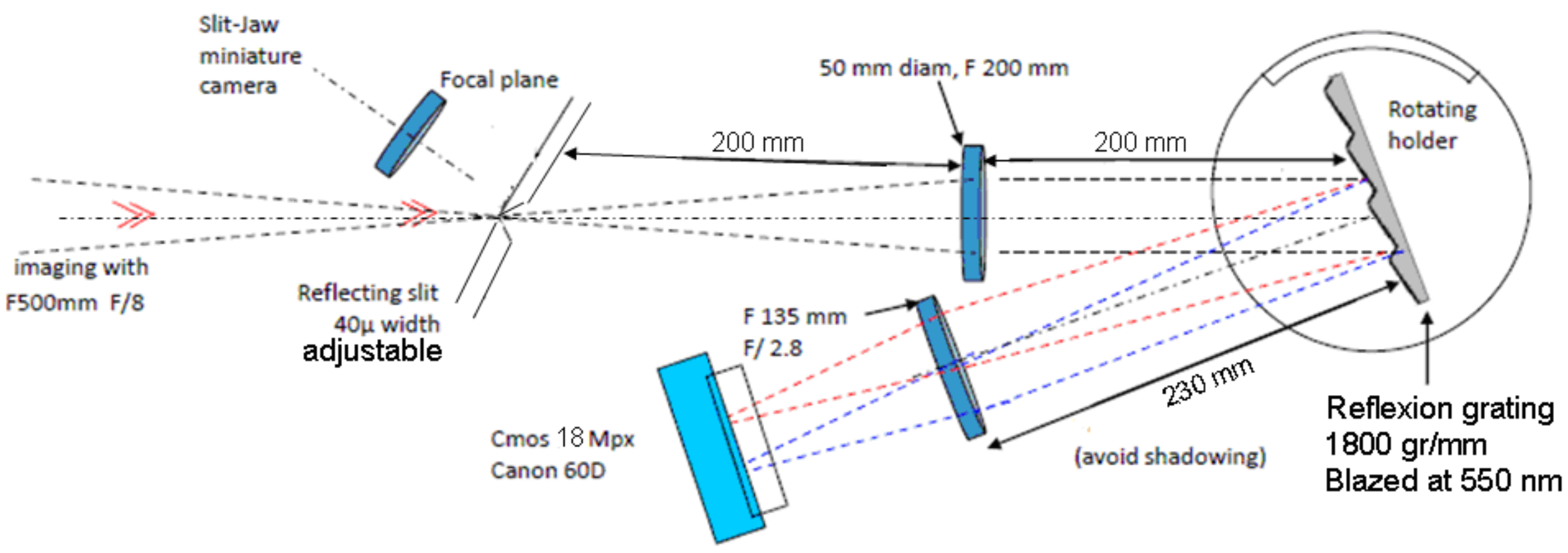

**Figure II-1-6:** *Schéma du spectrographe à fente ajustable utilisé à l'éclipse du 13 Novembre 2012 en Australie. Les acquisitions effectuées proches des contacts ont été de 10 ms. Les formats des images de l'appareil photo numérique images du Canon 60 D, chip CMos de 18 Mégapixels sont en 14 bit*

Des spectres éclair allant de la chromosphère à la couronne solaire ont été obtenus juste après le second contact, au moyen d'un spectrographe à fente. La réponse instrumentale a été sommairement évaluée, et une fente de 40 µm de large a été utilisée. La largeur des profils de raies affectée par la fonction instrumentale dépendait aussi de la mise au point, mais celle-ci était optimale pour le vert autour de 5300 ± 80 Å avec un profil gaussien. La FWHM instrumentale produisait un élargissement au plus égal à 1.0Å sur une raie du Fe I optiquement mince, produite par une lampe à basse pression au Fe I. Une autre mesure de fonction instrumentale a été tentée sur une raie tellurique (en absorption) aux mêmes longueurs d'onde que le Fe I avant la totalité, voir annexe 34 où des valeurs inférieures à 1Å ont été trouvées.

Les corrections de réponse instrumentale n'ont pas encore été effectuées dans toutes les analyses du profil de la raie verte et du continu étudiés au chapitre IV-4, car des étalonnages et vérifications « labo » sont encore nécessaires.
Une rafale d'images a été effectuée quelques secondes après le second contact, au début de la totalité, et dans cette séquence, les trois premières images ont été exploitées, car la durée totale de ces images était estimée inférieure à 0.7 s, le temps de pose étant de 1/100 s. Ceci représente plusieurs avantages, car le mouvement diurne n'a pas produit de « bougé » important sur les spectres, et le mouvement différentiel de la Lune sur le Soleil a été de 0.35 '', ce qui correspond à une variation d'altitude sur le limbe solaire de 760*0.35 = 220 km au total. L'échelle spatiale est de 1.9 Mm/pixel et la dispersion spectrale est de 0.177 Å/pixel. Les détails et méthodes d'obtention de ces paramètres sont décrits au chapitre III-5. L'objectif de cette méthode de spectrographie à fente est d'apporter une analyse quantitative et photométrique des spectres. Ces analyses spectroscopiques nécéssitent des images en lumière blanche réalisées simultanément durant les phases de totalité.

## II-3) Les images en lumière blanche aux éclipses totales de Soleil, un complément indispensable.

Les images en lumière blanche obtenues pendant la totalité des éclipses de Soleil permettent d'observer les structures fines associées à la diffusion électronique dans couronne solaire. Ce type d'observation est très important pour le diagnostic des densités, en analysant des profils d'intensité radiales. Un exemple d'image de la couronne solaire en lumière blanche obtenue à l'éclipse totale du 11 Juillet 2010 est présentée en figure II-1-7, où les structurations dans la couronne solaire ont été renforcées pour améliorer leur visibilité.

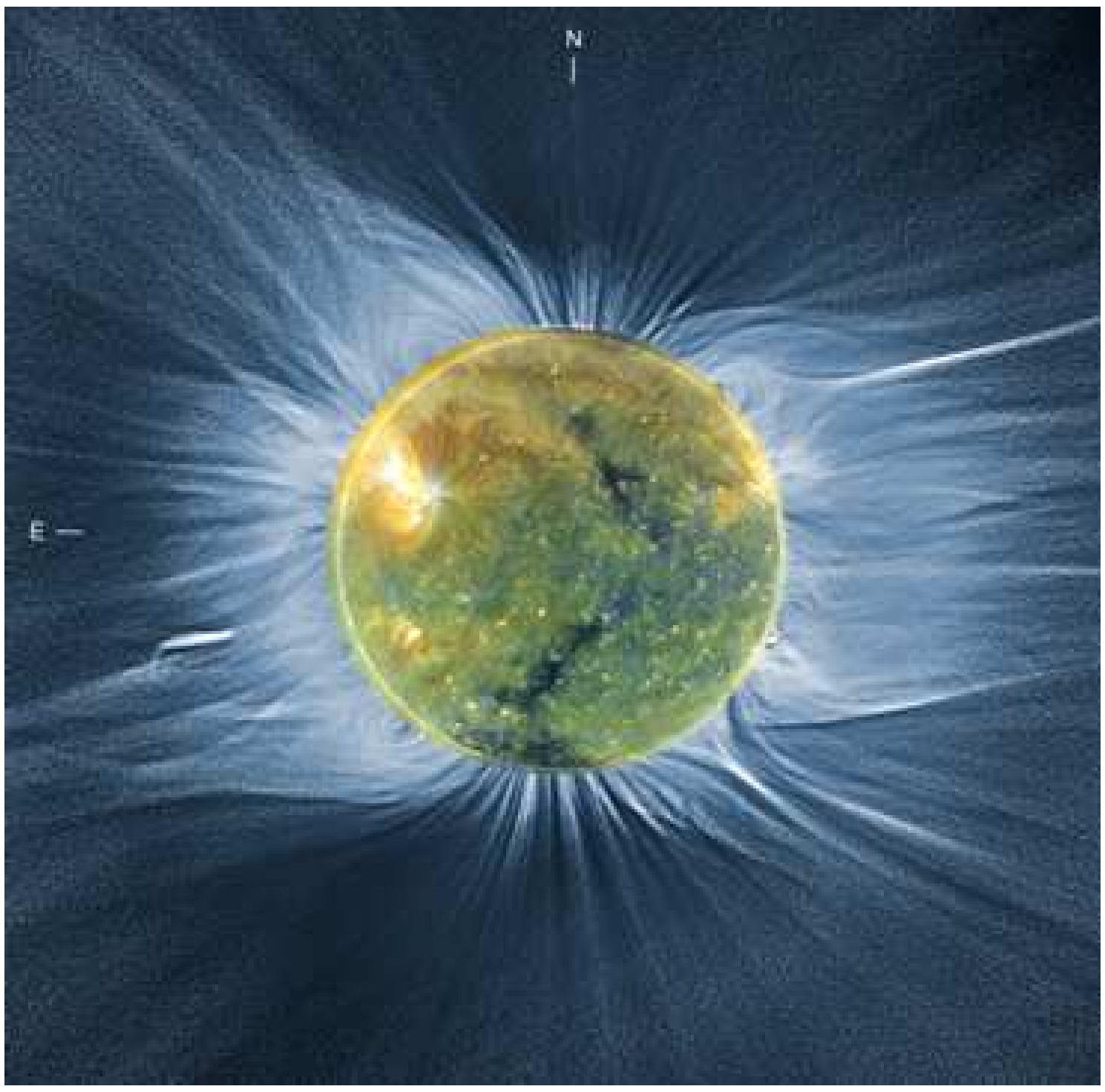

**Figure II-3-1:** *Image en lumière blanche de la couronne réalisée par Jean Mouette à l'éclipse totale du 10 Juillet 2010 à Hao en Polynésie Française. Superposition du disque des images 171Å, 193Å de AIA/SDO.*

Ces images en lumière blanche, révélant les inhomogénéités de densité électronique, ont été comparées avec les images en EUV de SDO/AIA pour l’analyse des protubérances et cavités coronales, et analyser des profils d’intensité radiales, en faisant l’hypothèse de températures hydrostatiques, voir Annexe 1.
Dans le cadre de cette thèse, des profils photométriques d’intensité *I(h)* ont été réalisés sur les spectres à fente de l’éclipse du 13 Novembre 2012, sur la raie du Fer XIV, voir chapitre IV-4. Le but est d’analyser les gradients du continu chromosphérique et coronal.
Le tableau II-3-2 résume les observations spectrales réalisées durant 5 éclipses totales de Soleil, qui sont analysées et discutées aux chapitres III et IV:

| Dates des éclipses totales | Lieu d'obser-vation Coordonnées altitude | Heure précise des contacts T.U. | Domaine spectral (nm), et Disp-ersion | Nombre de raies Identifiées et intensités en unités du disque solaire | Type de spectrographe utilisé et imageur | Type de caméra CCD, cadence et logiciel d'acquisition utilisé Nbre de pixels | Qualité du ciel ; hauteur du Soleil en degrés (ciel) | Remarques |
|---|---|---|---|---|---|---|---|---|
| 29 Mars 2006 | Side, Turquie 36°45' 52" N 31°23'55" E Alt 2m | C2 : 12h57' 11.1s C3 : 12h59' 34.6s | 540- 430 Chromo couronne 1.3 A/px | 25 $5*10^{-4}$ à $10^{-6}$ | Réseau-objectif 600 tr/mm par reflexion diam 5 mm et F=100 mm | Camescope CCD TRV 6 mini DV pixel = 5.4µ 300 K pixels | Très bonne + 53.7° | 2nd contact; 1ère tentative qualitative Acquisition 8bit |
| 1er Août 2008 | Ongaday, Siberie 50.7463° N 86.1355 E Alt : 836 m | C2 : 17h53' 12.9s C3 : 17h55' 22.8s | 460-475 Chromo 0.12 A/px | 22 $10^{-3}$ à $5*10^{-7}$ | Réseau-objectif 600 tr/mm par transmission diam 50 mm; F=600 mm | CCD video Watec 120 N+ Camescope mini DV pixel = 9µ 700 K pixels | Bonne Au 2nd contact couvert au 3ème contact + 27.5° | Mise en évidence de la seconde enveloppe d'hélium ionisé He II 468.6 nm α P Acquisition 8bit |
| 22 Juillet 2009 | Tianhuanping, Chine 30°28.1' N 119° 35.4' E Alt : 890 m | C2 : 1h33' 01.7s C3 : 1h38' 39.6s | 435-455 Chromo & Protu-bérances boucles 0.12 A/px | 46 $10^{-3}$ à $5*10^{-6}$ | Réseau-objectif 600 tr/mm par transmission diam 50 mm; F=600 mm | CCD Skynyx Lumenera Iris pixel = 4.6µ 1.2 M pixels | Voilé +54.3° | Acquisition en 8 bit Boucle-protu He I 4471 Macro-spicules, He I 4388 BaII, TiII… |
| 11 Juillet 2010 | Hao, Polynésie Française 18°06.18' S 140° 54.358 O Alt : 1m | C2 : 18h41' 33.1s C3 : 18h45' 02.6s | a)460-475 Chromo & Protus 0.12 A/px 1.1Mm/px b) 460-600 0.12 A/px | 22 $10^{-2}$ à $10^{-6}$ 44 $10^{-2}$ à $10^{-6}$ | a) Réseau-objectif 600 tr/mm par transmission diam 50 mm et F=600 mm b)Réseau-objectif 1200 tr/mm par transmission diam 30 mm et F=300 mm | a) CCD Skynyx Lumenera Lucam-Recorder pixel = 4.6µ 1.2 M pixels b) Canon 40D Pixel = 6µ 4 M pixels | Voilé +32.9° | a)Acquisition en 12 bit Protubérances He II 4686 Enveloppes Hélium, continu b)Acquisition 14 bit Continu coronal, region He I 4471 et H I 4340 |
| 13 Nov. 2012 | Howe, Australie, Queen's Land 17° 5.518' S 145° 24.812' E 556 m | C2: 20h39' 0.1s C3 : 20h40' 23.8s | 465-545 Chromo & Couronne 0.17 A/px | 28 $10^{-3}$ à $5*10^{-7}$ | Spectrographe à fente 40µm type Littrow, 1800 tr/mm diam 50 mm et F = 500 mm | Canon Cmos 60 D pixel = 6µ 5 M pixels | Bonne +13.7° | Acquisition 14 bit |

**Tableau II-3-2 :** *description et présentation des conditions des observations des précédentes éclipses totales de Soleil et les principaux objectifs objectifs atteints.*

Les images dans les spectres éclairs sont comparées aux images obtenues depuis l'espace dans les UV et EUV, pour réaliser une étude plus complète des interfaces photosphère-chromosphère/couronne, depuis les régions les plus proches du limbe solaire.
La particularité de ce travail de thèse en physique solaire, est d'avoir effectué des observations sur le bord du Soleil dans des altitudes très basses, accessibles seulement grâce aux éclipses totales, ce qui permet de résoudre et analyser les altitudes dans l'atmosphère solaire.

## II-4) Observations spatiales modernes du limbe solaire à partir d'images EUV avec les missions spatiales TRACE et Hinode

Les observations spatiales sont utilisées dans le but de comparer la structuration des couches de l'atmosphère solaire examinées dans plusieures raies d'émission correspondant à des températures d'ionisation, comme le calcium une fois ionisé, raie H du Ca II et la raie du carbone trois fois ionisé CIV. Les couches de l'atmosphère solaire proche du limbe présentent des embrillancements. Les altitudes où se produisent les maxima d'émission sont comparées aux régions où les petites raies d'émission low FIP observées aux éclipses sont produites, ainsi que les altitudes où apparaissent les raies de l'hélium neutre et une fois ionisé qui sont des raies plus sensibles à la couronne, et se présentent sous forme d'enveloppes, voir chapitres IV-1, IV-5. Les modèles à une dimension 1-D ne sont plus valables aux altitudes supérieures à 1000 km, à cause de la présence de structures fines et dynamiques comme les spicules, macrospicules, jets. Ces altitudes supérieures à 1000 km correspondent aux embrillancements observés, voir figure II-4-1.
Des régions inhomogènes liées aux spicules analysées grâce aux spectres flash sont préséntés dans cette thèse, et avec une meilleure résolution grâce aux images de la mission spatiale TRACE.
Ces travaux ont servi aussi à analyser la structure de la chromosphère, de compléter des analyses de l'ovalisation de la chromosphère (voir thèse J. Vilinga, et Filippov, Koutchmy, Vilinga 2007), et de comprendre à quelles altitudes ces enveloppes se forment, quels processus pourraient être responsables de l'ionisation de l'hélium comme high FIP, dans les raies optiquement minces, observées dans l'interface entre la photosphère et la chromosphère. L'analyse des embrillancements du limbe solaire dans les EUV est donc très importante pour comparer avec l'étude des enveloppes d'hélium dans les raies visibles, et aussi la contribution et le rôle des spicules pour définir les couches profondes de l'atmosphère solaire, entre 500 et 2000 kilomètres au dessus du bord solaire. Les figures II-4-1 à II-4-13 décrivent les profils d'intensité radiales du limbe solaire pour les raies Ly $\alpha$ et H du Ca II.

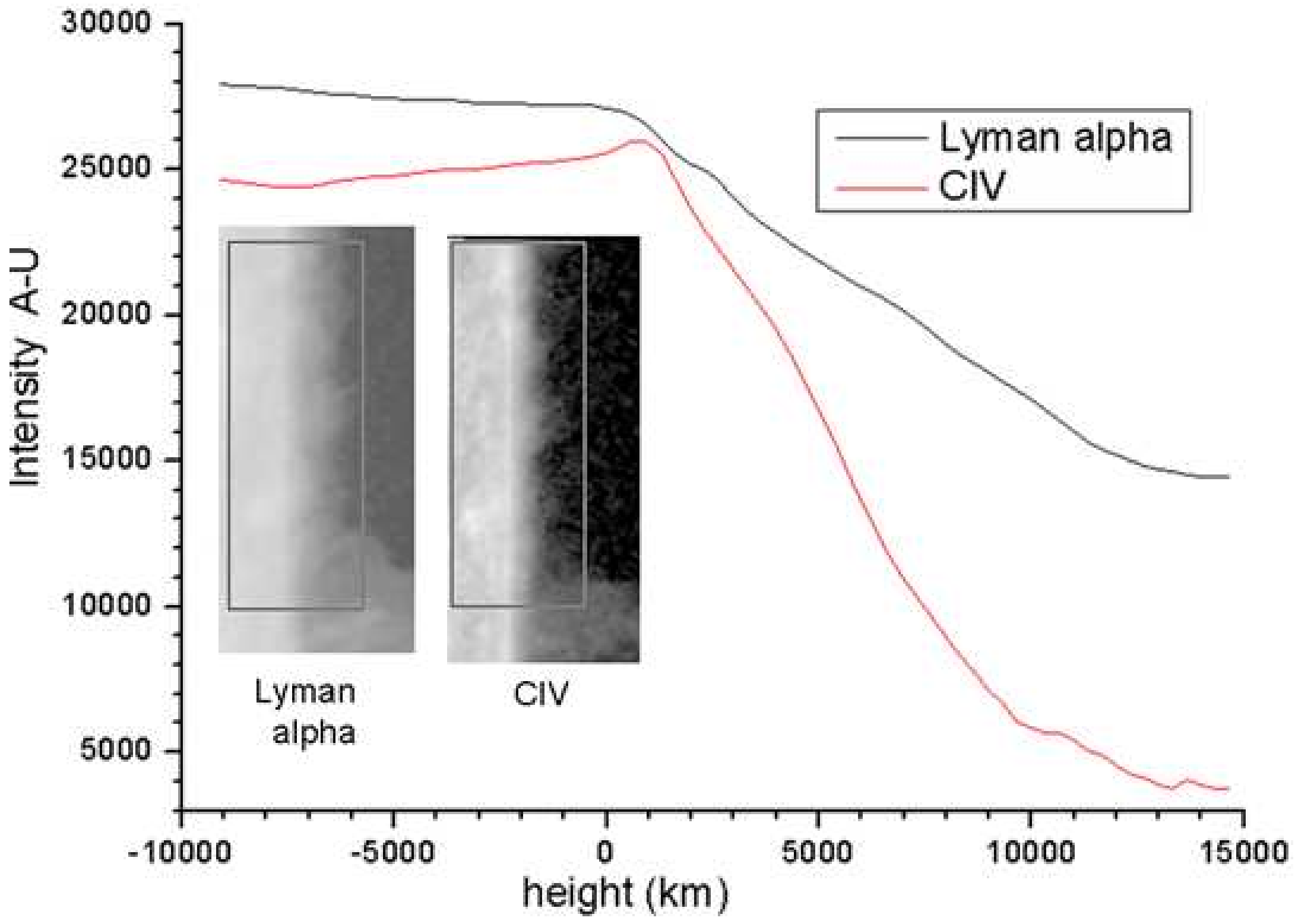

**Figure II-4-1:** *Aspect des profils des limbes avec TRACE en Lyman α et dans la raie du carbone CIV dans une région calme le 1^er Août 2008. Noter que sur le disque le signal en Lyα et en CIV n'est pas pur à cause de la fonction de transmission des filtres qui inclue du rayonnement provenant de nombreuses raies d'émission UV aux voisinage de CIV et Lyα, voir Annexe 10, figure A-10-2 pour Lyα.*

L'embrillancement du limbe est bien visible dans la raie du C IV à 1548.2 Å tandis qu'il n'est pas visible en Lyα à 1210 Å. Cette raie du C IV a une température électronique de 100000 K, voir annexe 33, et Annexe 10 sont indiquées des altitudes auxquelles cette raie est formée dans la chromosphère comme les raies O III et O IV, mais Lyα est une raie plus froide comme He II 304 Å ayant une température voisine de 30000 K.
Nous avons complété cette étude avec d'autres observations par imagerie avec le Solar Optical Telescope, (SOT) de Hinode dans la raie H du Ca II à 3970Å qui est une raie « froide » (autour de 10000 K) et la résolution spatiale est de l'ordre de 85 km/pixel sur le limbe solaire. Les spicules sont bien résolues, mais avec nos spectres éclairs (2008, 2009, 2010 et 2012), ils n'étaient pas résolus dans les raies « low FIP » et ni la raie de l'hélium neutre He I 4471Å. Cependant nous avions pu observer les macrospicules notamment lors des contacts juste après la totalité de l'éclipse du 22 Juillet 2009 dans la raie intense de l'hélium neutre He I 4471Å qui a été analysée en détails au chapitre IV-1.
Les régions suivantes sont extraites d'une image de Hinode dans H du Ca II prise le 10 Novembre 2007, après avoir sommé 13 images, et linéarisé, afin de réaliser des profils d'intensités dans cette raie avec un rapport signal/bruit suffisant. Cette date du 10 Novembre 2007 ne correspond pas aux dates des éclipses totales observées, mais s'inscrit dans une longue période de Soleil « calme ». Il est à noter que les éclipses totales de 2008, 2009, 2010 analysées dans cette thèse ont eu lieu aussi en période de minimum d'activité solaire. L'éclipse du 13 Novembre 2012 a eu lieu au début de reprise d'activité solaire. Les zones avec les profils d'intensités radiales sont indiquées, selon les régions avec spicules ou en dehors des spicules.

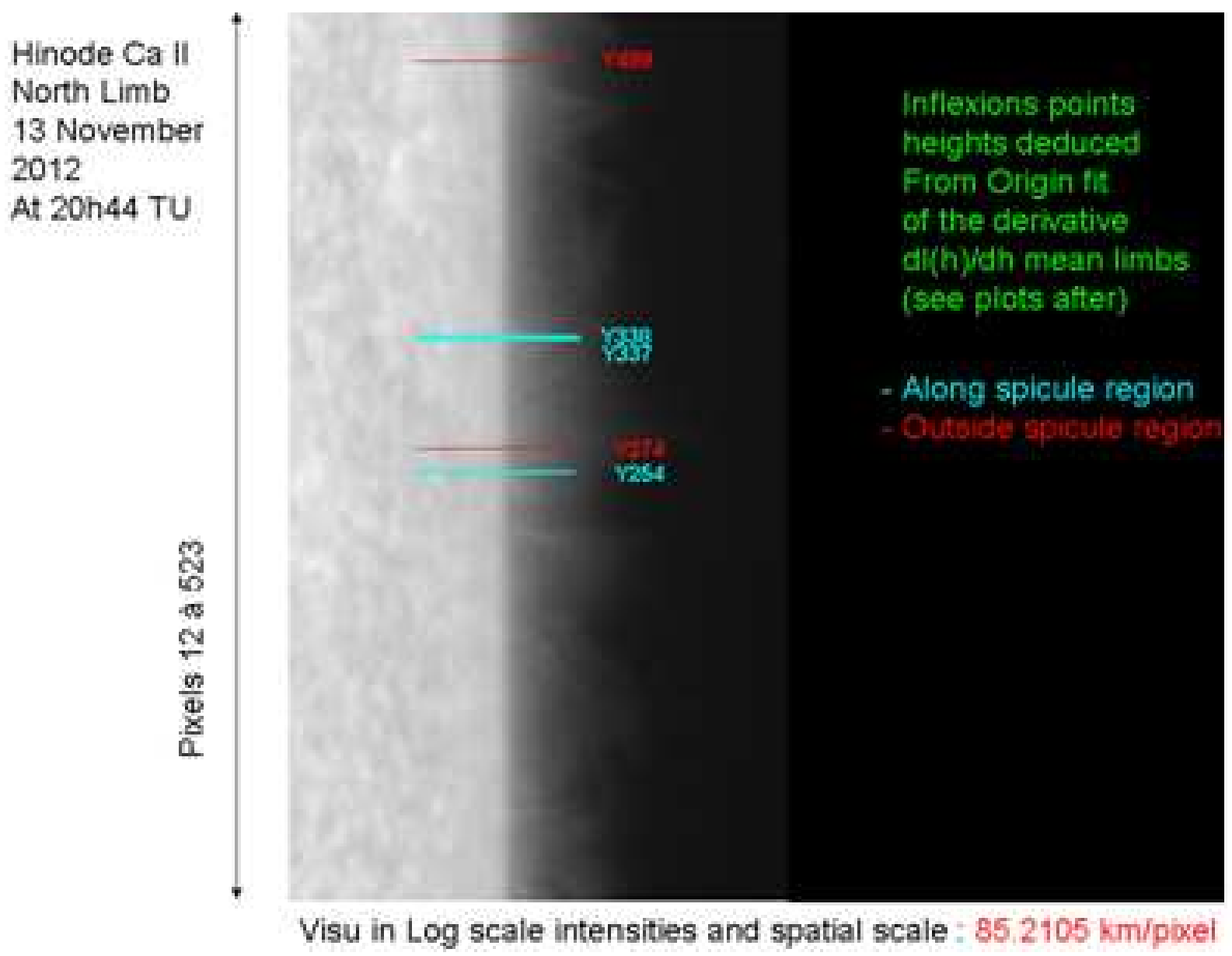

**Figure II-4-2:** *Image Hinode dans la raie H du Ca II redressée le long et en dehors des spicules du 13/11/2012.*

L’image figure II-4-3 est le résultat du renforcement de l’image figure II-4-2, afin de rehausser la visibilité des spicules:

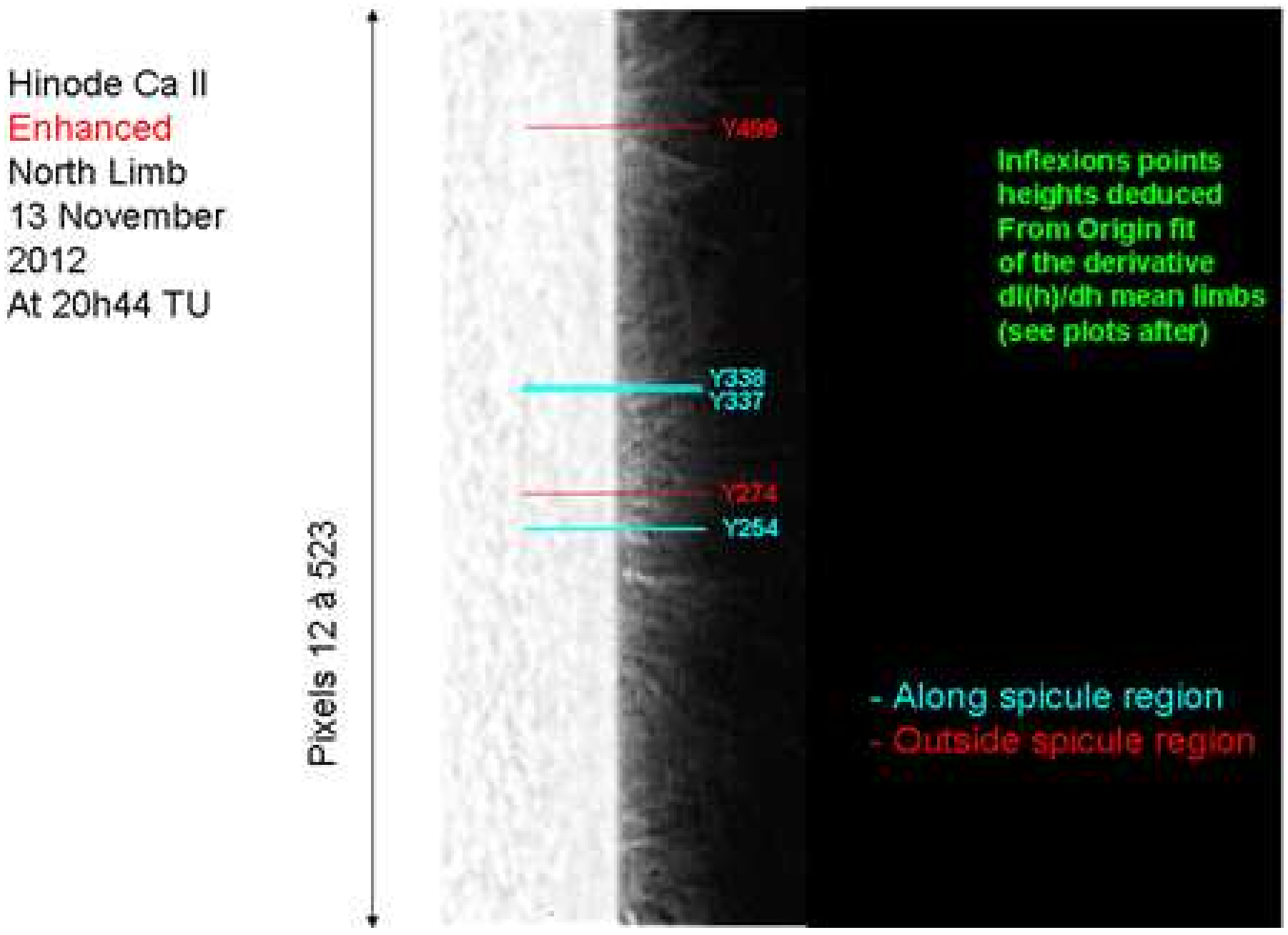

**Figure II-4-3:** *limbe Hinode dans la raie H du Ca II du 13/11/2012 redressé après traitement pour renforcer les contrastes. Les spicules inhomogènes sont visibles.*

Les profils d’intensité en fonction de l’altitude ont été obtenus d’après les images Hinode en H du Ca II redressées mais non renforcées selon les régions en dehors ou le long des spicules:

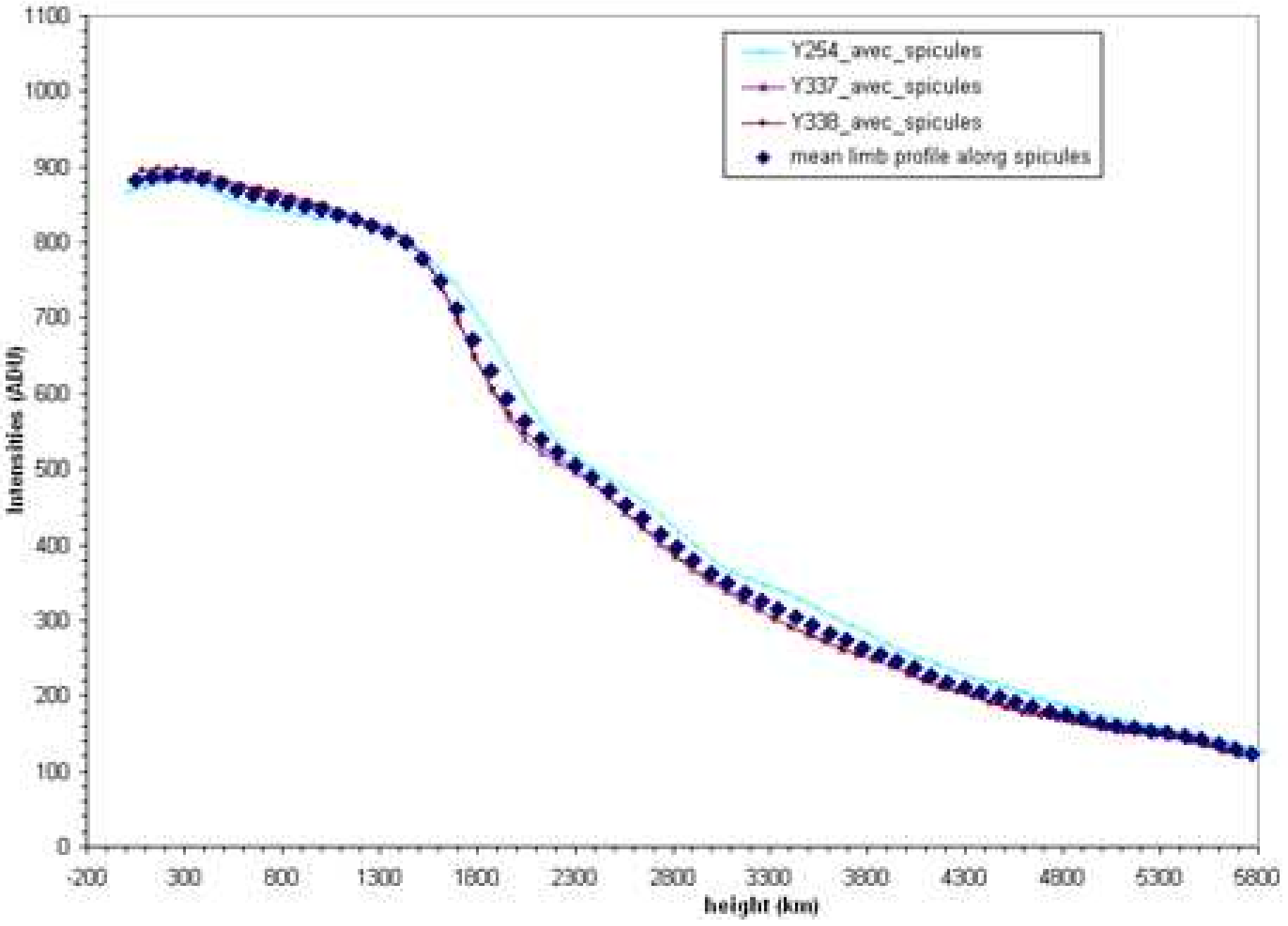

**Figure II-4-4:** *Profils d'intensité radiale le long des spicules inclinés réalisés sur 3 zones et moyenne de ces profils du 13/11/2012. Image dans la raie H du Ca II.*

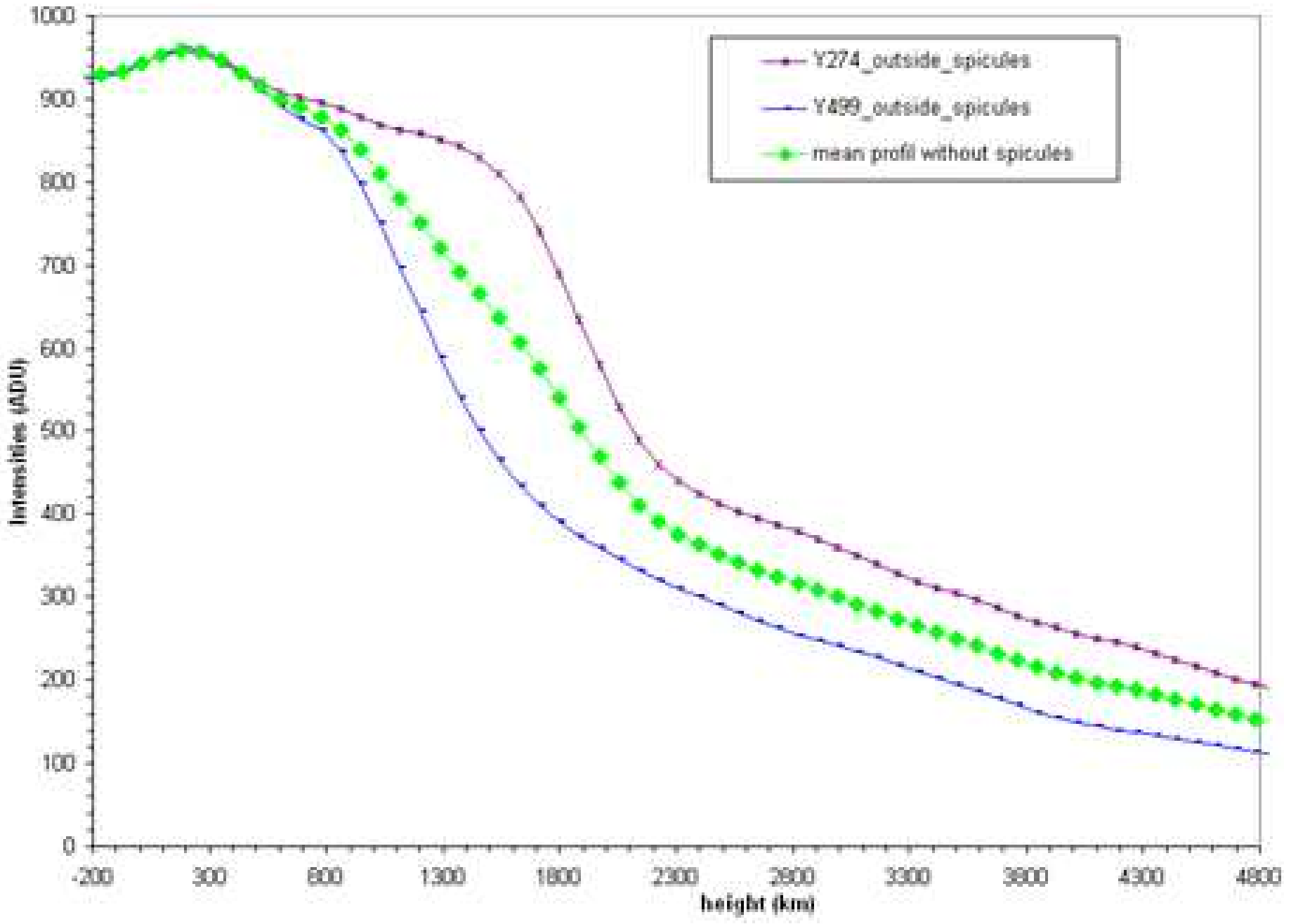

**Figure II-4-5:** *Profils d'intensité en dehors des spicules réalisés sur 2 zones bien choisies et la moyenne de ces profils, du 13/11/2012*

Les profils d'intensité en figures II-4-4 et II-4-5 présentent un embrillancement à 200 km étendu sur ± 200 km selon les régions le long ou en dehors des spicules, où ont été pris les profils radiaux. Le profil noté Y499 figure II-4-5 est plutôt caractéristique d'un profil sans spicules, tandis que le profil Y 274 peut être affecté par des spicules en arrière plan du disque. Le graphique figure II-4-6 est le résultat de la dérivée par rapport aux altitudes, - *dI/dh* du profil moyen pris en dehors des spicules en vue de déduire des profils d'émissivités des couches :

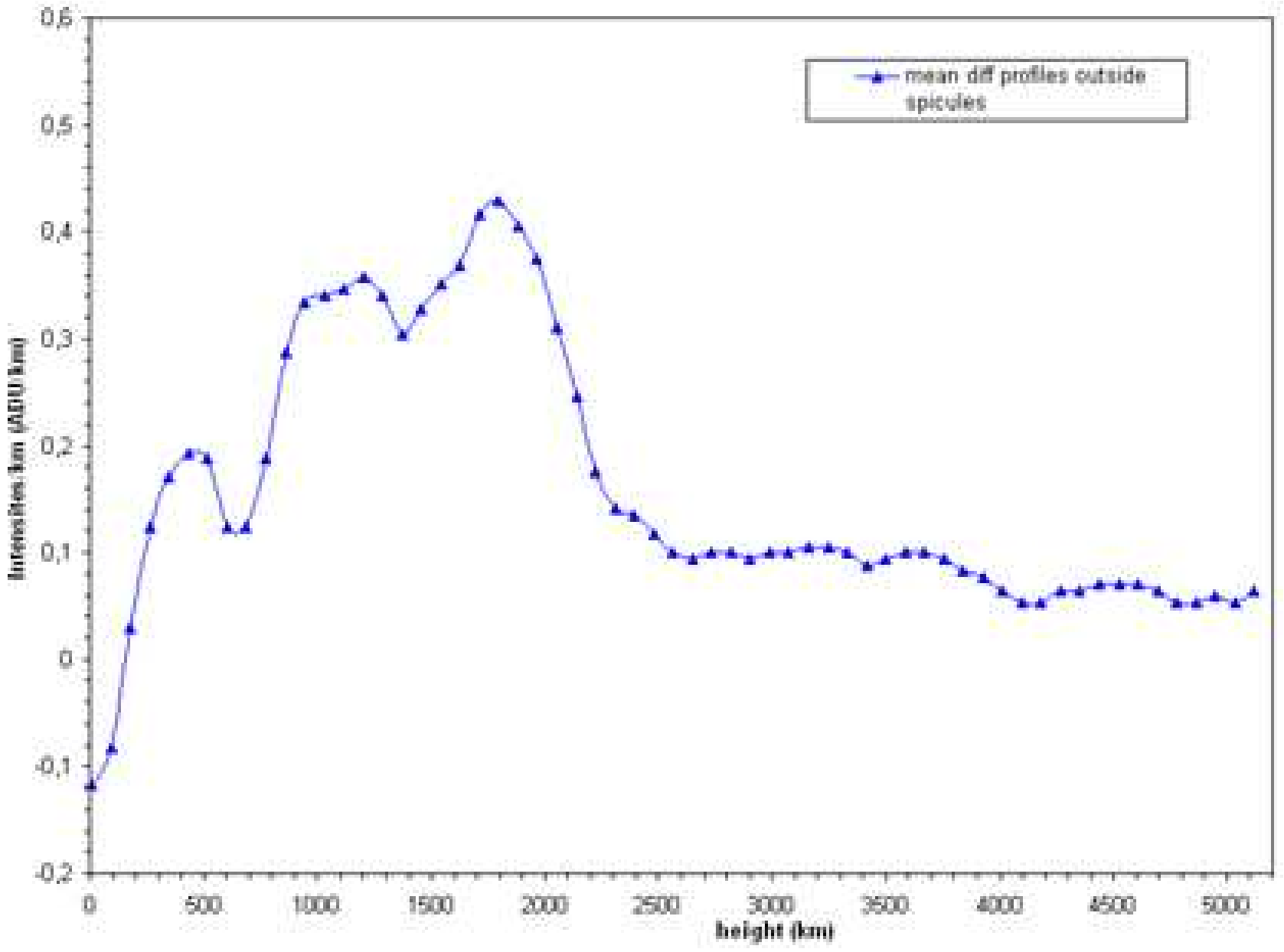

**Figure II-4-6:** *dérivée première des intensités par rapport aux altitudes des profils d'intensité pris en dehors des spicules, du 13/11/2012 d'après le profil moyen en figure II-4-5*

Le premier pic du graphique figure II-4-6 indique un maximum de brillance situé à des altitudes autour de 400 à 450 km, avec une largeur assez étendue sur ± 100 km. Il correspond relativement au limbe solaire à 380 km mesuré par Lites et al 1983. Cependant cet élargissement mesuré peut être toutefois lié à des spicules ayant des intensités plus faibles qui ont été prises en compte et ont apporté un embrillancement supplémentaire dans les profils d'intensité. Les autres modulations avec des maxima vers 1000 km puis vers 1700 km correspondent à la couche chromosphérique située au dessus du limbe.
Le graphique II-4-7 est le résultat de la dérivée par rapport aux altitudes du profil d'intensité radiale pris dans les spicules:

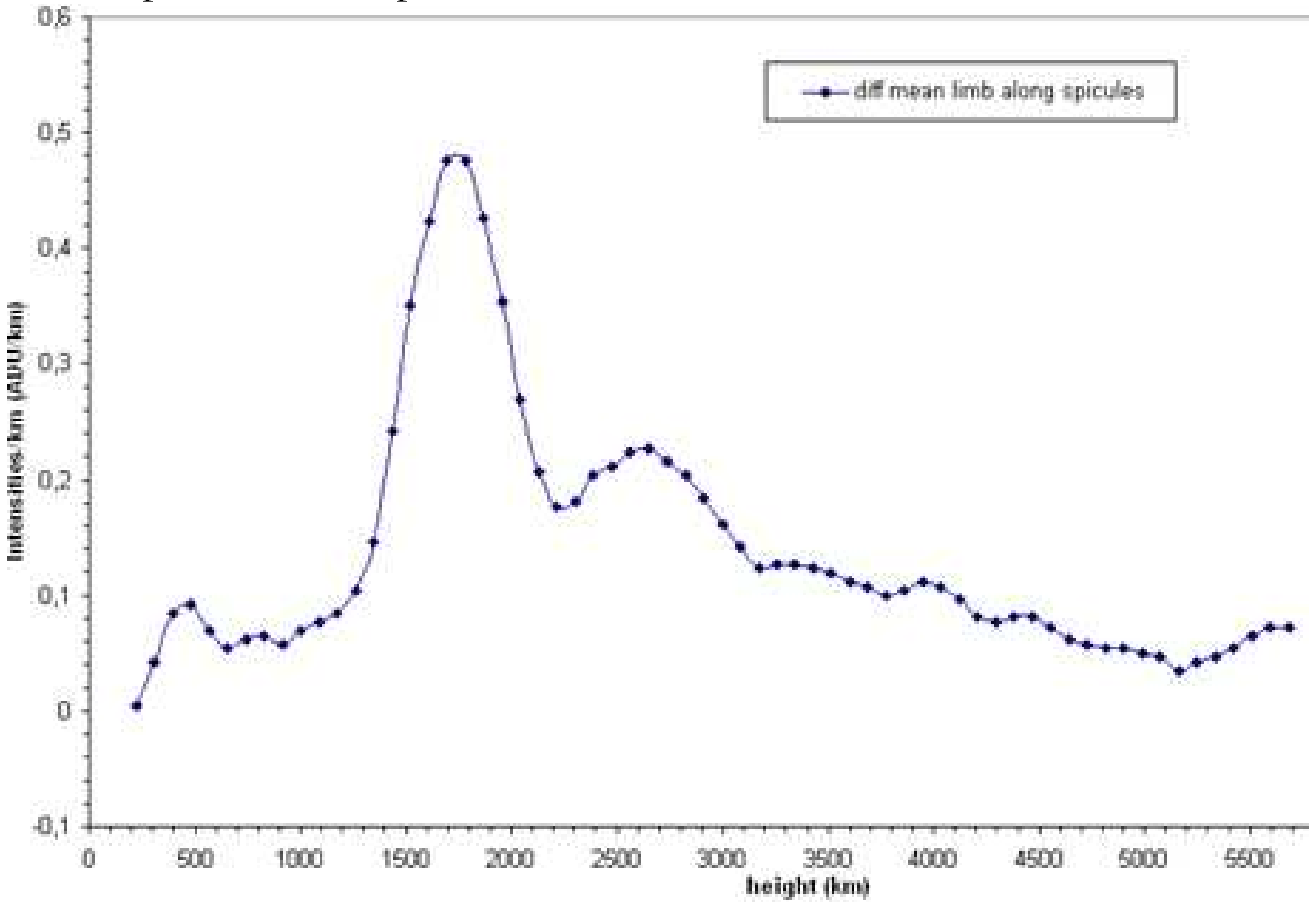

**Figure II-4-7:** *dérivée première des profils d'intensité pris le long des spicules, du 13/11/2012 d'après le profil moyen donné en figure II-4-4*

La position du limbe est mesurée à 400 $\pm$100 km avec les profils pris en dehors des spicules et cette position devient 450 $\pm$100 km avec des profils pris dans les spicules, ce qui est plutôt en assez bon accord avec les résultats obtenus par Lites et al malgré la dispersion des mesures. Les mesures avec le SOT (Hinode) sont plus précises sur les images car la résolution avec SOT (Hinode) est de 85 km/pixel, c'est-à-dire 4 fois meilleure qu'avec TRACE en C IV.

Ces résultats montrent que le limbe s'étend jusqu'à 2000 km sans tenir compte des spicules, ce qui correspond aux descriptions des modèles VAL, avant la brusque remontée de température au-delà de 2000 km entre la chromosphère et la couronne. Lorsque l'on considère et prend en compte les spicules, le modèle VAL n'est plus adapté.

Des images de Hinode à plus haute cadence ont été effectuées le 10 Novembre 2007. Il a été nécéssaire d'analyser les limbes plus en détails et d'effectuer des renforcements comme le montre l'image extraite figure II-4-8, où 25 images ont été sommées sur une durée de 2 minutes durant lesquelles les spicules ont très peu varié.

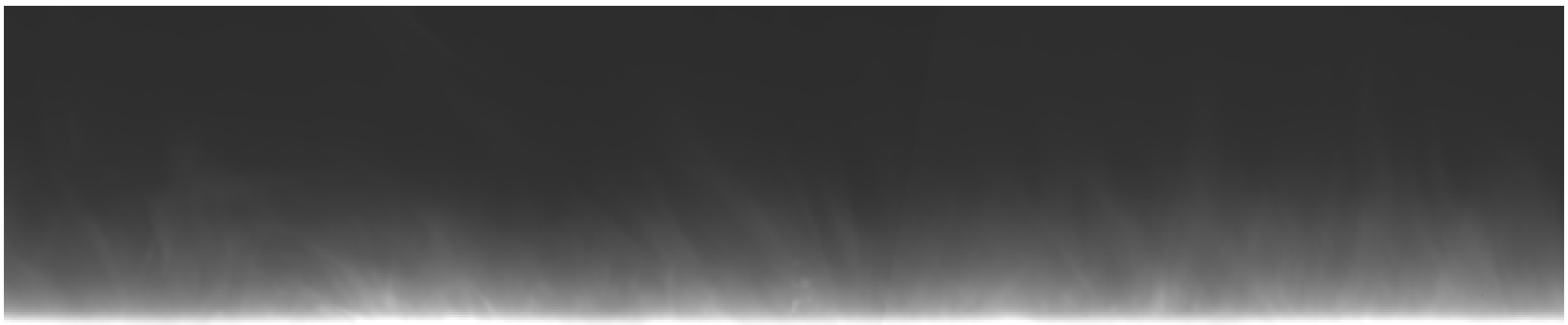

**Figure II-4-8:** *zone sélectionnée du limbe au pôle solaire sur l'image dans la raie H du Ca II par le Télescope Solaire Optique (SOT) de Hinode du 10 Novembre 2007 où 25 images ont été sommées sur 2 minutes d'observation.*

Le choix de cet intervalle de temps et cadence courts a permis de sommer suffisamment d'images, sans que les spicules ne se soient trop modifiés, afin de mieux étudier ensuite leur effet sur l'embrillancement du limbe. La figure II-4-9 a été obtenue à partir du masque flou depuis le logiciel Iris  Christian Buil- Centre National d'Etudes Spatiales - CNES:

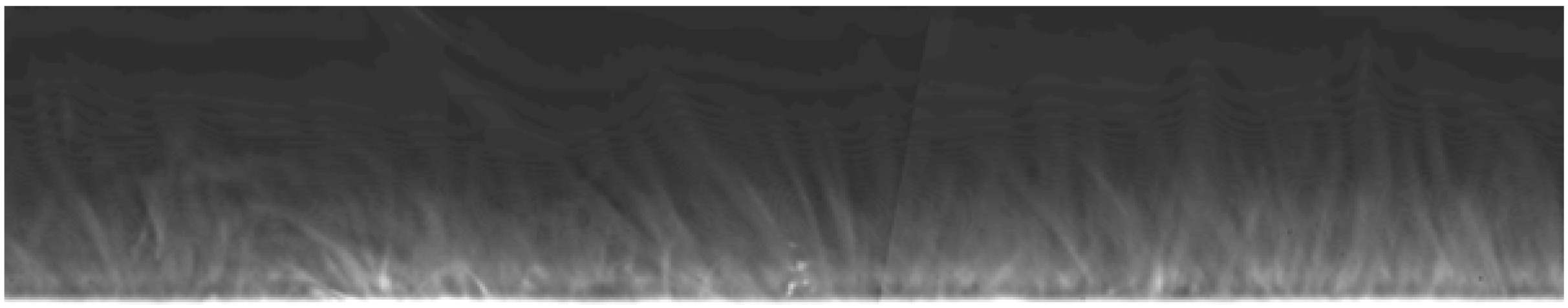

**Figure II-4-9:** *zone extraite du limbe au pôle solaire sur lequel un renforcement avec un masque flou réalisé sur 4.7 pixels a été effectué, afin de montrer certains spicules vus sur le disque, d'autres en arrière plan et d'autres en avant plan sur le limbe. Une légère absorption sur la ligne de visée, est visible juste au dessus du limbe (sur-exposé).*

Après ce premier traitement de masque flou avec le logiciel IRIS, les spicules sont renforcés, ce qui permet de mieux discerner leur base très proche du limbe; on distingue ceux qui sont en arrière plan et en avant plan et cela est important pour définir les altitudes où se forment les spicules, où l'on commence à voir apparaître leurs « pieds ».

Un second traitement au masque flou avec les mêmes paramètres dans le logiciel IRIS a été appliqué sur l'image figure II-4-9, afin de renforcer encore d'avantage les spicules à leur base et de révéler plus précisément leur structuration et l'image figure II-4-10 a été obtenue:

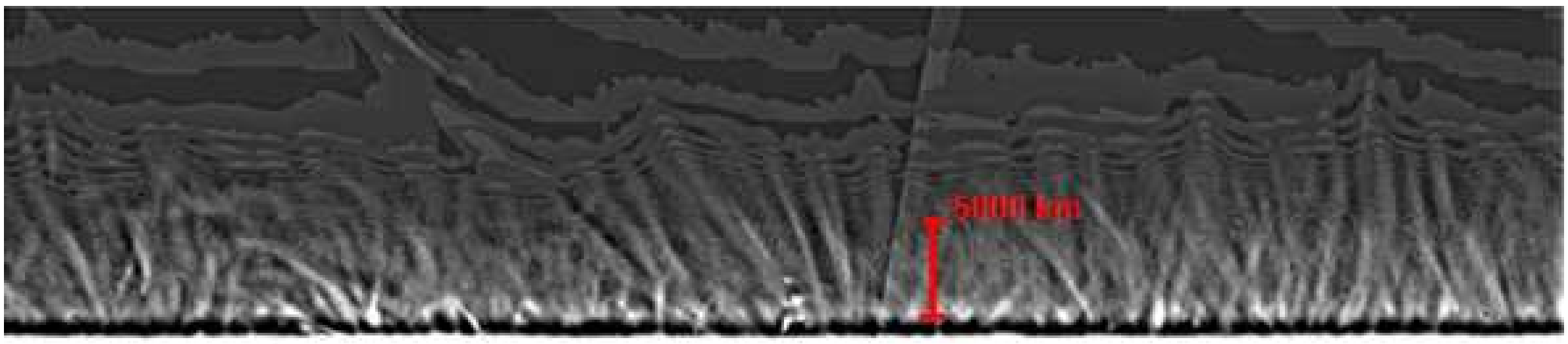

**Figure II-4-10:** *zone extraite du limbe au pôle solaire sur lequel des renforcements avec 2 masques flous consécutifs (réalisés l'un après l'autre) sur 4.7 pixels ont été effectués, afin de montrer certains spicules vus sur le disque, d'autres en arrière plan et d'autres en avant plan sur le limbe. La zone noire qui correspond à une légère absorption sur la ligne de visée, et qui a été renforcée après le second traitement mais peut être aussi affecté par des effets de pixels d'intensités négatives produits par le traitement. L'échelle est indiquée en liaison avec l'Annexe 7 des profils du limbe* $H\alpha$ .

Ce type de procédé de renforcement a permis d'obtenir des résultats assez comparables à ceux obtenus avec l'algorithme Madmax plus sophistiqué (O. Koutchmy, 1988) et utilisé par Tavabi et al 2012. Ces images avec le renforcement des spicules montrent que ces structures sont inhomogènes, turbulentes, et qu'elles n'occupent pas tout l'espace et qu'une région entre les spicules apparaît. D'autres analyses de profils radiaux sur les extraits d'images de Hinode dans H du Ca II ont été effectuées d'après les images en figure II-4-11, pour rechercher plus précisément la position du limbe solaire en tenant compte des spicules, à partir de profils d'intensités radiales effectuées sur les images brutes, où le courant d'obscurité, offset ont été soustraits, ainsi que la réduction du « flat field ».

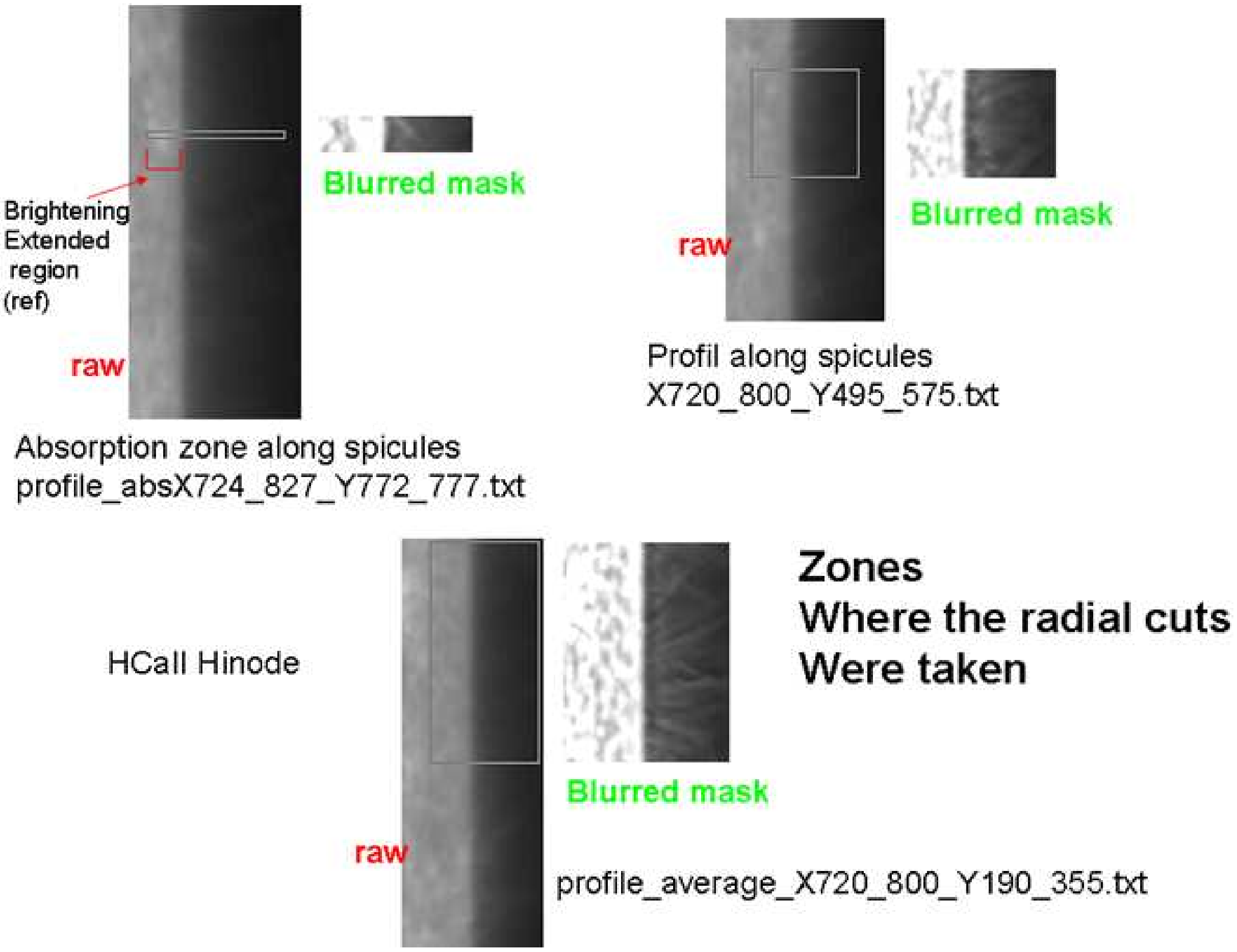

**Figure II-4-11:** *Extraits des zones d'où ont été effectuées des coupes photométriques sur les images du SOT de Hinode du 10 Novembre 2007, dans la raie H du Ca II en tenant compte de la localisation des spicules afin d'identifier le limbe sur lequel sont superposés les spicules. L'échelle est 1 pixel = 85.2 km. La région étudiée est située au pôle solaire et a été redressée.*

Les profils radiaux, figure II-4-12, ont été obtenus après intégration et moyenne sur 165 lignes de pixels, des limbes précédents redressés. Les positions du limbe, profil moyen, spicules et macrospicules sont indiqués. Ces analyses servent à montrer les contributions du vrai limbe solaire et les altitudes à partir desquelles les spicules dominent, en ayant intégré suffisamment d'images pour obtenir un bon rapport signal sur bruit.

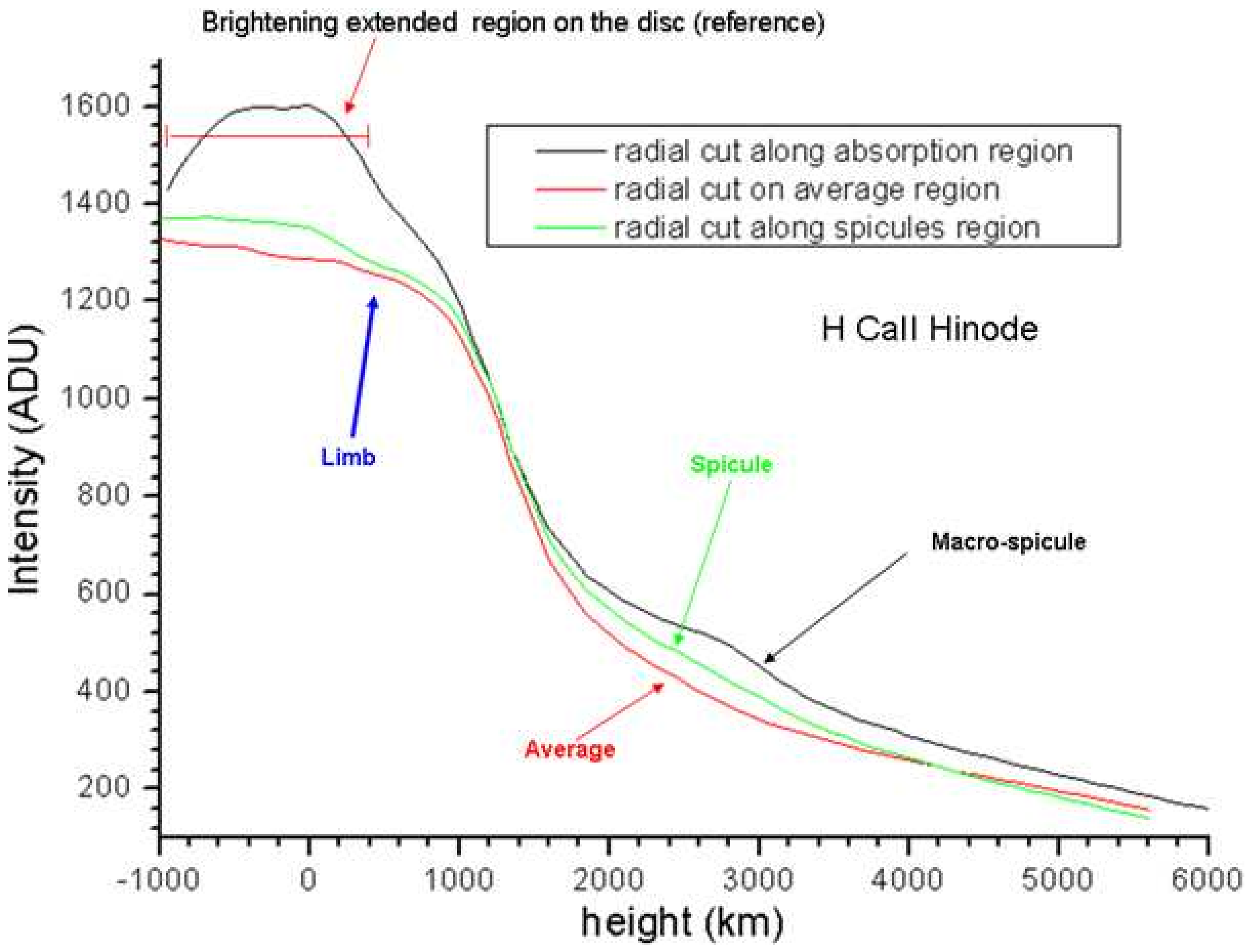

**Figure II-4-12:** *profils d'intensités en fonction de la hauteur réalisés dans différentes régions du limbe solaire d'après les images SOT-Hinode réalisées le 10 Novembre 2007, afin de différencier les contributions des spicules, macrospicules dans la structuration de l'atmosphère solaire et tenter de définir où se situe l'altitude h = 0 sur le limbe.*

Ces profils d'intensité montrent la contribution des spicules à partir de 2000 km d'altitude et au delà. L'émission provenant depuis l'arrière plan du limbe, est atténuée par l'absorption des couches chromosphériques avant de parvenir à l'observateur. Ceci peut expliquer le léger déficit ou absorption observée dans le profil du limbe et noté « absorption région » sur les figures II-4-9, II-4-10 et II-4-11 avec les extraits « blurred mask » et qui peuvent aussi être traduits par « unsharpmasking ».

A partir de ces profils d'intensités radiales, $I = f(h)$, des dérivées en fonction des altitudes sont effectués en figure II-4-13 afin de définir où se situe la référence $h = 0$. Cette méthode a été utilisée par Hirayama 1984 où la référence h = 0 correspondait aux gradients les plus élevés. Dans notre cas, nous avons effectué la dérivée de la courbe $I = f(h)$ par rapport à la hauteur, et nous avons considéré le limbe $h = 0$ *km* situé avant le point d'inflexion de la courbe I (h) avant l'apparition des embrillancements dus aux spicules.

Dans cette thèse nous montrerons qu'en réalité ce point d'inflexion est du à la lumière diffusée par les filtres et composants optiques de la mission du SOT- Hinode, et ayant pour origine la lumière parasite provenant du disque solaire observé hors conditions d'éclipses.

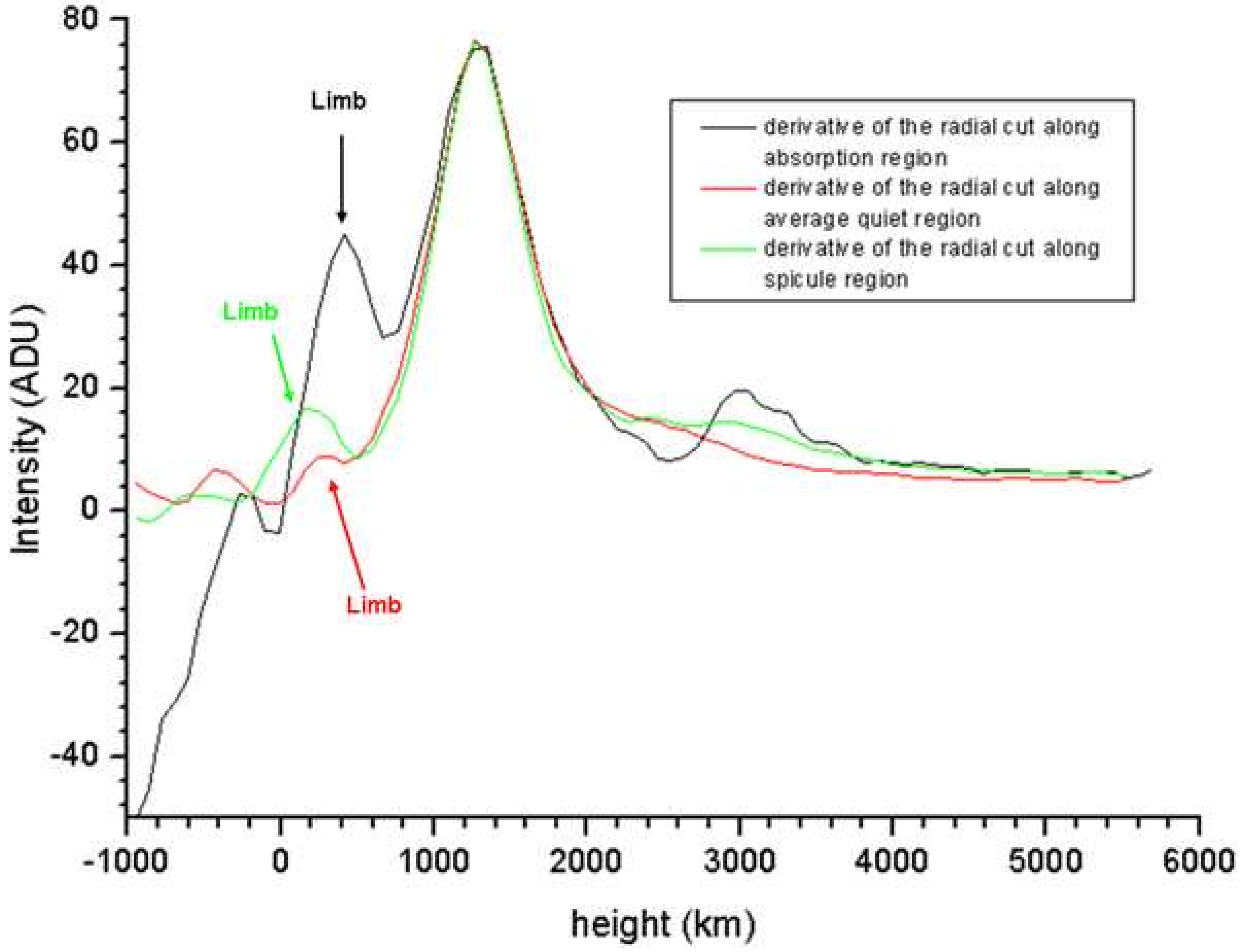

**Figure II-4-13:** *Profils dérivés –dI/dh, d'après les profils d'intensités effectués sur la figure II-4-12.*

Ces analyses aves les images de Hinode en H de Ca II montrent que le limbe se situe autour de 200 ± 50 km. Ces profils radiaux obtenus d'après les images H-Ca II de Hinode (hors éclipses) depuis l'espace, ont permis l'étude de la position du limbe solaire. Ils sont comparés avec les courbes de lumière dans le continu et raies low FIP obtenues aux spectres éclairs lors des éclipses.

Les profils radiaux dans la raie H – Ca II ont montré que les couches au dessus du limbe sont afféctées par l'embrillancement du aux spicules, qui sont inhomogènes et dynamiques, ayant une forme tubulaire, et traduisent l'émergence du champ magnétique, où un modèle hydrostatique, homogène et stratifié n'est pas adapté pour décrire ces régions. D'autre part, la région interspiculaire est mal connue, et l'analyse des spectres éclair dans les raies optiquement minces, contribue à mieux comprendre la nature de cette région (voir la fin du chapitre IV-2-2), ainsique les courbes de lumière des raies low FIP.

Cependant, pour essayer d'introduire les modèles d'atmosphère, des approximations sont effectuées en négligeant le champ magnétique en dessus de 2000km. Les auteurs en Annexe 24 utilisent les mécanismes de transfert de rayonnement.

Nous nous intéressons aux raies d'émission optiquement minces dans les spectres éclair (obtenus par spectrographie avec ou sans fente) dans le domaine visible. Les raies « low FIP » correspondent à la myriade de petites raies d'émission. Les raies « high FIP » associées à l'hydrogène neutre et hélium neutre et une fois ionisé sont moins nombreuses, et les raies He I 4713 Å et He II 4686 Å ont l'avantage d'être à des longueurs d'onde voisines pour être étudiées simultanément.

L'intérêt des éclipses totales de Soleil est de permettre d'étudier les interface photosphère-couronne d'une part et d'autre part la chromosphère dominée par les raies dites « high FIP » comme l'hélium et l'hydrogène.

L'occultation du disque solaire durant les éclipses totales a lieu dans l'espace, hors atmosphère terrestre, et il n'y a plus de problèmes de lumière parasite provenant du disque solaire très intense. Un aspect important que nous montrerons comme résultat nouveau concerne une équivalence entre l'interface photosphère -couronne qui serait de même nature que l'interface protubérance -couronne solaire (Prominence to Corona Transition Region). Nous avons trouvé la présence des raies « low FIP », aussi bien dans les spectres de protubérances que dans les basses couches de la région de transition photosphère-couronne solaire des spectres éclairs.

Ces observations de spectres éclairs sans fente ont été réalisées à l'éclipse totale du 29 Mars 2006, puis améliorées et renouvellées aux éclipses totales de 2008, 2009 et 2010. Ces observations sont chacune décrites chronologiquement dans le chapitre III et j'ai aussi utilisé les résultats de spectres éclairs obtenus lors de l'éclipse du 13 Novembre 2012, avec un spectrographe à fente.

# Chapitre III) Observations CCD des éclipses totales: Spectres éclairs, images au sol et espace simultanées

## III-1) Premières études préliminaires d'avant projet à l'éclipse du 29 Mars 2006

Cette éclipse totale de Soleil du 29 Mars 2006 a été observée en Turquie, dans la partie sud, proche d'Antalia.

Un site d'observation a été choisi le plus près possible de la zone centrale de la bande de totalité, afin de bénéficier de la durée d'éclipse la plus longue, et ainsi être dans les meilleures conditions d'assombrissement (aucune lumière parasite diffusée du disque, entièrement occulté par la Lune), surtout pour les instants de contact, 10 secondes avant et après la totalité qui nous intéressent pour cette thèse.

L'expérience de spectrographe sans fente de 2006 a été qualitative, afin d'évaluer, mettre au point et développer une méthode originale d'acquisition de spectres éclair avec un réseau-objectif, en ayant conçu de façon préliminaire une petite expérience transportable, relativement simple à mettre en œuvre.

Un réseau de 50 mm de diamètre par réflexion a été placé devant un objectif de 100 mm de focale diaphragmé à 5 mm. Cet objectif a été fixé sur un caméscope numérique modèle « TRV6 » afin d'enregistrer directement les spectres sur une mini-cassette de type mini DV, et dont la cadence était de 25 images/seconde. Ce montage est décrit en figure III-1-1 où le réseau de diffraction par réflexion était orienté pour obtenir le spectre à l'ordre 1 sur le CCD du caméscope enregistreur.

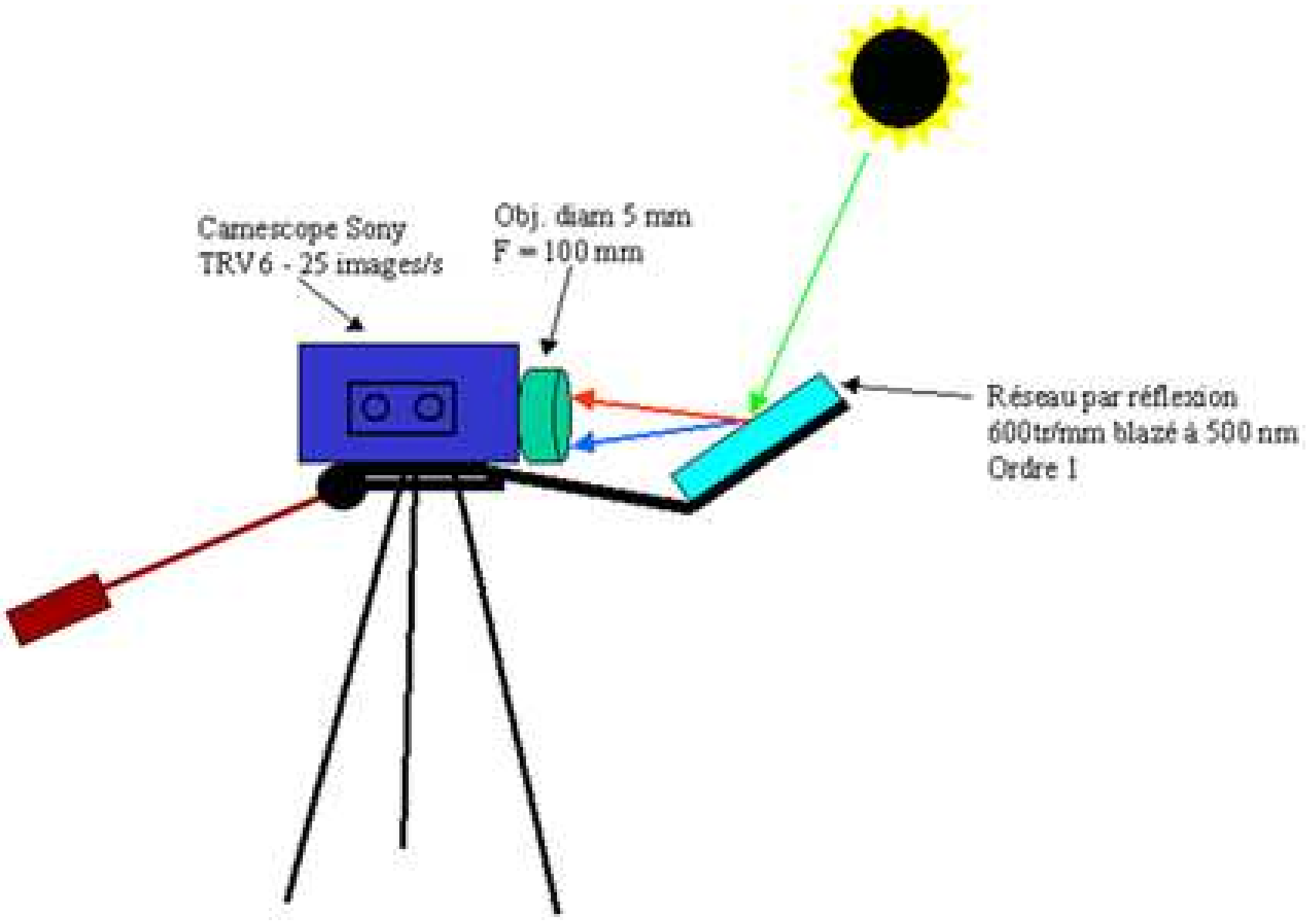

**Figure III-1-1:** *Schéma du réseau-objectif par réflexion avec caméscope enregistreur. Les données des spectres éclair au 1^er^ ordre étaient enregistrées sur cassette mini DV.*

Ce montage était très rudimentaire, car le trépied ne pouvait assuer qu'une orientation horizontale et verticale de l'expérience, et les degrés de liberté étaient ainsi limités. Cependant, différentes orientations de l'ensemble caméscope + réseau-objectif ont été effectuées le long du spectre ainsi formé, en vue d'examiner plusieurs intervalles de longueur d'onde, dont notamment les séquences des contacts. Celles-ci ont été réalisées au début dans la partie bleue-violette incluant les raies H et K du Ca II et quelques raies de l'hydrogène neutre H I, situées avant la discontinuité de Balmer, voir figure III-1-2.

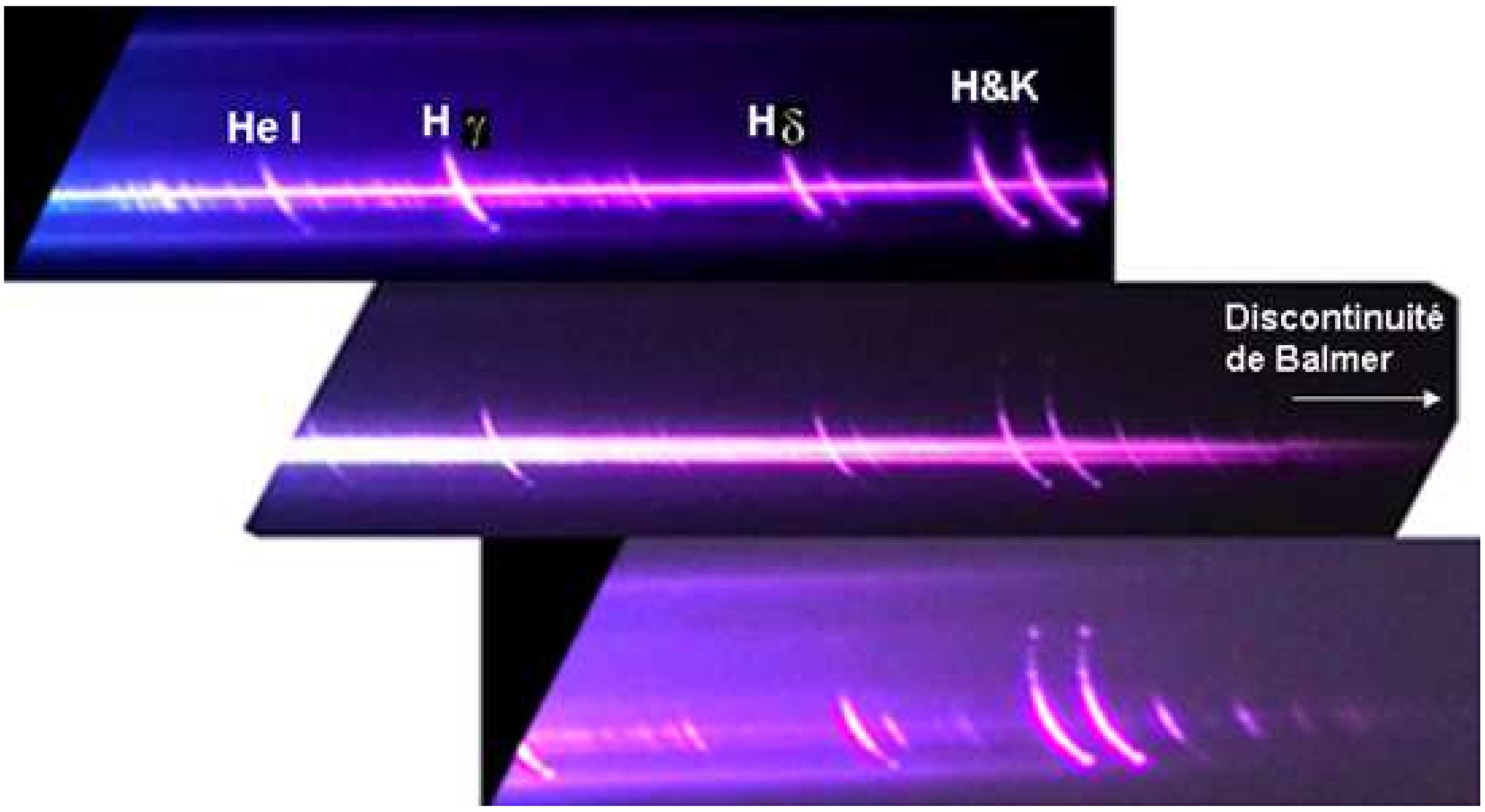

**Figure III-1-2:** *Extrait de séquence de spectres éclairs au second contact montrant les raies H&K de 3960 et 3930Å jusqu'à la partie UV du spectre avec les raies de l'hydrogène H I les plus intenses. La discontinuité de Balmer n'est pas visible. 3 spectres ont été superposés et sommés au second contact avec 0.04 secondes d'exposition.*

Durant les phases de totalité, un balayage en longueurs d'onde a été effectué en changeant l'angle d'inclinaison de l'expérience par rapport au disque solaire éclipsé, sans perdre les spectres dans le champ du camescope. Le spectre de la basse couronne faiblement résolue dans la raie verte du Fe XIV à 5303Å a été centré sur le CCD du camescope, et où des séquences de plus de 30 secondes ont été réalisées durant la totalité. De plus, au milieu de la totalité, l'inclinaison a été modifiée de telle façon à pointer l'ordre zéro du spectre, c'est-à-dire l'image du disque éclipsé sans spectroscopie, pour obtenir une image de la couronne en lumière blanche. La figure III-1-3 montre les images stabilisées obtenues depuis cette expérience. Une fois l'image de la couronne à l'ordre zéro acquise durant 10 secondes, le spectre à l'ordre 1 a été répointé sur la raie verte du Fe XIV 5303Å.

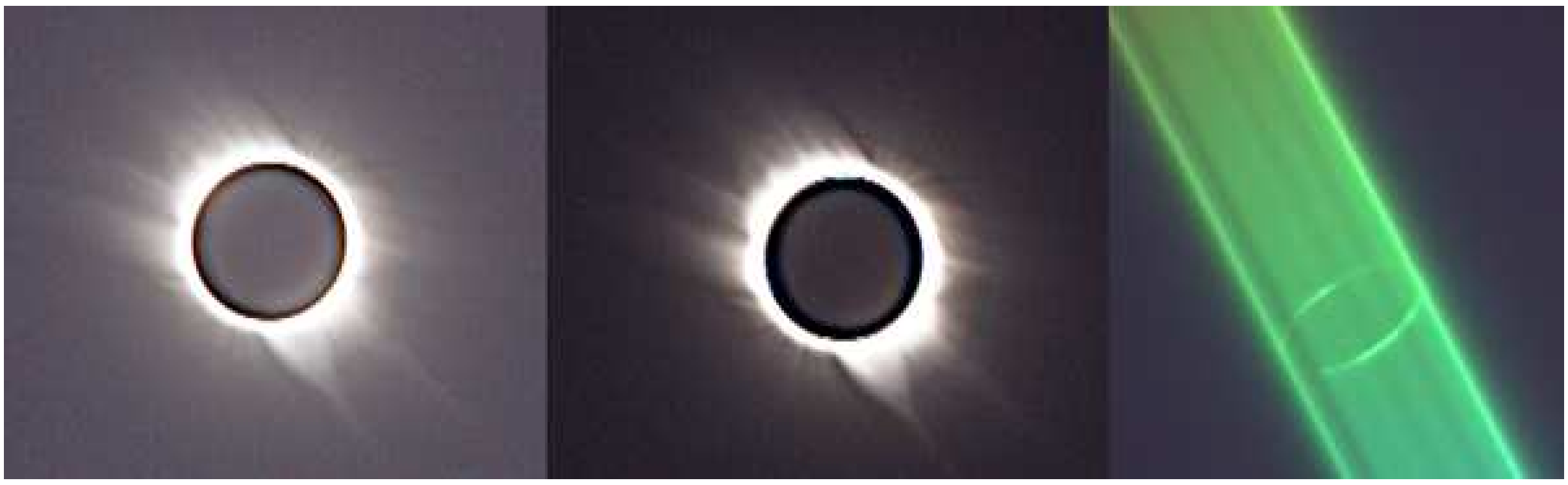

**Figure III-1-3:** *Extrait de séquences où le réseau a été placé dans l'ordre zéro du spectre pour obtenir quelques images en lumière blanche. Le réseau a été ensuite placé à nouveau à l'ordre 1 sur la raie verte afin de comparer à basse résolution, la structuration de la raie verte et du continu coronal, par comparaison avec la couronne blanche. La dispersion est de 1.3Å/pixel. Les images sont dans la même orientation. Les modulations dans le continu du spectre correspondent aux structures. Les images à l'ordre zéro ont subi un traitement de conversion des intensités en échelles logarithmique puis un masque flou d'une finesse de 5 pixels.*

La forme de la raie verte allongée et ovalisée est due à l'orientation du réseau, c'est-à-dire que le plan des faisceaux incidents et spectre réfléchi n'étaient pas perpendiculaires à la surface du réseau, et donc ont produit cet effet. Malgré la résolution spatiale insuffisante, quelques structures de grande échelle et plus étendues étaient visibles, avec quelques modulations dans le continu coronal.

La figure III-1-4 représente une courte séquence stabilisée de spectres éclair dans la région de $H\beta$, et au-delà du triplet b1, b2, b3 du magnésium neutre Mg I. Les variations des flux des raies et surtout du continu du bord photosphérique au troisième contact sont remarquées. Les images étaient stabilisées durant environ une seconde. Les images sont extraites individuellement et non sommées. Les spectres éclair du second contact et une partie de ceux du troisième contact n'ont pas pu être observés convenablement ni analysés en détails, car des vibrations d'images importantes se produisaient à cause du trépied mal conçu, et trop instable mécaniquement. Le maniement de ce trépied au cours des repointages produisait des bougers importants, et les raies dans de nombreux spectres ne pouvaient pas étre observées distinctement. De nombreux spectres n'ont pas été exploités. La résolution spatiale était insuffisante pour une analyse quantitative. Cependant cette expérience a servi à mettre en évidence les problèmes de stabilisation, choisir et définir les temps d'exposition et cadence d'images pour observer des raies d'émission plus faibles, mettre au point la méthode de pointage et acquisition des spectres, l'emploi d'un filtre amovible devant le réseau-objectif, pour ne pas saturer longtemps le CCD avant les phases de totalité des éclipses. Toutesfois cette expérience a permi d'observer la raie verte coronale du Fe XIV a basse résolution.

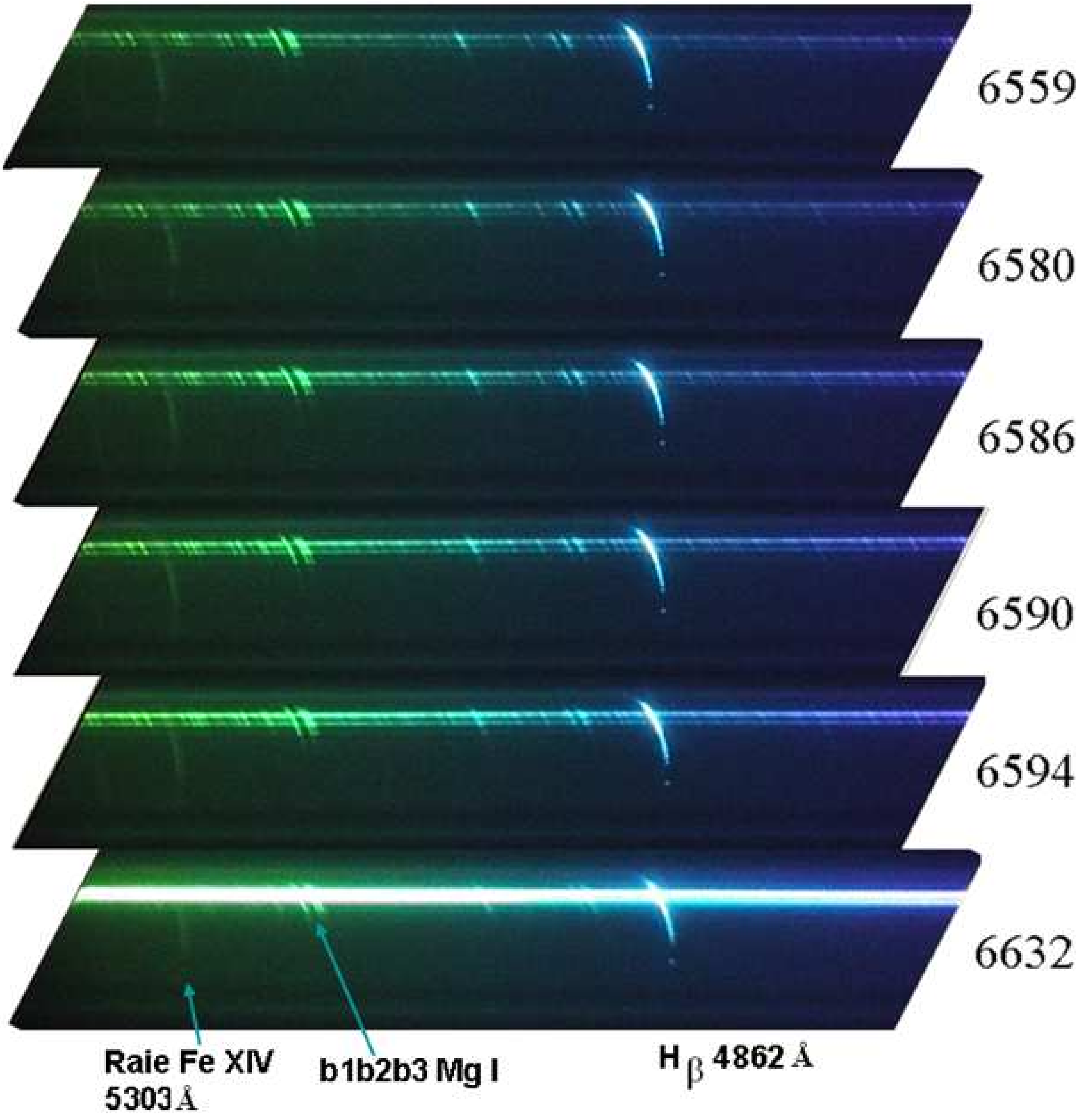

**Figure III-1-4:** *séquence des spectres éclairs au troisième contact de l'éclipse du 29 Mars 2006. La numérotation permet d'indiquer l'instant du contact, et peut être utilisée pour tenter de définir où est le limbe. La cadence était de 25 images/s, mais la dispersion et résolution étaient quelque peu insuffisantes pour effectuer une analyse quantitative et photométrique.*

La réapparition au troisième contact des raies chromosphériques, avec la raie H$\beta$ intense sur la partie bleue du spectre, puis le triplet b1 b2 b3 du Mg I, peut être comparée avec le continu entre ces raies. La raie verte coronale du Fe XIV à 5303Å est visible plus faiblement à gauche. Quelques autres raies plus faibles en émission sont aussi visibles, mais la résolution avec cette petite expérience était insuffisante pour analyser leurs profils, avec les modulations du relief du bord de la Lune. La raie de l'hélium ionisé He II 4686 Å n'était pas visible sur un seul spectre. Même après sommation de 30 images cette raie He II 4686 Å  a été détéctée avec faible un rapport signal sur bruit.
Cette première expérience dédiée àl'observation des spectres éclair a été ensuite améliorée pour l'éclipse du 1er Août 2008. L'utilisation d'un objectif de 50 mm d'ouverture et 600 mm de focale a permis d'atteindre une résolution spectrale 10 fois meilleure que celle utilisée en 2006 et surtout l'utilisation d'une monture équatoriale rigide, dont le maniement est effectué au moyen de flexibles (en ascension droite et déclinaison) a évité les bougers dans les spectres et de perdre des images. Le maniement de cette monture était manuel, plus aisé et confortable. Un réseau par transmission de 600 traits/mm de 50 par 60 mm sur la surface optique utile a été mis en place et conçu de telle manière à tourner autour de l'axe optique de la lunette 50/600 mm, ce qui a facilité l'orientation des spectres éclairs. Un enregistrement avec CCD à cadence élevée (15 à 25 images: seconde) a permis une analyse quantitative à l'éclipse de 2008. La méthode d'imagerie rapide des séquences de spectres éclairs, est présentée succintement sur un extrait en figure III-1-4. Elle permet de définir une procédure d'analyse de chaque spectre éclair extrait qui sera utilisée aux éclipses totales suivantes (2008, 2009, 2010..). Les spectres éclairs obtenus du 29 Mars 2006 avaient cependant été réalisés dans des conditions météorologiques très favorables, avec un ciel sans nuages. Les résultats de ces quelques spectres éclair de l'éclipse de 2006 n'ont pas été analysés en détails par les méthodes photométriques, pour être présentés dans cette thèse. L'expérience a été beaucoup améliorée pour l'éclipse de 2008, mieux préparée, répétée, testée au préalable sur le spectre du croissant lunaire pour adapter le réglage du gain et temps d'exposition de la caméra CCD de type « Watec 120 N+ » de sensibilité élevée 10 µLux. Le choix de sensibilité est adaptée aux faibles intensités des raies de He I 4713 Å et He II 4686 Å. Une meilleure résolution a été obtenue à l'éclipse totale de Soleil du 1er Août 2008, et des analyses photométrques sur les spectres ont pu être réalisées.

## III-2) Spectres éclairs à l'éclipse du 1er Août 2008 en Sibérie

Cette expérience de spectres éclairs a été réalisée en Sibérie dans la région de l'Altaï, voisine de la Mongolie. Les conditions météorologiques étaient prévues mitigées le jour de l'éclipse sur le site de Biisk, et j'avais prévu l'expérience transportable du réseau-objectif comprenant la lunette 50/600 mm munie de la caméra CCD Watec 120 N+ au foyer. D'avantage d'informations sont données en Annexe 16-2, où un caméscope numériseur en autonomie a permis d'enregistrer les données. En effet, par anticipation de risques de ciel couvert le jour de l'éclipse, un site meilleur avait été choisi sur un terrain d'une station météo dans la ville d'Ongudaï, plus proche de la frontière avec la Mongolie à l'est de l'Altaï, où les conditions météorologiques se sont avérées plus favorables que sur le site de Biisk, où d'autres expériences avaient été installées. En effet, sur le site d'Ongudaï, il a été possible d'observer l'éclipse jusqu'au début  de la totalité, avant l'arrivée des nuages. Des spectres éclair ont pu être enregistrés au second contact avec une cadence de 25 spectres/seconde. Une trouée entre 2 nuages avec le fond bleu du ciel s'était présentée sur ce site à l'instant du second contact jusqu'au début de la totalité, et la transparence dans cette zone du ciel était relativement bonne. Cependant, les nuages se sont déplacés après le début de la totalité de l'éclipse, et ensuite ont totalement couvert l'éclipse jusqu'à la fin, et il n'a pas été possible d'observer le troisième contact. La figure III-2-1 représente le profil du relief du bord de la Lune avec une résolution de l'ordre de une seconde d'arc. Les indications des positions des instants de contacts C2 avant la totalité et C3 juste après la totalité sont indiquées. Les instants et positions des circonstances de l'éclipse ont été calculés pour notre site à Ongudaï. La direction du zénith est indiquée par une flèche rose en haut de la figure III-2-1.

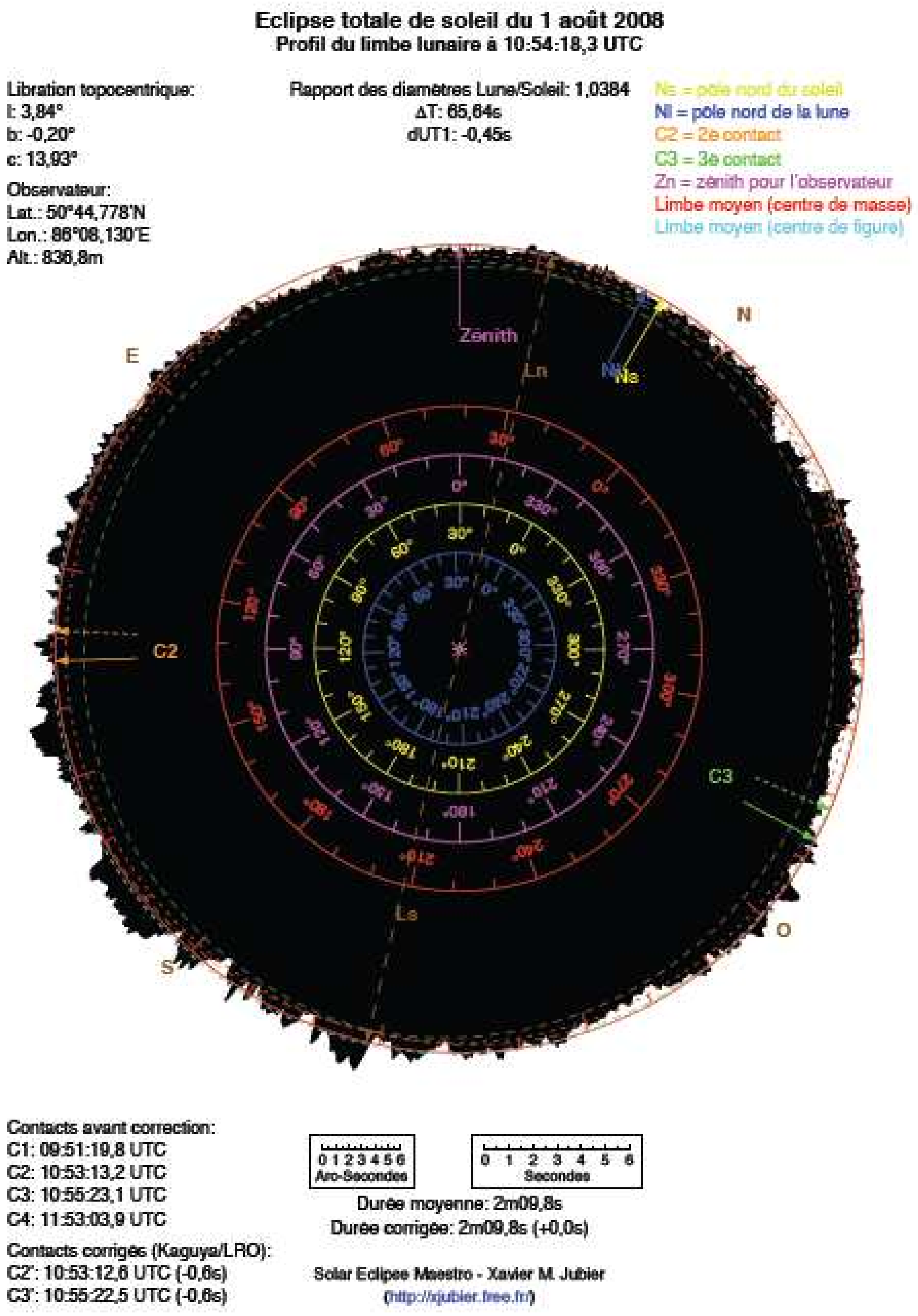

**Figure III-2-1:** *calcul des instants et position des contacts sur la carte du relief lunaire, pour l'éclipse totale du 1er Août 2008, sur le site d'Ongudaï, proche de la frontière avec la Mongolie. C2 et C3 représentent les contacts avec un profil lunaire calculé sans relief lunaire et C'2 et C'3 sont les contacts (pointillés) corrigés avec le relief lunaire.*

Nous étions plusieurs observateurs sur le site d'Ongudaï. Des images des contacts, juste avant la totalité ont été réalisées, figure III-2-2. Celles-ci m'ont servi pour l'orientation des spectres éclair et l'identification des modulations dans les spectres du continu des grains de Baily.

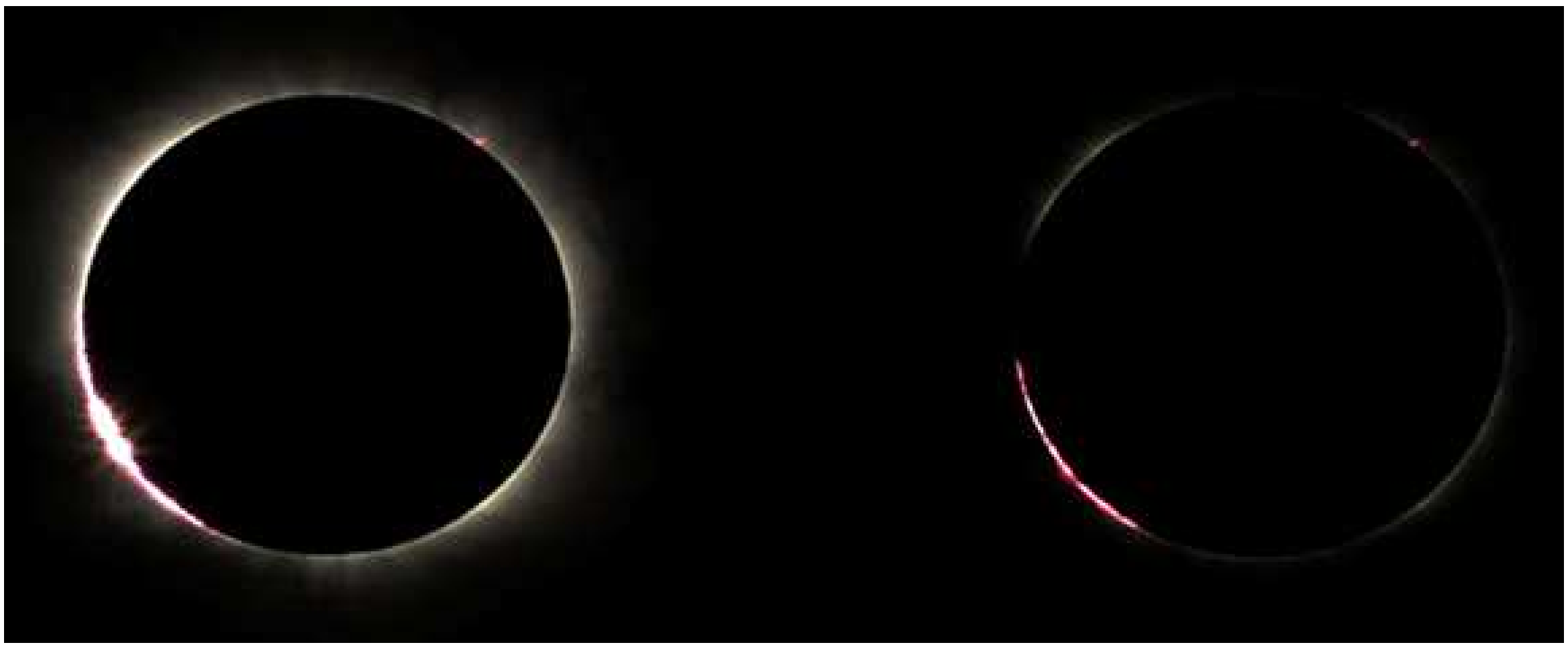

**Figure III-2-2 :** *vues du second contact à l'éclipse de 2008 en lumière blanche où simultanément les spectres flash ont été enregistrés. Image réalisée par Thierry Legault avec un appareil photo-numérique Canon EOS SD Mark II, poses* de *10 ms à 400 iso. 3 à 4 secondes séparent ces 2 images.*

Ces images en lumière blanche, notamment celle de droite, montre le mélange et superposition du continu de la haute photosphère, avec la frange rosée de la chromosphère qui s'ajoutent sur la ligne de visée. La couleur rosée de la chromosphère est due aux raies de l'hydrogène H$\alpha$ à 6562.8 Å, la raie D3 de l'hélium à 5876Å, et raie H$\beta$ à 4860Å. Le mélange de celles-ci conduit à cette colaration. Les modulations sur le limbe vu comme un arc, sont dues au relief lunaire, où le bord de la Lune présente des montagnes et vallées et qui se découpent sur le bord du Soleil et produisent les grains de Baily lors des contacts. L'image de droite présente un arc limité de la frange chromosphérique et où la Lune a complétement occulté le bord photosphérique, les grains de Baily ont disparu. Des spectres éclair en enregistrement continu avec une cadence de 25 spectres- images/seconde ont été obtenus au même moment que ces images en lumière blanche, figure III-2-2. Chaque spectre individuel a été analysé en relevant l'intensité dans le continu correspondant aux spectres des grains de Baily, dans certaines raies d'émission comme le Fe II 4629 Å, He I 4713 Å, He II 4686 Å. Pour améliorer le rapport signal sur bruit dans les raies d'hélium de plus faible intensité que la raie du Fe II, des sommations de 10 spectres tous les 5 spectres ont été effectués. Les séquences sont présentées en figure III-2-3 pour illustrer les variations rapides dans les spectres éclair. La première image de spectre en figure III-2-3 est obtenue 20 secondes avant l'éjection des filtres de densité neutre, montrant les raies d'absorption telluriques et de Fraunhofer, comme la raie du Mg I 4702.8 Å qui est plus large et profonde. Le temps de pose est de 40 millisecondes et reste le même durant toute la séquence d'acquisition en continu. La raie de Fraunhofer du Mg I 4702.8 Å correspond aux couches de la haute photosphère solaire et les spectres éclair de la figure III-2-3 sont obtenus après l'éjection de la densité neutre.

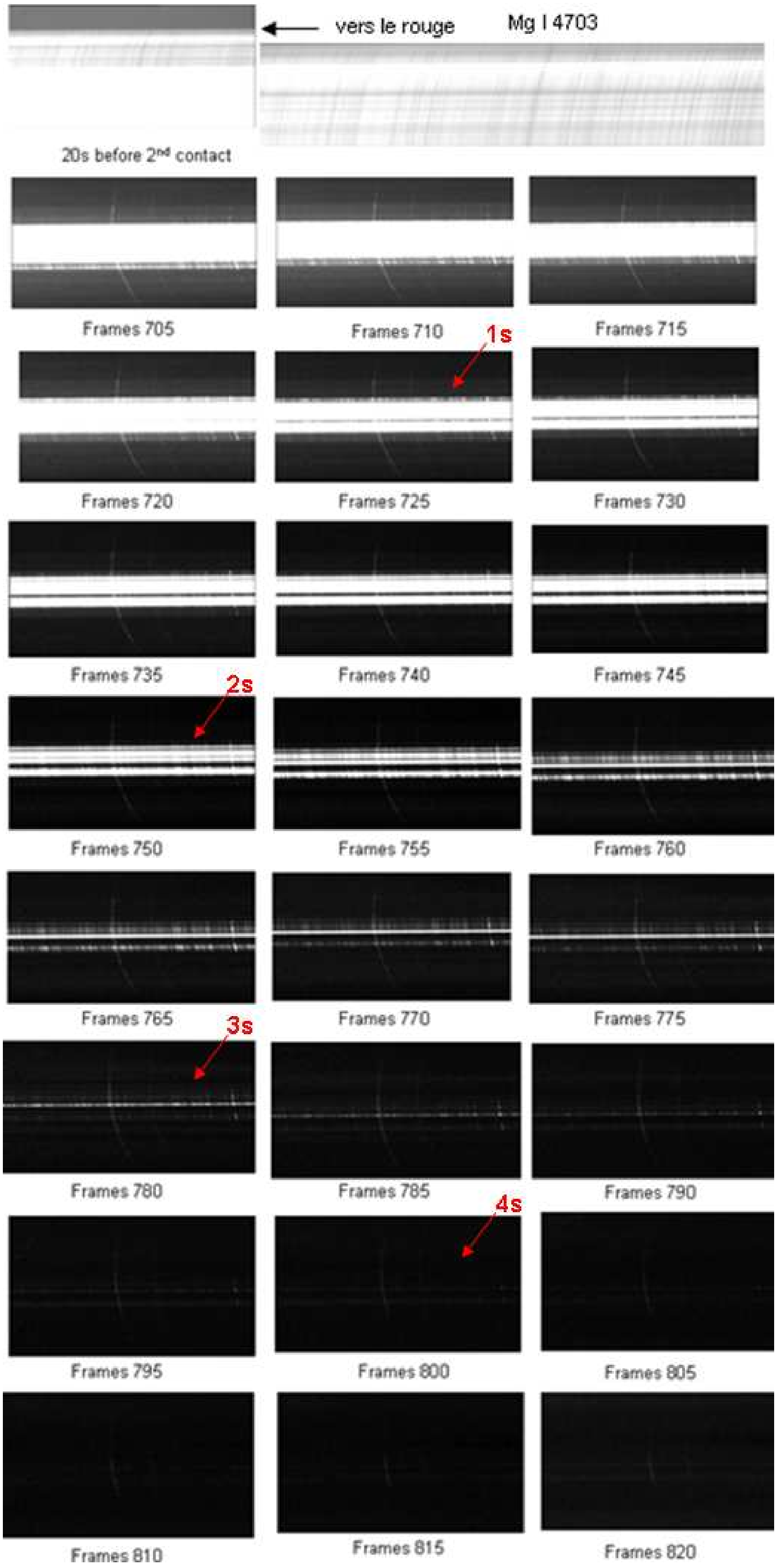

**Figure III-2-3:** *Montage d'une séquence de spectres éclairs au moment du second contact C2 de l'éclipse du 1er Août 2008. Durant 20s avec le filtre densité neutre avant le début du second contact, et après éjection du filtre, séquences des spectres éclairs révélantles basses couches de l'atmosphère solaire au-delà du profil de la Lune. Entre les spectres N° 705 et N° 820, 5 secondes se sont écoulées. Le temps de pose est de 40 milli-secondes.*

La figure III-2-4 montre le résultat d'un empilement de 40 spectres pour montrer la présence d'une seconde enveloppe d'hélium ionisé dans la raie He II 4686 Å.

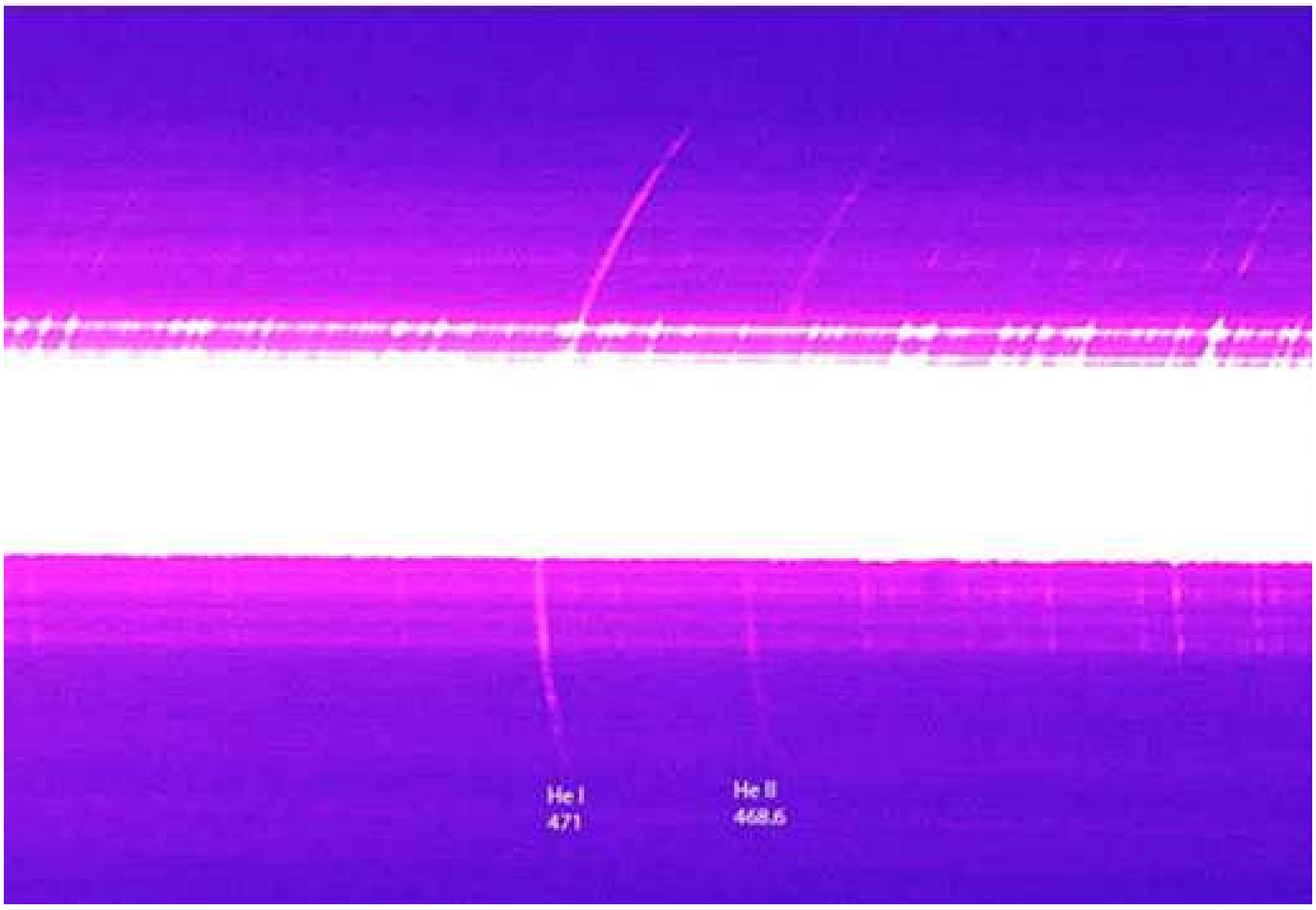

**Figure III-2-4:** *Extrait de la séquence spectrale obtenue au voisinage du second contact de l'éclipse totale de Soleil du 1er Août 2008. La bande blanche correspond à la partie surexposée de la haute photosphère, où les raies sont toujours mesurées en absorption. En simultané sur une plus grande étendue de la chromosphère autour du Soleil, les raies de l'hélium neutre 4713 Å, et de l'hélium ionisé 4686 Å se présentent sous forme d'enveloppes entourant le Soleil vues au-delà du bord lunaire. Les échancrures du limbe lunaire (montagnes et vallées) modulent fortement les intensités du spectre des enveloppes d'hélium. Les intensités de la 2ème raie d'hélium varient d'une façon plutôt monotone sur l'extension du croissant, à cause de la sur-occultation du Soleil par la Lune. Spectres éclairs obtenus avec le réseau-objectif de 600 tr/mm et un réfracteur achromatique de 600 mm de focale, avec un CCD au foyer. La vitesse de lecture originale était de 25 images/s. Moyenne de 10 spectres après soustraction du courant d'obscurité et le biais.*

Ces spectres montrent les raies d'émission d'hélium neutre et ionisé, séparées de 27 Å, qui ont l'apparence d'une enveloppe. La raie d'hélium ionisé He II 4686Å vue comme une enveloppe, est probablement un résultat nouveau, car il était difficile autrefois de l'observer comme une enveloppe, pour les raies optiquement minces. D'autre part les modulations dans le profil des enveloppes traduisent une corrélation avec le relief lunaire. Le continu chromosphérique et le continu coronal sont visibles entre les enveloppes d'hélium, sous forme de bandes dans le sens de dispersion spectrale.
Le continu photosphérique est un moyen aussi de déduire l'étendue de la « mésosphère » située juste au dessus de la haute photosphère où les intensités des raies d'émission « low FIP » et des grains de Baily commencent à décroître. Ces spectres du continu modulés par le relief lunaire correspondent aux grains de Baily comme sur l'image de gauche figure III-2-2. La diminution du flux de ce continu constitue la fin du bord photosphérique dans les vallées lunaires. Aux altitudes plus élevées, où le continu photosphérique a disparu, il reste le continu chromosphérique d'intensité 100 fois plus faible, comme le montrent les extraits des figures III-2-5, III-2-6 et III-2-7 de spectres éclairs obtenus à l’éclipse totale du 1er Août 2008.

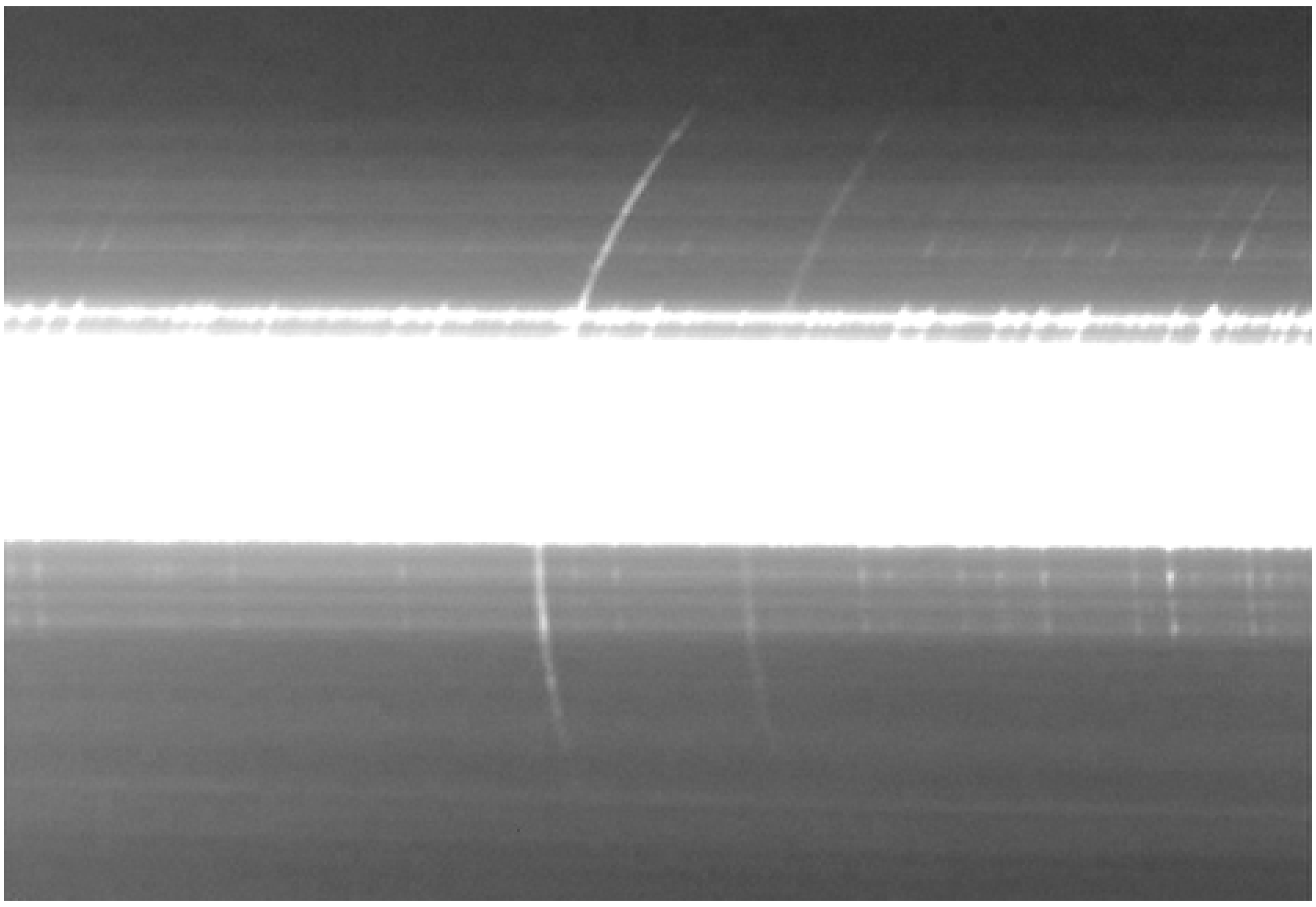

**Figure III-2-5:** *sommation de 10 spectres éclairs obtenus le 1er Août 2008, début du second contact, converti en échelle logarithmique. Cet extrait montre les 2 enveloppes d’hélium neutre He I 4713 Å et sans doute pour la première fois celle de l’hélium He II 4686 Å, et le continu photosphérique correspondant aux spectres des grains de Baily sur cet extrait.*

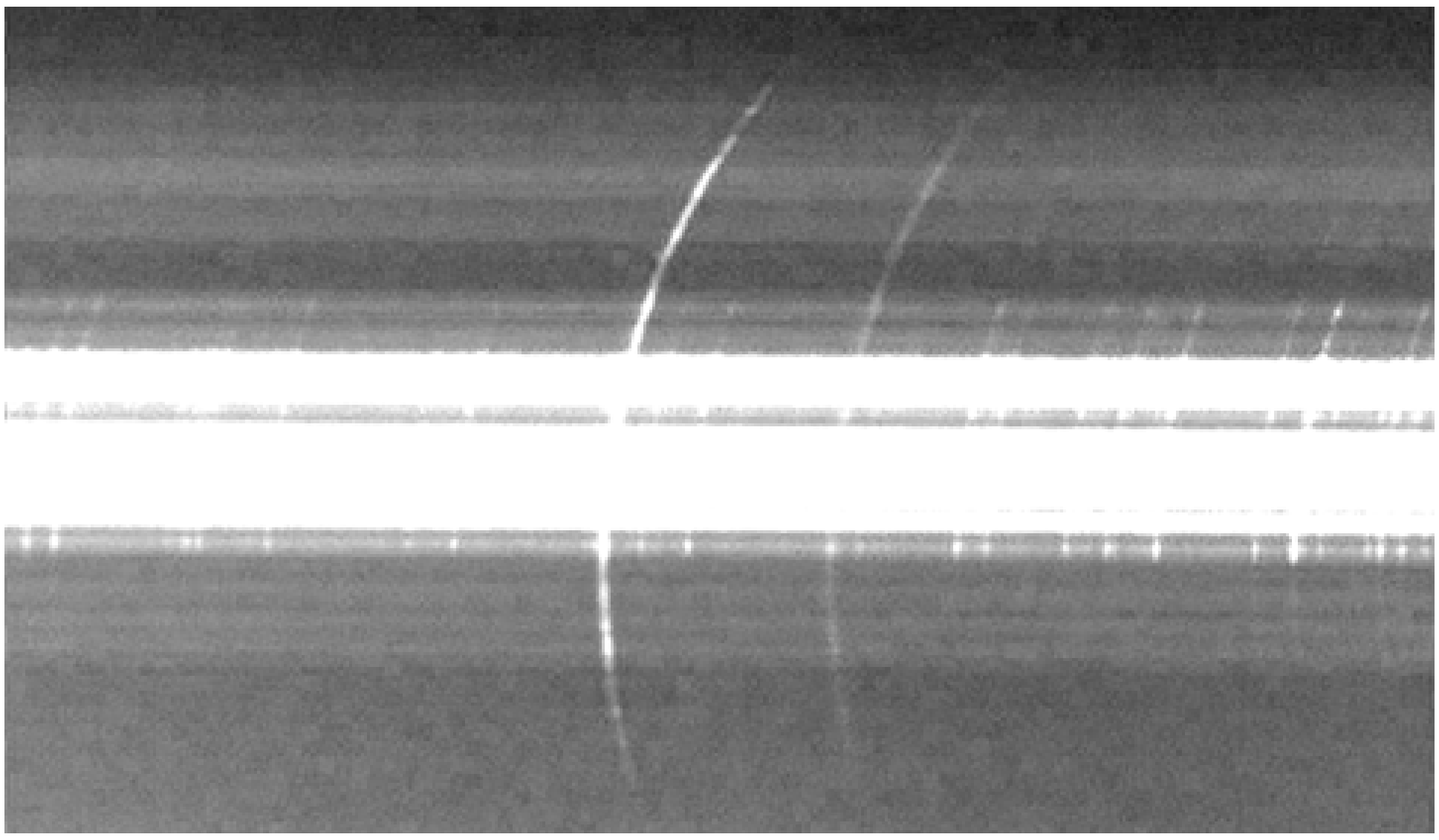

**Figure III-2-6:** *sommation de 10 spectres éclairs obtenu le 1er Août 2008, au milieu du second contact, converti en échelle logarithmique, montrant les 2 enveloppes d’hélium neutre He I 4713 et He II 4686 Å, et le continu photosphérique correspondant aux spectres des grains de Baily.*

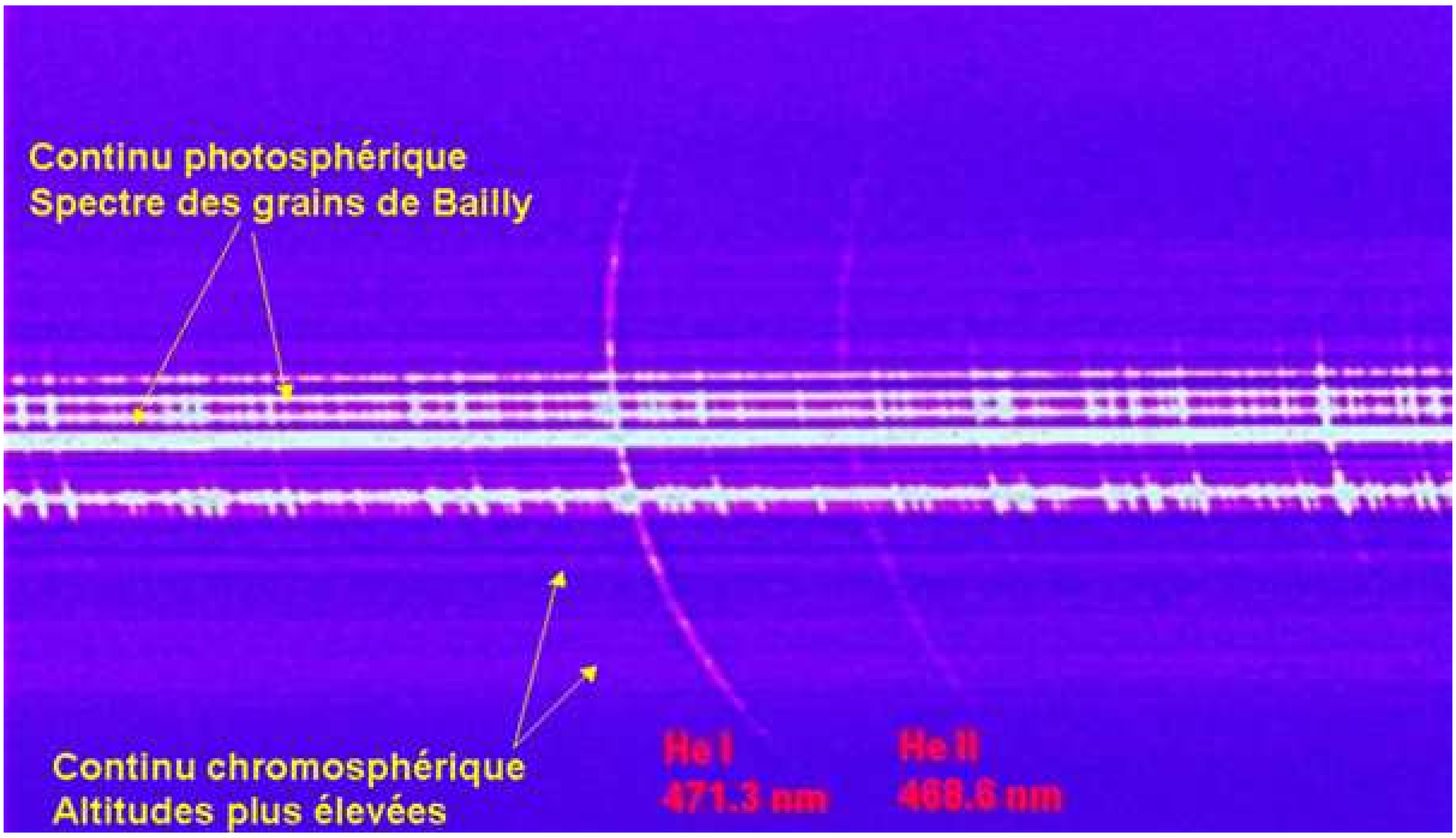

**Figure III-2-7:** *Extrait d'une séquence de spectres éclairs obtenue à la fin du second contact de l'éclipse du 1er Août 2008, dans le domaine spectral 4700 ± 60 Å, avec les nombreuses raies en émission entre les quelles se situent les continus photosphériques et chromosphériques, modulés par le relief lunaire. Le continu le plus intense (grains de Baily) correspond à la référence où l'altitude maximale est prise pour l'instant de contact.*

Le bord de la lune module le profil des raies, et le spectre continu le plus intense (grains de Baily) correspond à la référence où l'altitude maximale est prise pour définir l'instant de contact. Les numéros affectés à chaque spectre individuel sont ensuite convertis en altitudes et la référence $h = 0$ km est prise pour les flux les plus intenses, située dans la vallée lunaire la plus profonde, au centre du croissant, comme indiqué par la flèche C2 et C3 sur les figures III-2-8 et III-2-9. A partir des séquences de spectres éclairs numérotés et des relevés de variations des flux du continu, il a été possible de tracer la correspondance numéros de spectres avec les altitudes correspondantes prises au dessus du limbe. La vallée lunaire la plus profonde correspond au point de contact, comme indiqué en figure III-2-10. L'intervalle d'altitude entre 2 spectres consécutifs était de 16 km, compte-tenu de la cadence d'acquisition de 25 spectres/ seconde et du mouvement naturel du bord de la Lune sur le disque solaire de 0.52'' d'arc / seconde.

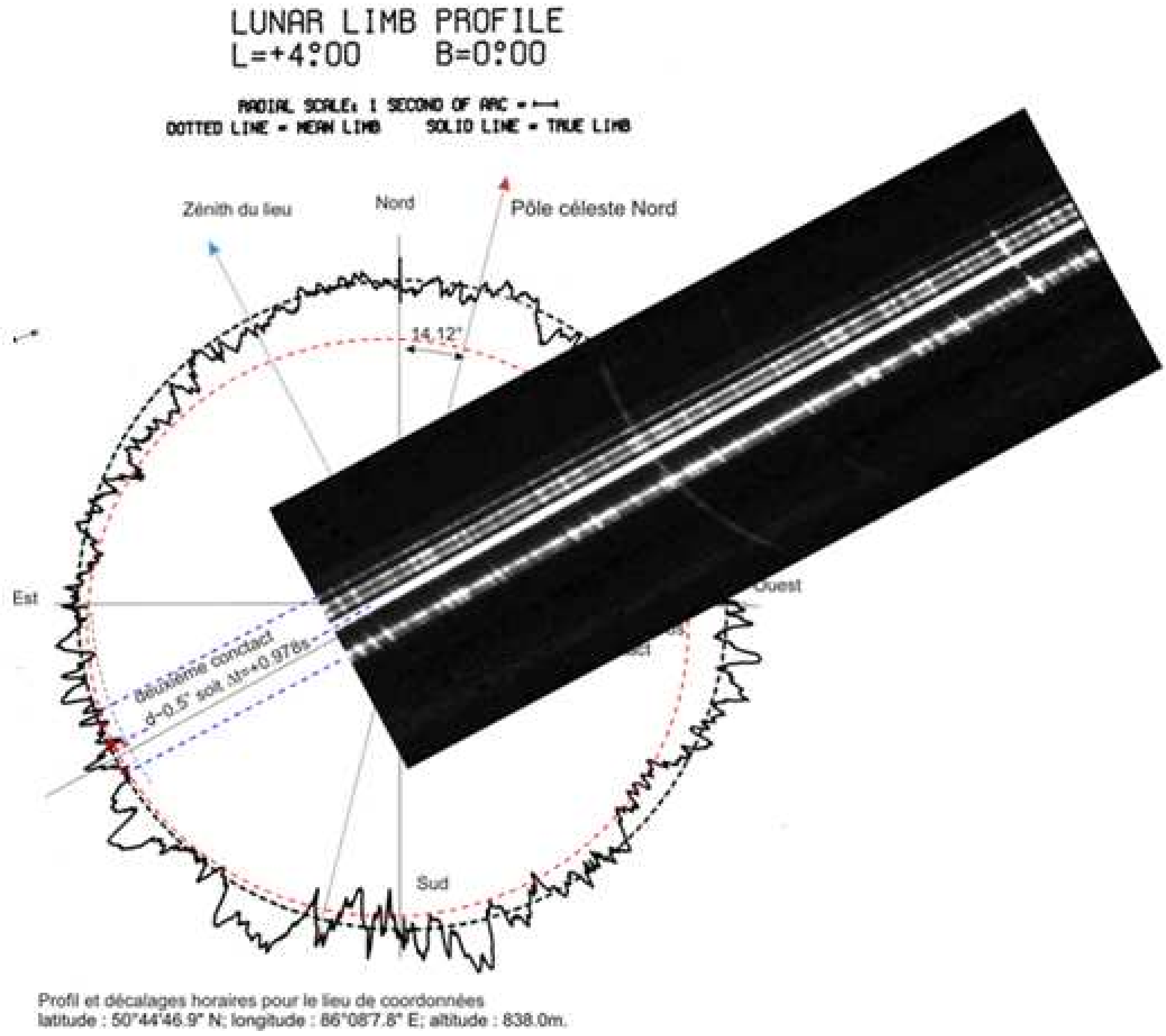

**Figure III-2-8:** *correspondance des spectres éclairs et des grains de Baily avec le profil lunaire de Watts. Second contact à l'éclipse du 1er Août 2008.*

La figure III-2-9 donne le profil lunaire détaillé sur la région du profil lunaire où est indiqué par une flèche orange le contact:

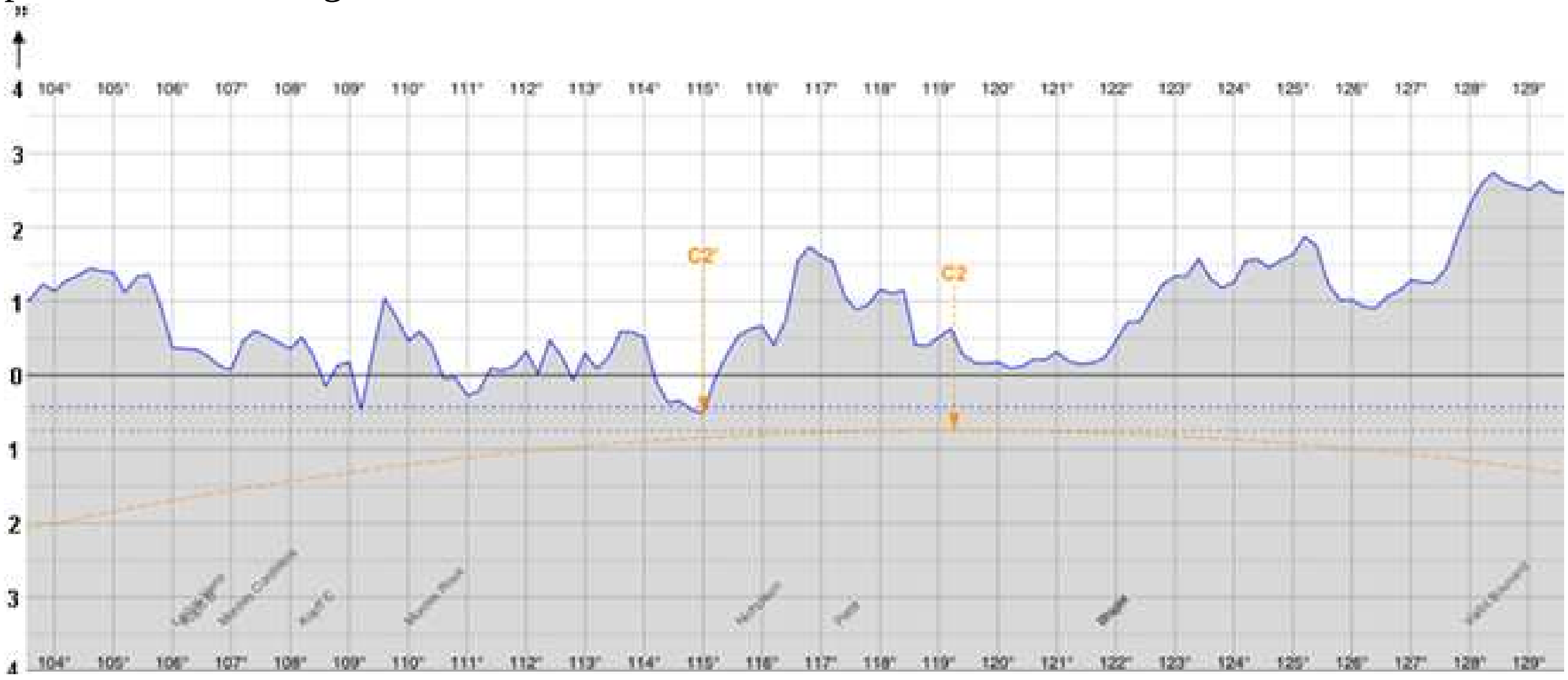

**Figure III-2-9:** *détail du relief lunaire à l'instant du second contact à l'éclipse totale de Soleil du 1er Août 2008 à Onguday, où ont été enregistrés les spectres éclair. Les vallées lunaires correspondent aux grains de Bailly dont les spectres ont été réalisés. L'échelle entre 2 graduations verticales correspond à 0.5 secondes d'arc.*

Les coordonnées du site où nous avons observé le 1er Août 2008, à Ongudaï, proche frontière Mongolie-Altaï étaient: 50,7463° Nord; 86,13550° Est. L'altitude était de 836 m au dessus du niveau de la mer.
A partir des premières analyses effectuées sur les spectres éclair, les correspondances altitudes et numéros d'images ont été établies dans le graphique de la figure III-2-10 pour seulement le second contact qui a été observé à l'éclipse du 1^er^ Août 2008 :

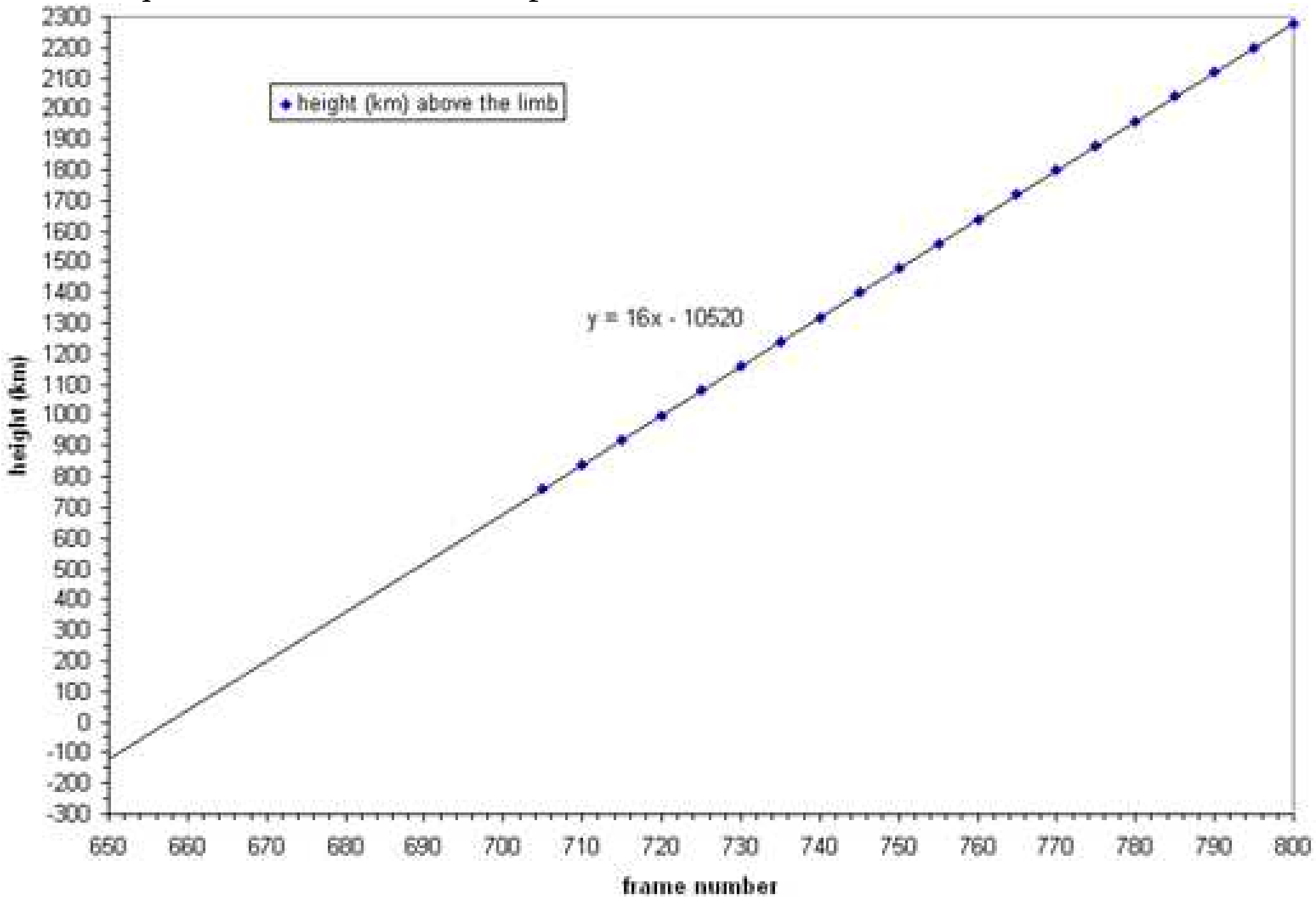

**Figure III-2-10:** *Altitudes estimées des images (position moyenne du bord lunaire) en fonction du numéro d'image au 2$^{ème}$ contact C2 à l'éclipse du 1$^{er}$ Août 2008*

Ces correspondances ont été établiées pour la partie la plus intense des spectres de grains de Baily. La caméra CCD Watec 120 N+ qui est limitée en 8 bit de dynamique, saturait pour les grains de Baily. Pour les mêmes altitudes, la caméra Lumenera Skynyx 2.1 M de 12 bit de dynamique utilisée en 2010 n'était pas saturée. Les grains de Baily montraient les raies d'absorption de Fraunhöfer, avec simultanément une myriade de raies d'émission.
Les altitudes des extensions des enveloppes d'hélium neutre He I 4713 Å et une fois ionisé He II 4686 Å ont pu être évaluées sur cette éclipse du 1$^{er}$ Août 2008. Les courbes de lumière sont données dans le chapitre IV-1.
La caméra utilisée en 2008 était plus sensible, Gain 60 % . Le capteur est constitué de pixels de 9 x 9 µ et 8 bits de dynamique. D'avantage d'informations techniques sur cette caméra Watec 120 N+ sont données en partie 2) de l'Annexe 16. La caméra utilisée en 2010 avait un gain de 2, les pixels font 4.6x4.6 µm ce qui a permis une meilleure résolution, 1150 km/pixels au lieu de 1800 km/pixels en 2008. Enfin, les régions sur le limbe solaire observées sont différentes entre 2008 et 2010. L'étalonnage a été réalisé avec les mêmes filtres et densité neutre suivi d'un filtre interférentiel centré sur 4700 Å. Une saturation importante à l'éclipse de 2008 a empéché de faire des analyses des raies d'émission d'hélium à des altitudes inférieures à 600 km au desus du limbe solaire occulté par le bord lunaire.
Nous avons observé par conséquent l'étendue de cette raie d'hélium He I 4388 Å jusqu'à 3500 kilomètres au dessus du limbe, tandis que Mictchell et al à l'éclipse de 1930 ne l'ont observée pas plus loin que 2000 km au dessus du limbe. Cela s'explique en partie à cause des limitations des plaques photographiques utilisées à cette époque.

## III-3) Spectres éclair à l’éclipse du 22 Juillet 2009 en Chine

Les expériences des spectres éclairs du 1er Août 2008 ont été reproduites lors de l’éclipse totale du 22 Juillet 2009 en Chine. Cette éclipse totale présentait un intérêt notable: sa durée était exceptionellement très longue, de plus de 5 minutes sur notre site à Tianhuanping en Chine, et plus de 6 minutes sur les Iles Marshall proches du Japon. Les autres éclipses avaient une durée plus courte:
Environ 4 minutes pour l’éclipse de 2006, 2 minutes pour celle de 2008, 2 minutes 30 secondes pour 2010 et moins de 2 minutes pour celle de 2012.
L’intérêt d’une éclipse totale de plus de 5 minutes de durée est que de plus nombreuses expériences et acquisitions plus longues peuvent être tentées durant la totalité, et surtout, le fond du ciel est encore beaucoup plus sombre que pour des éclipses totales ayant des durées plus courtes, car l’occultation du disque solaire par la Lune est plus importante.
Cependant la météorologie a hélas été décevante sur de très nombreux sites en Chine et en Asie, car la date du 22 Juillet 2009 était située dans la période de la saison des pluies.
Nous avons bénéficié d’un ciel voilé, mais heureusement suffisamment transparent à travers lequel il a été possible de voir l’éclipse et réaliser de nombreux spectres éclair.
Nous étions hébergés et installés dans dans un hôtel de montagne à Tianhuanping à 150 km à l’Est de Shanghaï, dans la région d’Anji au Zejiang. A côté de l’hôtel nous avions accès à un terrain dégagé sur lequel nous avions préparé nos expériences.
Nous étions aux coordonnées suivantes:
Lat.= 30°28.1'N ou plus précisément 30° 28' et 5,14"
Long.=119°35.4'E ou plus précisément 119° 35' et 22,5"
Elevation = 890.0 m
La figure III-3-1, comme pour l’éclipse de 2008, représente le profil du relief lunaire agrandi avec les indications des positions des instants de contacts C2 avant la totalité et C3 juste après la totalité, avec une échelle en secondes d’arc (arcseconds).

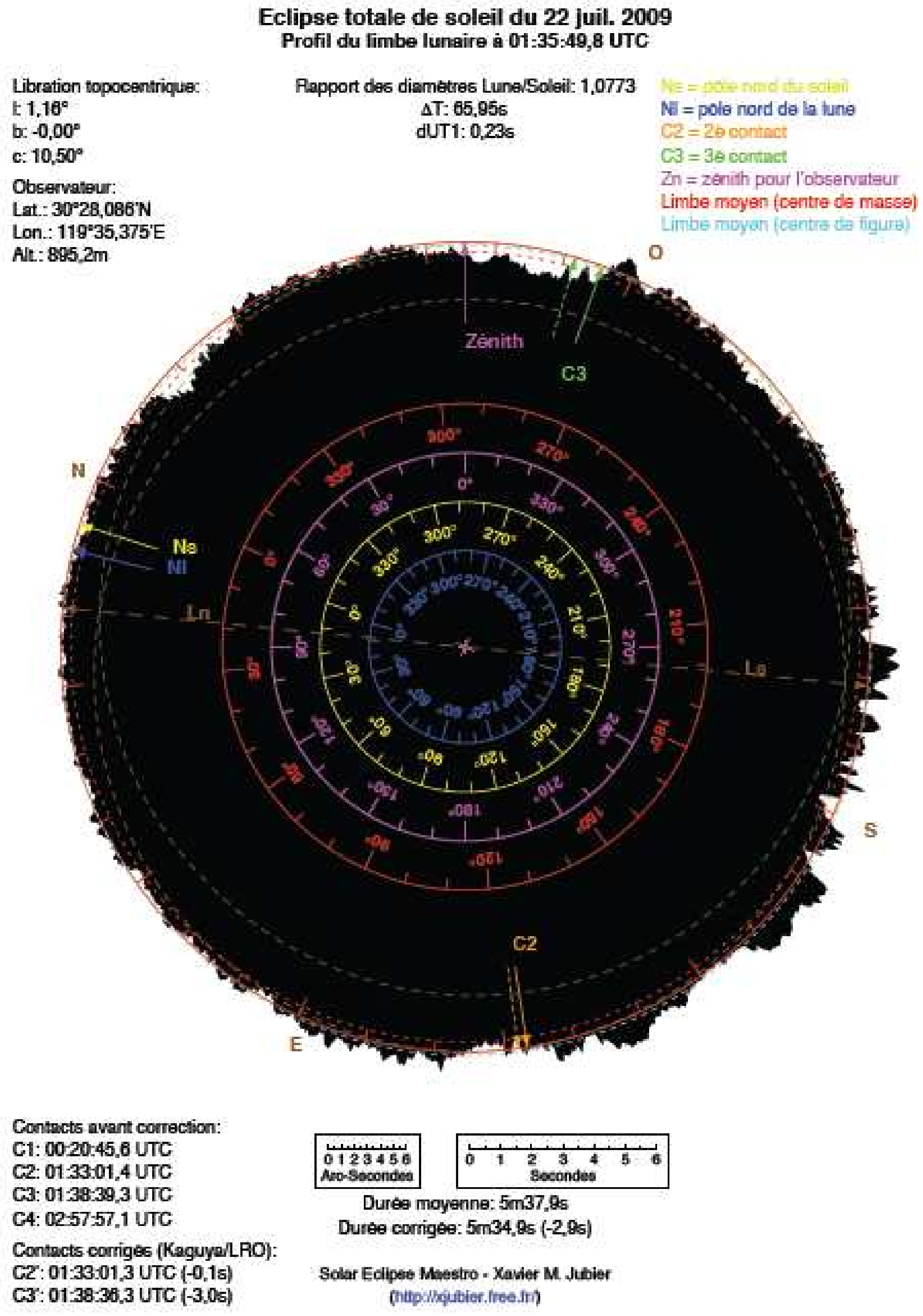

**Figure III-3-1:** *calcul des instants et position des contacts sur la carte du relief lunaire, pour l'éclipse totale du 22 Juillet 2009, sur le site de Tianhuanping en Chine au Nord Ouest de Shangaï. C2 et C3 représentent les contacts avec un profil lunaire calculé sans relief lunaire et C'2 et C'3 sont les contacts (pointillés) corrigés avec le relief lunaire.*

Le continu le plus intense (grains de Baily) sur la partie centrale correspond à la référence où l'altitude maximale est prise pour définir l'instant de contact et ensuite déduire la référence des altitudes $h$ = 0 km. Le montage extrait de la séquence du second contact de 2009, figure

III-3-2 permet d'identifier les correlations entre une image en lumière blanche et le spectre éclair correspondant réalisé aux mêmes instants:

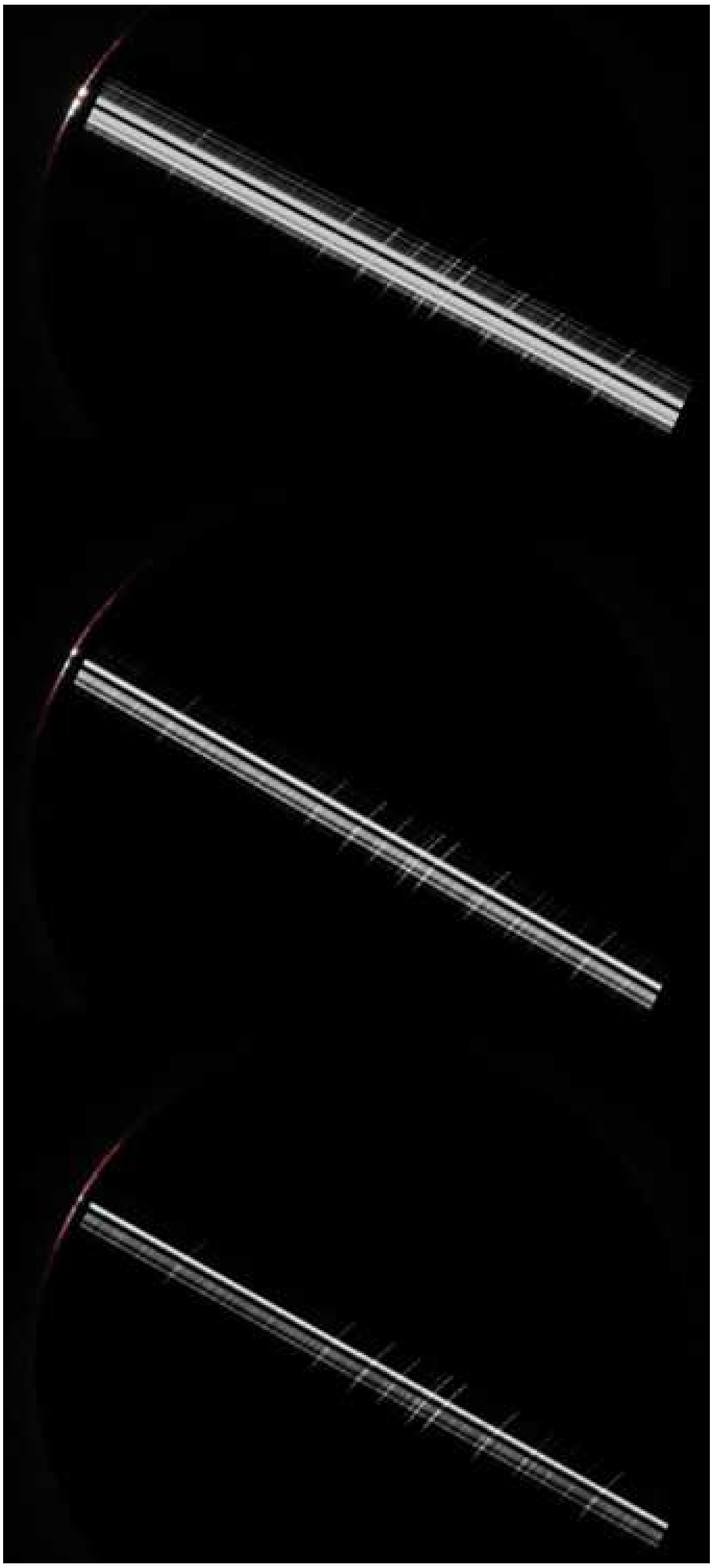

**Figure III-3-2 :** *Images des spectres éclair N° 60, 70 et 75 obtenus au second contact de l'éclipse du 22 Juillet 2009 à la cadence de 15 images/seconde, associés à une image en lumière blanche à la même échelle (arc rouge à gauche). Des grains de Baily sont obtenus avec un objectif de 500 mm de focale avec un appareil photo-numérique Canon, poses* de *1 ms à 400 iso avec un appareil photo numérique Canon EOS SD Mark II par Jean-Marie Munier.*

Lors de l’éclipse du 22 Juillet 2009, le logiciel « Lucam Recorder » n’était pas installé sur l’ordinateur ayant servi aux acquisitions, et celles-ci ont été effectuées au moyen du logiciel Iris, qui pouvait effectuer des acquisitions pour des caméras de type « Skynyx- 2.1 M Lumenera », en générant un fichier de format AVI. La dynamique de 8 bit a été utilisée et la cadence était de 15.1 images par seconde quel que soit le débit de données. La saturation des spectres lors des contacts (spectres de grains de Baily encore intenses après l’éjection des filtres amovibles) n’a pas ralenti la cadence d’acquisition, ce qui a été un avantage considérable, et où le processus d’acquisition n’a pas été interrompu.

A partir des numéros d’images des séquences, il a été possible d’étalonner la correspondance numéro d’image et altitude pour déterminer les altitudes, voir figure III-3-3 et III-3-4.

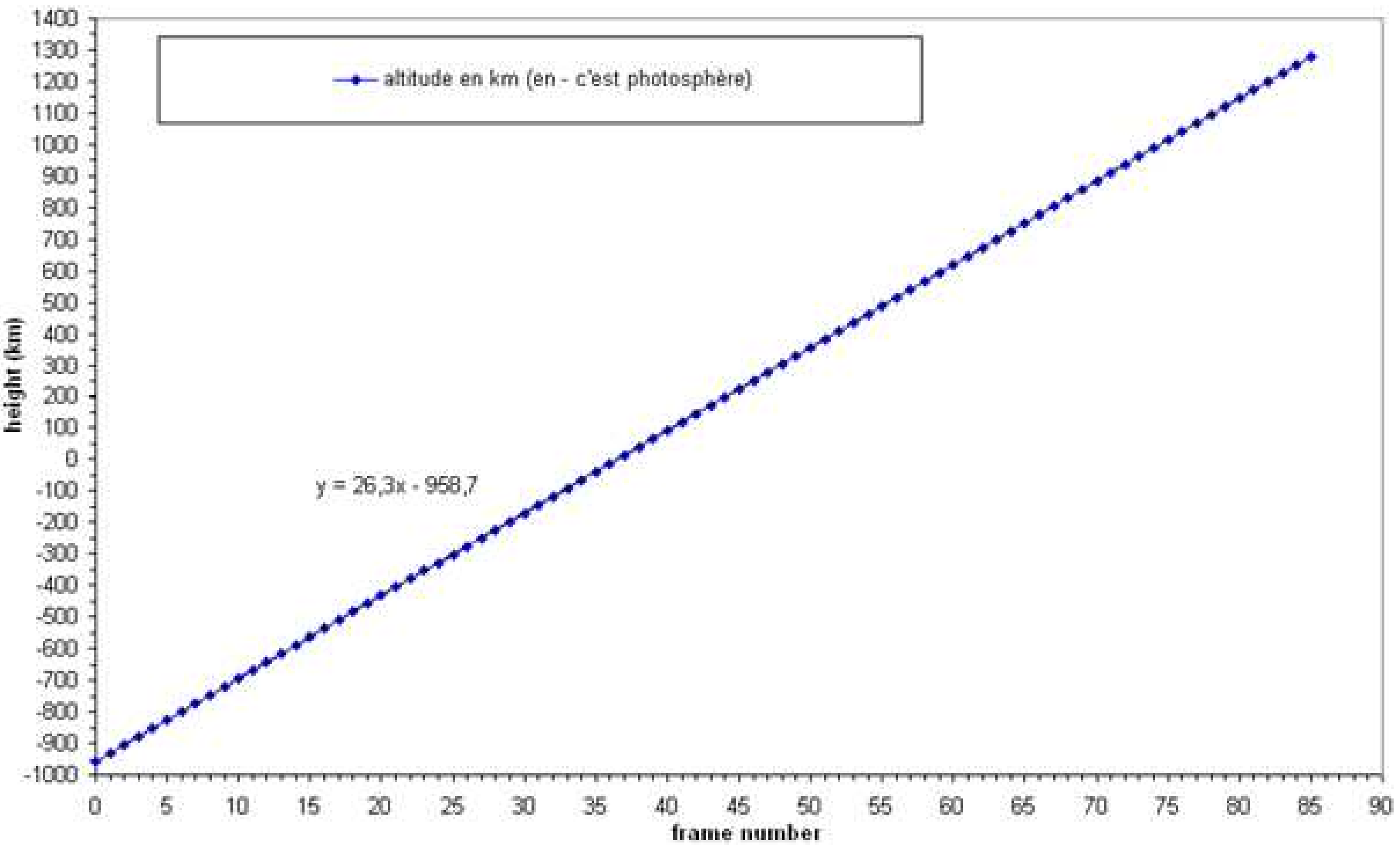

**Figure III-3-3:** *courbe d’étalonnage des altitudes en fonction du numéro d’image au 2$^{ième}$ contact de l’éclipse du 22 Juillet 2009.*

Les hauteurs ont été étalonnées avec les numéros d’image, compte-tenu de la cadence de 15 images par seconde et du mouvement différentiel de la Lune sur le Soleil et en prenant en compte le spectre du dernier grain de Baily le plus intense au centre du croissant au second contact. De la même façon, le premier spectre du grain de Baily a été relevé avant le troisième contact à la fin de la totalité, comme représenté en figure III-3-4:

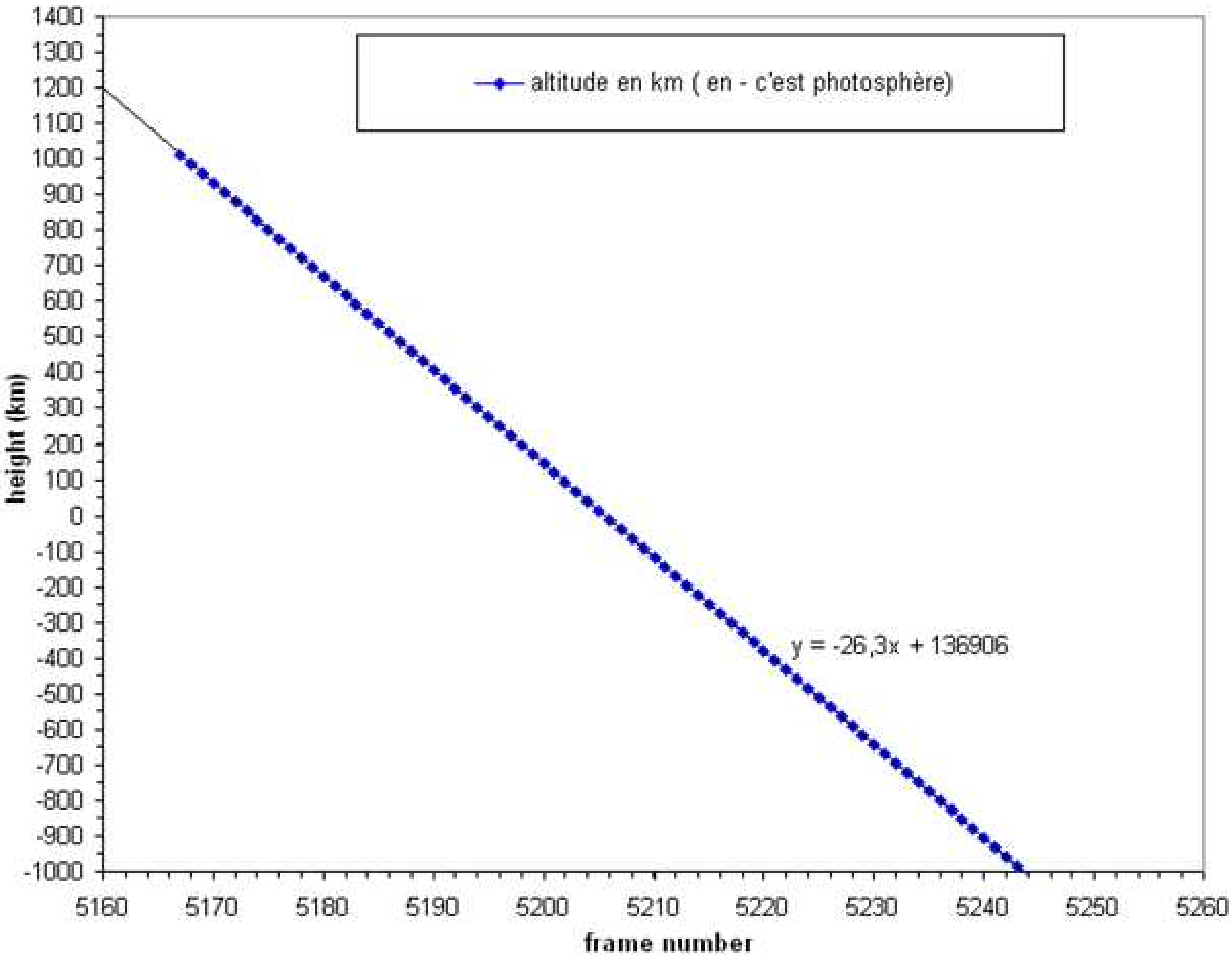

**Figure III-3-4:** *courbe d'étalonnage des altitudes en fonction du numéro d'image au 3ième contact de l'éclipse du 22 Juillet 2009.*

Une vérification avec le profil de la Lune est effectuée afin de distinguer des corrélations d'origine solaires au second et troisième contact de l'éclipse du 22 Juillet 2009, voir figures III-3-5 à III-3-9. Le bord de la Lune est aussi utilisé pour analyser les basses couches de l'atmosphère solaire, grâce à son mouvement naturel de l'ordre de 0.5 secondes d'arc par seconde de temps avec la caméra CCD Lumenera à la cadence de 15 images/seconde. L'intervalle d'altitude entre 2 spectres éclairs consécutifs limités par le mouvement naturel de la Lune était de 26.3 kilomètres, dans l'atmosphère solaire au dessus du limbe.

La région de transition photosphère - couronne solaire était autrefois appelée couche renversante car elle correspond à une inversion, où les raies d'absorption de Fraunhöfer une fois le disque occulté sont vues en émission, sur une très fine couche d'étendue inférieure à 500 km qu'il était très difficile de résoudre avant l'apparition des caméras CCD rapides. Le continu entre ces raies est aussi un élément important qu'il faut considérer pour appréhender le « vrai bord » du Soleil défini pris en dehors des raies d'émission qui produisent un embrillancement sur le limbe. Ce continu situé entre des raies d'émission est le prolongement du continu photosphérique superposé au continu chromosphérique lorsqu'on intègre sur la ligne de visée.
Nos résultats de spectres éclairs nous ont conduits à définir une nouvelle couche, la « mésosphère », située dans les parties les plus profondes des interfaces photosphère-chromosphère. La figure V-3-4 décrit ces couches de mésosphère, où elle est est associée à la mésogranulation par analogie. La couche mésosphérique est définie comme une partie de l'atmosphère solaire située au dessus de la haute photosphère (proche du minimum de température, comme on le verra plus loin dans les analyses des courbes de lumière), et qui correspond au continu sur les spectres éclairs. Dans la mésosphère, la couronne n'est pas dominante. Elle est constituée d’un mélange de plasma composé d’atomes excités et une fois ionisés (en dessous de la chromosphère plus dynamique).
La référence donnée à cette nouvelle couche mésosphérique a fait l'objet d'un article paru dans Journal of Advanced Research -JARE suite à la conférence IAGA III, Bazin Koutchmy 2012, voir Annexe 2. Cette couche de mésophère est aussi discutée au chapitre IV-3.
La figure III-3-5 montre un extrait de séquence de spectre éclair pour mieux décrire la mésosphère, mais réalisé dans un domaine de longueur d'onde situé autour de 4550Å, où une partie des raies d’émission les plus intenses ont été identifiées, et où ont été indiqués les potentiels d’excitation en électrons-volts d’atomes neutres et une fois ionisés (low FIP).
Ce potentiel,  pour n’importe quelle raie est l’énergie en électron-volt qui est requise pour élever l’atome depuis son plus bas niveau d’énergie vers l’état excité. D’après l’Atlas de Moore, 1966.

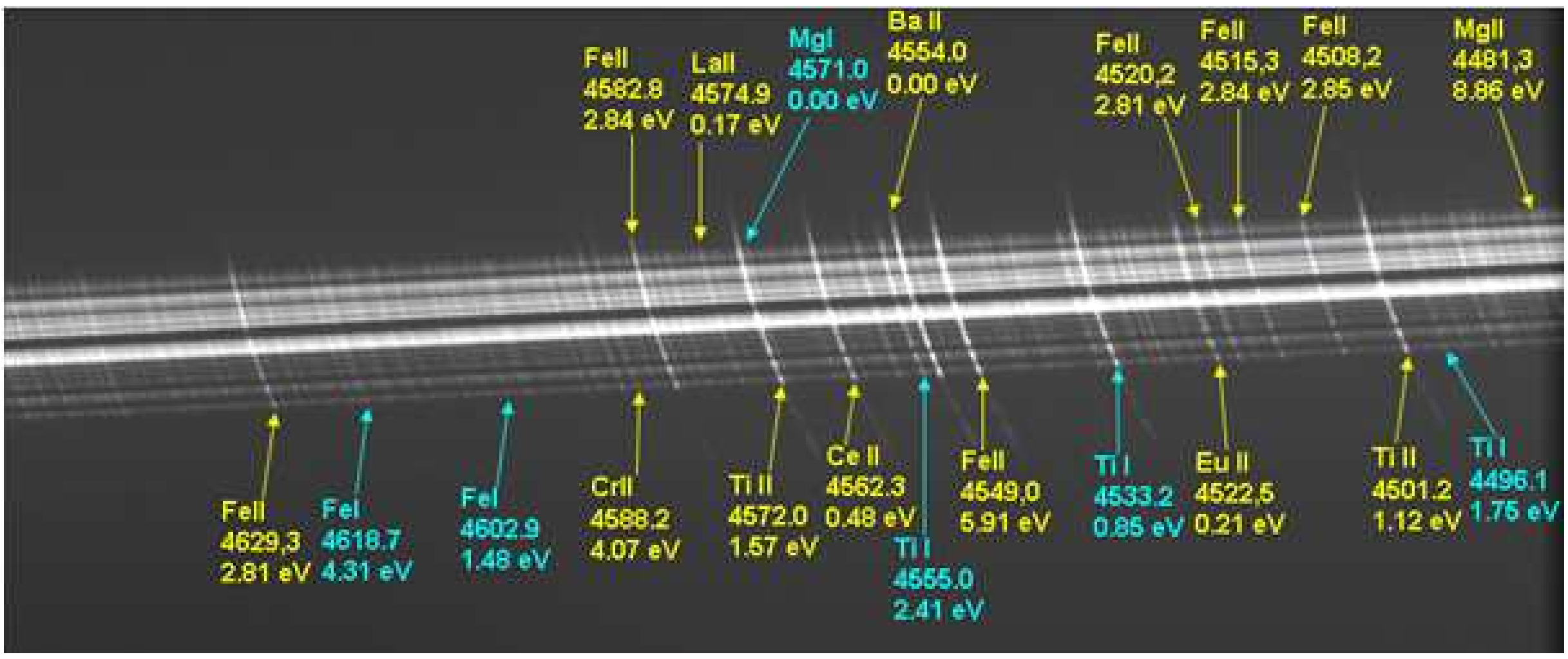

**Figure III-3-5:** *extrait de spectre éclair lors du second contact à l'éclipse du 22 Juillet 2009, montrant les myriades de raies low FIP dans le domaine spectral de  4480 à 4640 Å. Le continu du spectre du grain de Baily central plus intense est visible, et correspond à la fin du continu photosphérique où commence la mésosphère. La trace blanche correspondant au continu intense (grains de Baily) correspond à la référence où l’altitude maximale est prise. Les potentiels d’excitation sont indiqués en eV.*

Parmi les raies d’émission enregistrées aux spectres éclair, les raies du Barium Ba II 4553.8 Å Fe II 4629.3 et Ti II seront étudiées en détail dans le chapitre IV-2, avec des courbes de lumière, $I = f(h)$. En effet ces raies et surtout le Barium présentent des propriétés de faible potentiel d’excitation.
La raie du Fe II 4629.3 Å est en fait non résolue et le mélange de 3 raies:
Ti I, Co I .D’après les tables de Dunn et al 1968, ces 3 raies ont les mêmes intensités pour chaque élément. L’expérience des spectres éclair n’a pas permis de les différentier.
Une raie intense a été obtenue au troisième contact dans l’intervalle spectral 4380 Å à 4540 Å. Il s’agit de la raie de l’hélium neutre He I 4471 Å qui est une raie environ 100 fois plus brillante que la raie He I 4713 mais 3 à 4 fois moins brillante que la raie D3 de He I à 5876 Å. Cette raie He I 4471 Å a été observée jusqu’à des altitudes supérieures à 8000 km au dessus du limbe solaire à la fin de la totalité de l’éclipse du 22 Juillet 2009, et juste avant le troisième contact. Cependant les voiles nuageux ont atténué les domaines plus intenses des courbes de lumière des raies d'hélium neutre He I 4471 Å et He I 4387.9 Å, voir figures IV-2-1-9 et IV-5-1. Des correlations entre la raie He I 4471 « high FIP » et le relief lunaire d'une part puis avec le Ti II « low FIP » sont présentés sur la figure III-3-6.

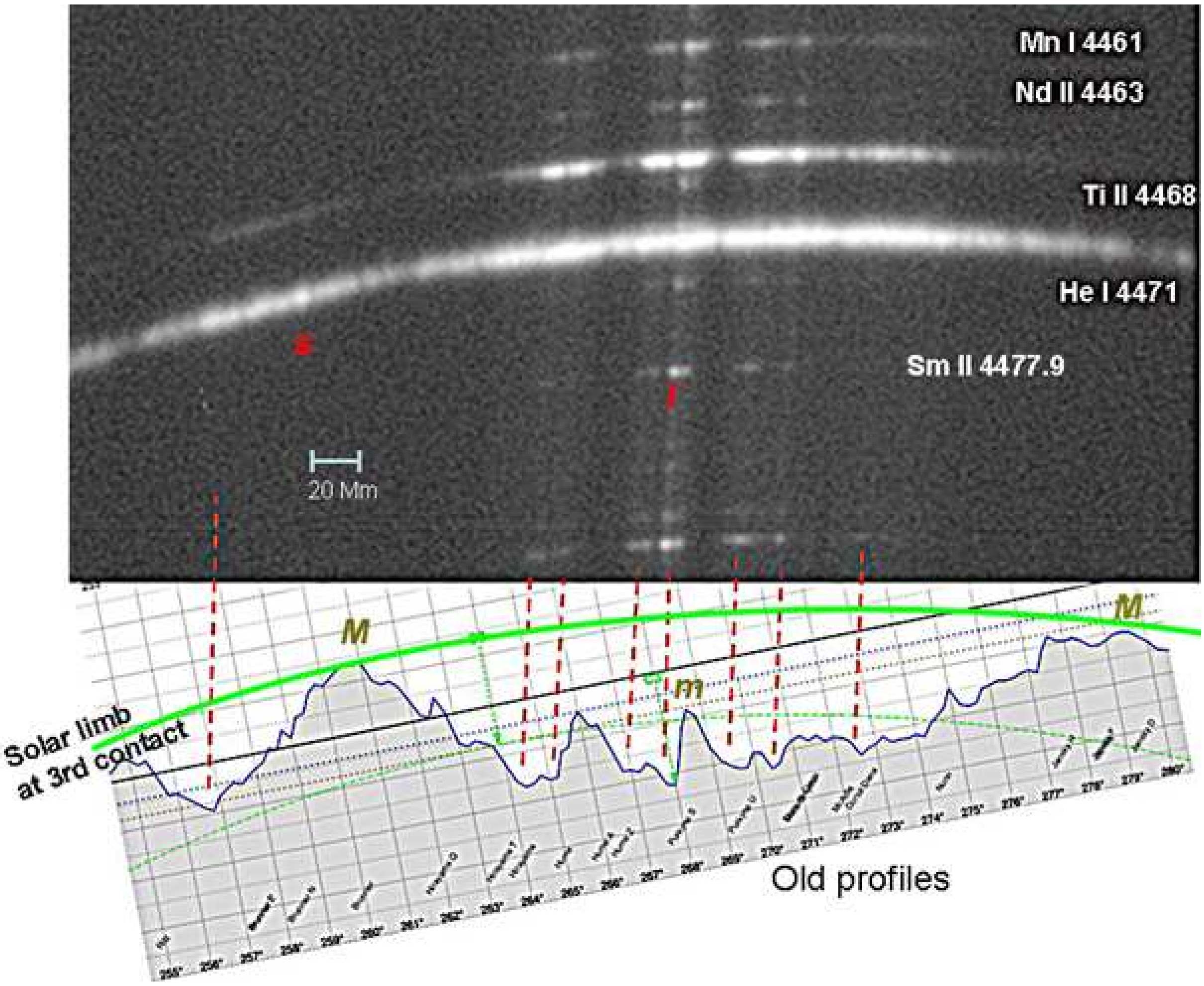

**Figure III-3-6:** *Corrélations entre le relief Lunaire (Calculé par Xavier Jubier) et les profils agrandis des spectres des raies low FIP, enveloppe d'hélium He I 4471 Å, et continu de 18 spectres éclairs sommés non redressés N°2474 à 2482. Les instants étaient entre 1h 38 min et 29.5 seconde à 1h 38 min et 30.0 seconde, qui correspondent aux altitudes de 1776 à 1555 km avant le 3$^{ième}$ contact de l'éclipse totale du 22 Juillet 2009.*

Les embrillancements dans les raies « low FIP » et He I 4471 Å sont corrélés avec le profil du bord de la Lune.
Ces résultats montrent que les interruptions les plus larges (10 à 20 Mm) sont liées au relief de la Lune. La diminution d'intensité produite par la petite montagne notée ***m*** est faible sur la raie He I 4471 Å. Elle devient plus importante sur la raie du Ti II 4468 Å, et encore plus importante pour les ions Mn I et Nd II. Ceci s'explique par le fait que ces raies Nd II, Mn I, et Ti II sont produites autour de 600 km d'altitude dans la région du minimum de température dans l'atmosphère solaire au dessus du limbe. Ces raies low FIP ont une étendue limitée contrairement à la raie He I qui s'étend radialement sur plusieurs milliers de kilomètres (enveloppe étendue), d'où cette modulation profonde par cette petite montagne ***m*** par le bas.
Le ***M***, qui est une montagne lunaire étendue, se traduit par une diminution d'intensité plus étendue sur la raie de He I 4471, et la raie du Ti II 4468 Å est très atténuée, par rapport aux régions sur la partie centrale des croissants.

Le ***l*** indiqué en rouge sur la raie du Samarium une fois ionisé Sm II 4477.9 Å du  spectre de la figure III-3-6 indique un point brillant qui semble anti-corrélé avec le profil lunaire. Ce point est environ 1.6 fois plus brillant que la partie juste à côté, mais sur la raie de l'hélium He I ceci est difficile à voir car cette raie est beaucoup plus intense que la raie du Sm II. La partie centrale des raies Mn I 4461 Å, Nd II 4463 Å, Ti II 4468 Å et Sm II 4477.9 Å semble un peu surexposée. Le ***S*** indique un embrillancement qui ressemble à un « Surge » tel que décrit dans l'article de Georgakilas et al 2001, et qui pourrait correspondre à une éruption.

Les images figure III-3-7 montrent cette même région spectrale agrandie 2 fois par rapport à la figure III-3-6, mais ce sont des images originales qui ne résultent pas d'une sommation d'images. Au début de l'instant du troisième  contact, où le continu chromosphérique commence à s'intensifier, les raies s'embrillancent également.

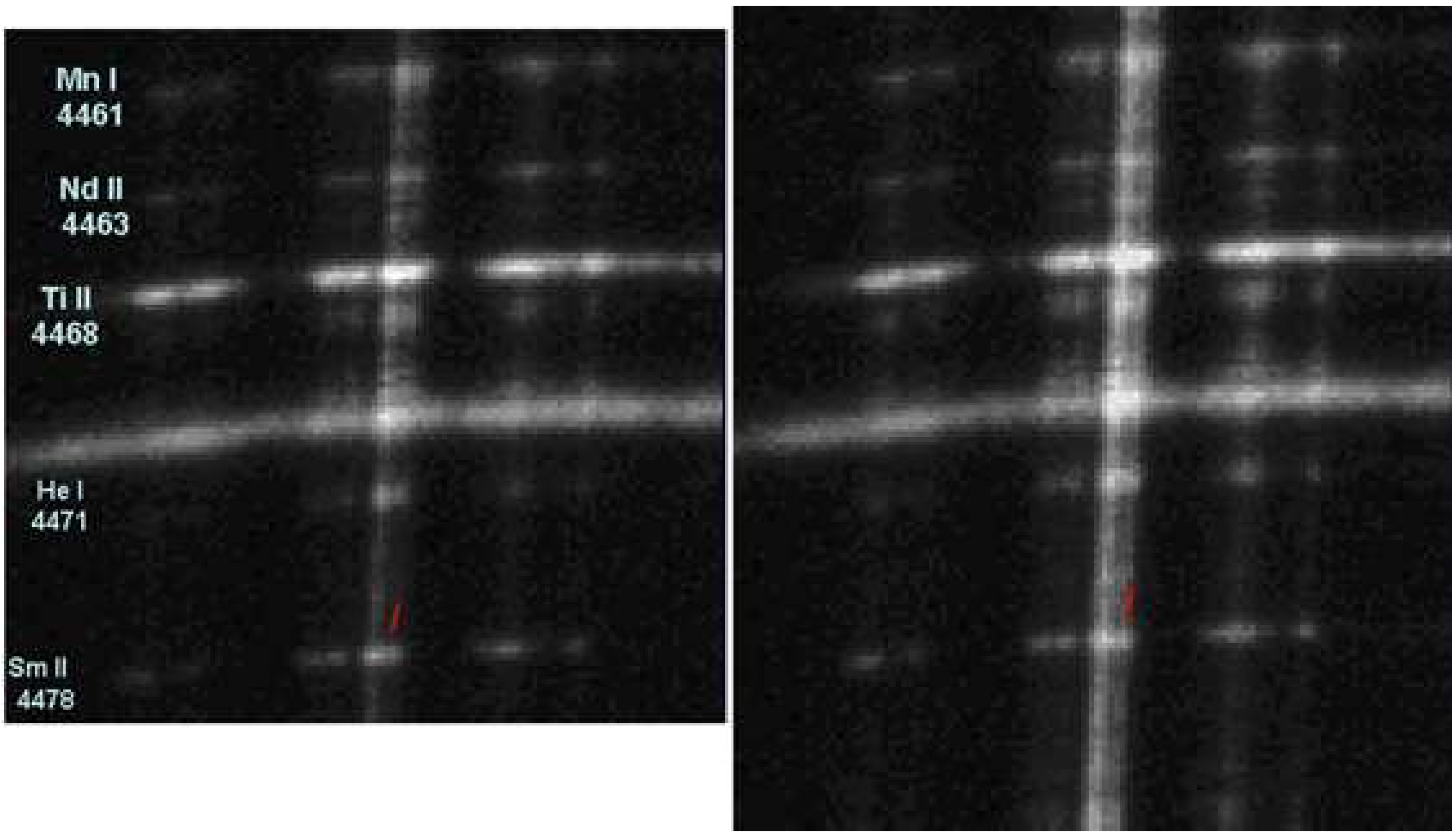

**Figure III-3-7:** *Extraits de spectres originaux (non sommés) agrandis 2 fois à l'instant du troisième contact où les raies ne sont pas surexposées. 330 milli- secondes  séparent  ces 2 images. Le renforcement (point brillant)  semble bien corrélé entre les raies low FIP et la raie He I 4471 Å. Frames N° 2500 à 1h 38 min et 31.2s,  et frame N° 2505 à 1h 38 min et 31.6 secondes, correspondant à des altitudes comprises entre 850 et 1000 km au dessus du limbe. La cadence était de 15.1 images/seconde.*

D'après ces résultats, il est possible de dire que cet embrillancement est présent dans les couches plus basses où sont formées les raies low FIP et Terres rares. La raie de l'hélium neutre He I 4471Å est observée à des altitudes plus élevées et au dessus des raies low FIP. L'interprétation demande une étude plus ciblée avec des documents complémentaires (observations au sol et espace) qui sont donnés dans le chapitre V.

D'autres corrélations sont bien visibles entre le continu et les raies low FIP dans une vallée lunaire où ces 2 types d'émission s'ajoutent en observant sur la ligne de visée, mais se différencient lorsqu'on regarde vers l'extrêmité et prolongement des croissants, où le continu a disparu, tandisque les raies d'émission low FIP sont encore visibles. Ceci confirme que le continu provient des structures des couches plus profondes au-dessus de la photosphère, de la mésosphère et de la chromosphère qui se superposent sur la ligne de visée.

Cette même étude des corrélations a été effectuée au second contact de cette éclipse de 2009, et nous avons utilisé 60 images sommées autour du Ba II 4553.8 Å et les mêmes notations que précédemment sont utilisées, voir les figures III-3-8 et III-3-9.

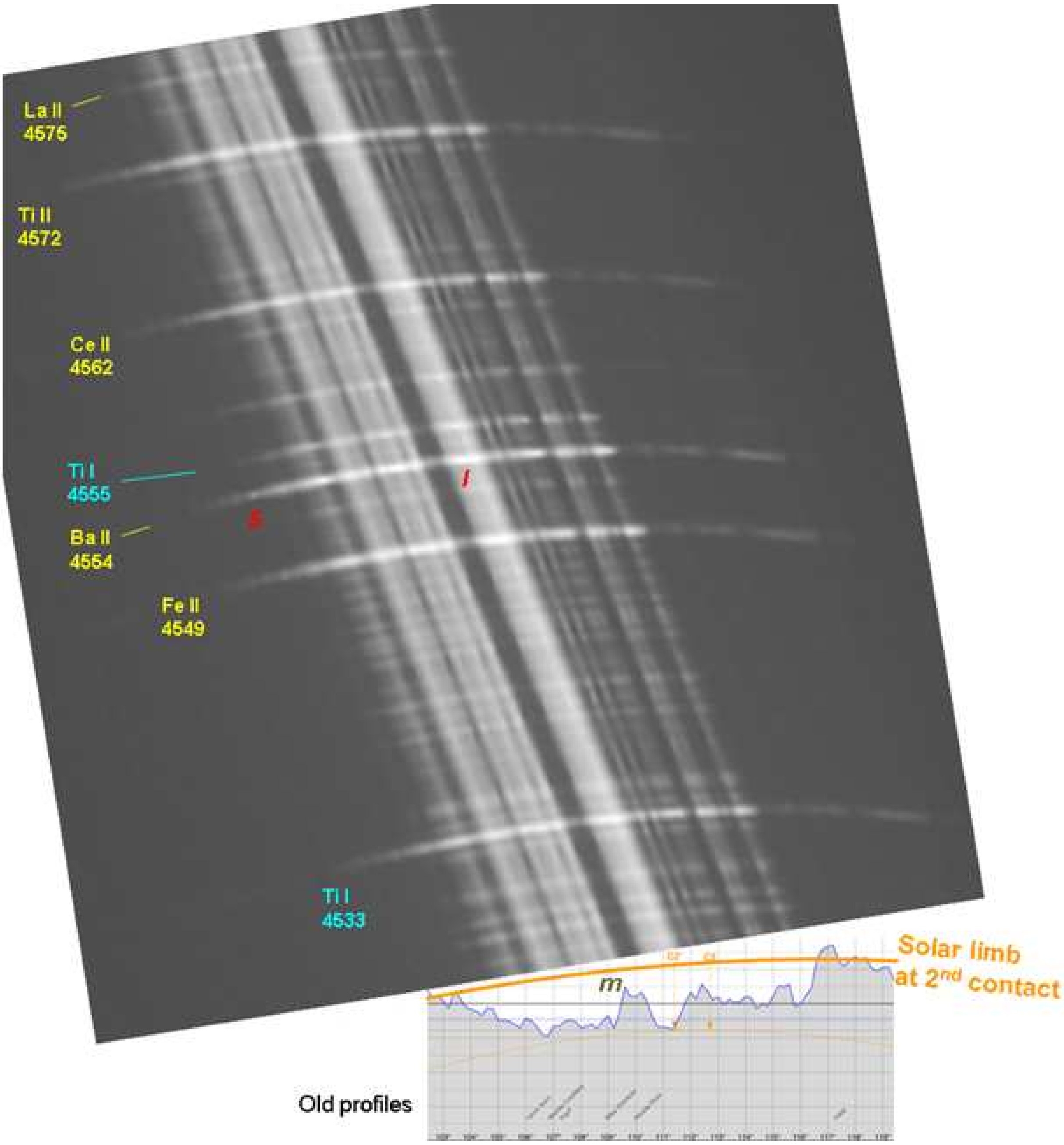

**Figure III-3-8:** *Corrélations entre le relief Lunaire (Calculé par Xavier Jubier) et les profils agrandis 4 fois des spectres des raies low FIP, et continu. 60 spectres éclairs sommés non redressés quelques secondes après le deuxième contact avant le début de la totalité, de N° 60: 1h 32 min et 34.4 seconde à N°120: 1h 32 min et 38.4 seconde. Les altitudes correspondantes sont de 619 à 2197 km au dessus du limbe. Eclipse totale du 22 Juillet 2009.*

Les corrélations entre le continu et les raies low FIP dans les vallées lunaires, sont confirmées sur cet extrait de spectre au second contact en ayant ajusté la taille du profil lunaire avec l'extrait des spectres éclairs, où les embrillancements dus aux spectres des grains de Baily coïncident avec les profils des vallées lunaires calculés. Les 2 extraits de spectres éclairs agrandis 4 fois en figure III-3-9 du second contact, servent à comparer les modulations plus faibles dans les images des raies d'émission. Par exemple, les raies du Ba II et Fe II qui ont des intensités voisines ont des modulations correlées dans leurs embrillancements.

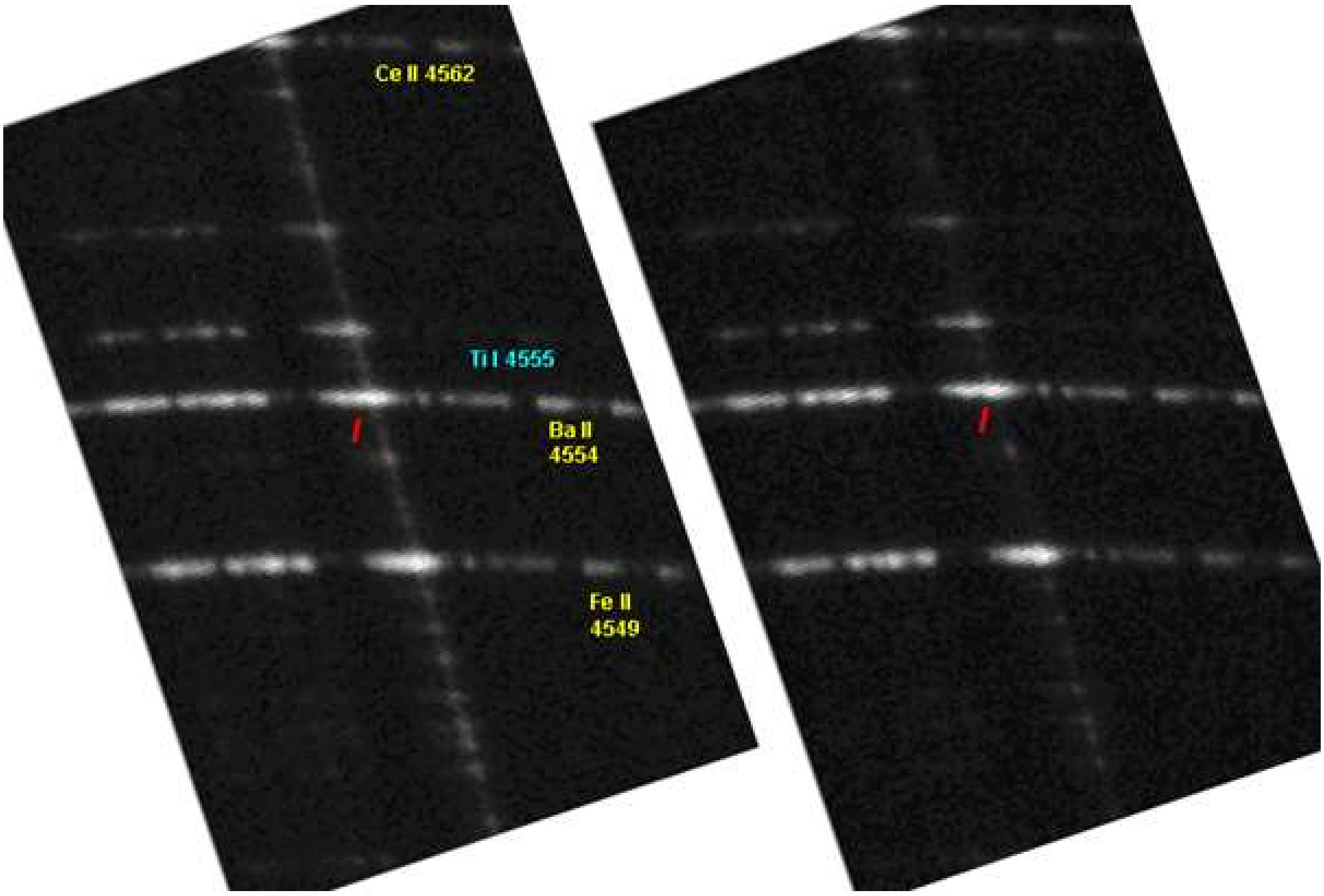

**Figure III-3-9 :** *Spectres non sommés agrandis 4 fois. 330 milli- secondes séparent ces 2 images. Le renforcement (point brillant) semble bien corrélé entre les raies low FIP du Ba II, Fe II, Ti I. Frames N° 100 : 1h 32 min 37.1 seconde soit h=1671 km et N° 105:1h 32 min 37.4 s soit h =1806 km. Cadence de 15.1 images/seconde. Second contact de l'éclipse du 22 Juillet 2009.*

Les spectres éclair obtenus à l'éclipse totale du 22 Juillet 2009 ont été analysés plus en détails, à partir de mesures d'intensités en fonction de l'altitude. Elles ont été réalisées sur les zones indiquées par la lettre « l » en rouge figure III-3-9 pour quelques raies d'émission low FIP comme le Fe II, Ti II, Ba II, dans le continu et la raie high FIP de l'hélium He I. Ces relevés d'intensité ont été répétés pour chaque spectre, afin de construire les courbes de lumière. La limitation ayant affecté les spectres éclair de 2009 provient de l'utilisation de la caméra Lumenera avec une dynamique limitée à 8 bits (256 niveaux d'intensité). Il n'a pas été possible d'observer les raies d'absorption de Fraunofer sur les grains de Baily encore intenses, car ceux-ci étaient saturés. Ce problème a été résolu en utilisant un autre logiciel, Lucam Recorder, qui a permis d'utiliser cette même caméra, avec une dynamique de 12 bits (4096 niveaux d'intensité), lors de l'éclipse totale du 11 Juillet 2010.

## III-4) Spectres éclair à l’éclipse du 11 Juillet 2010

Cette expérience de spectres éclairs a été réalisée en Polynésie Française, sur l’atoll de Hao dans les Iles Tuamotu à 1000 km à l’ouest de Papeete. La même expérience transportable du réseau-objectif avec la lunette 50/600 mm munie de la caméra CCD Lumenera Skynyx 2.1 M a été utilisée. La table récapitulative II-1-8 indique les principaux paramètres. Nous étions situés au bord de l’océan vers l’Est où avait lieu l’éclipse, et nous avions le courant électrique. Les conditions météorologiques étaient voilées au moment de l’éclipse sur l’emplacement où nous étions mais cela n’a pas empéché de réussir nos observations. Une expérience du CNES de mesure photométrique des courbes de lumière lors de l’éclipse a été réalisée, et décrite dans l’article Koutchmy et al 2011 MTPR 10. Cette expérience a consisté à utiliser plusieurs photométres répartis sur plusieurs sites dans la bande de totalité, et un des photomètres a été utilisé sur notre site de Hao. Cependant des spectres éclair ont pu être enregistrés au second et troisième contact avec une cadence moyenne de 10 spectres/seconde, et nous avons pu comparer les chronodatations des photomètres CNES avec les chronodatations GPS. Une autre expérience de spectrographe sans fente, avec réseau-objectif a été assurée par Serge Koutchmy avec un objectif de 300 mm de focale précédé par un réseau en transmission de 1200 traits/mm, et au foyer un appareil photo numérique Canon 60 D a été utilisé.
La figure III-4-1, comme pour les éclipses de 2008 et 2009, représente le profil du relief lunaire agrandi avec les positions des instants de contacts C2 avant la totalité et C3 juste après la totalité, et la flèche rose indiquant la direction du zénit est indiquée en haut.

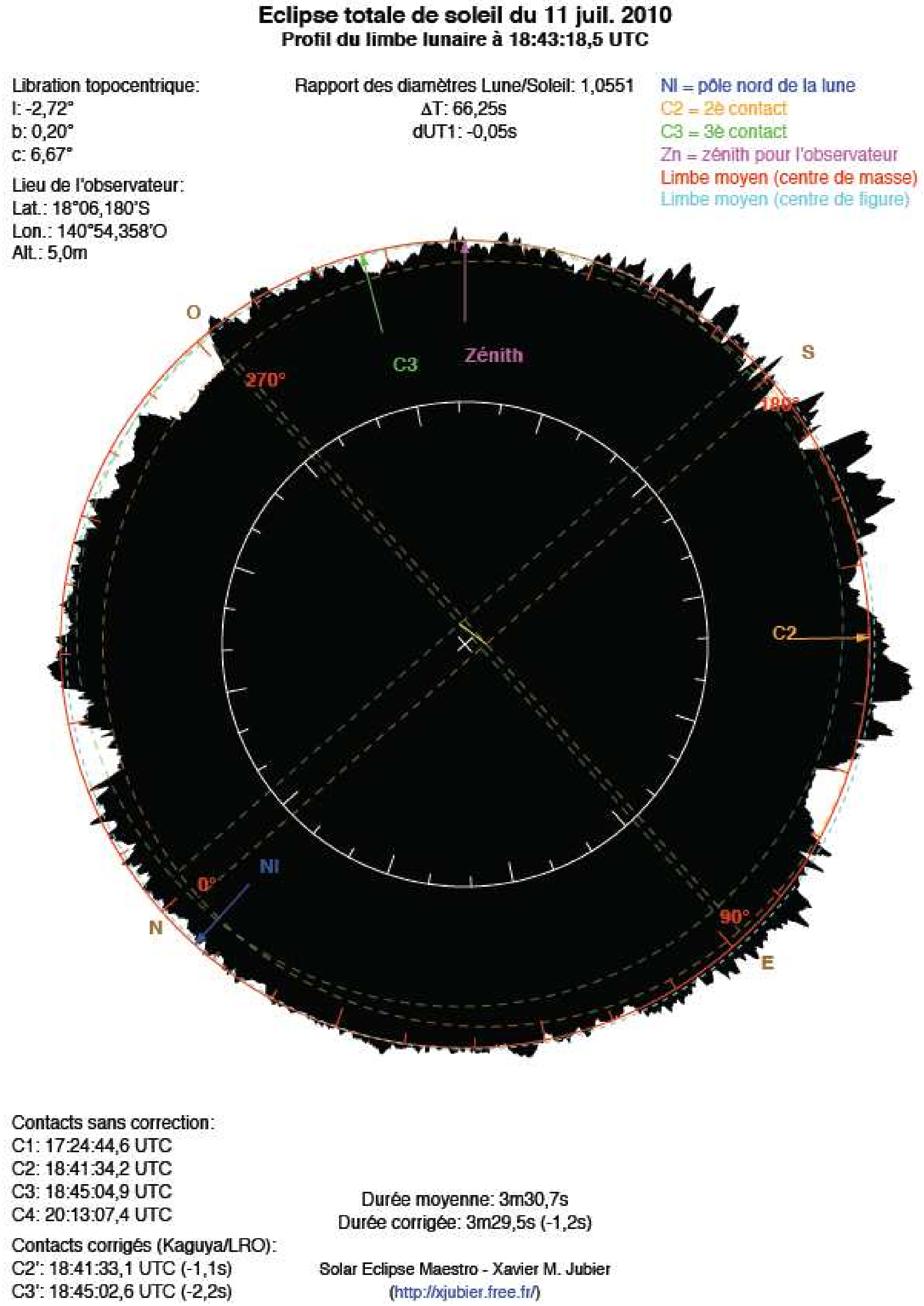

**Figure III-4-1 :** *calculs du relief lunaire et instants des contacts de l'éclipse totale du 11 Juillet 2010 pour l'atoll de Hao en Polynésie Française, d'après Xavier Jubier*

Une séquence de spectres éclair a été obtenue sur une petite expérience auxiliaire au moyen d'un réseau 1200 tr/mm devant un objectif F = 300 mm et de 40 mm de diamètre utile, voir figure III-4-2, incluant le même domaine spectral que celui utilisé en 2009.

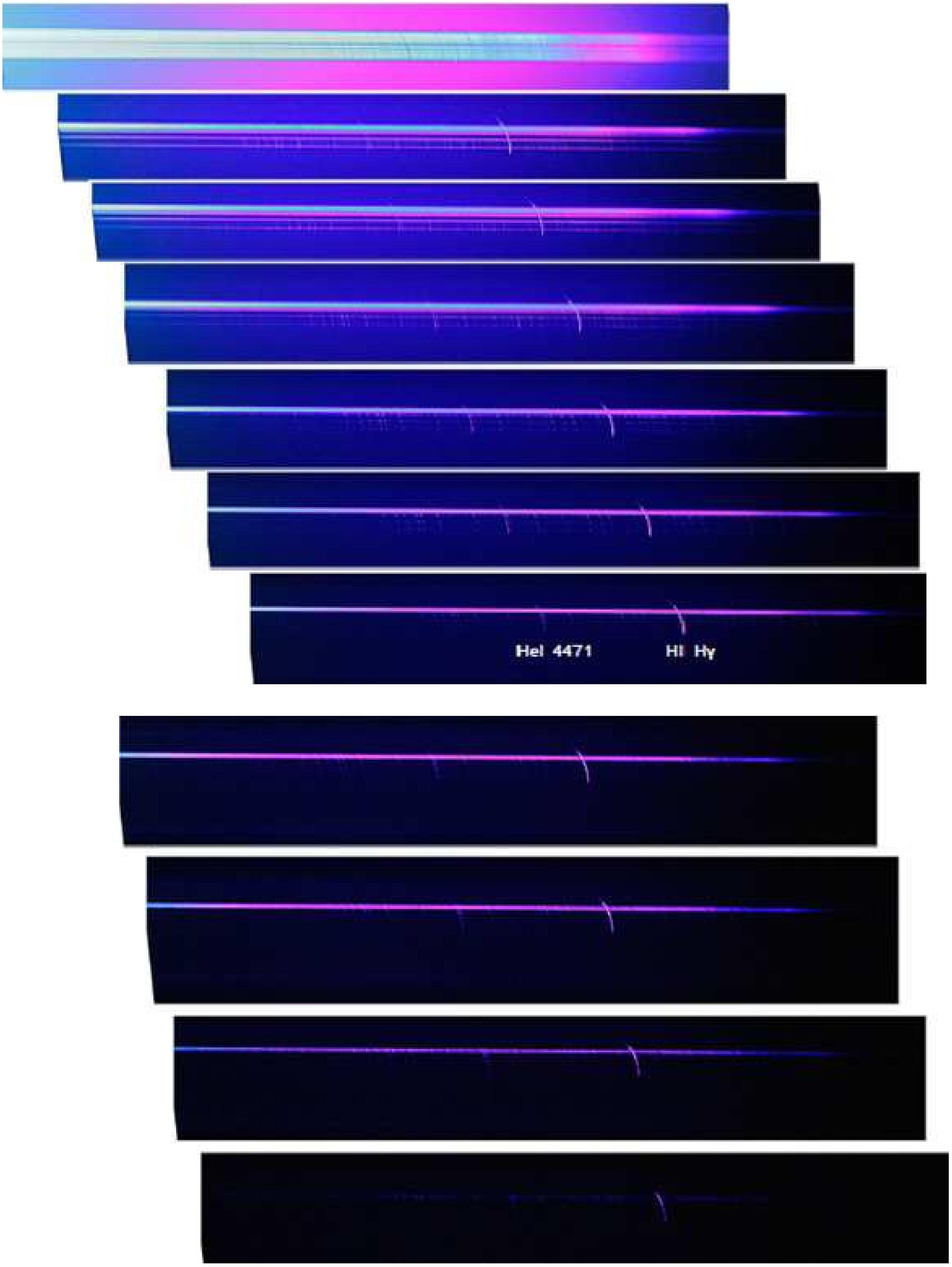

**Figure III-4-2:** *séquence de spectres éclair, cadence de 2 spectres/seconde au 2nd contact de l'éclipse du 11 Juillet 2010, région des raies H$\gamma$ 4340Å et raie de l'hélium neutre He I 4471 Å. Le premier spectre avant le contact, est atténué avec une densité neutre. Canon 60D.*

Les continus chromosphérique et coronal sont visibles sur ces spectres éclair, ainsi qu'une myriade de petites raies en émission, et les raies de l'hélium neutre He I 4471 Å et H I $\gamma$ 4340 Å, plus intenses. Le premier spectre réalisé quelques secondes avant le second contact et avant l'éjection de la densité neutre, présente les raies d'absorption de Fraunhofer. L'extrême limbe de la haute photosphère intense était encore visible, avant le second contact.

Les photomètres du CNES situés dans différents sites, dont certains en bordure de la bande de totalité, ont permis d'enregistrer des courbes de lumière. Les résultats obtenus sur le site de Hao Nord (atoll Polynésie Française) où nous avions installé nos expériences, sont indiqués dans les figures III-4-3, III-4-4 et III-4-5. Les contacts sont indiqués par « C2 », « C3 ».

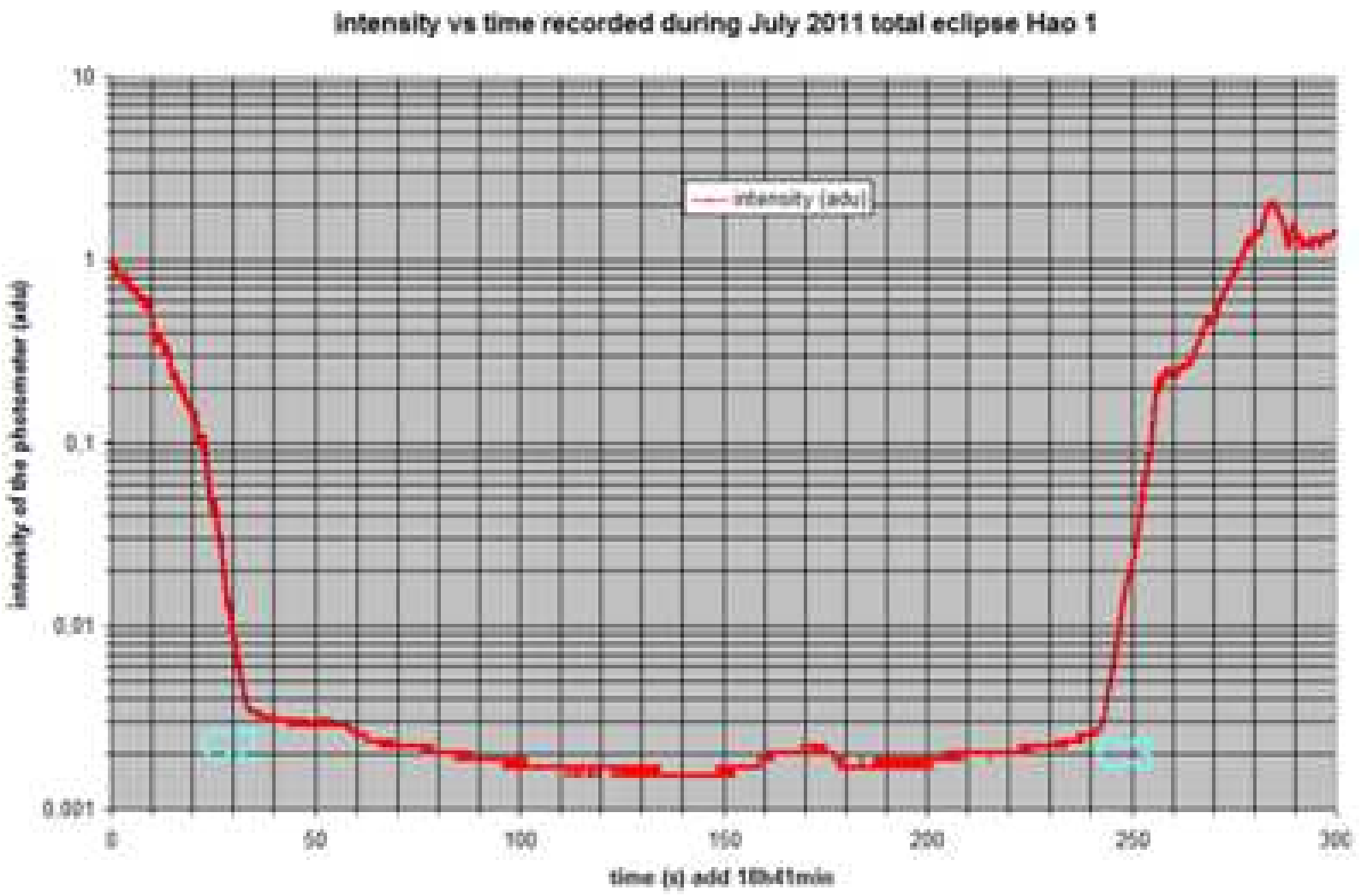

**Figure III-4-3:** *courbes de lumière obtenues avec le photomètre du CNES à raison de 100 points/seconde en contribution à la mission spatiale Picard. Les fluctuations peuvent être dues aux voiles nuageux. Les indications C2 et C3 indiquent le second et troisième contact.*

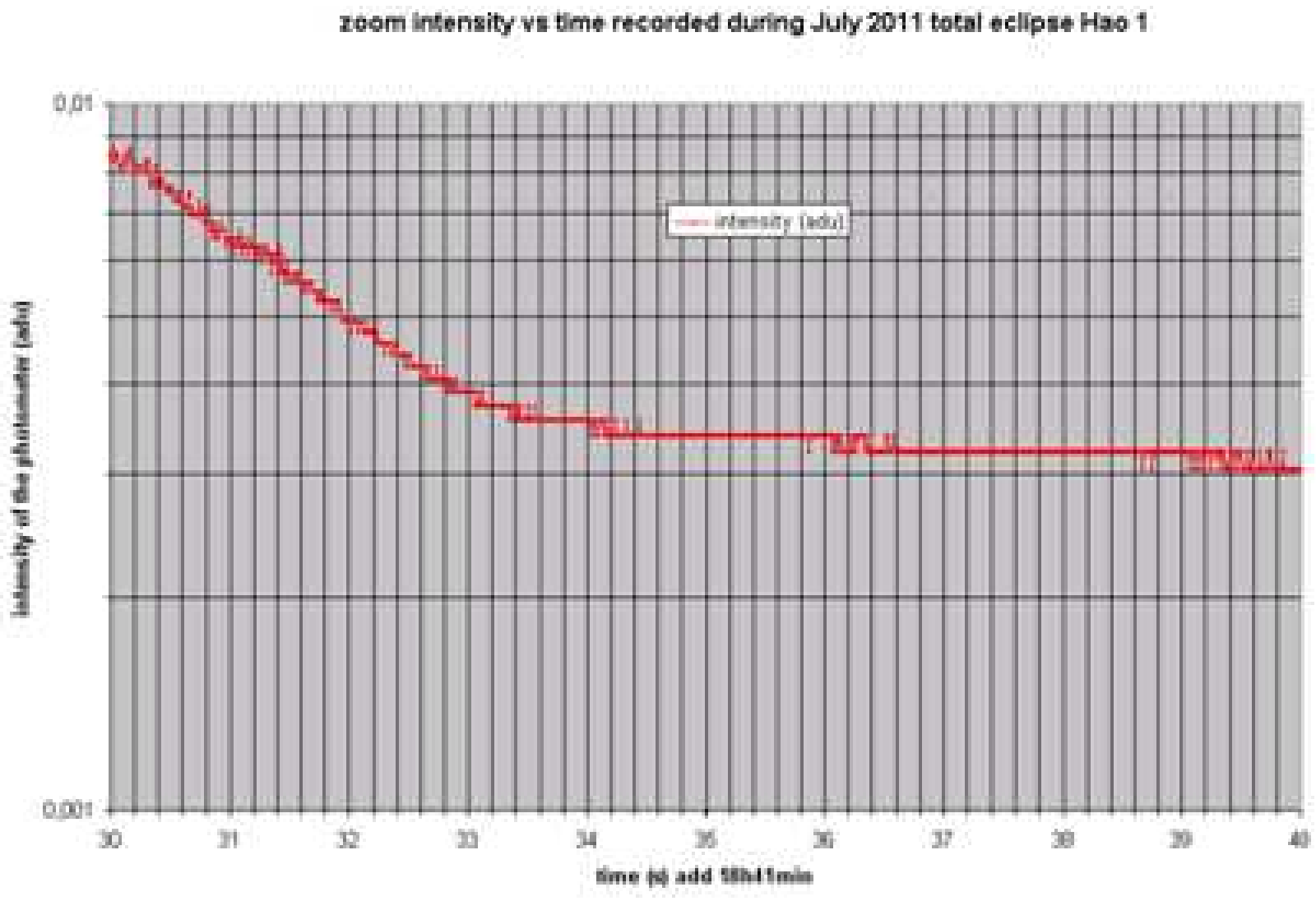

**Figure III-4-4:** *zone agrandie de la figure III-5-3 autour du second contact C2 estimé à 18h 41min et 33.6 s, montrant le changement de pente des variations du flux.*

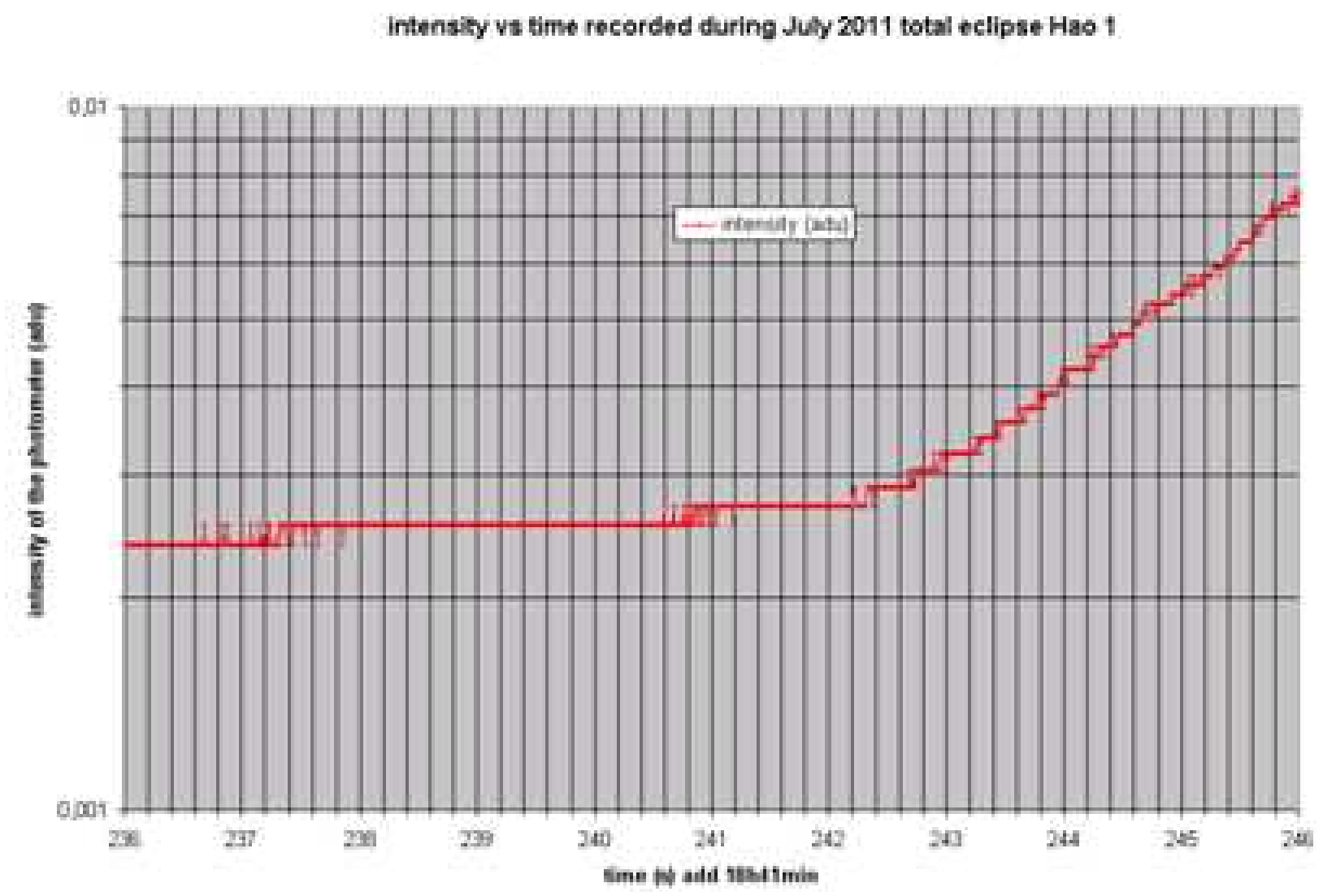

**Figure III-4-5:** *zone agrandie de la figure III-5-3 autour du troisième contact C3, montrant le changement de pente des variations du flux. Cet instant est estimé à 18h 45min et 2.6 s*

Ces courbes de lumière ont servi à évaluer les instants de contact, et estimer les fluctuations du fond de ciel durant la totalité. Ils ont permis de comparer et vérifier les instants des contacts déterminés par nos expériences de spectres éclair et imagerie W-L. Les résultats de chronodatations étaient en bon accord avec des éccarts inférieurs à une seconde. Ces résultats ont été déterminés en analysant chaque spectre éclair chronodaté, et pour lesquels des mesures d'intensité sur la partie continue du spectre a été réalisée.
L'extrait de spectre éclair III-4-6 illustre les corrélations entre le relief lunaire (vallée lunaire larges) avec le continu et les raies d'émission low FIP. Les raies He II 4686 Å et He I 4713 Å sont à droite en bordure de champ.

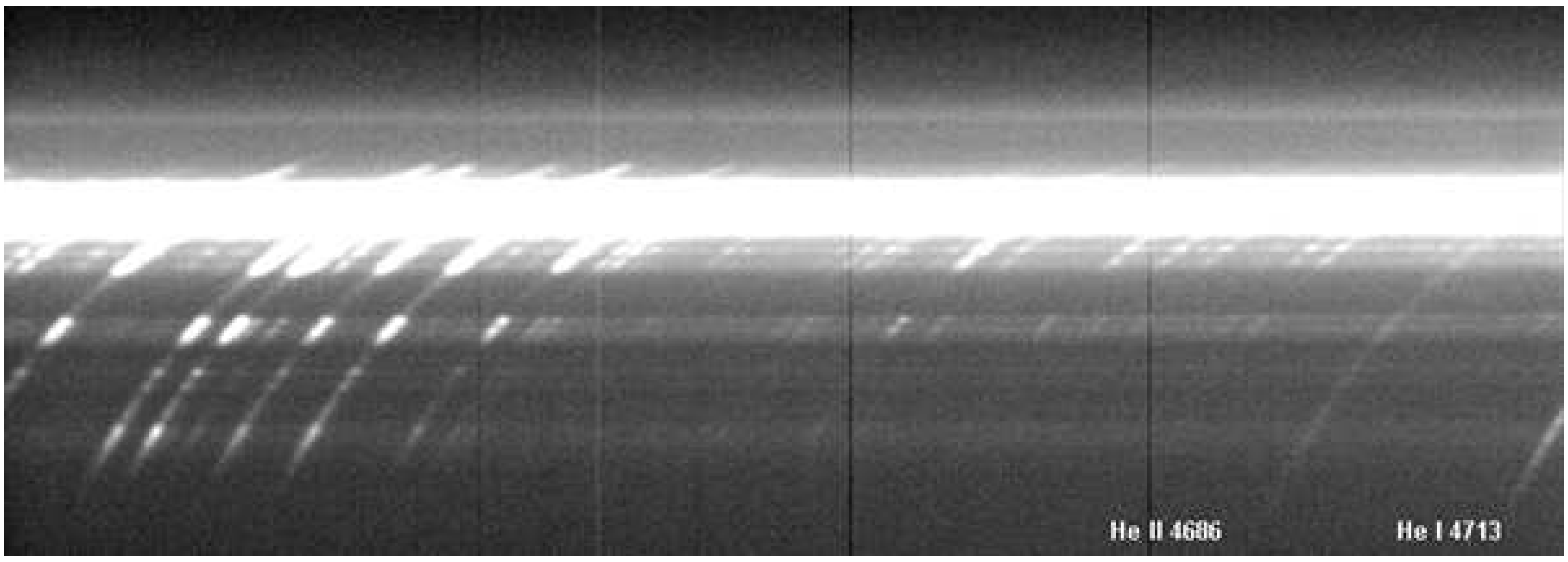

**Figure III-4-6 :** *spectre éclair au second contact, avec la raie He II 4686 Å au bord du champ et la raie He I 4713 Å en bordure du champ. 23 images sommées de N° 2793 à 2816. Deuxième contact lors de l'éclipse totale du 11 Juillet 2010.*

Les parties les plus intenses correspondent aux embrillancements produits dans les vallées lunaires qui laissent passer une partie encore intense du rayonnement provenant des hautes couches de la photosphère encore très intenses. Ces embrillancements diminuent en s'éloignant vers l'extrémité des croissants dans l'image des raies d'émission, où le bord de la Lune occulte des altitudes de plus en plus élevées de la chromosphère (Soleil plus petit) comme sur le spectre éclair en figure III-4-7.

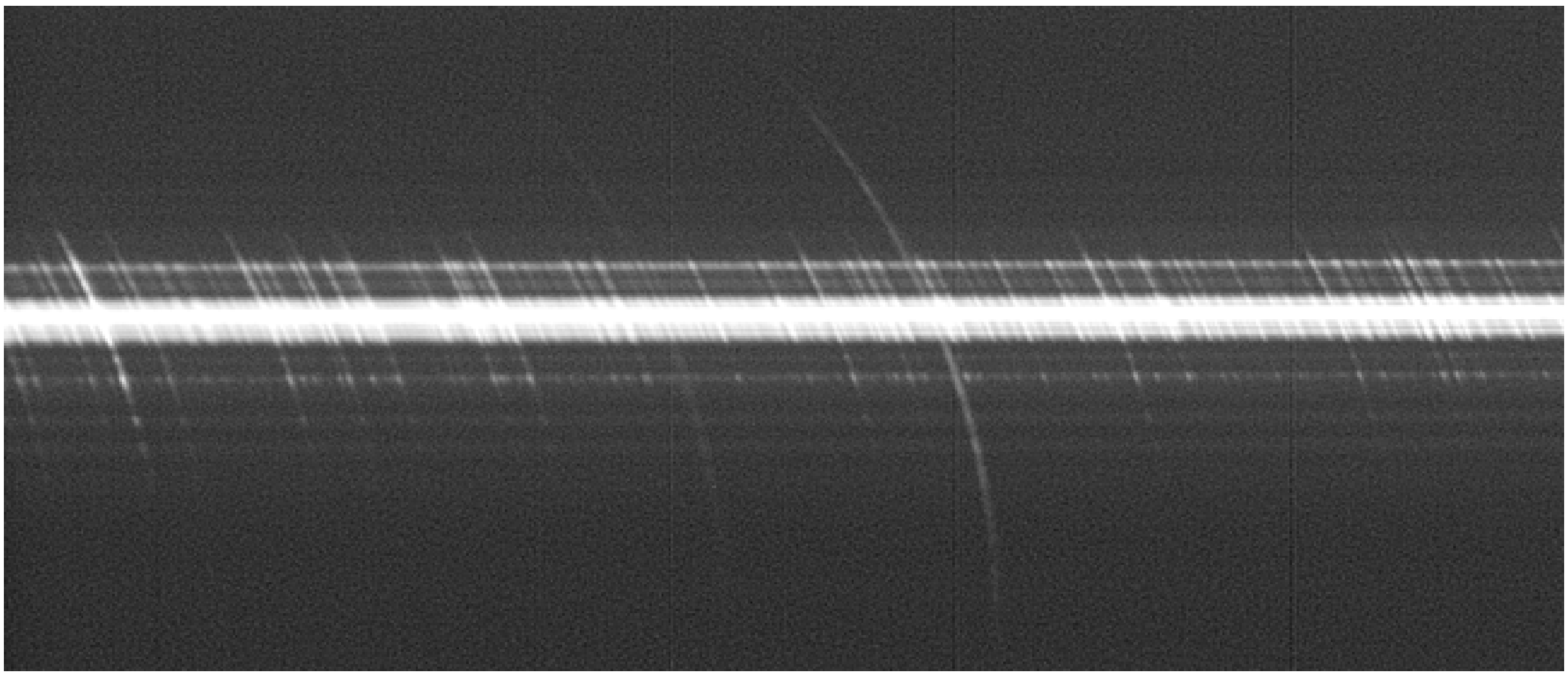

**Figure III-4-7:** *Image résultant de 26 spectres sommés (images 5120 à 5146) correspondant à une altitude moyenne estimée à 400 km au dessus du limbe photosphérique une seconde avant le 3ième contact. Ces spectres ont été convertis en échelle logarithme puis un masque flou de 2 pixels a été appliqué, contraste 2 pour faire apparaître plus de détails et les faibles intensités. Une myriade de petites raies low FIP ainsi que les 2 enveloppes d'hélium neutre He I 4713 Å et hélium ionisé He II 4686 Å apparaissent au milieu du spectre comme des arcs étroits et étendus.*

L'extension du continu entre les raies low FIP est bien visible dans les grains de Baily et il est modulé par le relief lunaire. Ceci montre une bonne corrélation entre le spectre du continu et les raies d'émission « low FIP » qui sont formées au dessus du continu.
Par ailleurs, ces spectres mettent en évidence que les enveloppes d'hélium apparaissent plus élevées que les raies low FIP. En effet, le bord de la Lune n'occulte pas entièrement les enveloppes d'hélium plus étendues. Taandisque les arcs correspondant aux raies low FIP sont moins étendus et plus étroits et plus affèctés par les modulations du relief lunaire. Les raies « low FIP » sont formées plus proches du limbe au dessus de la photosphère, et où commence la mésosphère. Cette structuration et formation des raies « low FIP » au dessus de la photosphère est observée sur la figure III-4-8 où les raies d'absorption de Fraunhofer sont observées simultanément en absorption avec les raies « low FIP » en émission. Une évaluation des spectres éclairs d'après la figure III-4-7 permet de déterminer à partir des extensions, et étendues des croissants des raies d'émission, que les raies d'hélium neutre et ionisée sont formées à des altitudes au dessus de 700 km par rapport à la myriade de raies low FIP qui apparaît dans la région du minimum de température en dessus de 600 km, au dessus de la haute photosphère voir figure IV-3-3-1 des courbes de lumière de 2010.

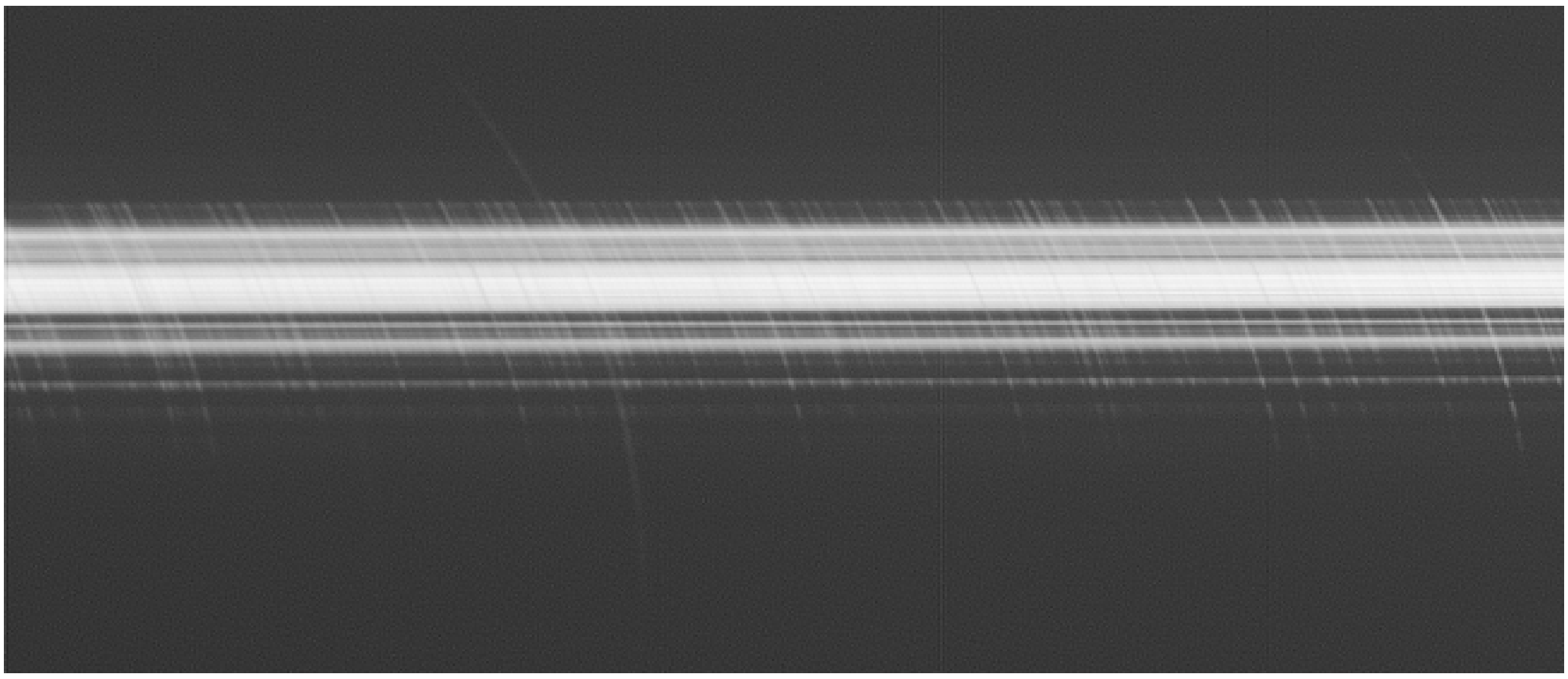

**Figure III-4-8:** *Comme sur la figure III-5-7, 36 spectres sommés (N°5126 à 5166) au troisième contact de l'éclipse du 11 Juillet 2010 montrant à la fois les myriades de petites raies low FIP en émission ainsi que le spectre du continu révélant simultanément les raies de Fraunhöfer vues en absorption.*

Les modulations la térales des spectres du continu révèlent les raies d'absorption dans la partie centrale, où le continu est le plus intense, et correspond aux spectres des grains de Baily.
Le montage III-4-9 décrit une séquence de spectres avec leurs hauteurs maximales moyennes attribuées <h>, au moment de la disparition des grains de Baily. Les raies d'émission « low FIP » présentent une décroissance en flux moins rapide que celle du continu correspondant aux grains de Baily. Cette observation contribue à définir ce qu'est le bord du Soleil sans lumière parasite venant du disque solaire, en séparant les raies « low FIP » du continu, car celles-ci ajoutent un certain embrillancement qui ne correspond pas au « vrai » bord du Soleil. Ainsi en prenant en compte le continu en dehors des raies low FIP, cela permet de différencier les natures différentes des basses couches de l'atmosphère solaire.
Ainsi, la couche renversante a été résolue spatialement et temporellement avec une cadence de 15 images/seconde, lors de l'éclipse totale du 22 Juillet 2009 et à 10 images/seconde à l'éclipse du 11 Juillet 2010, ce qui permet de sonder les basses couches de l'atmosphère solaire avec un échantillonnage de 25 km sur les pas en hauteur. L'apport des nouvelles caméras CCD numériques, où la dynamique de $2^{12}$ niveaux et plus, permet d'analyser les faibles variations de l'extrême-limbe solaire avec une dynamique suffisante en l'absence de lumière parasite, où sont indiquées les altitudes et chrono-datations des instants en figure III-4-9. Ces observations sont totalement impossibles à obtenir en dehors des éclipses, même avec de très grands coronographes.

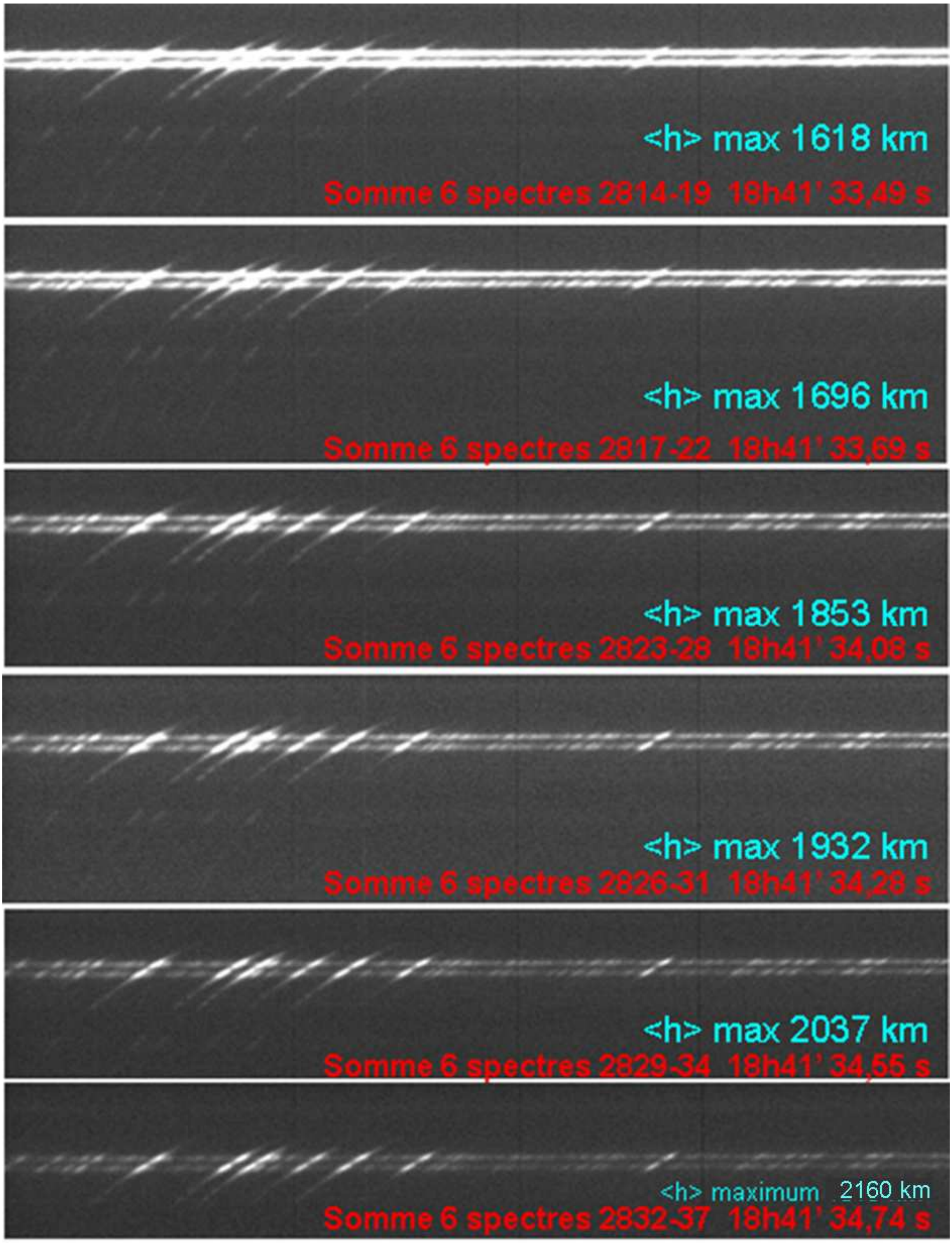

**Figure III-4-9:** *Extrait de séquences des spectres éclair CCD du second contact de l'éclipse de 2010 pour montrer la diminution d'intensité dans les vallées lunaires, les raies d'émission et la transition du continu photosphérique au continu chromosphérique. La sommation de 6 spectres a permis d'améliorer le rapport signal sur bruit et de distinguer de plus faibles détails dans le continu et les raies d'émission.*

Cette séquence montre que le continu photosphérique et le continu chromosphérique se raccordent ou se prolongent de manière monotone, définissant la limite du bord solaire photosphérique et où se situe la mésosphère.
Le second et troisième contact de l'éclipse du 11 Juillet 2010 sont données dans le tableau III-4-10. Les correspondances entre les numéros d'images avec l'altitude sont indiquées ainsique l'heure de l'horloge interne de l'ordinateur (PC) ayant été utilisé pour les acquisitions.
:

| Frames N° | PC time | height (km) | | Frames N° | PC time | height (km) |
|---|---|---|---|---|---|---|
| 2772-78 | 31,324 | 479,626 | | 5065-70 | 58,726 | 2130 |
| 2775-81 | 31,59 | 832,956 | | 5067-72 | 59,022 | 2008 |
| 2778-84 | 31,785 | 1120,781 | | 5070-75 | 59,451 | 1832 |
| 2781-87 | 31,983 | 1330 | | 5073-78 | 59,649 | 1751 |
| 2784-90 | 32,214 | 1539,616 | | 5076-81 | 60,077 | 1575 |
| 2787-93 | 32,675 | 1618,222 | | 5079-84 | 60,275 | 1494 |
| 2790-96 | 32,939 | 1696,431 | | 5082-85 | 60,407 | 1440 |
| 2793-95 | 33,272 | 1775,037 | | 5087-92 | 61,133 | 1142 |
| 2796-2802 | 33,632 | 1853,643 | | 5090-95 | 61,33 | 1061 |
| 2802-2807 | 33,995 | 1932,249 | | 5093-98 | 61,528 | 979 |
| 2805-2810 | 34,193 | 2037,057 | | 5096-5101 | 61,726 | 898 |
| 2808-2813 | 34,72 | 2115,266 | | 5099-5104 | 61,924 | 817 |
| 2811-2816 | 35,248 | 2220,074 | | 5102-5107 | 62,122 | 735 |
| 2814-2819 | 35,446 | 2508,296 | | 5105-5110 | 62,452 | 600 |
| 2817-2822 | 35,643 | 2874,727 | | 5108-5113 | 62,649 | 520 |

**Tableau III-4-10 :** *correspondances entre les numéros d'images sommées, les altitudes et l' heure de l'horloge interne de l'ordinateur ayant effectué les acquisitions, avant les corrections du temps GPS.*

Dans le procédé utilisé pour les acquisitions des spectres éclair, la caméra CCD Lumenera continuait à enregistrer en continu après les phases de totalité de l'éclipse. La caméra a été ensuite retirée de son logement au foyer du réseau-objectif, et le capteur a été protégé de la lumière ambiante, avant d'enregistrer les images de l'écran du GPS.
Un petit objectif de 35 millimètres de focale a été aussitôt placé et inséré devant la caméra CCD, et pour enregistrer les images de l'écran du GPS affichant l'heure, les minutes et secondes, sans avoir interrompu l'enregistrement depuis le début avant la totalité de l'éclipse.
Ce procédé a permis d'effectuer la correspondance entre heure du PC et heure GPS. Sur chaque image enregistrée l'heure du GPS est visible. Ces mêmes images contiennent la datation effectuée par le PC. En annexe N°12 sont donnés plus de détails sur le procédé.
Les altitudes ont été calculées avec la moyenne sur les images sommées voir table III-4-11.
L'étalonnage entre l'heure de l'ordinateur et du GPS est donné avec la dispersion suivante:

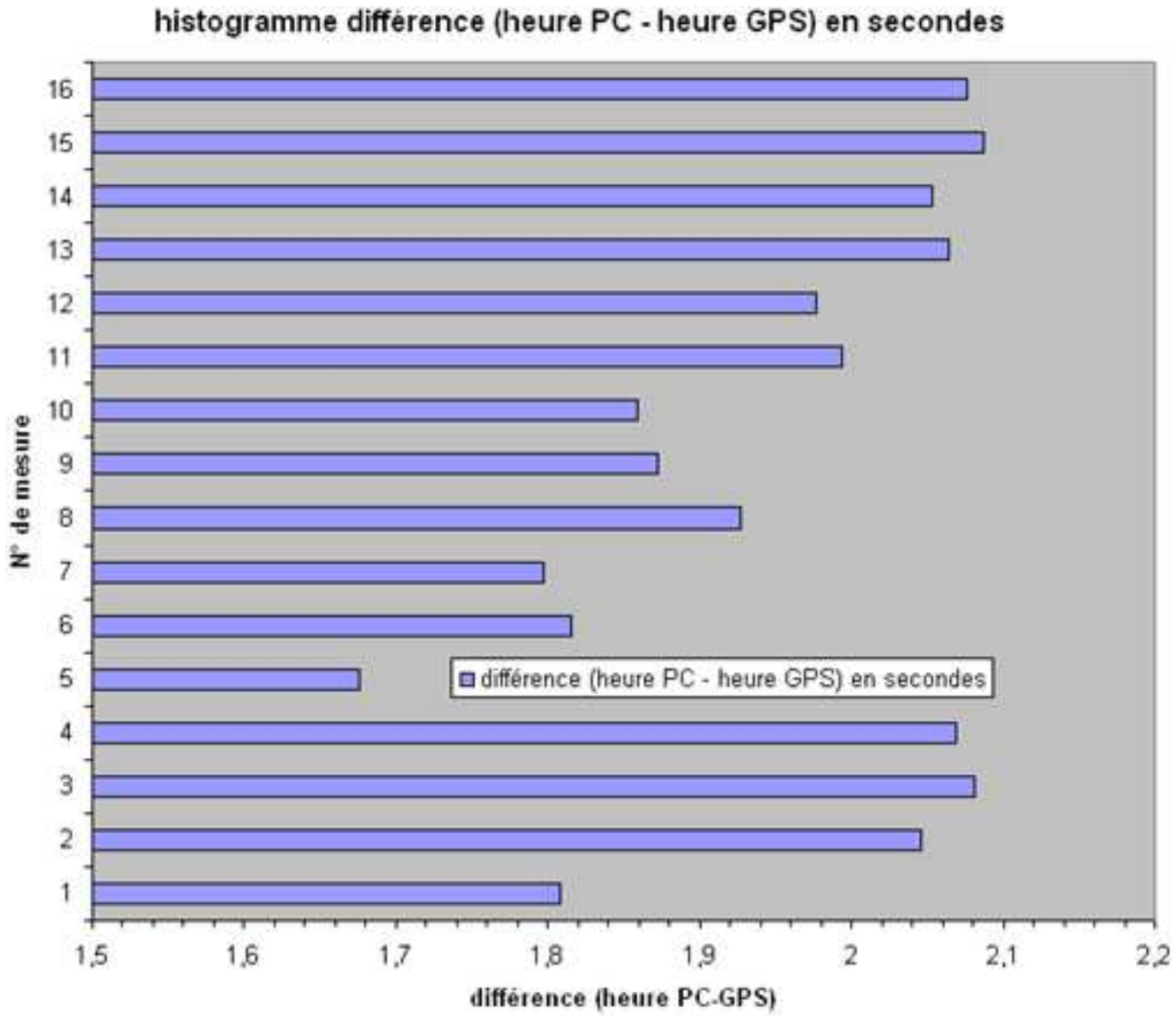

| différence (heure PC - heure GPS) en secondes | |
|---|---|
| 1,808 | |
| 2,046 | |
| 2,081 | |
| 2,069 | |
| 1,676 | |
| 1,816 | |
| 1,797 | |
| 1,927 | |
| 1,872 | |
| 1,859 | |
| 1,994 | |
| 1,977 | |
| 2,064 | |
| 2,053 | |
| 2,087 | |
| 2,076 | |
| 1,950125 | Moyenne |
| 0,125735772 | eccart-type |

**Figure III-4-11:** *calcul des différences de l'heure lue au GPS et indiquée sur la datation du PC. Ces relevés ont été effectués à partir d'images filmées du GPS, au moment du changement de digit des secondes, et à partir du temps affiché dans l'entête des images*

Des mesures de courbes de lumière ont étés réalisées le long de la raie Fe II 4629 Å, et du continu relevé entre les raies d'émission pour prendre en compte le vrai flux du continu correspondant au vrai bord solaire, simultanément dans une vallée lunaire. Les intensités ont été relevées dans l'étendue limitée par la largeur de la vallée lunaire. Les mesures des raies d'émission et du continu ont été réalisées dans les mêmes conditions sur chaque spectre pris individuellement.

Les figures III-4-12 et III-4-13 montrent un extrait de séquence de spectre éclair agrandie 10 fois pour montrer les fluctuations du vrai continu et des raies d'émission le long du relief lunaire et comment prendre en compte le relief lunaire pour analyser le « vrai » bord solaire.

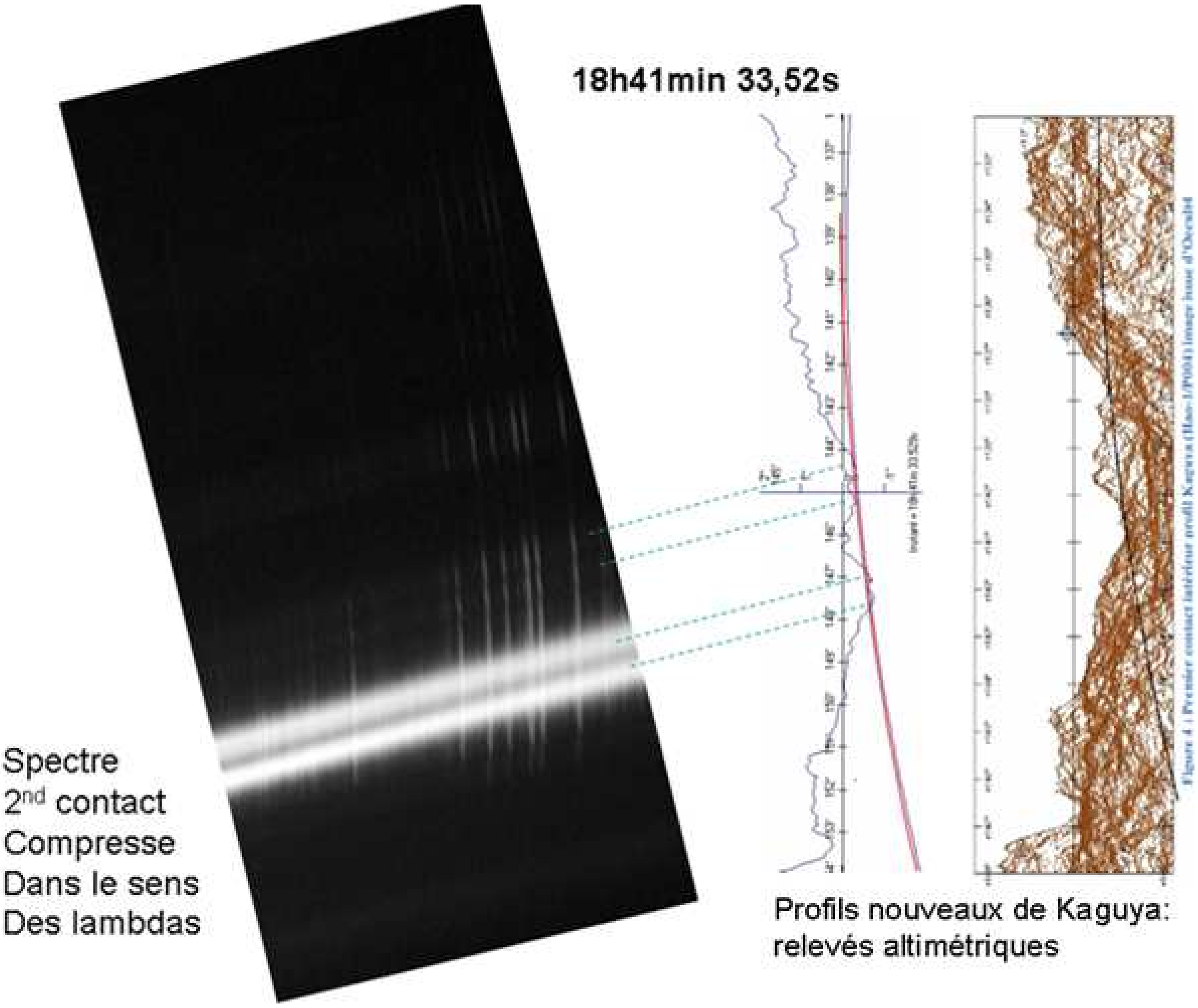

**Figure III-4-12:** *correspondance entre les spectres éclair modulés par le relief lunaire et le profil de la Lune (données nouvelles de Kaguya, et calculs de Patrick Rocher, IMCCE)*

Le schéma figure III-4-13 représente comment peut être structurée une vallée lunaire et effectuer des tranches succéssives pour l'évaluation des flux relevés sur le bord du Soleil lors des contacts d'éclipses. Cette analyse a fait l'objet d'un article dans les actes du colloque de la Société Français d'Astronomie et d'Astrophysique SF2A, Bazin, Koutchmy, Rocher 2012, voir Annexe 4. Elle présente quelques analyses du bord du Soleil aux éclipses affectées par les reliefs du bord de la lune, en l'abscence de lumière parasite et d'autre part une étude de lumière diffractée du bord solaire sur un mur vertical à 150 m du lieu d'observation.

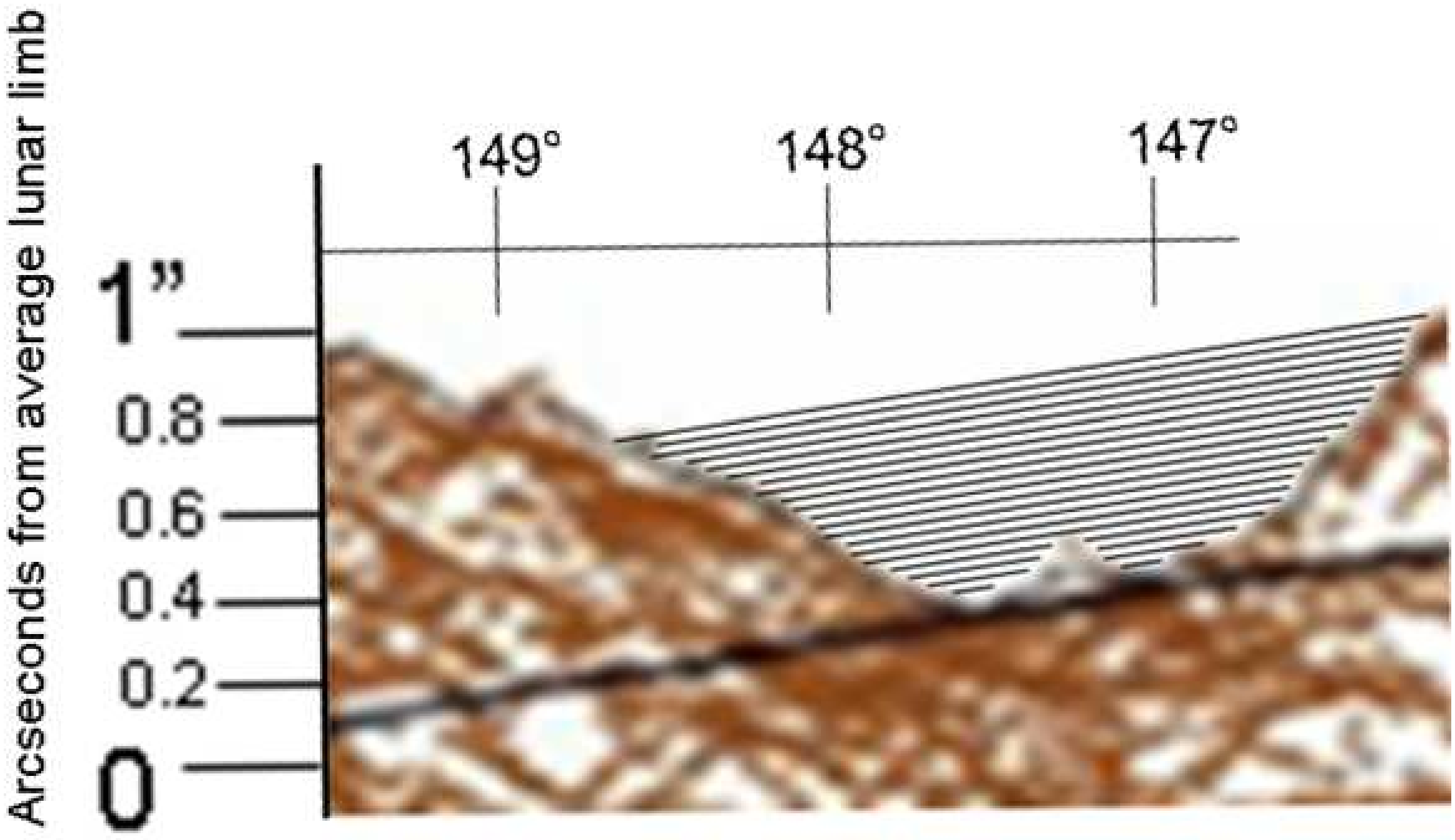

**Figure III-4-13:** *Profil agrandi 10 fois de Kaguya de la vallée à l'angle AA-147° et 148 durant le second contact C2 de l'éclipse totale du 11 Juillet 2010. Les grains de Baily sont supposés être représentés par des niveaux successifs qui s'additionnent sur la ligne de visée à travers cette vallée. L'aire dans la coupe de la vallée est divisée en 20 sous-segments équidistants de 20 milli-secondes d'arc et correspondent à des intervalles de temps de 50 milli-secondes. Durant les instants des contacts avant et après la totalité, chaque coupe est remplie avec différentes sous-couches du limbe solaire à intervalles de temps constants: une seconde pour tout ce grain au total.*

La méthode d'analyse des grains de Baily en lumière blanche est décrite l'article Sigismondi, Raponi, Bazin, et Nugent, 2011 voir Annexe 3. Les 50 millisecondes indiquées dans la légende de la figure III-4-13 sont liées à la cadence d'acquisition de 15 images/seconde de la caméra CCD Lumenera utilisée à l'éclipse de 2010 où le temps d'exposition était de 55 millisecondes (Annexe 17).
Cette méthode est utilisée pour l'analyse des spectres des grains de Baily, notamment pour les décroissances du spectre continu mesuré lors des contacts d'éclipses. Une partie du continu est relevée entre la myriade de raies d'émission pour définir le « vrai bord » du Soleil. Cela sert à distinguer les contributions des raies d'émission « low FIP » qui se superposent au « vrai bord » du Soleil considéré sans la contribution des raies d'émission. Cette structuration relativement complexe est représentée dans le montage figure III-4-14 avec un spectre éclair extrait et agrandi provenant de l'éclipse de 2010 (avec spectrographe sans fente) auquel les profils d'intensité et relief lunaire ont été ajustés pour indiquer les corrélations.

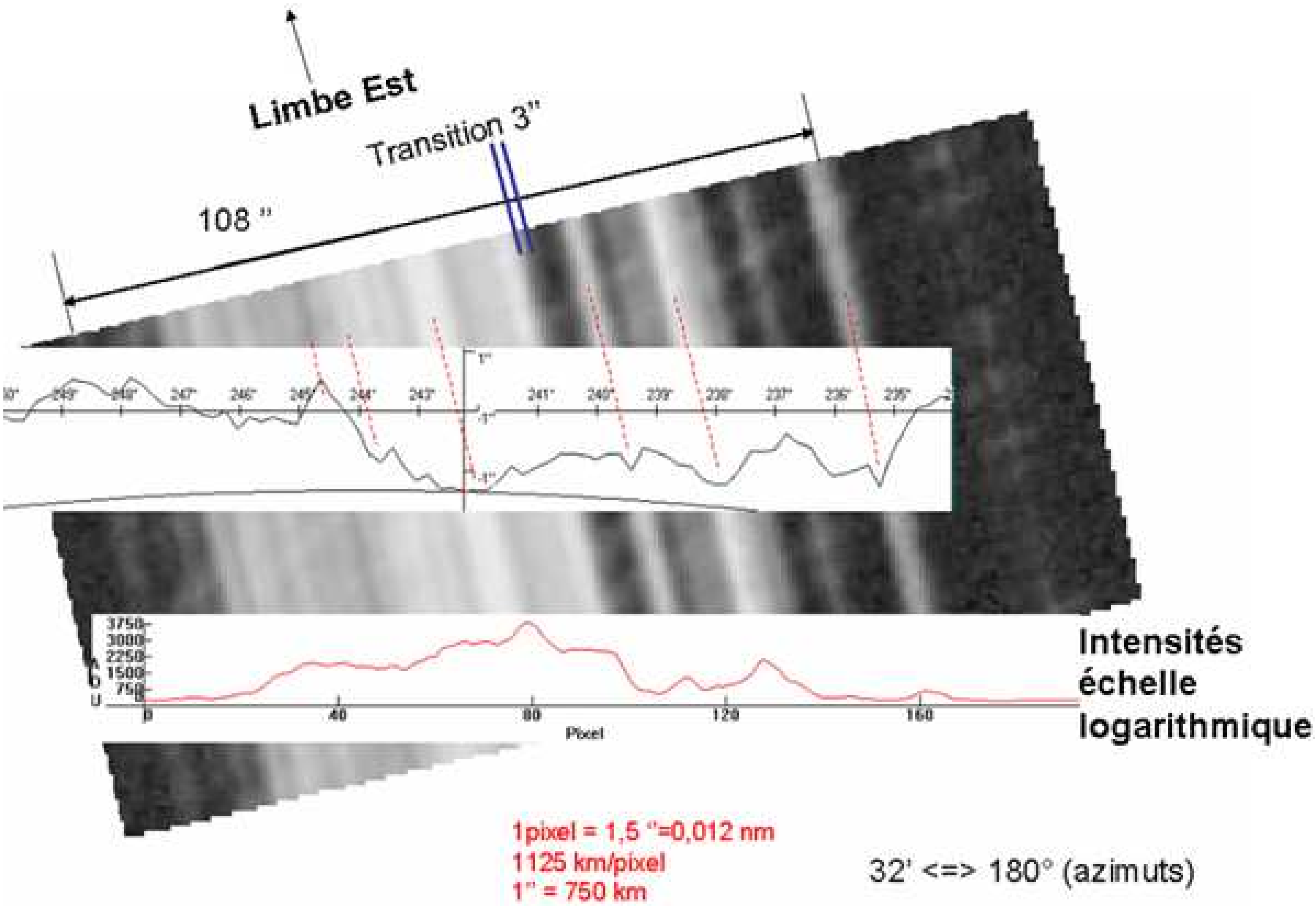

**Figure III-4-14:** *vue agrandie d'un extrait des spectres des grains de Baily à l'éclipse de 2010 montrant au troisième contact C3, la structuration complexe des raies d'émission, du continu, et des raies d'absorption séparées, constituant le vrai bord solaire, appelé autrefois « couche renversante », avec mention des échelles.*

Cette figure permet d'illustrer, après un agrandissement de 20 fois, une observation des couches mésosphériques plus en détail, lors de l'éclipse du 11 Juillet 2010. Dans les vallées lunaires, le fin liseret du continu photosphérique se superpose avec la myriade de petites raies d'émission le long de la ligne de visée, où ces raies apparaissent distinctement au dessus du continu. Cette transition entre le spectre continu et l'apparition des raies d'émission peut correspondre aux régions (200 à 600 km) où pourraient se former les spicules, où le champ magnétique est dominant ($\beta_{plasma} < 1$) et qui correspond au minimum de température.

Les spectres sans fente ont permis d'utiliser comme « fente naturelle », le mouvement apparent de la Lune sur le limbe solaire pour étudier structurations des couches de l'atmosphère solaire et où il est possible d'analyser les flux des raies d'émission sur chaque spectre éclair. Ce type d'analyse permet d'étudier que les altitudes transversales, limitées seulement par le bord accidenté de la Lune. Cette étude est restreinte car on ne peut pas évaluer les distances prises radialement à partir du limbe solaire. C'est pourquoi, j'ai utilisé les résultats de spectres obtenus à l'éclipse totale du 13 Novembre 2012, où un spectrographe à fente a été utilisé au voisinage de l'intervalle spectral contenant la raie verte coronale du Fer XIV et du triplet b1, b2, b3 du magnésium neutre Mg I. Ces spectres obtenus avec un spectrographe à fente à l'éclipse de 2012 apportent des observations complémentaires du bord solaire, par rapport aux observations réalisées aux éclipses de 2008, 2009 et 2010 utilisant un spectrographe sans fente. Ces expériences sur le bord Solaire entrent dans le contexte de l'expérience Picard.

## III-5) Spectres effectués sur la raie verte du fer Fe XIV à 5303, Eclipse du 13 Novembre 2012

Les observations ont été réalisées en Australie dans la partie Nord, proche de la barrière de corail, appelée Queensland. Des expériences ont été préparées, et installées sur plusieurs sites autour de la bande de totalité, afin d'augmenter les chances d'obtenir un ciel dégagé au moment de la totalité, car les conditions météorologiques étaient prévues peu favorables en cette période de Novembre au Nord de l'Australie.
Les expériences de Serge Koutchmy, Jaime Villinga et Jonathan Cirtain ont été réalisées sur le site de Howe Plantation, dont les coordonnées sont indiquées sur la figure III-5-1. Les instants et positions du second et troisième contact sont indiqués pour ce site. Les conditions météorologiques ont été favorables et de nombreuses expériences ont réussi.
Dans ce chapitre, nous nous intéréssons exclusivement aux résultats obtenus avec le spectrographe à fente, et pour un intervalle spectral, situé autour de la raie verte coronale du Fe XIV à 5303Å.

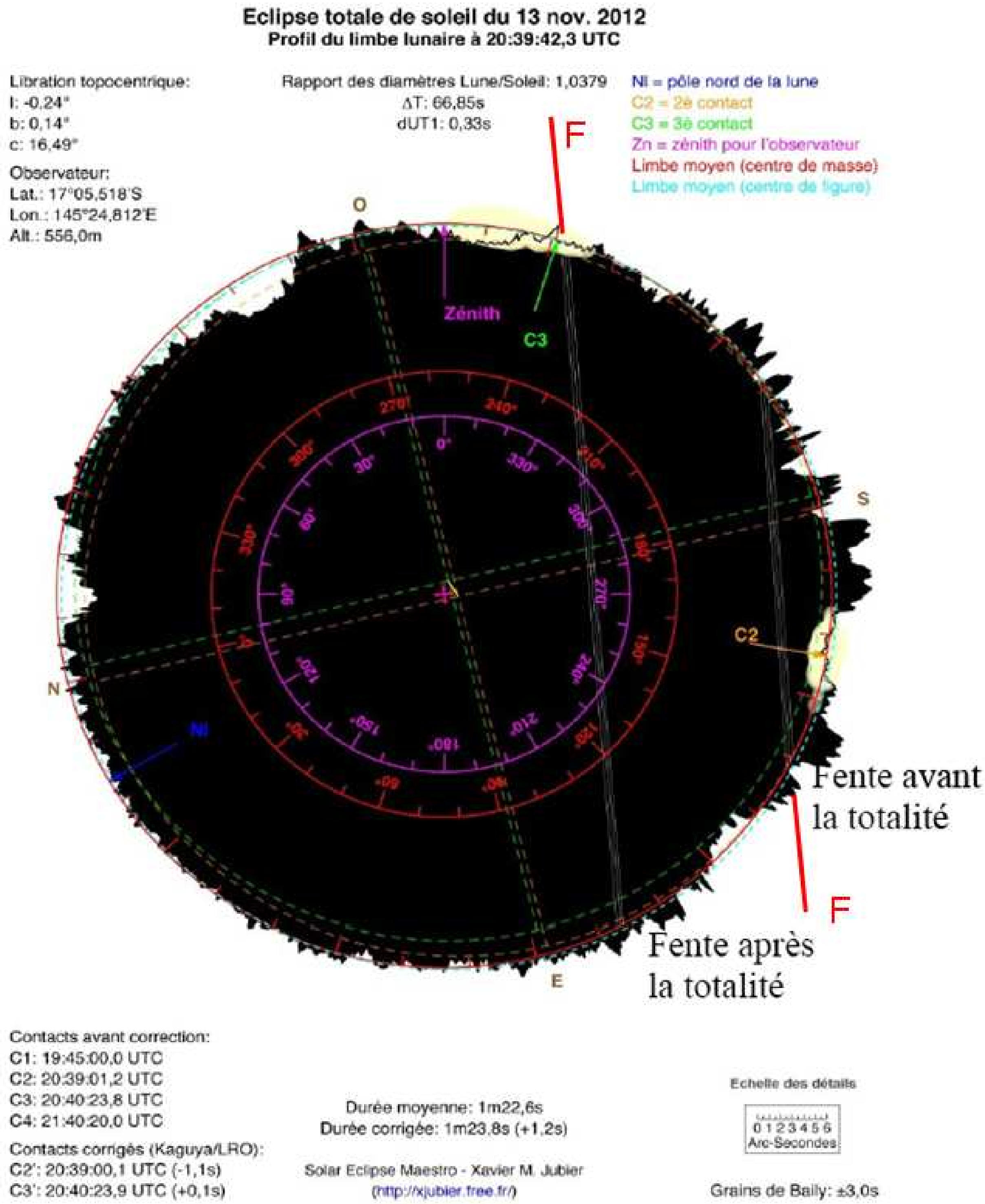

**Figure III-5-1:** *Profils lunaires calculés pour le lieu d'observation des spectres discutés dans ce sous chapitre (Howe Plantation) par Xavier Jubier, montrant les instants des contacts. Les orientations du disque solaire et les coordonnées précises du site d'observation de l'éclipse totale de Soleil du 13 Novembre 2012 sont indiqués. Les positions relatives de la fente du spectrographe avant et après la totalité, sont indiquées à partir des prises de vue des images réfléchies par la fente d'entrée.*

L'expérience du spectrographe miniature a globalement bien réussi et une vingtaine de spectres éclairs ont été obtenus au total, durant les instants du second contact et pendant les phases de totalité. Quelques séquences de spectres éclairs ont été perdues au début du second contact, car d'une part la caméra « slit jaw » voir schéma figure II-1-6 a été paramétrée avec un gain automatique et son temps de réponse était ralenti à cause du débit plus important lié aux images saturées au début du second contact, et d'autre part, les déclanchements des prises de vues des spectres éclairs avec l'appareil photographique numérique ont été réalisés plus tardivement au milieu et à la fin du second contact.

Je me suis intéressé aux 3 premiers spectres éclairs consécutifs obtenus par rafale quelques secondes après le second contact au début de la totalité. Ces spectres montrent simultanément le continu de l'interface chromosphère-couronne, avec de nombreuses raies « low FIP ». Le choix initial de l'intervalle de longueurs d'ondes entre 4650 et 5450 Å dans le champ du CCD du Canon, a permis l'obtention de spectres de  la raie verte coronale dont l'émission est renforcée dans la région d'interface chromosphère-couronne. 3 spectres correspondant à la fin du second contact juste avant la totalité ont été additionnés et ajustés. Le spectre résultant de ce traitement est indiqué en figure III-5-2. En effet l'observation de la raie verte verte du Fe XIV à 5303 Å dans ces régions plus basses, permet d'étudier si la couronne chaude entre 1 et 2 M K, peut être présente (mais sans dominer) dans l'interface photosphère-couronne solaire. La raie verte du Fe XIV est une raie coronale interdite et qui a pour origine des atomes hautement ionisés (plus de la moitié des électrons d'atomes de fer ou nickel arrachés). Ces atomes hautement ionisés dans la couronne ont une température électronique supérieure à 1 million de Kelvins (Evans 1963, Shklovskii, 1965).

La compréhension des mécanismes d'ionisation et d'identification des raies coronales est due aux travaux de Grotrian 1939 qui a montré que les longueurs d'onde des transitions entre les composantes du Fe IX et du Fe X étaient proches des longueurs d'onde coronales, comme le Fe X à 6374 Å et Fe X 7892 Å. C'est ensuite Edlen 1936 et 1942 qui a montré en laboratoire que le fer ayant perdu 13 électrons produisait la raie verte à 5303Å.

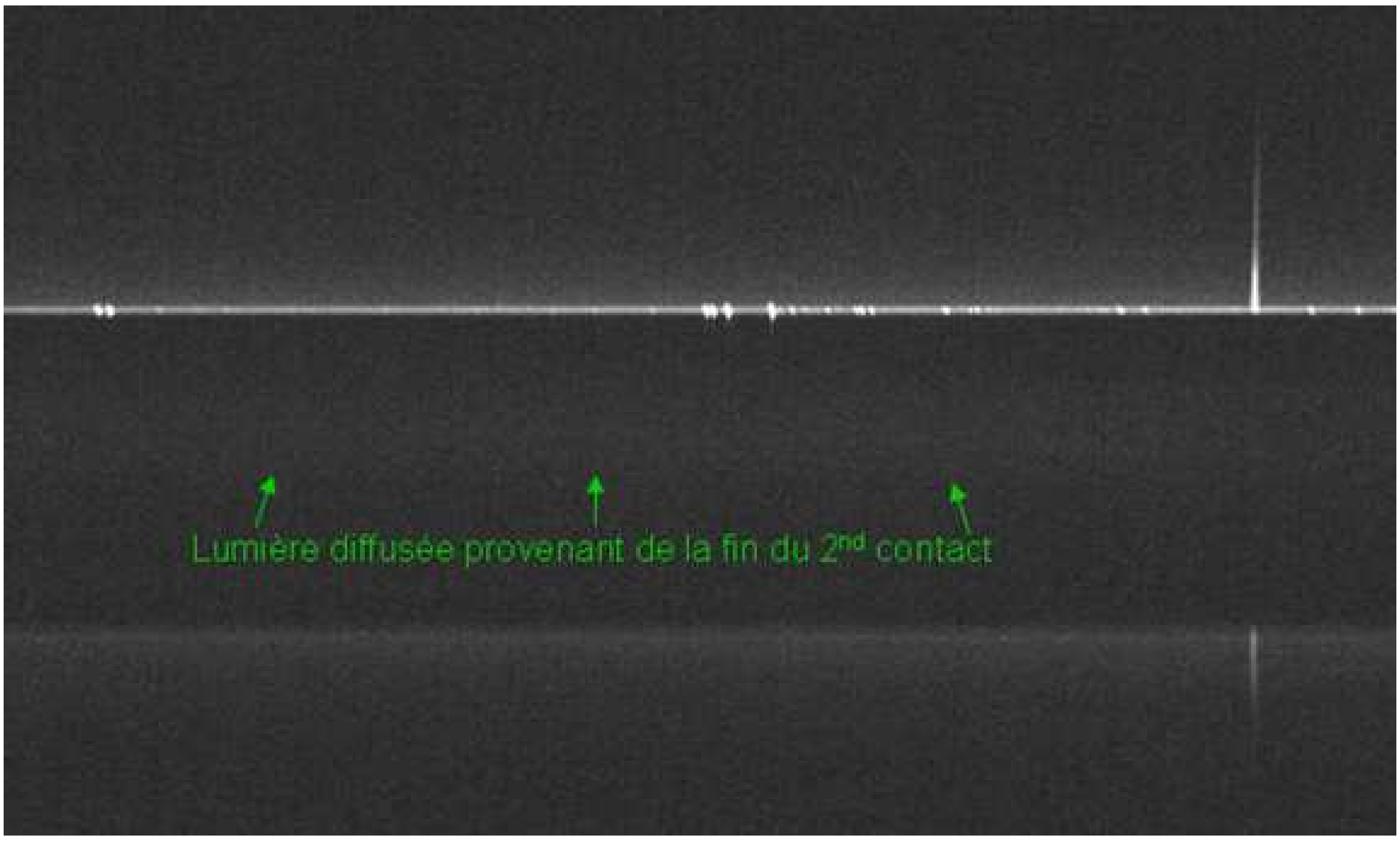

**Figure III-5-2:** *Somme de 3 spectres obtenus le 13 Novembre 2012 par Serge Koutchmy et Jonathan Cirtain au moyen d'un spectrographe à fente. Le temps d'exposition est de 10 ms, obtenus au moyen d'un appareil  numérique Canon 60D, à chip CMos de 18 Mpixels numérisé sur 14 bit (format RAW). Le bord au Sud-Est solaire est dans la partie supérieure et est situé un peu sur le côté du contact C2.*

La fente de 40µm interceptait une bande de 24.5’’ de large sur l’image du disque solaire éclipsé. La partie un peu plus brillante correspond à de la lumière résiduelle située entre la haute photosphère et la basse couronne, juste au début de la totalité. Les raies d’émission de l’interface photosphère-chromosphère/couronne sont identifiées figure III-5-6 et répertoriées dans la table III-5-5. La détermination des altitudes des couches de l’atmosphère solaire est déterminée et calculée à partir de la figure III-5-3 qui présente des extraits de la séquence d’images de la caméra « slit jaw » montrant quand et où les spectres précédents ont été réalisés:

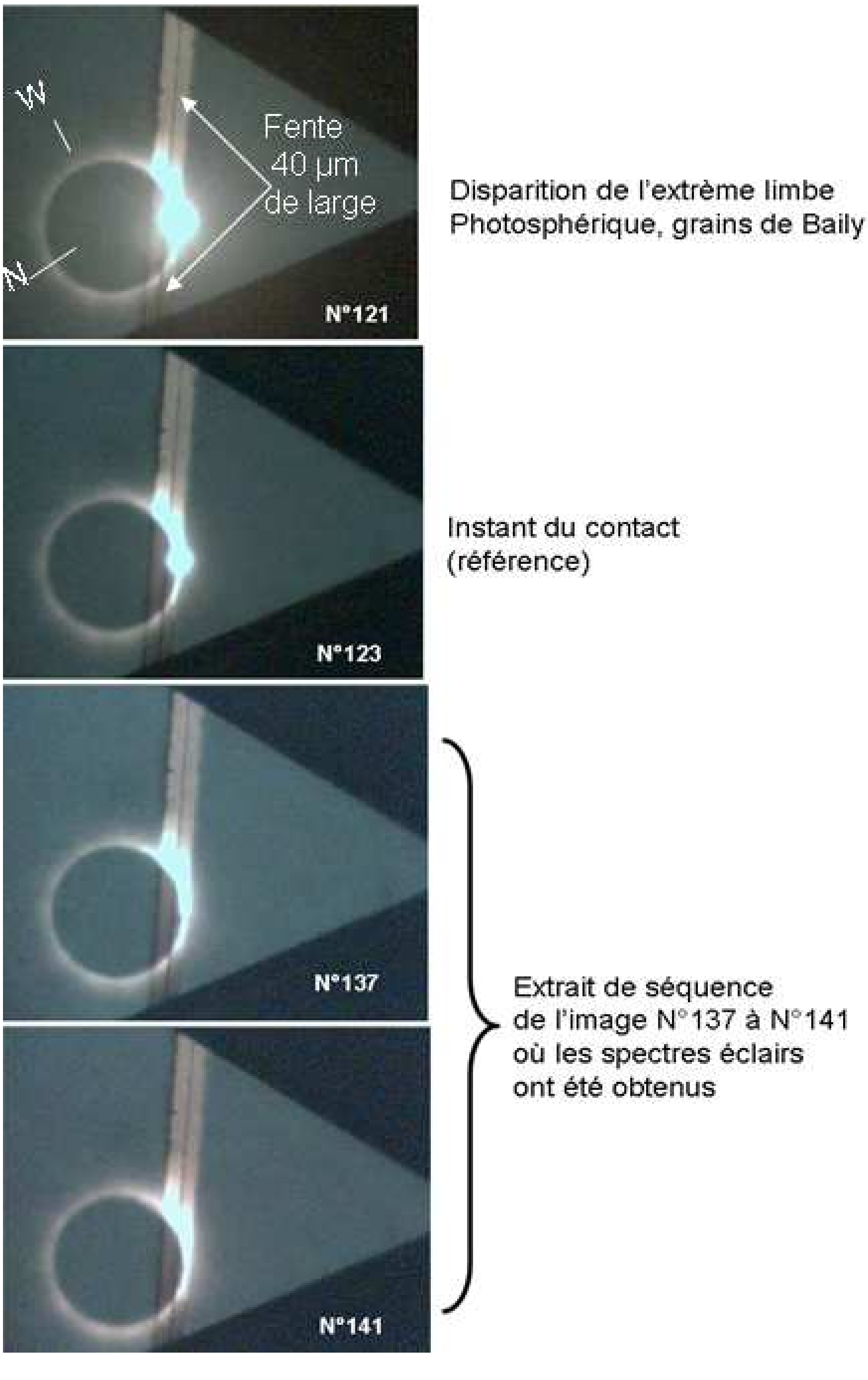

**Figure III-5-3:** *extraits des images de la web-cam « slit jaw » montrant la position de la fente sur le limbe solaire, à partir desquelles les spectres en rafale ont été réalisés.*

D’après les images figure III-5-3, la fente a coupé le disque solaire éclipsé sur une corde et le mouvement apparent du Soleil a produit un léger déplacement vers le bas de la fente sur la dernière image. Ce qui signifie que le bord ouest Solaire devrait être situé sur la partie supérieure du disque. L’obtention des positions précises des limbes ouest, est, nord et sud est délicate, et les positions calculées par Xavier Jubier nous ont aidé.
Cette méthode d’enregistrement d’images du disque solaire éclipsé sur la fente d’entrée du spectrographe au moyen d’une caméra webcam (slitjaw) est indispensable pour contrôler les positions et les altitudes dans les couches de l’interface photosphère - couronne. Elle permet aussi de mieux établir les orientations sur le disque, afin d’assurér les étalonnages des altitudes des spectres éclairs étudiés. La fente était située en dehors de la zone du contact très intense, comme indiqué sur le montage en figure III-5-4. L’évaluation des altitudes correspondant à la position du limbe lunaire dans l’atmosphère solaire était limité par la largeur de la fente d’analyse. Nous présentons une méthode pour déterminer les altitudes, où les évaluations ont été difficiles à cause de lègères vibrations lors des déclanchements des vues et de la résolution limitée de l’image obtenue avec la web-cam « slit-jaw ». La cadence d’acquisition des images de la webcam était de 5 images/s mais cette valeur reste une estimation, car les fortes variations des flux au second contact, après l’éjection du filtre de densité neutre, ont conduit à une augmentation du débit de données et la cadence a été ralentie à cause de la saturation de la webcam. D’après les numérotations des images de la webcam, trois secondes se seraient écoulées entre l’instant de contact de l’image N° 123 et l’image N°137 où le premier spectre a été réalisé. Ces 3 secondes d’écart permettent une évaluation de l’altitude moyenne limitée par le bord de la Lune. En une seconde de temps, la Lune s’est déplacée de 0.5’’ d’arc environ. Or une seconde d’arc correspond à environ 760 km d’altitude sur le limbe solaire. Donc en 3 secondes l’altitude explorée est de 0.5*3*760 = 1140 km près du contact sur limbe solaire. Or la fente du spectrographe n’était pas située exactement au point de contact, mais coupait l’atmosphère solaire sur une corde comme le montre le schéma figure III-5-4:

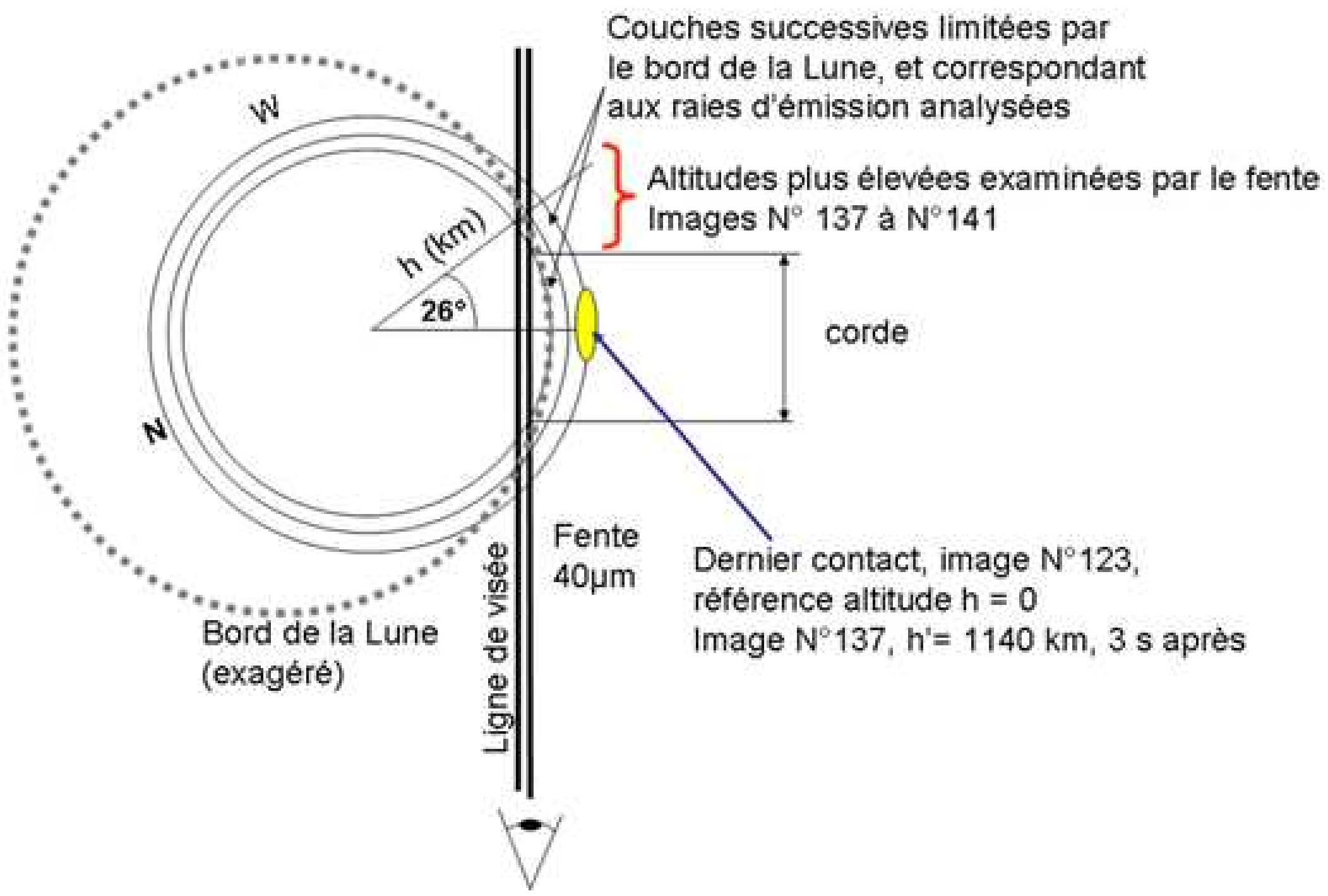

**Figure III-5-4:** *Schéma du disque solaire projeté sur la fente d’entrée pour expliquer comment définir les altitudes correspondant aux spectres analysés, et l’inclinaison des raies.*

A partir du schéma III-5-4, l'élévation correspondant à la position de la fente sur la partie la plus basse dans l'atmosphère solaire dont le spectre est effectué, correspond à un angle de 26° entre le rayon pris perpendiculairement à la fente et les premières couches de l'atmosphère solaire interceptées par le prolongement supérieur de la fente avant la totalité de l'éclipse. L'altitude prise comme référence sans tenir compte de la fente est de 1140 km au dessus du limbe solaire, en tenant compte des 3 secondes écoulées après le second contact. A cette altitude s'ajoute une certaine tranche d'altitude correspondant à la largeur de la fente, et de son inclinaison de 26° par rapport au rayon solaire. L'élévation résultante est alors de: 1140 + sin(26°)*1140 = 1640 km au dessus du limbe solaire. A partir du schéma optique utilisé pour le spectrographe miniature (voir figure II-1-6), un diamètre solaire voisin de 0.009 radians au foyer primaire de la lunette de 500 mm fait un angle de 32' et son diamètre moyen fait 500*tan(32')= 4.6 $\pm$ 0.3 mm sur la fente d'entrée du spectrographe. Avec le système collimateur utilisé, un objectif de 200 mm est suivi d'un objectif de 135 mm. La taille de l'image au foyer CCD du spectrographe fait 4.6*135/200 = 3.1 $\pm$ 0.2 mm. La dimension exacte des pixels de l'appareil photo numérique Canon 60 D couleur est de 4.7*4.3 µm. Donc un diamètre solaire moyen au foyer du spectrographe correspond à 3100/4.3 = 720 pixels. En prenant le diamètre solaire égal à 1392 Mm ce calcul aboutit finalement à une échelle de 1.9 Mm/pixel et cette valeur a été adoptée par la suite. Une fois les évaluations des altitudes déterminées et les échelles, les spectres ont été analysés. De nombreuses raies d'émission sont visibles (voir figure III-5-6) et sur une plus grande étendue au-delà de la chromosphère, on distingue la raie verte du Fe XIV à 5303 Å qui s'étend dans la couronne adjacente bien plus haut. Les identifications des raies d'émission « low FIP » ont été effectuées à partir des tables de R.B. Dunn. Les raies caractéristiques du triplet du Mg I b1 5183.60Å, b2 5172.68Å et b3 5167.32Å, sont situées à gauche du spectre, voir Figure III-5-6 et la position de la raie verte à 5303 Å, à droite, a permis d'établir une dispersion de 0.177 Å/pixel. Les longueurs d'onde des autres raies ont pu être mesurées et les éléments identifiés, comme par exemple la raie « low FIP » du Fe II 5169 Å à côté de la raie b3 du Mg I.
L'inclinaison des raies dans la couche chromosphérique peut s'expliquer soit parce que la fente coupe une tranche de chromosphère d'étendue limitée par la bord de la Lune avec un angle de 26° par rapport au rayon du Soleil, soit par une éventuelle erreur de parallélisme entre le sens des sillons du réseau et le sens de la longueur de la fente. Ces effets conjugués pourraient être à l'origine de l'inclinaison des raies « low FIP » par rapport au sens de dispersion. Au-delà de la chromosphère, où commence la couronne, le profil de la raie verte n'est pas incliné. Une fois les raies identifiées, il a été utile de comparer les potentiels d'excitation et d'ionisation pour quelques raies identifiées, afin de les comparer avec ceux des raies « low FIP » observées aux éclipses de 2008, 2009 et 2010.

Le tableau figure III-5-5 indique pour quelques raies sur le spectre figure III-5-6, les potentiels d'excitation et de première ionisation des éléments chimiques correspondants.

| $\lambda$ (Å) | 5169 | 5170 | 5188 | 5204 | 5205 | 5208 | 5226 | 5234 | 5237 | 5261 | 5303 | 5316 | 5328 | 5337 |
|---|---|---|---|---|---|---|---|---|---|---|---|---|---|---|
| element | Fe II | Mg I | Ti II | Fe I | Y II | Cr I | Ti II | Fe II | Cr II | Ca I | Fe XIV | Fe II | Cr I | Ti II |
| Potentiel de première ionisation (eV) | 7.86 | 7.64 | 6.83 | 7.86 | 6. | 6.74 | 6.83 | 7.86 | 6.74 | 6.11 | 7.86 | 7.86 | 6.74 | 6.83 |
| Potentiel d'excitation (eV) | 2.87 | 2.71 | 1.57 | 0.09 | 0.92 | 0.93 | 1.55 | 3.2 | 4.05 | 2.51 | 355 | 3.20 | 2.90 | 1.57 |

**Tableau III-5-5 :** *potentiels d'excitation et de première ionisation pour certaines raies étudiées dans les spectres éclair de l'éclipse totale du 13 Novembre 2012. D'après "The Solar spectrum 2935 Å to 8770 Å, C Moore et al 1966.*

Les valeurs des potentiels d'excitation des raies low FIP sont inférieurs à 10 eV, et sont comparables à ceux obtenus dans les régions spectrales de 4500Å et 4700Å, des éclipses totales antérieures.
En Annexe 18 est indiquée une table d'où proviennent les valeurs des premiers potentiels d'ionisation des éléments chimiques observés.

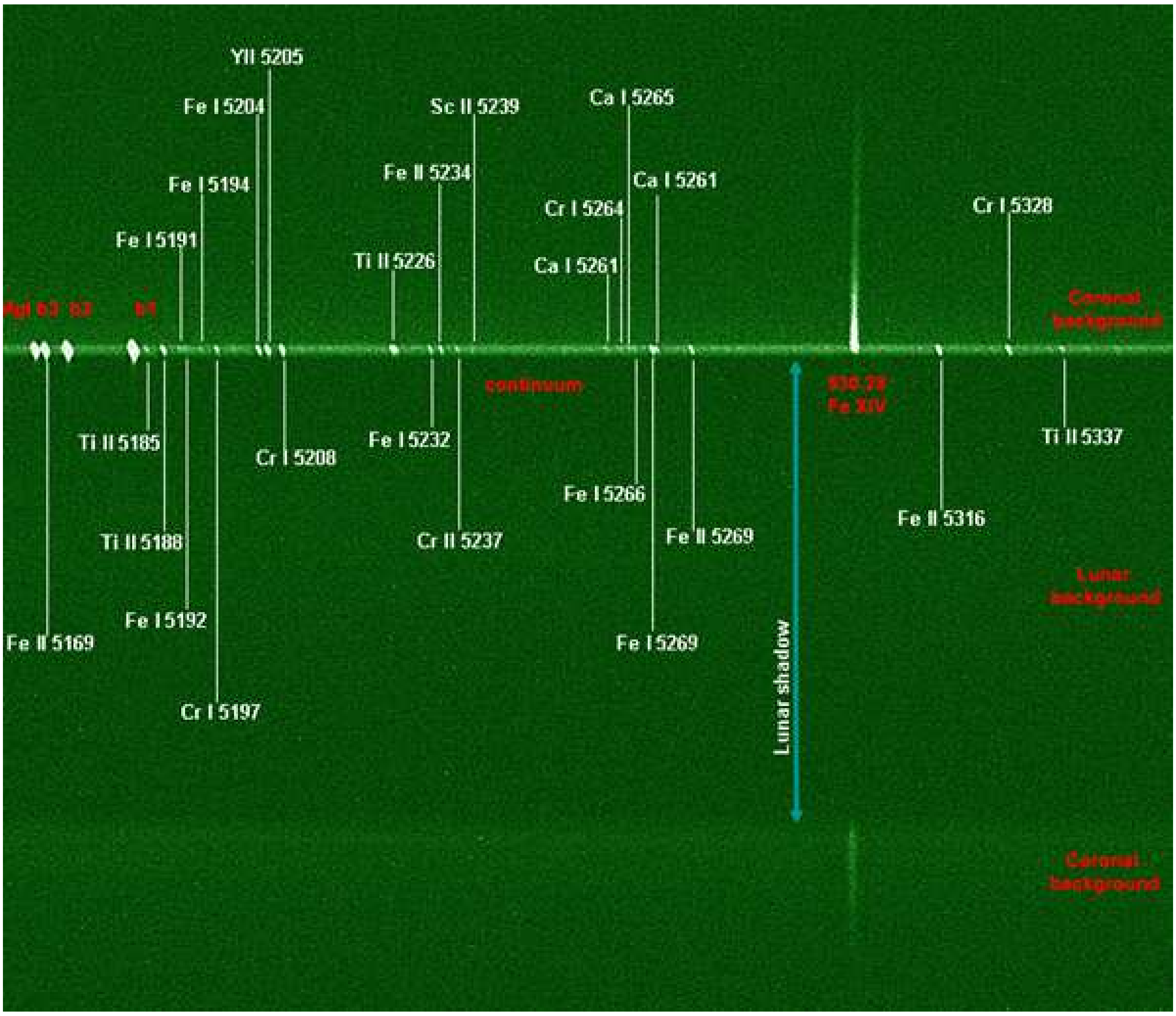

**Figure III-5-6:** *image traitée, renforcée et colorisée avec identification des raies chromosphériques low FIP encore visibles à des altitudes supérieures à 1690 km au dessus du limbe solaire. L'identification des espèces chimiques a été réalisée avec les tables de C. Moore, Revised Rowland Spectra 1966.*

Dans ce chapitre III, nous avons décrit et présenté les principaux résultats des spectres éclair sans fente obtenus à l'éclipse totale de 2006, puis améliorés en 2008, 2009 et 2010 avec une meilleure résolution spatiale et dispersion spectrale. Ces nombreux spectres éclair, 100 à 150 par contact, ont été analysés un par un, pour bien montrer les variations importantes d'intensité. Et enfin les résultats de spectres éclairs avec spectrographe à fente en 2012 ont été décrits dans le but de présenter 2 méthodes complémentaires d'observations et d'analyses.
Des analyses des profils de la raie verte où elle a été observée dans les régions de l'interface photosphère-couronne ont été tentées en ayant retranché le continu coronal.

# Chapitre IV) : Analyses des résultats de chaque éclipse totale

## IV-1) Eclipse du 1er Août 2008

Les séquences des spectres éclair obtenues à l'éclipse totale du 1er Août 2008 ont été extraites, et chaque spectre a été visualisé un par un afin de vérifier le nombre de spectres exploitables, à partir desquels les intensités maximales sont relevées dans des zones des spectres correspondant au continu, et à quelques raies d'émission du Fe II, He I, He II.. qui ont été identifiées. Des relevés ont été effectués dans les grains de Baily et continu en dehors des raies. Pour les étalonnages en intensités du disque solaire moyen en Arbitrary Digital Unit – ADU, 1 ADU = $5.76923 \times 10^{-6}$ unités du disque solaire moyen, lors du second contact de l'éclipse du 1er Août 2008. Un tableau des étalonnages des caméras utilisées pour chaque éclipse est donné en Annexe 17-3. Les coefficients de transmission des filtres sont indiqués en figure II-1-5 et le choix de l'unité en unités du disque solaire est précisé en Annexes 28 et 29. Les intensités ADU mesurés par la caméra Watec 120 N+, sont codées sur 256 niveaux. Selon le profil du limbe lunaire, des variations dans les profils des courbes de lumière (relevés sur les spectres éclairs) pour le continu et pour la raie du Fe II 4629 Å peuvent être constatées: vallées larges et peu profondes, ou vallées étroites et profondes, comme le montrent les courbes de lumière tracées et ajustées avec le tableur Excel, voir figures IV-1-1 et IV-1-2:

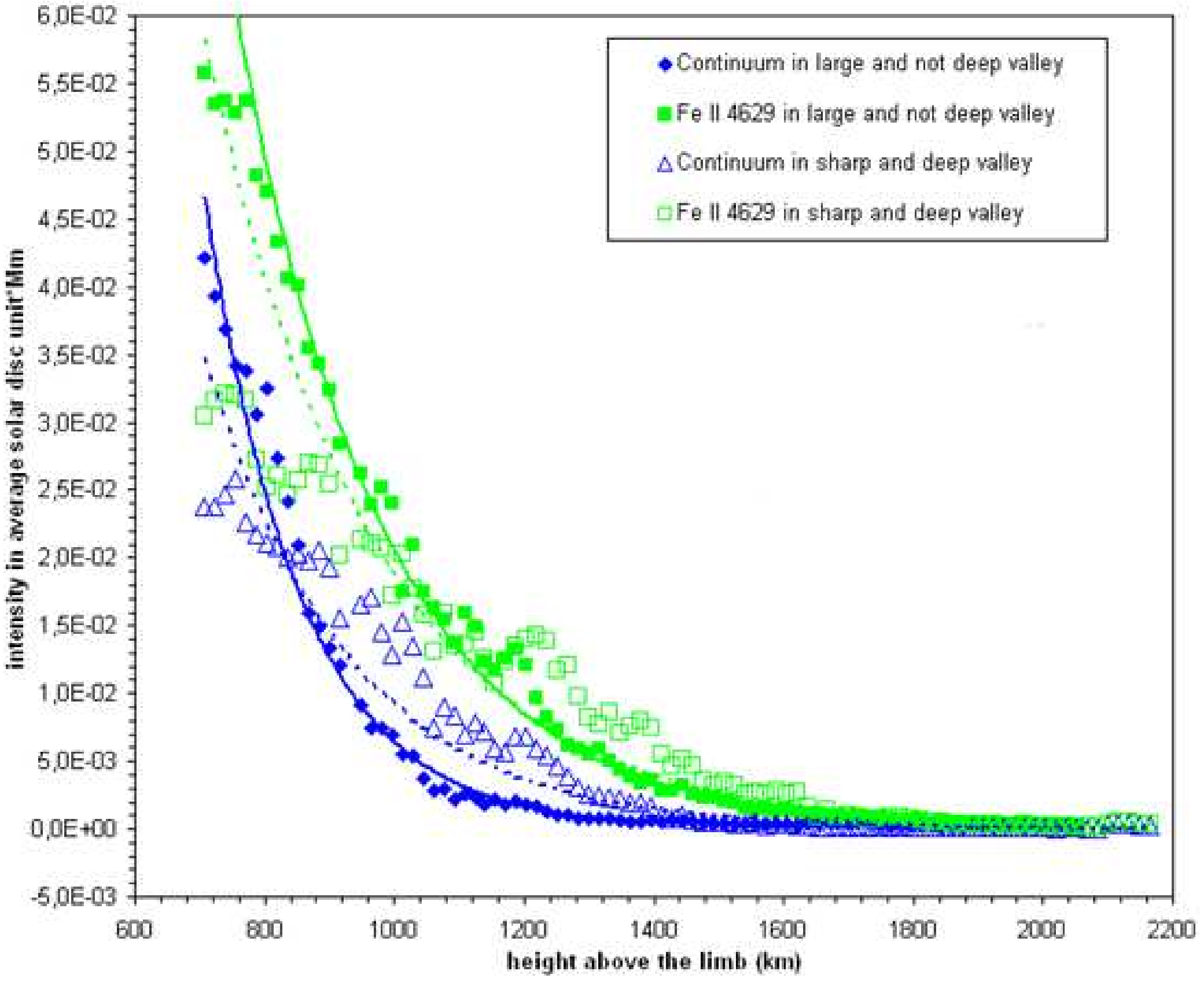

**Figure IV-1-1:** *Intensités linéiques relevées dans le continu en dehors des raies d'émission et relevées sur la raie du Fer II 4629 Å (« low FIP »), prises dans deux types de profils de vallée lunaire sur une étendue de plusieurs pixels en considérant i/ vallée étroite profonde et ii/ vallée large et peu profonde. Second contact C2 de l'éclipse du 1er Août 2008.*

Ces premières courbes servent à montrer l'effet du relief lunaire sur les mesures des courbes de lumière, et où les fluctuations peuvent être importantes, autour des courbes d'ajustement. Des premiers résultats avec des courbes d'ajustements ont été effectués, et les exponentielles décroissantes sont adaptées sur l'étendue des courbes $I = f(h)$ expérimentales.
Les écarts entre les valeurs de brillance de la courbe d'ajustement et les points de mesures sont de $\pm$ 0.003 unités du disque solaire*Mm, pour une brillance de $7*10^{-3}$ unités du disque solaire moyen*Mm autour de 1200 km d'altitude.
Les unités de ces relevés en « unités du disque solaire moyen*Mm » proviennent du procédé où chaque mesure correspond à un relevé d'intensités prises dans toute l'étendue de la vallée lunaire (largeur en Mm), et multiplié par les intensités moyennées relevée dans cette même vallée. Ces valeurs peuvent être assimilées à une brillance linéique. Les équations des courbes d'ajustement ont été appliquées sur les courbes de lumière et données ci-dessous :

Equation fit courbe de lumière Fe II Y243 : $I(h) = 15.16194*e^{-0.004400322\ h}$

Equation fit courbe de lumière Fe II Y277 : $I(h) = 6.255083*e^{-0.003871599\ h}$

Equation fit courbe de lumière Continu Y277 : $I(h) = 8.36297*e^{-0.004539491\ h}$

Equation fit courbe de lumière continu Y243 : $I(h) = 5.667979*e^{-0.006779284\ h}$

Ces équations montrent des différences pouvant atteindre un facteur 2 pour le terme pré-exponentiel, ce qui traduit l'effet de la nature du relief lunaire (topologie, vallées étroites ou profondes) sur les courbes de lumière du limbe solaire sans lumière parasite.
A partir des courbes de lumière obtenues et de leurs courbes d'ajustement, il est possible de déduire un autre paramètre, l'échelle de hauteur *H*, en faisant l'hypothèse d'un milieu homogène, hydrostatique et stratifié.
Les méthodes de R. Wildt pour le calcul des échelles de hauteurs, voir Introduction Chapitre I-3 ont été utilisées. Il suffit de prendre l'inverse du coefficient dans le terme exponentiel, précédant la variable *h*, pour déduire directement *H* en kilomètres. Une première estimation d'échelles de hauteur apparente déduite des courbes de lumière pour l'ion Fe II 4629 Å donne 227$\pm$ 50 km pour une vallée large et peu profonde, et devient 258 $\pm$ 50 km pour les relevés dans une vallée étroite et profonde. Pour le continu, ces résultats conduisent à 220 km pour une vallée étroite et profonde, et 147 km pour une vallée large et peu profonde.
Pour vérifier ces résultats, nous avons refait les mêmes analyses mais en utilisant le tableur Origin. Les résultats sont présentés en figure IV-1-2. Il a été possible d'ajuster le profil du continu dans la vallée large et peu profonde.

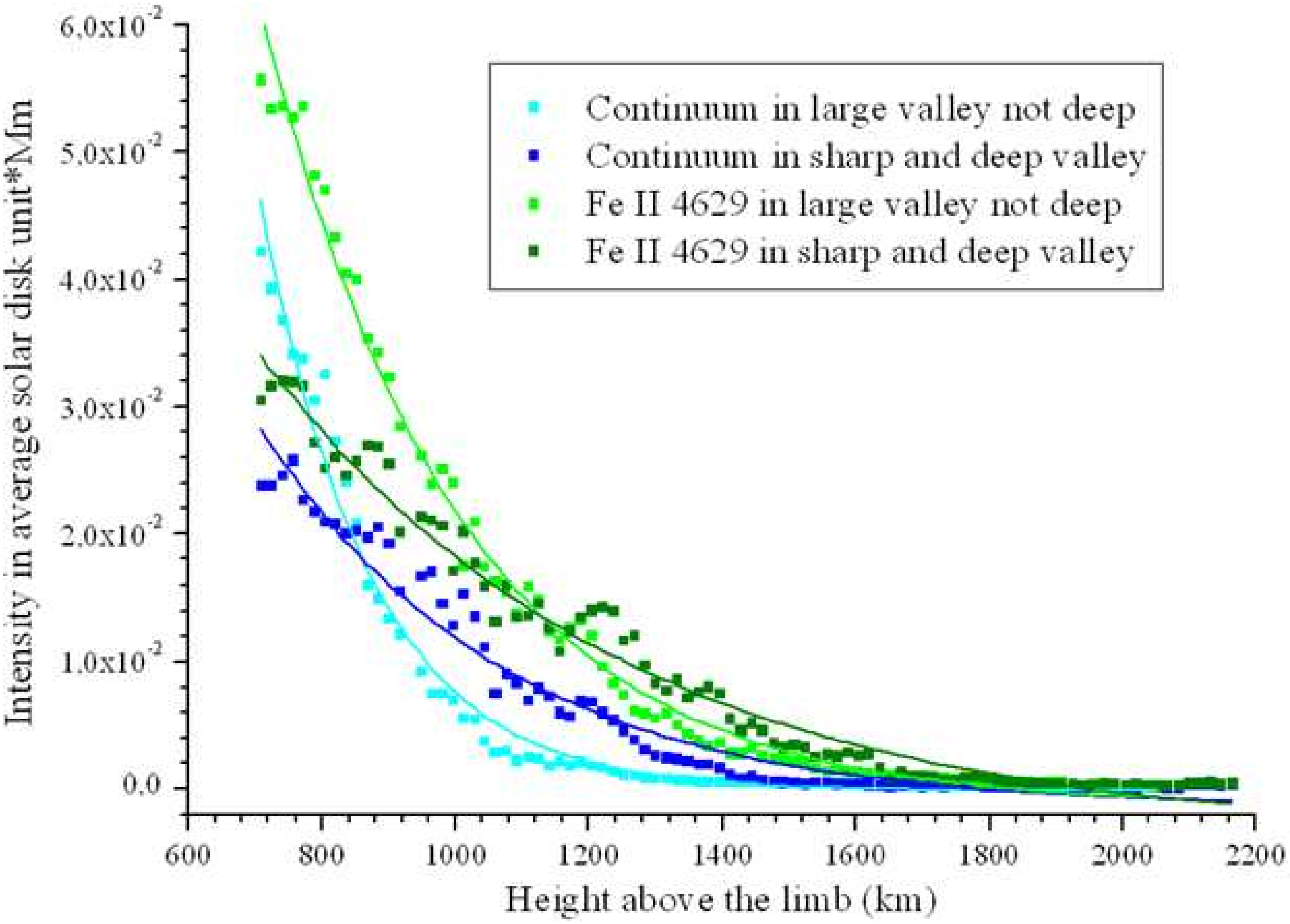

**Figure IV-1-2:** *Profils des courbes de lumière selon le type de vallée lunaire considéré, pour le continu relevé entre les raies et l'intensité dans la raie du Fe II 4629 Å, obtenu avec le Tableur Origin.*

D'après ces résultats les termes constants des exponentielles décroissantes d'ajustement permettent de déduire directement les échelles de hauteur, *H*, dans les équations types des exponentielles décroissantes:

$$I(h) = A{*}e^{h/H}$$

Le tableau IV-1-3 indique les différentes valeurs de *H* selon le type de vallée pour lesquelles les courbes de lumière des raies et continu ont été effectués et relevées selon le type de tableur utilisé, Excel 2003 et Origin 7, afin de comparer les résultats obtenus par les 2 logiciels.

| | |
|---|---|
| Echelles de hauteur du continu dans une vallée lunaire large et peu profonde | **147 ± 20 km** |
| Echelles de hauteur du continu dans une vallée lunaire étroite et profonde | **220 ± 50 km** |
| Echelles de hauteur du Fe II 4629 dans une vallée lunaire large et peu profonde | **227 ± 50 km** |
| Echelles de hauteur du Fe II 4629 dans une vallée lunaire étroite et profonde | **258 ± 50 km** |

**Tableau IV-1-3:** *Résultats des mesures d'échelles de hauteur apparente du continu et des intensités du Fe II 4629Å selon le type de vallée lunaire considéré, et le type de logiciel utilisé.*

Les barres d'erreurs sont de l'ordre de ± 50 km sur chaque estimation d'échelles de hauteur du Fe II, mais deviennent 2 fois plus faible pour une vallée large et peu profonde pour le continu. Ces résultats montrent que le relief lunaire a une influence sur l'estimation de l'échelle de hauteur, selon topologie (la largeur et profondeur des vallées lunaire). Les coefficients de corrélation se situent entre 95 et 98 % dans tous les cas, ce qui indique un bon ajustement. Les échelles de hauteurs sont plus élevées dans les vallées plus profondes.
Les dispersions des points pour cette courbe de lumière du Fe II et du continu sont de $\pm 3*10^{-3}$ unités du disque solaire*Mm. Les fluctuations de l'ordre de $5*10^{-3}$ unités du disque solaire moyen situées à 1200 km peuvent être liées aussi à la météo.
Le tableau IV-1-4 indique les températures hydrostatiques déduites à partir des premières évaluations des échelles de hauteur. L'expression $T_{hydro} = \frac{RH}{\mu_{ion} * G_{sol}}$ a été utilisée. $R$ = 8.32 $J*K^{-1}$ est la constante des gaz parfaits, $H$ est l'échelle de hauteurs en kilomètres, $\mu_{ion}$ est la masse atomique de l'espèce considérée, et $G_{sol}$ est la constante de gravitation à la surface du Soleil qui vaut 273 $m*s^{-2}$.

| Raie étudiée C2/2008 | longueur d'onde (Å) | masse ion g/moles | Températures hydrostatiques (modèle hydrostatique) |
|---|---|---|---|
| H I continu | 4700 | 1.0 | **4820 ± 500 K large peu profonde** |
| Fe II | 4629 | 55.8 | 410000 K large et peu profonde |
| H I continu | 4700 | 1.0 | **7220 ± 500 K étroite et profonde** |
| Fe II | 4629 | 55.8 | 470000 K étroite et profonde |

**Tableau IV-1-4:** *températures hydrostatiques du continu et du Fer II selon le relief lunaire, déduites des échelles de hauteurs mesurée et présentées au tableau IV-1-3.*

Les températures pour les ions associés aux raies « low FIP » sont représentées en grisé, car elles apparaissent 50 à 100 fois plus élevées que les températures évaluées pour le continu. Ces valeurs n'ont pas de sens dans ces régions proches du minimum de température.
Ces résultats montrent que le modèle hydrostatique stratifié ne permet pas de déduire les températures des ions associés aux raies « low FIP » dans les couches profondes de la région de transition, située au dessus de la photosphère.

D’autres raies d’émission à « low FIP » ont été observées, mais n’ont pas été étudiées en détail. Les raies d’émission de l’hélium neutre He I 4713 Å et He II 4686 Å qui sont des raies dites « high FIP » (c'est-à-dire que leur potentiel de première ionisation est supérieur à 10 eV) ont été enregistrées sur chaque spectre éclair. Les variations de ces raies ont été relevées pour chaque spectre sur une même région équatoriale. Les intensités dans les raies ont été mesurées au moyen du logiciel Iris, en utilisant le curseur pour effectuer les mesures d’intensité maximales prises dans un seul pixel à la fois, aux mêmes coordonnées dans les croissants des raies étudiées du spectre éclair (et ici ces mesures ne sont pas effectuées dans l’aire limitée par une vallée lunaire, car ces raies d’hélium sont formées plus haut que le continu et raies « low FIP »).
Ces mesures ont été répétées plusieurs centaines de fois, et en ayant tenu compte du mouvement du spectre. Les altitudes définies sur les courbes de lumière, intensité en fonction de l’altitude, $I = f(h)$, des raies d’hélium ont été déduites à partir des altitudes exprimées dans les courbes de lumière mesurées sur les spectres du continu. Les courbes de lumière du continu servent à définir la référence $h = 0$. Ces altitudes correspondent au prolongement de la diminution des flux des spectres des grains de Baily de la haute photosphère, lors des instants de contacts. Les enveloppes d’hélium (raies optiquement minces) « high FIP » ont un potentiel de première ionisation supérieur à 10 eV. Les courbes de lumière des raies de l’hélium neutre, et ionisé ainsi que le continu sont données sur la figure IV-1-4 en unités du disque solaire moyen où cette unité a été utilisée pour étalonner la brillance de la couronne voir cette méthode en Annexes 28 et 29.

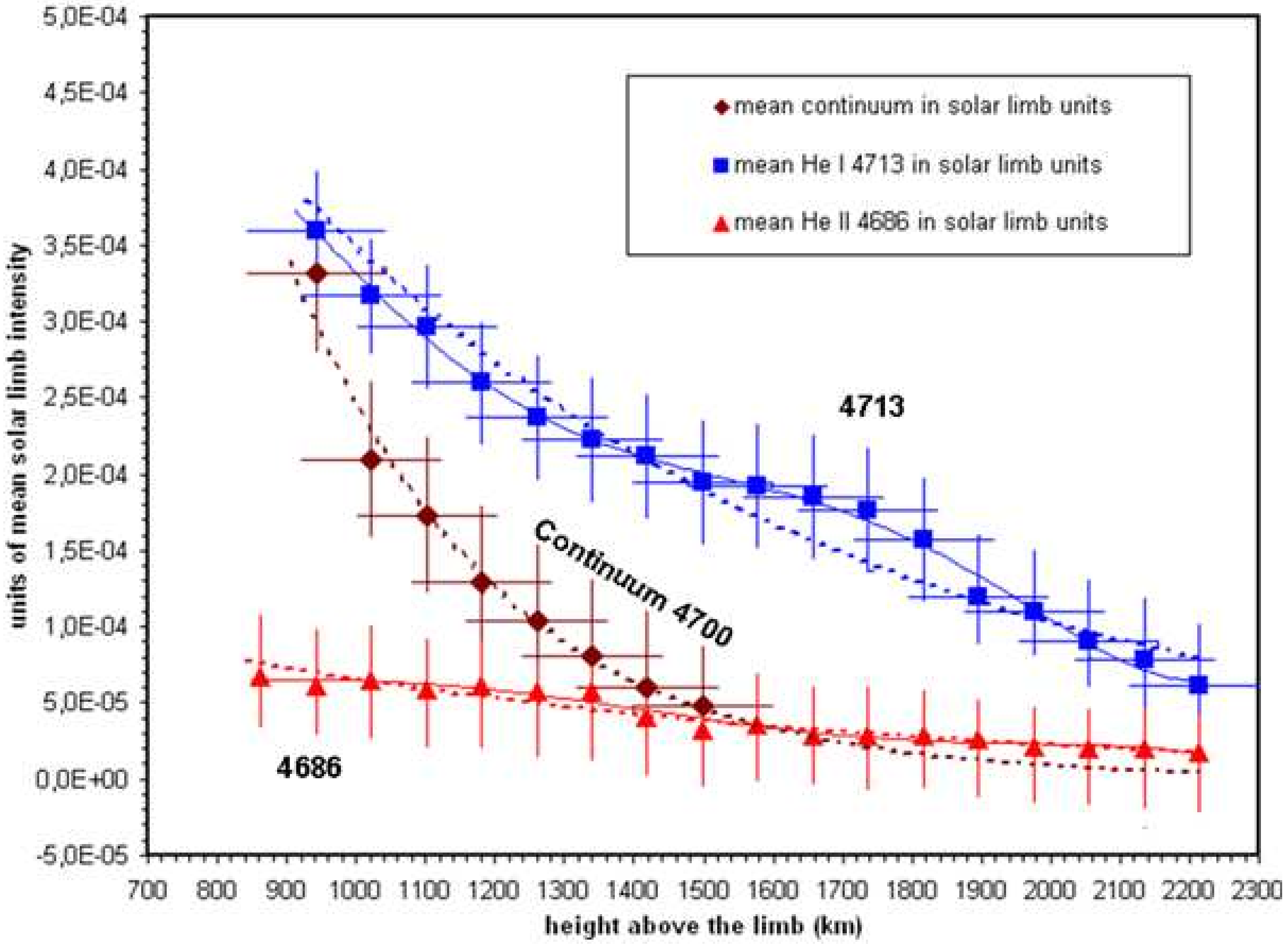

**Figure IV-1-4:** *Profils d'intensités des courbes de lumière obtenues à l'éclipse du 1^er^ Août 2008 au second contact. L'incertitude des hauteurs est de ±100 km, avec un niveau de confiance de 1 sigma. Remarquer la grande différence de gradients continu et intensité des raies d’hélium.*

Les courbes d'ajustement des enveloppes d'hélium peuvent être effectuées avec des polynômes de degré 5 ou 6 si on souhaite étudier les modulations de plus faible amplitude des enveloppes. Mais aussi, des exponentielles décroissantes conviennent si on ne prend pas en compte des variations plus faibles (émissivités des enveloppes), pour déduire les échelles de hauteur et effectuer ensuite les inversions d'intégrale d'Abel.

L'équation de la courbe d'ajustement sur la courbe de lumière *I(h)* du continu est une exponentielle décroissante en négligeant les effets d'enveloppe :

$I(h)_{continu} = 6.925051*10^{-3}*e^{-h/300}$ *H = 300 km*

avec *h*, altitude au dessus du limbe en kilomètres. La barre d'erreur est de $\pm\ 5*10^{-5}$ unités du disque solaire moyen. Le coefficient de corrélation entre les points expérimentaux et la courbe d'ajustement est :

$R^2 = 0.9925457$

Ceci montre un bon accord entre les points expérimentaux et la courbe d'ajustement.
De même pour les courbes de lumière des raies d'émission d'hélium neutre He I et ionisé He II, les équations d'ajustements type « exponentielle décroissante » sont :

$I(h)_{He_I_4713} = 1.179298*10^{-3}*e^{-h/821}$ *H = 821 km*

Les barres d'erreurs sont de $\pm\ 4*10^{-5}$ unités du disque solaire moyen
Le coefficient de corrélation entre les points expérimentaux et la courbe d'ajustement est :

$R^2 = 0.9433009$

$I(h)_{He_II_4686} = 1.935862*10^{-4}*e^{-h/926}$ *H = 926 km*

La barre d'erreur est de $\pm\ 6*10^{-5}$ unités du disque solaire moyen et le coefficient de correlation pour cette courbe d'ajustement est :

$R^2 = 0.9496437$

Le coefficient de corrélation des profils exponentiels est meilleur pour la courbe de lumière du continu que pour les courbes de lumière des raies d'hélium neutre et ionisé, à cause de la « bosse » qui correspond à l'effet « enveloppe » sur les raies d'hélium neutre et ionisé et introduit un éccart.
Pour les enveloppes d'hélium, les ajustements avec des courbes polynômiales de degré 6 ont été effectués. Ce type d'ajustement est adapté car un polynôme de degré 6 a permis de prendre en compte la « bosse » autour de 1700 km pour le profil de l'hélium neutre He I 4713 Å. L'analyse suivante a consisté à utiliser des méthodes de calculs d'inversion d'intégrales d'Abel pour déduire les échelles de hauteur et ensuite à partir des mêmes courbes de lumière, rechercher les altitudes des maximum d'émissivité une fois l'inversion effectuée. Les méthodes de calculs, formules sont décrites en Annexe 30. La méthode utilisée a consisté à effectuer des simplifications selon les étapes suivantes :

- application d’une courbe d’ajustement de type exponentielle décroissante sur chaque courbe de lumière
- application d’une courbe d’ajustement de type polynômial sur chaque courbe de lumière
- Ensuite j’ai effectué la dérivée par rapport aux altitudes pour chaque courbe d’ajustement dans les 2 types d’ajustement et pour chaque courbe de lumière.
- J’ai ensuite divisé chaque point des courbes dérivées par la valeur de la corde correspondante à chaque altitude pour tenir compte des effets de ligne de visée
- J’ai ensuite intégré par approximations rectangulaires et où j’ai appliqué les calculs sur les lignes des colonnes dans le tableur Excel (Annexe 30)
- A partir des courbes obtenues, il a été possible d’obtenir des échelles de hauteur

Les courbes figure IV-1-5 indiquent le tracé des émissivités après l’inversion d’intégrale d’Abel effectuée d’après les équations des courbes d’ajustements (fit) avec l’approximation des exponentielles décroissantes, sans tenir compte de l’effet d’enveloppe. Ces résultats correspondent aux analyses du second contact avant la totalité de l’éclipse du 1er Août 2008.

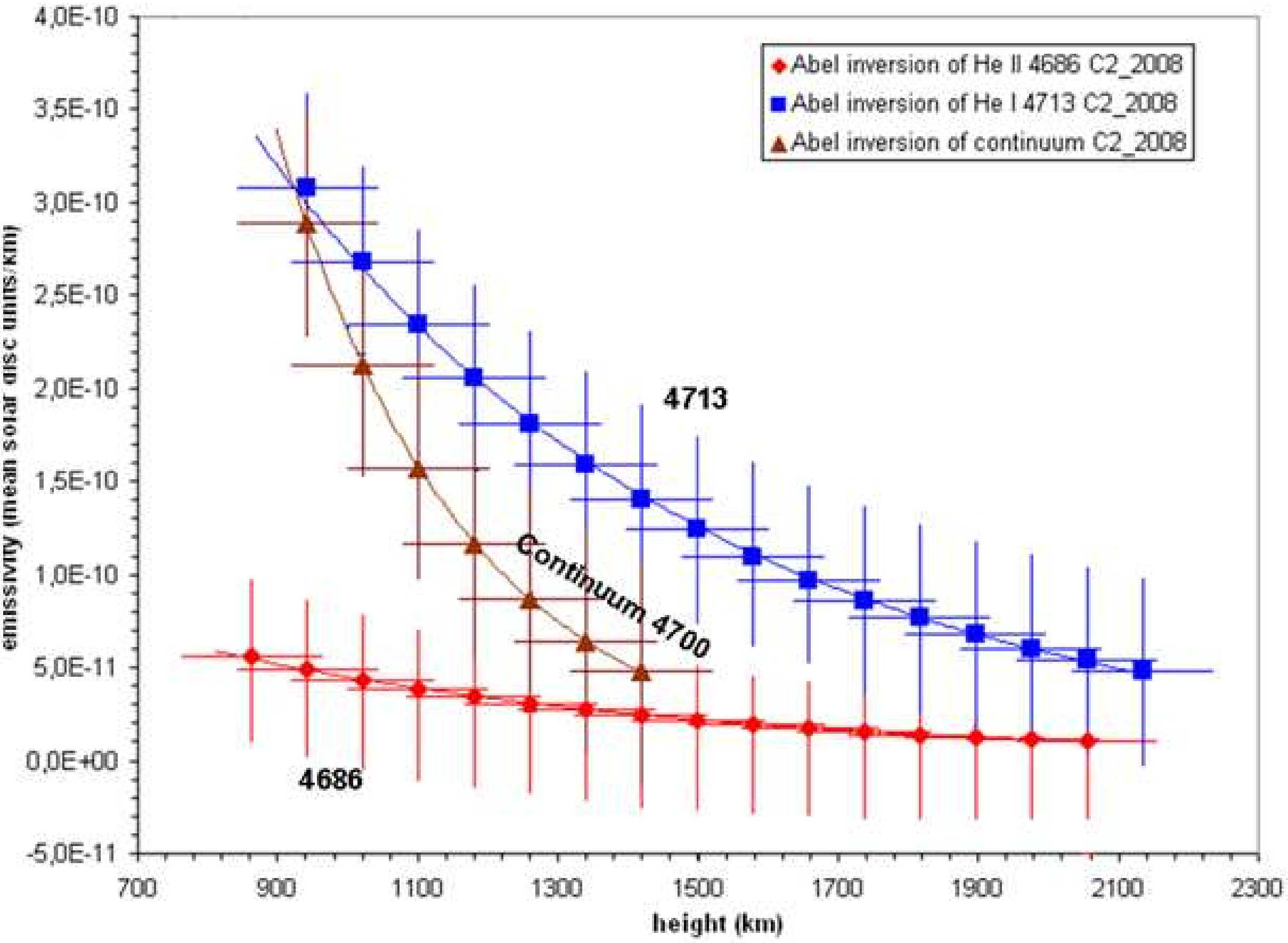

**Figure IV-1-5:** *émissivités déduites par inversion d’intégrale d’Abel d’après les courbes d’ajustements avec approximations d’exponentielles décroissantes des courbes de lumière au second contact de l’éclipse du 1er Août 2008.*

Les échelles de hauteur peuvent être déduites sur les courbes en décroissances exponentielles, car l’inversion d’intégrale d’Abel appliquée conserve l’allure exponentielle et ne modifie pas ou très peu les valeurs des échelles de hauteurs. Celles-ci peuvent être estimées aussi bien sur les courbes d'émissivité en fonction de la hauteur comme sur les courbes d'ajustement *I = f(h)*.
Les équations des courbes d'ajustements déduites des graphiques d'inversions d'intégrales d'Abel au second contact de l'éclipse de 2008 sont indiquées pour le continu et les raies de He I 4713Å et He II 4686Å:

Continu à 4700Å: $E(h) = 9.969444 * 10^{-9} * e^{-h/265}$ $\quad H = 265 \text{ km}$

Les valeurs en précision des barres d'erreurs sont $\pm\ 6*10^{-11}$ unités du disque solaire moyen/km. Les unités sont exprimées en unités du disque solaire moyen par kilomètre, même si ce type d'unité n'est pas usuel pour exprimer des émissivités. Ces émissivités évaluées ici sont exprimées en fonction de la brillance du disque solaire moyen, pris comme référence.

He I 4713: $E(h) = 1.292673 * 10^{-9} * e^{-h/643}$ $\quad H = 643 \text{ km}$

Les valeurs en précision des barres d'erreurs sont $\pm\ 5*10^{-11}$ unités du disque solaire moyen/km

He II 4686: $E(h) = 1.880803 * 10^{-10} * e^{-h/696}$ $\quad H = 696 \text{ km}$
Les valeurs en précision des barres d'erreurs sont $\pm\ 6*10^{-11}$ unités du disque solaire moyen/km.
D'après la courbe d'émissivité figure IV-1-5 et les équations d'ajustement déduites, l'échelle de hauteur de 265 km est mesurée pour le continu associé à l'ion H I de l'hydrogène neutre (Hiei 1962). Cette valeur conduit à une température hydrostatique de 8685 K qui est excessive car les altitudes correspondent aux régions du minimum de température (T< 5000 K). Cette valeur exscissive peut s'expliquer des façons suivantes : i/ instrumentales, où la caméra manquait de dynamique, seulement en 8 bit, et les spectres étaient saturés dans les couches profondes, ce qui rendait les mesures difficiles à des échelles de hauteurs plus faibles. ii/ le relief lunaire dans le quel le continu a été mesuré correspond à des vallées étroites et profondes, ce qui conduit à des valeurs plus élevées des échelles de hauteurs, voir tableau IV-1-3.
A partir des échelles de hauteur, il est possible de déduire des températures, en faisant l'hypothèse de l'équilibre hydrostatique, et que les couches sont homogènes et stratifiées.
Le tableau IV-1-6 indique les températures *hydrostatiques*, déduites à partir des mesures des échelles de hauteur après l'inversion d'intégrales d'Abel.

| Raie étudiée C2/2008 | longueur d'onde (Å) | masse ion g/mole | Température hydrostatique (modèle hydrostatique) |
|---|---|---|---|
| H I continu | 4700 | 1.0 | **8685 $\pm$ 656 K** |
| | | | |
| He I | 4713 | 4.0 | 84400 K |
| He II | 4686 | 4.0 | 91350 K |

**Tableau IV-1-6 :** *températures hydrostatiques déduites des échelles de hauteur à partir des courbes d'émissivité des ions de raies He I et He II et une raie « low FIP » Fe II 4629* Å.

Les températures indiquées en grisé, pour les ions associés aux raies « low FIP » n'ont pas de sens dans ces régions si proches du minimum de température où celle-ci ne peut excéder 10000 K.

Ces résultats montrent que l'hypothèse de modèle hydrostatique stratifié est inadaptée pour déduire les températures des ions associés aux raies « low FIP » dans les couches profondes de la région Ph-C/I, située au dessus de la photosphère jusque dans la chromosphère.
Les courbes Figure IV-1-7 donnent l'inversion d'intégrale d'Abel en ayant pris des ajustements polynomiaux qui tiennent compte des variations dues à l'enveloppe vers 1700 km:

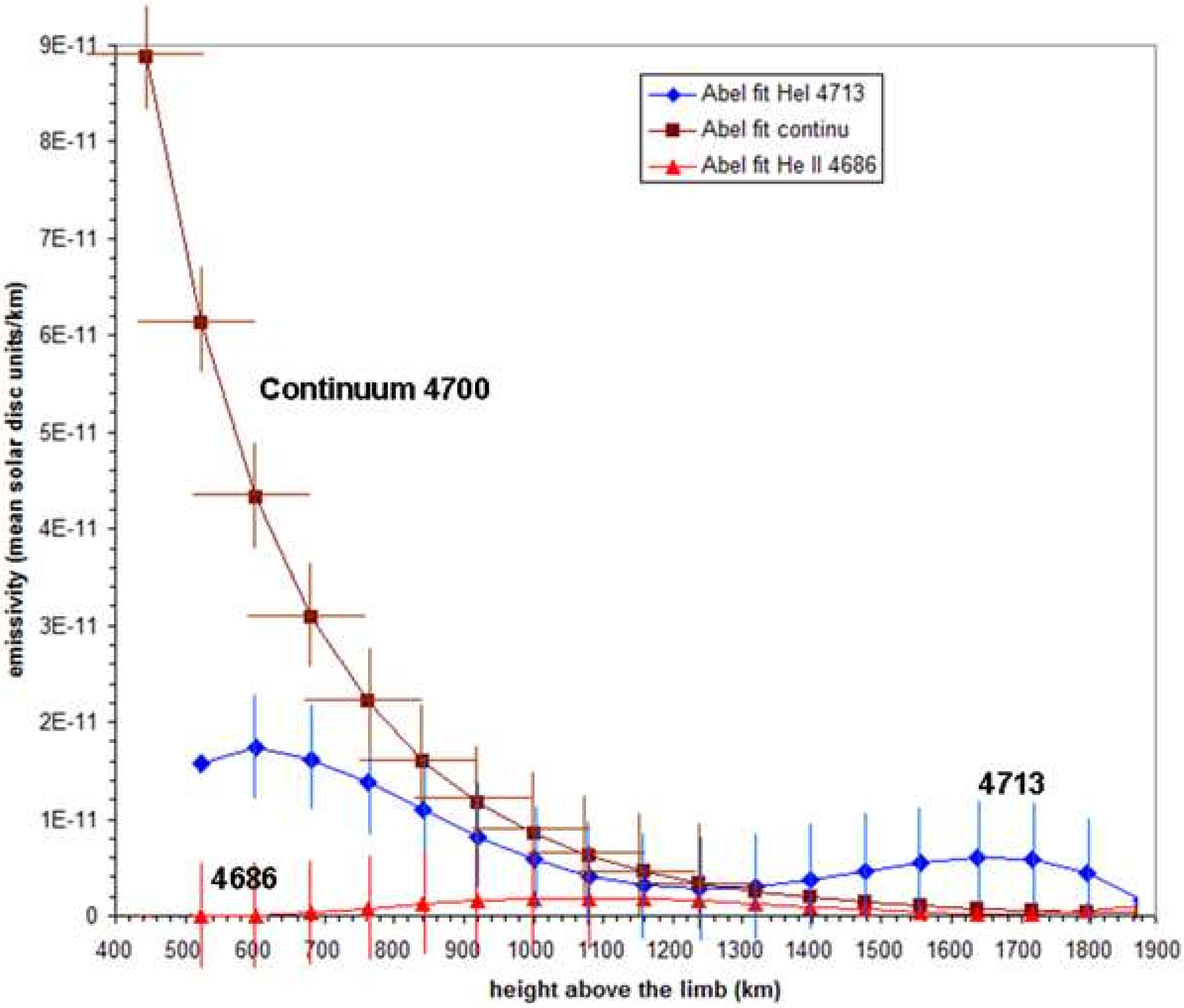

**Figure IV-1-7:** *émissivités déduites par inversion d'intégrale d'Abel d'après les courbes d'ajustements de type polynomiales des courbes de lumière au second contact de l'éclipse de 2008.*

Les équations des courbes d'ajustements polynômiales pour les courbes de lumière, $I = f(h)$ *avec* $h$ en km, de l'hélium neutre He I 4713 Å et hélium ionisé He II 4686 Å sont les suivantes :

He I 4713 Å : $I(h) = 7{,}382085*10^{-19}*h^5 - 4{,}139790*10^{-15}*h^4 + 8{,}630895*10^{-12}*h^3 - 8{,}121657*10^{-9}*h^2 + 3{,}055533*10^{-6}*h + 5{,}570210*10^{-5}$

He II 4686 Å : $y = -1{,}32786410^{-19}*h^5 + 6{,}989322*10^{-16}*h^4 - 1{,}332817*10^{-12}*h^3 + 1{,}106825*10^{-9}*h^2 - 4{,}131881*10^{-7}*h + 1{,}221720*10^{-4}$

La courbe du continu ne montre pas de point d'inflexion. Ce résultat est important pour la définition du bord solaire, où les modèles prévoyaient un point d'inflexion. Ce point d'inflexion observé hors conditions d'éclipses est du à l'étalement de la lumière parasite provenant du disque solaire, et des effets instrumentaux.
Les courbes de lumière des raies d'hélium neutre 4713 Å et hélium ionisé 4686 Å obtenues en 2008, ont été à nouveau reproduites à partir des résultats de l'éclipse du 11 Juillet 2010, où ces mêmes raies ont été observées. Des nouvelles courbes de lumière ont été obtenues pour ces mêmes raies. Cela a permis de comparer les résultats obtenus lors de l'éclipse du 1$^{er}$ Août 2008 avec ceux du 11 Juillet 2010 sur les mêmes raies. Le maximum d'émissivité pour la raie de l'hélium He I 4713Å se situe autour de 1670 km. Le maximum d'émissivité de l'enveloppe de l'hélium ionisé He II 4686 a été difficile à mesurer à cette éclipse de 2008.
Les 2 méthodes d'inversion, l'une avec un ajustement de type décroissance exponentielle et l'autre par ajustement polynomial, permettent d'obtenir des paramètres complémentaires (échelles de hauteurs ou maxima d'émissivité des enveloppes) en faisant des approximations. En conclusion, l'ajustement de type « exponentielle décroissante » ne tient pas compte des enveloppes, et il est possible de déduire des échelles de hauteurs. L'ajustement de type polynômial permet d'accentuer les variations des émissivités dans les enveloppes d'hélium, mais le profil du bord de la Lune introduit des fluctuations, ce qui rend difficile les mesures de l'enveloppe d'hélium ionisé 4686Å obtenue en 2008.
Par ailleurs d'autres raies d'hélium neutre He I 4471 Å et He I 4387,9 Å ont été observées lors de l'éclipse du 22 Juillet 2009, avec la même expérience mais sur un intervalle spectral différent et centré sur 4450 ± 120Å.
Les résultats des analyses des spectres éclair obtenus à l'éclipse du 22 Juillet 2009 sont présentés dans le chapitre IV-2, où des effets d'enveloppe ont été bien observés dans la raie de l'hélium neutre He I 4471Å et aussi constatés dans la raie du Baryum une fois ionisé Ba II 4554 Å, et plus faiblement dans une raie du Fe II 4541Å.

# IV-2) Analyses des résultats de l'éclipse du 22 Juillet 2009

## IV-2-1) Analyses des courbes de lumière des raies low et high FIP, et du continu

Le contenu de ce chapitre IV-2-1 a pour objectif de présenter les analyses des profils des courbes d'émissivités, ou de brillance à partir des courbes de lumière des spectres éclairs de 2009, pour en déduire les échelles de hauteur pour plusieurs raies d'émission observées. Ce chapitre a pour objectif de rechercher des corrélations à partir des raies « low FIP » et « high FIP » des spectres éclair de l'éclipse de 2009 entre l'interface photosphère-couronne et l'interface protubérance-couronne, pour les ions correspondant aux éléments « low FIP » (Fer, Titane) et « high FIP » He I 4471Å. Des analyses d'extensions des macrospicules dans la chromosphère et jusqu'à la couronne sont effectuées, où des condensations sont observées dans la raie de He I 4471Å après traitement, aux limites de l'enveloppe, dans l'interface macrospicule-couronne. La mésosphère solaire ou atmosphère intermédiaire est introduite, en vue d'apporter plus de détails sur la couche située au-dessus de la photosphère, qui était appelée autrefois « couche renversante » et qui avait une étendue de 500 km. Grâce aux spectres éclairs CCD et avec la résolution atteinte de 25 km dans l'atmosphère, cette couche mésosphérique, constituée de la myriade de raies d'émission « low FIP » et située dans la région du minimum de température, et très bas dans l'interface Ph-C/I a pu être analysée plus en détails à partir des courbes de lumière analysées sur quelques raies « low FIP ».

Les spectres éclair obtenus à l'éclipse du 22 Juillet 2009 ont été analysés un par un et des courbes de lumière, intensité dans les raies en fonction de l'altitude, sont données, ainsi que les évaluations d'échelles de hauteur, d'après les courbes d'ajustement de type exponentielles décroissantes et polynômiales. Les résultats sont présentés pour quelques raies « low FIP » (Ti II, Ba II, Fe II), et « high FIP » hélium neutre He I. Quelques calculs et tracés de des variations de l'émissivité en fonction de l'altitude ont été effectués, après inversion d'intégrale d'Abel, afin de renforcer la visibilité et extentions des enveloppes.
La figure IV-2-1-1 montre les profils d'intensité de quelques raies lors du second contact de l'éclipse du 22 Juillet 2009. Le choix sur l'axe des ordonnées des unités du disque solaire moyen des intensités des raies low FIP dans les courbes de lumière, reprend la méthode décrite aux Annexes 28 et 29.

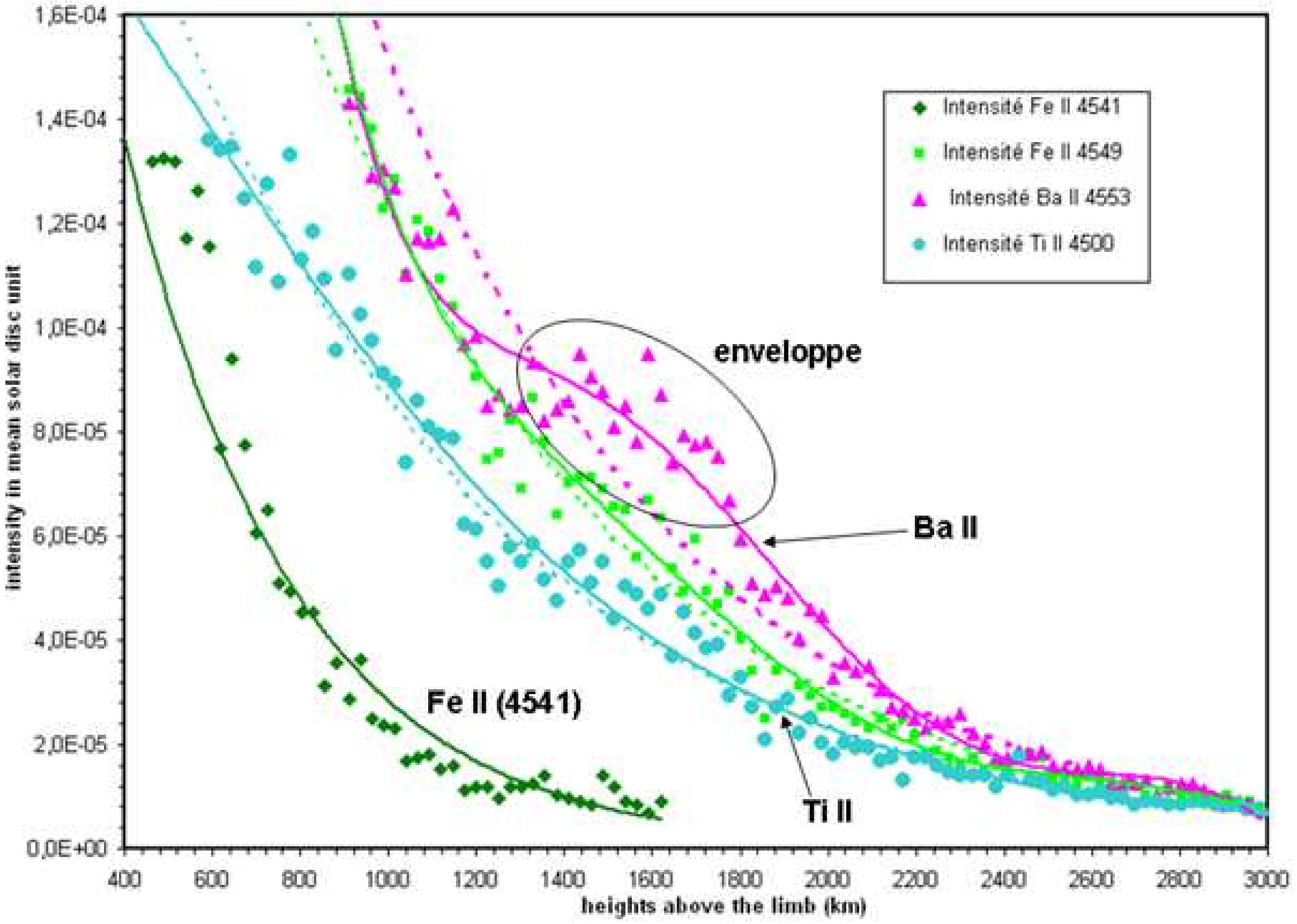

**Figure IV-2-1-1:** *profils des intensités des courbes de lumière obtenues par les mesures des intensités dans les raies du Fe II, Ti II, et Ba II lors du second contact C2 à l'éclipse totale de Soleil du 22 Juillet 2009. Des courbes d'ajustements polynomiaux sont effectuées en traits pleins, et avec des exponentielles décroissantes représentées en pointillés.*

Les altitudes représentées en abcsisse sont déduites des étalonnages donnés en figure III-4-3. Les formules des courbes d'ajustements polynômiaux du Ba II 4554 Å et Fe II 4549 Å sont les suivantes:

Ba II 4554 Å: $I(h) = 4{,}029954 \cdot 10^{-23} \cdot h^6 - 5{,}584986 \cdot 10^{-19} \cdot h^5 + 3{,}102089 \cdot 10^{-15} \cdot h^4 - 8{,}817311 \cdot 10^{-12} \cdot h^3 + 1{,}350853 \cdot 10^{-8} \cdot h^2 - 1{,}064299 \cdot 10^{-5} \cdot h + 3{,}491768 \cdot 10^{-3}$

Fe II 4549 Å: $y = 3{,}319240 \cdot 10^{-23} \cdot h^6 - 4{,}276436 \cdot 10^{-19} \cdot h^5 + 2{,}23660510^{-15} \cdot h^4 - 6{,}072323 \cdot 10^{-12} \cdot h^3 + 9{,}053776 \cdot 10^{-9}\ h^2 - 7{,}135072 \cdot 10^{-6} \cdot h + 2{,}436966 \cdot 10^{-3}$

Les formules des exponentielles décroissantes d'ajustement pour les courbes de lumière des 4 éléments chimiques sont les suivantes :

Ba II 4554 : y = 6,288087E-04e-1,425572E-03x

Fe II 4549 Å : $I(h) = 5{,}009585*10^{-4}*e^{-0.001403227h}$

Fe II 4541 Å : $I(h) = 3{,}869843*10^{-4}*e^{-0.002612912h}$

Ti II 4500 Å : $I(h) = 3{,}169448*10^{-4}*e^{-0.001300093h}$

La courbe du Ba II montre clairement une distribution due à l'effet d'enveloppe entre 1250 et 1750 km, ce qui est un résulat nouveau.
Pour le Ba II 4554 Å et Fe II 4549 Å. Les précisions des différentes raies analysées sont présentées dans le tableau suivant et servent à évaluer les barres d'erreurs dans les courbes d'émissivités présentées plus loin:

| Raie étudiée | longueur d'onde (Å) | Précision sur les mesures d'intensités autour de h = 1400 km au dessus du limbe solaire |
|---|---|---|
| Fe II | 4541 | $1.0*10^{-5} \pm 0.9*10^{-5}$ unités du disque solaire moyen |
| Fe II | 4549 | $7.8*10^{-5} \pm 1.2*10^{-5}$ unités du disque solaire moyen |
| Ba II | 4554 | $9.4*10^{-5} \pm 1.0*10^{-5}$ unités du disque solaire moyen |
| Ti II | 4500 | $5.0*10^{-5} \pm 1.6*10^{-5}$ unités du disque solaire moyen |

**Tableau IV-2-1-2 :** *dispersions des mesures d'intensité dans les relevés des courbes de lumière pour le Fe II, Ba II, Ti II pour une altitude 1400 km au dessus du limbe*

Les dispersions sur les points de mesures dans les courbes de lumière incluent les fluctuations dues aux conditions météorologiques qui sont 5 à 10 fois plus importantes que les fluctuations autour de la linéarité lors des étalonnages de la caméra Lumenera Skynyx utilisée en 8 bit à cette éclipse (Voir annexe 17.1, figure 17.1.1).
Les courbes figures IV-2-1-3, IV-2-1-4 et IV-2-1-5 donnent l'émissivité déduite par inversion d'intégrales d'Abel des courbes d'ajustements effectuées sur les courbes de lumière précédentes avec un ajustement de type exponentielle décroissante. Les méthodes de calcul d'inversions d'intégrales d'Abel et simplifications, à partir des tableaux de données sont indiquées en annexe 30.

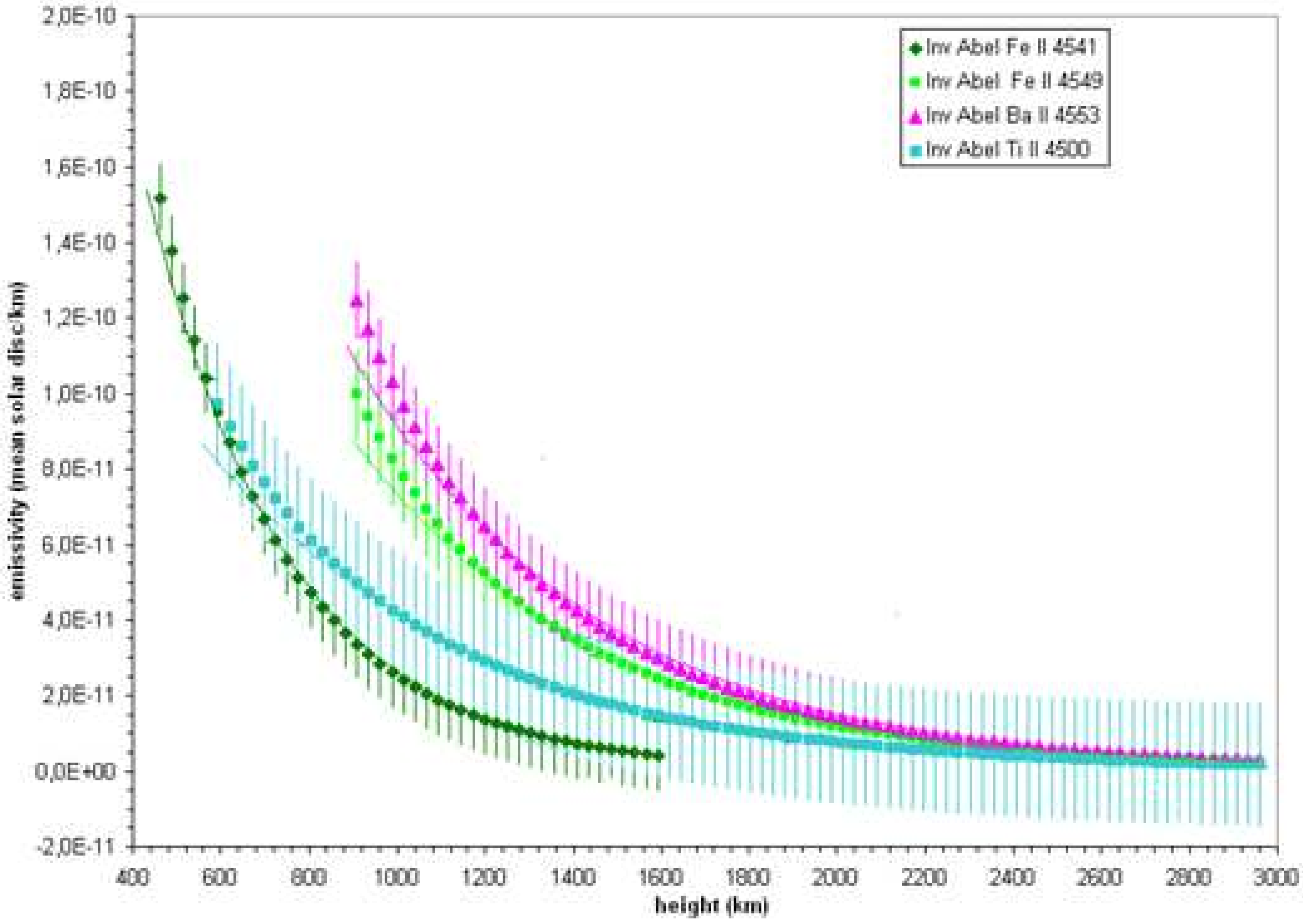

**Figure IV-2-1-3 :** *émissivités (ou brillances) des raies du Fe II 4549 Å, Fe II 4541 Å, Ba II 4553.8 Å et Ti II 4500 Å déduites des courbes de lumière précédente du second contact C3, à partir d'ajustements d'exponentielles décroissantes, sans tenir compte des effets d'enveloppes.*

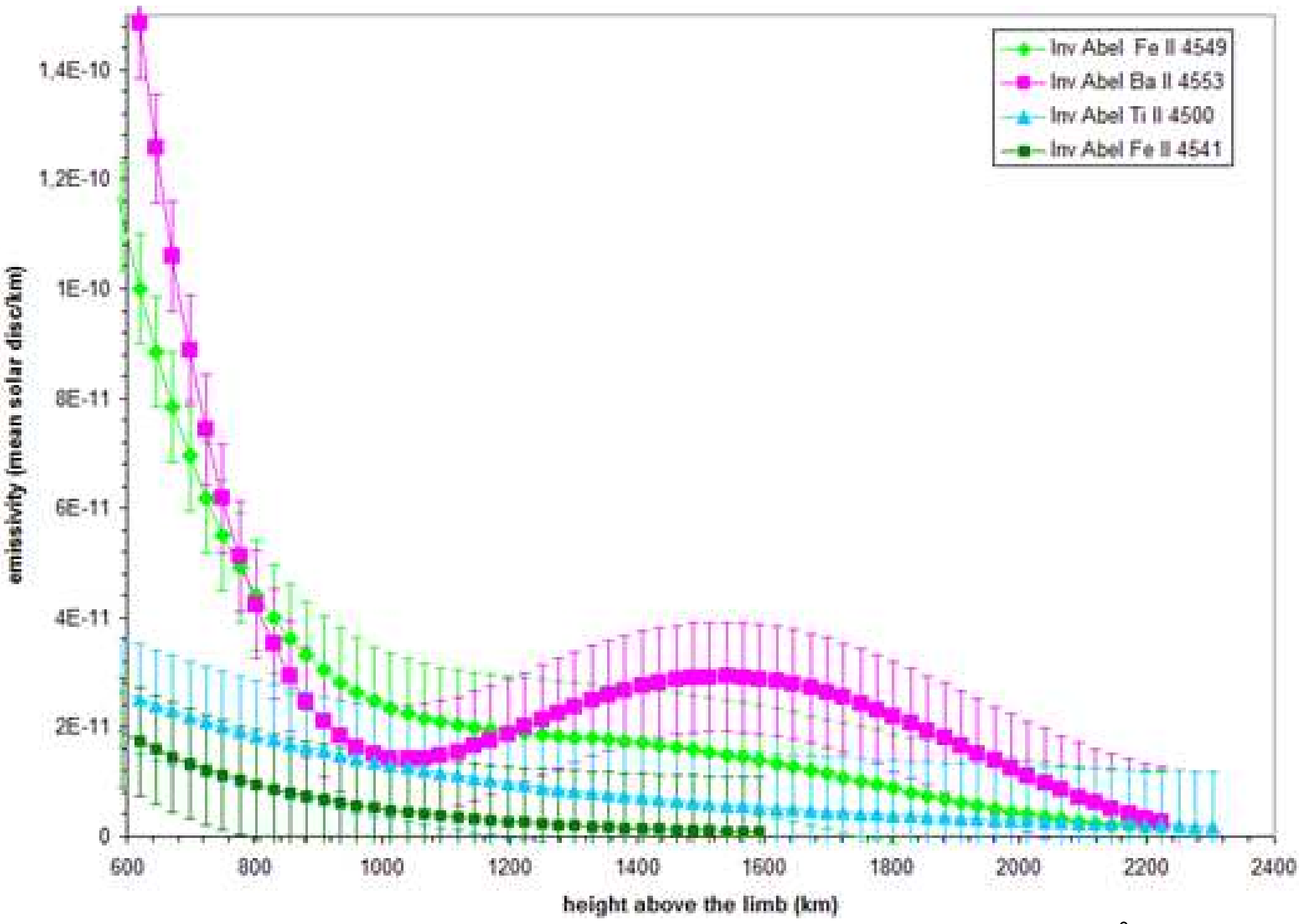

**Figure IV-2-1-4:** *émissivités (ou brillances) des raies du Fe II 4549 Å, Fe II 4541 Å, Ba II 4553.8 Å et Ti II 4500 Å déduites des courbes de lumière précédentes, en ayant utilisé les ajustements polynomiaux pour prendre en compte les enveloppes. Résultats après calcul d'inversion d'intégrales d'Abel du second contact C2 de l'éclipse totale de Soleil du 22 Juillet 2009.*

Le maximum d'émissivité du Barium Ba II 4553.8 Å se situe autour de 1550 km d'altitude au dessus du limbe solaire. Cette altitude correspond aussi au maximum de brillance des enveloppes d'hélium neutre He I 4713Å, observées en 2008 et 2010. Les courbes figure IV-2-1-4 sont ensuite converties en échelle log pour mieux évaluer l'enveloppe plus faible de la courbe d'émissivité du Fe II 4549 Å en figure IV-2-1-5.

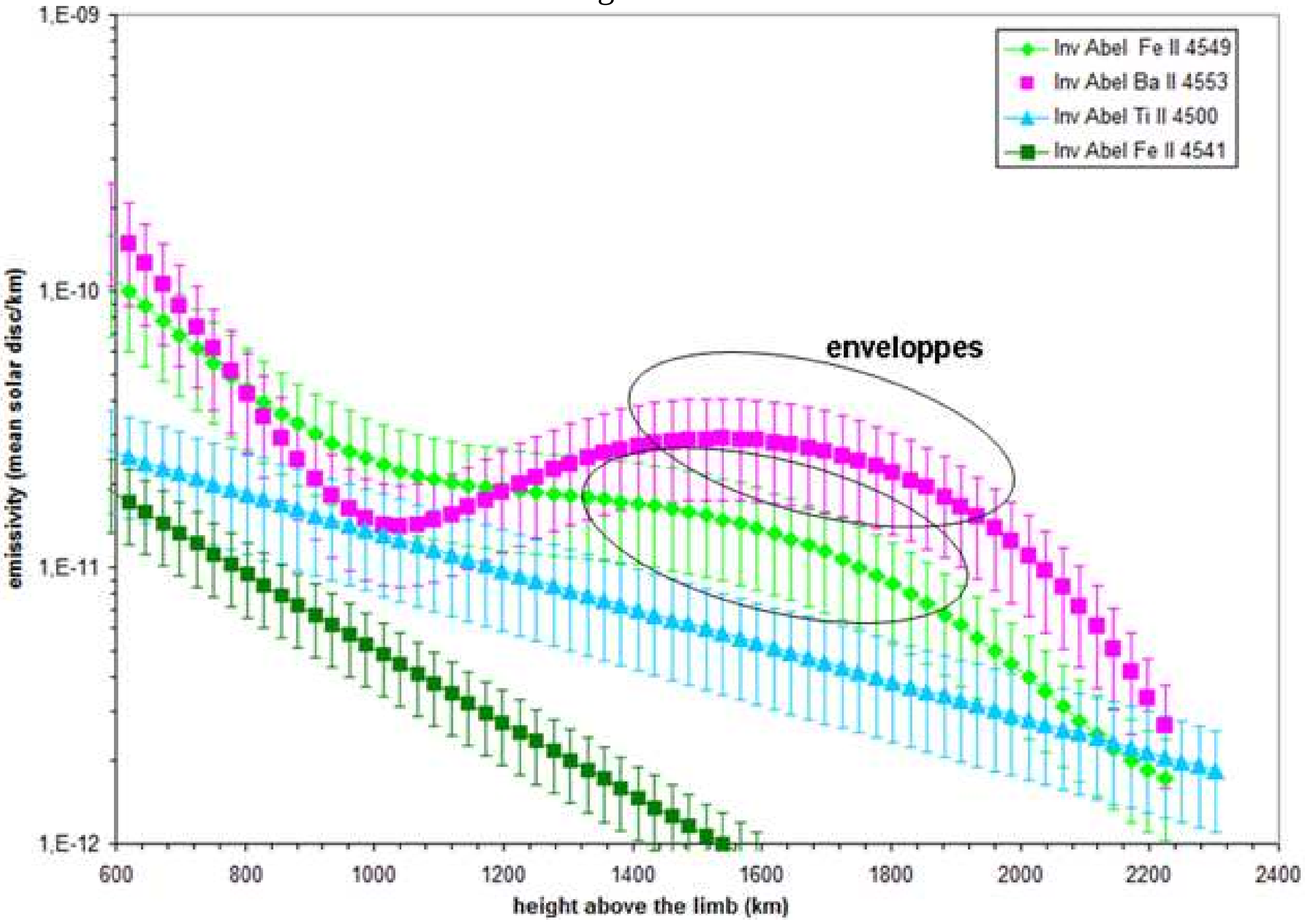

**Figure IV-2-1-5:** *émissivités (ou brillances) déduites des courbes de la figure précédente après conversion en échelle logarithmique des raies du Fe II 4549 Å, Fe II 4541 Å, Ba II 4553.8 Å et Ti II 4500 Å.*

Les courbes d'ajustement appliquées sont des polynômes de degré 6, pour les courbes de lumière (avant inversion d'intégrale d'Abel) du Ba II 4553.8 Å et du Fe II 4549 Å. C'est à partir de ces équations des courbes d'ajustements que les calculs d'inversions d'intégrales d'Abel ont été effectués.

Le calcul d'inversion d'intégrale d'Abel a permis de renforcer la visibilité des enveloppes du Ba II 4553.8Å et surtout du Fe II 4549Å afin de mieux déduire l'étendue de l'enveloppe du Barium une fois ionisé, mais dont l'existence et la vraisemblance sont discutables à ces altitudes, car bien que ces enveloppes soient observées pour les raies He I « high FIP », ces enveloppes ne sont pas observées pour les autres raies du Ti II 4500 Å et Fe II 4541 Å. En effet, la raie du Ba II est sensible à la raie Lyα très intense, et cette raie UV augmente la visibilité de l'enveloppe de Ba II voir Landman 1985.

Cependant, l'ajustement avec un polynôme de degrés 5 et 6 a été difficile à effectuer pour les courbes de lumière de Ti II 4500 Å et Fe II 4541 Å car les intensités relevées sur les courbes de lumière étaient plus faibles que ceux du Ba II 4553.8 Å et Fe II 4549 Å.

Nous avons effectué un ajustement avec une exponentielle décroissante pour les raies du Ti II 4500 Å et Fe II 4541 Å, car l'ajustement de type polynomial ne permettait pas de distinguer des effets d'enveloppes sur ces raies.

D’après ces courbes d’émissivités obtenues par les approximations d’exponentielles décroissantes, sans tenir compte des effets d’enveloppes, les équations des courbes d’ajustement obtenues pour les différents éléments au second contact de 2009 sont les suivantes :

Fe II 4541 Å : $5.988131*10^{-10} * e^{-0.00313542\ h}$

Fe II 4549 Å : $4.354371*10^{-10} * e^{-0.00178146\ h}$

Ba II 4553 Å : $5.551069*10^{-10} * e^{-0.001803805\ h}$

Ti II 4500 Å : $2.174547*10^{-10} * e^{-0.001646436\ h}$

Le tableau IV-2-1-6 présente les échelles de hauteur déduites à partir des équations des courbes d’ajustement précédentes:

| Raie étudiée C2/2009 | longueur d’onde (Å) | Echelles de hauteur |
|---|---|---|
| Fe II | 4541 | 319 ± 50 km |
| Fe II | 4549 | 561 ± 60 km |
| Ba II | 4554 | 554 ± 60 km |
| Ti II | 4500 | 607 ± 60 km |

**Tableau IV-2-1-6 :** *Echelles de hauteurs déduites d’après les courbes d’émissivité, après inversion d’intégrale d’Abel.*

A partir de l’équation suivante exprimant la relation linéaire entre l’échelle de hauteur et la température hydrostatique, il est possible de déduire des températures en faisant l’hypothèse d’un modèle hydrostatique stratifié.

$$H = \frac{RT_{hydrostatique}}{\mu_{ion} G_{solaire}}$$

R= 8.32 $JK^{-1}$, $G_{solaire}$ = 273 $ms^{-2}$

Nous déduisons des températures hydrostatiques pour les émissivités déduites des courbes de lumière des différents ions, à partir des échelles de hauteur mesurées.

| Raie étudiée C2/2009 | longueur d’onde (Å) | masse ion g/moles | Températures hydrostatiques (modèle hydrostatique) |
|---|---|---|---|
| Fe II | 4541 | 55.8 | 0.58 MK |
| Fe II | 4549 | 55.8 | 1.02 MK |
| Ba II | 4553 | 137.3 | 2.50 MK |
| Ti II | 4500 | 47.9 | 0.95 MK |

**Tableau IV-2-1-7:** *températures hydrostatiques déduites des échelles de hauteur de rayonnement à partir des courbes d’émissivité des ions de raies low FIP.*

Ces résultats montrent que l’hypothèse d’un modèle hydrostatique stratifié est inadaptée pour déduire les températures des ions associés aux raies « low FIP » dans les couches profondes de la région de transition, située au dessus de la haute photosphère. Les courbes de lumière obtenues au troisième contact de l’éclipse de 2009, ont permis de comparer les profils obtenus pour des raies « low FIP », et simultanément avec la raie de l’hélium neutre He I « high FIP ».

Les courbes de lumière figure IV-2-1-8 ont été obtenues à l'éclipse du 22 Juillet 2009 en Chine, lors du troisième contact. Les altitudes ont été déduites à partir de la figure III-4-4.

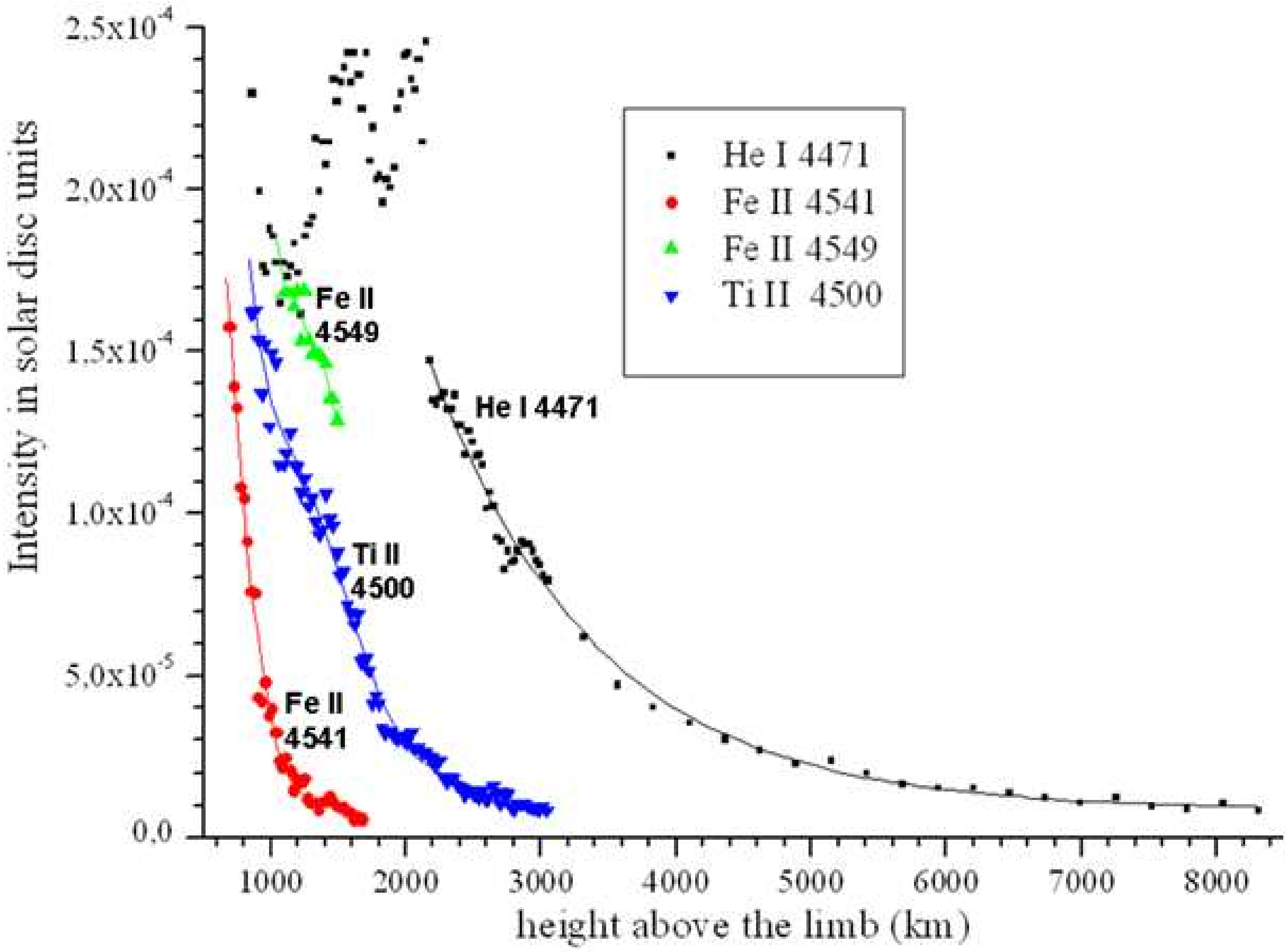

**Figure IV-2-1-8:** *profils des intensités des courbes de lumière obtenues par les mesures des intensités dans les raies du Fe II, Ti II (« low FIP »), et He I (« high FIP ») lors du troisième contact à l'éclipse totale de Soleil du 22 Juillet 2009.*

En dessous de 2500 km, les profils présentent des fluctuations qui sont difficiles à expliquer: elles pourraient être dues aux instabilités météorologiques (à cause des voiles nuageux pendant l'éclipse), mais les maxima constatés sur la courbe de He I 4471Å correspondent aux altitudes où l'intensité de l'enveloppe est maximale en dessous de 2000 km, ce qui a été observé également pour la raie He I 4713Å aux éclipses de 2008 et 2010, voir chapitres IV-1, IV-3-3, IV-5 et figures IV-1-4, IV-3-3-1, IV-3-3-2 et IV-5-1.
Les précisions des mesures d'intensité sur les profils des courbes de lumière obtenues sur la figure IV-2-1-8 sont les suivantes :

Pour la raie He I 4471 Å c'est $9.0*10^{-4} \pm 1.0*10^{-5}$ à 3000 km
Pour la raie Fe II 4541 Å c'est $6.0*10^{-5} \pm 1.5*10^{-5}$ à 900 km
Pour la raie Fe II 4549 Å c'est $1.7*10^{-4} \pm 0.8*10^{-5}$ à 1200 km
Pour la raie du Ti II 4500 Å c'est $1.4*10^{-4} \pm 1.2*10^{-5}$ à 1200 km

Les fluctuations importantes d'intensité sont constatées pour toutes les courbes de lumière. Afin d'améliorer la lecture des fluctuations sur 3 ordres de grandeurs, les courbes de lumière dans les raies du Fe II, Ti II et He II sont tracées en échelle logarithmique voir figure IV-2-1-

9. Cette méthode de tracé permet d'améliorer la lecture afin de déceler si des corrélations existeraient entre les courbes de lumière des différents éléments chimiques. Il est possible qu'autour de 1600 km d'altitude où le premier maximum apparaît pour He I 4471, un maximum de modulation soit aussi constaté pour la courbe de lumière du Ti II. Cette constatation reste difficile à confirmer sur les courbes affectées par les fluctuations d'origine météorologiques. Ces résultats confirmeraient des effets d'enveloppe comparables à ceux de la raie He I 4713 Å dans la courbe de lumière figure IV-1-4 et d'émissivité IV-1-7. Les graphiques en figure IV-2-1-9 ont été tracés en échelle Logarithme et la raie du Fe II 4549Å n'a pas été représentéée (car le nombre de ses points est trop limité).

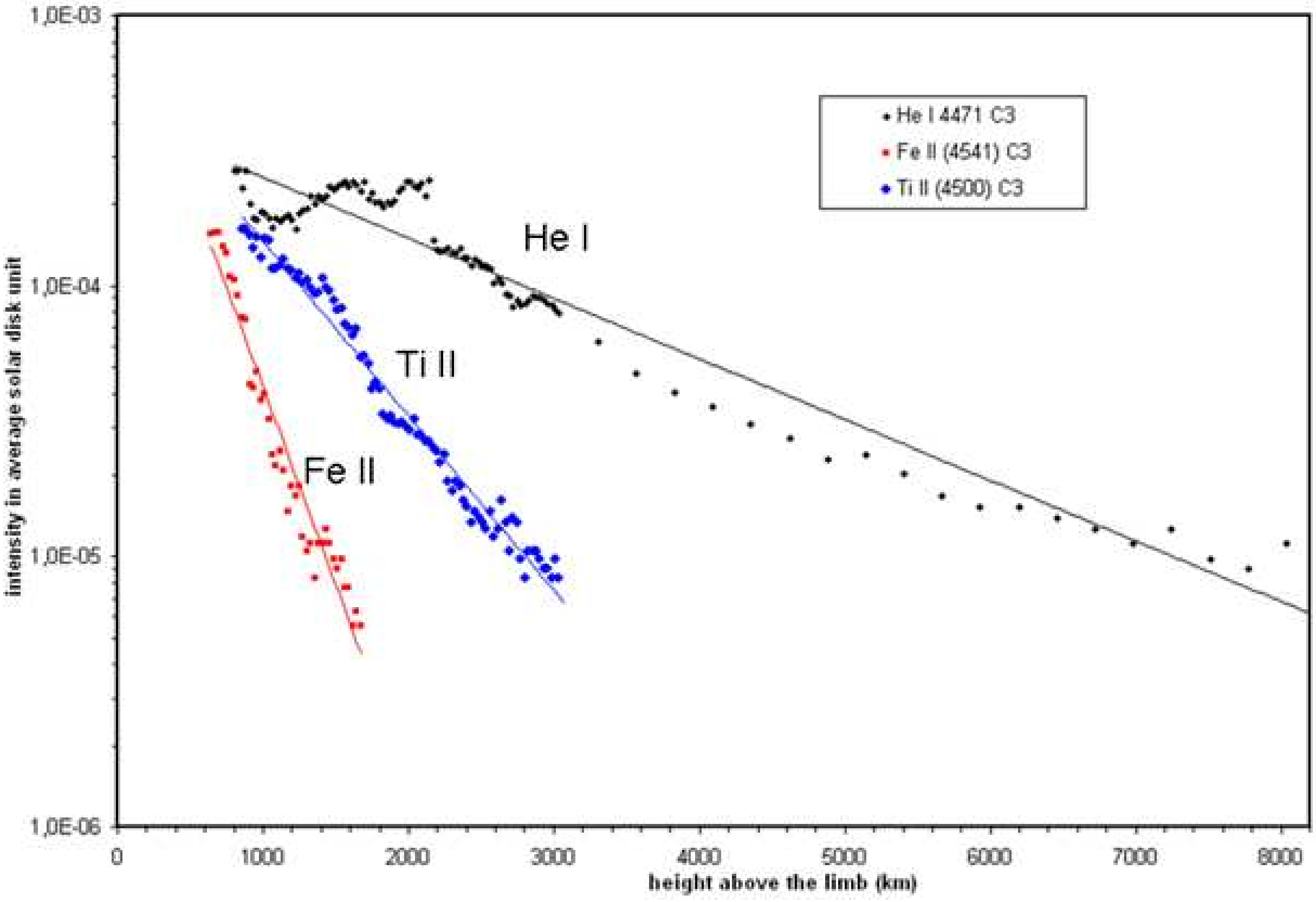

**Figure IV-2-1-9:** *profils des intensités des courbes de lumière tracés en échelle logarithme, obtenues par les mesures des intensités dans les raies du Fe II, Ti II, et He I lors du troisième contact C3 à l'éclipse totale de Soleil du 22 Juillet 2009, et ajustements avec courbes exponentielles décroissantes, afin d'évaluer les modulations.*

Les fluctuations autour des courbes d'ajustement (exponentielles décroissantes) sont évaluées pour distinguer quelques corrélations autour de 1700 km entre le Ti II 4500 Å et He II 4471 Å qui pouraient être des effets d'enveloppes. Les altitudes commencent à 330 km au dessus du bord solaire, avec l'épaisseur optique $\tau_{5000} = 1$ définie pour la haute photosphère, voir Annexe 24. Les couches de la haute photosphère émettent du rayonnement continu encore intense, et ce rayonnement est relevé dans les spectres sur la partie continue, prise entre les raies d'émission.

Les résultats des relevés dans le continu des spectres éclair (mesuré entre les raies d'émission « low FIP ») ont permis de définir le bord solaire à partir de courbes de lumière, voir figure IV-2-1-10. Ces résultats reproductibles par rapport à ceux de 2008 et 2010 ont permis de montrer qu'il n'y a pas de point d'inflexion et des échelles de hauteur situées entre 100 et 200 km ont été déduites. Ces valeurs sont en très bonne adéquation avec les modèles hydrostatiques stratifiées et homogènes. Ces courbes du continu ont été relevées avec une résolution suffisante aux altitudes proches du minimum de température.
Les profils des continus sont pris dans différentes vallées lunaires, pour les 2 contacts C2 et C3, sur ce même graphique (voir figure IV-2-1-10) afin d'évaluer les dispersions sur les profils.

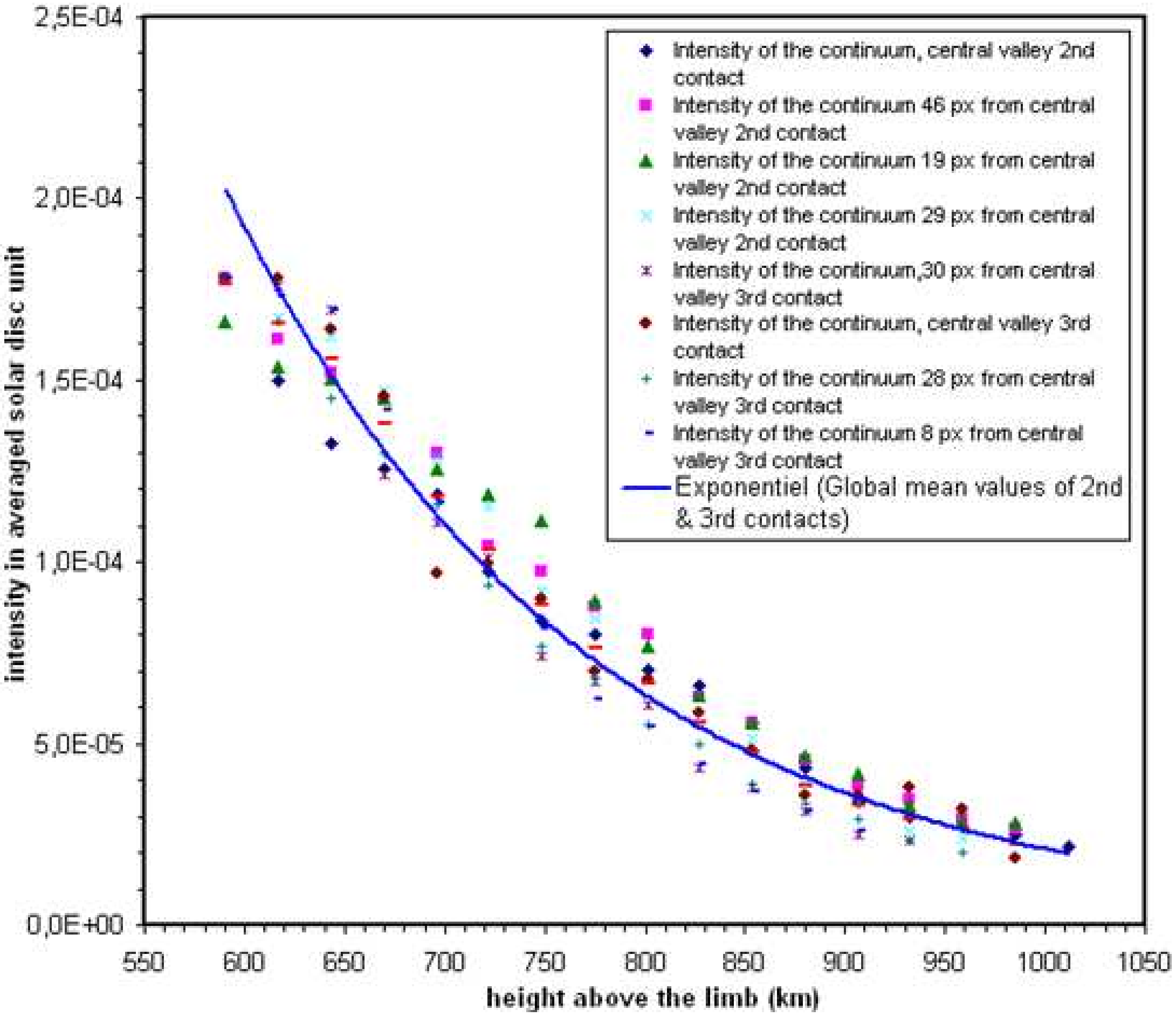

**Figure IV-2-1-10:** *Intensités du continu entre les raies mesuré aux second et troisième contacts de l'éclipse de Soleil du 22 Juillet 2009, et avec la valeur moyenne du continu à 4500Å, et courbe d'ajustement. Etalonné en unités de disque solaire moyen.*

L'appréciation du continu entre les raies se trouve améliorée dans la région 4450 ± 80 Å plutôt qu'à 4700 Å. Cela montre que le "vrai" continu du bord solaire est mieux évalué ou défini en allant vers les UV, mais où les raies low FIP sont plus abondantes.
A partir de la figure IV-2-1-10 sur les variations d'intensités du continu en fonction de l'altitude, il est possible d'interpréter l'allure de la courbe moyenne aux altitudes inférieures à 1000 km, comme étant la fin du continu photosphérique et du début du continu chromosphérique, et où les contributions se superposent sur la ligne de visée sans discontinuité apparente, et sans point d'inflexion.
Pour les étalonnages en intensité, 1 ADU = $6.9745 \times 10^{-7}$ unités du disque solaire avec la caméra Lumenera utilisée en 8 bit de dynamique lors de l'éclipse du 22 Juillet 2009. Le tableau Annexe 17-3 indique les étalonnages des intensités et compte-tenu des filtres utilisés pour cette éclipse. La caméra (modèle Skynyx 2.1 M - Lumenera) fonctionnait à raison de 15 images/seconde soit un pas de hauteur équivalent à 26 km dans l'atmosphère solaire. Les relevés obtenus en figure IV-2-1-10 au 2nd et 3ième contact sont comparables et ont été tracés sur un même graphique afin de comparer les profils obtenus dans les 2 cas.
La précision des mesures des dispersions d'intensités des profils moyennés du continu a été évaluée pour l'altitude h =750 km. L'intensité moyenne correspondante mesurée est de $9*10^{-5} \pm 2*10^{-5}$ unités du disque solaire moyen.

Les équations des courbes d'ajustement des courbes de lumière *I(h)*, des différents ions et du continu moyenné sont données ci-après pour les 3 éléments présentés en figure IV-2-1-9. Le terme au dénominateur dans l'exponentielle indique une valeur de l'échelle de hauteur en kilomètres, avant l'inversion d'intégrale d'Abel.

He I 4471 Å : $I(h) = 8.2*10^{-4} * e^{-h/1219}$ *H = 1219 km*
Fe II 4541 Å : $I(h) = 1.2218*10^{-3} * e^{-h/297}$ *H = 297 km*
Ti II 4500 Å : $I(h) = 6.428*10^{-4} * e^{-h/674}$ *H = 674 km*
Continu : $I(h) = 5.368*10^{-3} * e^{-h/180}$ *H = 180 km*

A partir des équations des courbes d'ajustement des ions précédents, il est possible d'en effectuer l'inversion d'intégrale d'Abel pour déduire l'émissivité en fonction des altitudes, réprésentées sur la figure IV-2-1-1:

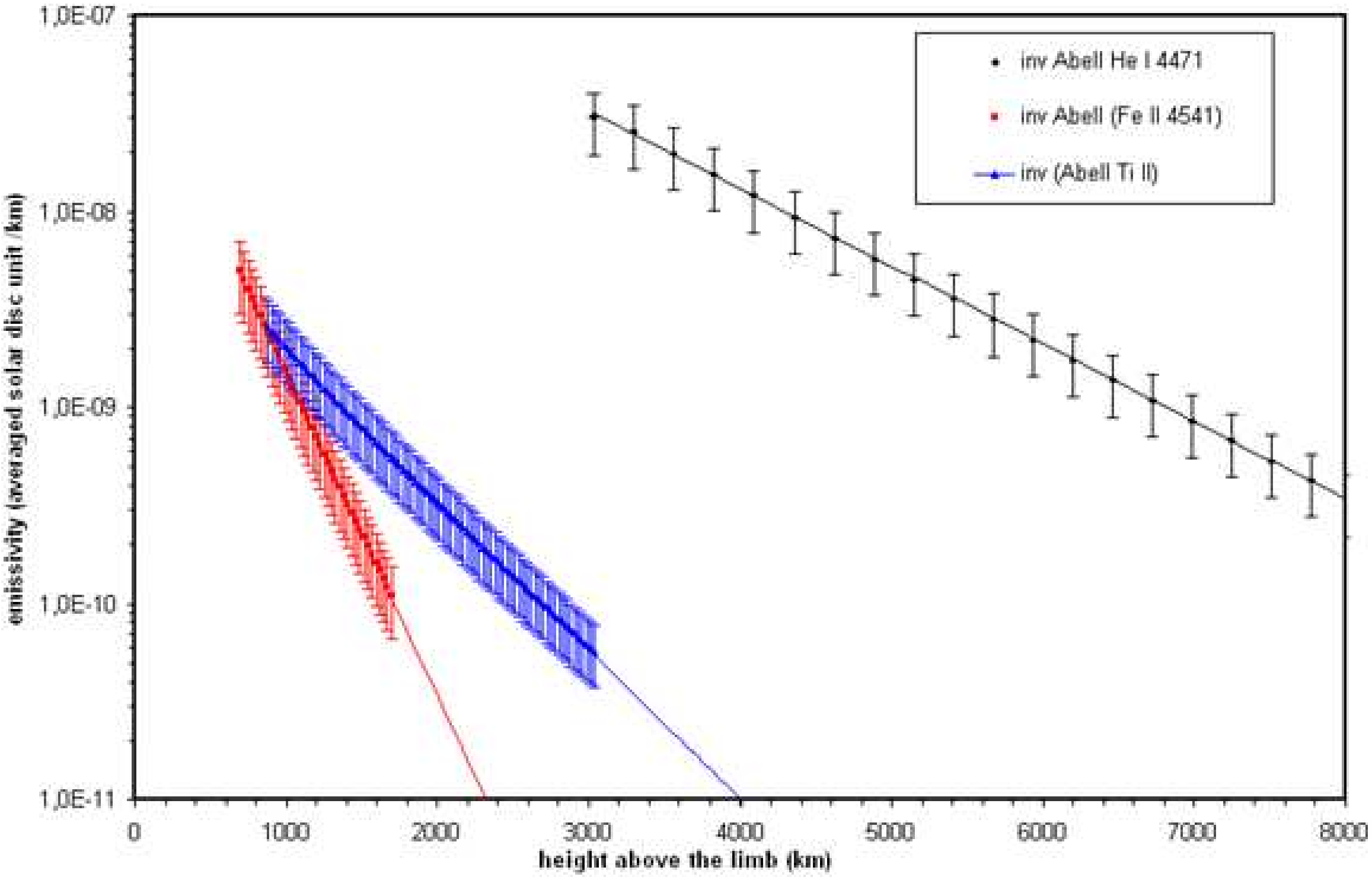

**Figure IV-2-1-11 :** *émissivités (ou brillances) des raies de l'hélium neutre 4471 Å, Fe II 4541 Å et Ti II 4500 Å déduites des courbes de lumière précédente du troisième contact C3, avec la méthode décrite au chapitre VI-2.*

Concernant l'émissivité de l'hélium He I 4471 Å indiquée sur la figure IV-2-1-11, les valeurs situées en dessous de 3000 km n'ont pas été représentées à cause de fluctuations trop importantes, et dont l'interprétation est difficile.

Les équations suivantes pour quelques raies proviennent des courbes d'ajustement effectuées sur les courbes d'inversions d'intégrales d'Abel figure IV-2-1-11 et pour le continu IV-2-1-13 au 3[ième] contact de l'éclipse de 2009:

He I 4471 Å : $E(h) = 4.956411*10^{-7} * e^{-0.0009103888h}$

Fe II 4541 Å : $E(h) = 6.803320*10^{-8} * e^{-0.003800614h}$

Ti II 4500 Å : $E(h) = 1.132399*10^{-8} * e^{-0.001756678h}$

Pour l'hélium neutre l'échelle de hauteur est : 1098 ± 50 km
Pour le Fer II 4541 l'échelle de hauteur est : 263 ± 60 km
Pour le Ti II 4500 l'échelle de hauteur est : 569 ± 60 km
Pour le continu de Paschen $\alpha$ associé à l'hydrogène neutre H I (voir Hiei 1962), l'échelle de hauteur est : 161 ± 10 km
Ces valeurs d'échelle de hauteur déduites après inversion d'intégrale d'Abel, sont inférieures à celles déduites dans les équations d'ajustement des courbes de lumière.

La différence entre l'échelle de hauteur mesurée directement sur les courbes de lumière pour les raies low FIP et par méthode d'inversion d'intégrale d'Abel est de 110 ± 10 km c'est-à-dire : $H_{courbedelumière} = H_{inv.\mathrm{int}\, egraleAbel}$ +110 km
Nous appliquons l'hypothèse d'un modèle homogène, hydrostatique stratifié pour déduire des températures *hydrostatiques* à partir des mesures des échelles de hauteur des courbes de lumière et émissivités. Le tableau IV-2-1-12 présente les températures déduites:

| Raie étudiée C3/2009 | longueur d'onde (Å) | masse ion g/moles | Températures hydrostatiques (modèle hydrostatique) |
|---|---|---|---|
| Fe II | 4541 | 55.8 | 0.48 MK |
| H I continu | 4500 | 1.0 | **5282 K** |
| He I | 4471 | 4.0 | 0.14 MK |
| Ti II | 4500 | 47.9 | 0.89 MK |

**Tableau IV-2-1-12:** *températures hydrostatiques déduites des échelles de hauteur à partir des courbes d'émissivité des ions de raies low FIP.*

Bien que ce résultat soit attendu pour les raies low FIP, les températures indiquées indiquées en grisé n'ont pas de sens pour les ions Ti II, Fe II, He I dans ces régions proches du minimum de température, mais situées au dessus de la haute photosphère, où commencent la mésosphère et la basse chromosphère.
Ces résultats montrent que le modèle hydrostatique stratifié est à nouveau inadapté pour déduire les températures des ions associés aux raies low FIP dans les couches profondes de la région de transition, située au dessus de la photosphère, pour les résultats de cette éclipse.
Par contre la température de 5282 K pour le continu autour de 4500 Å associé H I (P$\alpha$) a un sens et le modèle hydrostatique stratifié est adapté pour ces altitudes situées autour du minimum de température, et au dessus de la haute photosphère. En effet, la température de 5282 K est en adéquation avec la région du minimum de température, inférieure à 6000 K, et pour les altitudes situées entre 500 et 1000 km, l'émission à 4500 Å et 4700Å est directement associée à la diffusion éléctronique et la température électronique est de 5750 K pour *h = 500 km* d'après Athay, Menzel, Pecker, Thomas 1955.
La courbe IV-2-1-13 indique la courbe résultant de l'inversion de l'intégrale d'Abel, pour les continus moyennés aux seconds et troisièmes contacts de l'éclipse de 2009 à partir de laquelle l'échelle de hauteur de 161 km et la température hydrostatique de 5282 K sont déduites.

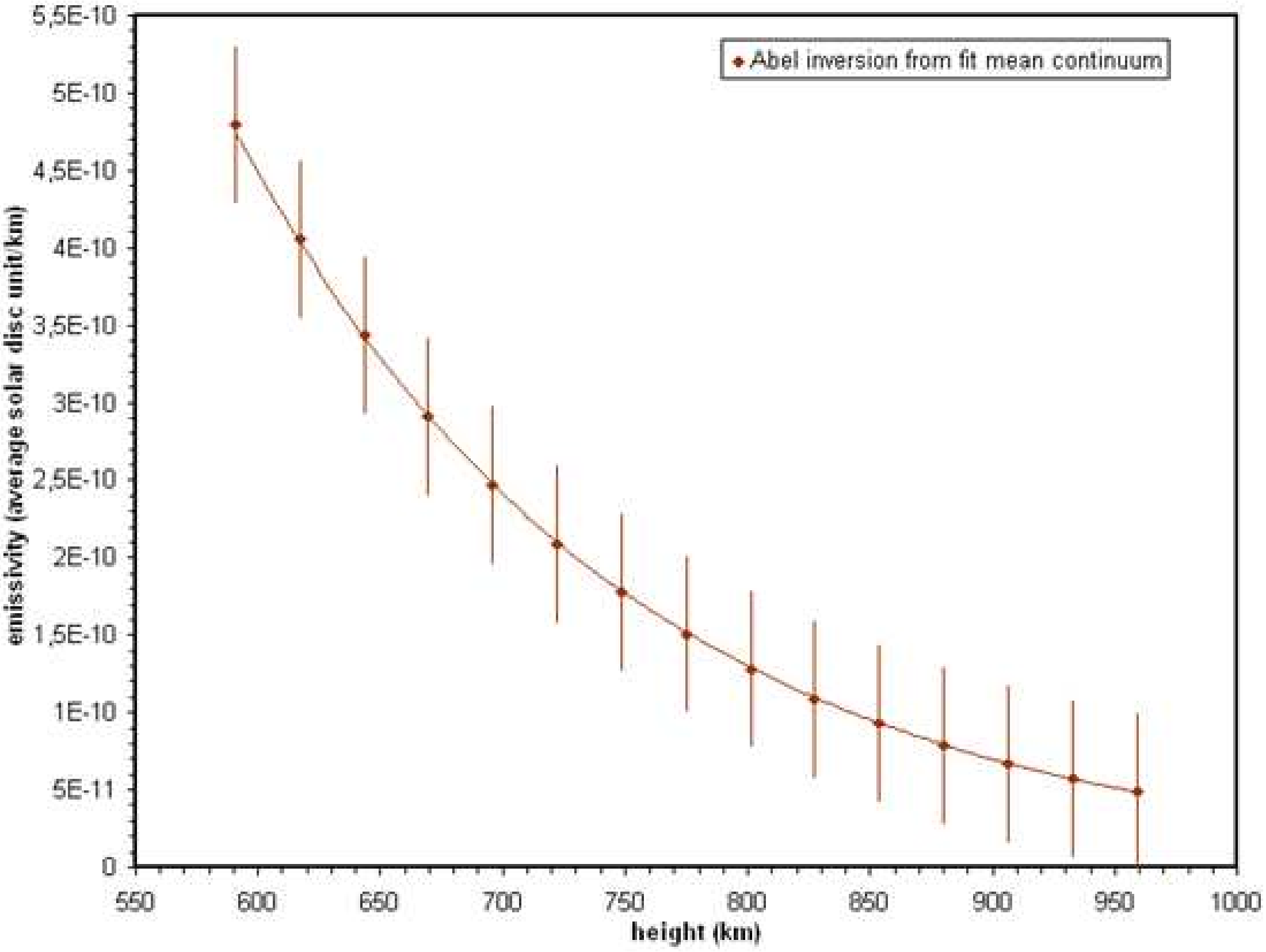

**Figure IV-2-1-13:** *émissivités des courbes du continu aux 2ième et 3ième contacts de l'éclipse de 2009, obtenus par inversions d'intégrale d'Abel sur les équations des courbes d'intensités des courbes de lumière du continus I = f(h).*

L'équation de la courbe d'ajustement pour l'émissivité moyenne du continu en 2009 ($2^{nd}$ et 3ièmes contacts) est:

$$E(h) = 1.85161*10^{-8}*e^{-0.006201321h}$$

L'échelle de hauteur déduite est: 161 km pour une longueur d'onde moyenne de 4500 Å.
Pour la décroissance exponentielle appliquée aux courbes de lumière du continu, la valeur de l'échelle de hauteur se conserve, après inversion d'intégrale d'Abel. La température hydrostatique associée à cette échelle de hauteur est de 5286 K. Par contre les échelles de hauteur ont une différence de 100 km entre les approximations d'exponentielles d'écroissantes ajustées aux courbes de lumière et après inversion d'intégrales d'Abel pour les raies « low FIP » comme les ions Fe II et Ti II. Cette différence peut s'expliquer par la nature inhomogène des basses couches de l'atmosphère solaire où se forment ces raies « low FIP ». Ces basses couches sont aussi le siège du champ magnétique concentré qui émerge, et cela entraine des instabilités dans le plasma.

Le chapitre IV-2-3 décrit d'autres études sur une protubérance en boucle observée dans les spectres éclairs de 2009. Cette structure est comparée avec d'autres observations réalisées depuis la mission spatiale SoHO, aux mêmes horaires qu'à l'éclipse du 22 Juillet 2009. Ce type d’observation a été aussi complété avec d’autres spectres éclair qui ont révélé des condensations dans l’interface chromosphère-couronne, dans le prolongement des spicules.

## IV-2-2) analyse d’une petite protubérance en boucle et interface chromosphère-couronne à l’éclipse du 22 Juillet 2009

Une petite protubérance en boucle, notée « pp-b » est visible dans le prolongement de raie de l’hélium neutre He I 4471Å et indiquée comme « loop-prominence » sur la figure IV-2-2-1. Elle est située sur la partie supérieure du spectre. A l’altitude de 1500 km au dessus du bord solaire, les nombreuses raies « low FIP » de la mésosphère ont disparu. Les raies d’émission ayant des extensions plus élevées jusque dans la chromosphère, et les modulations du continu chromosphérique et coronal sont encore visibles.

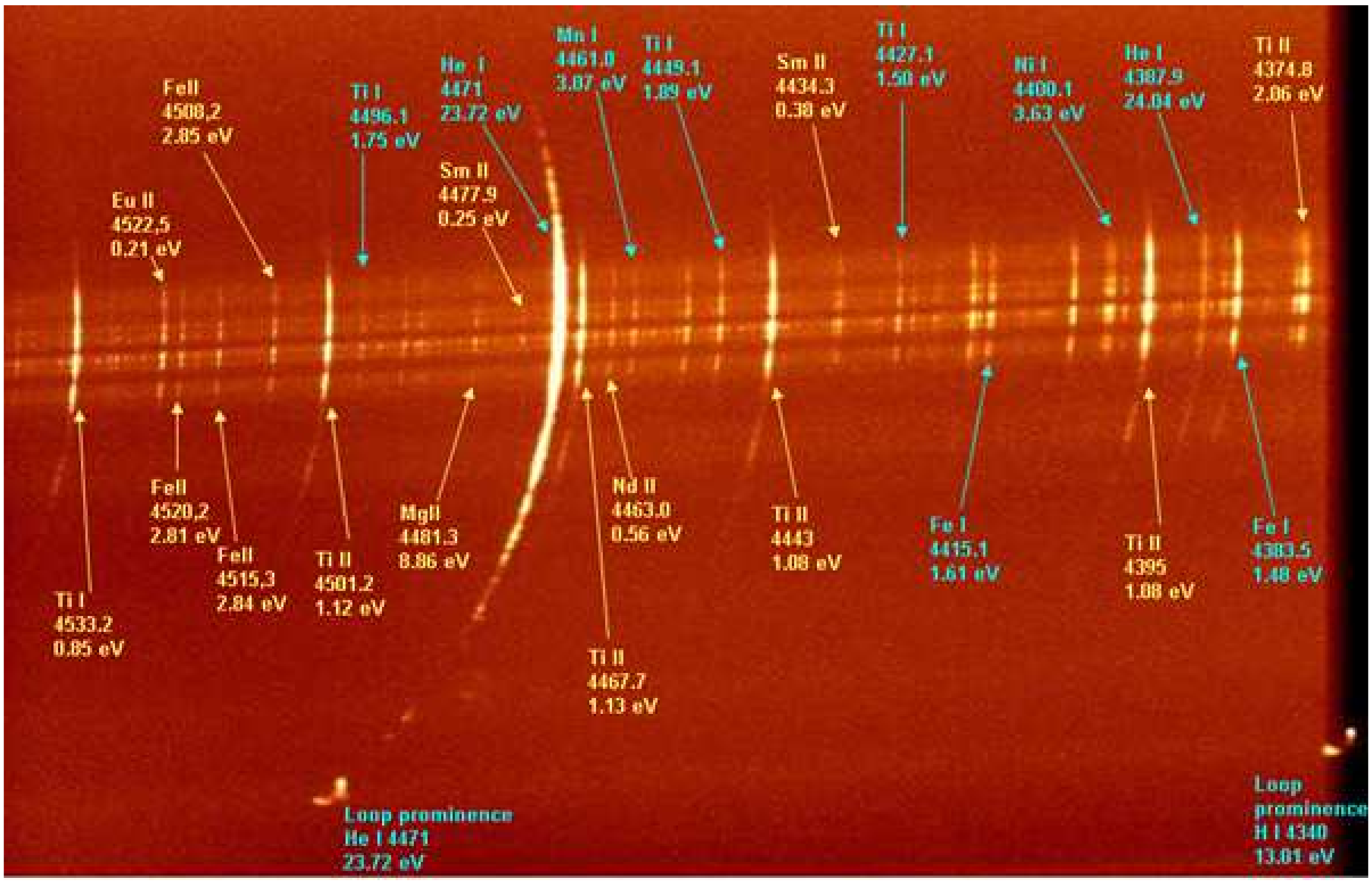

**Figure IV-2-2-1:** *Sommation de 90 spectres éclairs quelques secondes après C3 à 1h 38min et 28 s TU pour montrer notamment la structuration du profil de la raie de l'hélium neutre He I 4471 Å avec des groupes de macro-spicules, et les modulations du continu chromosphérique et coronal. Les potentiels d'excitation sont indiqués en eV sous les longueurs d’onde des raies (Moore et al 1966, The solar spectrum*2935 Å to 8770 Å*). Remarquer aussi les images de protubérances.*

Les 90 images de spectres éclairs sommées précédentes prises environ 10 secondes avant le troisième contact à la fin de la totalité à 1h38 39 s TU, révèlent des fines structures avec des embrillancements le long du croissant de la raie de l’hélium neutre He I 4471Å limitée par le bord de la Lune. Les grandes amplitudes de cette structuration voir figure IV-2-2-1 sont liées en partie aux motifs du profil du bord de la Lune. Les autres petits embrillancements sont dus à la présence de macrospicules. Ces structures de spicules et macrospicules en forme de tubes sont décrits dans les ouvrages de Athay 1961. Le matériel chromosphérique serait concentré dans ces structures (Koutchmy Stellmacher 1976**).** Des modèles ont été proposés par Lorrain

Koutchmy 1996. L’orientation de ce spectre a été choisie dans le sens de défilement du spectre de la gauche vers la droite, longueurs d’onde décroissantes. La pp-b dans l’hydrogène neutre H I$\gamma$ 4340 Å disparaît en bordure de champ, à mesure que la séquence de spectres se rapproche du troisième contact C3.
Les images des figures IV-2-2-2 et IV-2-2-3 représentent la comparaison d’une image SoHO dans l’hélium 304Å et d’une image en lumière blanche réalisée par Sylvain Weiller aux mêmes échelles, quelques instants avant le troisième contact de l’éclipse du 22 Juillet 2009. Ces images indiquent la position de la protubérance sur le limbe sud suest d’où elle a été observée simultanément sur les spectres éclairs voir figure IV-2-2-1. L’analyse de cette boucle de plasma « froid » consiste à rechercher la signature d’une absorption par du plasma, qui pourrait être observée sur la ligne de visée et mesurée à ses pieds. Par ailleurs un autre objectif consiste à étudier l’environnement coronal de cette boucle, c'est-à-dire l’interface entre cette pp-b « froide » et la couronne « chaude » ambiante à des altitudes supérieures à 1700 km au dessus du limbe.

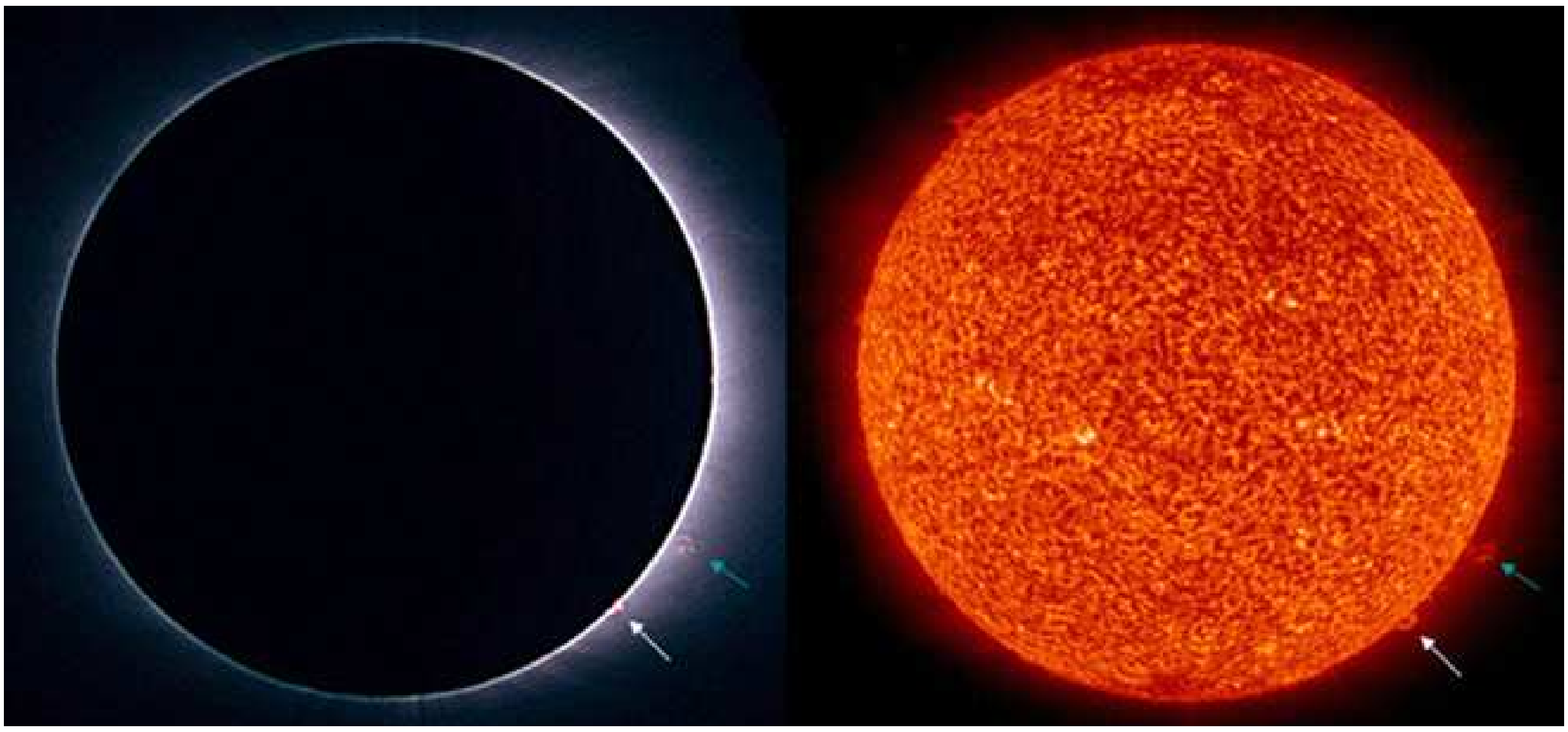

**Figure IV-2-2-2:** *image en He II 304 Å SoHO le 22 Juillet 2009 à 2h36 TU à droite et à gauche image en lumière blanche réalisée par Sylvain Weiller quelques instants avant le 3 ième contact vers la fin de la totalité de l’éclipse totale. La flèche blanche indique la pp-b.*

La figure IV-2-2-2 permet de montrer les disques solaires en entier de l’éclipse totale en lumière blanche, et où les protubérances apparaissent de coloration rosée liée à la raie H $\alpha$ de l’hydrogène neutre HI et à la même échelle pour l’image en entier obtenue par la mission SoHO dans la raie optiquement épaisse de l’hélium He II 304Å. La figure IV-2-2-3 est un extrait de la figure IV-2-2-2 sur la région de la pp-b agrandie 5 fois au moment de l’éclipse totale du 22 Juillet 2009.

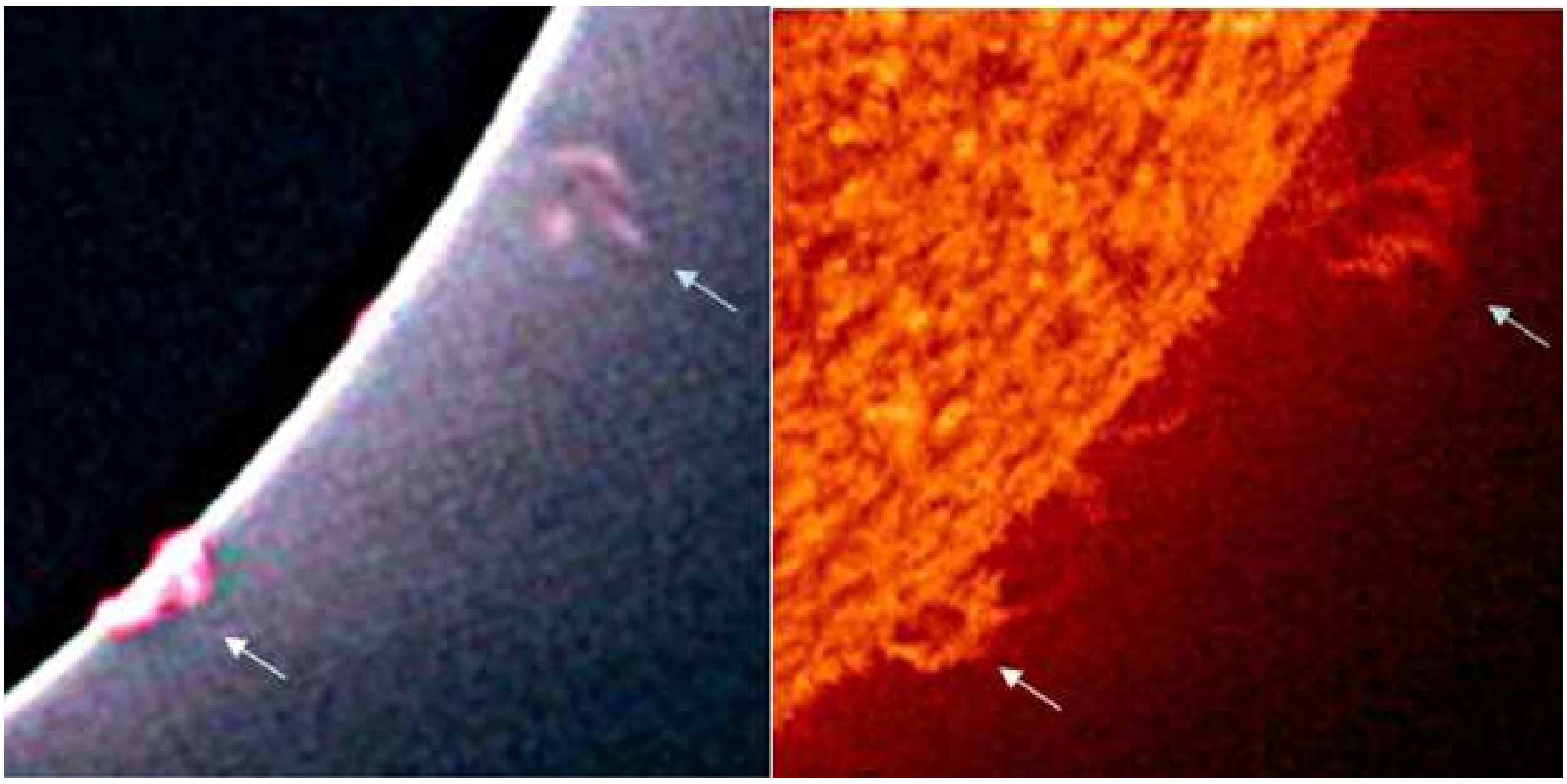

**Figure IV-2-2-3:** *Montage d'une image W-L réalisée par Sylvain Weiller et d'une image SOHO dans l'hélium He II 304 Å montrant la petite protubérance en boucle et son orientation ; dans le prolongement, la flèche blanche indique sa position au limbe Sud-Ouest. Une autre protubérance plus faible et plus élevée est visible un peu plus haut sur l'image et indiquée par la flèche bleue.*

Les images de la pp-b en H$\alpha$ sont optiquement minces tandis que l'image de la pp-b dans la raie de l'hélium He II 304Å est optiquement épaisse et correspond à des températures du plasma inférieures à 30000 K. Cette pp-b est associée à du plasma « froid » et ne correspond pas à une boucle coronale.
Les vues figure IV-2-2-4 et IV-2-2-5 sont extraites d'une image EIT en 304 Å traitée, à bord de la mission SoHO Delaboudinère 1995, afin d'étudier les contours de l'interface de la pp-b avec la couronne environnante.

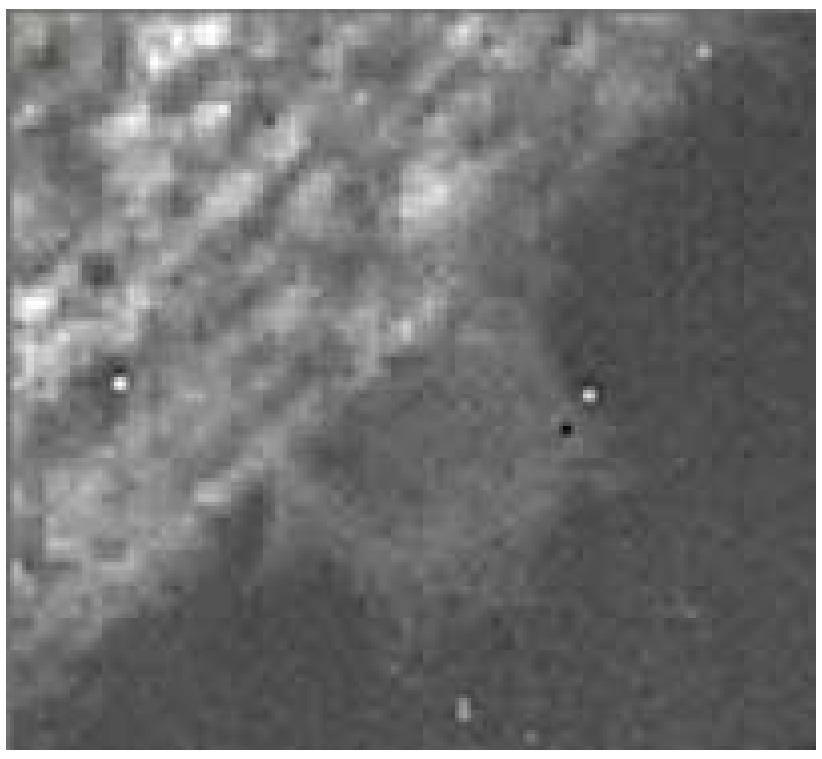

**Figure IV-2-2-4:** *extrait d'une somme brute de 2 images EIT (SOHO) en 304 Å sur la région de la protubérance en boucle, dans la raie He II 304 (30000 K) du limbe sud ouest.*

La pp-b n'apparaît pas en absorption aux températures coronales, probablement à cause de la présence de plasma coronal en émission sur la ligne de visée. Cette pp-b est aussi observée dans les raies « low FIP » comme le Titane ionisé Ti II 4501 Å, Ti II 4443 Å, Ti II 4395 Å voir en figure IV-2-3-1, sans doute pour les parties les plus denses.

Par ailleurs, des images de la couronne vue dans les rayons X provenant de la mission Hinode XRT prises 30 minutes avant l'éclipse ont été analysées afin de localiser le plasma froid vu en absorption associé aux protubérances. Il a été possible d'analyser la structuration complexe de la couronne environnante autour de la pp-b, après sommation de 6 images converties en échelle logarithmique pour une meilleure visibilité, voir figure IV-2-2-5.

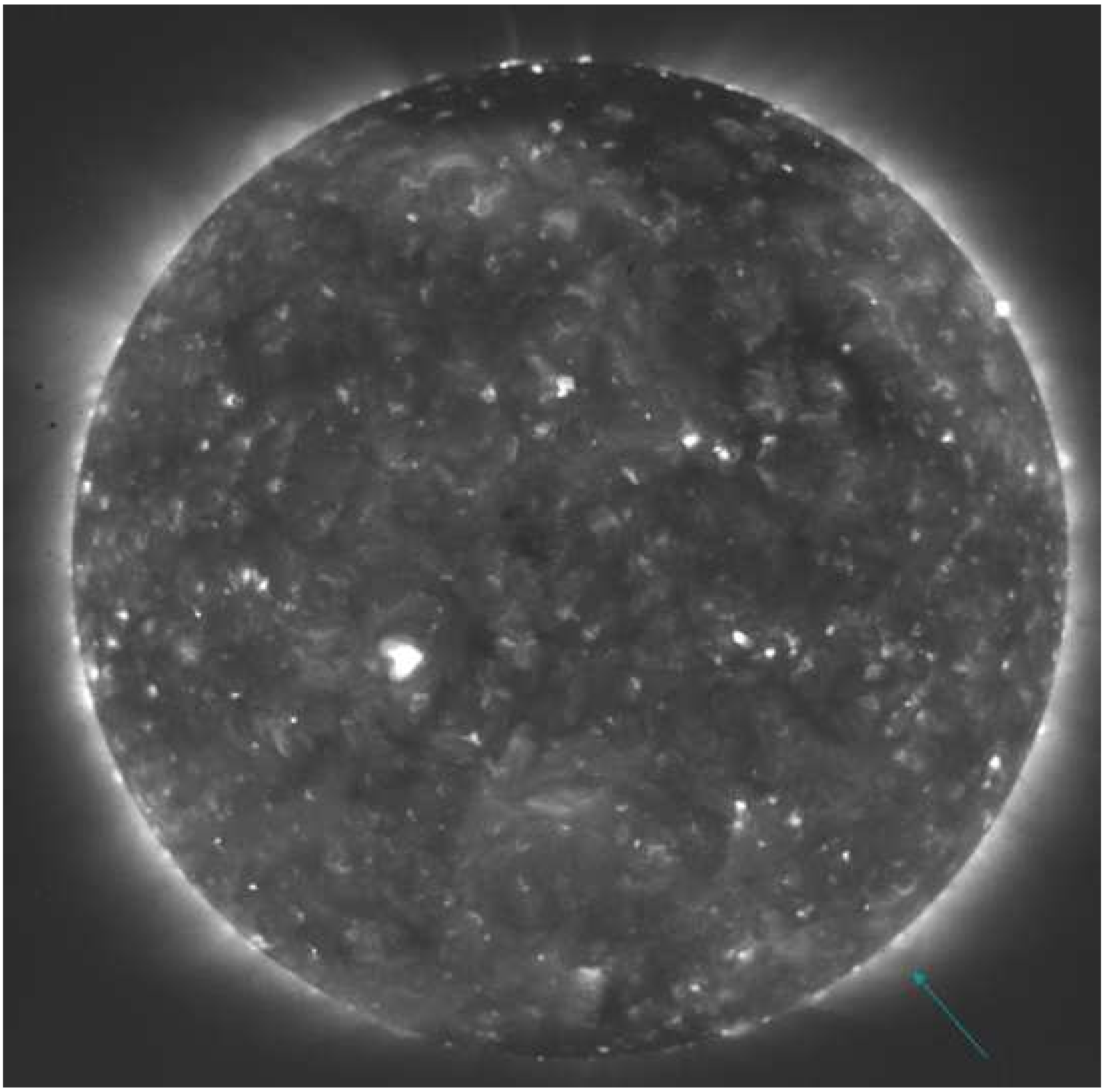

**Figure IV-2-2-5:** *Somme de 6 images de Hinode SXR avec filtre Ti/poly correspondant à des températures supérieures à 2 millions de Kelvins pour des longueurs d'onde entre 5 et 40 Å. Image convertie en échelle Logarithme, prises entre 1h 08 et 1h 10 TU, soit une demi-heure avant l'éclipse totale du 22 Juillet 2009. La flèche bleue indique la région correspondante où se situe la protubérance en boucle observée simultanément dans le domaine visible.*

L'image du disque solaire dans les EUV, révèle un embrillancement du limbe qui sera étudié plus en détails au chapitre IV-7. La petite protubérance en boucle n'est pas visible en absorption sur l'image figure IV-2-2-5, au bord du disque indiqué par la flèche bleue. Un agrandissement d'environ 10 fois autour de cette région a été effectué en figure IV-2-2-6 pour montrer cependant un léger déficit de plasma.

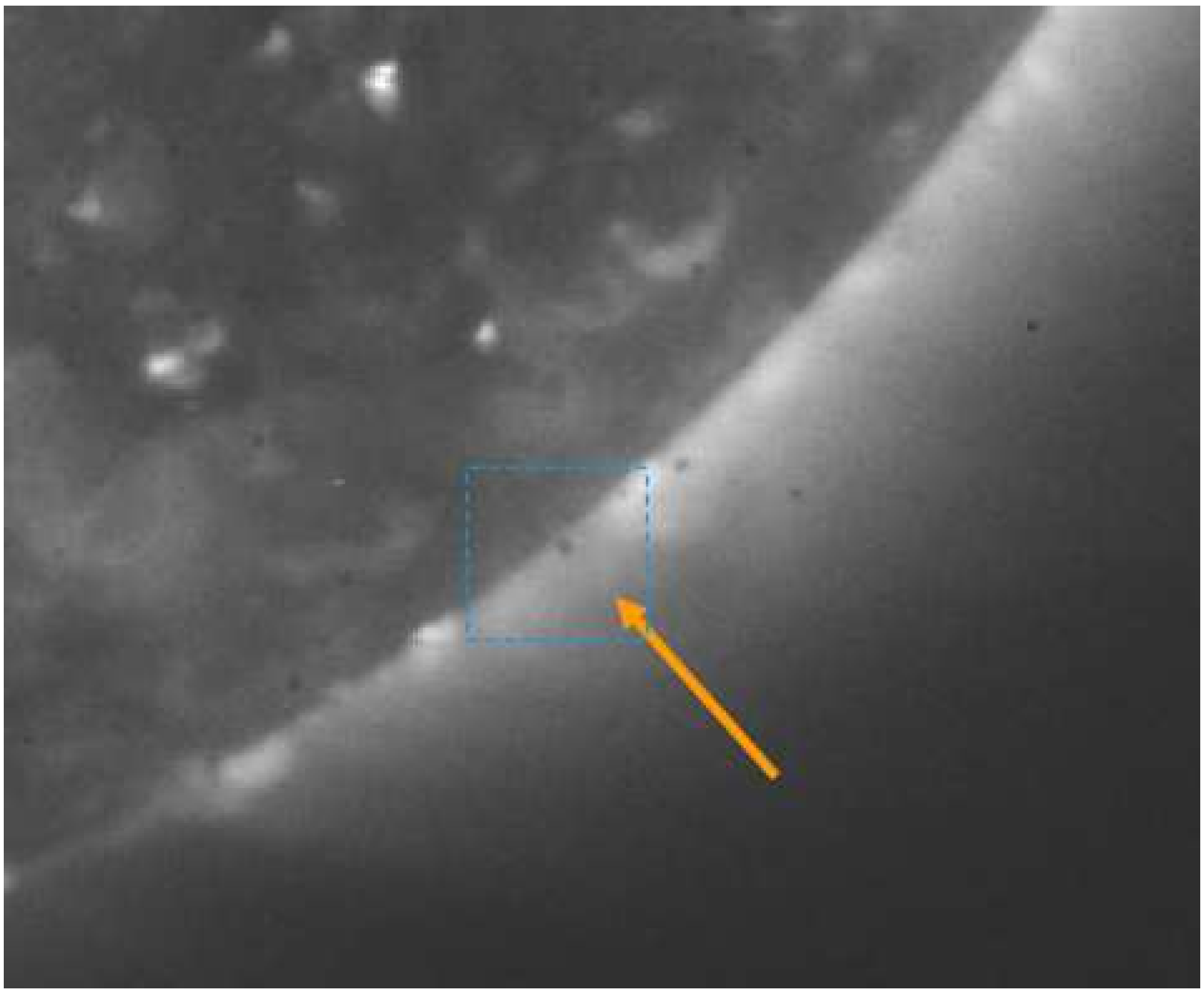

**Figure IV-2-2-6:** *Extrait sur la région autour de la protubérance en boucle à partir de l'image de la figure IV-2-3-5, avec 6 images sommées espacées de 30 secondes, convertie en échelle Logarithme pour une meilleure visibilité des détails plus faibles. Le carré bleu en pointillés indique le champ des vues précédentes examinées avec SoHO et correspondent à l'emplacement de la petite pp-b, et où un faible déficit de plasma est constaté.*

Cette figure IV-2-2-6 avec la zone carrée indique la présence de plasma coronal en émission et sur une plus grande étendue, que pour le plasma « froid » correspondant à la pp-b.
Bien que l'agrandissement de 10 fois ait été effectué, la pp-b n'est pas visible en absorption, même après sommation d'images où le rapport signal sur bruit a été amélioré. Cela peut s'expliquer par le fait qu'il n'y ait pas assez d'hélium He II susceptible de s'ioniser en He III.

Cette pp-b de plasma est étudiée plus en détail dans l'image de la raie de l'hélium He I 4471Å où elle est visible dans cette raie optiquement mince et « froide », dans le but de rechercher des zones d'absorption dans ses pieds ou aux altitudes plus basses, limitées par le bord de la Lune, et aussi analyser les variations d'intensité en considérant l'interface entre cette protubérance en boucle et la couronne ambiante. Une interface de transition entre cette pp-b et la couronne ambiante est considérée comme au chapitre V-1 à venir. Un extrait a été effectué, sur une image résultant de la somme de 30 images, puis a été agrandi et une analyse avec des isophotes a été effectuée en figure IV-2-2-7 au moyen du logiciel DS9, pour rechercher des faibles variations d'intensité autour de la pp-b.

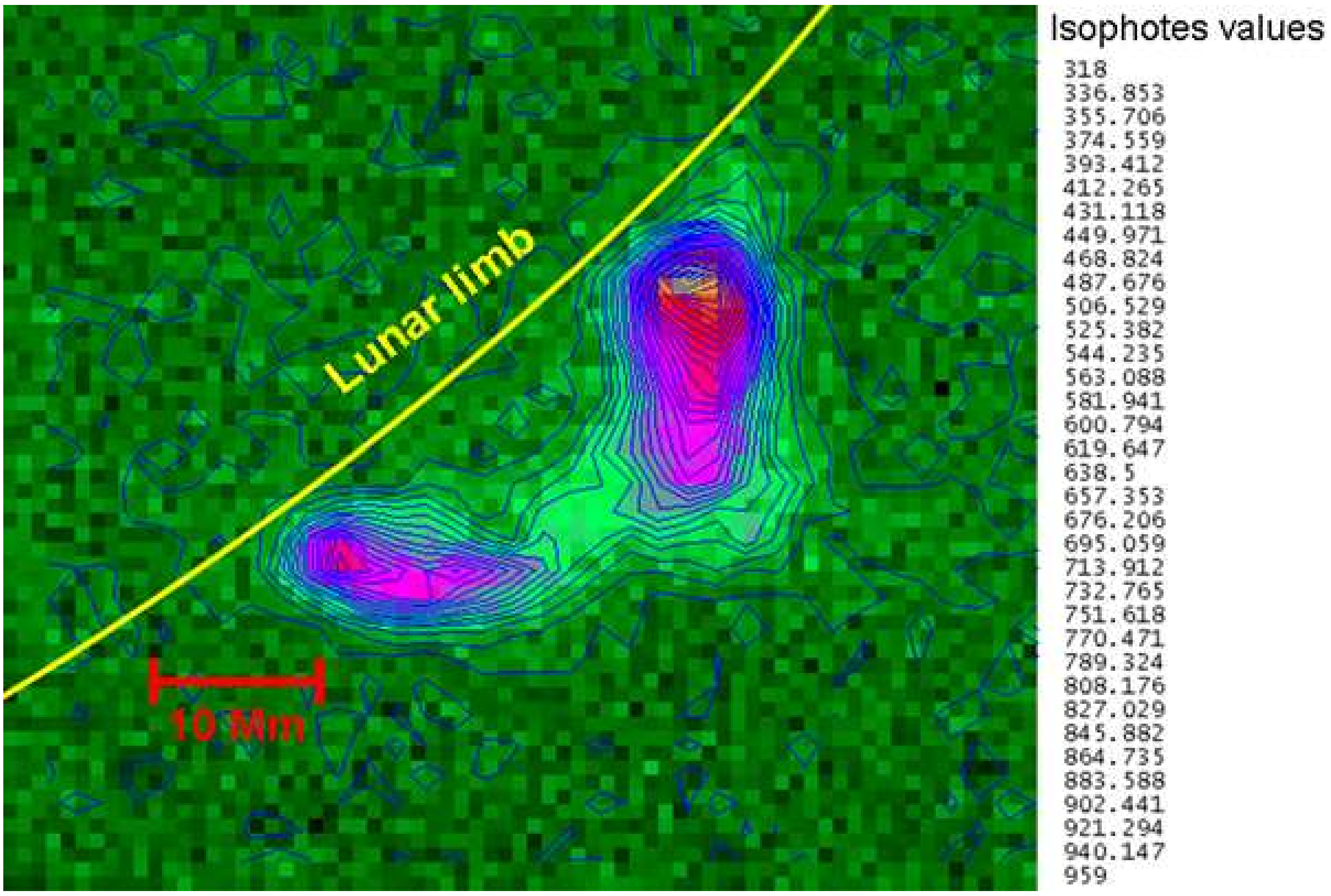

**Figure IV-2-2-7:** *Image agrandie extraite du spectre éclair figure IV-2-2-1, centrée sur la boucle de plasma à 4471Å sur laquelle une cartographie d'isophotes a été effectuée.*

Les lignes d'isophotes, ou isodensités révèlent un « étranglement » ou rétrécissement à la base des pieds où se situe le limbe solaire. Cependant il est difficile de déceler l'absorption même faible, liée à la chromosphère, car il faudrait pouvoir observer en détail la frange de la raie plus proche du limbe solaire. Or ces vues sommées correspondent à des altitudes supérieures à 1600 km. En dessous de cette altitude de 1600 km, les spectres sont trop intenses (parfois saturés) pour distinguer des détails dans la pp-b à 4471 Å, pour les spectres éclairs se rapprochant du 3$^{ième}$ contact. La dynamique de la caméra CCD était limitée à 256 niveaux d'intensité (8 bit), ce qui a limité l'analyse.

Il est difficile de conclure si une quelconque absorption pourrait être observée au pied de cette boucle, avec la limitation donnée par le limbe lunaire au dessus du limbe solaire.
Par ailleurs, aux altitudes plus élevées 7930 à 7457 km au dessus du limbe, cette boucle a bien été observée simultanément dans les raies de l'hélium neutre He I 4471 Å et H$\gamma$ de l'hydrogène neutre HI, avec des condensations coronales comme le montre l'extrait de 18 images sommées, voir figures IV-2-3-1 et IV-2-3-2 au chapitre suivant IV-2-3.

## IV-2-3) Analogies interfaces photosphère-chromosphère/couronne, condensations aux extrêmités des macrospicules et continu coronal à l'éclipse du 22 Juillet 2009

Les spectres éclairs obtenus au 3$^{ième}$ contact de l'éclipse de 2009 ont été additionnés à raison de 6 spectres espacés de 3 spectres afin d'améliorer le rapport signal sur bruit. Les spectres sommés ont permis de distinguer des structures dans la raie de l'hélium He I 4471 Å. Ces structures apparaissent sous forme de condensations situées dans la haute chromosphére. Ces structures ont un intérêt pour notre étude des interfaces, car elles pourraient être situées au niveau de la transition entre la fin des macrospicules qui constituent la raie He I 4471 Å et le début de la couronne solaire où se termine la chromosphère.
Les macro-spicules ont été étudiés dans le passé sur les images de la mission Skylab à partir de spectrohéliogrammes dans la raie de l'hélium à 304 (Bohlin, Vogel, Purcell, et al 1975). Elles ont la forme de structures en forme de « spikes ». Elles ont été analysées relativement plus récemment en période de minimum d'activité avec le satellite Yohkoh, où des phénomènes éruptifs polaires et macrospicules ont été observés aux régions polaires par A.A. Georgakilas, Koutchmy et Christopoulou, 2001. L'intérêt de ces structures de macrospicules vient de ce qu'elles peuvent être associées à des éruptions dans les rayons X de type flares, et ceci a été observé pour la première fois par Moore et al, 1977.
Les macrospicules ont été aussi analysées au pôle sud solaire avec une lunette doublet de 200 mm de diamètre et 3 m de focale, avec du film Ektachrome sur une plaque de 18x24 cm, durant l'éclipse totale du 30 Juin 1973, par Koutchmy, S. et Stellmacher, G. 1976. Les auteurs avaient donné le nom de « spikes » pour désigner les spicules ou macrospicules. Ils avaient mesuré que leurs largeurs étaient de 1.6'' et d'une densité de l'ordre de $10^{10}/cm^3$. D'avantage d'informations sont données en Annexe 27.

Les structurations de plus petites échelles dans la raie de l'hélium He I 4471 Å aux spectres éclairs 2009, sont liées aux spicules et/ou macro-spicules dont la taille n'est pas résolue. Les images sont quasi monochromatiques. On ne peut guère voir un décalage produit par une vitesse dans une protubérance. En effet, un décalage correspondant à 1 pixel correspond à 0.125 Å à 4471 Å, ce qui se traduit par une vitesse qui correspondrait à $v_r = c\frac{\delta\lambda}{\lambda} = 3*10^5*\frac{0.125}{4471} = 8$ km/s. Or les vitesses dans les protubérances sont de l'ordre de 10 km/s dans l'interface protubérance – couronne (voir Labrosse 2010), mais les résolutions spectrales et spatiales sont insuffisantes sur ces spectres pour évaluer de façon plus précise ces faibles décalages, sur 2 pixels.
La protubérance en boucle est également observée dans les raies du titane une fois ionisé low FIP Ti II, comme le montrent les images sommées figure IV-2-3-1, ce qui permet d'établir une analogie entre l'interface de transition photosphère-chromosphère et protubérance-couronne où ces mêmes raies sont présentes.

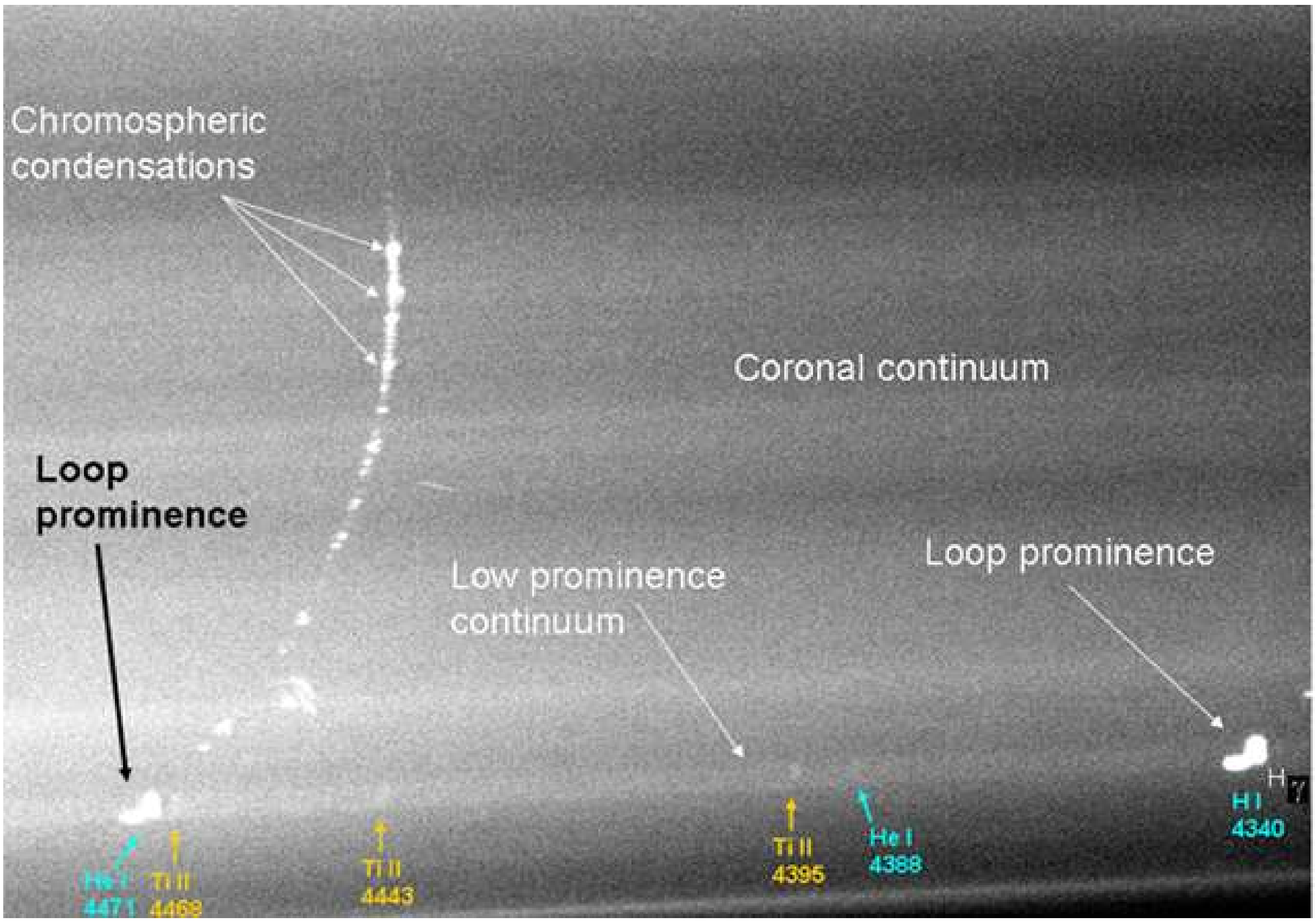

**Figure IV-2-3-1:** *Identification des images monochromatiques de la protubérance en boucle observée dans les raies du titane une fois ionisé. Somme de 50 images (N° 2240 à N° 2290) correspondant aux altitudes voisines de 7000 km par rapport au bord lunaire moyen au dessus du limbe solaire, et avant le troisième contact de l'éclipse du 22 Juillet 2009.*

Les modulations du continu coronal sont apparentes sur la figure IV-2-3-1 et correspondent à la diffusion Thomson des électrons libres dans la couronne. Les renforcements d'intensité des modulations montrent une corrélation avec les protubérances vues en émission dans les raies « low FIP » et de la même manière qu'à l'éclipse de 2008, ces modulations sont aussi corrélées avec les embrillancements dans les structures chromosphériques, comme les petites protubérances dans la raie de l'hélium neutre He I 4471Å. Ces résultats ne montrent pas de façon évidente des corrélations du continu coronal avec les régions de macrospicules observés dans la raie de l'hélium neutre 4471 Å. Malgré le ciel voilé, il a été possible d'améliorer les observations et de renforcer le continu coronal en sommant beaucoup d'images, ce qui a permis d'atteindre les faibles niveaux d'intensité en figure IV-2-3-1.
On distingue plus faiblement l'image monochromatique de la pp-b dans la raie d'émission de faible intensité de l'hélium neutre 4387.9 Å, malgré le continu de la protubérance. Cette image monochromatique de He I 4387.9 Å à a pu être identifiée et localisée car située entre les images monochromatiques dans les raies de l'hélium neutre He I 4471Å et H$\gamma$ 4340 Å « high FIP ». Ces images monochromatiques permettent ainsi une mesure de longueur d'onde plus précise. Les images monochromatiques de la pp-b dans le Ti II « low FIP » confirment l'équivalence et la présence du titane entre les 2 types d'interface de transition photosphère-chromosphère d'une part et d'autre part, l'interface protubérance-couronne ambiante.

L’image figure IV-2-3-2 a été obtenue en sommant 100 spectres, afin de montrer cette pp-b à des altitudes plus basses et quelques secondes avant le troisième contact où les autres raies d’émission des couches inférieures apparaissent en forme de croissants:

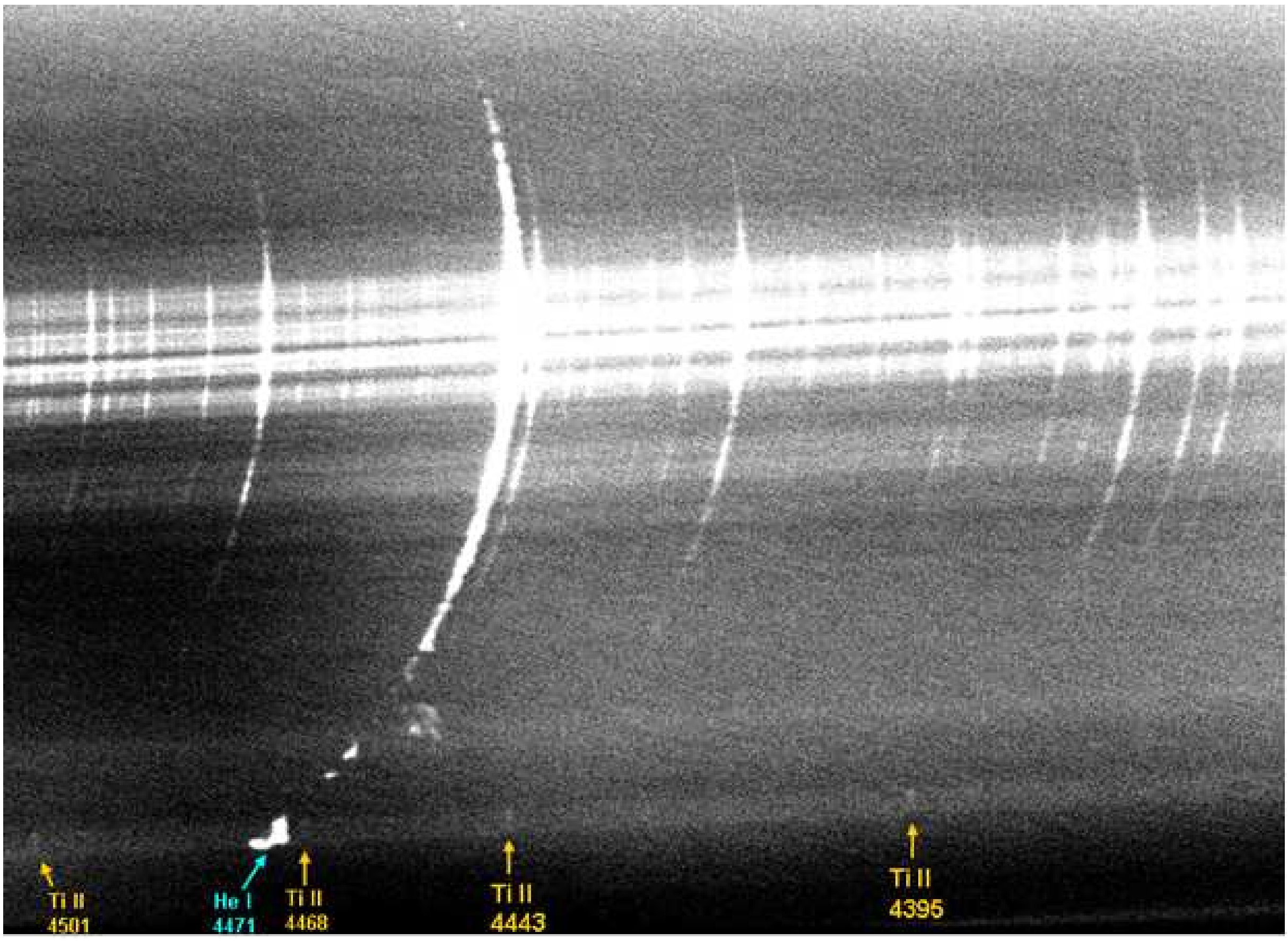

**Figure IV-2-3-2:** *Identification des images monochromatiques de la protubérance en boucle observée dans les raies du titane une fois ionisé. Somme de 100 images de spectres éclairs aux altitudes entre 2500 et 1500 km par rapport au bord lunaire moyen, et avant le troisième contact de l’éclipse du 22 Juillet 2009. La protubérance en He I 4388 Å n’est plus visible.*

Il apparaît que les images monochromatiques de la pp-b dans les raies du Ti II 4443Å et Ti II 4501Å se situent dans le prolongement de ces mêmes raies vues comme des croissant.
On ne voit plus l’image monochromatique de la protubérance en boucle dans la raie de l’hélium He I 4388 Å, car cette raie est très faible intensité, et peut être vue plus facilement dans la phase de totalité, avant le 3$^{ième}$ contact. Dans l’image des spectres éclairs sommés de la figure IV-2-3-2, les composantes continues chromosphériques et photosphériques s’intensifient, avec la réapparition des grains de Baily à la fin de la totalité sur la partie centrale des croissants. On retrouve les mêmes éléments chimiques présents dans les raies low FIP de la région d’interface photosphère-chromosphère, que dans les protubérances. Ce résultat et cette corrélation sont importants pour l’analyse des mécanismes d’interface, où l’on retrouve notamment la présence du Ti II « low FIP » dans l’interface photosphère-couronne et dans l’interface protubérance-couronne.

Une étude approfondie et détaillée est effectuée sur la structuration des spectres des raies du Ti II et He I en fonction de l’altitude. Il s’agit de 2 raies voisines en longueur d’onde, l’une est « low FIP » et l’autre « high FIP », et des extraits de séquences de spectres sommés sont assemblés pour montrer l'évolution sur un même montage. Nous avons sommé 6 spectres pris tout les 3 spectres et pour étudier les profils, les modifications dans la structuration des images dans les raies juste avant le 3$^{ième}$ contact (éclipse 2009). Comme ces images ont une forme de croissant (bord lunaire), il a été utile de linéariser en vue de réaliser ensuite des profils d'intensité. La figure IV-2-3-3 montre une image résultante de 6 spectres sommés, puis redressée ensuite, à partir de laquelle des analyses détaillées des profils d'intensité sont tracés dans le sens transverse des raies He I 4471 Å et Ti II 4468 Å:

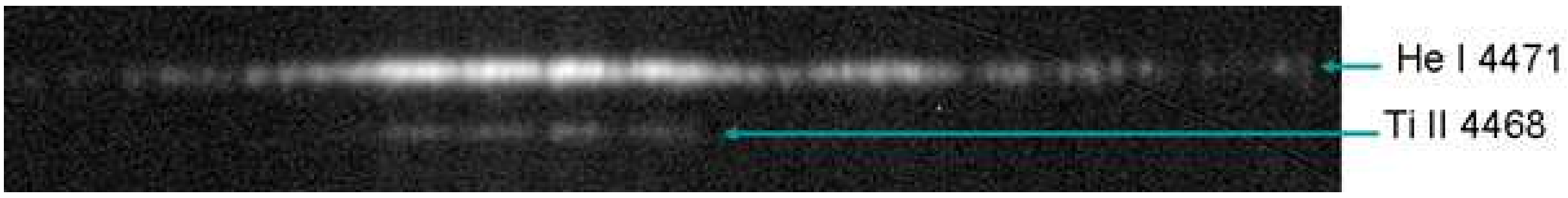

**Figure IV-2-3-3:** *vue agrandie « redressée » des spicules et macro-spicules dans la raie de l'hélium neutre He I 4471 Å. Somme de 6 spectres tous les 3 spectres, pour respecter le critère de Shannon. Echelle Logarithme. Altitude moyenne 2749 km par rapport au bord lunaire moyen et au dessus du limbe solaire.*

Les embrillancements très rapprochés et leurs décalages comme des « stries » correspondent aux effets Doppler dans les macrospicules, modulés par le relief du bord lunaire. Ces effets Doppler observés dans les spicules, sont décrits en Annexe N°7, profils du limbe en H$\alpha$ .

Les altitudes ont été déterminées d'après la figure III-4-3 qui exprime l'altitude en fonction du numéro d'image. D'autre part l'instant de contact a été relevé et a été pris comme référence. La cadence d'images était constante à raison de 15.1 images/seconde. L'altitude d'espacement dans l'atmosphère solaire entre 2 spectres consécutifs limité par le bord de la Lune est de 26.3 km. Mais en sommant 6 spectres éclair tous les 3 spectres, cela revient à un espacement résultant d'environ 80 km.
A l'altitude de 2749.7 km (par rapport au bord moyen de la Lune), l'extrait figure IV-2-3-3 présente une raie « low FIP » du Ti II 4467.7 Å qui est visible sous celle de l'hélium He I 4471Å après la linéarisation. Les montages en tranches regroupés de façon séquentielle, les uns en dessous des autres, aux figures IV-2-3-4 à IV-2-3-9 servent à montrer l'évolution puis disparition des condensations coronales limitées par le bord de la Lune vers 7930 km. Ces condensations se situent dans le prolongement des macrospicules vues par effet Doppler. Ces condensations s'affaiblissent aux altitudes de 5000 km et ne sont plus guère visibles aux altitudes inférieures, où les macrospicules s'intensifient, avant le troisième contact. Cette méthode d'analyse visuelle, grâce au mouvement naturel du bord de la Lune, s'apparente un peu à celle d'un « scanner ». Elle montre les variations de la structuration de ces 2 raies (He I 4471Å « high FIP » et Ti II 4468Å « low FIP ») par altitudes décroissantes, (avant le contact C3) de l'éclipse du 22 Juillet 2009. Cette méthode permet aussi une analyse très précise des variations d'intensité dans les macrospicules directement dans l'image des raies, pour étudier leurs extensions, les corrélations, et l'interface entre le sommet des macrospicules et la couronne ambiante.

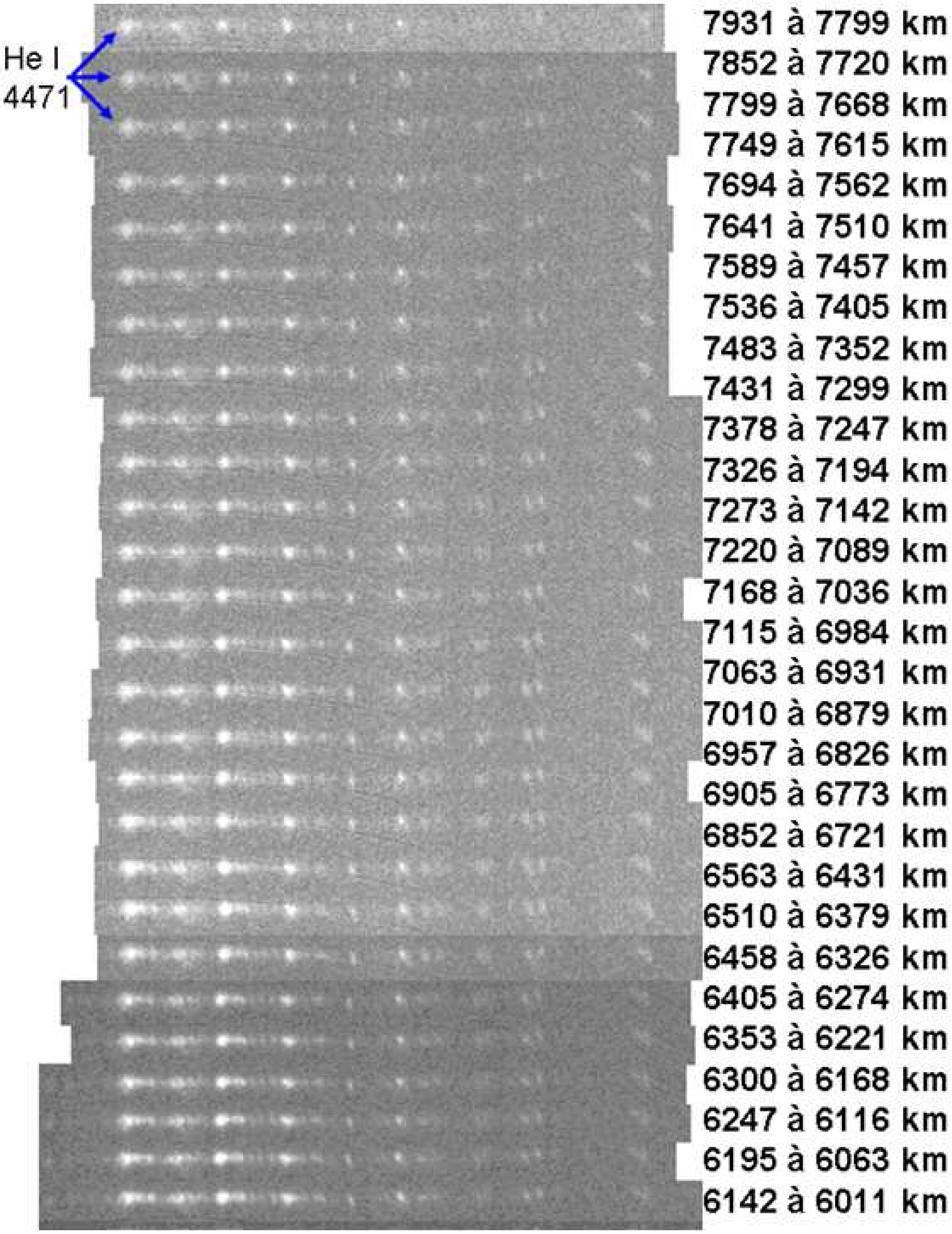

**Figure IV-2-3-4:** *Aspect des spectres dans les parties supérieures des condensations des macro-spicules une fois les profils des raies He I 4471Å (Le Ti II 4468Å n'est pas encore visible dans cette série) redressées, avec les altitudes correspondantes, en échelle linéaire pour les intensités. Altitudes 6011 à 7931 km par rapport au bord lunaire moyen, au dessus du limbe solaire. Le sens de dispersion spectral est aussi dans le sens vertical.*

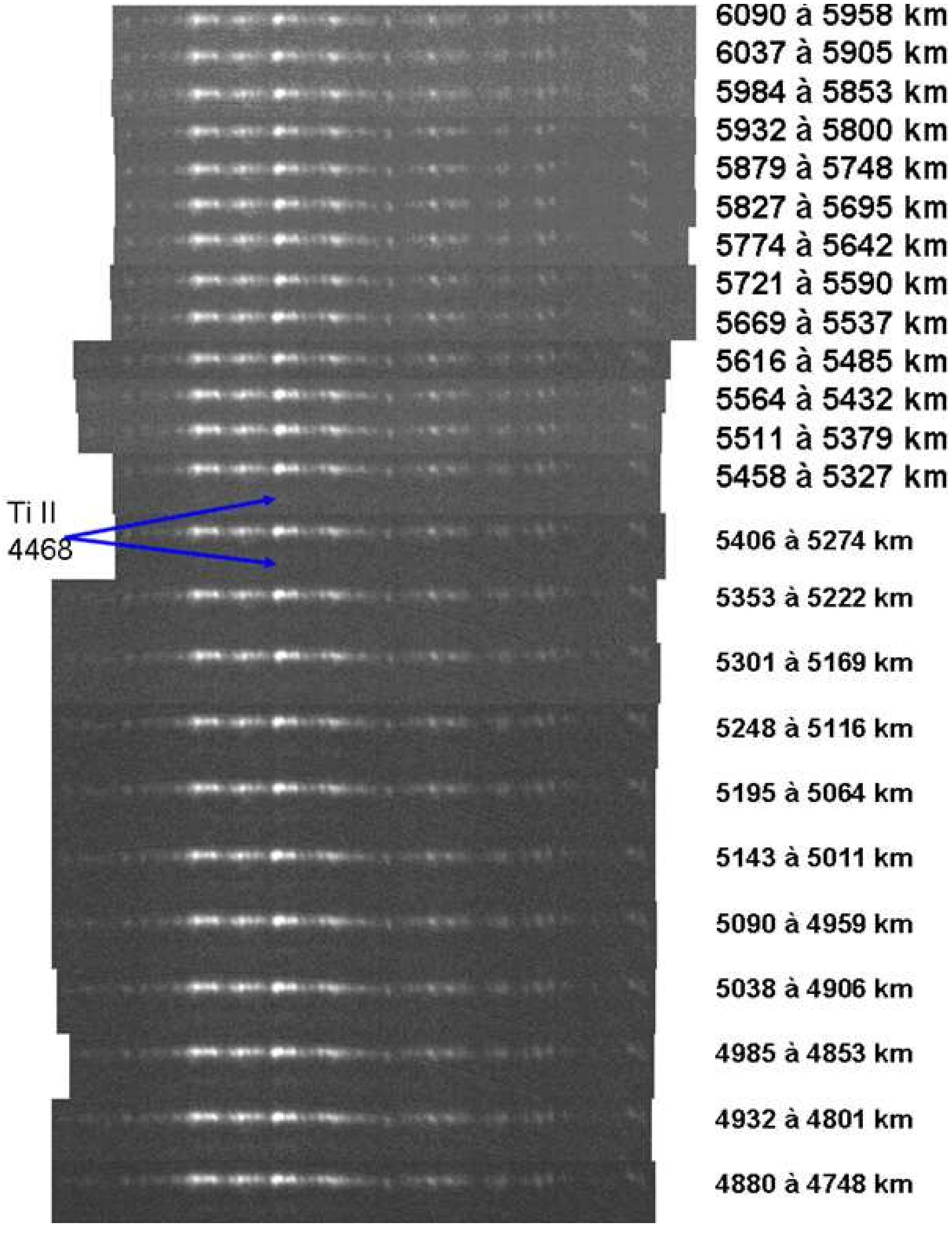

**Figure IV-2-3-5:** *Remarquer l'apparition de la raie du Ti II 4468 Å à partir de 5200 km sur le tirage papier mais elle est décelée dès 5400 km sur les images visualisées avec le logiciel Iris. Altitudes 4748 à 6090 km par rapport au bord lunaire moyen, au dessus du limbe solaire. En dessous de 5400 km, les décalages Doppler du aux macrospicules commencent à être visibles.*

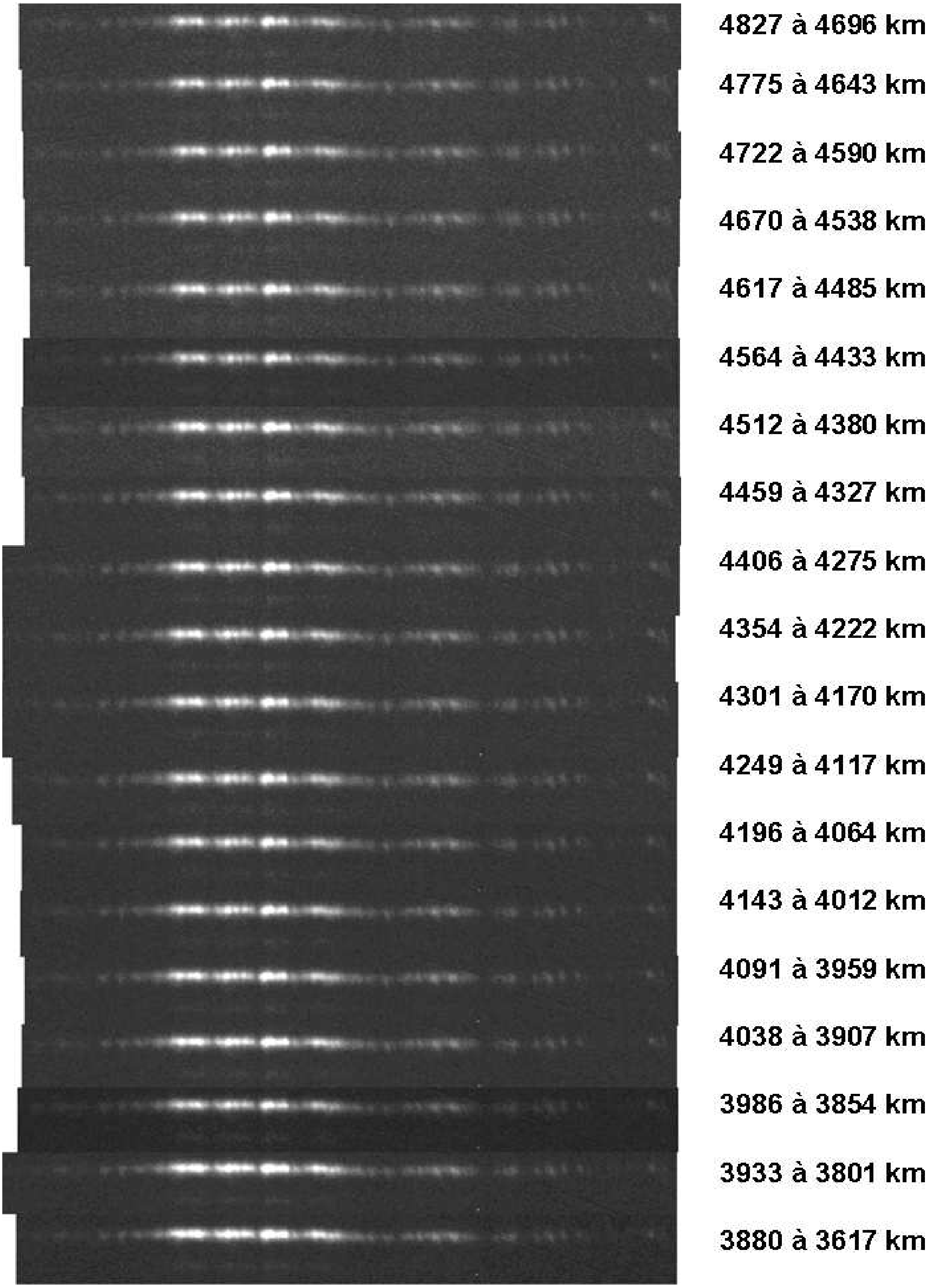

**Figure IV-2-3-6:** *altitudes de 3617 à 4827 km au dessus du limbe solaire par rapport au bord lunaire moyen.*

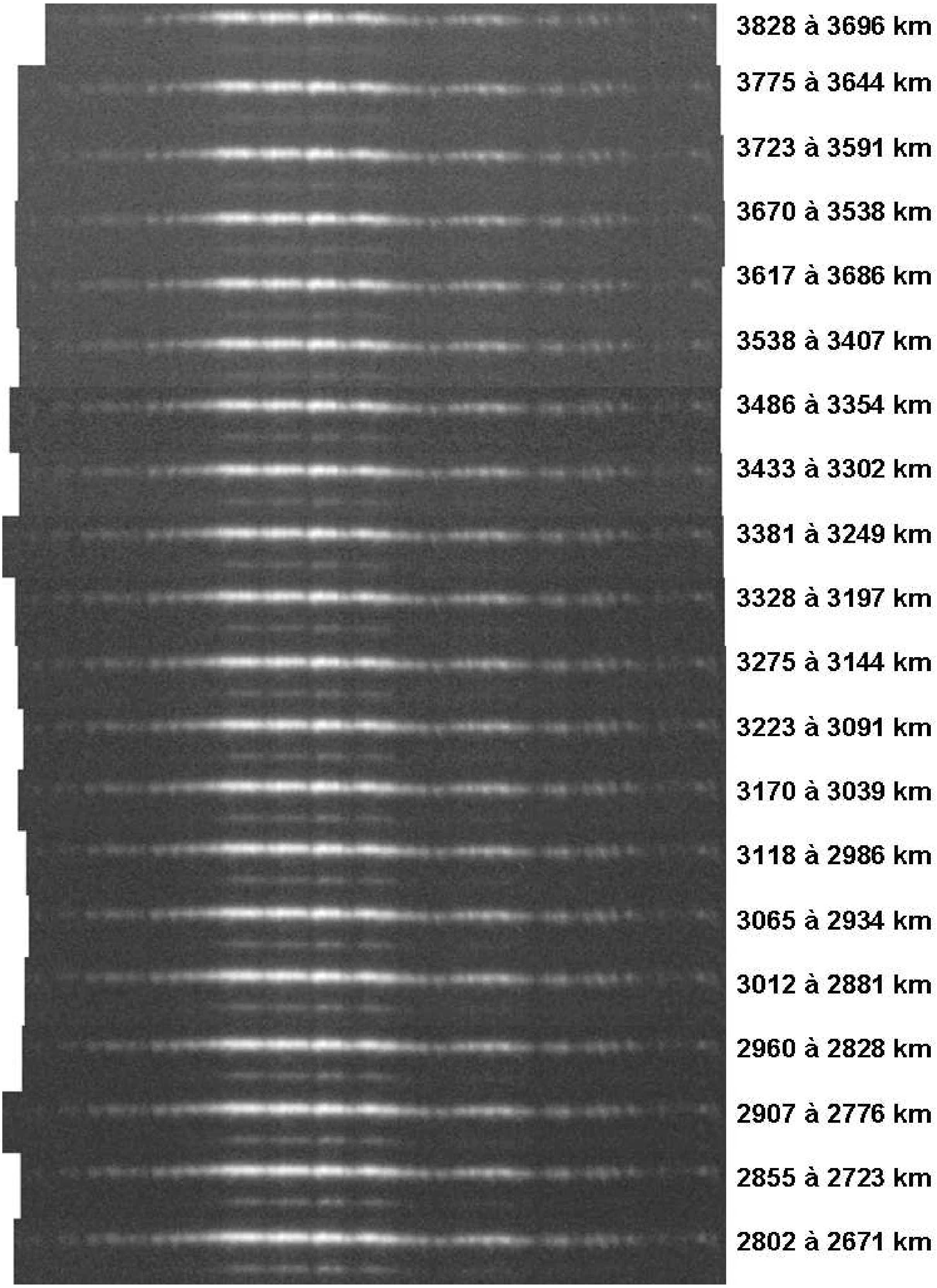

**Figure IV-2-3-7:** *altitudes 2671 à 3828 km au dessus du limbe solaire par rapport au bord lunaire moyen.*

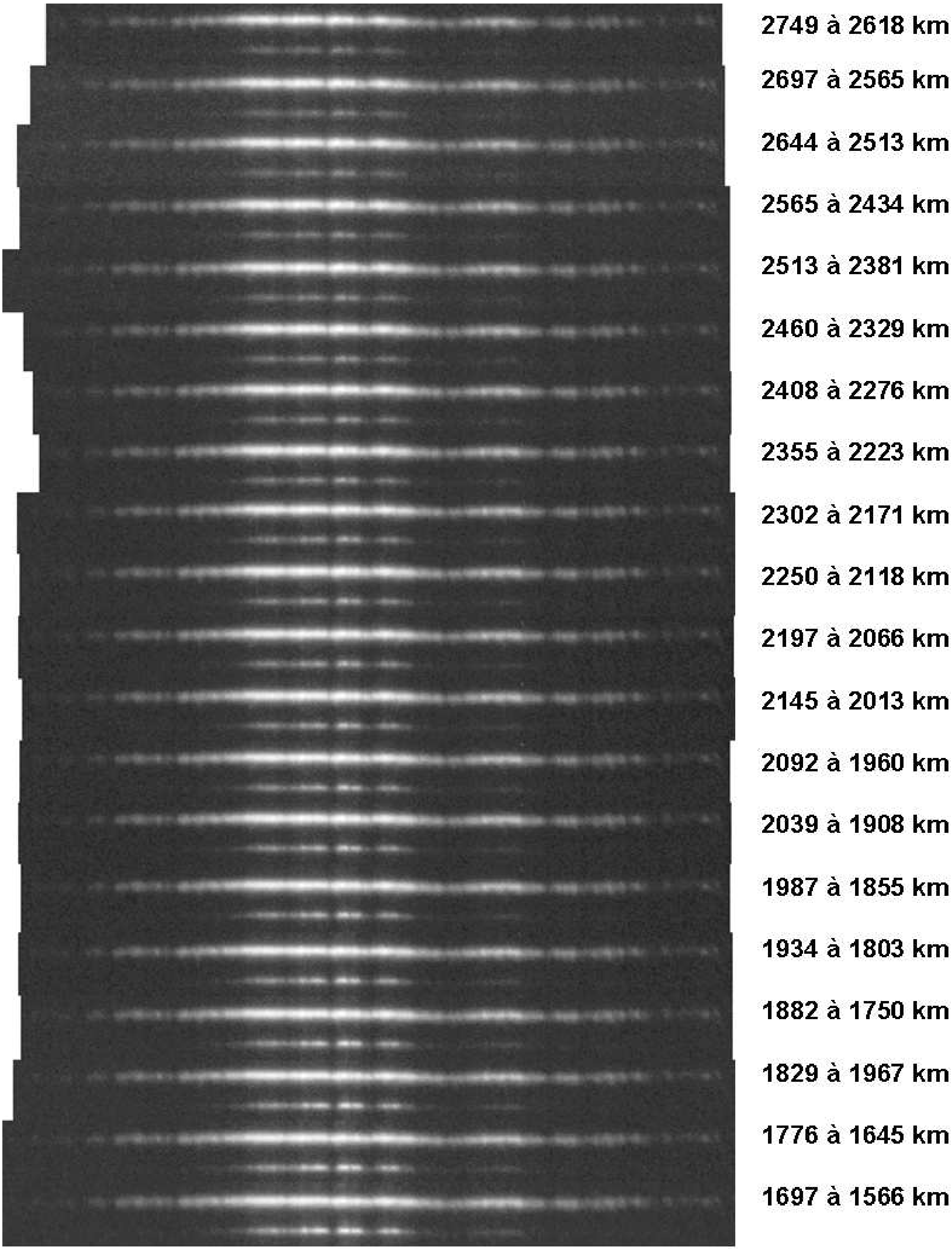

**Figure IV-2-3-8:** *altitudes 1566 à 2749 km au dessus du limbe solaire par rapport au bord lunaire moyen.*

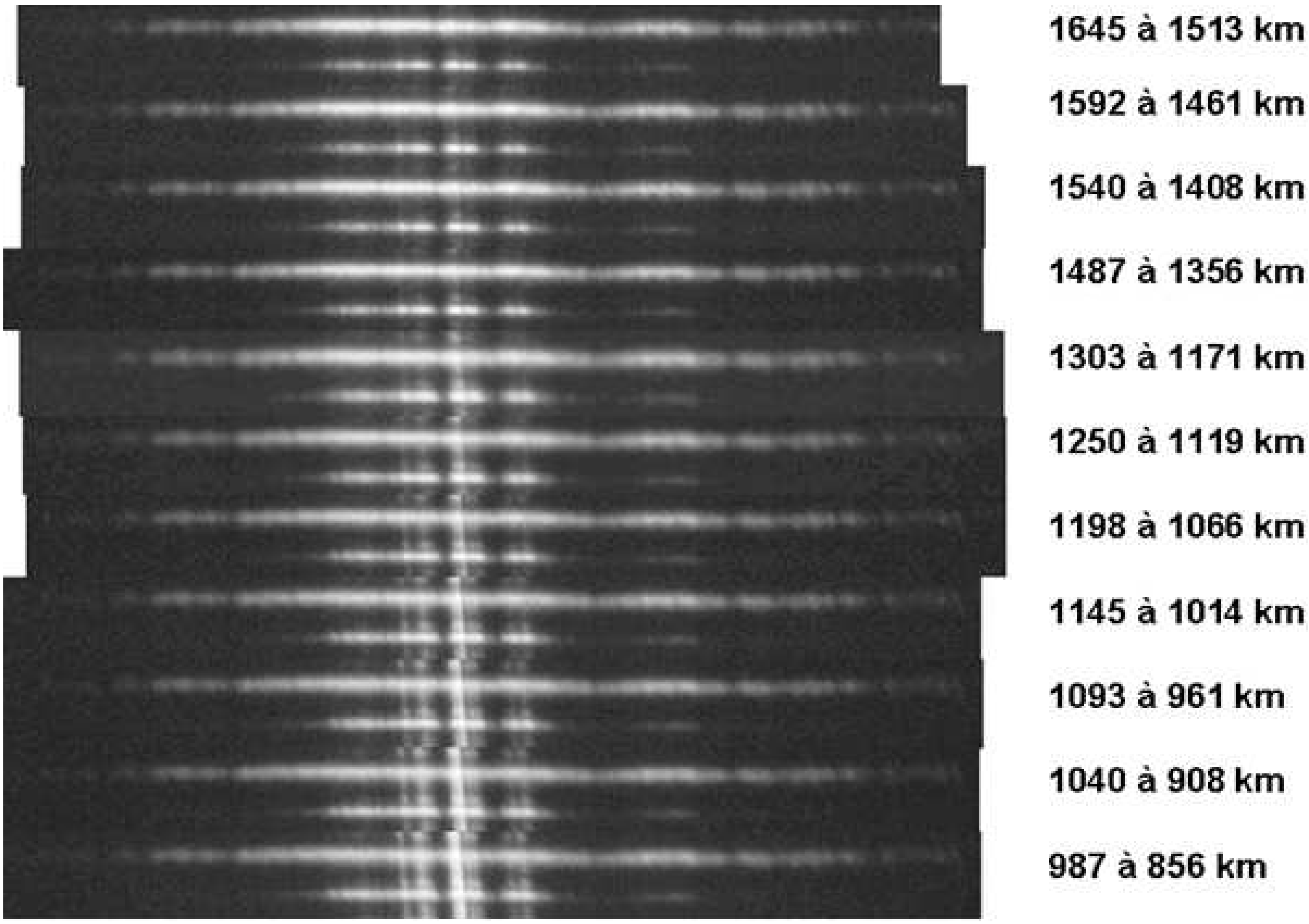

**Figure IV-2-3-9:** *altitudes 856 à 1645 km par rapport au bord lunaire moyen, au dessus du limbe solaire proches de l'instant du troisième contact, selon le spectre continu du grain de Baily qui s'intensifie plus au centre.*

Des structurations en forme de « grumeaux » sont visibles au dessus de la chromosphère, dans la région de transition chromosphère-couronne en figures IV-2-3-4 à IV-2-3-6. Ces « grumaux » qui sont assimilés à des condensations sont situés sur la partie supérieure de la raie He I 4471 Å, à sa terminaison. Cette raie comprend des inhomogénéités associées aux macrospicules. Ces structures apparaissent limitées par le bord de la Lune sur la ligne de visée et leurs extrêmités sont situées dans la région d'interface chromosphère-couronne où ils disparaissent ensuite lorsque la Lune les a recouverts au moment de la totalité de l'éclipse. Ces résultats pourraient en partie apporter un éclaircissement sur des structures similaires (bulles chromosphériques) observées en période de Soleil calme par Guy Simon et Zadig Mouradian, 1975.

Les sommations de 6 spectres tous les 3 spectres ont permis d'analyser les variations le long des profils d'intensités des raies de l'hélium neutre 4471Å et du Ti II 4468 Å. Cette sommation est utile pour analyser plus en détails la nature du plasma entre les macrospicules et le continu entre les raies He I et Ti II.

Des profils d'intensité ont été réalisés dans le sens transversal, afin d'étudier la structuration du milieu interspiculaire de la raie He I 4471Å, et étudier ce milieu encore mal connu à ce jour.

Les figures IV-2-3-10 montrent un tel profil d'intensité réalisé sur une largeur de 14 Mm et une longueur équivalente de 510 Mm avec la correspondance des images linéarisées dans la raie He I 4471 Å:

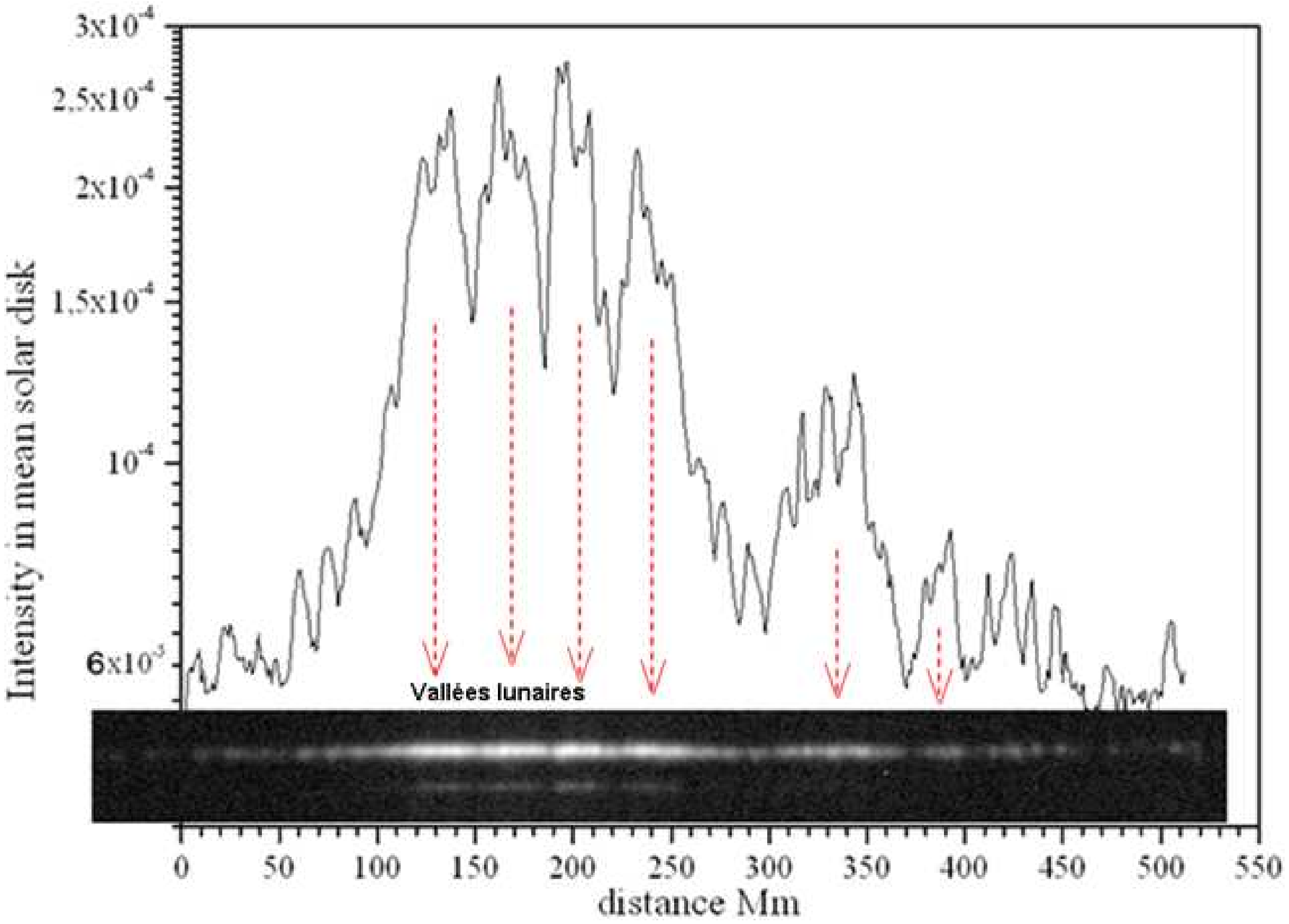

**Figure IV-2-3-10 :** *Profils d'intensités relevées sur la tranche à 2749.7 km d'altitude maximale (par rapport au bord lunaire moyen) dans la raie redressée de l'hélium neutre He I 4471 Å et correspondances avec le profil linéarisé. L'intégration est effectuée dans le sens de la raie de l'hélium neutre sur un rectangle de 14 Mm de largeur. Le niveau du continu n'a pas été soustrait et échelle Logarithme pour les intensités.*

Les modulations d'étendues supérieures à 30 Mm sont liées au relief lunaire, car on retrouve ces mêmes étendues de modulations sur les autres raies comme le Ti II 4468 Å.
Les maxima d'intensité les plus étroits correspondent à la contribution des macrospicules. La majeure partie de ceux-ci mesurent moins de 10 Mm de largeur à mi-hauteur.
D'autres sommations voir figure IV-2-3-11 ont été réalisées avec 18 spectres sommés espacés tous les 9 spectres, et ajustés pour améliorer le rapport signal sur bruit et représenter une séquence plus courte entre 1555 et 4040 km d'altitudes afin d'analyser si des corrélations existent entre les raies low FIP, leur altitude d'apparition avec la raie chromosphérique de l'hélium neutre 4471Å. Les altitudes sont indiquées à droite des figures en km.

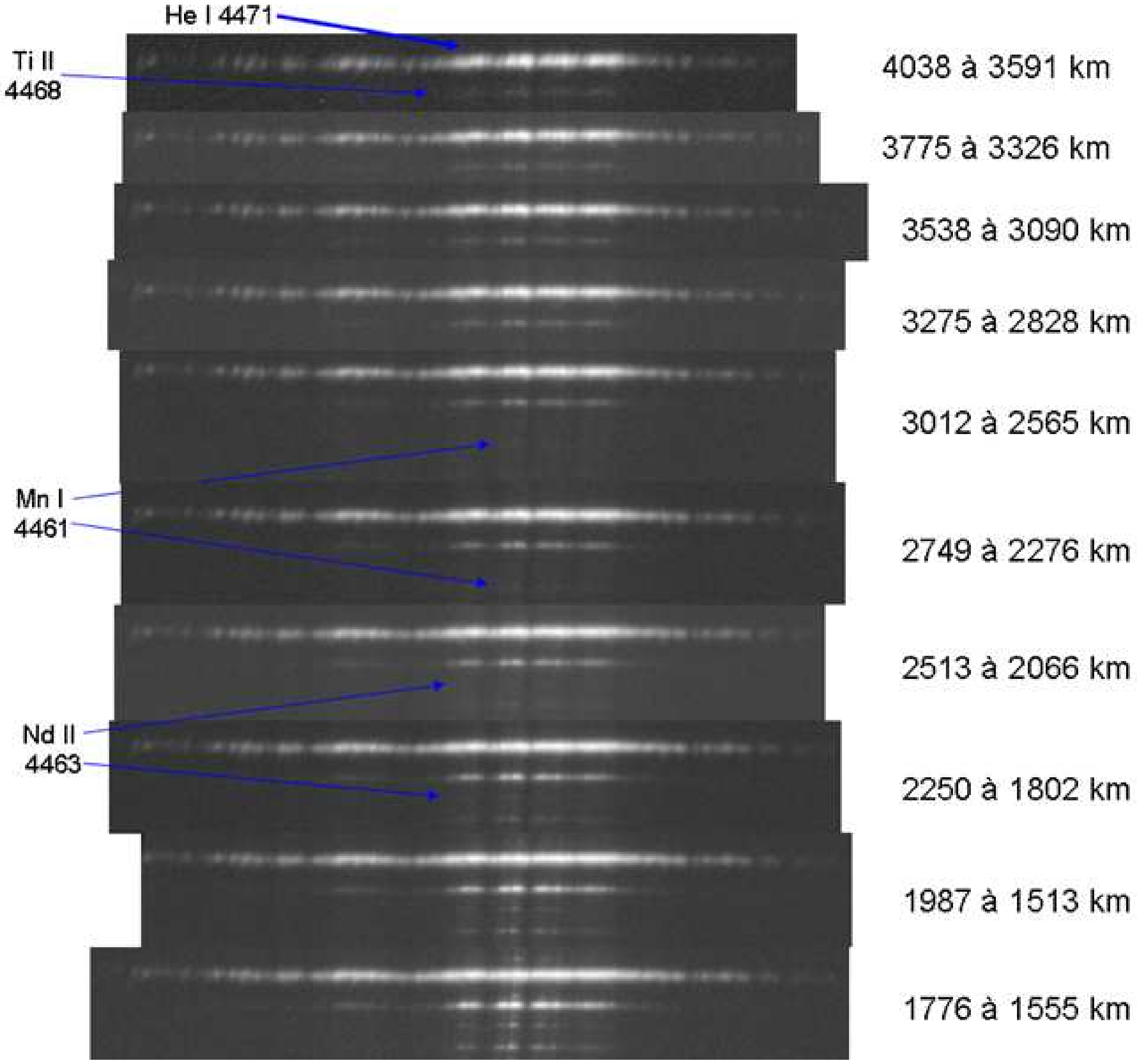

**Figure IV-2-3-11:** *Séquence de spectres résultant de sommes de 18 spectres tous les 9 spectres, redressés, puis convertis en échelle logarithmique pour une meilleure visualisation. L'apparition de raies low FIP (Mn I, Nd II) est indiquée à gauche. Les altitudes examinées sont indiquées à droite et par rapport au bord lunaire moyen.*

Les corrélations entre les raies « low FIP » et la raie de l'hélium He I 4471 Å sont d'origine solaire, c'est-à-dire que les déficits d'intensités entre la raie du Ti II 4468 Å et du Mn I 4461 Å coïncident et aussi avec la raie He I 4471 Å qui est formée plus haut que les raies low FIP (Mn I, Ti II, Fe II…).

Une interprétation de la forme de la structure des macrospicules observés peut être donnée par le schéma IV-2-3-12 où les tubes des macrospicules sont observés sur la ligne de visée et coupés par le bord de la Lune.

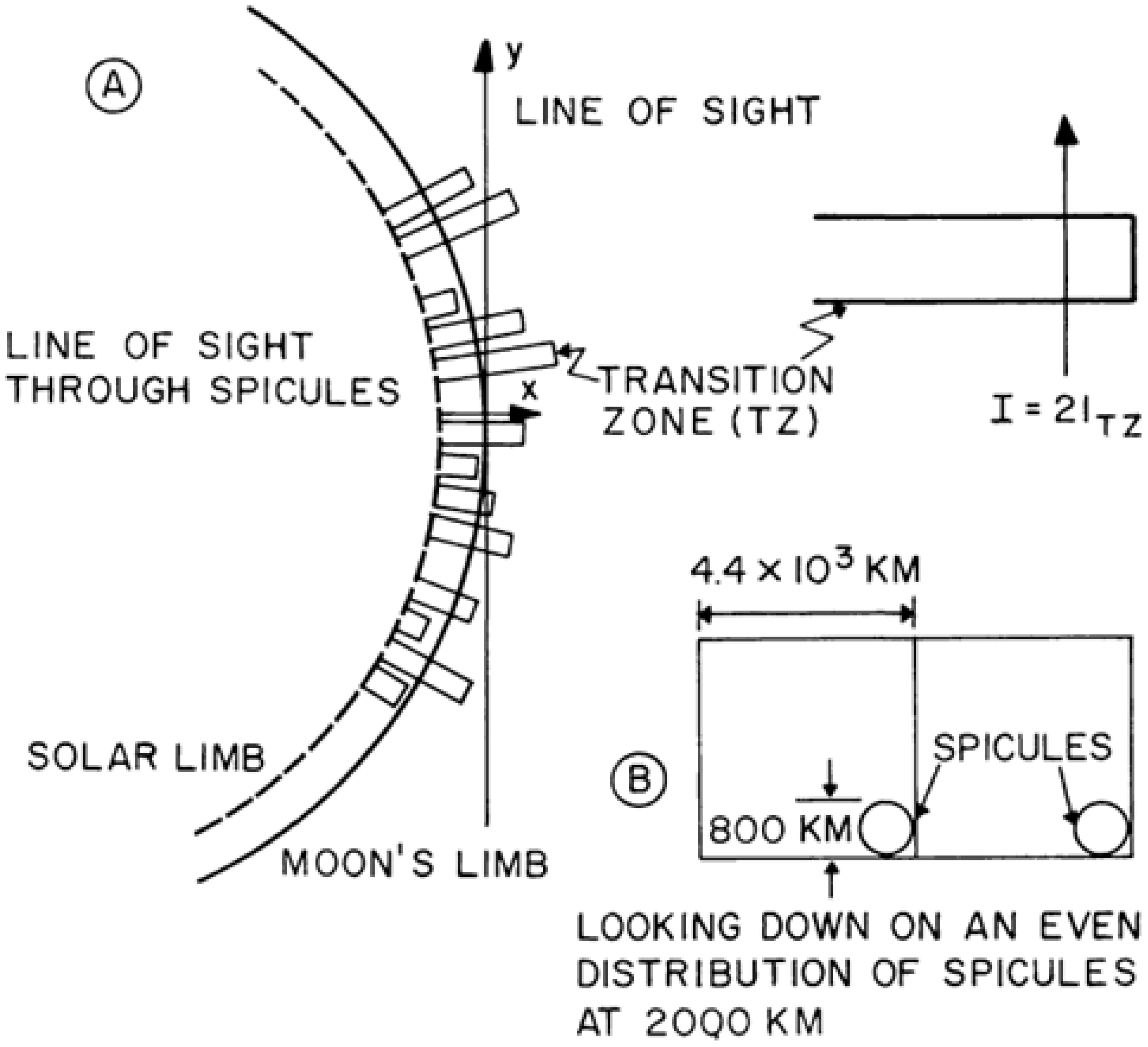

**Figure IV-2-3-12:** *Schéma selon la ligne de visée à travers les spicules en dehors du limbe. D'après Brueckner, Nicolas 1973, figure 5 page 311. Une fine zone de transition entoure chaque spicule. Partie B: Représentation schématique sur le disque montrant leur surface projetée à 2000 km au-dessus de la surface solaire à μ = 1.0. Dans ce modèle, les spicules sont placés sur des positions aléatoires dans chaque rectangle. Ces représentations ont été améliorées dans l'article Tavabi, Koutchmy, Ajabshirizadeh 2011.*

La résolution des spectres éclairs avec l'expérience transportable du réseau-objectif et caméra CCD Lumenera décrite aux figures II-1-3 et II-1-4 n'atteind pas une résolution spatiale suffisante pour résoudre les spicules dont les dimensions sont inférieures à 500 km (Tavabi et al 2011), et on peut considérer que le schéma de la figure IV-2-4-12 s'applique aussi pour les macrospicules observés à l'éclipse du 22 Juillet 2009.
Une analyse complémentaire est effectuée à partir des spectres éclair non redressés dans la raie de l'hélium neutre He I 4471 Å, où des embrillancements dans les macro-spicules ont été constatés sur le limbe à l'équateur ouest. Un ajustement est effectué pour identifier avec une image SoHO (dans la raie de l'hélium He II 304 Å) que ces embrillancements coïncident avec une protubérance calme et de plus faible intensité.
Le ***S*** représenté en rouge sur la figure IV-2-4-13 des spectres sommés avec le profil lunaire au 3$^{ième}$ contact de l'éclipse de 2009, indique la présence d'un renforcement en intensité ou « spike ». Il a été localisé sur le limbe ouest solaire, et par comparaison avec l'image de l'hélium He II 304 Å de SoHO. Les correspondances sont indiquées sur la figure IV-2-4-13 en traits pointillés bleus, entre les embrillancements de petite échelle sur une partie de la raie He I 4471 Å (quelques macrospicules) et la protubérance calme observée simultanément dans la raie He II Å 304 avec SoHO, sur le même limbe.

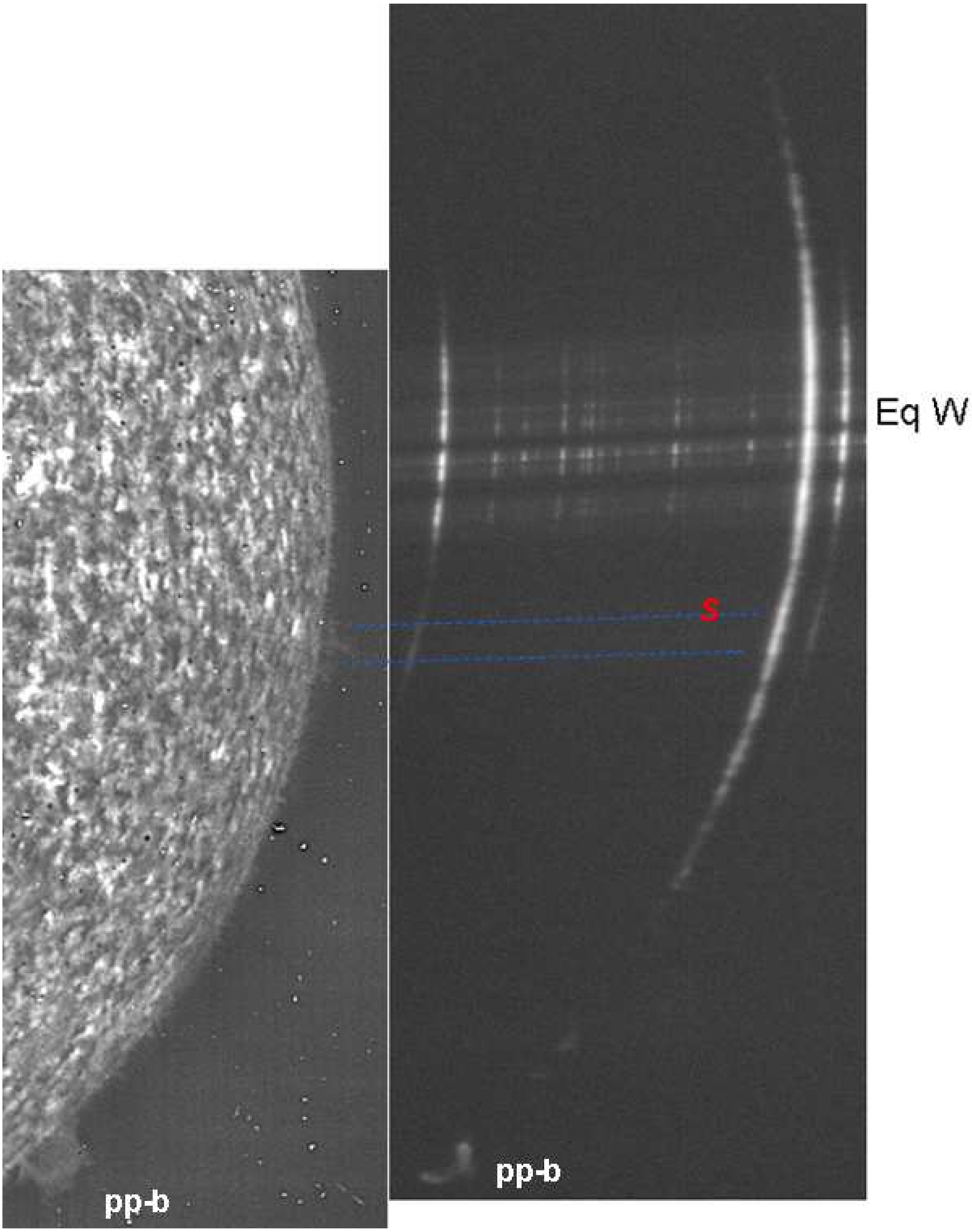

**Figure IV-2-3-13 :** *Analyse du renforcement ou « spike » dans les macrospicules par comparaison avec une image de SoHO en He II 304 Å et l'image monochromatique dans la raie He I 4471Å (spectre éclair) avant le troisième contact de l'éclipse totale du 22 Juillet 2009. L'image de SoHO est prise à 0h 36 TU, soit environ 1heure avant cet instant du 3ième contact. La pp-b visible en bas des images a servi pour l'orientation et a peu varié.*

Ces renforcements pourraient être probablement liés à un embrillancement de macrospicule. Le montage figure IV-2-3-13 montre que cette structure est associée à une faible protubérance calme observée en He II 304 Å. Ces observations semblent confirmer qu'un embrillancement de type « spike » observé dans quelques macro-spicules pourrait être lié ou corrélé avec l'apparition d'une protubérance. Ceci peut-être un autre moyen d'analyser les phénomènes similaires dans les interfaces chromosphère-couronne et protubérance-couronne.

Ce chapitre IV-2-3 a montré la structuration des macro-spicules dans la raie « froide » He I 4471Å, et des phénomènes d'interface chromosphère-couronne qui sont de même nature que l'interface protubérance-couronne, grâce aux correlations observées. Ce phénomène de même nature dans les interfaces se confirme aussi par la présence des raies « low FIP » comme le Titane Ti II 4468 Å présent dans l'interface protubérance-couronne et visible dans l'interface photosphère-couronne, dans le prolongement de l'image de ces raies en croissants. Ces résultats montrent que le titane est un élément chimique important comme « traceur » dans les interfaces photosphère-couronne et protubérance-couronne.
Bien que l'intervalle spectral autour de 4500 Å obtenu à l'éclipse de 2009 ait été différent de celui de l'hélium He II 4686 Å et He I 4713 Å aux éclipses de 2008 et 2010, ces observations de l'éclipse de 2009 ont été fort utiles et complémentaires, pour analyser les interfaces grâces aux éclipses totales.
Le chapitre IV-3 présente quelques analyses des résultats obtenus lors de l'éclipse totale du 11 Juillet 2010, où l'intervalle spectral était situé autour de 4700 Å avec simultanément les 2 raies de l'hélium He I 4713 et He II 4686 Å, ainsi que les raies « low FIP » du Fe II 4629 Å et Mg I 4702 Å. Les protubérances visibles dans ces mêmes raies d'hélium ainsi que dans le titane une fois ionisé Ti II sont discutées au chapitre V.

# IV-3) Analyses et interprétations des résultats de l'éclipse du 11 Juillet 2010. Vers une nouvelle définition du bord solaire

## IV-3-1) Definition, analyse de la structuration du bord solaire sans lumière parasite

Les résultats des spectres éclair obtenus à l'éclipse totale de Soleil du 11 Juillet 2010 ont permis d'aboutir à une nouvelle conception du bord du Soleil. Les analyses qui vont suivre présentent en détail la façon dont le bord du Soleil apparaît, en l'abscence de toute lumière parasite, avec une certaine complexité. Les modulations liées au bord de la Lune d'une part, et d'autre part l'enchevêtrement des raies d'émission low FIP et de raies de Fraunhofer en absorption se superposent au « vrai » continu qui définit le bord solaire (sans lumière parasite).
L'extrait agrandi 5 fois figure IV-3-1-1 provient de spectres éclair de la séquence au troisième contact de l'éclipse du 11 Juillet 2010. Les raies d'émission et d'absorption sont vues simultanément, avec un renversement, et ces spectres sont modulés par les vallées et montagnes du bord de la Lune.

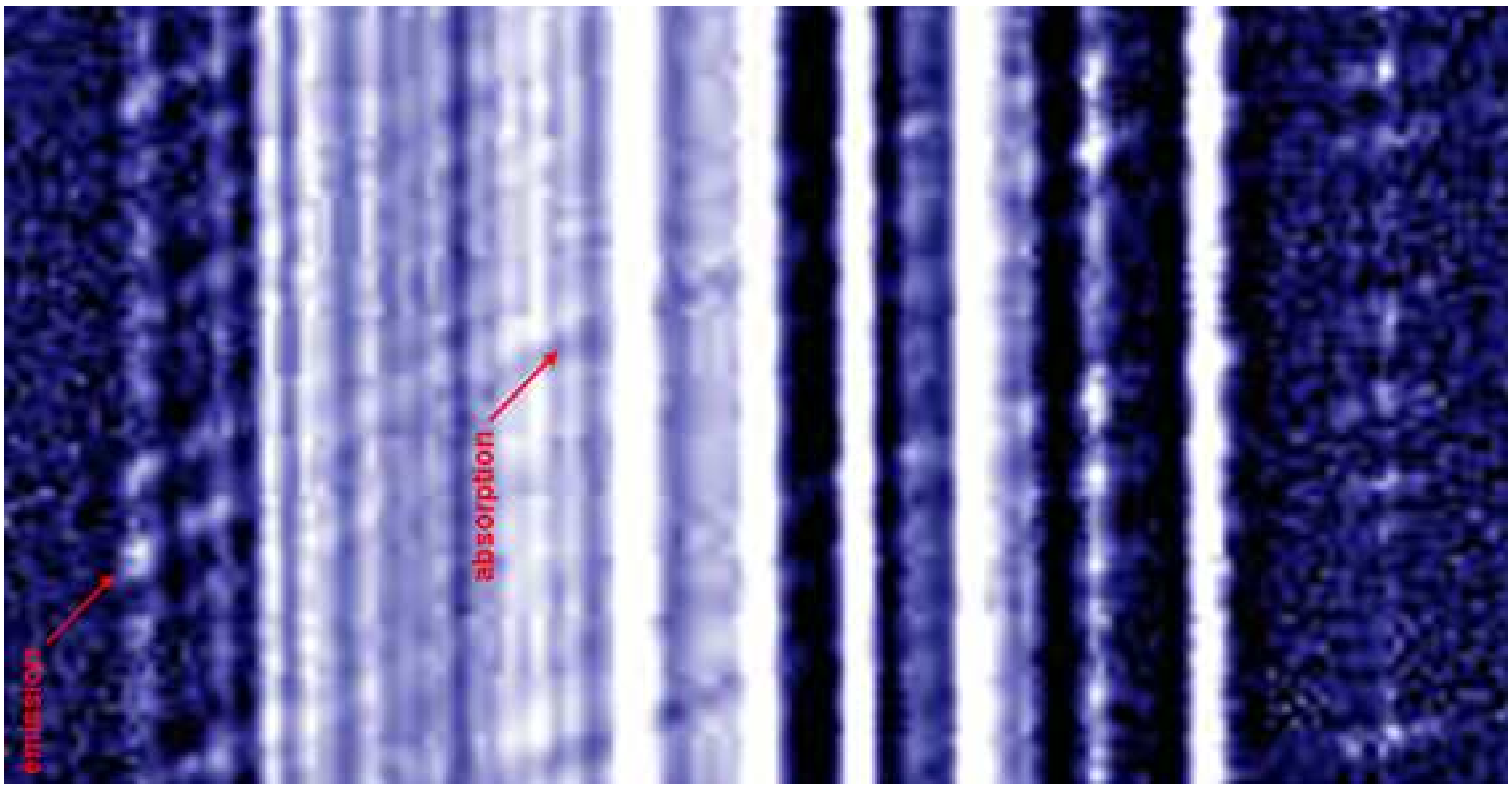

**Figure IV-3-1-1 :** *vue agrandie extrait des spectres précédents avec un renforcement des contrastes et colorisation, pour illustrer les observations simultanées de raies en absorption et de raies en émission avec le spectre continu du bord photosphérique*

Le renforcement des contrastes permet de mieux distinguer la structuration des raies en émission, simultanément avec les raies en absorption, au bord du Soleil sur un même spectre, et pour définir la région de la mésosphère comme introduite et discutée au chapitre III-3.
La distinction entre l'absence d'émission et l'absorption s'effectue dans un intervalle limité d'une même bande de spectre de grain de Baily sur le continu. Les raies en absorption conduisent à un déficit dans le continu, tandis que les raies d'émission qui s'ajoutent à ce continu, produisent une augmentation d'un facteur 2 en intensité. L'abscence d'emission, correspond au domaine spectral, pris en dehors des raies d'émission et d'absorption, et correspond au « vrai » bord solaire, c'est-à-dire sans superposition de raies.

Ces nouvelles observations avec la caméra CCD Lumenera de 12 bit de dynamique (4096 niveaux d'intensité), permettent d'apporter une nouvelle conception et définition de la couche renversante au moment des contacts des éclipses totales de Soleil à partir des séquences de spectres éclairs sans fente. Ces observations simultanées apportent ainsi une nouvelle façon de définir et traiter la question de la définition du bord du Soleil et de la mésosphère sans effets de lumière parasite, où le bord solaire est étudié en prenant en compte le spectre continu, relevé entre les raies d'émission.
Le montage Figure IV-3-1-2 décrit le procédé pour extraire une fine bande spectrale de 1.5 Å de large sur chaque spectre pour reconstruire l'extinction du limbe solaire par le bord de la lune à une seule longueur d'onde. Les raies d'absorption de Fraunhofer sont observées simultanément avec la myriade de petites raies faibles en émission grâce à la dynamique suffisante de 4095 niveaux (12 bits) de la caméra CCD Lumenera Skynyx 2.1 M.
Ces caméras permettent d'obtenir des images de qualité photométrique.

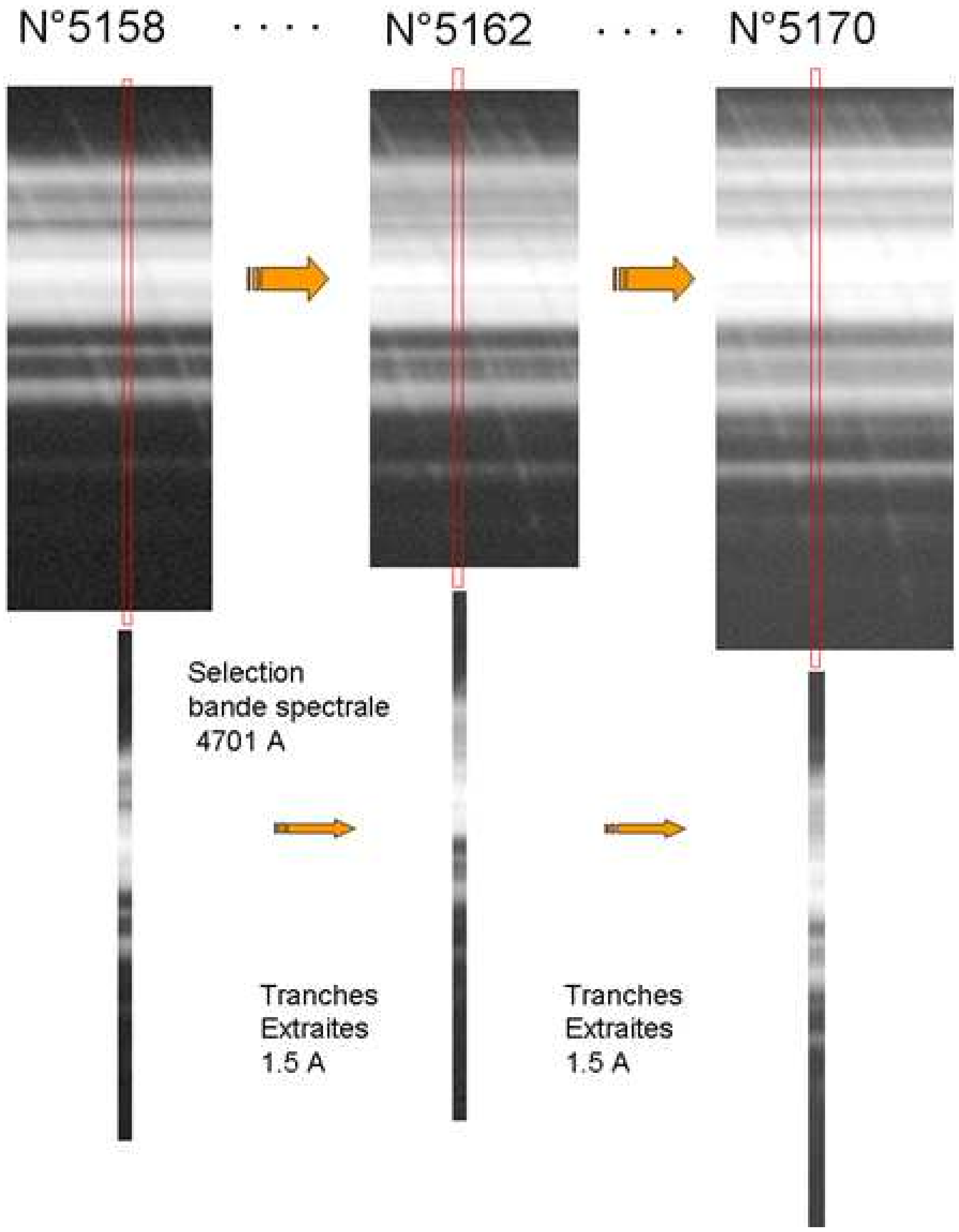

**Figure IV-3-1-2 :** *montage permettant de décrire chaque procédé pour extraire une par une les tranches de spectre dans les grains de Baily de 1.5 Å de large (rectangle rouge). Etapes préliminaires de reconstruction de la disparition du bord solaire (et sa réapparition), résolu spatialement, avec le mouvement du profil du bord de la Lune.*

Ces montages permettent de décrire le procédé d'extraction d'un rectangle noté « tranche extraite » sur la figure IV-3-1-2**,** de 1.5 Å de large. Chacune des tranches extraites est accolée et ajustée à la suivante, de façon à toutes les empiler. Cette reconstruction des profils en coupe du bord du Soleil résolue temporellement (spatialement) en altitude tient compte des modulations du relief lunaire. Cela est ensuite utilisé pour comparer ces reconstructions avec un profil d'intensités relevées transversalement dans les grains de Baily, voir figure IV-3-1-5. Les figures IV-3-1-3 et IV-3-1-4 décrivent une autre manière d'appréhender ce qu'est le vrai bord du Soleil, en ayant extrait des intervalles de coupes dans une tranche de spectre pris dans le continu des grains de Baily, selon le procédé figure IV-3-1-2 et reproduit pour les contacts précédant et succédant la phase de totalité.

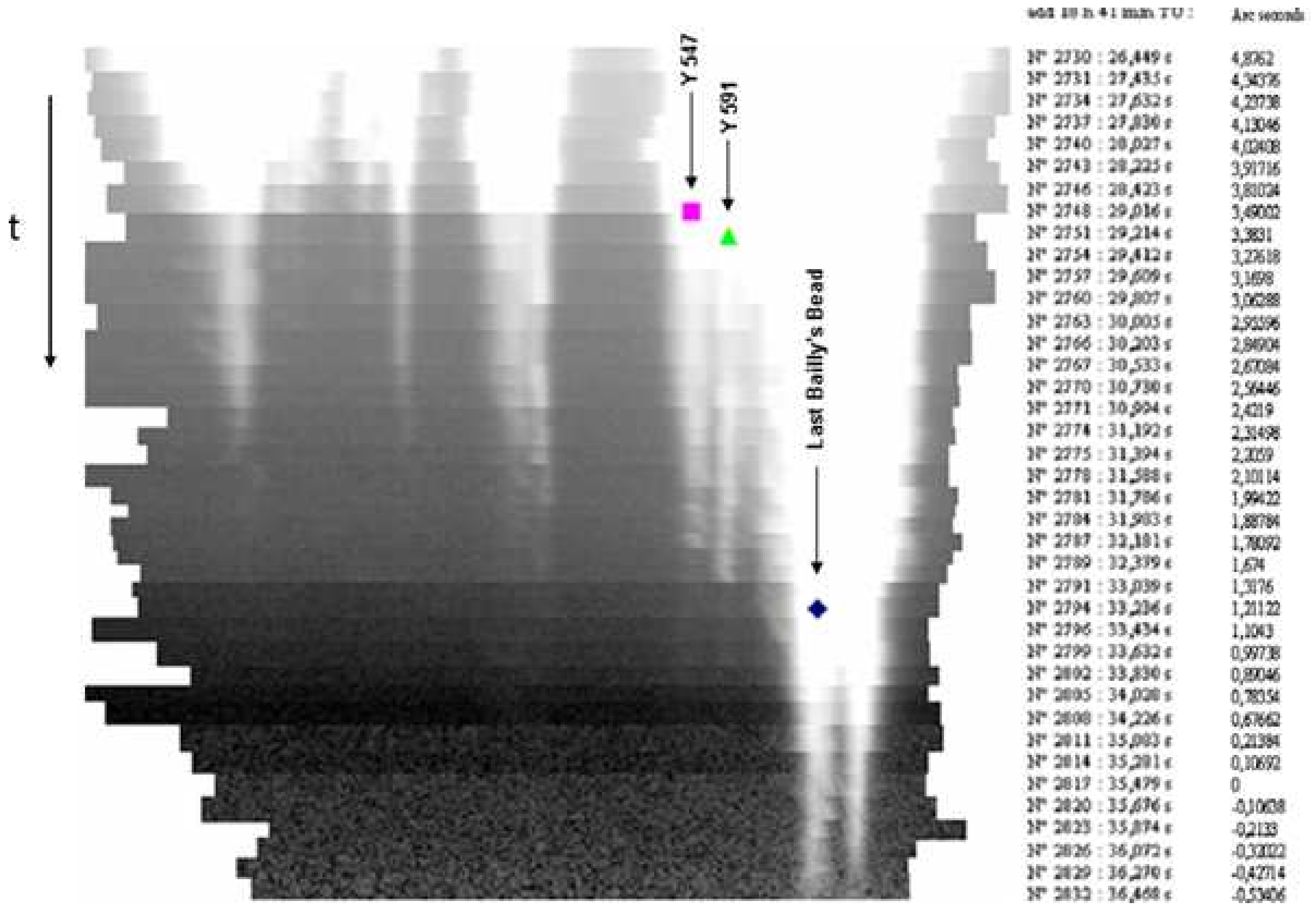

**Figure IV-3-1-3:** *structuration des profils par coupe de 1.5 Å de largeur agrandis, du vrai bord solaire observé, relevé dans le continu, afin de quantifier l'assombrissement puis l'extinction de l'extrême limbe solaire au moment du 2nd contact, juste avant la totalité de l'éclipse du 11 Juillet 2010. Les instants correspondant à chaque tranche sont indiqués à droite. Les symboles colorés indiquent les positions des grains de Baily étudiés pour lesquels des courbes de lumière ont été analysées.*

La différence entre le fond de ciel sur la tranche N° 2789 (32.379s) et N° 2791 (33.039) peut s'expliquer par l'ajustement des niveaux d'intensité permettant de mieux visualiser les plus faibles intensités dans les grains de Baily plus faibles.
De la même façon, cette analyse a été effectuée au 3ième contact, voir figure IV-3-1-4, montrant un profil avec la même méthode que la figure IV-3-1-3 du « vrai » bord solaire modulé par le relief lunaire au moment de la réapparition.

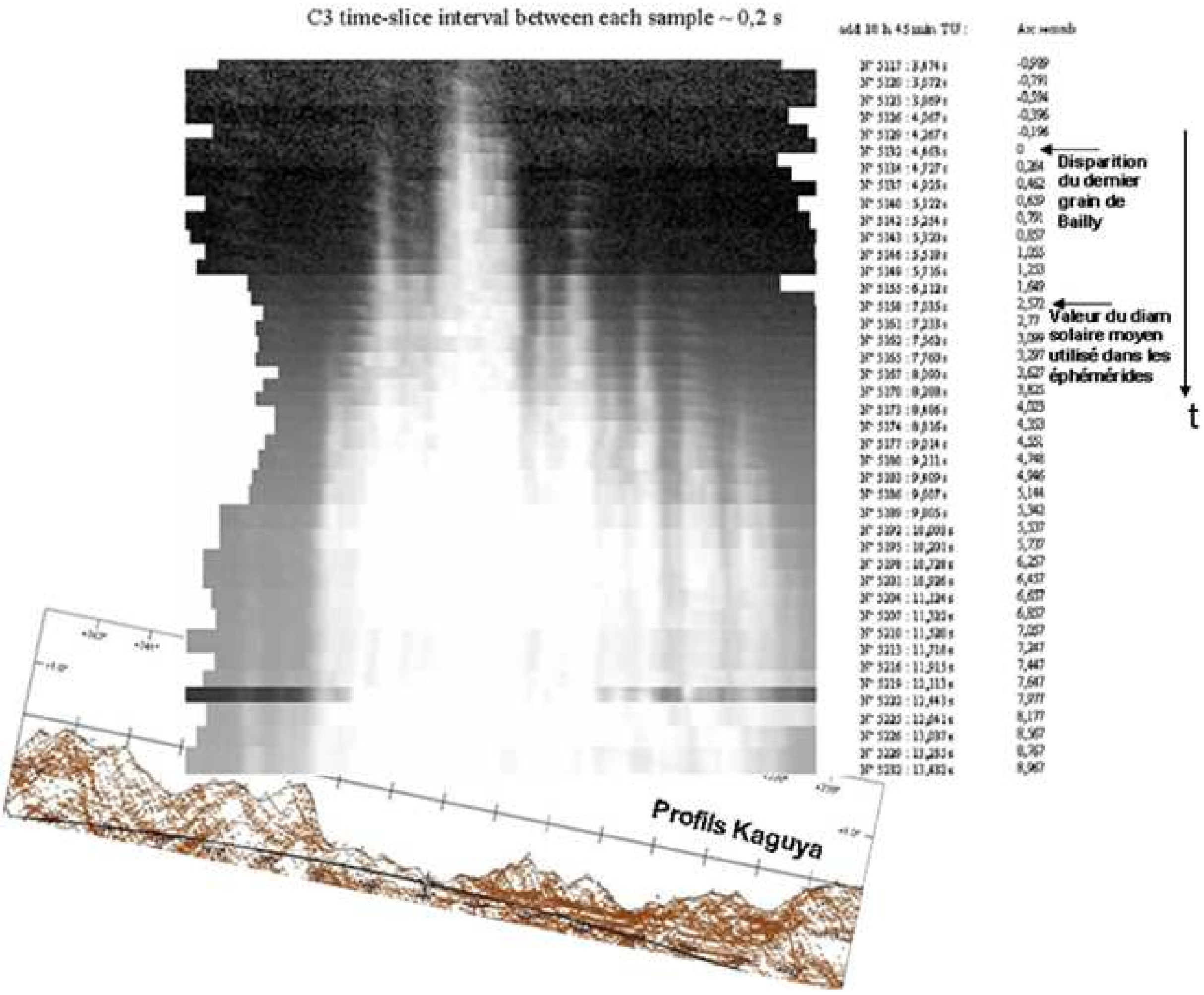

**Figure IV-3-1-4:** *empilement des profils en coupe de 1.5 Å du vrai bord solaire observé, relevé dans le continu (grains de Baily), afin de décrire les structurations correspondantes liées au profil lunaire (altimétrie Kaguya). L'assombrissement puis l'extinction de l'extrême limbe solaire au moment du 3$^{ième}$ contact apparaissent progressivement, juste après la totalité de l'éclipse du 11 Juillet 2010. Les instants correspondant à chaque tranche sont indiqués à droite.*

Ces coupes sont réalisées en vue d'expliciter et révéler la structuration complexe du « vrai » bord du soleil limité par le relief du bord lunaire en utilisant un procédé de reconstruction visuel, pour ensuite réaliser des profils d'intensité en 2 dimensions sur l'étendue des spectres des grains de Baily, voir figure IV-3-1-5. Cela permet de définir au moyen des courbes d'intensités les variations du profil d'intensité du bord du Soleil occulté et découpé par le relief du limbe lunaire.

Ce type d'analyse et de reconstruction du bord du Soleil est une approche nouvelle par rapport à ce qui a été publié sur le sujet. En effet, les courbes de lumière en figure IV-3-2-1 montrent une différence entre l'instant du contact défini par les éphémériques et la bifurcation qui pourrait correspondre à l'instant où le dernier grain de Baily disparaît, avec la raie d'émission du Fe II 4629Å qui apparaît et se distingue du spectre continu. Cette méthode permet de mieux définir le bord solaire lors des contacts d'éclipses. Les profils d'intensité sont une autre manière de le représenter. La figure IV-3-1-5 représente les profils d'intensités, au 3$^{ième}$ contact concernant les mêmes régions présentées en figure IV-3-1-4.

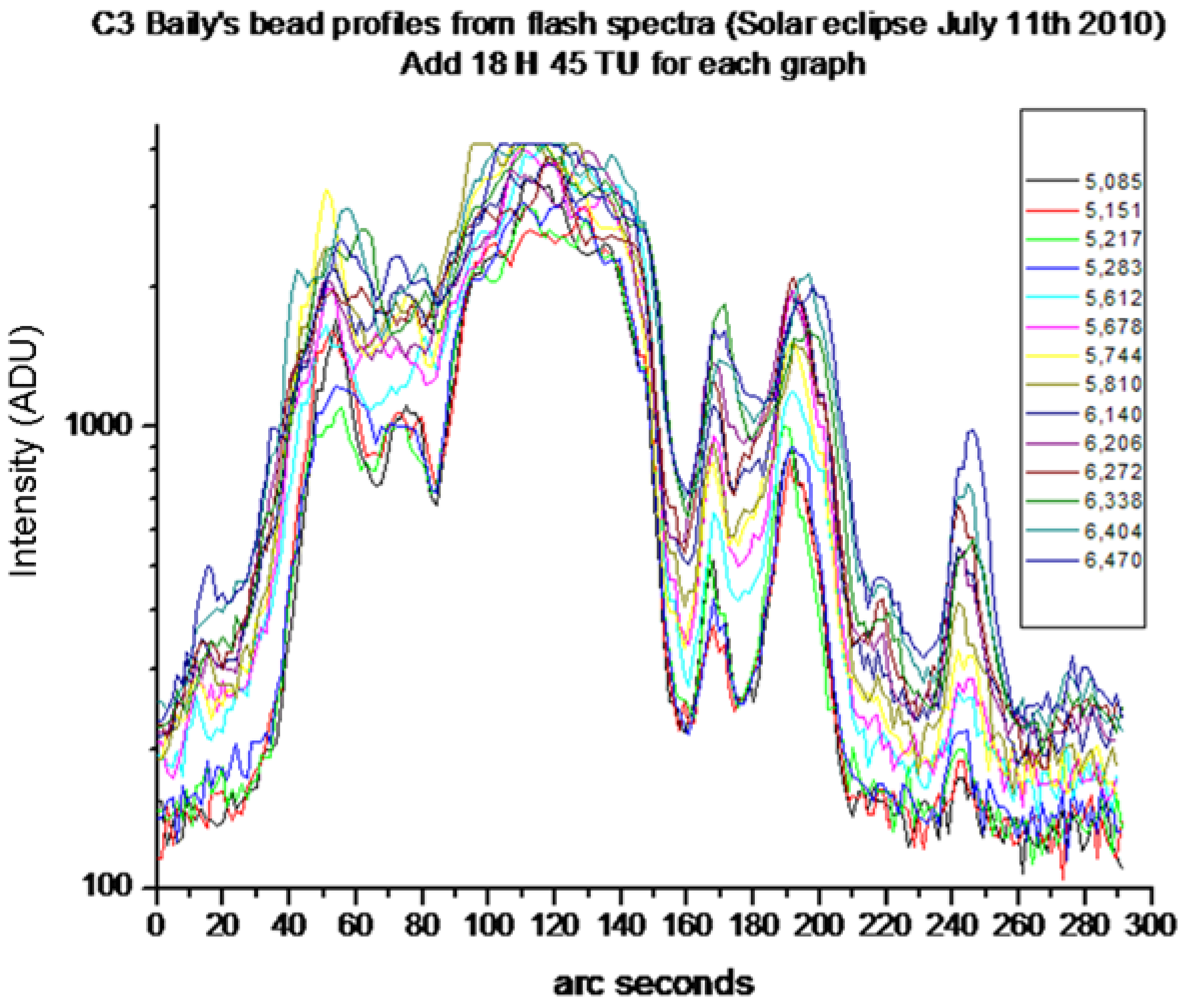

**Figure IV-3-1-5:** *profils d'intensité transverse relevés le long des tranches des spectres des grains de Bailly sur un intervalle spectral étroit 1Å à la longueur d'onde de 4700Å, avec un intervalle variant entre 0.1 et 0.3 s entre chaque spectre. Les intensités sont exprimées en ADU (Arbitrary Digital Unit), et au-delà de 4095 ADU il y a saturation.1 ADU équivaut à* $0.823*10^{-6}$ *unités du disque solaire moyen (voir Annexe 17-3).*

Ces résultats ont contribué à montrer ce que le satellite de la mission Picard aurait dû observer depuis l'espace pour définir le « vrai » bord du Soleil en l'absence de lumière parasite diffusée, en vue de mesurer ensuite son diamètre, une fois le bord ainsi défini à cette longueur d'onde 4700 Å. Ceci montre l'importance des éclipses totales, et de la réalisation des spectres éclair, en l'abscence de lumière parasite, pour appréhender le vrai bord du Soleil.
Cependant, le relief lunaire, montagnes et vallées rendent le travail difficile pour évaluer les altitudes et hauteurs au dessus du limbe solaire lors des contacts des éclipses totales. La détermination de la référence d'altitude $h = 0$ est difficile. Lites 1983 a tenté d'estimer des fluctuations du limbe solaire avec des modèles théoriques et des observations.
Dans sa publication, il indique clairement les inconvénients pour savoir où définir la référence $h = 0$ des hauteurs lors des observations d'éclipses. Cependant dans cette thèse nous avons considéré le bord comme étant la région où l'intensité présente la décroissance la plus abrupte, c'est-à-dire où le gradient d'intensité est le plus élevé.
Chaque grain de Baily dont le spectre est réalisé correspond en réalité à un bord du Soleil découpé dans une valée lunaire, dont il est possible de réaliser les profils d'intensité.

Cependant à cause des irrégularités du relief lunaire, ceux-ci sont décalés comme le montrent les courbes de lumière IV-3-1-6:

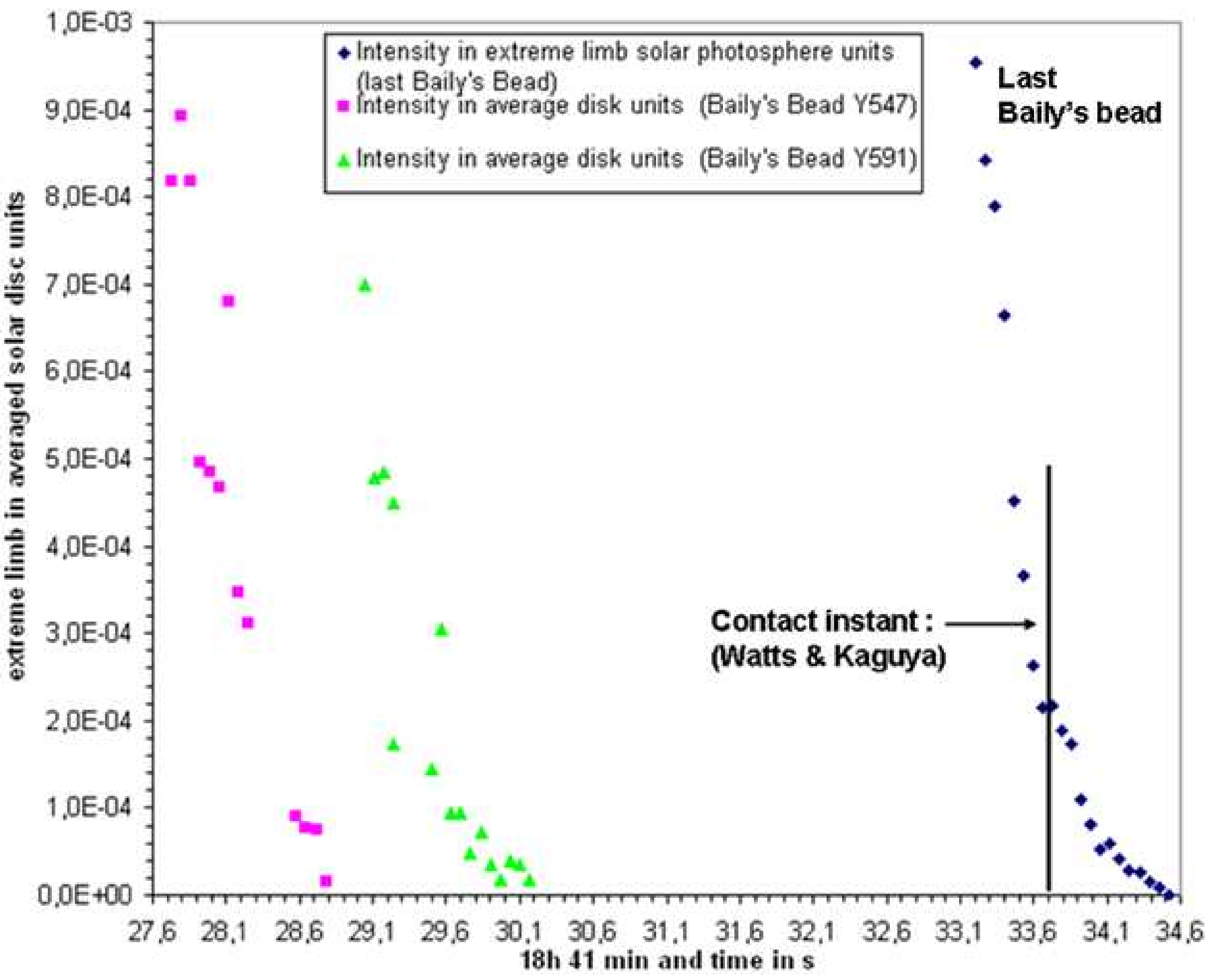

**Figure IV-3-1-6:** *courbes de lumière en fonction du temps pour 3 spectres de grains de Baily relevés dans des vallées lunaires différentes sur le bord de la Lune au second contact de l'éclipse de 2010. Les mentions Y 547 et Y 591 correspondent aux ordonnées pour lesquelles 2 des profils ont été pris d'après la figure précédente du 2$^{nd}$ contact. Le profil avec les « diamants» bleus correspond au dernier grain de Baily.*

Ces résultats montrent la difficulté de définir le bord du Soleil compte-tenu du relief accidenté du bord de la Lune. La chronodatation au moment des contacts de l'éclipse est indiquée en abscisses.

### VI – 3-2) Analyses des profils du continu correspondant au « vrai » bord solaire. Eclipse totale du 11 Juillet 2010. Inversion d'intégrale d'Abel pour déduire les brillances.

Les courbes figures IV-3-2-1 et IV-3-2-2 ont pour objectif d'approfondir les résultats obtenus sur les variations d'intensités des raies du Fer II et du continu, directement mesurées sur chaque spectre éclair, sans avoir lissé les courbes. Une comparaison avec les travaux antérieurs de Weart S.R. 1971 sur l'éclipse de 1970 dans le continu à 4815 Å est effectuée en vue comparer des méthodes pour définir le bord solaire et l'effet d'embrillancement du limbe dû aux raies d'émission low FIP situées très proches du limbe et au minimum de température. Les courbes de la figure IV-3-2-1 ont été déduites après la chronodatation au troisième contact de l'éclipse du 11 Juillet 2010. La correspondance avec les prédictions des éphémérides par Patrick Rocher (IMCCE), est indiquée, ainsi que l'instant d'insertion des filtres et instant de contact.

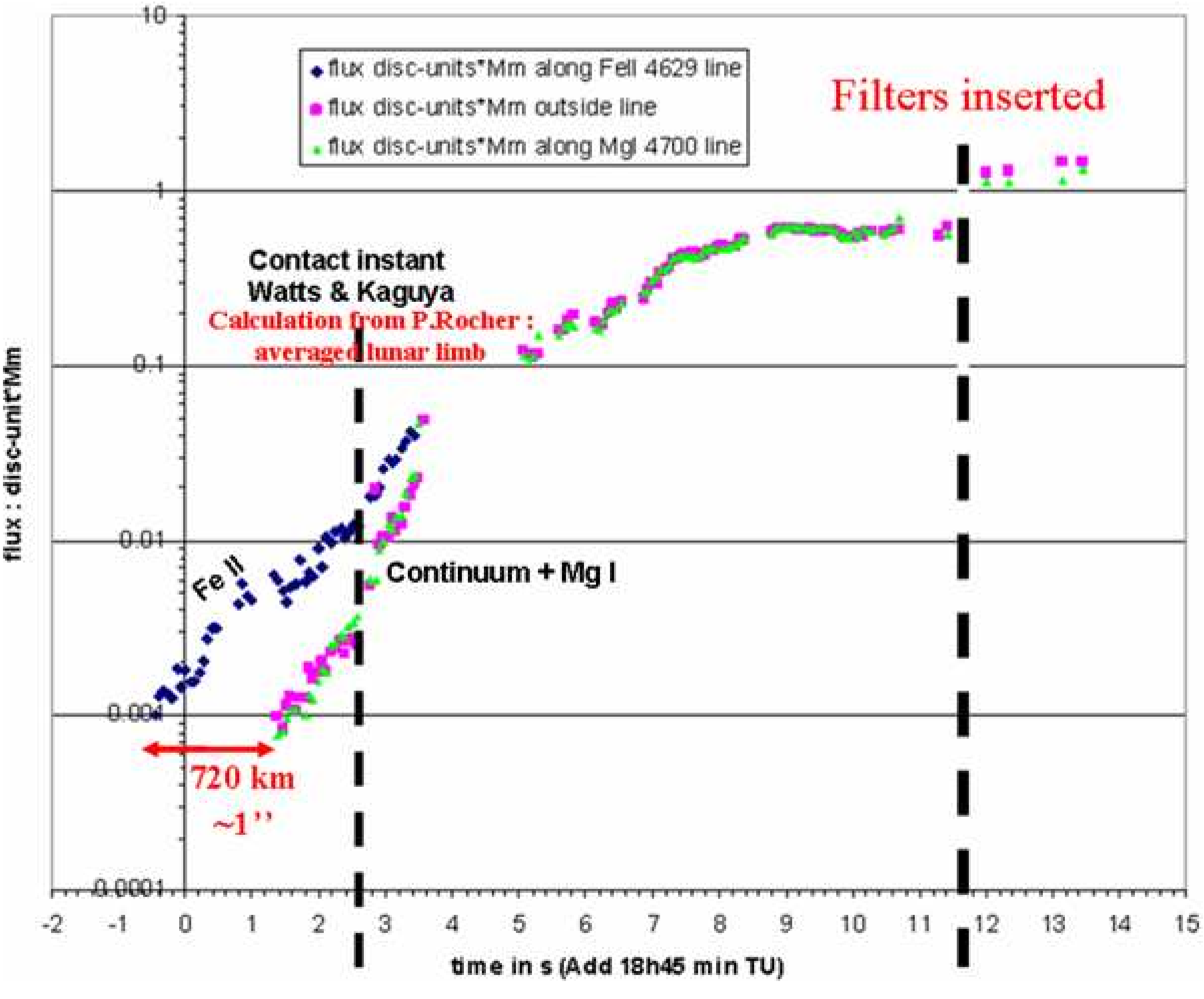

**Figure IV-3-2-1:** *analyse du bord solaire au troisième contact à l'éclipse du 11 Juillet 2010, (sans lumière parasite) et comparaison des flux relevés dans une raie d'absorption du Mg I, avec les flux pris dans une raie du Fe II, et dans le continu relevé entre les raies. Les échelles sont indiquées. Relevés d'intensité moyenne effectués dans une vallée lunaires.*

Les unités en flux « disc-unit*Mm » découlent du même procédé de mesure de brillance dans une vallée lunaire, comme décrit au début du chapitre IV-1 et des Annexes 28 et 29.

Ces graphiques IV-3-2-1 permettent d'analyser chaque séquence lors du troisième contact où le bord solaire est défini malgré un manque de certains points de relevés à cause du ralentissement de la cadence d'acquisition des spectres liés à la saturation de la CCD et débit limité en 12 bits de dynamique. La figure IV-3-2-2 reprend les mesures des flux de la figure IV-3-2-1 où sont indiquées les altitudes (à la place des temps de la figure IV-3-2-1) et en ayant reporté les relevés réalisés par Weart à l'éclipse de 1970, sur le même graphique.
Nous avions eu des voiles nuageux lors de l'éclipse totale du 11 Juillet 2010 qui ont atténué les flux (d'un facteur 3 à 5), ce qui pourait expliquer en partie les différences entre les courbes, mais la cadence du nombre de relevés effectués par Weart à l'éclipse du 7 Mars 1970 et publiés (Weart, 1971) était faible comparée à celle de nos caméras CCD numériques.

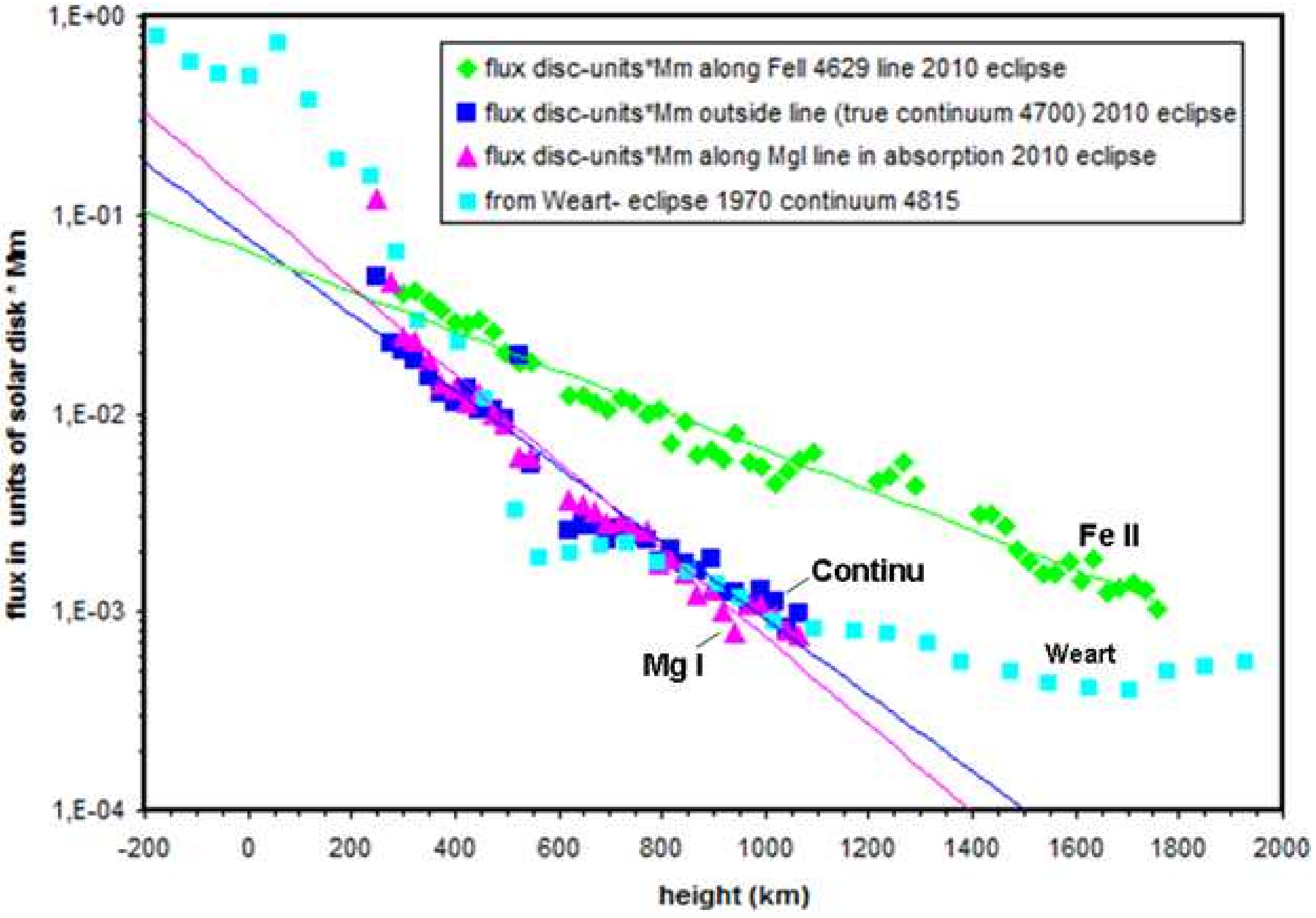

**Figure IV-3-2-2:** *courbes brutes en échelle logarithme au troisième contact- C3 de l'éclipse du 11 Juillet 2010, montrant le vrai continu pris dans une vallée lunaire, en dehors des raies d'émission, et par comparaison, le flux de la raie du Fe II 4629 Å. En bleu clair, est représentée la courbe du continu à 4815 Å réalisée par Weart et analysée-publiée par Weart 1971, d'après l'éclipse du 7 Mars 1970. Relevés d'intensités moyennes effectués dans une vallée lunaire.*

Les courbes de lumière de Weart, S.R., obtenues sous un ciel bien dégagé du 7 Mars 1970, ne montrent pas de point d'inflexion autour de l'altitude $h = 0$, définissant le bord du Soleil pris dans le continu. L'avantage des relevés pris sur l'étendue des valées lunaires « aires dans les vallées lunaires » est de pouvoir moyenner. Tandis que le relevé de l'intensité maximal dans un seul pixel, permet de rendre compte d'une mesure plus localisée dans le relief lunaire.

La courbe de lumière du Fe II  montre une plus grande étendue que celle du continu sur le bord solaire lorsque ces 2 composantes ne sont pas prises séparément grâce à la séparation des composantes.
Ces courbes de lumière expérimentales, relevées sur les spectres éclair, sont ensuite utilisées pour déduire les émissivités par la méthode d'inversion d'intégrale d'Abel, avec une méthode simplifiée décrite en Annexe 30.
Les analyses des enveloppes d'hélium et des raies « low FIP » sont effectuées pour des altitudes inférieures à 1000 km.  Ceci est une avancée par rapport aux observations antérieures où il était très difficile d'observer en dessous de 1000 km, voir chapitre I-1-6.
Les variations d'intensité constatées sur le continu et pour les mêmes raies de l'hélium neutre 4713 Å et ionisé 4686 Å observées aux éclipses totales de 2008 et 2010, sont dues à une transmission atmosphérique différente: en 2008, le second contact a eu lieu entre 2 nuages, le fond de ciel était bleu autour de l'éclipse, tandis qu'en 2010, ce ciel était totalement voilé au moment des phases de la totalité, mais on voyait l'éclipse.

## IV-3-3) Analyses des courbes de lumière et inversions d'intégrales d'Abel des données de 2010

Dans ce chapitre, nous avons tenté de réaliser les courbes d'ajustement sur les courbes de lumière où les premières mesures commencent vers 450 km d'altitude, voir figure IV-3-3-1, car les dynamiques de 8 bits et surtout 12 bits des caméras CCD nous ont permis d'examiner les intensités des raies d'émission à ces altitudes inférieures à 600 km. C'est ensuite à partir des profils d'intensité ajustés, que des inversions d'intégrale d'Abel sont ensuite tentées, en considérant le cas de raies optiquement minces (épaisseurs optiques très inférieures à 1, « transparent »). Ce chapitre décrit les méthodes d'obtention du calcul d'inversion d'intégrale d'Abel, et en tenant compte du relief lunaire, après l'analyse des courbes *I = f(h).*

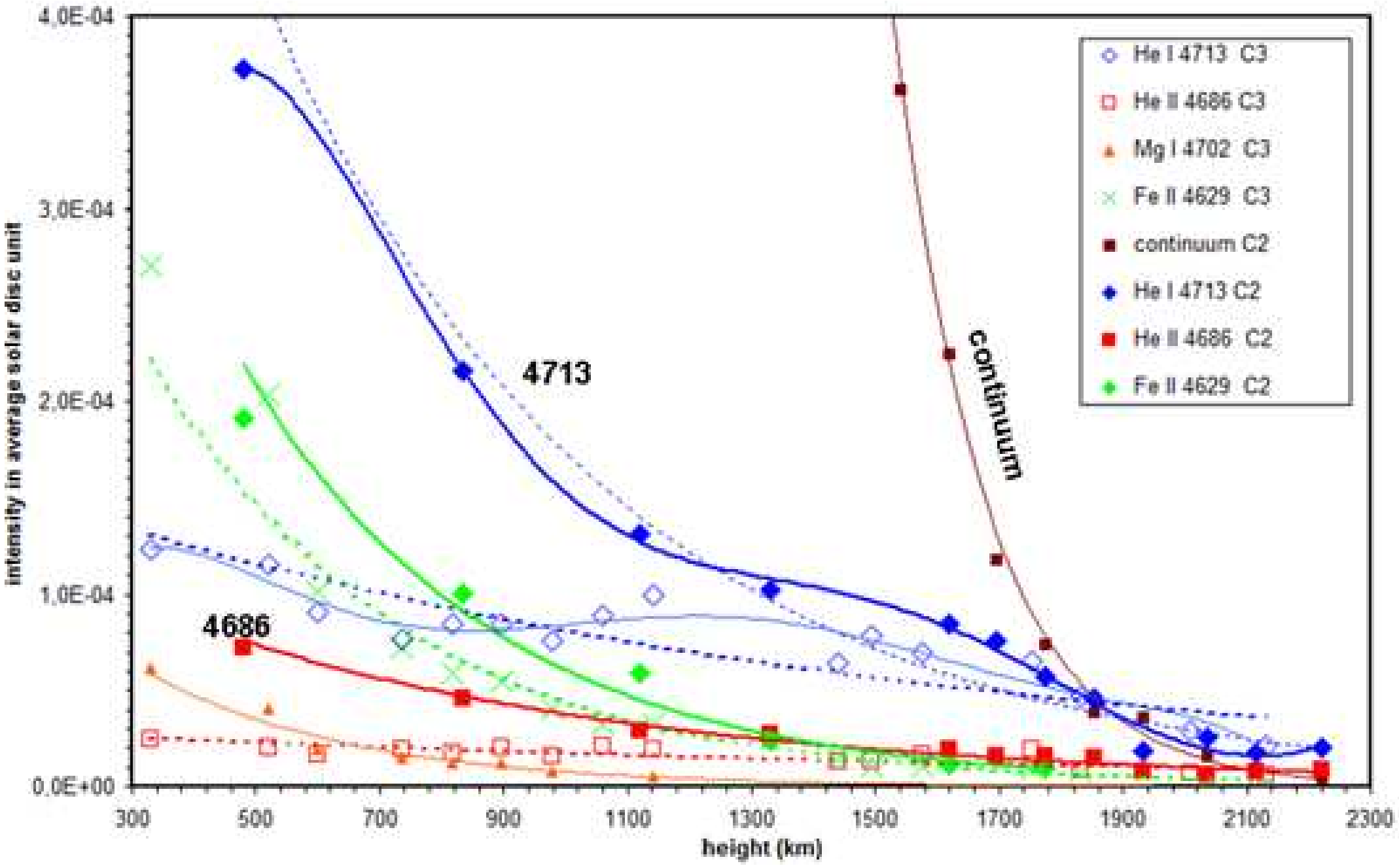

**Figure IV-3-3-1:** *Courbes de lumière I = f(h)  au 2$^{ième}$  et 3$^{ième}$ contacts de l'éclipse de 2010. Les courbes d'ajustement en traits sont un ajustement polynomial, et les courbes en pointillés un ajustement en exponentielles décroissantes. Relevés des valeurs d'intensité moyenne dans une vallée lunaire, en prenant le pixel d'intensité maximale.*

L'analyse des courbes de lumière de l'hélium He II 4686 Å montre des variations et dont les variations montrent une enveloppe entre 1200 et 1700 km. La résulution numérique de l' inversion d'intégrale d'Abel n'a pas été nécéssaire dans ce cas où une interpolation est effectuée sur la distribution des brillances radiales (voir Wildt 1947, page 39, chapitre relations géométriques). Bien que les relevés aient été dispersés autour de la courbe de lumière mesurée d'après les résultats des spectres éclairs, une interpolation polynomiale a été effectuée, et a permis une estimation des altitudes des maxima de brillance des enveloppes. Le maximum d'émissivité de l'hélium une fois ionisé He II 4686Å a été évalué autour de 1600 km, voir figure IV-3-3-2:

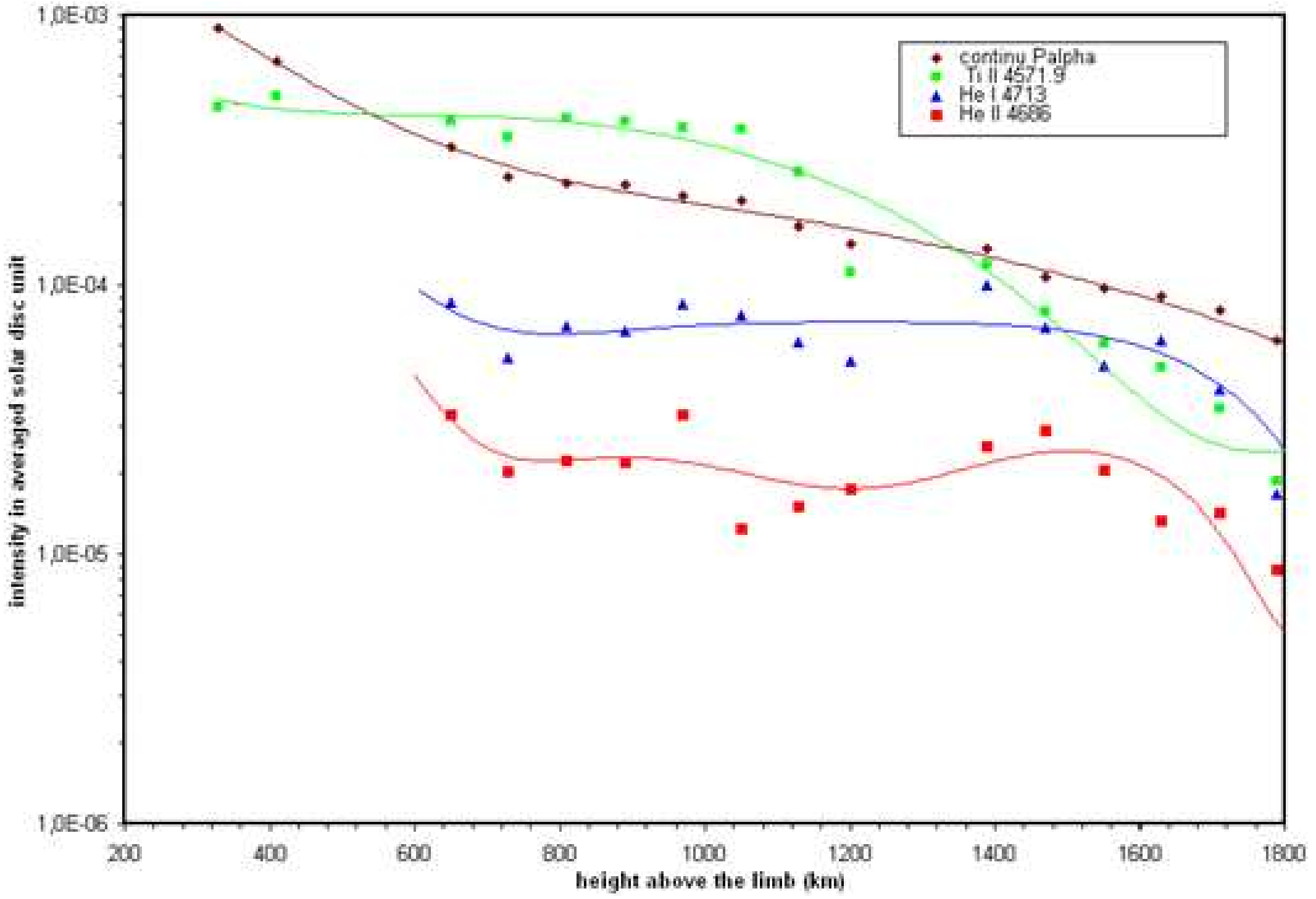

**Figure IV-3-3-2: C***ourbes de lumière I = f(h) au 2$^{ième}$ contact de l'éclipse de 2010. Les courbes d'ajustement sont de type polynomial. Relevés des valeurs d'intensité dans une vallée lunaire, en prenant le pixel d'intensité maximale.*

Ces courbes en échelle Logarithmique permettent une évaluation plus précise des positions des altitudes des maxima d'émissivité qui traduisent des effets d'enveloppes pour les raies d'hélium mais aussi des fluctuations sont constatées sur la raie du Fe II 4629, tandis que le profil du continu est plus régulier. Les courbes du second contact sont plus significatives pour les raies d'hélium car les valeurs d'intensités sont plus élevées. Cependant, après avoir appliqué une courbe d'ajustement sur ces points, il apparaît que le maximum d'émissivité de l'hélium neutre He I 4713 Å se situe vers 1300 km au troisième contact. Une « bosse » est aussi visible au second contact vers 1550 km pour les 2 raies He I 4713Å et He II 4686Å.
L'intérêt de ces analyses sur les raies « low FIP » est d'avoir analysé une raie comme le Fe II 4629 Å et aussi une raie du Mg I 4702.8 Å qui a été observée en absorption sur la figure III-2-3 du Chapitre III-2, et où cette même raie a été étudié sur la courbe de lumière en figure IV-3-3-1 et dont l'émissivité est déduite en figure IV-3-3-5.
Des analyses de fluctuations sont réalisées sur certains relevés pour évaluer les barres d'erreurs. Le tableau IV-3-3-3 décrit les valeurs de flux et les précisions évaluées pour les courbes de lumière de la figure IV-3-3-1 et IV-3-3-2.

| Raie étudiée longueur d'onde (Å) | Précision sur les mesures d'émissivités autour de h = 1100 km au dessus du limbe solaire (sauf pour le continu) |
|---|---|
| He I C2 4713 | $1.7*10^{-4} \pm 1.2*10^{-5}$ unités du disque solaire moyen |
| He II C2 4686 | $3.0*10^{-5} \pm 1.2*10^{-5}$ unités du disque solaire moyen |
| Fe II C2 4629 | $8.0*10^{-5} \pm 1.0*10^{-5}$ unités du disque solaire moyen |
| Continu C2 (1900 km) 4700 | $6.0*10^{-5} \pm 1.0*10^{-5}$ unités du disque solaire moyen |
| He I C3 4713 | $1.3*10^{-4} \pm 1.3*10^{-5}$ unités du disque solaire moyen |
| He II C3 4686 | $2.0*10^{-5} \pm 1.3*10^{-5}$ unités du disque solaire moyen |
| Fe II C3 4629 | $4.0*10^{-5} \pm 1.1*10^{-5}$ unités du disque solaire moyen |
| Mg I C3 4702 | $5.0*10^{-6} \pm 0.8*10^{-5}$ unités du disque solaire moyen |
| Continu C3 | Difficultés des relevés |

**Tableau IV-3-3-3:** *relevé des brillances en unités du disque solaire moyen et leurs précisions relevées sur la courbe de lumière figure IV-3-3-1 pour les raies de l'hélium, du fer du magnésium et du continu, au second et troisième contact lors de l'éclipse totale de Soleil du 11 Juillet 2010.*

Le continu au troisième contact a été difficile à relever, car les intensités étaient plus faibles qu'au second contact. Une fois les courbes obtenues les barres d'erreurs sont estimées par la dispersion des points, puis les courbes ajustées avec une exponentielle décroissante, voir figures IV-3-3-4 et IV-3-3-5.

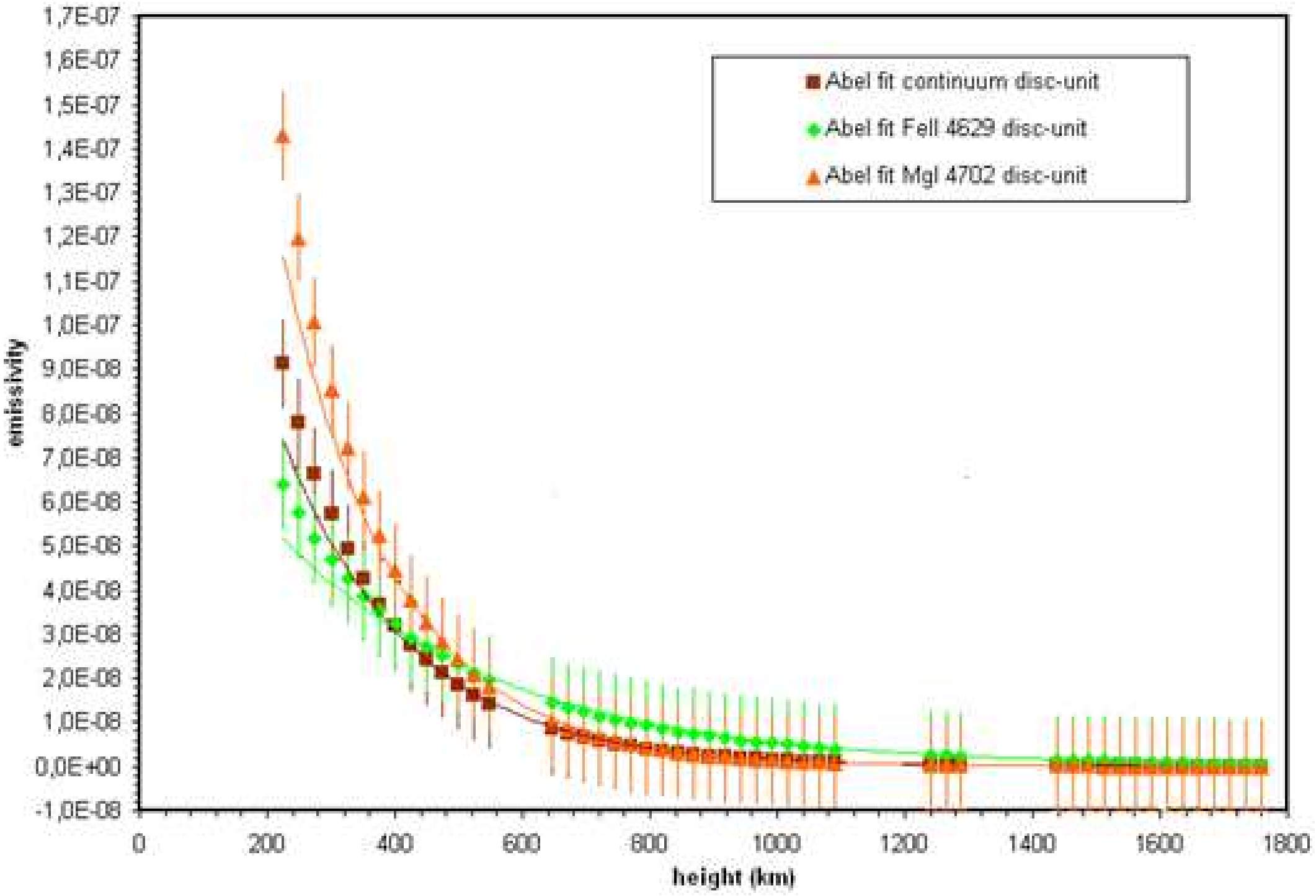

**Figure IV-3-3-4:** *Tracés des inversions d'intégrale d'Abel à parir des* ***relevés d'intensités moyennes dans les vallées lunaires*** *en échelles linéaire pour les brillances dans le continu au troisième contact C3 de l'éclipse du 11 Juillet 2010 et la raie du fer une fois ionisée. Les courbes d'ajustements sont des exponentielles décroissantes, et les nuages de points permettent d'évaluer la dispersion des mesures.*

Les relevés des intensités ont été effectués sur une étendue de plusieurs pixels correspondant à l'étendue d'une vallée lunaire qui correspondent à une surface étendue de plusieurs milliers de km sur le bord du Soleil.
Ces courbes, en décroissances exponentielles, montrent des pentes différentes en échelle logarithme, ce qui suggère des températures plus élevées pour l'ion Fe II que pour le Mg I et le continu.

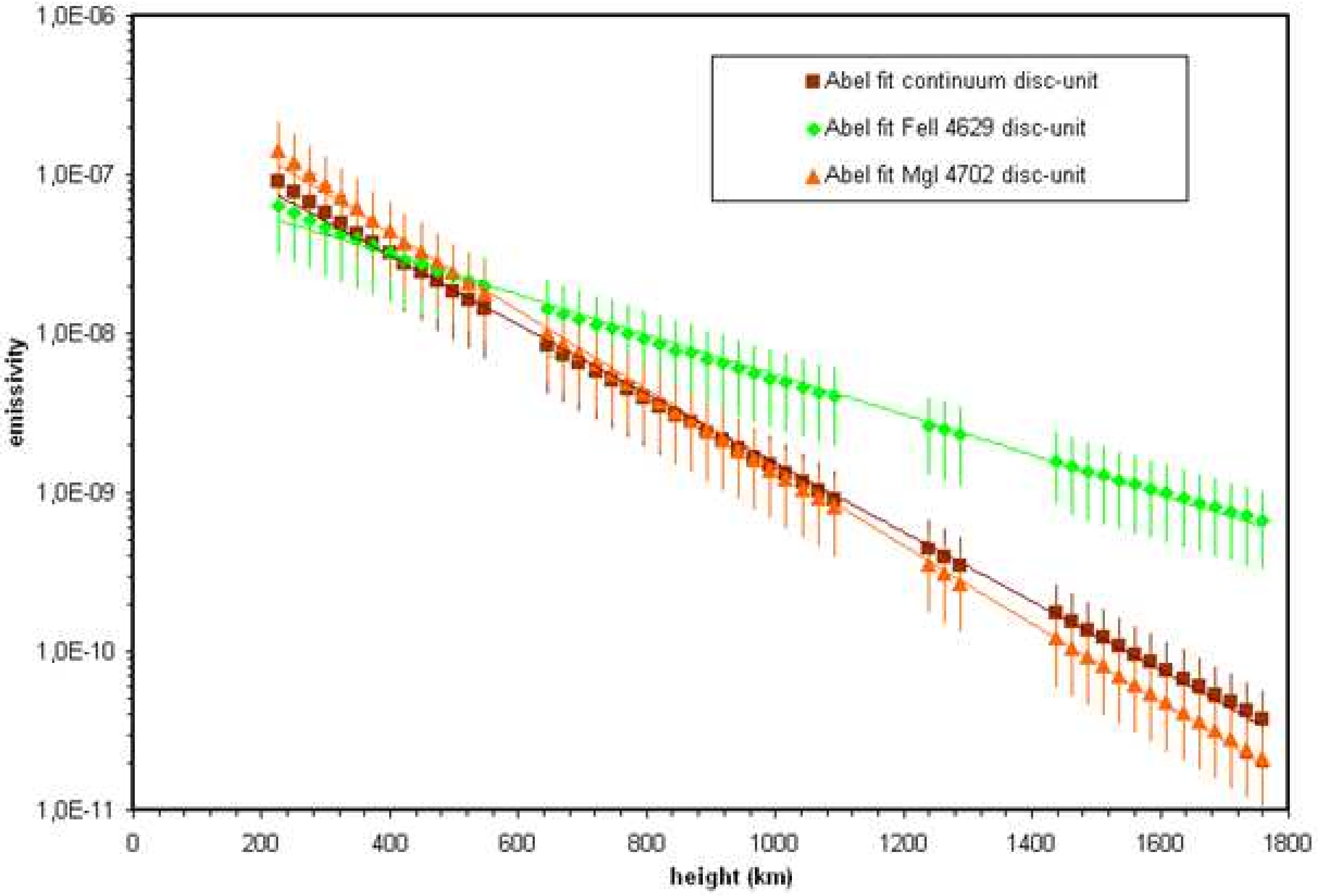

**Figure IV-3-3-5:** *Tracés des inversions d'intégrale d'Abel à partir des équations d'ajustement des courbes de lumière (relevé d'intensité moyenne dans une vallée lunaire). Echelles logarithmique. Brillances dans le continu au troisième contact C3 et des raies du Fe II et Mg I. Les courbes d'ajustements sont des exponentielles, et les nuages de points permettent d'évaluer la dispersion des mesures.*

Les premiers points à 200 km ont été difficiles à évaluer, pour une altitude aussi basse.
La méthode d'estimation des altitudes aux contacts d'éclipses reste une taché difficile.
L'inversion d'intégrale d'Abel a permis de tenter d'évaluer les flux aux basses couches de l'atmosphère solaire immédiatement au dessus de la photosphère.
Les barres d'erreurs proviennent des estimations des fluctuations réalisées sur les courbes de lumière avant inversion d'Abel.
Une fois les courbes d'ajustements de type « exponentielle décroissante » effectués, il faut prendre le terme constant devant la variable *h* dans l'exponentielle, et prendre son inverse, exprimé en kilomètres. Cela permet de déterminer les échelles de hauteurs en première approximation. Les tableaux suivants donnent les relevés des équations des courbes d'ajustement et les valeurs des échelles de hauteurs. Les incertitudes sont déduites de la dispersion des points de mesures. L'unité d'émissivité, indiquée dans les graphiques figures IV-3-3-5 et IV-3-3-6, équivaut à une unité de disque solaire moyen par kilomètre, comme mentionné à la figure IV-1-5 du chapitre IV-1.

## VI – 3-4) Détermination des échelles de hauteur d'après les inversions d'Abel du continu et de la raie du Fer II au second contact de l'éclipse du 11 Juillet 2010

Le tableau IV-3-4-1 donne les équations de courbe d'ajustements sur l'inversion d'intégrale d'Abel du continu et Fe II et les échelles de hauteurs sont déduites à partir du terme dans l'exponentielle:

| | Equation fit inversion d'Abel | Echelles hauteur estimées |
|---|---|---|
| Continu 4700 Å (éclipse 2010) | $E(h) = 2.279195*10^{-7}*e^{-0.005003543h}$ | **199.8 ± 30 km** |
| Continu 4815 Å (éclipse 1970) | $E(h) = 2.324851*e^{-0.0048h}$ | 208 ± 60 km |
| Fe II 4629 Å (éclipse 2010) | $E(h) = 9.933578*10^{-8}*e^{-0.002894446h}$ | **345 ± 40 km** (384 avant) |
| Mg I 4702 Å (éclipse 2010) | $E(h) = 4.145651*10^{-7}*e^{-0.005664498h}$ | **176 ± 30 km** |

**Tableau IV-3-4-1**: *équations d'ajustement et des échelles de hauteurs déduites des relevés effectués dans une vallée lunaire au 3$^{ième}$ contact (relevés des aires dans les vallées lunaires) pour le continu et la raie du Fe II 4629 Å. A l'éclipse de 2010.*

Les courbes d'ajustements pour les données de l'éclipse de 1970 sont données en Annexe 13, en ayant utilisé un autre procédé d'ajustement moins précis, mais permettant d'évaluer l'échelle de hauteur pour le calcul d'émissivité dans le continu à 4815 Å.

Les dispersions des points de mesure sur les échelles de hauteur du continu peuvent provenir des voiles nuageux que nous avons eus à l'éclipse de 2010 et qui ont atténué les intensités des raies et du continu.
Il apparaît toutefois que l'échelle de hauteur pour la raie du fer ionisé Fe II 4629Å est plus importante que celle du continu ce qui signifie que malgré son faible potentiel d'ionisation (faible FIP), sa température reste plus élevée que le continu.

Le choix de cette raie du fer ionisé à Fe II 4629 Å est important car c'est une raie froide à faible valeur de potentiel de première ionisation 2.81 eV, inférieur à 10 eV, et est un exemple parmi la myriade de petites raies d'émission, car ces raies de type low FIP sont la signature des couches mésosphériques juste au dessus de la photosphère où se formeraient les spicules, plus précisément leur pieds, et dans ces très basses couches où le champ magnétique est très concentré et dominant. L'abondance de cet élément aussi se retrouve dans les couches plus élevées de la région de transition. Dans le cas d'une atmosphère isotherme, et d'une atmosphère hydrostatique stratifiée et homogène,

$$\boldsymbol{P = P_0 * e^{\frac{-G\mu}{RT}h}}$$

G est l'accélération de pesanteur, $\mu$, le poids moléculaire, $R$ la constante des gaz, $T$ la température, et $P_0$ la pression au point $h=0$

L'échelle de hauteur est définie comme étant la variation de la pression d'une quantité 1/e, hauteur caractéristique. Elle vaut H = $\frac{RT}{\mu G}$ , Pecker Schatzman, 1959.

## VI-3-5) Détermination des échelles de hauteur du continu, des raies de l'hélium et fer ionisé (Fe II) au second contact de l'éclipse du 11 Juillet 2010

Les courbes de lumière figures IV-3-5-1 et IV-3-5-2 ont été obtenues à partir des relevés effectués sur les spectres flash du 11 Juillet 2010. La méthode utilisée pour les relevés sur les courbes de lumière consiste à considérer seulement le pixel dont la valeur d'intensité est maximale sur l'étendue de la vallée lunaire où le profil de la raie est le plus intense limité par la dimension du pixel (1.1x1.1 Mm sur le disque solaire). A partir de ces courbes de lumière, les inversions d'Abel ont été effectuées comme le montrent les figures suivantes, avec des approximations rectangulaires pour les calculs d'intégration :

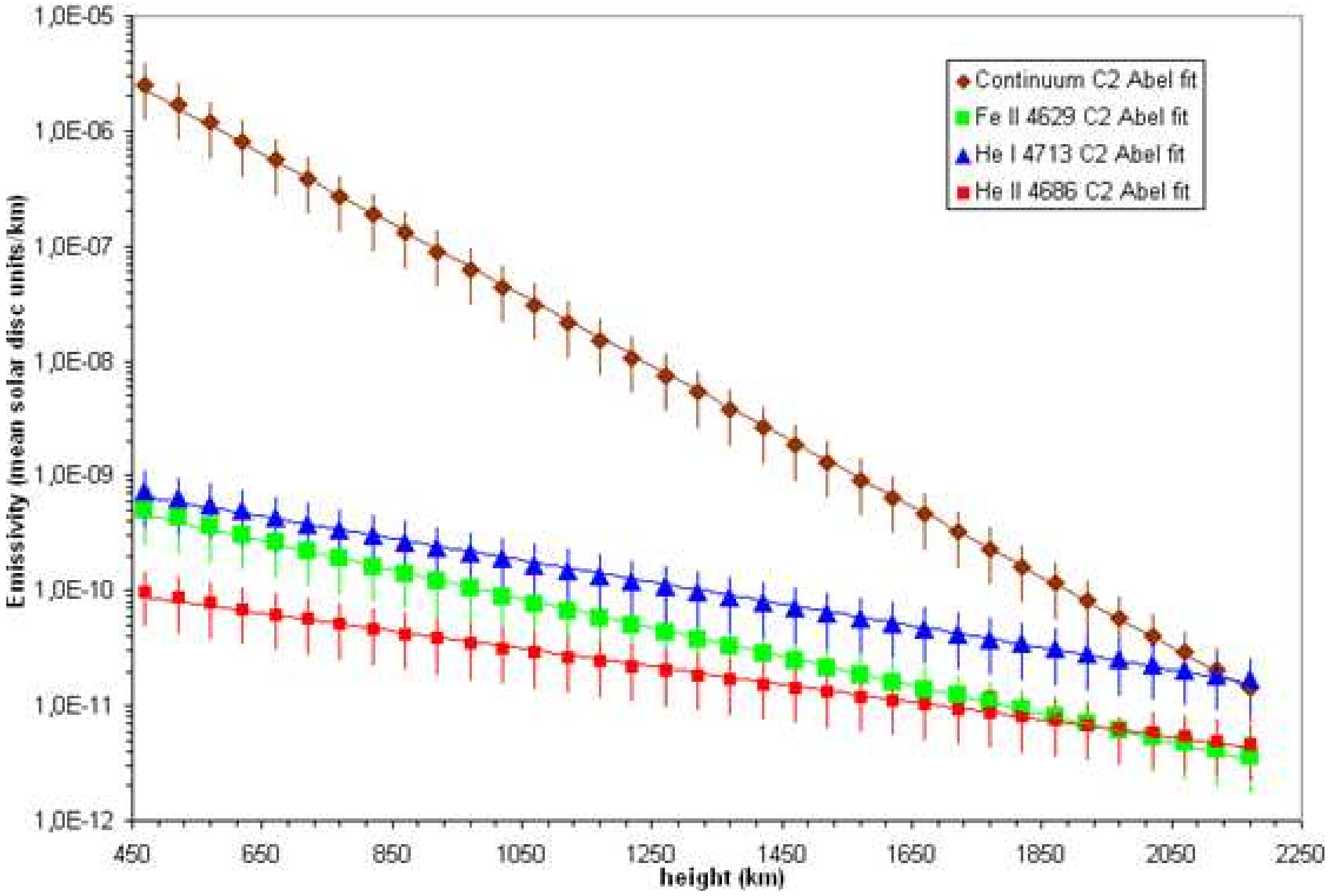

**Figure IV-3-5-1:** *inversions d'Abel des courbes de lumière en échelle logarithme pour déduire la brillance ou émissivité, pour le 2$^{ième}$ contact pour le continu relevé entre les raies, les raies He I 4713Å et He II 4686Å. Relevés de l'intensité maximale (courbe de lumière I(h)) sur 1 seul pixel (cf courbe de lumière figure* IV-3-3-1).

On constate que la pente du continu est plus abrupte que celle de l'hélium neutre 4713 Å, et que la pente de l'hélium ionisé 4686 Å est légèrement plus faible que celle de l'hélium neutre 4713 Å, pour ces mêmes altitudes. Cela traduit l'effet de la température, c'est à dire que la température de l'hélium ionisé est plus élevée que celle de l'hélium neutre. La pente du Fe II est plus importante que celle des raies d'hélium, ce qui montre que la température de cette raie est plus faible que celles de l'hélium optiquement mince. Pour le continu en figure IV-3-5-1, il s'agit de la diffusion des électrons libre liés de HI et associé aussi au continu de Paschen. La différence entre l'émissivité du continu représenté sur la figure IV-3-5-1 et IV-3-3-5, peut provenir de la nature différente entre les 2 contacts avant et après la totalité où, le profil du relief du bord lunaire est différent, le mouvement relatif du limbe lunaire par rapport au bord solaire est différent entre ces 2 contacts aussi parce qu'à cette éclipse, nous n'étions pas exactement au milieu de la bande de totalité. La variation de la pente des profils du continu représentée en échelles logarithmiques, entre 450 km et 900 km diminue ainsi d'un facteur 30 sur la figure IV-3-5-1 tandis que la pente du continu diminue seulement d'un facteur 15 entre 450 et 900 km pour la figure IV-3-3-5. Par contre pour le Fe II 4629Å, les différences sur les pentes sont voisines sur ces mêmes intervalles d'altitudes:
Une diminution d'un facteur 3 sur la figure IV-3-3-5 et une diminution d'un facteur de 2.5 pour la figure IV-3-5-1. Par contre les valeurs d'émissivités sont 100 fois plus faibles sur la figure IV-3-5-1 que sur la figure IV-3-3-5 car les procédés de relevés d'intensités ont été différents sur les courbes de lumière, afin d'obtenir le rapport signal sur bruit le plus élevé possible dans les raies et continu.
Les mesures du continu intégré dans une vallée lunaire réalisées précédemment permettent de mieux estimer ses valeurs et variations en fonction de l'altitude, en commençant à des altitudes de 300 à 400 km au dessus du bord solaire. Le choix de cette méthode de relevé dans un seul pixel 1100x1100 km sur le limbe solaire permet de tenir compte des faibles variations des flux localisées, associées aux brillances, et d'assimiler une vallée lunaire à un seul pixel.
Par ailleurs, nous avons relevé les équations des courbes d'ajustement sur les courbes de lumière. A partir de ces équations, nous avons tracé les profils d'émissivité déduits après inversions d'intégrale d'Abel et indiqué les barres d'erreurs déduites des profils précédents:

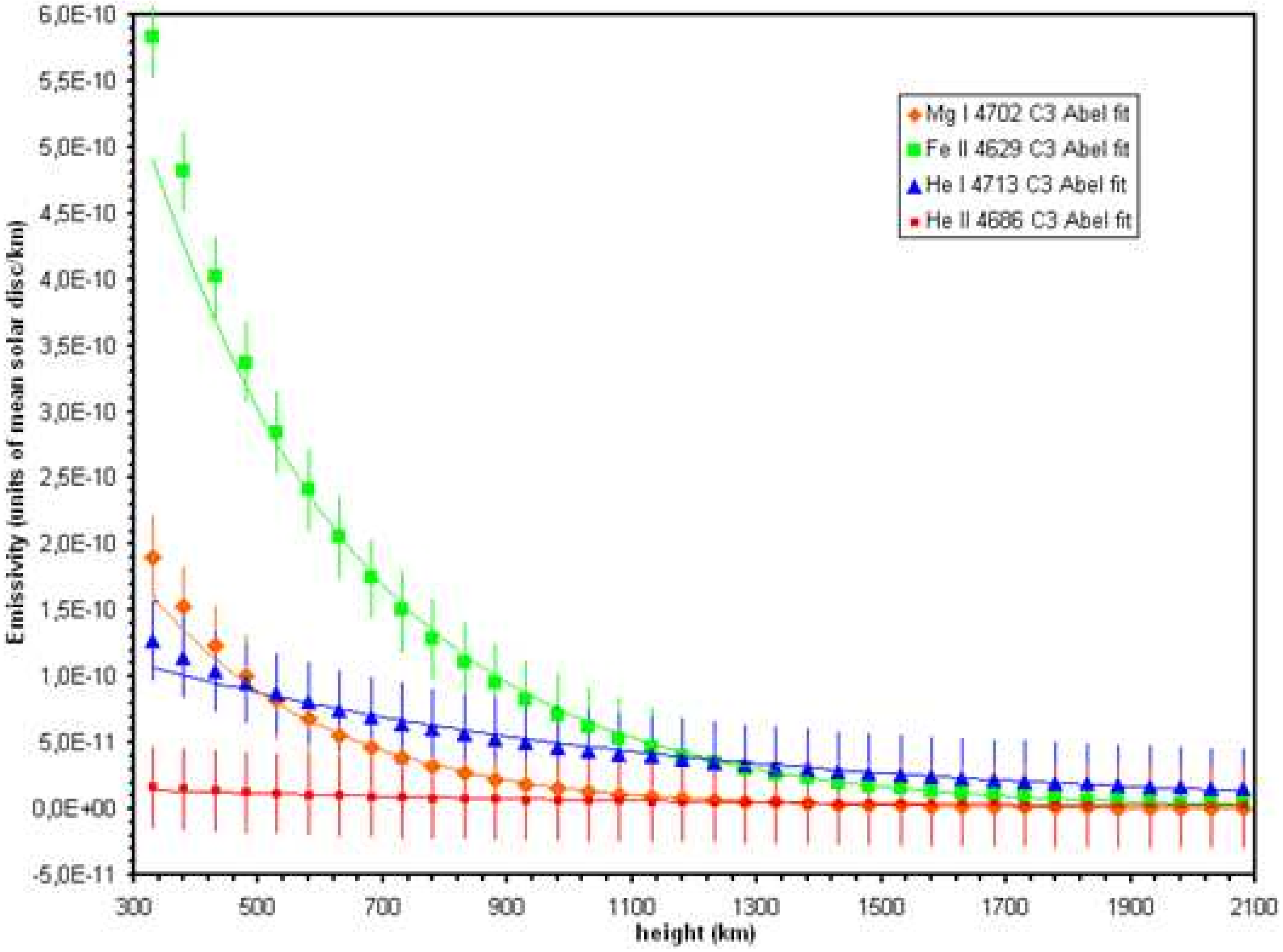

**Figure IV-3-5-2:** *profils d'émissivité ou brillance, de l'hélium neutre 4713 Å, hélium ionisé 4686 Å, Fe II 4629 Å et Mg I 4702Å calculés par inversion d'intégrales d'Abel à partir des équations des courbes d'ajustements exponentiels effectués sur les courbes de lumière déduites des spectres éclairs au troisième contact de 2010. Relevé d'intensité maximale sur 1 seul pixel (cf courbe de lumière figure IV-3-3-1).*

Les graphiques ont été ensuite représentés en échelle logarithme pour comparer plus facilement les valeurs des émissivités aux altitudes plus élevées :

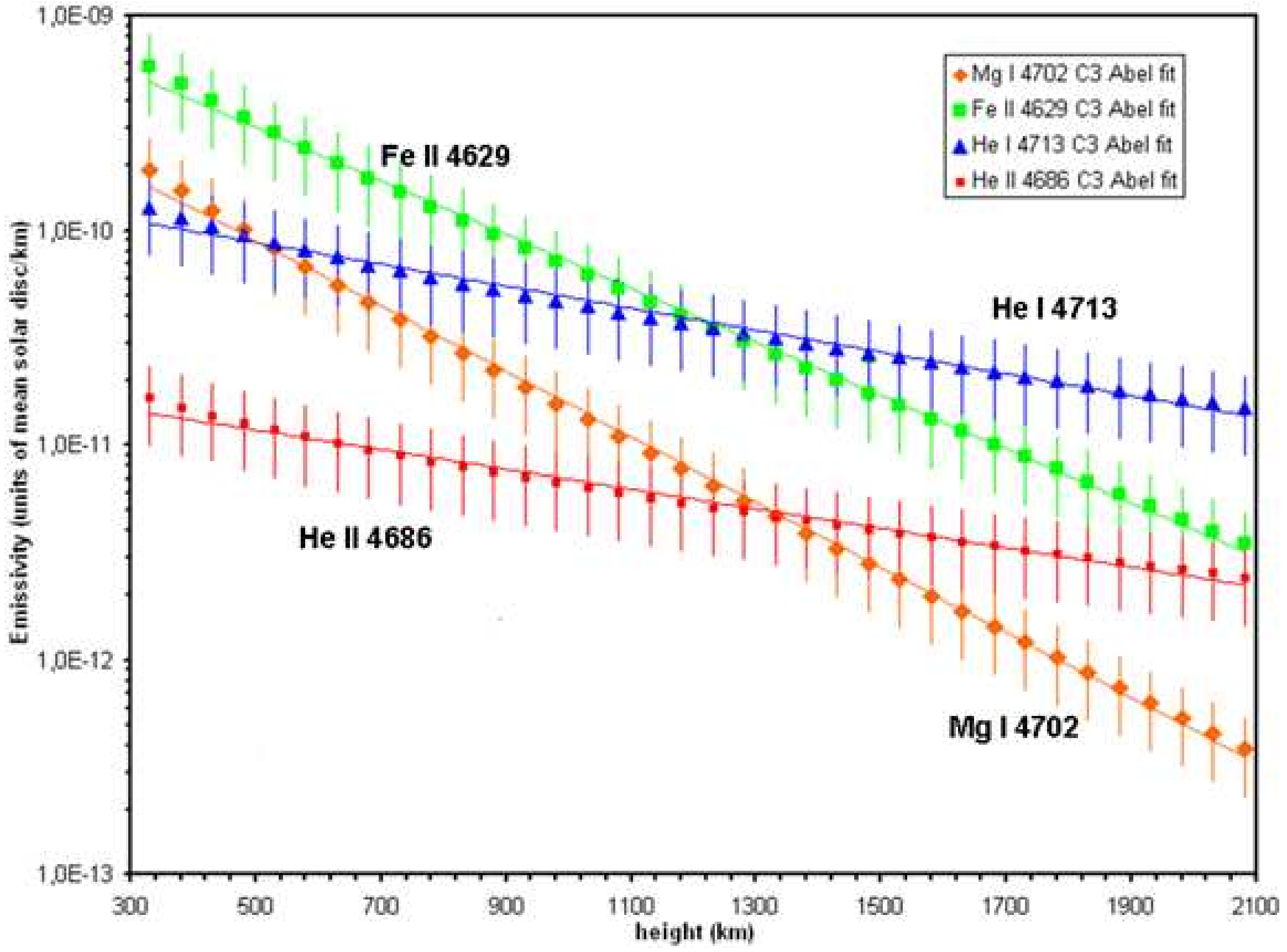

**Figure IV-3-5-3:** *profils d'émissivités en échelle logarithmique, de l'hélium neutre 4713 Å, hélium ionisé 4686 Å, Fe II 4629 Å et Mg I 4702 Å, calculés par inversion d'intégrales d'Abell à partir des équations des courbes d'ajustements effectués sur les courbes de lumière déduites des spectres éclairs au 3$^{ième}$ contact de 2010. Relevés des intensités initiales sur 1 seul pixel (cf courbe de lumière figure IV-3-5-2)*

Les mesures d'émissivité restent toutefois difficiles à ces altitudes aussi basses au dessus de la photosphère. Les gradients sont différents selon la nature de l'ion considéré. Les figures IV-3-5-2 et IV-3-3-4 ont été comparées, et les figures IV-3-5-3 et IV-3-3-5 ont permis une meilleure évaluation des emissivités aux faibles intensités en échelle logarithmique. Le tableau suivant a permis de comparer les valeurs d'émissivités pour les ions Fe II et Mg I à 2 altitudes 300 et 900 km pour comparer les gradients, selon la méthode de mesure indiquées dans les figures IV-3-5-2 et IV-3-3-4.

| Méthode de mesure | h | 300 km | 900 km | Gradients $\frac{I_{300}}{I_{900}}$ |
|---|---|---|---|---|
| $I_{max}$ dans 1 pixels (pris dans 1 vallée lunaire) | Mg I 4702 Å (fig IV-3-5-2) | $1.9*10^{-10}$ | $2.1*10^{-11}$ | 9.0 |
| $I_{max}$ dans 1 pixels (pris dans 1 vallée lunaire) | Fe II 4629 Å (fig IV-3-5-2) | $5.9*10^{-10}$ | $9.5*10^{-11}$ | 6.2 |
| $<\bar{I}>$ dans une vallée lunaire | Mg I 4702 Å (fig IV-3-3-4) | $8.3*10^{-8}$ | $2.3*10^{-9}$ | 34.7 |
| $<\bar{I}>$ dans une vallée lunaire | Fe II 4629 Å (fig IV-3-3-4) | $4.7*10^{-8}$ | $7*10^{-9}$ | 6.7 |

**Tableau IV-3-5-4:** *Comparaison des émissivités pour 2 raies « low FIP » Fe II 4629Å et Mg I 4702Å, selon les méthodes de relevés d'intensité : intensité maximale dans un seul pixel et intensité moyenne dans une vallée lunaire, et en fonction des altitudes pour comparer les gradients.*

Ces résultats montrent que pour le Fe II 4629Å, les gradients sont semblables indépendament de la méthode de mesure. Par contre il y a une différence de l'ordre d'un facteur 100 entre la valeur moyennée dans une vallée lunaire. Pour le Mg I 4702Å, on trouve un gradient pratiquement 4 fois plus élevé avec l'intensité moyennée que pour l'intensité maximale dans un seul pixel, mais les mesures ont été difficiles à réaliser à 900 km au dessus du limbe pour cette raie du Mg I et la valeur de $2.11*10^{-11}$ unités du disque solaire/km est faible. On peut déduire de ces résultats que les gradients du Fe II 4629Å sont plus faibles que ceux du Mg I 4702Å, ce qui suppose que la température de la raie du Fe II est plus élevée que celle du Mg I. Par ailleurs, les pentes du Fe II et Mg I qui sont des raies « low FIP » sont plus élevées et correspondent à des températures plus basses que celles de l'hélium He I et He II qui sont des raies dites « high FIP » sur la figure IV-3-5-3. Les différents ions des éléments « Low FIP » (inférieurs à 10 eV) permettent d'identifier les couches inférieures à 1000 km dans lesquelles ces ions sont formés, ainsi que leur extension dans les régions correspondant au minimum de température.

Une analyse plus approfondie a été réalisée en vue de mesurer plus précisément l'extension des enveloppes d'hélium qui ont des potentiels d'ionisation plus élevés, 24 eV pour He I 4713 Å et 54 eV pour He II 4686 Å. Les mesures des extensions des enveloppes d'hélium ont été réalisées en ayant redressé les spectres éclair, puis ont été analysées pour les différents instants des séquences des contacts. Ces séquences correspondent à des altitudes dans l'atmosphère solaire, situées dans les basses couches de la région de transition aux altitudes supérieures à 400 km au dessus du limbe solaire.

Le tableau IV-3-5-5 résume les équations des courbes d'ajustements déduites des précédents graphiques et inversions d'Abel, pour le continu, le Fe II, hélium neutre et ionisé. Les échelles de hauteur sont déduites à partir du terme dans l'exponentielle.

La présence des raies d'hélium aux altitudes inférieures à 1000 km permet d'apporter de nouveaux diagnostics sur la stratification des enveloppes d'hélium au dessus des altitudes correspondant au minimum de température ($T_{min}$), et que ces raies « high FIP » sont situées au dessus des altitudes proches du $T_{min}$ où les raies « low FIP » sont formées.

| | Equation fit inversion d'Abel | Echelles hauteur estimées |
|---|---|---|
| Continu 2nd contact 2010 | $y = 6{,}210533 \cdot 10^{-5}\, e^{-0.007066255h}$ - | 141 ± 30 km |
| Fe II 4629 Å 2nd contact 2010 | $y = 0.7529 \cdot 10^{-9} \cdot e^{-0.002882739h}$ | 346 ± 60 km |
| He I 4713 Å 2nd contact 2010 | $y=1.832821 \cdot 10^{-9} \cdot e^{-0.00219089h}$ | 456 ± 60 km |
| He II 4686 Å 2nd contact 2010 | $y=1.994827 \cdot 10^{-10} \cdot e^{-0.001766141h}$ | 566 ± 80 km |

**Tableau IV-3-5-5 :** *équations exponentielles des courbes d'ajustements et estimations des échelles de hauteur correspondantes pour les raies de l'hélium et du continu au 2nd contact de l'éclipse du 11 Juillet 2010.*

Ces valeurs montrent que l'échelle de hauteur radiative de l'hélium ionisé He II 4686 Å est 1.2 fois plus importante que celle de l'hélium neutre He I 4713 Å. Ces raies sont sensibles aux raies coronales plus chaudes. Le tableau IV-3-5-6 indique les températures hydrostatiques, où seule la température du continu H I a un sens physique.

| Raie étudiée C2/2010 | longueur d'onde (Å) | masse ion g/moles | Températures hydrostatiques (modèle hydrostatique) |
|---|---|---|---|
| H I | 4700 | 1.0 | **4626 ± 656 K** |
| Fe II | 4629 | 55.8 | 0.37 MK |
| He I | 4713 | 4.0 | 59850 K |
| He II | 4686 | 4.0 | 74280 K |

**Tableau IV-3-5-6:** *températures hydrostatiques déduites des échelles de hauteur à partir des courbes d'émissivité des ions de raies low FIP au second contact de l'éclipse totale du 11 Juillet 2010.*

Les valeurs grisées indiquées dans le tableau représentent les températures hydrostatiques calculées pour les ions He I, He II, Fe II. La température calculée d'après les mesures d'échelles de hauteur (voir tableau IV-3-5-5) de l'He II 4686Å et plus importante que celle de He I 4713 Å, ce qui a un sens compte tenu de l'état d'ionisation de l'hélium. Il apparaît que la température de l'ion Fe II est 5 à 6 fois plus élevée que les températures des raies He I et He II. Le sens de ces résultats est difficile à interpréter, il semblerait que la masse ionique puisse avoir une incidence sur la température, mais ceci n'est pas établi car le milieu est dynamique, avec des mouvements de plasma, vitesses.

Le modèle hydrostatique stratifié n'est valable que pour le continu en faisant l'hypothèse qu'il est associé à l'H I pour un calcul de température hydrostatique, et correspondant au continu de la haute photosphère.

Ces résultats montrent que le modèle hydrostatique stratifié est inadapté pour déduire les températures des ions associés aux raies « low FIP » à partir des mesures des échelles de hauteurs radiatives, dans les basses couches de l'interface photosphère- couronne, et juste au dessus de la photosphère.

## VI-3-6) Détermination des échelles de hauteur du continu et raies de l'hélium et raie du Fe II au troisième contact de l'éclipse du 11 Juillet 2010

Les courbes obtenues après l'inversion d'intégrale d'Abel suivantes ont été calculées pour le troisième contact C3 mais le voile nuageux étant devenu plus épais, les mesures ont été plus difficiles à réaliser. Cependant, certains relevés ont pu être effectués à des altitudes plus basses où les intensités des raies low FIP sont plus élevées, afin de mieux comparer les variations des intensités des raies dans les couches plus profondes.
On retrouve aussi la présence des raies d'hélium optiquement minces dans les couches aux altitudes inférieures à 800 km, comme au second contact de l'éclipse de 2010, et les mesures sont relativement proches, compte-tenu des conditions de fluctuations météorologiques.
Des relevés ont été effectués pour le $3^{ième}$ contact et présentés dans le tableau ci-dessous:

| | Equation fit inversion d'Abel | Echelles de hauteur radiatives estimées |
|---|---|---|
| Fe II 4629 Å $3^{ième}$ contact 2010 | $y = 1.2691*10^{-9}*e^{-0.002879952h}$ | 347 ± 70 km |
| Mg I 4702 Å $3^{ième}$ contact 2010 | $y = 5.093788*10^{-10}*e^{-0.003498937h}$ | 285 ± 70 km |
| He I 4713 Å $3^{ième}$ contact 2010 | $y=1.581236*10^{-10}*e^{-0.001177199h}$ | 849 ± 80 km |
| He II 4686 Å $3^{ième}$ contact 2010 | $y= 0.976206*10^{-11}*e^{-0.001048976h}$ | 953 ± 90 km |

**Tableau IV-3-6-1 :** *équations des courbes d'ajustement exponentielles et des mesures d'échelles de hauteurs pour quelques raies au $3^{ième}$ contact de l'éclipse du 11 Juillet 2010.*

Des différences de mesures d'échelles de hauteurs sont constatées entre les mesures prises dans les aires des vallées lunaires et prises sur un seul pixel. Ces différences de l'ordre d'un facteur 2 entre les échelles de hauteurs de l'hélium neutre et ionisé (high FIP) au second et troisième contact (tableaux IV-3-5-4 et IV-3-6-1) peuvent s'expliquer à cause de la méthode de dispersion utilisée lors des contacts, qui était différente. Le second contact présentait une dispersion avec un angle de 26° comme le montre la figure IV-3-6-2 suivante :

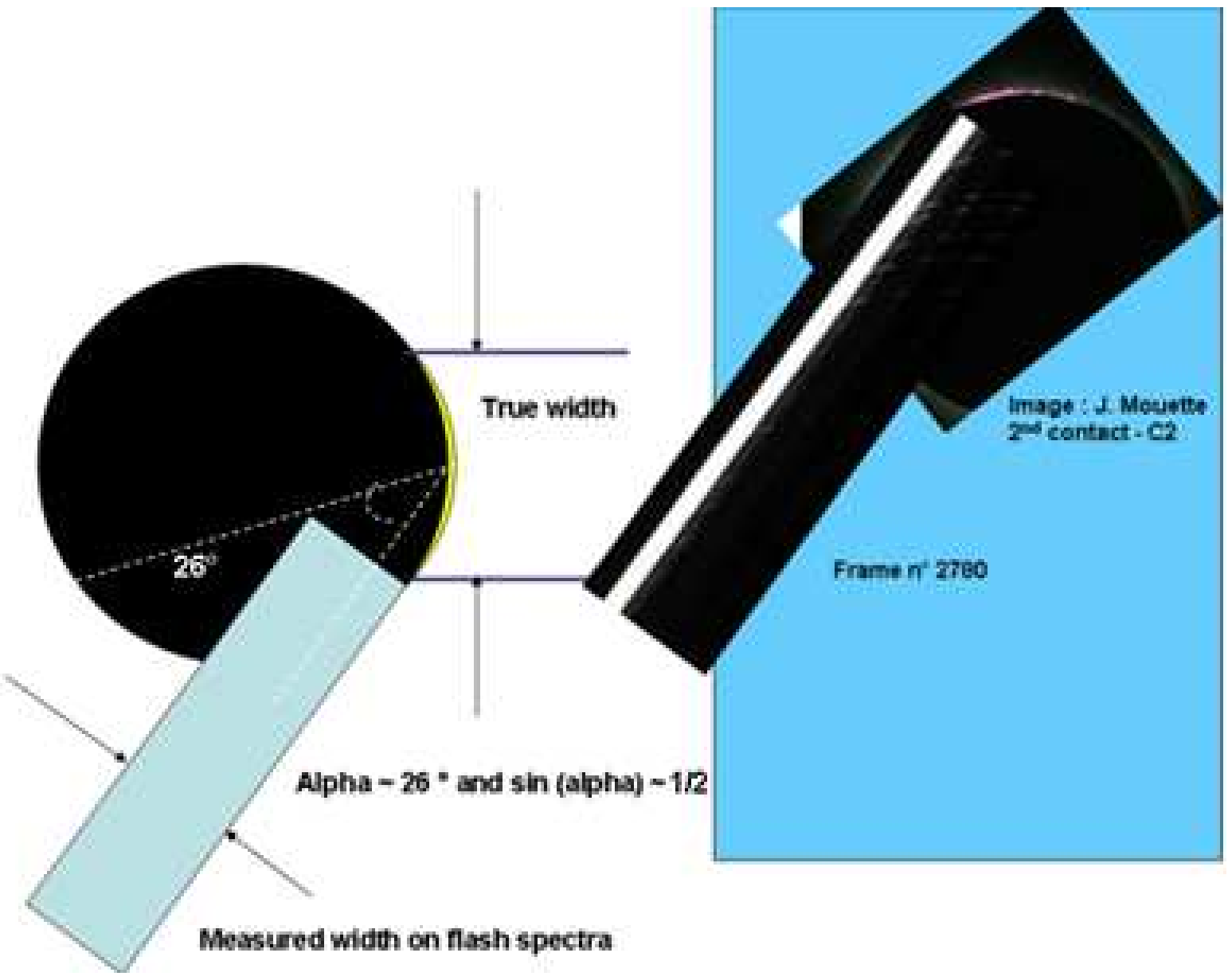

**Figure IV-3-6-2 :** *montage indiquant le sens de dispersion des spectres éclairs au second contact, avec orientation du sens de dispersion faisant un angle de 26° avec l'axe du croissant solaire occulté par la lune lors du second contact de l'éclipse totale du 11 Juillet 2010.*

Cette indication permet de dire que les largeurs des spectres des croissants sont sous estimées d'un facteur de l'ordre de ½ à cause de l'angle de 26° entre l'axe de dispersion spectral, et le croissant photosphérique lors du second contact C2. Au troisième contact C3, le sens de dispersion et l'axe du croissant étaient confondus, ce qui n'a pas entraîné d'effets d'inclinaison des raies à C3. Ces remarques peuvent expliquer les différences sur les spectres éclairs avec les inclinaisons des raies à C2 et pas d'inclinaison à C3. Nous n'étions pas situés au milieu de la bande de totalité, où les enveloppes ont été observées sur une durée plus longue à C3 qu'à C2, et dans des régions plus proches des régions du pôle sud solaire à C2 plus proche de l'équateur à C3.
La figure III-4-1 du sous chapitre III-4 montre en effet, que le second contact avait eu lieu entre l'équateur est et le pôle sud, tandisque le troisième contact avait eu lieu plus proche de l'équateur ouest. Ces régions différentes traduisent aussi des phénomènes d'occultations du bord solaire par le bord lunaire différentes. Cependant cela n'a pas eu d'incidences sur l'échelle de hauteur radiative du Fe II 4629Å, car on retrouve la même valeur de 346 km entre C2 et C3.
Un autre aspect concerne la topologie différente du relief lunaire entre C2 et C3, où les figures IV-3-1-3 et IV-3-1-4 représentent le bord solaire modulé par le relief lunaire durant toute les phases des contacts. Les grains de Baily semblent moins nombreux au second contact qu'au troisième contact. Cela peut entraîner des variations d'intensités plus importantes au troisième contact, dont la durée a été plus longue que celle du second contact, et conduire à des différences d'un facteur 2 sur les valeurs des échelles de hauteur pour les raies d'hélium.
Les fluctutations atmosphériques du ciel voilé entre le second et le troisième contact peuvent avoir changé, mais sur une moins grande ampleur. La topologie du relief lunaire représente un paramètre de plus grande influence, car selon une vallée large et peu profonde et une vallée étroite et profonde des différences d'échelles de hauteurs avaient été constatées, voir figure IV-1-1, IV-1-2, et tableau IV-1-3. Ces différentes irrégularités du profil du limbe lunaire qui n'étaient pas les même entre C2 et C3 peuvent expliquer ces différences sur les échelles de hauteurs mesurées pour les raies He II 4686Å et He I 4713Å.

Les différences constatées pour les autres échelles de hauteurs concernant les raies low FIP peuvent aussi s'expliquer selon les points évoqués précédemment.
Le tableau IV-3-6-3 donne les températures hydrostatiques déduites du tableau IV-3-6-1 pour discuter sur le sens des interprétations en termes de températures.

| Raie étudiée C3/2010 | longueur d'onde (Å) | masse ion g/moles | Températures hydrostatiques (modèle hydrostatique) |
|---|---|---|---|
| Fe II | 4629 | 55.8 | **0.36 MK** |
| Mg I | 4702 | 24.3 | 0.22 MK |
| He I | 4713 | 4.0 | 0.11 MK |
| He II | 4686 | 4.0 | 0.12 MK |

**Tableau IV-3-6-3:** *températures hydrostatiques déduites des échelles de hauteur à partir des courbes d'émissivité des ions de raies low FIP.*

Ces températures pour les ions associés aux raies « low FIP » n'ont pas de sens dans ces régions proches du minimum de température. Les valeurs grisées indiquées dans le tableau représentent les températures hydrostatiques calculées pour les ions He I, He II, Fe II, Mg I à partir des échelles de hauteur radiatives.
Ces résultats sont très éloignés des valeurs de gradients mesurés dans les modèles empiriques de la région de transition de Mariska 1992, voir Annexe 10 figures A-10-7 et A-10-8. En effet, l'hypothèse stationnaire est utilisée. En réalité il y a des vitesses, mais le modèle des vitesses est inconnu, même en essayant d'introduire une pression « dynamique » $P_{dyn}$ proportionnelle à la vitesse $V$ : $P_{dyn} = mV$, où $m$ est la masse d'un ion. En effet, ces basses couches de l'atmosphère solaire sont composées de flots de plasma le long des lignes de champ magnétique, dont certains sont montants, d'autres descendants, et tout est dynamique, et c'est difficile à modéliser. Bien que le modèle hydrostatique stratifié soit utilisé, il est inadapté pour déduire les températures des ions associés aux raies « low FIP » dans les couches profondes de la région de transition, située au dessus de la photosphère.

Ces résultats permettent de discuter sur la structuration complexe des couches profondes de la région d'interface photosphère-couronne, où des échelles de hauteurs de quelques centaines de kilomètres au dessus du bord solaire sont considérées, et où les gradients sont importants. Notamment le bord solaire défini pour le continu est pris comme référence où les gradients sont les plus élevés. Les cas optiquement mince est considéré.
Plus de précisions concernant la définition de cette épaisseur optique pour décrire les basses couches de la haute photosphère sont décrites en Annexe N°24.
Les courbes figure IV-3-6-4 donnent un récapitulatif des différentes inversions d'intégrales d'Abel pour les courbes du continu comparés aux éclipses de 2008, 2009 et 2010. Il est important de noter que les relevés sur les spectres éclairs pour le continu, ont été effectués entre les raies d'émission, ce qui permet de définir le « vrai » bord solaire.

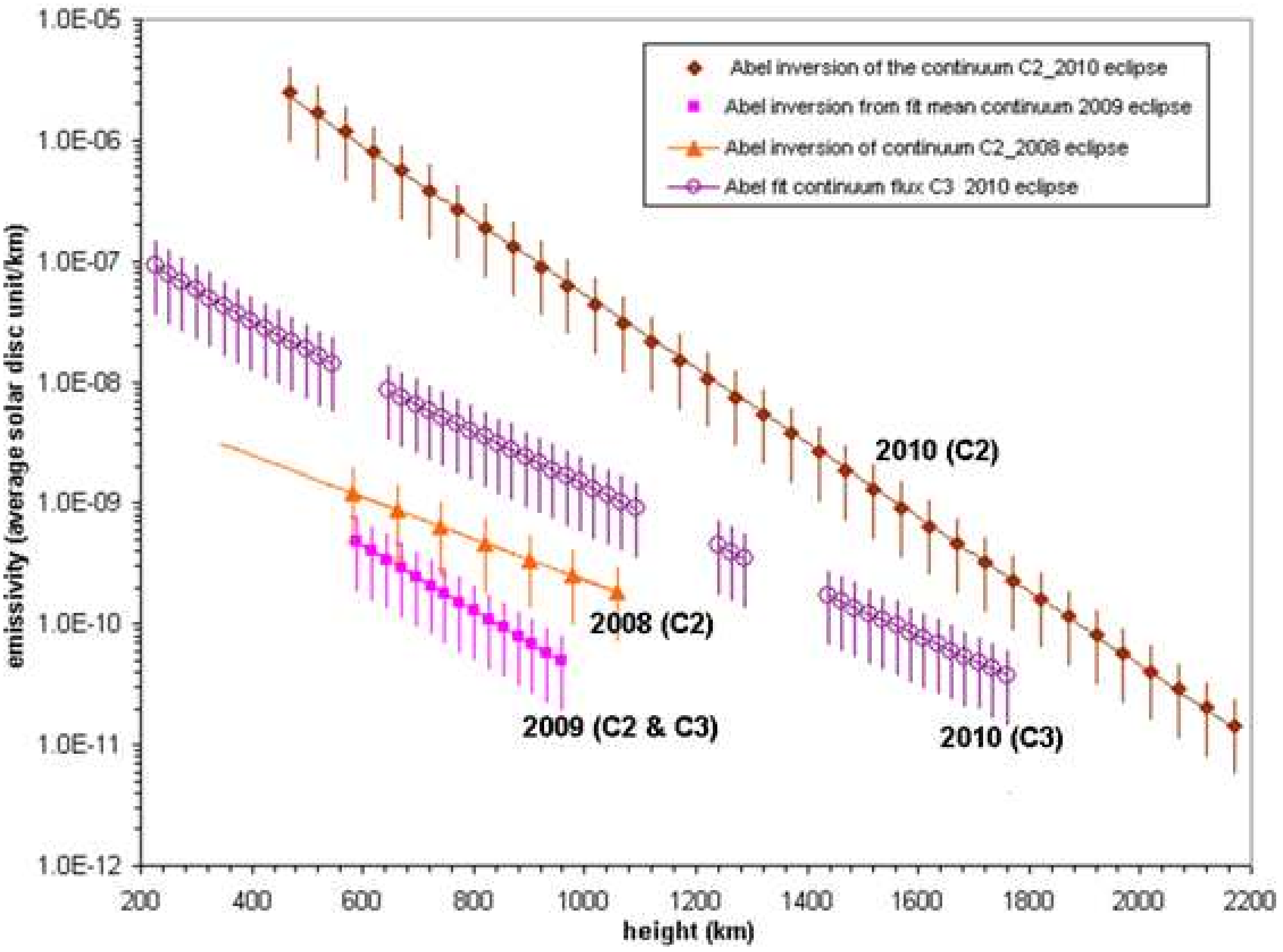

**Figure IV-3-6-4:** *comparaisons des émissivités du continu pour les 3 éclipses de 2008, 2009 et 2010. Après inversion d'intégrale d'Abel des courbes de lumière du continu de chaque éclipse.*

Deux méthodes différentes ont été utilisées pour mesurer les courbes de lumière I = f(h) du continu pour C2 et C3 de 2010, avant d'effectuer l'inversion d'intégrale d'Abel pour déduire les émissivités. La méthode utilisée au second contact a consisté à mesurer l'intensité maximale prise dans un seul pixel dans la valée lunaire, où le brain de Baily était relevé, tandis qu'au troisième contact, l'intensité moyenne a été relevée en prenant en compte l'étendue de la valée lunaire dans le grain de Baily. L'emploi de 2 méthodes différentes se justifie sur l'obtention du rapport signal sur bruit le plus élevé possible.
Les équations des courbes d'ajustements pour les émissivités des continus en fonction des altitudes *h* sont les suivantes et les valeurs des échelles de hauteurs ***H*** sont indiquées:

$E(h) = 1.209257*10^{-8}*e^{-0.003960736\ h}$ pour le continu C2 de 2008 ***H* = 252 km**

$E(h) = 1.8516*10^{-8}*e^{-0.006201321h}$ pour le continu moyen C2 & C3 de 2009 ***H* = 161 km**

$E(h) = 2.279195*10^{-7}*\ e^{-0.005003543\ h}$ pour le continu C3 de 2010 ***H* = 199.8 km**

$E(h) = 6.210533*10^{-5}*\ e^{-0.007066255\ h}$ pour le continu C2 de 2010 ***H* = 141.5 km**

Les échelles de hauteur déduites des différentes courbes d'émissivité conduisent à une valeur moyenne de 188 km. La dispersion des valeurs peut être associée aux irrégularités du relief lumaire, qui « découpe » le bord du Soleil, et où des motifs (valées, montagnes) non résolus provoquent des fluctuations d'intensités dans les spectres des grains de Baily, lors des contacts. Ces dispersions ne viennent pas du « seeing », voir Raponi 2011 et Sigismondi 2011. Il est possible par ailleurs que des effets de scintillation aient lieu et conduisant a des variations de brillance d'un facteur 2, mais ce n'est pas établi.
Cette valeur est relativement élevée par rapport à une valeur théorique (modèles VAL) autour de 100 km qui est attendue pour les couches les plus élevées de la haute photosphère.
L'échelle de hauteur de 141 km est celle qui se rapproche le plus des résultats attendus.

## IV-4) Analyses des profils de la raie verte de l'éclipse du 13 Novembre 2012 et du continu coronal avoisinant, en fonction de la hauteur (issus du spectrographe à fente)

Une autre expérience utilisant un spectrographe à fente a été utilisé lors de l'éclipsse du 13 Novembre 2012, à lafin du second contact juste avant la totalité. Des spectres sur la raie verte du Fe  XIV à 5303 Å ont permis de compléter l'étude de l'interface photosphère-couronne et des mesures photométriques ont été réalisées en vue d'analyser les gradients d'intensité.
Cette raie du Fe XIV est la signature de l'atome de fer ayant perdu 13 électrons, et dont le potentiel d'ionisation correspondant est de 355 eV soit environ 100 fois plus élevé que celui du Fe II 4629 Å , qui a perdu un seul électron. L'intérêt de ces analyses est de traiter le problème de la surabondance relative du fer dans la couronne, qui est établi depuis longtemps, et cette thèse montre que le fer et d'autres éléments « low FIP » comme le titane, sont abondants dans les basses couches de cette interface photosphère-couronne. Les éléments « low fIP » dans les couches profondes de l'interface photosphère-couronne alimenteraient en permanence la couronne en fer et autres ions comme le titane, et le champ magnétique contribue à ce phénomène.
Le but de ce chapitre IV-4 est de donner des résultats d'analyses des spectres  de la raie verte du Fe XIV à 5302.86 Å. Cette raie a été observée dans les couches plus basses de l'interface photosphère-couronne solaire, ce qui est difficilement observable si proche du limbe solaire en dehors des éclipses. Les paramètres des profils de la raie verte sont analysés pour en tirer des renseignements physiques sur la nature de cet embrillancement dans son environnement, et les variations des gradients d'intensité avec l'altitude, en compte l'intégration sur la ligne de visée.

Les profils d'intensité dans la raie verte ont été relevés parallèlement au limbe solaire, et effectués perpendiculairement le long de la raie verte, par tranches d'altitudes croissantes. L'objectif est d'obtenir les profils de la raie verte pour chaque tranche d'altitude, en vue d'étudier la largeur à mi-hauteur de la raie verte en fonction de la distance radiale, la longueur d'onde centrale et maximale de la raie verte  en fonction de la distance radiale. Un profil gaussien a été utilisé plutôt qu'un profil lorentzien, car l'ajustement gaussien correspondait mieux aux profils expérimentaux mesurés sur les spectres de la raie verte. Une fois la position de la fente en partie identifiée, voir chapitre III-5, les analyses en figure IV-4-1 sur les profils de la raie verte et du continu coronal ont été obtenus.  Ces analyses ont pour objectif d'étudier les variations des profils d'intensité de la raie verte avec la distance radiale au-delà du bord solaire, pour ensuite effectuer des analyses sur la nature du plasma, après réduction du continu coronal, dans la seconde région de transition entre la chromosphère et la couronne solaire.

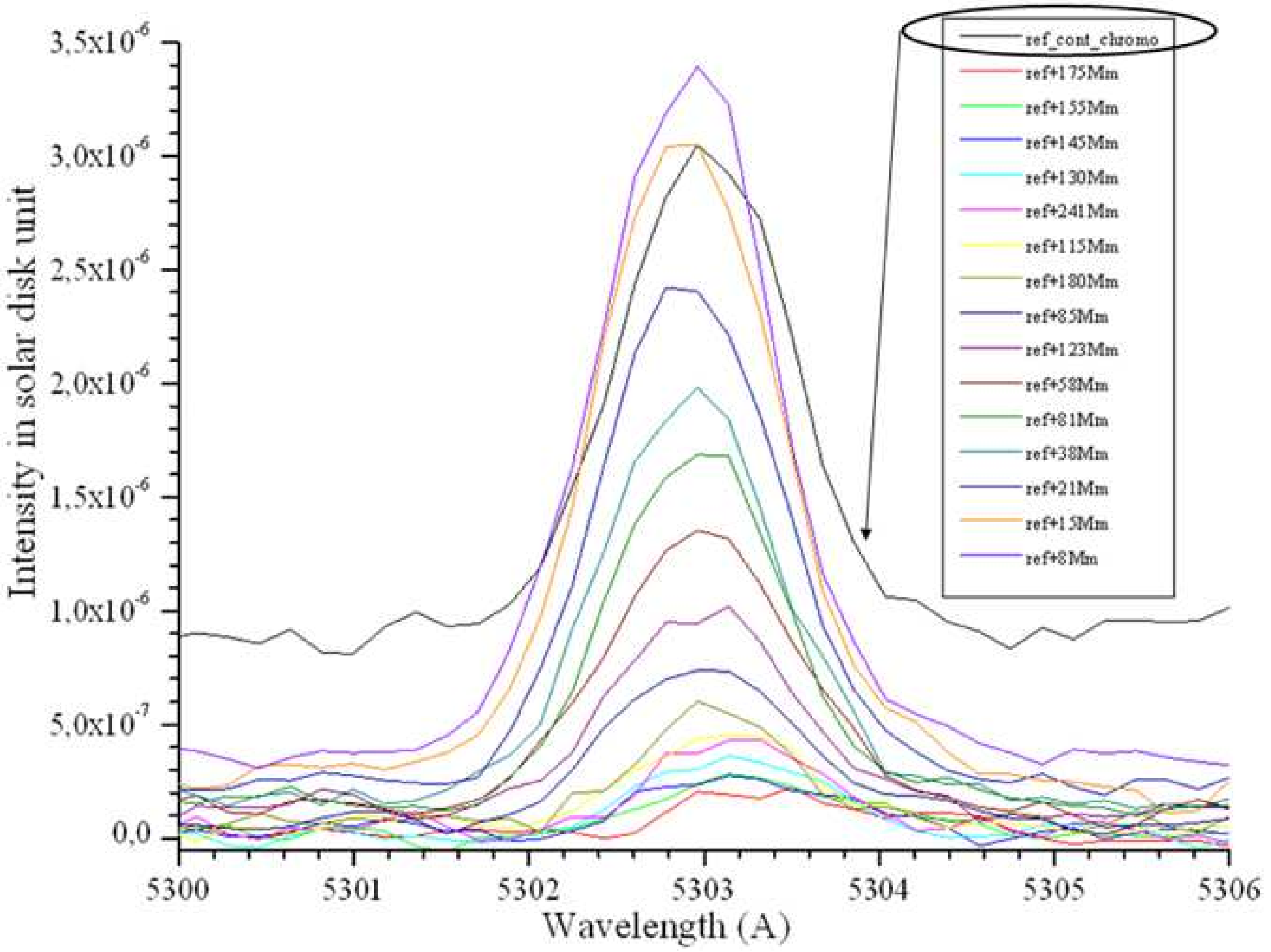

**Figure IV-4-1:** *analyse des FWHM des profils gaussiens de la raie verte relevés radialement, avec l'altitude en ayant pris le continu chromosphérique comme référence.*

Les mesures obtenues sur la figure IV-4-1 ont été effectuées à partir des 3 spectres sommés et alignés pour améliorer le rapport signal sur bruit. Le pas entre chaque profil est de 7 à 15 Mm afin de montrer les variations.
Le continu coronal a été soustrait pour chaque profil. Les étalonnages en unités du disque solaire des profils gaussiens précédents ont été effectués en prenant le modèle des couronnes K (de plasma) et F (de poussières) de Koutchmy-Lamy 1985. Seule le modèle de la couronne K de plasma est prise en compte, et correspond au ions Fe XIV sous forme de plasma et dont nous étudions les intensités. La méthode utilisée a consisté à relever la valeur de l'intensité du disque solaire à 1.1 rayon solaire et où l'intensité correspond à $1.8991.10^{-6}$ unités d'intensité du disque solaire.
Voir diagramme des brillances Koutchmy, S. et al, (1974), Lebecq, C. et al (1985) en Annexes N°28 et N°29.
A cette altitude de 1.1 rayons solaires, le flux mesuré correspond à 577 ADU, et les valeurs d'intensités en ADU ont pu être converties en unités du disque solaire de la façon suivante :

$$I_{Arbitrary_Digital_Unit} = I_{unites_disque_solaire} * \frac{1.8991.10^{-6}}{577}$$

A partir de ces profils, les paramètres FWHM, Intensité maximale, longueur d'onde centrale, et continu peuvent être déterminés pour chaque relevé le long de la raie verte.

Ces profils préliminaires sur la raie verte ont permis de définir ensuite un pas plus étroit et régulier de de 2 Mm afin de réaliser des mesures plus précises de la longueur centrale, du maximum d'intensité. Les variations de la FWHM en fonction de la distance radiale depuis le bord du Soleil  jusqu'à des extensions de 380 Mm ont été évaluées. Ces mesures réalisées avec le spectrographe à fente, permettent une résolution des distances radiales grâce à la fente, tandis que le spectrographe sans fente, utilisant le seul mouvement du bord de la Lune ne permet pas ce type d'analyse. Les 2 méthodes, spectrographe sans fente et spectrographe à fente sont complémentaires. Les analyses effectuées à partir des spectres de la raie verte avec le spectrographe à fente servent à montrer avec une résolution suffisante, l'interpénétration de la couronne chaude (1 à 2 MK), dans la région d'interface chromosphère-couronne solaire, sur la ligne de visée. C'est une analyse importante, car dans les conditions d'éclipse, sans lumière parasite, il est possible d'examiner très près du bord du Soleil, dans les couches les plus basses de son atmosphère, et la fente d'analyse utilisée radialement permet d'obtenir des étalonnages complémentaires des altitudes au dessus du limbe solaire.
Le logiciel Maxim DL a été utilisé pour relever chaque rectangle le long de la raie verte avec un pas régulier de 4 Mm et un chevauchement de 2 Mm pour l'échantillonnage, en respectant le critère de Nyquist.
Le logiciel Origin a été ensuite utilisé pour obtenir les profils gaussiens pour effectuer ensuite les relevés des paramètres à chaque pas de hauteur. Les coupes ont été prises suffisamment loin dans les ailes de la raie verte. A chaque profil de raie verte brut a été soustrait le continu coronal, et ceci a été répété pour chaque altitude. Les flux de la raie verte ont été mesurés sous l'aire des profils de la raie verte une fois réduits, et exprimés en unités du disque solaire moyen.
Cette méthode semble la plus adaptée par rapport à celle qui consisterait à prendre une coupe dans le sens radial, le long de la raie verte, à laquelle on soustrait 2 autre coupes moyennées prises de chaque côté en dehors de la raie verte.

## IV-4-1) Analyses des profils des flux de la raie verte du Fer XIV à 5302.86 Å

La figure IV-1-1 montre un extrait du profil de la raie verte agrandi sur lequel des isophotes ont étés réalisés dans la région de transition pour définir les paramètres d'analyse autour de celle-ci et qui ont servi à choisir les pas en hauteurs:

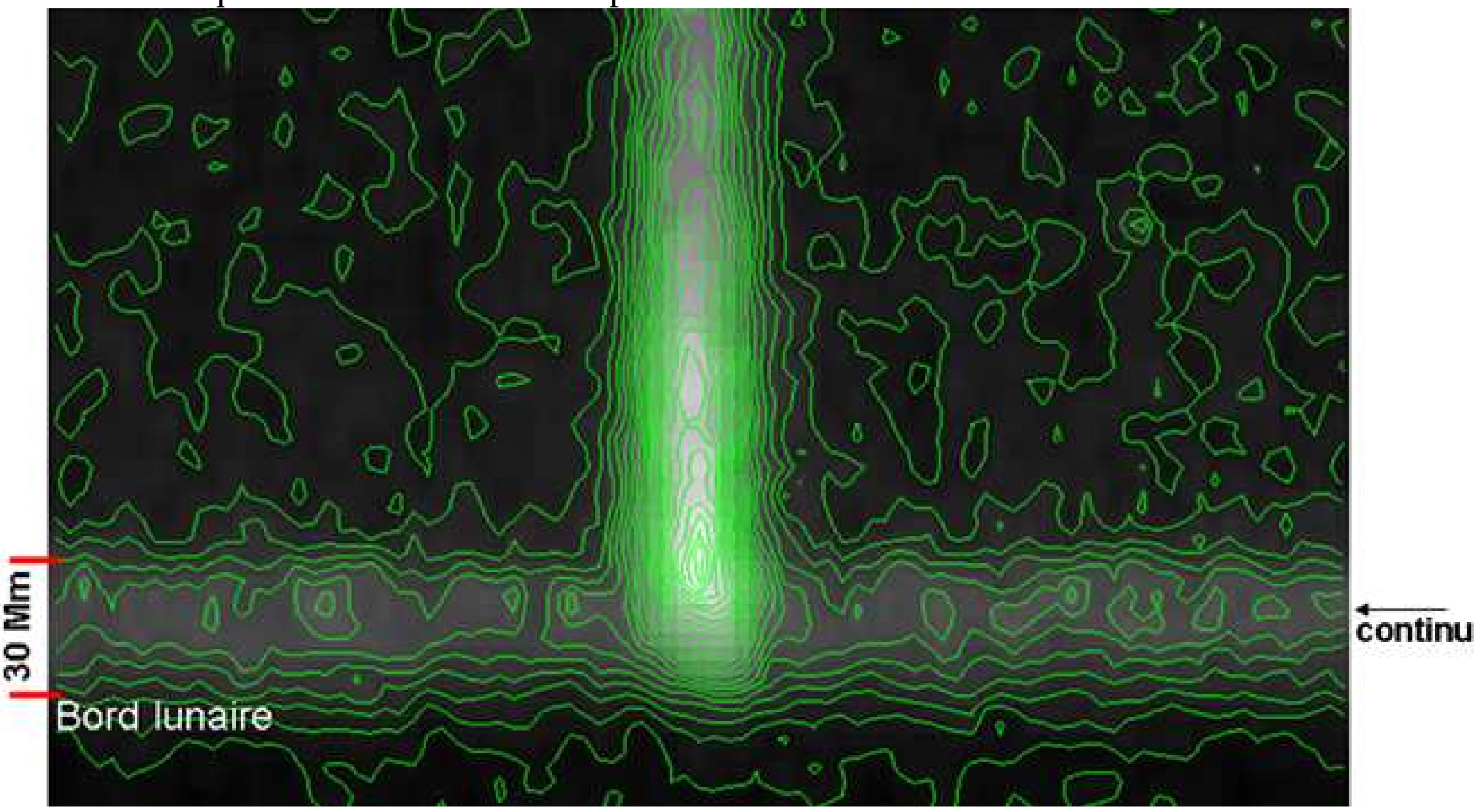

**Figure IV-4-1-1:** *courbes d'isophotes sur un extrait autour de la raie verte à partir du bord de la Lune. La valeur minimale est de 16 adu et la valeur maximale est de 1216 ADU. (smoothness 1). 24 isophotes ont été tracés avec DS9. Remarquer l'importance du continu aux faibles hauteurs.*

Cette analyse montre que la raie verte est observée jusque dans les couches de la chromosphère, et vient se superposer au continu limité par le bord de la Lune.
Le continu observé horizontalement pour cet intervalle spectral autour de 5303 Å peut correspondre à plusieurs contributions:
Le continu chromosphérique dû à HI, le continu Thomson, et le continu de Paschen.
Ces 2 derniers ont des intensités de l'ordre de $10^{-6}$ à $10^{-8}$ unités du disque solaire moyen. Mais comme les spectres ont été enregistrés bien après le second contact de l'éclipse du 13 Novembre 2012, le continu chromosphérique au niveau du bord de la Lune est vu à une altitude d'environ 1600 km au dessus du bord du Soleil, et donc l'intensité des H I est plus faible. Cette intensité est comparable à celle obtenue aux spectres flash pour ces mêmes altitudes, mais aux longueurs d'onde de 4500Å et 4700 Å.

La présence de raies d'absorption est à discuter, car des zones plus brillantes peuvent aussi correspondre à des raies chromosphériques et/ou « low FIP » de plus faible intensité à ces altitudes. 24 pas d'isophotes ont été relevés avec les niveaux d'intensités figure IV-4-1-1:

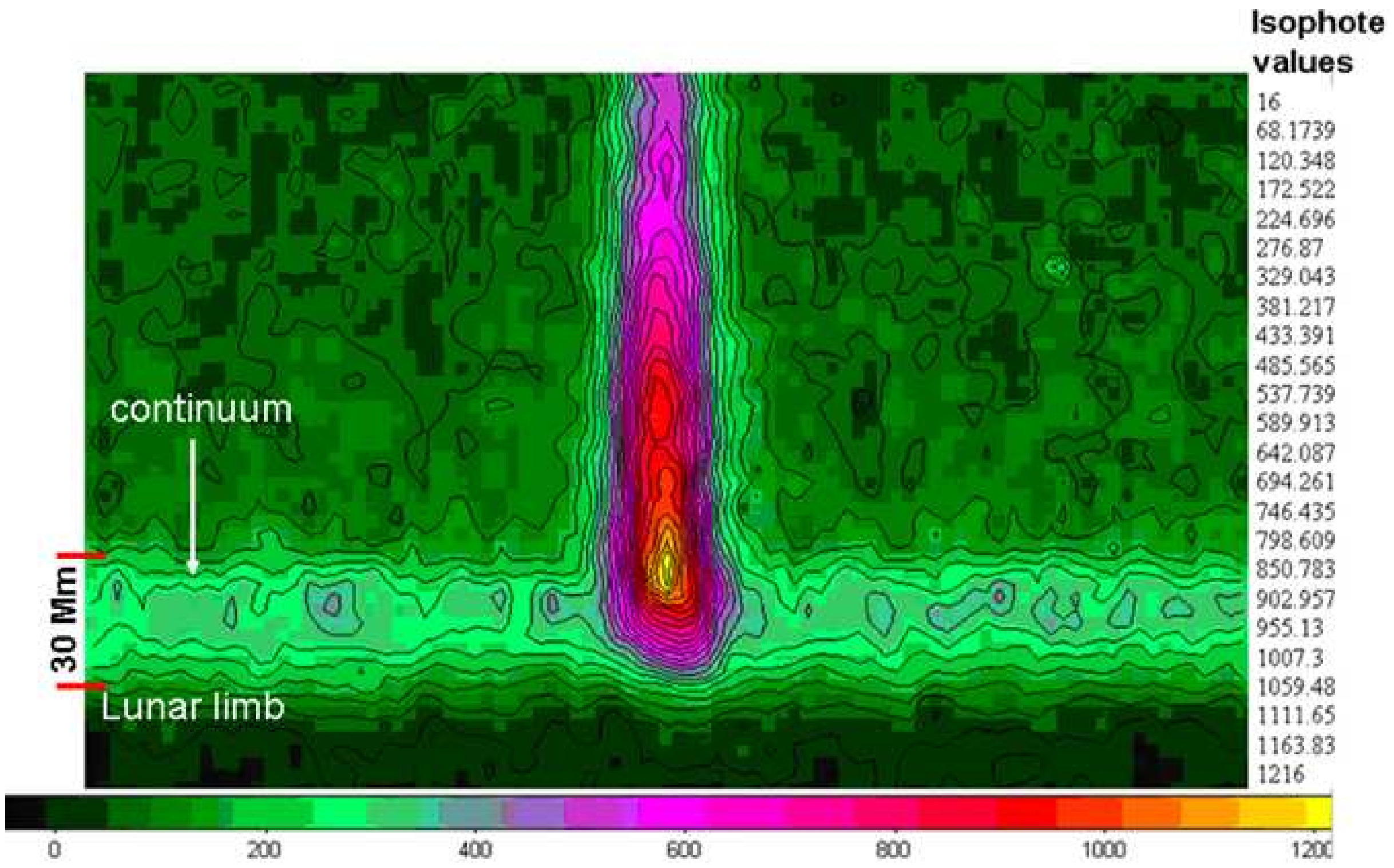

**Figure IV-4-1-2:** *rehaussement des contrastes et des couleurs à partir de la carte d'isophotes précédente, pour mieux montrer les contours et le bord de la Lune. L'échelle de couleurs horizontale indique les niveaux d'isophotes.*

Les gradients sont relativement élevés sur les contours de la raie verte d'après la figure IV-4-1-2, par rapport au continu. A partir de cette représentation avec les isophotes, un pas de 4 Mm est choisi pour tracer les profils pris le long de la raie verte, dans la région de transition chromosphère-couronne. Ces profils sont représentés sur IV-4-1-3:

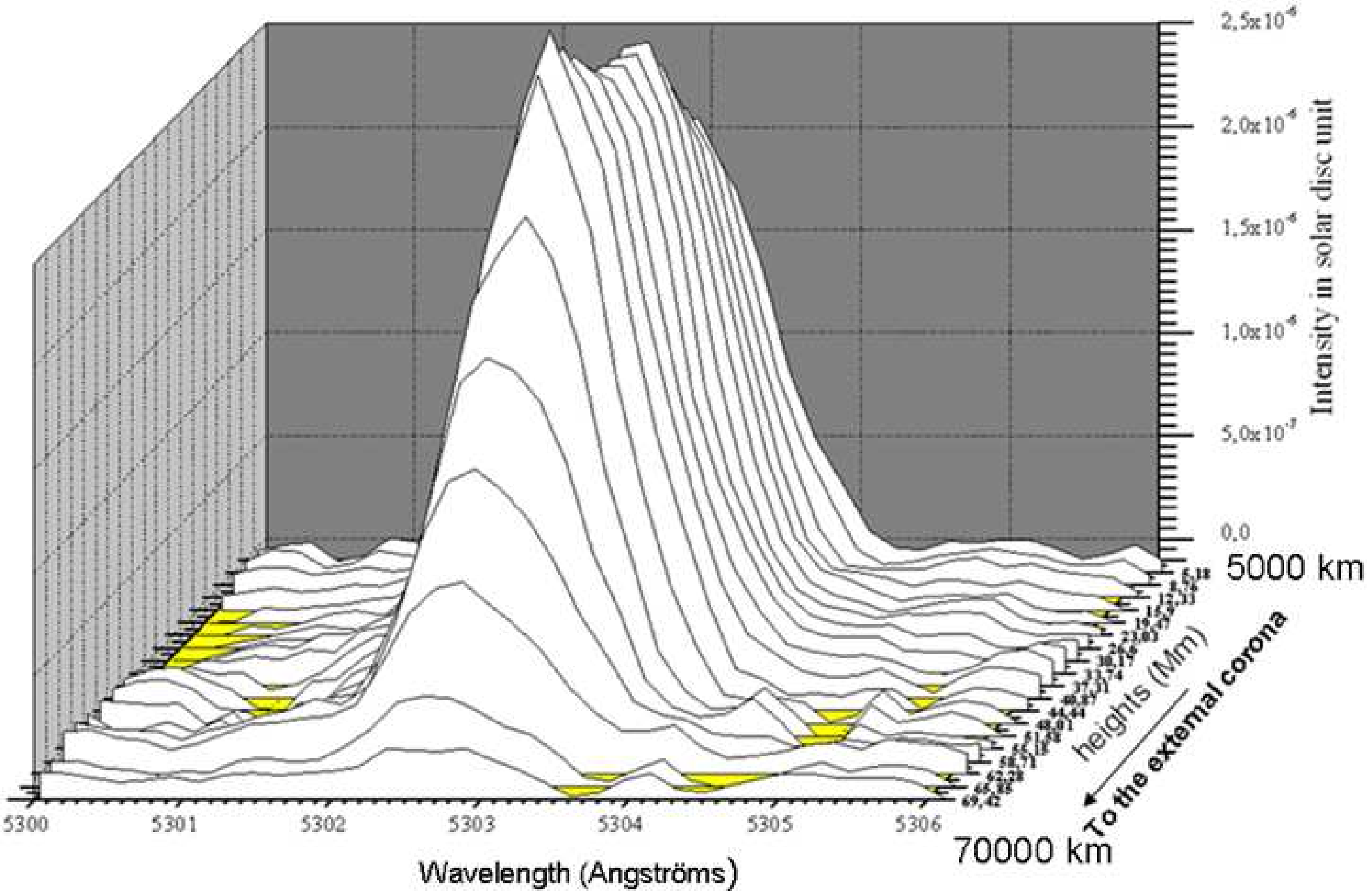

**Figure IV-4-1-3:** *Profils d'intensités en fonction de la longueur d'onde et reproduits le long de la hauteur de la raie verte du Fe XIV à 5302.86Å, avec un pas de 4 Mm, afin de montrer les variations des profils dans les altitudesde 5000 à 70000 km correspondant à la région de transition chromosphère-couronne solaire. Le premier profil est réalisé à l'altitude de 5.18 Mm au dessus du limbe et le dernier est à 69.42 Mm qui est dans la couronne.*

Les profils IV-4-1-4 ont été obtenus en ayant mesuré l'aire prise sous le profil de la raie verte, avec un pas régulier de 2 Mm.
La position de la fente par rapport au disque et l'angle donné en figure III-6-3, ont été pris en compte pour le calcul des hauteurs.
Le graphique IV-4-1-4 représente les variations des aires mesurées sous la raie verte et étalonnées en unités d'intensité du disque solaire multipliée par des Angströms, et les altitudes sont représentées en Mm. 1 rayon solaire vaut 696 Mm.

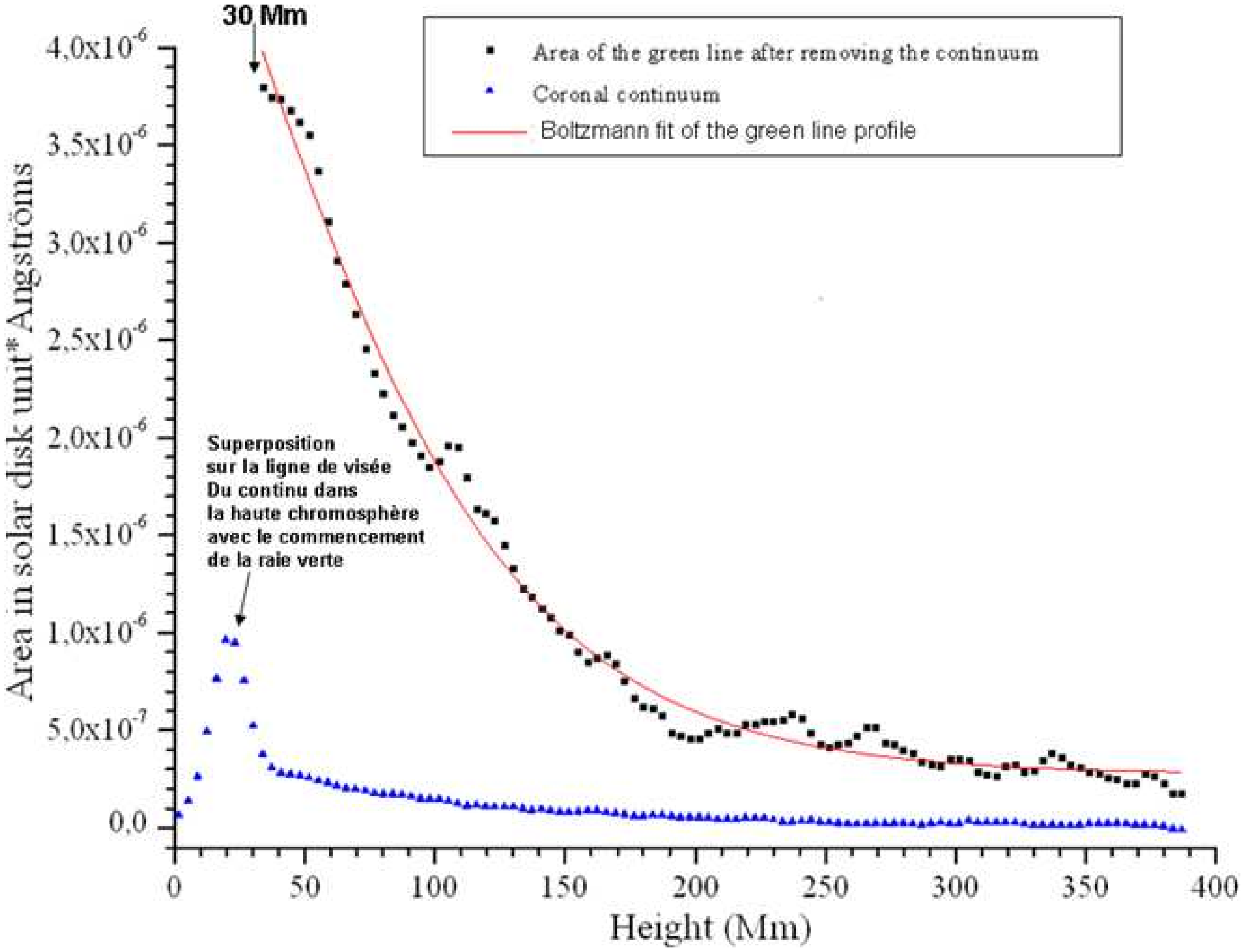

**Figure IV-4-1-4:** *Variations de l'aire mesurée sous le profil de la raie verte en ayant soustrait le continu, et le graphique du continu, tracés à la même échelle.*

Une courbe d'ajustement de type Boltzmann $I(h)=A_2+\frac{A_1-A_2}{1+e^{(h-h_0)/C}}$ a été effectuée, en bon accord avec les valeurs expérimentales. A, B, et C et $h_0$ sont des constantes.
Ces courbes figure IV-4-1-4 représentent les variations de l'aire corrigée sous la raie verte, c'est à dire son flux. Les paramètres et termes constants sont donnés avec l'équation de la de la courbe d'ajustement:
I(h) = A2 + (A1-A2)/(1 + exp((h-$h_0$)/C)) de l'aire de la raie verte sont les suivants :
A1 = 9.3258E-6 $\pm$ 1.7519E-6
A2 = 2.7448E-7 $\pm$ 2.1192E-8
$h_0$ = 12.97606 $\pm$ 18.66484
C = 56.64451 $\pm$ 3.64384

Les points de la courbe de la raie verte en dessous de 30 Mm n'ont pas été pris en compte pour réaliser la courbe d'ajustement à cause des effets de superposition sur la ligne de visée. Un changement de pente ou de gradient est mis en évidence à partir de l'altitude 200 ± 20 Mm sur la courbe de flux de la raie verte, voir figure IV-4-1-4. Ceci traduit des échelles de hauteur différentes, et un changement de température ou de densité électronique ou de fonction de remplissage dans la couronne à partir de cette altitude de 200 Mm.

## IV-4 –2) Analyses du continu coronal dans les couches chromosphériques et basse couronne obtenus sur les spectres de l'éclipse de 2012

L'étendue de cet embrillancement chromosphérique est mesurée dans la courbe figure IV-4-2-1 où un profil de type Lorentzien a été utilisé comme ajustement et les échelles sont exprimées en rayons solaires. La valeur 1 correspond au bord solaire:

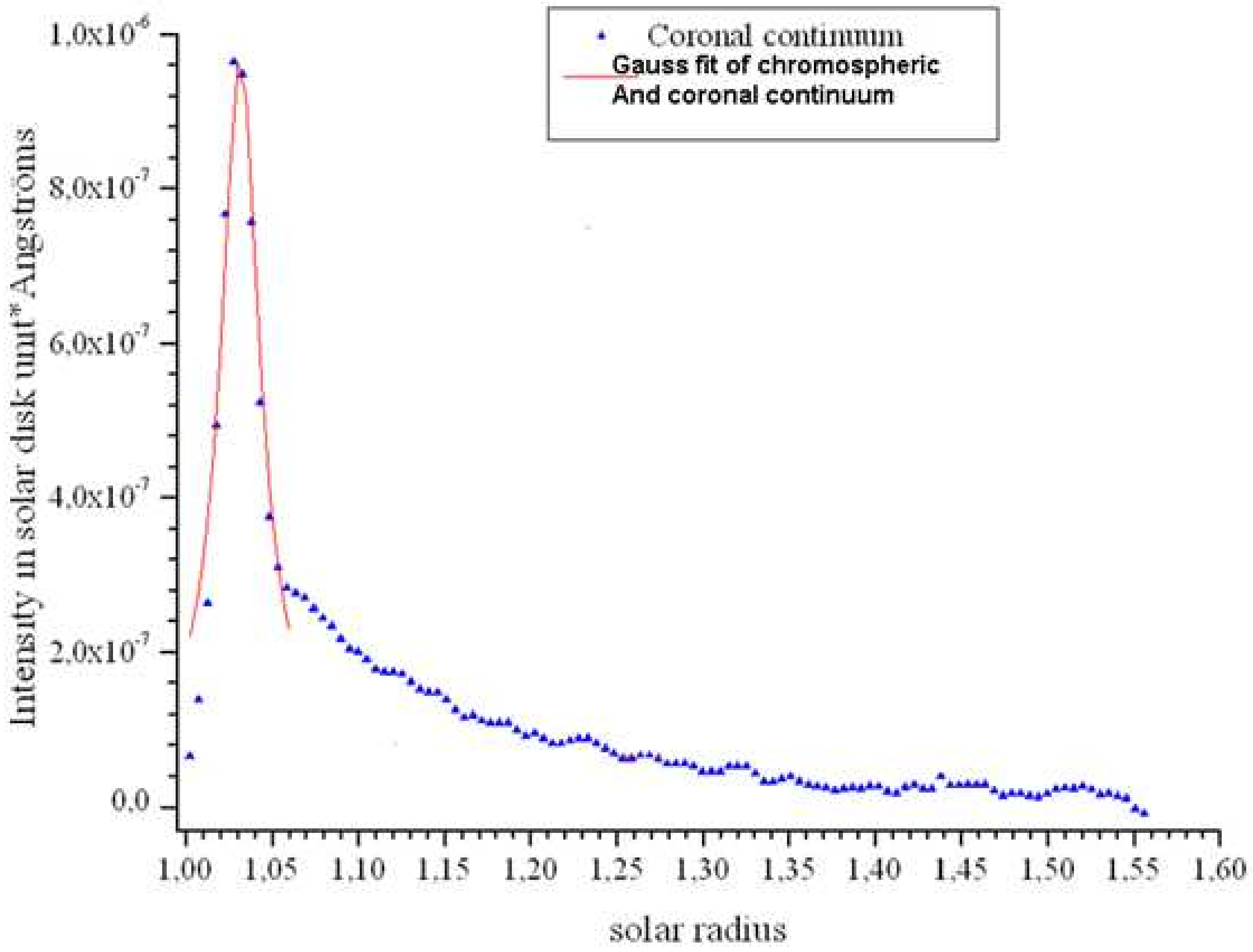

**Figure IV-4-2-1:** *mesure de coupe radiale en rayon solaire, de l'embrillancement chromosphérique correspondant à la région d'interface de la chromosphère à la couronne solaire pour le continu coronal.*

La largeur à mi-hauteur estimée de cet embrillancement mesure 0.02698 ± 0.00177 rayons solaires soit 18.77 ± 1.23 Mm. Une valeur de 19000 km peut être admise. Aucune échelle de hauteur ne peut être associée à cette hauteur car la raie 5303Å est produite vers 2 MK, avec une échelle de hauteur correspondant à 100 Mm. Cette enveloppe sera étudiée plus en détails par comparaison avec l'analyse de l'embrillancement observé par SDO/AIA dans les raies coronales EUV au dessus de la surface, pour rechercher la nature de la lumière chromosphérique diffusée, et aussi savoir si les embrillancements des limbes constatés sur les images de AIA/SDO pourraient provenir de ce même continu qui serait émis aussi dans les EUV pour des étendues et altitudes comparables aux observations d'éclipse dans le visible. Cette enveloppe brillante d'environ 20 Mm d'épaisseur pourrait révéler le chauffage des pieds de boucles coronales avec un processus « d'évaporation » dû à l'impact vers le bas de particules énergétiques.
Cet anneau brillant devrait être visible dans tous les continus et raies chromosphériques, comme par exemple Lyman $\alpha$ qui est à la fois chromosphérique et coronal.
Pour la suite de notre étude, le pic d'embrillancement du continu coronal entre 1.00 et 1.05 rayons solaires a été ignoré par la suite des analyses de profils, pour effectuer le calcul des coefficients du polynôme d'ajustement avec l'équation suivante :
$I(\rho) = P_1 * \rho^{-17} + P_2 * \rho^{-7} + P_3 * \rho^{-2.5}$ où $\rho$ désigne l'altitude en rayons solaires (Solar radius) en abscisse du graphique figure IV-4-2-2. Ce polynôme résulte d'une étude et d'un calcul provenant d'un modèle utilisé par Baumbach, 1937, revu par November et Koutchmy 1996, pour mesurer la brillance en unités du disque solaire dans la basse couronne à des échelles de hauteurs inférieures à 1 rayon solaire, en vue de mesurer la température hydrostatique.
Le terme en $P_3 * \rho^{-2.5}$ correspondant aux intensités de la couronne lointaine, au-delà de 2 rayons solaires est négligé devant les termes $P_1 * \rho^{-17} + P_2 * \rho^{-7}$ qui correspondent aux altitudes basses et intérmédiaires au dessus du limbe. Ce sont seulement ces couches moins élevées de l'interface à la basse couronne qui sont étudiées pour cette thèse.
Avec le logiciel Origin, 200 itérations ont été effectuées et les coefficients du polynôme d'ajustement sont donnés dans le graphique suivant:

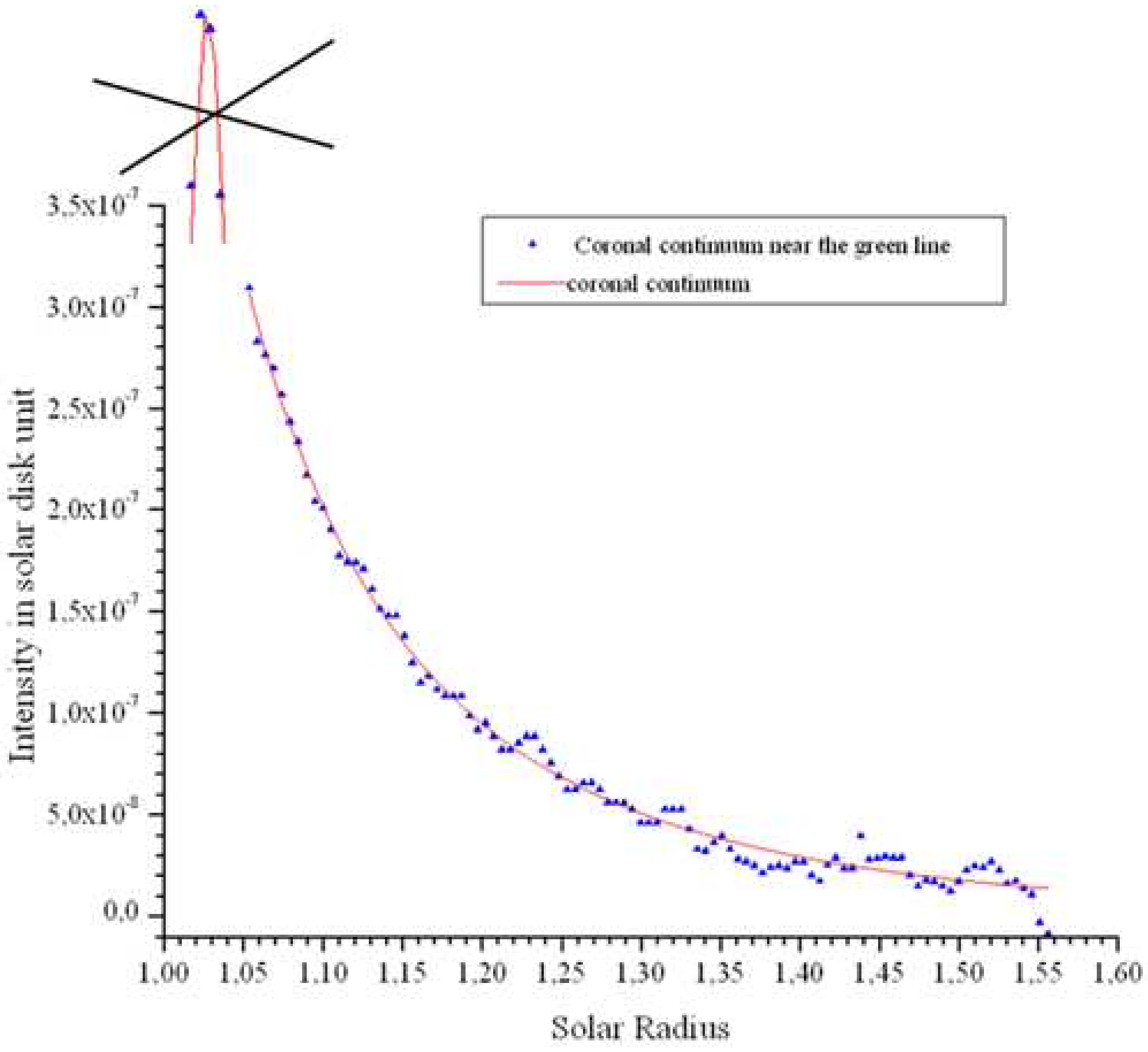

**Figure IV-4-2-2:** *Courbe d'ajustement obtenue sur le continu coronal dont les intensités sont exprimées en unités du disque solaire moyen.*

D’après cet ajustement, $P_1 = 2.3735*10^{-7}$ et $P_2 = 3.0086*10^{-7}$
L’équation de ce polynôme est donc:

$$I(\rho) = 2.3735*10^{-7}*\rho^{-17} + 3.0086*10^{-7}*\rho^{-7}$$

Les modulations observées peuvent être liées à des structures coronales,
comme par exemple les boucles et les grands jets de plasma magnétisé, dont la fente du spectrographe a probablement intercepté une tranche lors des acquisitions des spectres.
Ces équations des polynômes d’ajustement servent à déterminer les paramètres constants devant les termes en $\rho^{-7}$ et $\rho^{-17}$.
Les modulations observées le long de la courbe d’ajustement polynomiale sont liées à des structures coronales, et à mesure que l’on s’éloigne, le profil décroit et les points sont plus dispersés autour du polynôme d’ajustement.
Ces résultats permettent des analyses aux altitudes chromosphériques et confirmer ce qu’avaient mesuré Shlovskii 1965. Les observateurs de cette époque étaient gênés par la lumière parasite diffusée au-delà du limbe provenant du disque solaire occulté. Les premiers spectres obtenus hors éclipse voir annexe N°32, montrent la présence d’une frange continue brillante qui empêche d’observer la raie verte à des altitudes plus basses au dessus du limbe solaire, et n’avaient pas pu mesurer aussi bas dans la chromosphère, que lors de ces résultats d'éclipse du 13 Novembre 2012.
Le chapitre suivant a pour but d’examiner les variations des paramètres physiques selon la distance radiale, déduits des profils analysés autour de la raie verte avec le spectrographe à fente.

## IV– 4-3) Analyses sur les paramètres de la raie verte du Fer XIV en vue de diagnostiquer des corrélations, facteurs d'échelles et nature du plasma, à l’éclipse du 13 Novembre 2012

Ces résultats récents permettent de mieux analyser les variations de longueur d’onde, de vitesse radiales, et FWHM dans le profil de la raie verte du Fe XIV avec la distance radiale.
Par ailleurs, les coupes relevées le long de la raie verte pris à intervalles de distance radiale réguliers, ont permis de mesurer des effets de rétrécissements de la largeur de la raie verte avec une meilleure précision qu’aux éclipses totales de 1994, 1996, 2001 et 2006 (réalisées aussi avec un spectrographe à fente).
Les graphiques IV-4-3-1, IV-4-3-2 et IV-4-3-3 montrent les analyses effectuées après le second contact lors de l’éclipse du 13 Novembre 2012 en Australie. Les effets de décalage en longueur d’onde avec l’altitude ont été analysés et présentés sur la figure IV-4-3-1:

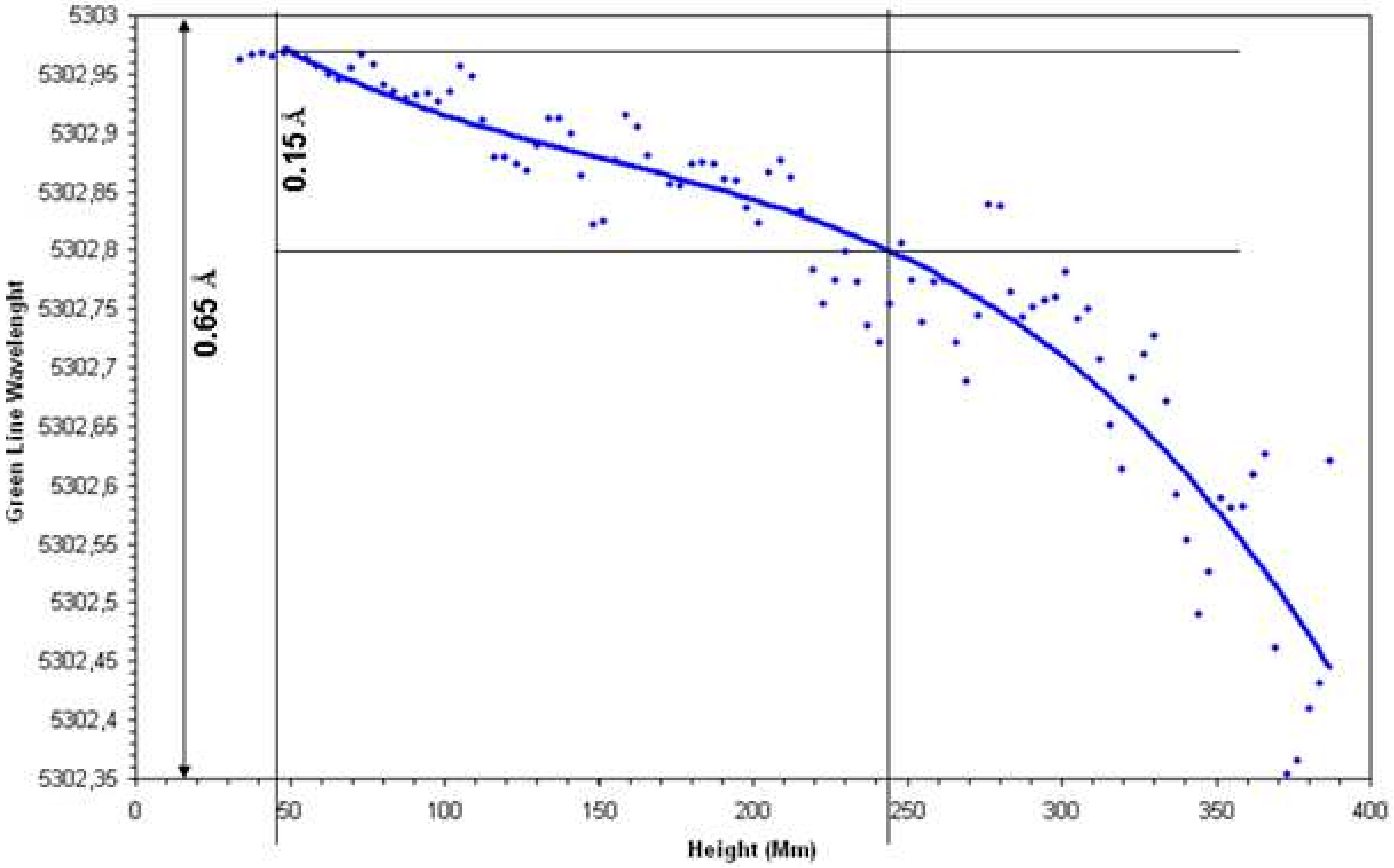

**Figure** IV-4-3-1**:** *variations des décalages de la longueur d'onde centrale de la raie verte en fonction de l'altitude.*

Un ajustement préliminaire avec un polynôme de degré 3 a été effectué pour « lisser » et mieux apprécier les modulations, et fluctuations des décalages compte-tenu de la dispersion des points autour de la courbe d'ajustement. L'équation du polynôme est :

$\lambda_{max} = -2*10^{-8} h^3 + 1*10^{-5} h^2 - 2.2*10^{-3} h + 5303.1$

Un écart de 0.15 Å est constaté entre 40 et 240 Mm, ce qui correspond à une vitesse radiale des jets coronaux en expansion : $\frac{\delta\lambda}{\lambda_0} = \frac{V_r}{c}$ , soit $Vr$ = 8 km/s (avec $\lambda_0$ = 5303 Å)

Puis cet écart devient 0.55 Å entre 240 et 380 Mm, soit $V_r$ = 31 km/s.

Sur toute l'étendue entre 40 et 380 Mm, $\delta\lambda$ = 0.65 Å, ce qui correspond à $Vr$ = 39 km/s.

Pour rappel, $c$ est la vitesse de la lumière dans le vide, $\lambda$ la longueur d'onde mesurée, $\lambda_0$ est la longueur d'onde de référence du Fe XIV, $\delta\lambda = \lambda_0 - \lambda$ et $V_r$ est la vitesse radiale en km/s.

Le graphique IV-4-3-2 donne les variations des vitesses radiales de la raie verte en fonction de la distance, d'après les relevés précédents, en considérant que le jet en arrière plan « s'éloigne » :

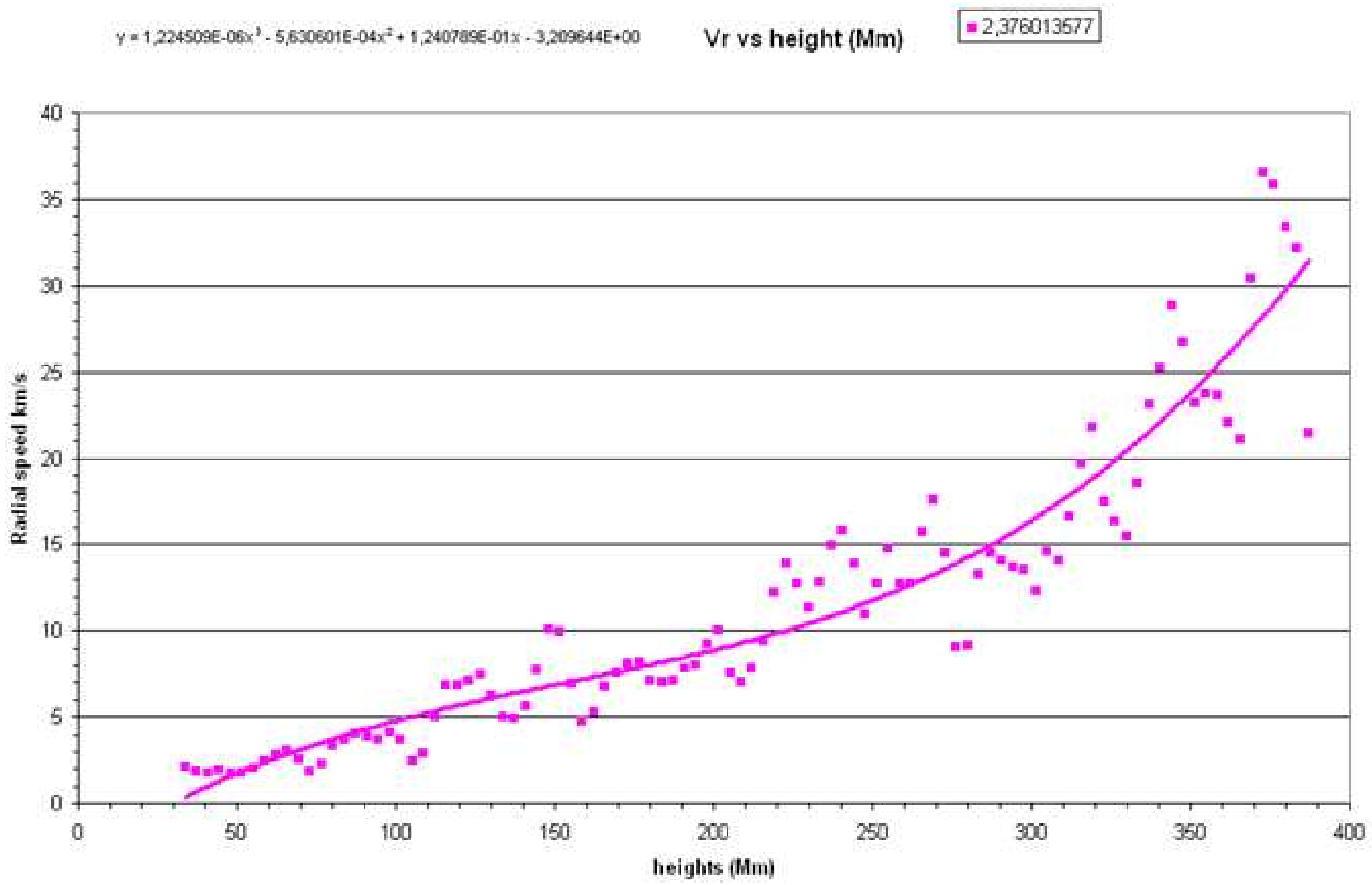

**Figure IV-4-3-2:** *variation des vitesses radiales déduites de la mesure du décalage de la position de la longueur d'onde centrale en km/s en fonction de la distance au bord solaire pour la raie verte du Fe XIV 5303 Å.*

Deux valeurs d'accélérations peuvent être déduites, traduisant une expansion du plasma. Les analyses plus approfondies sur le vent solaire et son origine qui en découlent, sont en dehors du cadre de cette thèse. Par ailleurs, les largeurs à mi-hauteur de la raie verte ont été mesurées en fonction de l'altitude, afin d'étudier les modulations de celle-ci comme indiqué dans le graphique IV-4-3-3:

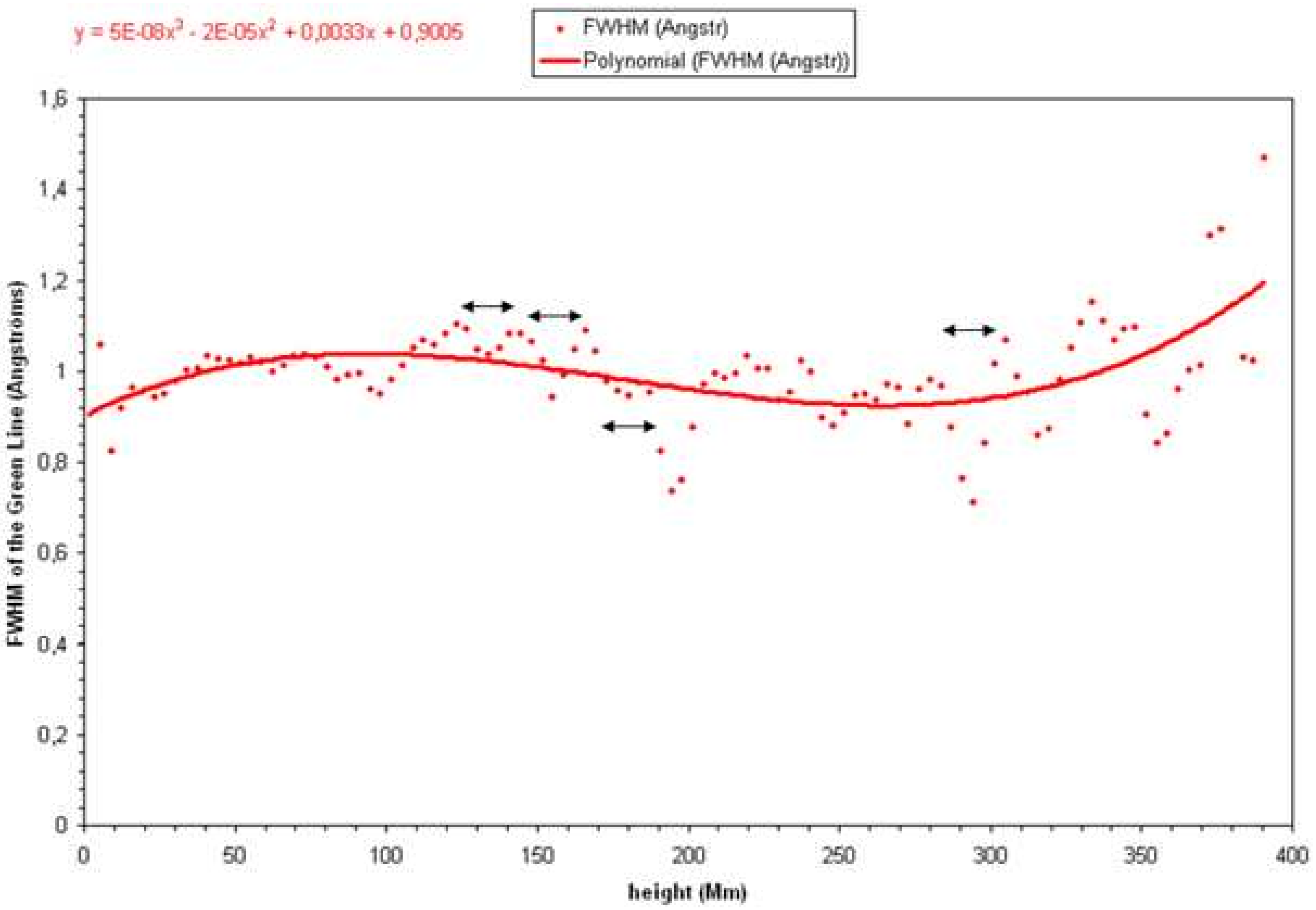

**Figure IV-4-3-3:** *variations de la largeur totale à mi-hauteur de la raie verte en fonction de l'altitude. Une courbe d'ajustement polynômiale préliminaire a été ajoutée pour « lisser » et mieux apprécier les modulations, et fluctuations des décalages*

Ces variations observées peuvent être liées à la vitesse turbulente non thermique dont la contribution joue un rôle dans les mécanismes de transport des ions, électrons dans l'interface chromosphère – couronne solaire. Les modulations de quelques dizaines de Mm des vitesses turbulentes pourraient être liées à  des structures plus étendues dans la couronne, comme les grands jets et modulations. Ces fluctuations sont beaucoup plus difficiles à observer dans l'interface photosphère – couronne solaire aux altitudes inférieures à 1 Mm.

La présence des ions de Fe XIV 5303Å à des hauteurs de plusieurs dizaines de Mm, et la présence de ce même élément comme le Fe II 4629 Å « low FIP » dans l'interface protubérance-couronne (voir chapitre V-1) et observé dans les spectres éclairs (interface photosphère-couronne), montre un mélange de ce même élément sur une gamme plus étendue d'échelles. Ces résultats d'observations donnent des indications sur un mécanisme possible d'alimentation de la couronne en éléments « low FIP », et où la sur-abondance en éléments « low FIP » dans la couronne est établie depuis longtemps voir Annexe 19, figures A-19-1 et A-19-2. L'origine de ces éléments « low FIP » qui alimenteraient la couronne en masse provient des basses couches de l'interface photosphère-couronne. C'est ce qu'ont permis de montrer les nouveaux spectres éclairs CCD obtenus aux éclipses de 2006, 2008, 2009, 2010 et 2012 et leur analyses présentés dans cette thèse.

## IV-5) Synthèse des analyses comparatives des courbes de lumière et enveloppes des raies d'hélium aux éclipses de 2008, 2009, et 2010

Les éclipses de 2008, 2009 et 2010 ont toutes permis d'observer et de comparer les raies d'hélium optiquement minces lors des contacts, et les analyses des courbes de lumière ont montré que ces raies commencent à être visibles pour des altitudes supérieures à 500 et 600 km au dessus du limbe, au dessus de la région du minimum de température. Ce sont des résultats importants pour ces travaux de thèse qui ont permis d'explorer ces régions difficilement accessibles, grâce aux éclipses totales.
Les graphiques IV-5-1 montrent une comparaison des raies d'hélium observées aux éclipses, et sur lesquels le profil d'intensité de la raie D3 de l'hélium neutre 5876 Å observé hors éclipse, par S. Koutchmy en 1991 à Sacramento Peak et ayant fait l'objet d'un DEA, a été ajouté:

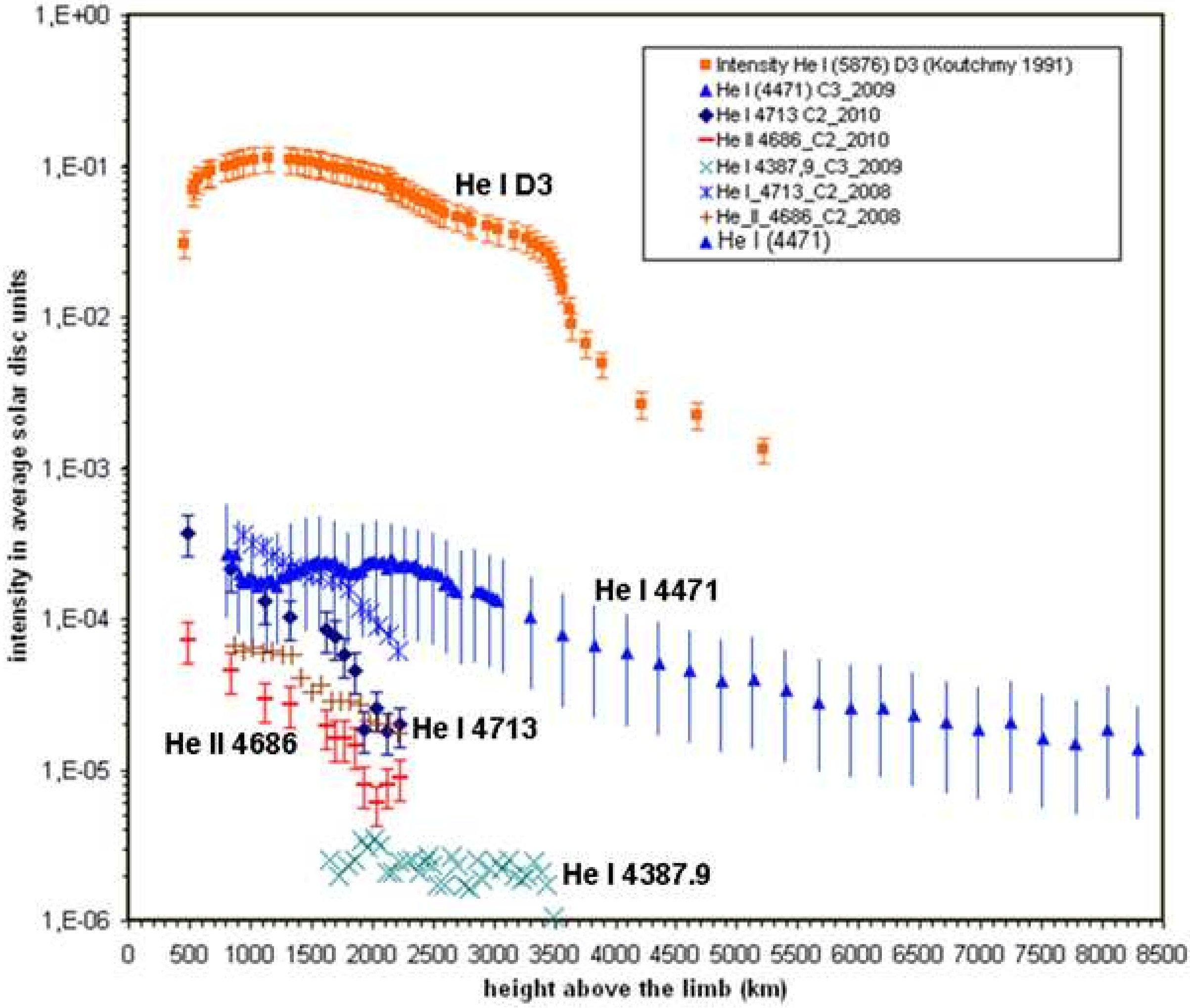

**Figure IV-5-1:** *Comparaison des courbes de lumière de différentes raies d'hélium obtenues aux éclipses de 2008, 2009 et 2010 et la courbe obtenue par observation directe de la raie intense D3 de He I à Sacramento Peak en 1991.*

Les émissivités des enveloppes d'hélium neutre et ionisée ont été déduites après le calcul d'inversion d'intégrale d'Abel à partir des courbes de lumière relevées dans ces raies lors des contacts d'éclipse.
Le tableau IV-5-2 récapitule les altitudes maximales des émissivités observées, selon les constacvts des éclipses.

| Dates et contact de l'éclipse étudiée | Altitude du maximum d'émissivité He I 4471Å | Altitude du maximum d'émissivité He I 4713Å | Altitude du maximum d'émissivité He II 4686Å | Altitude du maximum d'émissivité Ba II 4553Å |
|---|---|---|---|---|
| C2 de 2008 (ciel sans voiles) | - | 1670 ± 50 km | 1150 ± 50 km | - |
| C2 de 2009 (ciel voilé) | - | - | - | 1550 ± 70 km |
| C3 de 2009 (ciel voilé) | 1700 +/- 200 km | - | - | - |
| C2 de 2010 (ciel voilé) | - | 1550 ± 100 km | 1550 ± 100 km | - |
| C3 de 2010 (ciel voilé) | - | 1300 ± 100 km | - | - |

**Tableau IV-5-2:** altitudes des maxima des émissivités des courbes de lumière des raies He I 4471Å, He I 4713Å, He II 4686Å et Ba II 4553Å pour les éclipses totales de Soleil de 2008, 2009 et 2010

Les altitudes des maxima d'émissivité se situent autour de 1600 km, et correspondent aux enveloppes.
Les différences des résultats peuvent s'expliquer en partie à cause des fluctuations météorologiques et la présence de voiles nuageux. La raie de l'hélium He I 4471Å observée au troisième contact de l'éclipse totale du 22 Juillet 2009 est plus intense que les raies He I 4713 Å et He II 4686 Å. Cependant des fluctuations importantes ont été relevées dans sa courbe de lumière, et l'étendue de son émissivité est plus incertaine.

Les analyses suivantes ont pour objectif de montrer des différences d'extensions des enveloppes d'hélium neutre et ionisé, mesurées à partir des profils des raies en croissants des spectres éclairs. Les extensions des enveloppes d'hélium ont été mesurées d'après la largeur en pixels relevées sur les spectres des raies. Les largeurs mesurées des enveloppes sont limitées par le bord intérieur de la Lune et les enveloppes s'étendent au-delà du limbe avec une décroissance monotone, une FWHM a été prise, car elles se superposent au continu du fond coronal.
Des moyennes ont été réalisées sur les évaluations des extensions des enveloppes d'‘hélium car la présence de structures (macrospicules) produisent des fluctuations.
A l'éclipse du 22 Juillet 2009 une raie faible de l'hélium neutre He I 4387.9Å a été étudiée. Cette raie a été observée par Mitchell et al, 1933. Les altitudes où cette raie a été observée historiquement ne s'étendaient pas plus que de 2070 km. A cette époque, elle n'était pas vue comme une enveloppe. Nos résultats de l'éclipse de 2009, ont montré qu'elle s'étendait plus loin jusqqu'à 3500 km, à des extensions comparables à celles de l'hélium neutre 4713 Å et ionisé He II 4686 Å. La figure IV-5-3 montre des mesures des extensions:

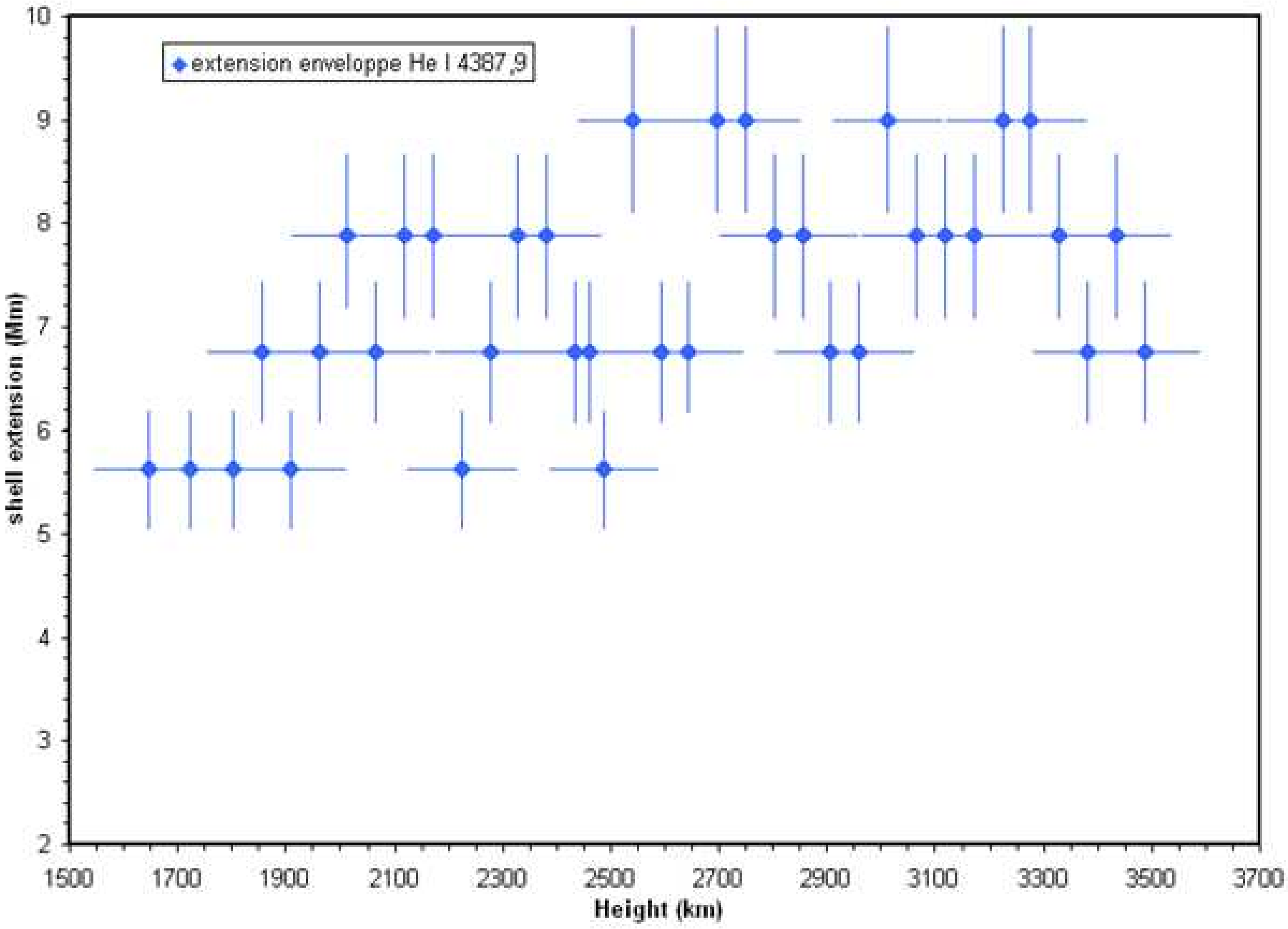

**Figure IV-5-3:** *mesure des extensions de l'enveloppe de l'hélium neutre 4387.9 Å observée au 3ième contact de l'éclipse du 22 Juillet 2009 avec l'altitude limitée par le bord lunaire.*

Cette raie He I 4387.9Å est de plus faible d'intensité que celle de He I 4713Å, un facteur 5 à 10 d'après la figure IV-5-1, et cette raie est vue aussi comme une enveloppe. On constate donc que toutes les raies d'hélium neutre optiquement minces sont vues comme des enveloppes, dont l'extension dépend de l'intensité de la raie et du degré d'ionisation. Pour l'éclipse du 11 Juillet 2010, nous avons comparé les extensions des enveloppes d'hélium neutre He I 4713 Å et He II 4686 Å, pour chaque contact et les extensions mesurées sont données dans la figure IV-5-4.

D'après les spectres éclair de l'éclipse 2010, des mesures ont été réalisées sur chaque spectre résultant de 6 spectres sommés tous les 3 spectres, afin d'améliorer le rapport signal sur bruit, et d'autre part de montrer les variations des extensions sur chaque spectre, correspondant au bord de la Lune définissant une altitude dans l'atmosphère solaire malgré ses irrégularités.

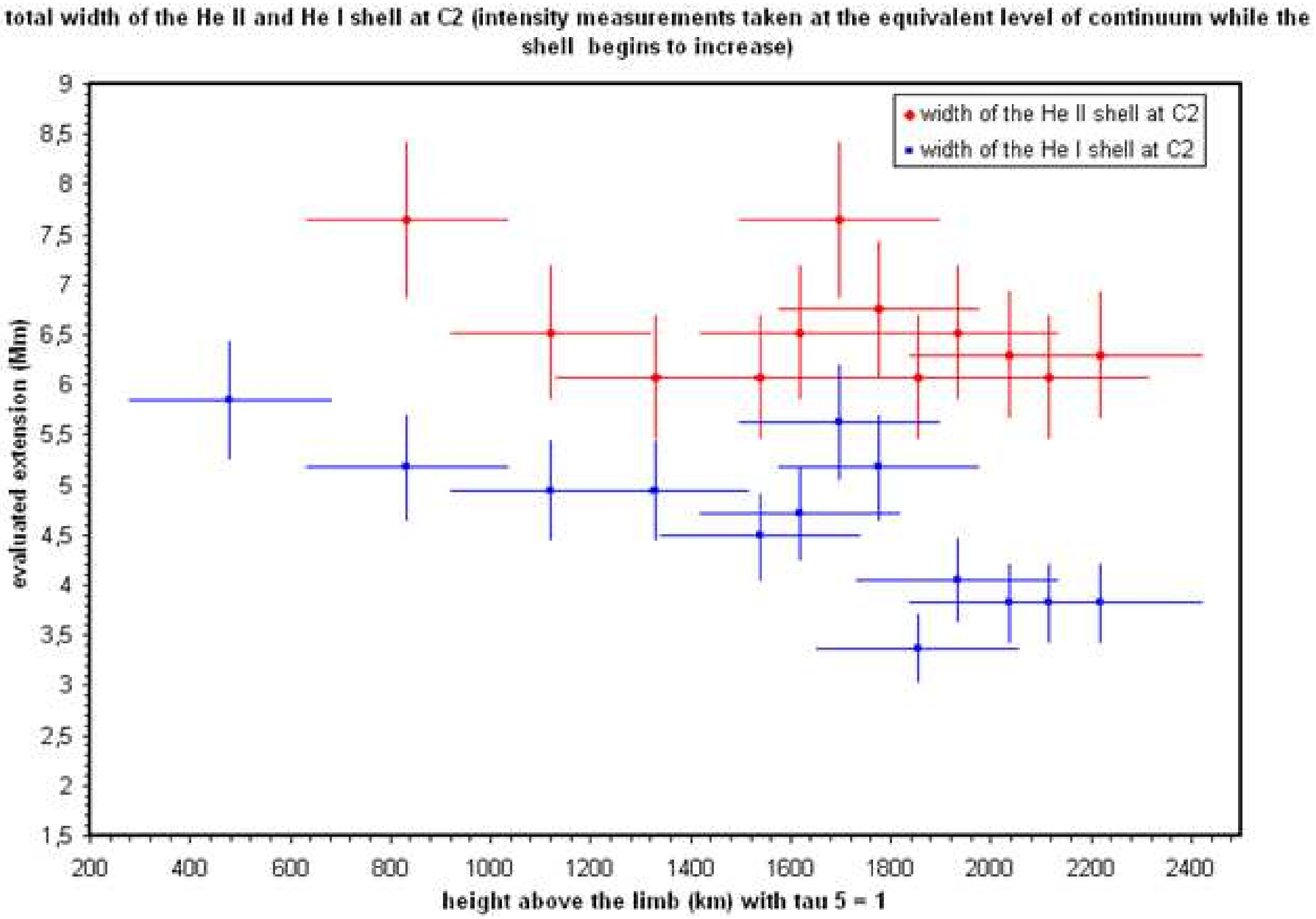

**Figure IV-5-4:** *mesures des extensions des enveloppes d'hélium neutre et ionisé au second contact C2 de l'éclipse du 11 Juillet 2010 avec l'altitude limitée par le bord lunaire*.

Les barres d'erreur des relevés des extensions des enveloppes d'hélium neutre et ionisé correspondent à une incertitude sur les altitudes prises comme référence pour mesurer les extensions, à cause du relief lunaire. Le relief lunaire, et les fluctuations des voiles nuageux peuvent être aussi responsables des modulations constatées.
L'enveloppe de l'hélium ionisé He II 4686 Å est 1.3 fois en moyenne plus étendue que celle de l'hélium neutre, et ces résultats se confirment au troisième contact comme le montrent les courbes de la figure IV-5-5:

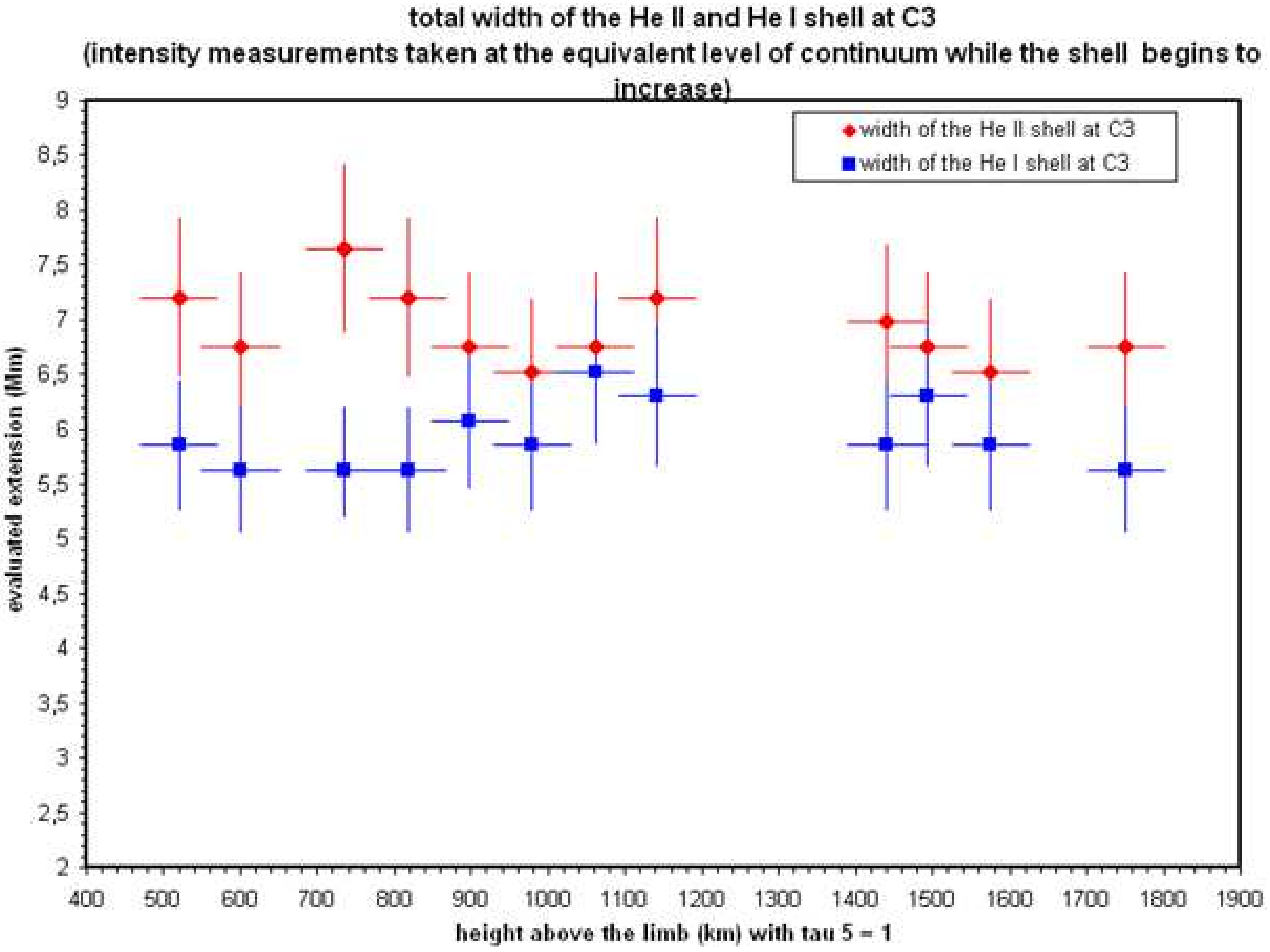

**Figure IV-5-5:** *mesures des extensions des enveloppes d'hélium neutre et ionisé au troisième contact C3 de l'éclipse du 11 Juillet 2010, avec l'altitude limitée par le bord lunaire.*

L'enveloppe de l'hélium ionisé est plus étendue que celle de l'hélium neutre dans des proportions comparables à celles mesurées lors du second contact. Ces extensions ont été mesurées sans l'utilisation des résultats d'inversion d'intégrales d'Abel. Cependant les altitudes au dessus du limbe des maxima d'émissivités des enveloppes d'hélium neutre et ionisé sont sensiblement voisines autour de 1700 km voir Figures IV-3-3-1 et IV-3-3-2 où les courbes de lumière des raies He I 4713 Å et He II 4686 Å des 2ieme et 3$^{\text{ième}}$ contacts de l'éclipse de 2010 sont comparés. Il est à noter que l'intensité de la raie de l'hélium ionisé à 4686 Å est de 3 à 4 fois plus faible que celle de l'hélium neutre 4713 Å.

A partir des inversions d'intégrale d'Abel effectuées sur les courbes de lumière des raies de He I 4713 Å et He II 4686 Å, des ajustements avec des courbes d'exponentielles décroissantes ont  été effectués, dans le but de mesurer des échelles de hauteurs et de comparer ces mesures pour les éclipses de 2008 et 2010 où elles ont été observées simultanément. Le tableau IV-5-6 présente une synthèse des résultats obtenus:

| Date et contact de l’éclipse | Echelle de hauteur déduite après inversion d’Abel<br><br>He I 4713 Å | Echelle de hauteur déduite après inversion d’Abel<br><br>He II 4686 Å |
|---|---|---|
| C2/ 2008 | 643 ± 70 | 696 ± 70 |
| C2/ 2010 | 456 ± 60 | 566 ± 80 |
| C3/ 2010 | 849 ± 80 | 953 ± 90 |

**Tableau IV-5-6:** *récapitulatif des échelles de hauteurs déduites sur les raies d’hélium, d’après les courbes d’ajustement en exponentielles décroissantes et après inversion d’intégrale d’Abel des courbes de lumière.*

Ces résultats montrent clairement que les échelles de hauteurs de l’hélium ionisé He II 4686 sont plus importantes que celles de l’hélium neutre He I 4713 Å, ce qui confirme le sens des mesures des extensions des enveloppes d’hélium réalisées spatialement en utilisant le bord de la Lune. Ces mesures d’échelles de hauteur confirment que l’extension de l’hélium ionisé He II 4686 Å est plus élevée que celle de l’hélium neutre He I 4713 Å.
Ces raies optiquement minces sont sensibles au rayonnement EUV émis par la couronne.
Les extensions des enveloppes d’hélium traduisent la pénétration des émissions EUV de la couronne dans ces couches profondes. Un article a été publié dans *Journal of Advanced Research,* Bazin, C, Koutchmy, S 2012, voir Annexe N°2. Par ailleurs, les différences d’échelles de hauteurs constatées pour une même raie s’expliquent par la géométrie du relief lunaire, où dans une vallée étroite et profonde d’avantage de flux de l’enveloppe est intégré sur la ligne de visée, que par rapport à une vallée large et peu profonde. Cela se traduit par des profils différents. Le relief lunaire est aussi différent selon la position des contacts sur le bord lunaire, car nous n’étions pas exactemment au milieu de la bande de centralité de l’éclipse de 2010, et cela s’est traduit par des phénomènes l’occultation différents selon le second et troisième contact par rapport aux limbes lunaire et solaire.

## IV-6) Synthèse d’analyses : embrillancement du limbe solaire

L’embrillancement du limbe solaire a été constaté avec les analyses des observations des spectres de la raie verte et du continu, obtenues avec le spectrographe à fente, et la pénétration de la raie verte dans les couches plus profondes peut aussi expliquer cet embrillancement associé à des effets de photoionisation par les raies EUV.
En effet, le renforcement d’intensité de la raie verte dans les couches plus profondes et plus proches du limbe solaire, pourrait être associé à la couronne chaude entre 1 et 2 MK. Le rayonnement EUV coronal provoque une photo-ionisation des couches plus basses au dessus de 1000 km comme les enveloppes d’hélium optiquement minces. Cependant cette observation montrant un maximum d’émission de la raie verte sur les altitudes chromosphérique peut être associée aux effets d’intégration sur la ligne de visée du rayonnement émis. Toute fois un effet d’enveloppe peut être considéré, avec des structures de petite échelle non résolues.
La présence des ions Fe XIV 5302.86 Å à des échelles de hauteur de plusieurs dizaines de méga-mètres et la présence de ce même élément Fe II  4629 Å  en « low FIP » montrent des degrés d’ionisation sur une gamme plus étendue d'échelles. Ce phénomène pourrait expliquer l’enrichissement de la couronne en éléments « low FIP » et dont l’origine des « low FIP » se situe aux altitudes inférieures à 1000 km, en dessous des enveloppes d’hélium.
La photo-ionisation des atomes par les rayonnements émis par les raies EUV, peut expliquer l’embrillancement du limbe solaire. Cet embrillancement qui est un effet d’opacité sur la ligne de visée est aussi constaté dans des observations dans l’UV avec TRACE, dans la raie du Carbone C IV, et où le carbone a perdu 3 électrons.

## IV-6-1) Embrillancement du limbe constaté avec TRACE en C IV pour diagnostiquer une correspondance entre les enveloppes d’hélium observées dans les raies optiquement minces

Les images de TRACE à partir desquelles le profil a été réalisé sont présentées en figure IV-6-1-1. Une image à 1550 Å correspondant au continu est soustraite de l’image à 1700 Å, pour mieux évaluer l’extension de l’embrillancement.

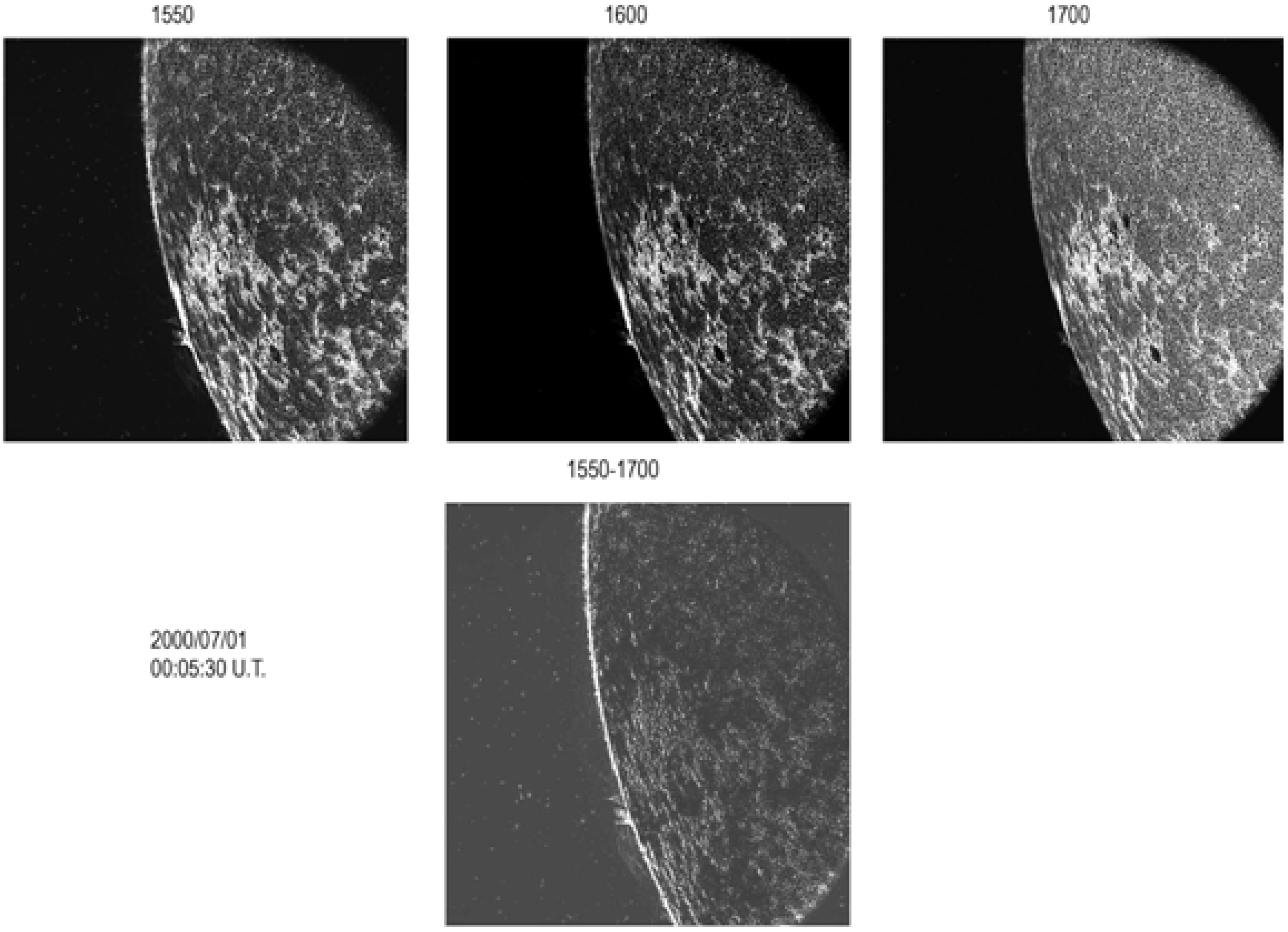

**Figure IV-6-1-1:** *Image dans la raie du carbone C IV où le continu a été soustrait : image à 1550 à laquelle l’image à 1700Å a été soustraite. L’embrillancement du limbe constitue une enveloppe avec Trace 1550 – 1700 Å*

Une fois l’image obtenue, des profils d’intensité radiaux sont effectuées sur le limbe, afin de révéler le profil de cet embrillancement, et effectuer une évaluation de sa largeur voir figures IV-6-1-2.

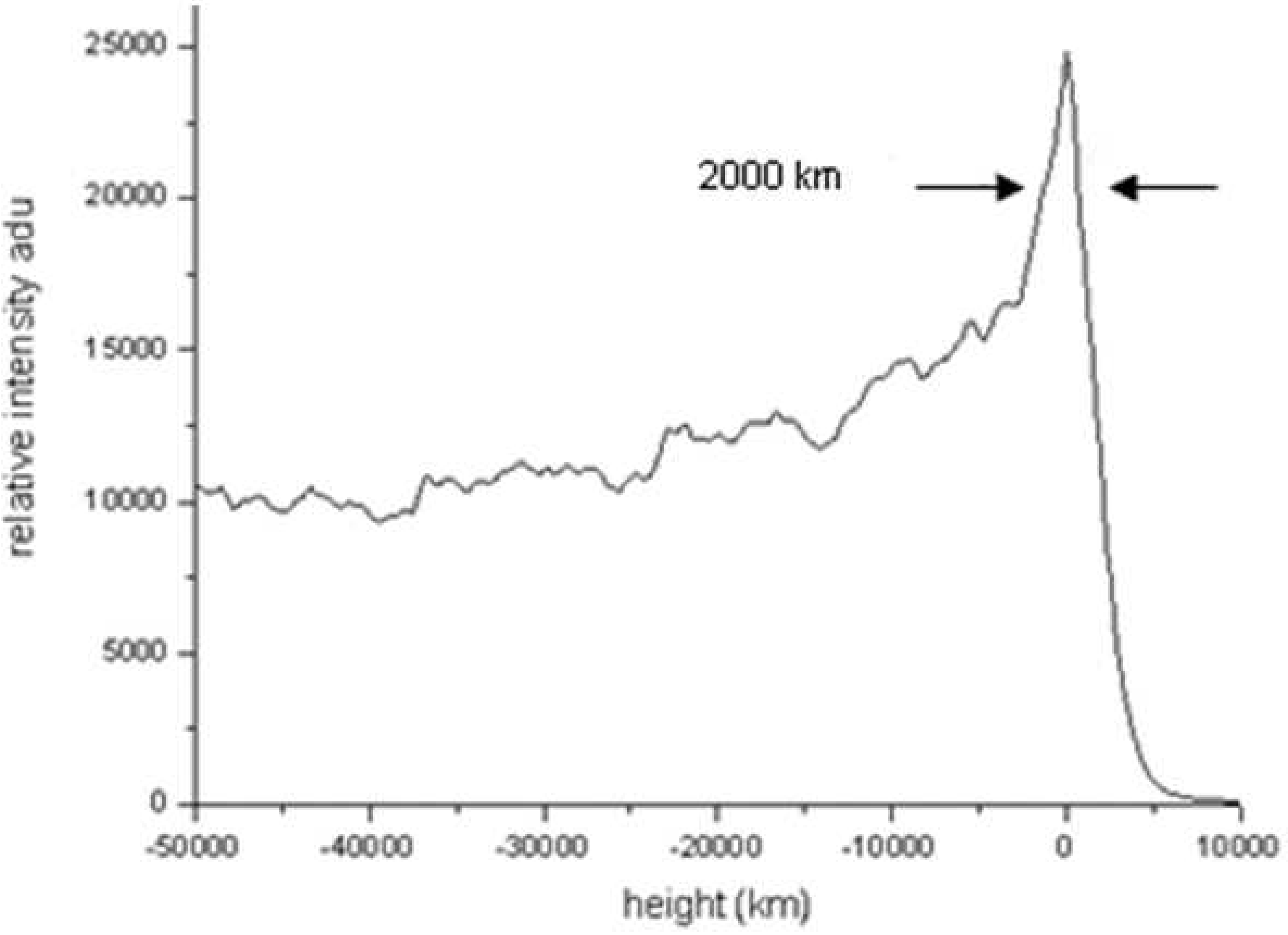

**Figure IV-6-1-2:** *profils d'intensités obtenu par une moyenne de 200 lignes intégrées radialement pour montrer l'embrillancement du limbe et sa taille dans la raie CIV correspondant aux mêmes couches où les raies de l'hélium He I 4713 Å et He II 4686 Å sont observées et aussi le continu chromosphérique.*

Le profil d'intensité radiale voir figure IV-6-2 montrant l'embrillancement, révéle une largeur et rapport d'intensité asssez comparable aux extensions des enveloppes d'hélium He I 4713Å et He II 4686Å voir chapitre IV-5. Dans l'article Vial 1980, des profils dans la raie de OVI, présentent un embrillancement d'amplitude comparable.
Par la suite une analyse de l'embrillancement du limbe est réalisée dans la raie du Fe IX/ Fe X, pour évaluer l'extension de l'embrillancement du limbe dans ces raies correspondant à des températures plus élevées 500000 K, et étudier l'effet de la température sur l'extension de l'enveloppe.

## IV-6-2) Comparaison de l’embrillancement avec SWAP Proba2, AIA/SDO dans les raies du Fe IX/X 171 Å, He II 304 Å et Fe XII 193 Å

Les nouvelles images du Soleil en extrême Ultraviolet EUV ont été obtenues par les missions AIA de SDO et une analyse des profils des embrillancements des limbes a été réalisée en vue de comparer les données des spectres éclairs obtenus aux éclipses totales de soleil dans le domaine visible. L'objectif est de comparer l'étendue des embrillancements et de mesurer l'étendue de l'atmosphère "froide", car les images EUV concernent la couronne dite "chaude" dont les températures sont de 0.7 MK pour la raie du Fe X à 171Å, et de à 2 MK pour la raie du Fe XII à 193 Å. Les images dans la raie 304 Å de l'hélium He II ont une température de l'ordre de 30000 K, ce qui correspond à des températures plus basses que les raies EUV coronales, dans la région de transition chromosphère-couronne. La raie de l’He II 304Å est 1000 à 10000 fois plus épaisse que la raie de He II 4686 Å, bien que ce soit le même élément. Des images de la mission SWAP Proba 2 ont été utilisées, et 20 images ont été sommées, puis redressées sur un secteur. La linéarisation a été effectuée avec le logiciel Iris, en paramétrant le rayon, et les ordonnées correspondant à l’intervalle où le bord solaire est linéarisé:

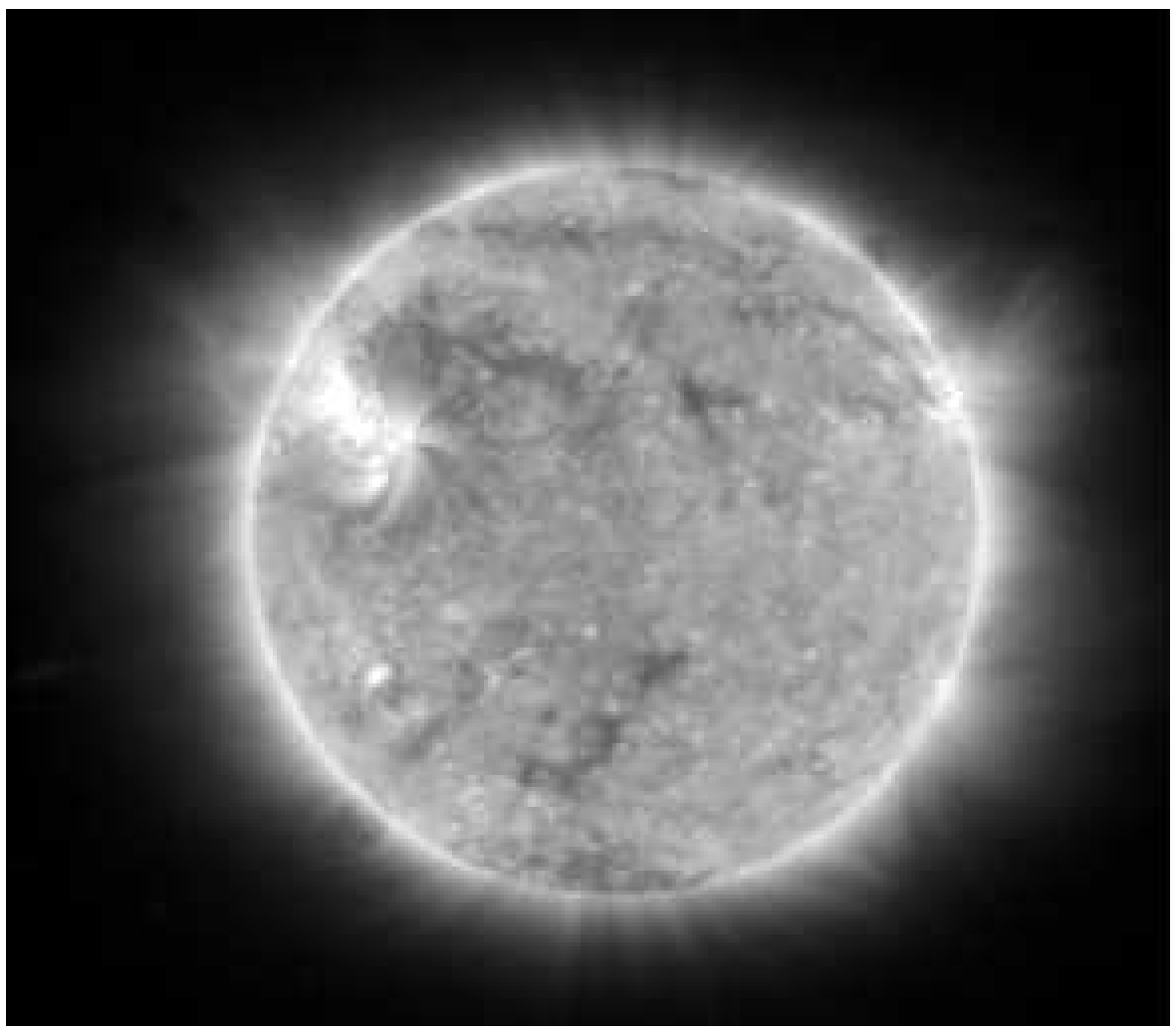
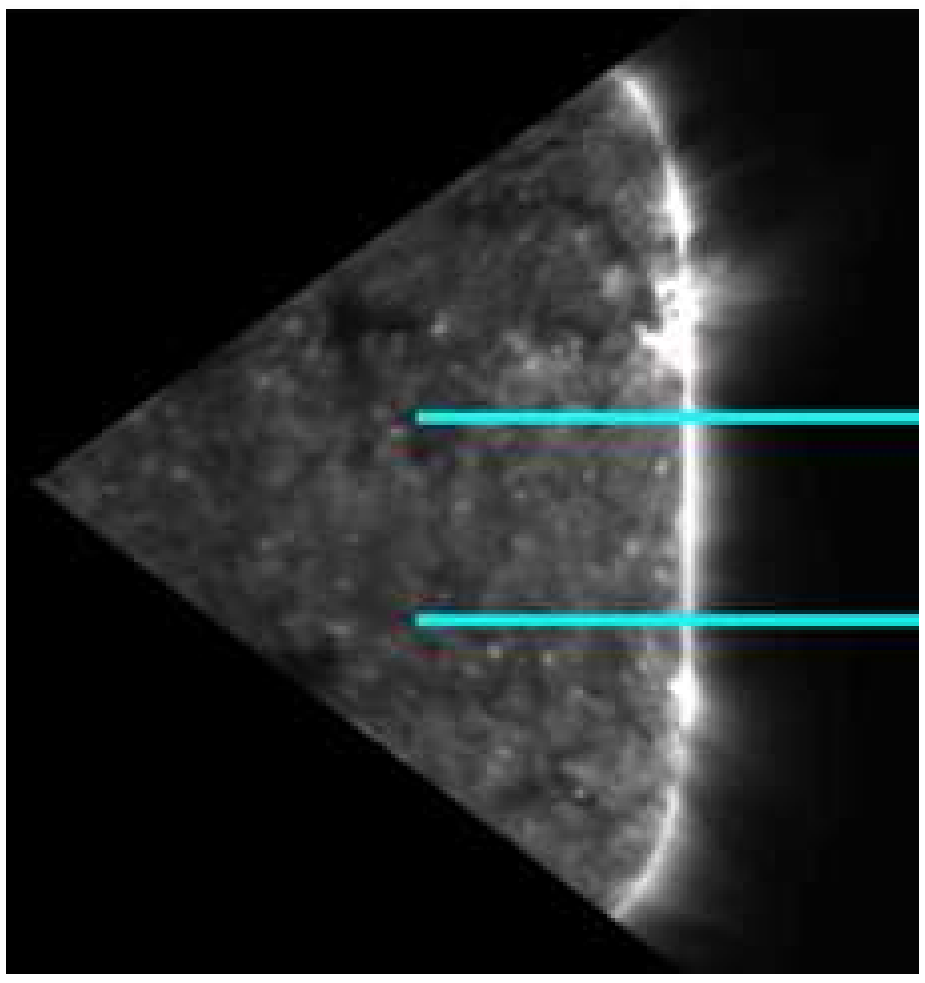

**Figure IV-6-2-1:** *Image de SWAP/Proba2 dans le Fe IX/FeX à 171 Å à gauche le 11 juillet 2010, pris dans une région calme. A droite, une partie de l’image redressée et indication de la zone intégrée sur 200 lignes, avec le logiciel Iris*

La figure IV-6-2-2 montre les variations d’intensité des profils, avec l’altitude:

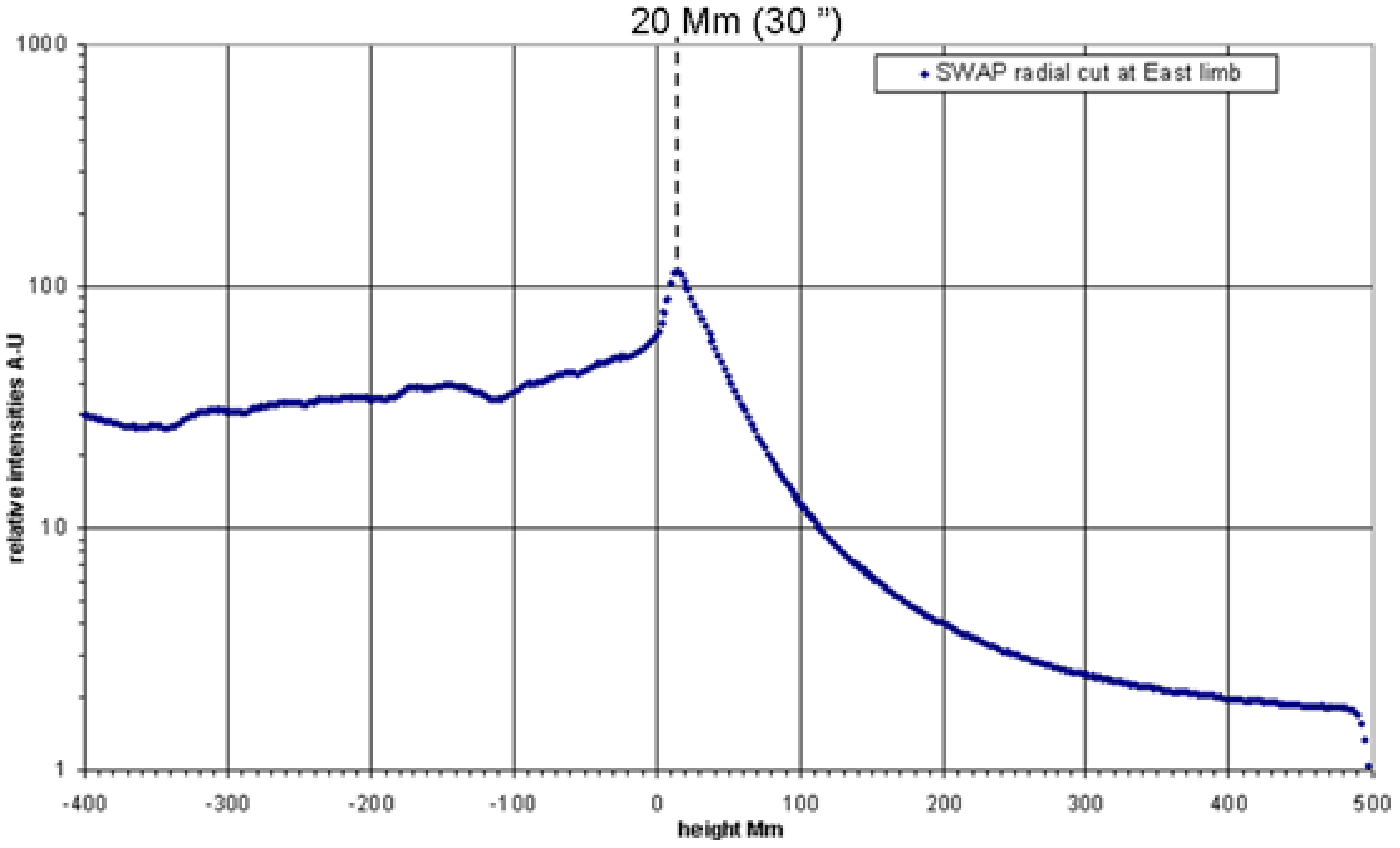

**Figure IV-6-2-2:** *profils d’intensités après intégration de 200 lignes sur les images de SWAP/Proba 2 montrant l’embrillancement du limbe dans les raies du Fe IX/X à 171 Å*

La largeur à mi-hauteur de l’embrillancement fait 25 ± 5 Mm et est située entre 20 et 30 Mm au dessus du limbe.

Cette largeur de 25 Mm s’avère plus élevée que celles mesurées dans les raies du C IV et He II 4686, et montre que plus la température d’ionisation est élevée, plus l’étendue de l’embrillancement du limbe est importante.
La sommation des images de SWAP a pour avantage d'améliorer le rapport signal sur bruit. Une étude de l'influence du nombre d'images sommées et du rapport signal sur bruit est donnée en Annexe N°21. Les profils d’intensité réalisés d’après les images de SWAP 171 Å en Fe IX/Fe X présentent une largeur d’embrillancement de l’ordre de 20 Mm. L’étalement peut être lié à la réponse du filtre de SWAP (voir Berghmans et al 2008) qui transmet d’autres raies d’émission EUV voisines de celles du Fe IX/ Fe X, et ces raies supplémentaires pourraient produire l’embrillancement, et celui-ci s’ajoute par effet d’intégration sur la ligne de visée sur le bord du disque.
L’autre explication possible consisterait à dire que les images SWAP montrent des émissions coronales correspondant à des parties plus chaudes (171 Å -174Å) et qui seraient renforcées par effets d’intégration sur la ligne de visée.
L’étude de l’embrillancement du limbe a été effectuée aussi avec les images de AIA/SDO, qui permettent d’analyser la basse couronne jusqu’aux températures de 2 MK.
Nous avions à disposition des données provenant de SDO, où les images étaient en format FITS qui avaient été rendues accessibles, et à partir desquelles des analyses approfondies en période de soleil calme ont été effectuées (voir chapitre V-5), comparablement aux éclipses de 2009, 2010 et 2012. Les réponses en températures des filtres de AIA sont indiquées figure IV-6-2-3.
Les images figures IV-6-2-4, IV-6-2-6 et IV-6-2-8 montrent des zones où ont été examinés les limbes sur le Soleil calme et dans les trous coronaux, pour la date du 14 Novembre 2011, où des données ont été disponibles, qui correspond à la suite de l'étude des limbes solaires, car des séquences temporelles ont été réalisées en ayant sommé 200 images.

Les graphiques IV-6-2-5, IV-6-2-7, IV-6-2-9 et IV-6-2-10 donnent la correspondance entre les longueurs d'onde des raies et les températures, selon les réponses des filtres de AIA, afin d'interpréter les températures pour analyser l'embrillancement du limbe solaire.

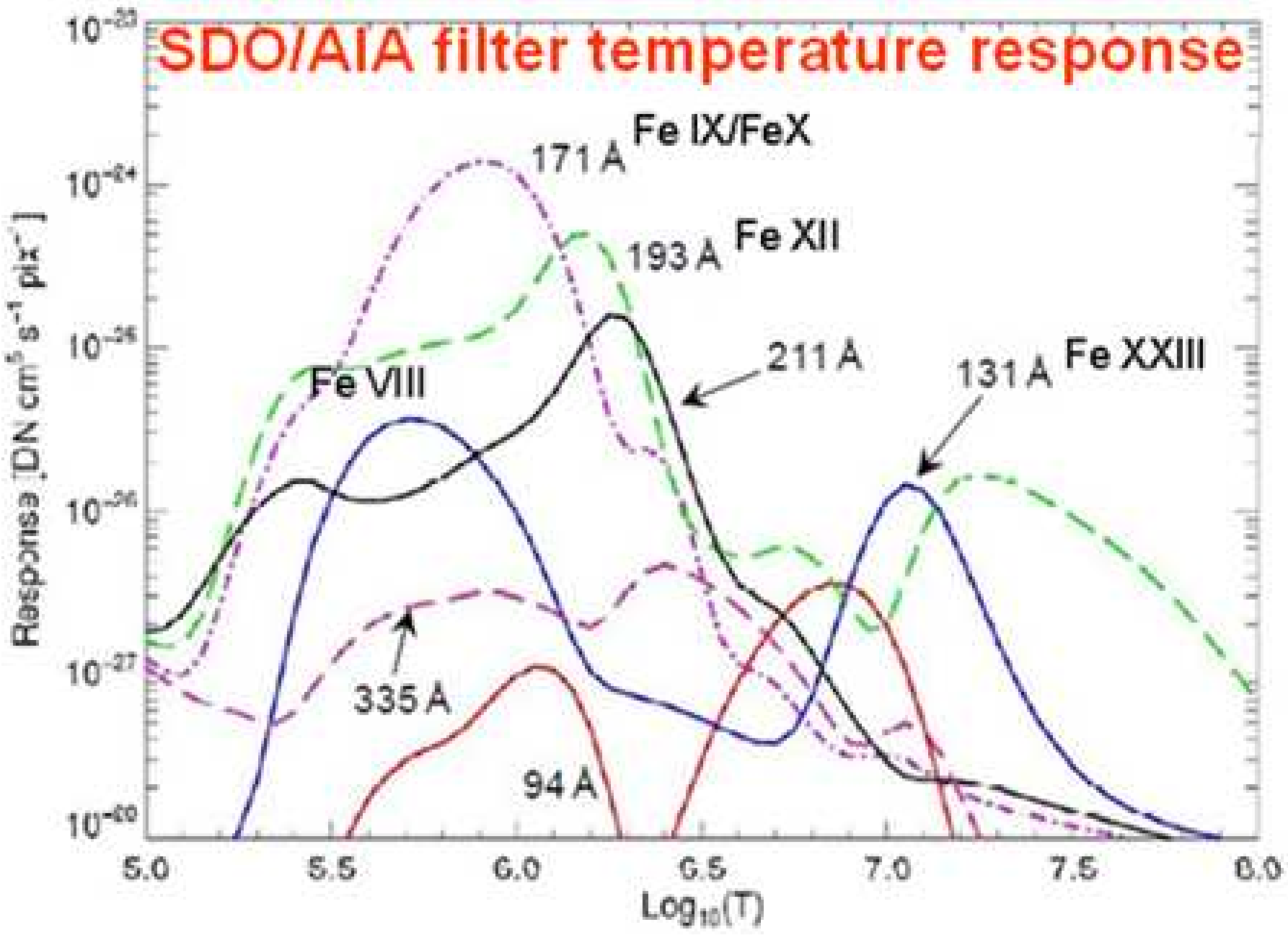

**Figure IV-6-2-3:** *Courbes des réponses en température des filtres de SDO/AIA, et pour les différents ions. D'après Boerner P. et al.*

La réponse des filtres dans la raie de l'hélium He II 304 n'est pas mentionnée, et la température de cette raie est supérieure 30000 K.

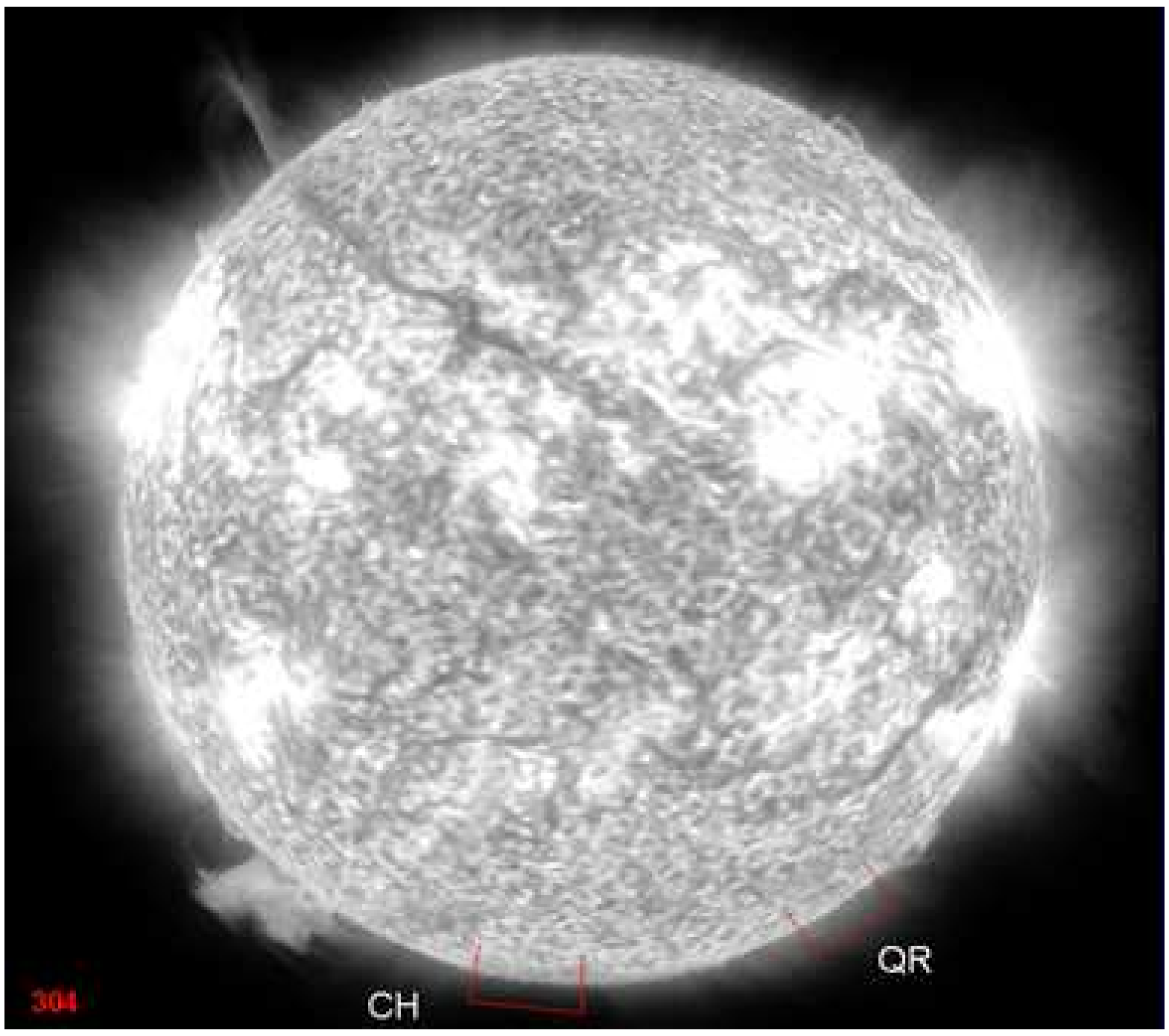

**Figure IV-6-2-4:** *Image du Soleil le 14 Nov. 2011 de 12h à 14h TU dans la raie He II 304 Å, avec les indications où les mesures des limbes ont été réalisées.*

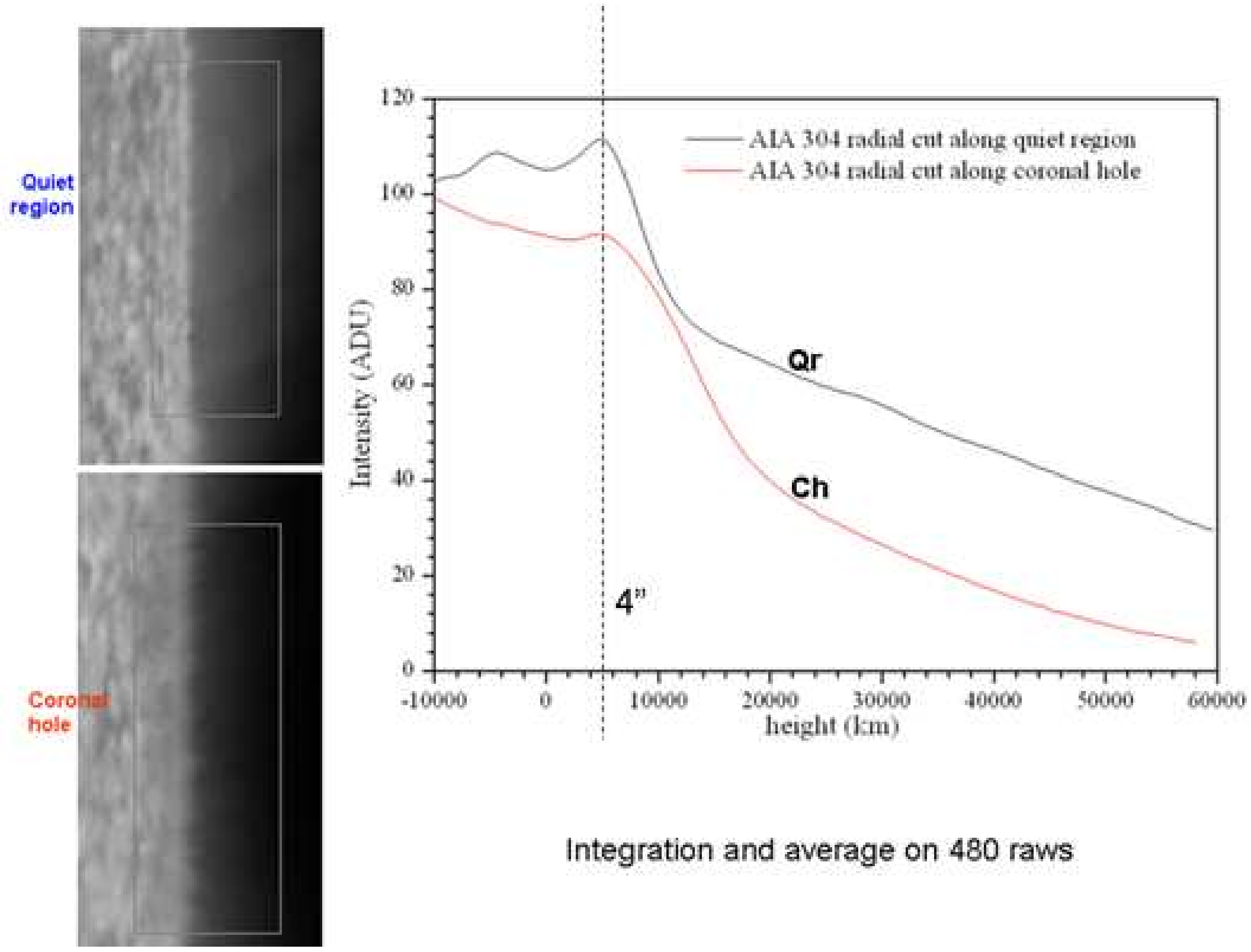

**Figure IV-6-2-5:** *Profils d'intensités des limbes en 304 Å, après linéarisation des limbes examinés et montrant un faible embrillancement dans les raies froides.*

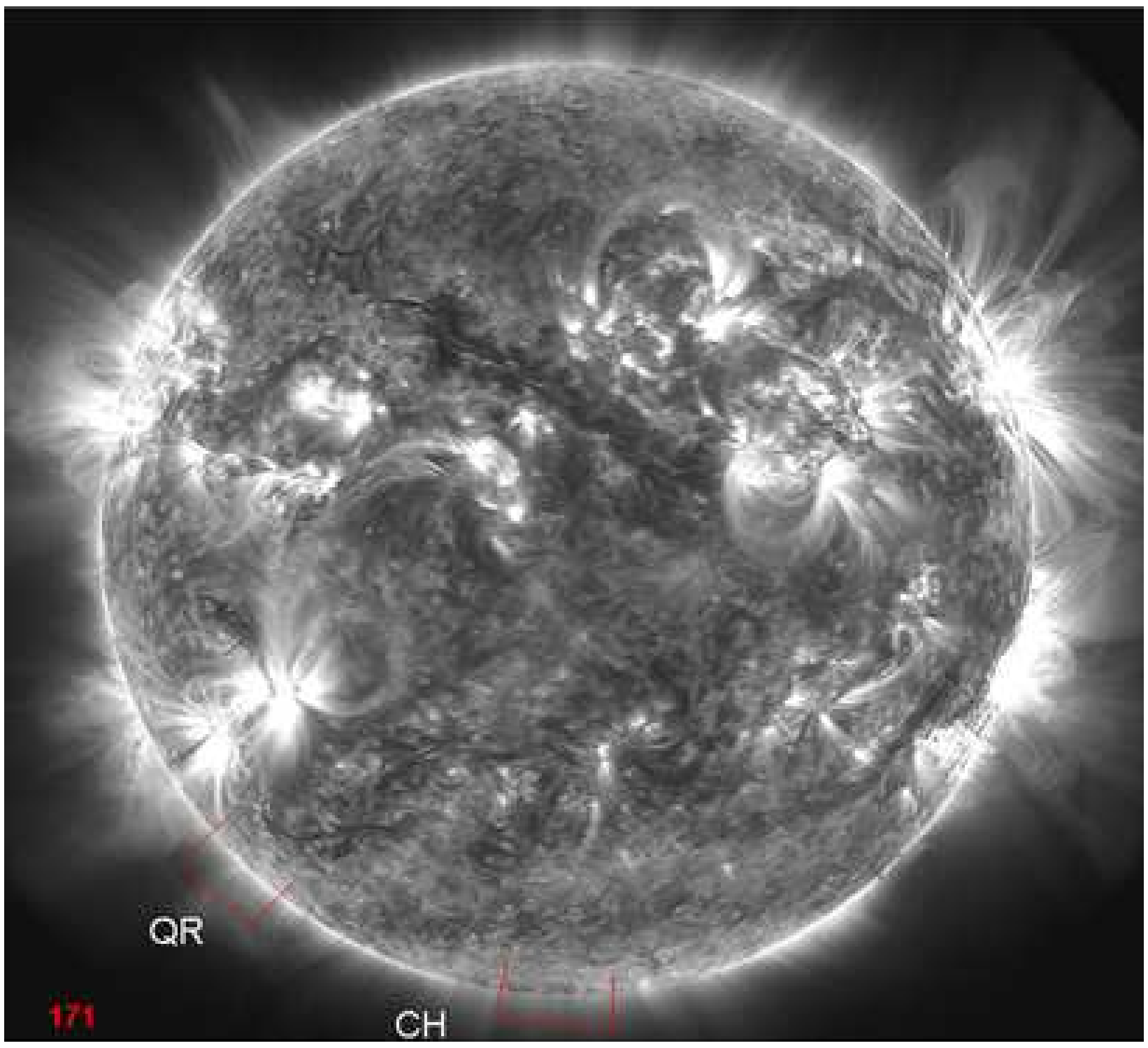

**Figure IV-6-2-6 :** *Image du Soleil le 14 Nov. 2011 de 12h à 14h TU dans la raie du Fer X à 171 Å AIA, avec les indications où les mesures des limbes ont été réalisées*

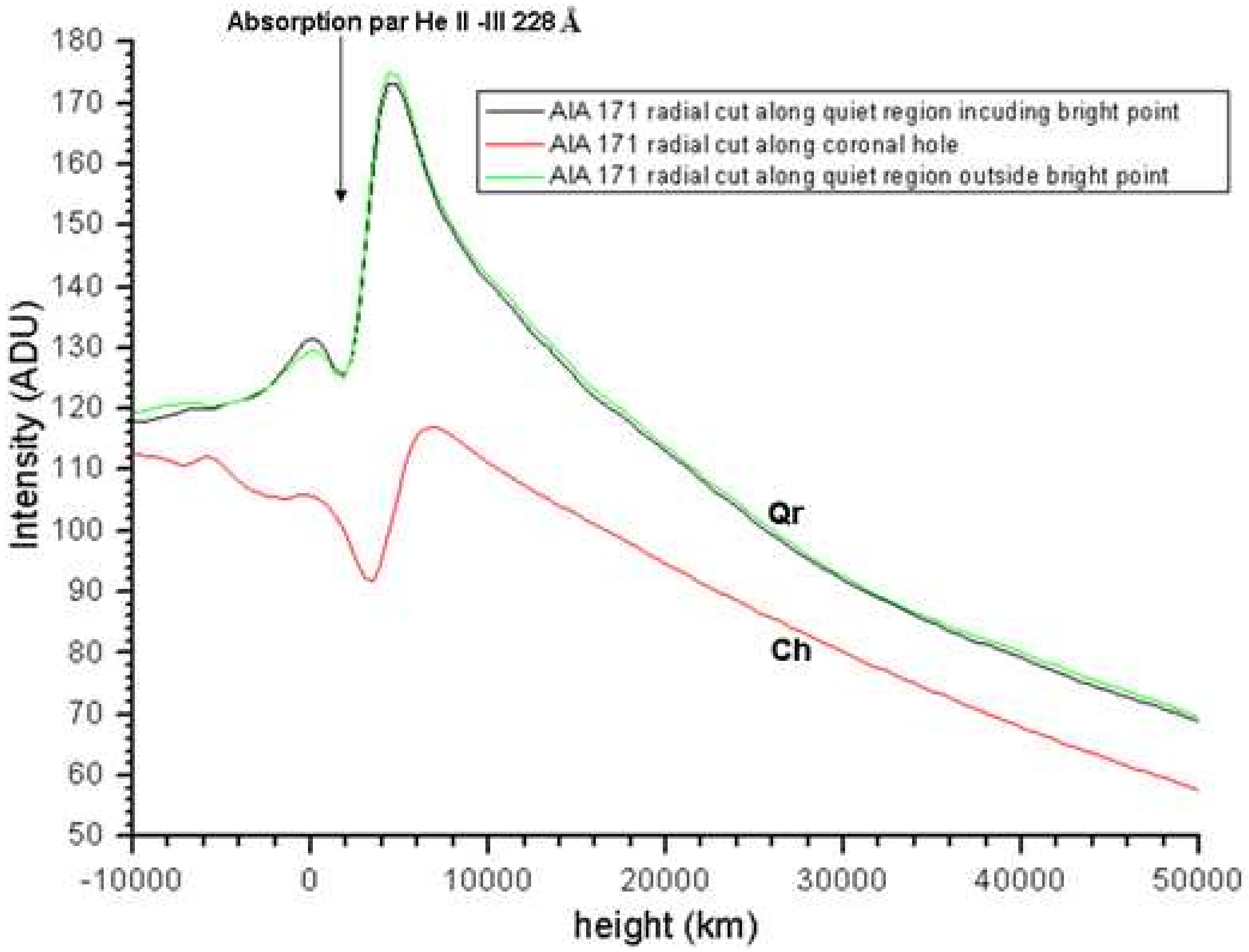

**Figure IV-6-2-7:** *Profils d'intensités des limbes à 171 Å en fonction de l'altitude. Intégrations sur 480 lignes de pixels, après linéarisation des secteurs des limbes examinés. Les embrillancements sont importants dans les raies plus chaudes.*

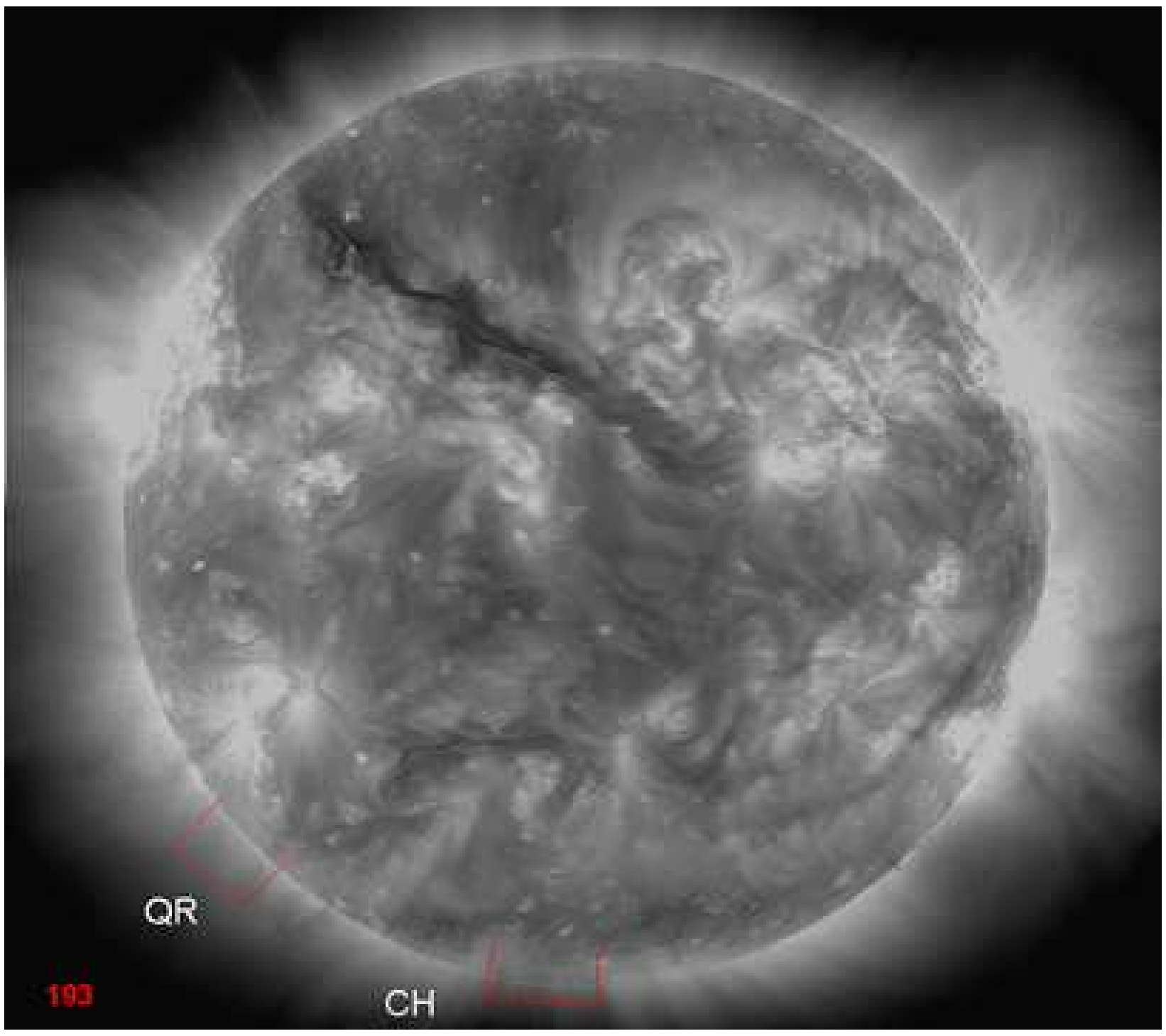

**Figure IV-6-2-8:** *Image du Soleil le 14 Nov. 2011 de 12h à 14h TU dans la raie du Fer XII à 193 Å, avec les indications où les mesures des limbes ont été réalisées*

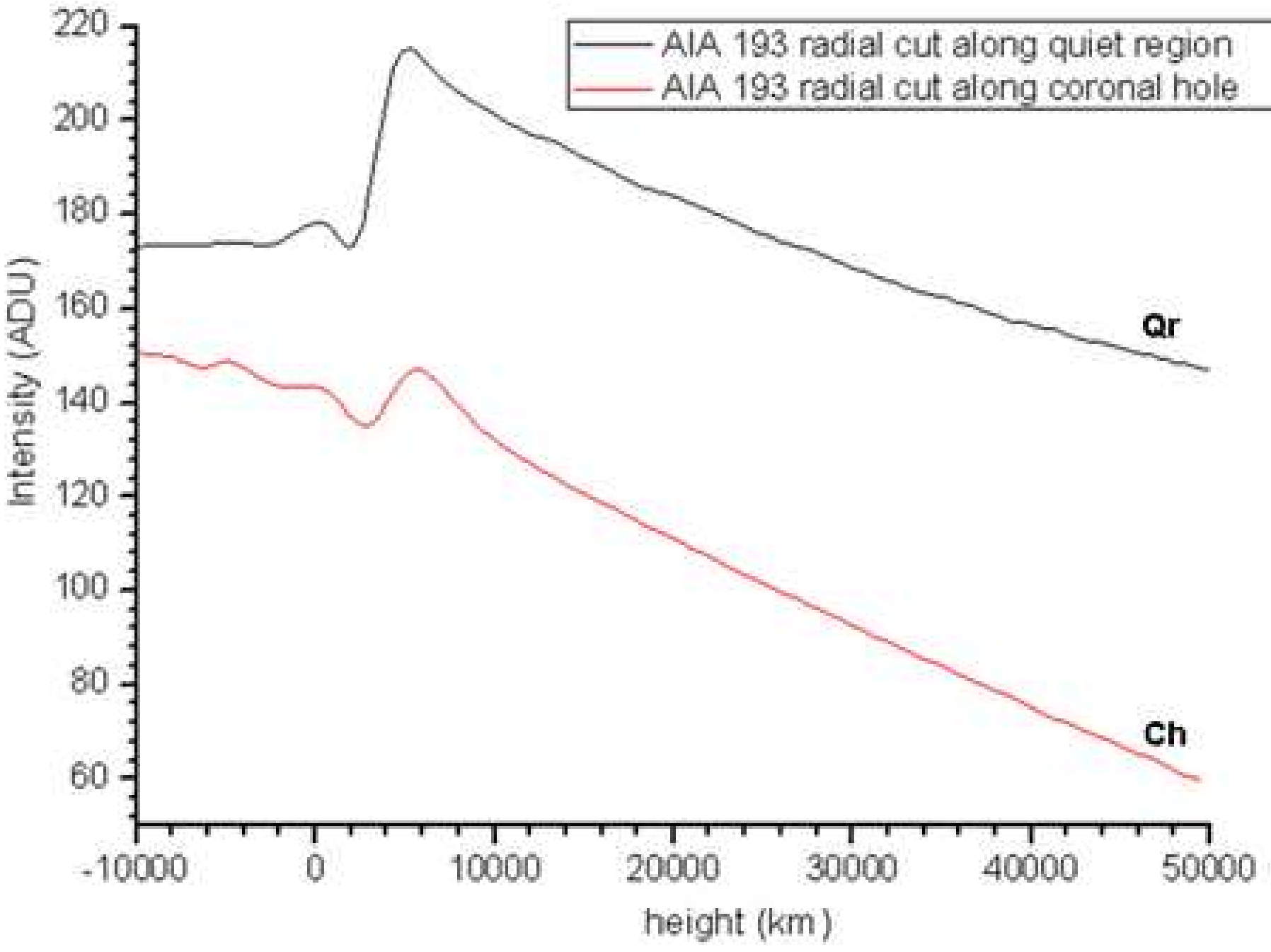

**Figure IV-6-2-9:** *Profils d'intensités des limbes à 193 Å en fonction de l'altitude. Intégrations sur 480 lignes de pixels, après linéarisation des secteurs des limbes examinés*

Les graphiques figure IV-6-2-10 résument tous les profils des limbes comparés ensemble afin d'étudier les étendues des embrillancements et les absorptions :

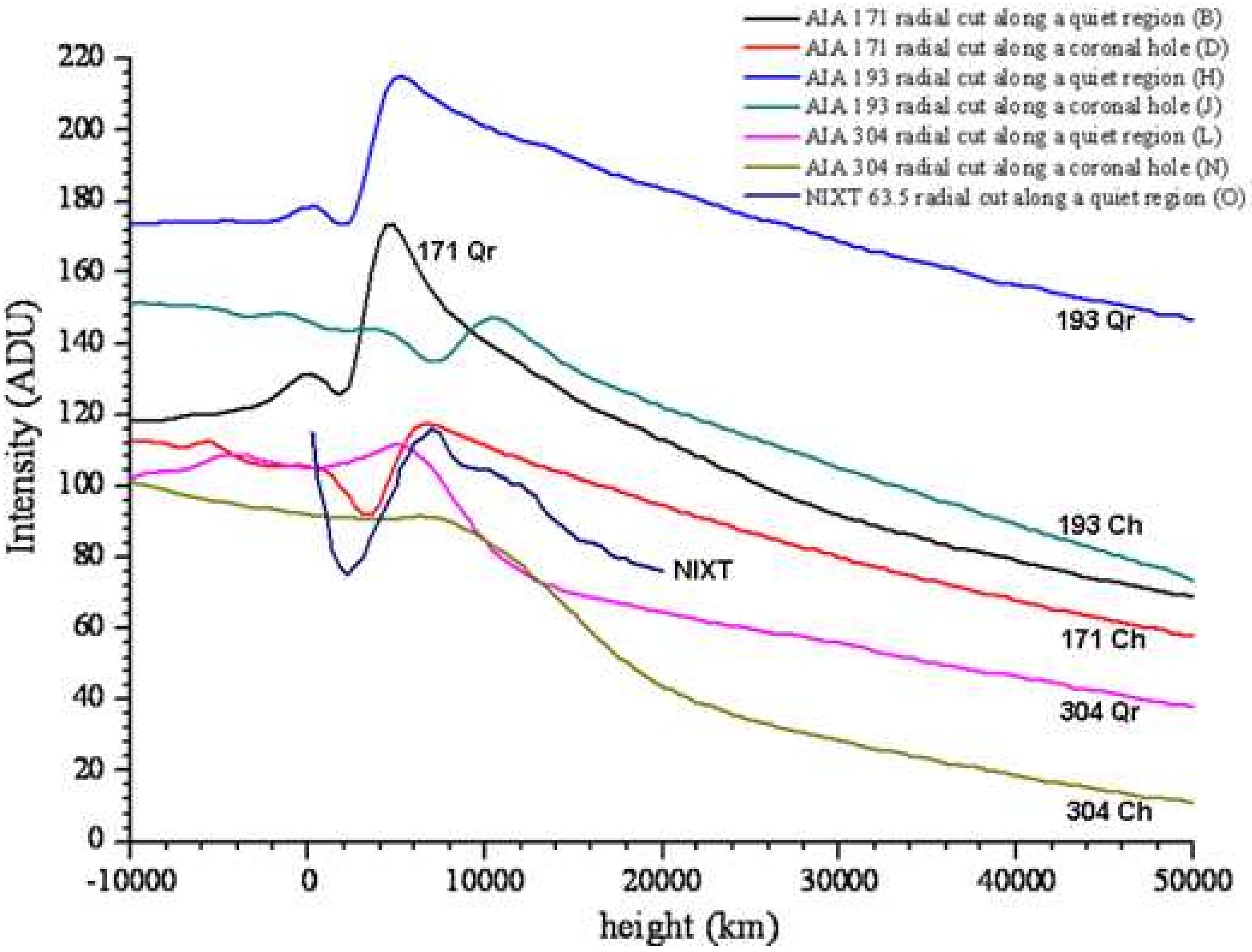

**Figure IV-6-2-10:** *graphiques comparés des profils d'embrillancements des limbes dans les EUV et avec un profil de NIXT à 63.4 Å par comparaison, mais réalisé à une date antérieure*

Les profils d'embrillacements sur les figures IV-6-2-7, IV-6-2-9, et IV-6-2-10 sont précédés d'une absorption dans les régions très proches du limbe. Ces absorptions proviennent du plasma « froid » plus dense absorbé par les raies d'émission coronales EUV, et aussi de l'absorption par He II – He III à 228Å, et recombinaison par les particules $\alpha$, voir Daw, Delucas, Golub 1995. Les profils obtenus sur la figure IV-6-2-5, dans la raie He II 304 Å optiquement épaisse, sont conformes à la région de transition chromosphère – couronne dans un trou coronal, voir Huber et al 1974, à partir des résultats de SKYLAB.
Les profils d'absorption correspondent aux altitudes comparables à celles de la mésosphère solaire, à du plasma dominé par les collisions associé aux raies optiquement minces « high FIP » de la chromosphère et de l'interface de transition photosphère-couronne. Ces profils radiaux montrent que l'étendue de l'embrillancement du limbe augmente avec la température selon le type de région. Des analyses supplémentaires avec des déconvolutions par masque flou ont été réalisées au chapitre V pour discuter de la correspondance des enveloppes d'hélium optiquement minces He I 4713Å et He II 4686 Å, pour les embrillancements observés, et mieux analyser les profils d'absorption situés juste au dessus du limbe, par analogie avec les spectres éclair dans le domaine visible et raies optiquement minces.

# Chapitre V) Interprétation et discussions des résultats obtenus aux éclipses et avec les données spatiales en EUV.

## V- 1) Interprétation des spectres éclair

Les spectres éclair montrent les modulations du profil des raies vues comme des croissants, à cause du relief lunaire, montagnes et vallées. Chaque raie indique la signature d'un élément chimique constituant les couches profondes de l'atmosphère solaire observée sur la ligne de visée. Pour la plupart des raies, les potentiels d'excitation en électron-volts sont indiqués. Chaque raie spectrale est supposée émise (ou absorbée) en liaison avec la transition d'un atome dans un niveau d'énergie à un autre différent, à la fréquence $\nu$ proportionnelle à ce changement d'énergie. Les niveaux ou états (distincts par leurs nombres quantiques) sont les composants des termes spectroscopiques, qui sont multiples. La différence entre deux niveaux d'énergie est mesurée en unités de fréquence et correspond au nombre d'onde (avec corrections du vide) de la raie spectrale résultant d'une transition électronique. Le potentiel d'excitation pour n'importe quelle raie donnée, est l'énergie $h\nu$ qui est requise pour élever l'atome de son plus bas niveau d'énergie $E_1$ vers l'état d'énergie plus élevé $E_2$ dans lequel il peut absorber cette enérgie $h\nu$, telle que $E_2 - E_1 = h\nu$. Les raies dont les niveaux les plus faibles sont caractérisés par une energie $E_1$ de quelques eV sont pratiquement exemptes d'autoabsorption, d'après St John, Moore, et al 1928. La figure V-1-1 montre les résultats des spectres éclairs obtenus en 2010 dans la même région spectrale qu'à l'éclipse de 2008, réseau objectif 600 traits/mm devant la lunette de 50 mm de diamètre et 600 mm de focale (Chapitre II-1).

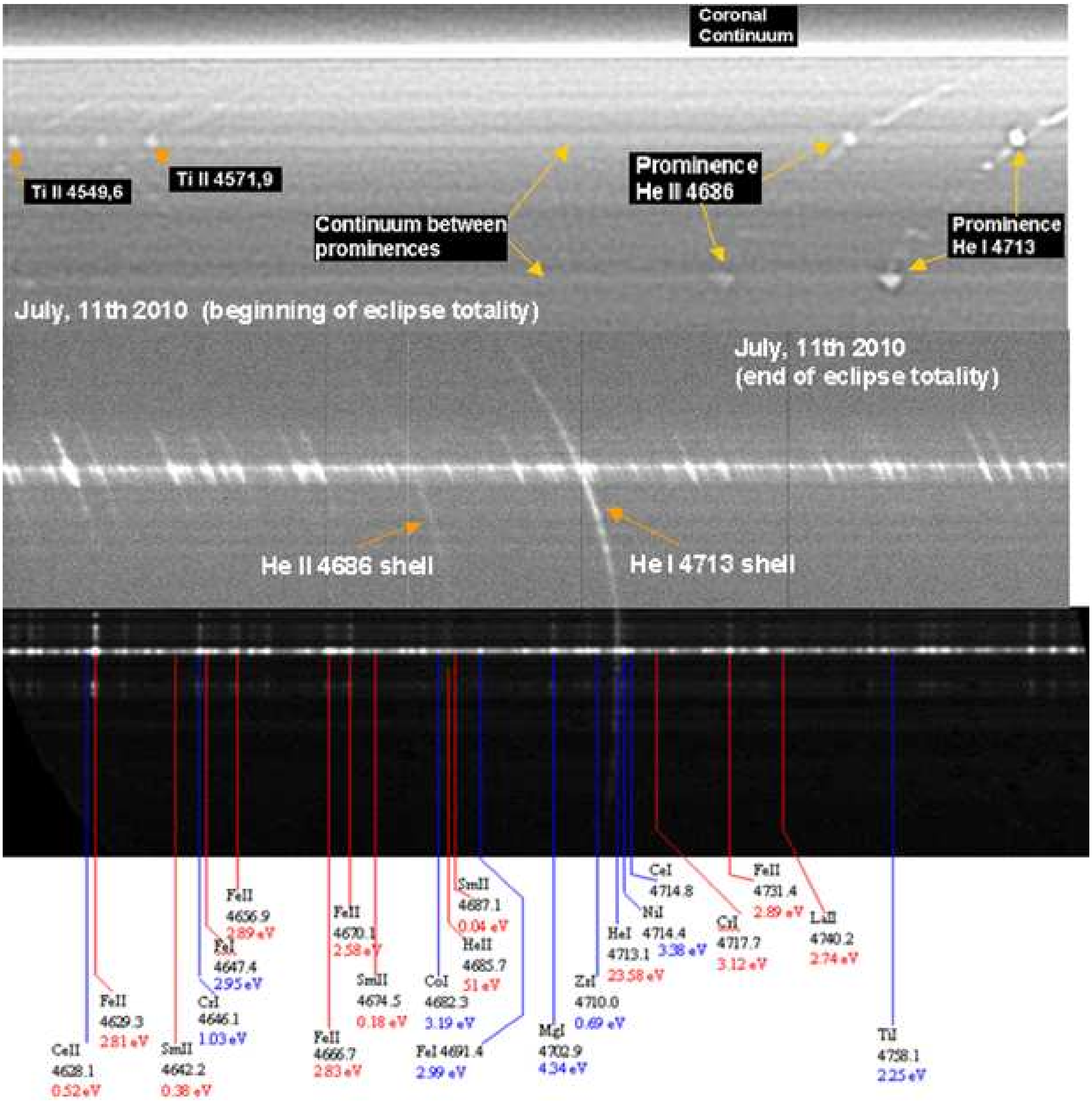

**Figure V-1-1:** *Extrait de séquences de spectres éclairs des 2 éclipses de 2008 et 2010 avec les identifications des raies, des protubérances et niveaux d'énergie correspondant aux faibles potentiels d'excitation. Seulement le spectre de 2008 (en bas) est montré après un redressement effectué par ordinateur. La dispersion spectrale est de 0.12 Å/pixel et la résolution spatiale est de 1.1 Mm/pixel.*

La majeure partie des éléments présents sont des métaux, dont les potentiels de première ionisation « low FIP » sont nettement inférieurs à 10 électrons-volts. Ces raies « low FIP » sont abondantes dans ces couches profondes de l'interface photosphère-couronne. On peut distinguer sur les extraits des spectres précédents que l'intensité des raies est la plus intense proche du milieu du croissant où la partie supérieure de la photosphère (spectre étroit continu) est encore visible dans une vallée lunaire. A ces altitudes très proches du limbe solaire, sur les spectres non saturés, la myriade de raies métaliques composant les couches profondes de l'atmosphère produit du rayonnement par émission. Le bord de la lune, par son mouvement naturel sur le bord du disque solaire, permet de masquer le reste du disque solaire, et ainsi des tranches d'atmosphère solaire, modulées par le relief lunaire, sont explorées avec séparation des raies d'émission et du spectre continu (grains de Baily). Ensuite à mesure que le bord de la Lune occulte par pas de 25 km successifs en altitude (acquisitions CCD rapide) les couches d'atmosphère solaire, les raies des spectres éclair diminuent en intensité. La quantité de rayonnement intégrée sur la ligne de visée décroit, avec l'avancement du bord de la lune et sur les extrémités des croissants d'atmosphère solaire, et les intensités des cornes des croissants sont plus faibles. La chute d'intensité continue est vertigineuse et le bord du Soleil est défini là où cette pente est la plus élevée.
Chaque croissant correspondant à une raie d'émission possède sa propre étendue, correspondant aux altitudes où est formée chaque raie. Dans le cadre de cette thèse, nous nous sommes intéressés à quelques raies, pour lesquelles nous avons analysé les profils des courbes de lumière, afin de déduire des variations dans la structuration des couches.
Nous avons étudié les variations des intensités des myriades de raies low FIP, et les rapports des intensités entre l'hélium neutre et ionisé dans les couches profondes de la région de transition et dans les protubérances.
Ces variations des rapports d'intensité peuvent être reliées aux variations du taux d'ionisation mais il est difficile de les quantifier directement d'après ces seules observations. Ces variations du taux d'ionisation des raies de l'hélium peuvent être aussi liées au brusque saut de température de 10000 K à 2 MK dans l'interface de transition entre la chromosphère et la couronne, et de façon similaire dans l'interface protubérance couronne, voir contribution Bazin, Koutchmy Tavabi 2012. Les interprétations nécéssitent un traitement Hors Equilibre Thermodynamique Local (HETL).

D'autres analyses ont été effectuées et des profils d'intensité le long d'une protubérance observée sur la fin des spectres éclairs au début de la totalité de l'éclipse de 2010. Les profils ont été obtenus selon un axe en pointillés comme le montre la figure V-1-2.

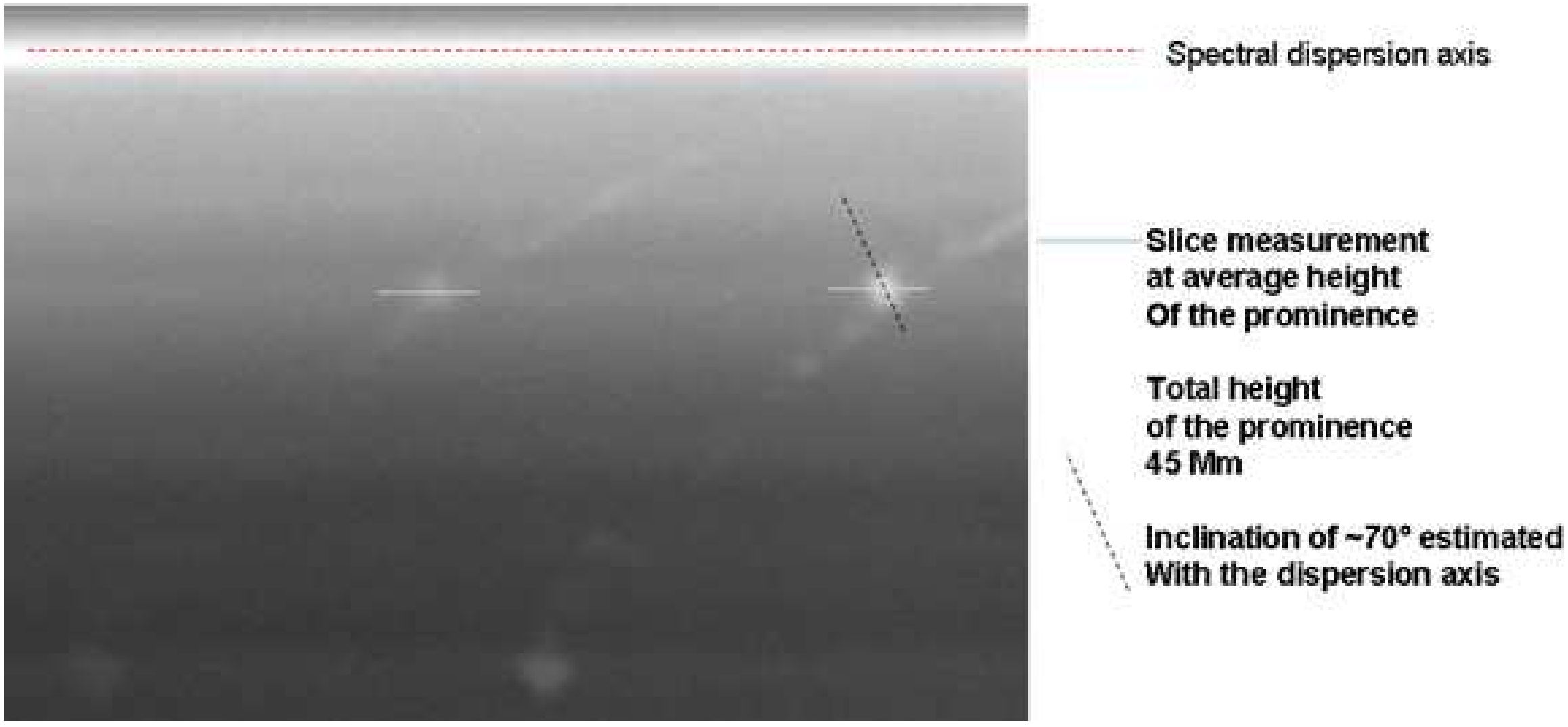

**Figure V-1-2:** *vues des protubérances en He I 4713 Å et He II 4686 Å sur lesquels les profils d'intensité sont tracés (pointillés noirs). 1 ADU = 0.823.10$^{-6}$ unités du disque solaire moyen et étalonnage sur le continu.*

La figure V-1-2 sert à montrer une méthode utilisée pour tracer des profils d'intensité le long des protubérances observées monochromatiquement dans les raies d'hélium optiquement minces, voir article Bazin Koutchmy, Tavabi 2012.

Les analyses des profils du continu dans l’interface photosphère-chromosphère aux altitudes proches et au-delà du minimum de température, peuvent être discutées et être le résultat des contributions suivantes :

- le continu de Paschen correspondant à l’hydrogène, et dont la composante dans le domaine visible est importante
- Les transitions libre lié des ions HI, au dessus de la photosphère, dans la mésophère et autour de la région du minimum de température, mais où les neutres dominent
- La diffusion Thomson des électrons dans la couronne, où les atomes du plasma sont complètement ionisés. Les flux sont de l’ordre du millionième de l’intensité du disque solaire moyen

En effet en intégrant sur la ligne de visée, et à la fin des contacts au début de la totalité ou au début du troisième contact avant la fin de la totalité, ces effets peuvent s’ajouter, avec l’expérience du réseau-objectif (spectrographe sans fente).
D’après l’article Vial Koutchmy 1992, les valeurs des flux d’émission dus à la diffusion Thomson sont 5 fois plus faibles que l’émission de Paschen, voir table V-1-3 extraite de la table N°2 de l’article Vial Koutchmy 1992:

| $n_e$ ($cm^{-3}$) | Emission Thomson (CGS) | Emission Paschen (CGS) | Emission Fer X 6374 (CGS) |
|---|---|---|---|
| $10^{11}$ | 63 | 320-640 | 1120 |
| $10^{10}$ | 6.3 | 32 | 56 |
| $10^{9}$ | 0.6 | 0.32 | 1 |

**Tableau V-1-3:** *Emissions Thomson, Paschen, et raie du Fe X à 6374 calculées pour différentes valeurs de densités électroniques. La température est supposée être de 7500 K pour l’émission de Paschen et elle est de 1.2*10 $^{6}$ K pour la raie rouge coronale du Fer X 6374.D’après les données Vial, Koutchmy 1992.*

Les mesures des échelles de hauteur pour le spectre du continu ont conduit à des valeurs de températures inférieures à 10000 K, ce qui montre que ce rayonnement observé dans les basses couches de la région de l’interface photosphère –couronne et mésosphère, est dû à de l’émission de Paschen, et dont l’intensité est proportionnelle à $\lambda^3$ (inversement proportionnel à la fréquence).
La nature de l’émission du continu est liée aussi à la faible ionisation, et donc des températures faibles. Les régions où les raies low FIP sont formées est un milieu faiblement collisionnel. Les raies « low FIP » dynamiques sont produites hors équilibre thermodynamique local, et sont sensibles au champ magnétique, avec la couronne qui pénètre dans les régions profonde. Les raies « low FIP » comme le Ti II sont présentes dans protubérances, ce qui est un résultat très important pour l’analyse de l’interface protubérance - couronne. Les raies « high FIP » comme l’hélium et l’hydrogène neutre appartenant à la chromosphère et apparaissent sous forme d’enveloppe. L’ionisation de l’hélium est complexe et peut être produite par photo-ionisation de ces raies optiquement minces par les photons EUV provenant de la couronne à des températures supérieures à 1 MK.
L’étude de l’interface protubérance-couronne est un aspect important, car comparable à l’interface photosphère –couronne. Le continu entre les protubérances est soustrait, afin d’analyser les variations d’intensité dans les images monochromatiques des protubérances et simultanément dans les raies de l’hélium neutre et ionisé et titane ionisé. Les profils en figure

V-1-4 indiquent les variations d’intensité exprimées en unités du disque solaire le long des protubérances, voir coupe figure V-1-2.

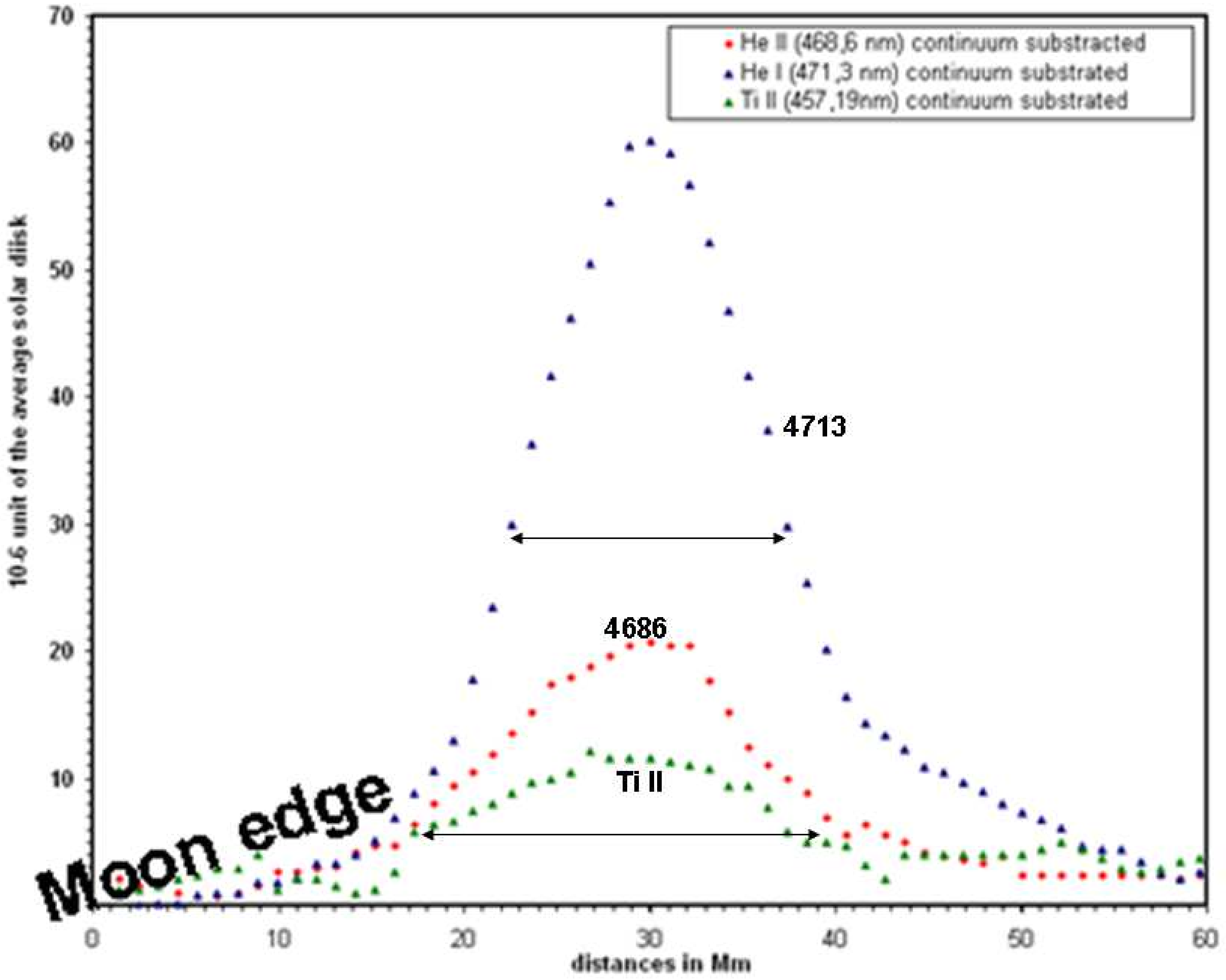

**Figure V-1-4:** *profils d’intensités le long de la protubérance prise dans les raies de l’hélium neutre He I 4713 Å, ionisé He II 4686 Å et dans la raie du Titane Ti II 4571.9 Å, après soustraction du fond coronal. D’après les spectres éclairs obtenus à l’éclipse totale du 11 Juillet 2010. Noter que le bord de la lune se situe du côté de des distances proches de zéro.*

Les profils d’intensité réalisés indiquent que l’image de la protubérance est beaucoup plus étendue dans le Titane que dans l’hélium neutre et une fois ionisé. Ce résultat est important pour utiliser le Titane ionisé comme indicateur ou traceur dans les interfaces, compte tenu de sa masse ionique de 47.9 g/moles qui est environ 12 fois supérieure à celle de l’hélium. Un rapport des intensités a été effectué pour tenter d’analyser soit des effets de variations du degré d’ionisation pour l’hélium He II 4686 Å et pour le Ti II 4572 Å par rapport à l’hélium neutre He I 4713 Å, soit des effets associés à la gravité, mais dont les effets sont limités. Ces rapports d’intensités sont donnés à la figure V-1-5.

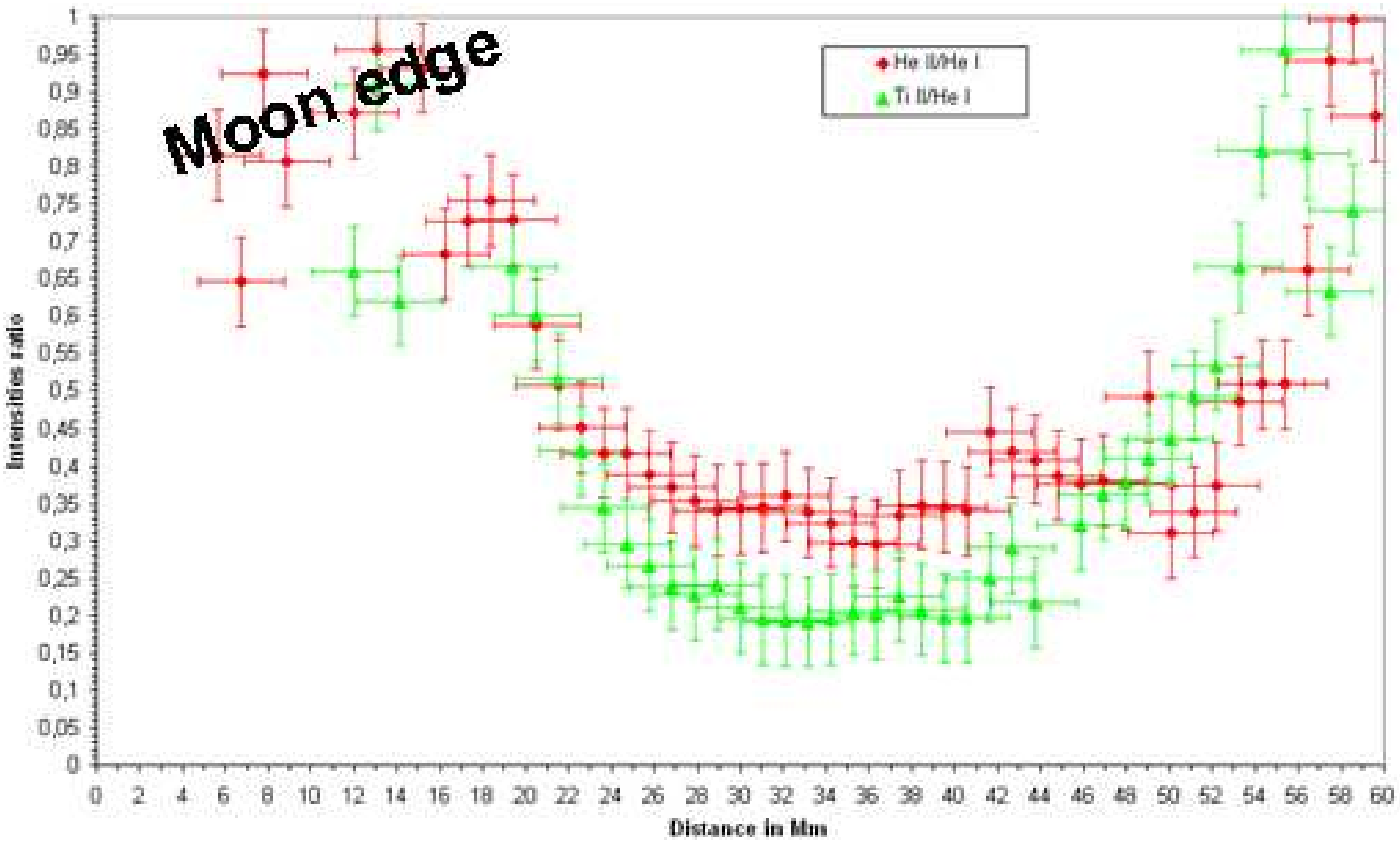

**Figure V-1-5:** *rapports d'intensités déduits d'après les profils FigureV-1-6 de He II/He I et Ti II/He I. Le bord de la Lune se situe à gauche aux distances proches de h = 0*

Ces variations des rapports d'intensité indiquent que le rapport d'intensités He II/HeI est moins important que celui de Ti II/He I, pris au centre de la protubérance, et pour une étendue de 20 Mm. Les pentes entre 18 et 26 Mm et entre 50 et 56 Mm sur la figure V-1-5 traduisent les gradients de température de l'interface protubérance-couronne.
Cette différence entre $I_{HeII} / I_{HeI}$ et $I_{TiII} / I_{HeI}$ peut être due à des effets de densité, d'abondance et de masse de l'ion dans la protubérance. En dehors de la protubérance les rapports sont proches de 1 qui correspond aux altitudes où la couronne domine.
Ces résultats montrent que les raies low FIP présentes dans les couches profondes de l'interface de transition photosphère-chromosphère, se trouvent également dans l'interface protubérance-couronne et l'on peut déduire que les processus d'ionisation sont semblables dans les deux cas. L'origine de la surabondance des éléments low FIP se situe dans les couches plus basses de l'interface photosphère-couronne.
Ces atmosphères sont considérées optiquement minces en considérant les raies dans le domaine visible. Bien que ces couches soient inhomogènes, il apparaît cependant une stratification, où les raies « low FIP » sont formées entre 200 et 600 km au dessus du bord solaire, et les enveloppes d'hélium, qui sont des raies « high FIP », commencent à apparaître à partir de 800 km du bord solaire, avec des maxima d'enbrillancements situés entre 1400 et 1700 km. Une analyse complémentaire avec les images EUV de AIA/SDO a été nécéssaire pour discuter des phénomènes d'interfaces aux altitudes comparables, entre les régions où les enveloppes d'hélium optiquement minces sont formées et l'analyse des extensions des limbes EUV, aux altitudes comparables.

## V-2) Analyses et discussions des enveloppes d'hélium par comparaison avec les images AIA

Les spectres d'éclipse dans le visible, sont comparés avec les images en Extrême- UV réalisées simultanément par la mission spatiale SDO/AIA. Cette mission produit des images de la couronne solaire dans une gamme étendue de températures comme indiquée en figure IV-6-2-3 et Annexe N° 22 où sont données les températures et les contributions des mesures d'émission. En effet ces images obtenues simultanément dans le domaine visible et le domaine Extreme- UV permettent une étude plus étendue des régions d'interface photosphère-chromosphère/couronne, où les structures observées dans les raies optiquement minces comme l'hélium neutre He I 4713 Å et une fois ionisé He II 4686 Å, sont comparées avec les raies coronales optiquement plus épaisses. Cette étude a pour objectif d'analyser comment la couronne chaude (1MK) peut pénétrer dans les couches plus profondes de l'interface de transition sur le limbe solaire. Elle a aussi pour but de comparer les analogies entre la nature des raies d'émission dans l'interface photosphère chromosphère, et dans l'interface protubérance-couronne solaire.
L'image du spectre éclair figure V-2-1 révèle la disparition du bord du Soleil lors du second contact de l'éclipse du 11 Juillet 2010 et simultanément une image du limbe avec AIA/SDO dans le Fe VIII à 131Å. Le continu du bord solaire, sur cette image correspond à la disparition des grains de Baily en lumière blanche. Les images en EUV après traitement montrent un embrillancement sur une étendue de l'ordre de 5 à 7 Mm. Ce résultat est en désaccord avec les échelles de hauteur de 50 à 100 Mm correspondant aux températures coronales supérieures ou égales à 1 MK dans l'hypothèse d'un modèle hydrostatique stratifié. L'embrillancement sur le bord est le résultat d'effets d'opacité sur la ligne de visée, mais aussi il peut être produit par des phénomènes dynamiques, liés à l'émergence du champ magnétique, jets, macrospicules, et associé aux enveloppes d'hélium neutre et ionisée et qui ne peuvent pas être décrits par des modèles hydrostatiques stratifiés.

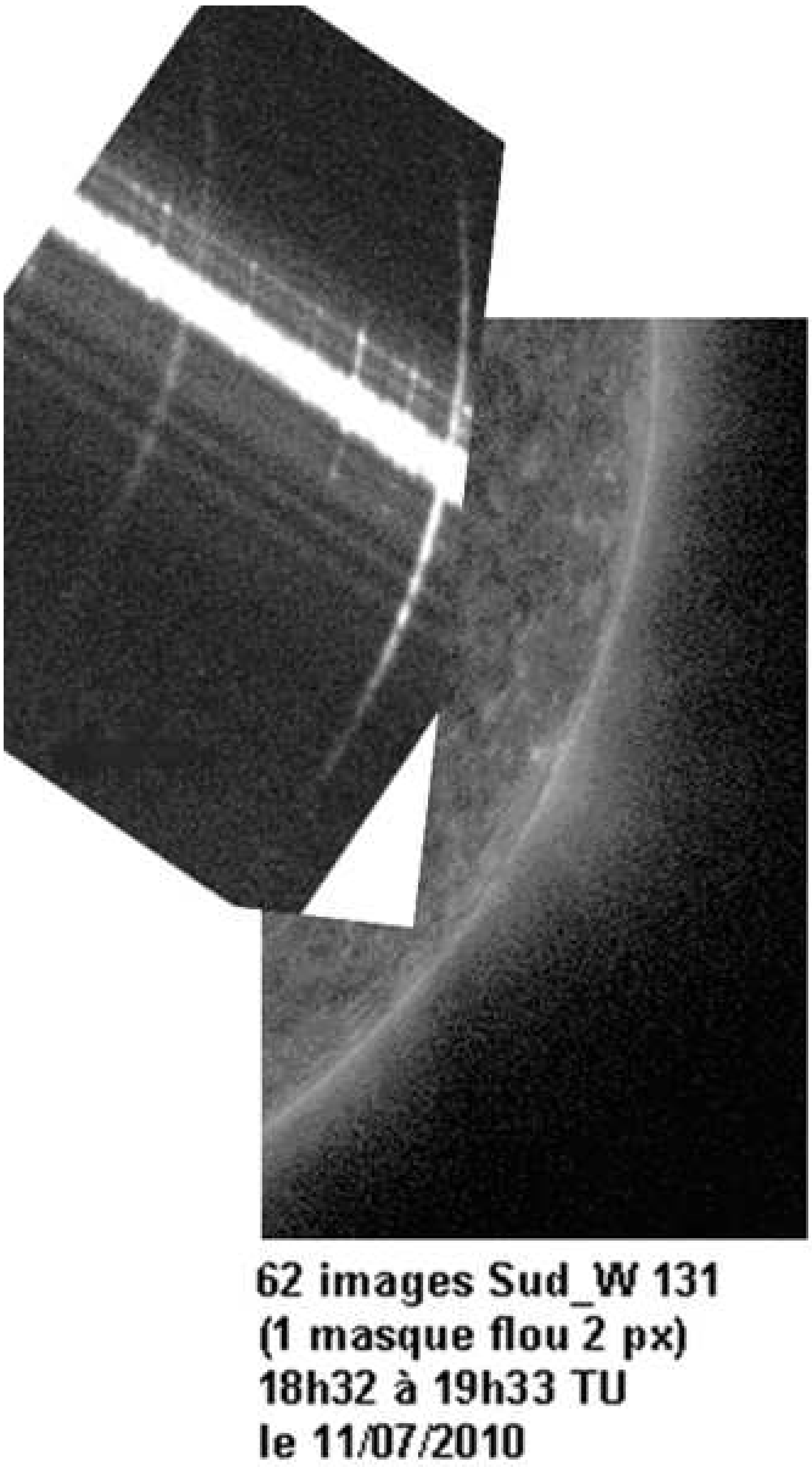

**Figure V-2-1:** *comparaison d'un extrait de spectre éclair (à gauche) comprenant les enveloppes d'hélium, avec le limbe en 131Å du Fer VIII superposé à droite au moment du troisième contact de l'éclipse du 11 Juillet 2010.*

Des phénomènes de transport d'ions et d'électrons ont lieu, où le champ magnétique est très concentré aux altitudes inférieures à 1 Mm sur le limbe solaire.
D'autres phénomènes inhomogènes comme la convection contribuent à ces mouvements de plasma aux petites échelles, mais ce type d'analyse par convection n'est pas développé ou approfondi dans cette thèse.

Les ions des petites raies low FIP peuplent ces altitudes correspondant au minimum de température.
Il est important de considérer le rôle du champ magnétique associé aux particules en mouvement dans les couches mésosphériques et au dessus, ce qui produit les inhomogénéités observées dans les couches supérieures à 1000 km, et où apparaissent les spicules. Le champ magnétique contribue au chauffage, au mouvement et transport des charges par diffusion ambipolaire le long des lignes de force.

## V-3) Discussion du champ magnétique dans les couches mésosphériques et minimum de température

Un lien peut être fait avec le champ magnétique concentré, où la pression magnétique domine : le béta du plasma est donné comme étant le rapport suivant :

$$\beta = \frac{pression_totale_du_plasma}{pression_magnetique}$$

le $\beta$ du plasma est inférieur à 1 dans ces régions autour du minimum de température, car la pression magnétique est dominante.

Le graphique V-3-1 est obtenu à partir des pressions calculées pour le modèle du VAL C.
A partir des valeurs des pressions qui varient selon une loi exponentielle décroissante, il a été possible ensuite de faire le calcul de correspondance des échelles de hauteur en fonction de la hauteur au dessus du limbe solaire.

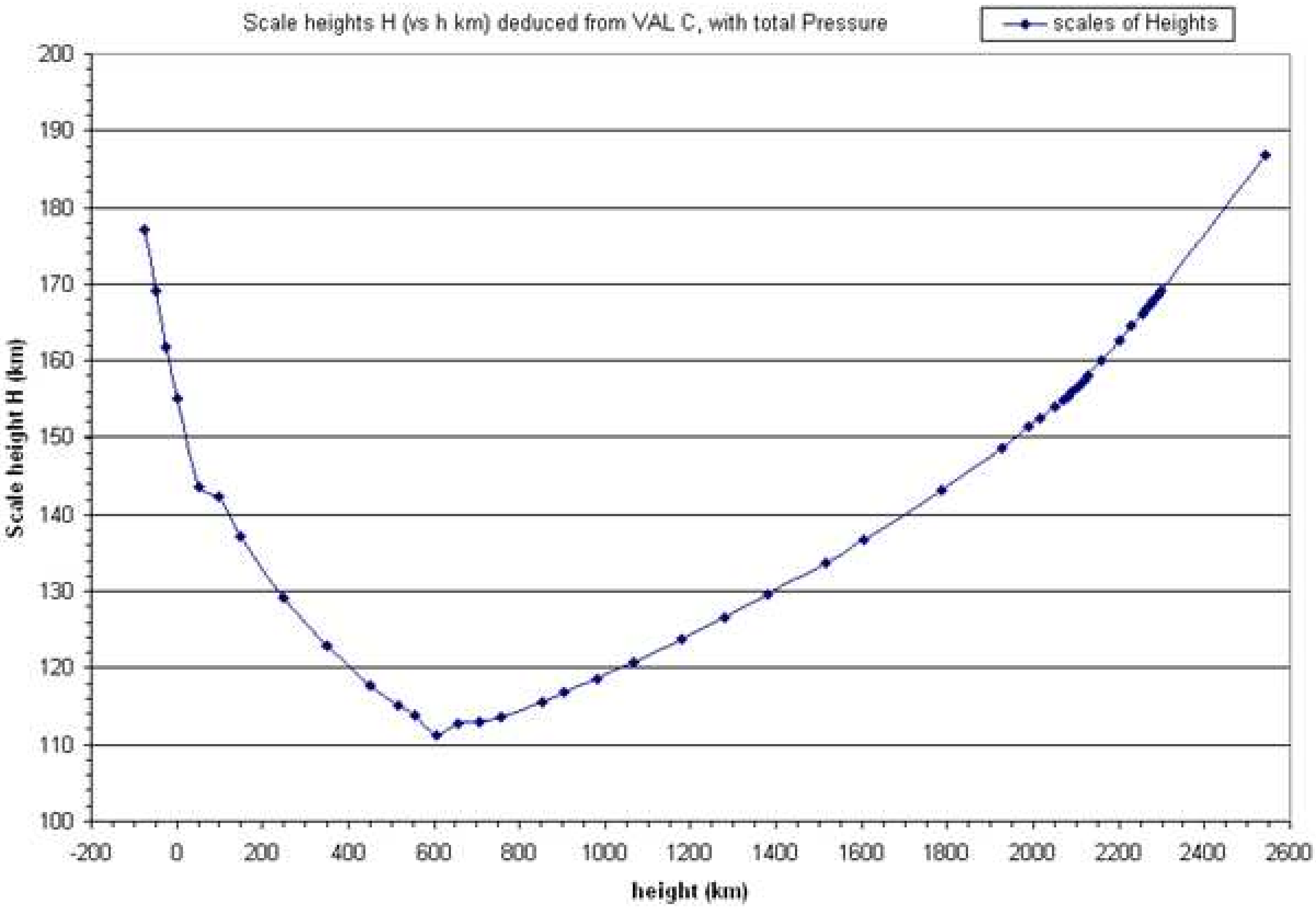

**Figure V-3-1:** *correspondances des échelles de hauteur en fonction des altitudes, déduites à partir des tableaux des pressions en fonction des altitudes hydrostatiques, issus du modèle VAL.*

Ces graphiques sont considérés par la suite pour les calculs d'échelles de hauteur et les correspondances, afin d'analyser des corrélations entre ces modèles et analyses des courbes de lumière et émissivités des spectres éclairs. D'autres graphiques issus de ce modèle permettent de comparer les profils des paramètres de densité, température, pressions, afin de mieux interpréter la structuration des couches profondes.

Les graphiques suivants ont été extraits du modèle de VAL, afin de comparer les différents paramètres avec les échelles de hauteurs précédemment évaluées. Ils sont effectués pour comparer les profils de densité avec ceux des courbes de lumière et émissivités des ions de low FIP comme le Fer et Magnésium, puis le continu (traduisant la diffusion Thomson des électrons et Paschen $\alpha$) et les enveloppes d'hélium neutre et ionisé. Les échelles sont ajustées par des coefficients d'échelles, afin de les représenter tous sur un même graphe. Ces graphiques V-3-2 et V-3-2 sont extraits des données du modèle VAL, pour montrer les variations des paramètres de densité, pression et température dans les altitudes correspondant au minimum de température, avec un modèle hydrostatique stratifié. Nous avons montré que ces modèles ne sont pas adaptés pour décrire les échelles de hauteurs et surtout les températures, déduites sur les courbes de lumière des raies low FIP lors des éclipses totales.

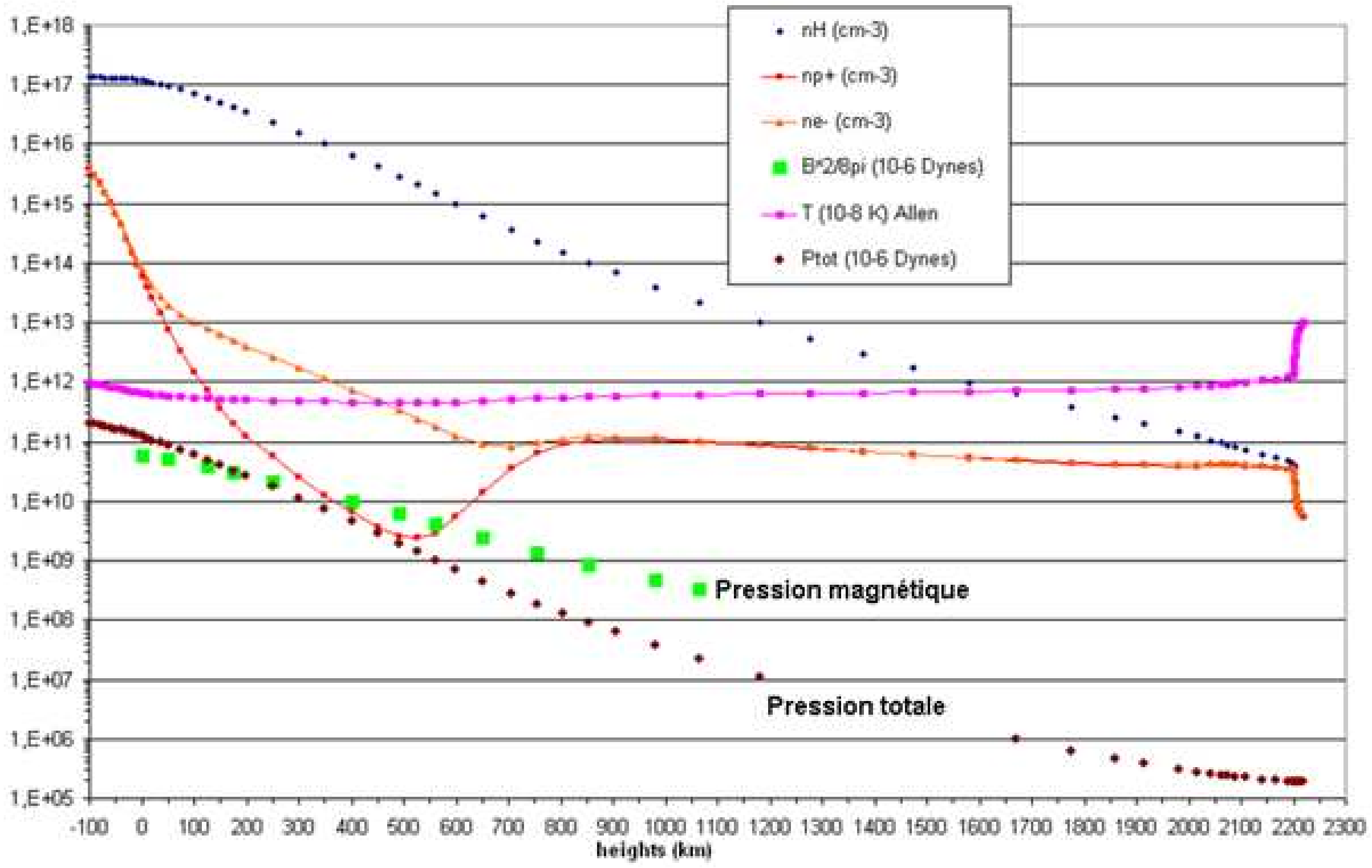

**Figure V-3-2:** *variations des paramètres de pression magnétique, pression totale, densités des neutres, des protons, électrons et température en fonction de la hauteur au dessus du bord solaire. D'après le modèle VAL.*

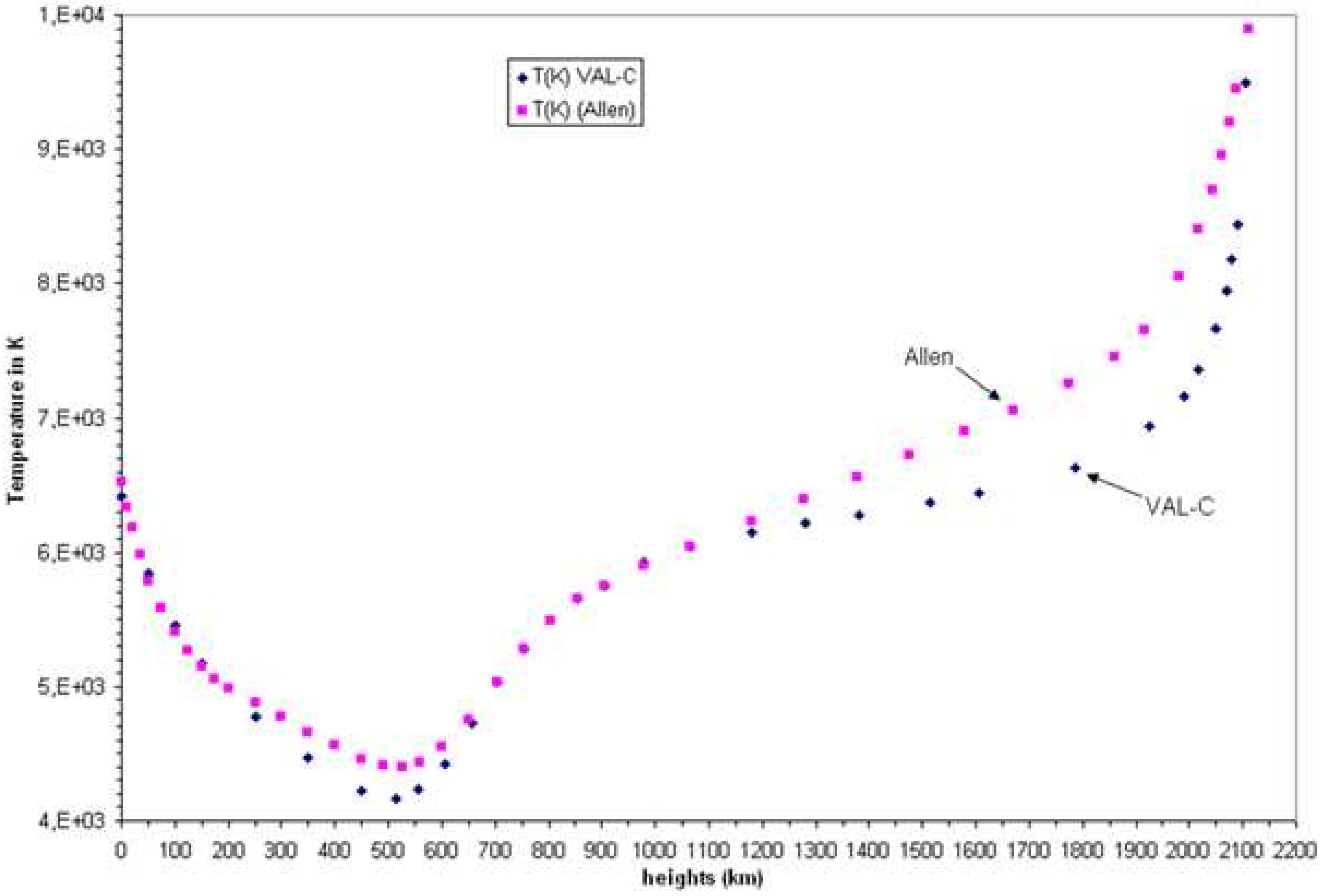

**Figure V-3-3:** *régions agrandies autour du minimum de température avec localisation des altitudes, en ayant comparé le modèle VAL, et les données Allen.*

Des minima de température et de densités sont représentés pour des altitudes comprises entre 150 et 700 km. Or d'après les analyses des données des spectres éclairs, il apparaît clairement que d'après les échelles de hauteur mesurées pour les raies low FIP comme le Fer, Magnésium, Titane, Barium (étudiés à l'éclipse de 2009, Chapitres IV et V), ces éléments correspondent et sont formés aux altitudes de ce minimum de température. Au dessus du minimum de température, nous avons mesuré d'après les courbes de lumière des spectres éclair, que les raies d'hélium He I 4713Å et He II 4686 Å sont formées au dessus du minimum de température, c'est-à-dire à partir de 700 à 800 kilomètres au dessus du limbe et où commence la chromosphère, dominée par les raies « high FIP » associées aux enveloppes.

L'altitude $h$ = 0 des modèles correspond à un bord défini par Thuillier et al 2011, Piau, L. et al 2010 avec un point d'inflexion. Les altitudes de formation des raies tiennent compte de l'épaisseur optique prise à $\tau_{5000}$ = 1 dans la haute photosphère (Annexe 24), et qui correspond à une altitude de 330 km prise au dessus du "vrai" bord solaire, c'est-à-dire du continu des spectres de grains de Baily, relevé entre les raies d'émission, et en l'absence de tourte lumière parasite, aux instants de contacts.

Le champ magnétique aux petites échelles de 160 à 300 km est discuté, où des observations supplémentaires ont été apportées à des altitudes restant inférieures à 500 km au dessus du bord solaire, depuis les expériences d'éclipses analysées par R-G Athay (éclipses de 1952 et 1970).

La densité des protons décroît avec l'altitude comme indiqué en figure V-3-2, et le degré d'ionisation des éléments low FIP comme le fer augmente dans l'interface de transition photosphère-chromosphère-couronne et perdent ensuite la moitié de leurs éléctrons comme indiqué en figure A-10-3 de l'Annexe 10 et figure I-1-3-1 sur l'interface protubérance-couronne. La densité des autres ions et des neutres décroit aussi avec l'altitude mais le degré d’ionisation augmente dans les interfaces, ce qui pourrait contribuer au brusque saut de température et aussi jouer un rôle dans les mécanismes du chauffage de la couronne qui sont encore mal connus et mal expliqués à ce jour. L’analogie entre les interfaces protubérances-couronne et interface photosphère-couronne, dans lesquelles le titane ionisé est présent est un point important sur l’enrichissement de ces régions en métaux, et pour l’action et de la pénétration de la couronne chaude qui ionise des couches de plasma plus froid.
C’est pourquoi, et pour des analyses plus étendues, nous avons aussi étudié les raies d’hélium neutre et ionisé grâce aux spectres éclair. Ces raies ont un potentiel d’ionisation supérieur à 10 eV, high FIP, et jouent un rôle d’interface, avec la couronne ambiante à 1 MK qui pourrait photo-ioniser ces raies plus froides. Le mécanisme aussi pourrait être des collisions dues à des éléctrons rapides en provenance de la couronne.

Nous avons évalué l’éxtension de l’enveloppe d’hélium ionisé à 4686 Å pour des couches optiquement minces à l'éclipse du 1er Août 2008. Ces raies de l’hélium neutre et ionisé ont l’avantage d’être émises à des longueurs d’onde très voisines sur un spectre effectué sur un intervalle allant de 4620 Å à 4750 Å, d’où la possibilité de les obtenir simultanément sur un même spectre éclair CCD. Ces raies d’hélium correspondent évidemment à des potentiels d’ionisation plus élevés que les raies low FIP: 24 eV pour He I à 4713 Å et 54 eV pour He II 4686 Å.
Cela permet d’apporter des éléments relatifs à l’embrillancement constaté sur le pourtour du limbe solaire qui est aussi observé dans la raie du Carbone C IV à 1600 Å, et aussi avec les images de SDO/AIA en absorption à 193 Å.
Par ailleurs, nous avons observé la raie verte du Fe XIV à 5302.86 Å à partir des spectres flash CCD lors de l’éclipse totale du 13 Novembre 2012. Nous avons décelé la présence de cette raie dans les couches correspondant aux altitudes de la région de transition chromosphère-couronne solaire. Les profils photométriques de celle-ci ont été ensuite analysés, une fois le continu coronal soustrait.
La myriade de raies low FIP peut être aussi associée à la pénétration de la couronne à 2 MK dans ces couches plus froides et profondes. Les mécanismes d’ionisation de l’hélium sont encore mal compris. Leur nature peut être de type collisionnel, par des faisceaux d’électrons, ou accélération de particules (ions, éléctrons) avec le champ magnétique dans la couronne, ou bien par photo-ionisation par les EUV en provenance des raies d’atomes ayant perdu plus de la moitié de leurs éléctrons dans la couronne. Mais ce problème est récurrent. D’où vient l’énergie pour arracher les électrons des atomes du plasma et porter sa température de 10000K à quelques millions de Kelvins sur une si courte distance ?
Le rôle du champ magnétique contribue aux mécanismes d’ionisation, mais il est difficile à mesurer. Cependant son action est détectée de manière indirecte, sur la manière dont le plasma est transporté par exemple. Dans cette thèse, nous avons analysé les répartitions des différents ions de masses atomiques différentes et évalué les échelles de hauteur pour tenter d’identifier des différences de températures et une structuration de ces couches et selon la masse atomique des éléments et de raies low FIP optiquement minces. Les courbes de lumière $I = f(h)$ pour chaque ion, se situent dans des altitudes relativement distinctes dans l’atmosphère solaire. Les raies « low FIP » sont situées dans le minimum de température entre 200 et 600 km, au dessus de la haute photosphère, et les raies d’hélium neutre et ionisé sont réparties au dessus du minimum de température à partir de 800 km.

Par ailleurs, le calcul de températures des raies de l'hélium neutre et ionisé provient des travaux réalisés par Hirayama, 1972, en supposant que les raies He II 4686Å et He I 4713 Å sont émises aux mêmes régions.
Les rapports d’intensités des raies ont été analysés, en utilisant la formule suivante qui ne fait pas intervenir les lois de Saha:

$$\frac{I(HeII_4686)}{I(HeI_4713)} = \frac{4.6*10^9}{N_e} \text{ pour } T_e = 10^4 \text{ K et}$$

$$\frac{I(HeII_4686)}{I(HeI_4713)} = \frac{12.7*10^9}{N_e} \text{ pour } T_e = 2*10^4 \text{ K}$$

$N_e$ , densité électronique, est comprise entre $10^{10}$ et $10^{11}/cm^3$ si les rapport d'intensité de He II sur he I sont compris entre 0.1 et 0.5

$T_e$ est la température électronique (Hirayama 1972).

Ces calculs sont effectués en considérant les intensités de rayonnement EUV, en dessous de 228Å, voir Hinteregger 1965, où les sections efficaces de photoionisation et coefficients d'absorption sont donnés pour les gaz rares dans le domaine EUV.

La température de radiation du continu de Lyman de l'hélium He II donnée par l'équation suivante est de $2*10^4$ K :

$$\frac{N_e n_{III}}{n_{II}} = 7.2*10^8 * \left(\frac{T_e}{10^8}\right)^{0.8}$$

où $n_{III}$ est la densité de He III, $n_{II}$ est la densité de He II

Les températures déduites des profils de raies sont inférieures à 10000 K, car les rapports d'intensité He II/He I sont compris entre 0.1 et 0.5 dans les protubérances voir les articles en Annexes 1 et 2. Les résultats des mesures d'échelles de hauteurs pour les courbes de lumière et émissivités de ces raies avaient conduit à des mesures de température 10 à 20 fois supérieures en considérant un modèle hydrostatique. Les résultats et calculs d'Hyrayama semblent plus appropriés à nos travaux pour évaluer les températures de ces raies.

L'expérience d'Hyrayama qui a été réalisée avec un spectrographe à fente n'a pas permis d'examiner les enveloppes d'hélium.

Dans notre cas, comme nous avons un réseau-objectif, nous ne pouvons pas déduire le profil des raies, mais seulement leur extension avec les altitudes dans les croissants, ce qui rend complémentaires ces travaux avec ceux effectués avec un spectrographe à fente (Hirayama 1972).

Les différents modèles VAL montrent quelques variations des profils sur les minima de température, voir figure V-3-3. Le minimum de température à 4300 K correspond aussi au minimum de densité des protons (hydrogène ionisé) entre 200 et 600 km. Or c'est à ces mêmes altitudes que sont formées les raies low FIP, associées aux métaux.

Le schéma V-3-4 permet de décrire de façon naïve ou intuitive la structuration des couches, et appréhender la manière dont le champ magnétique émerge depuis les régions profondes, juste au dessus de la photosphère, que l'on désignera comme mésosphère qui est décrite dans l'article Bazin, Koutchmy 2012, voir Annexe 2:

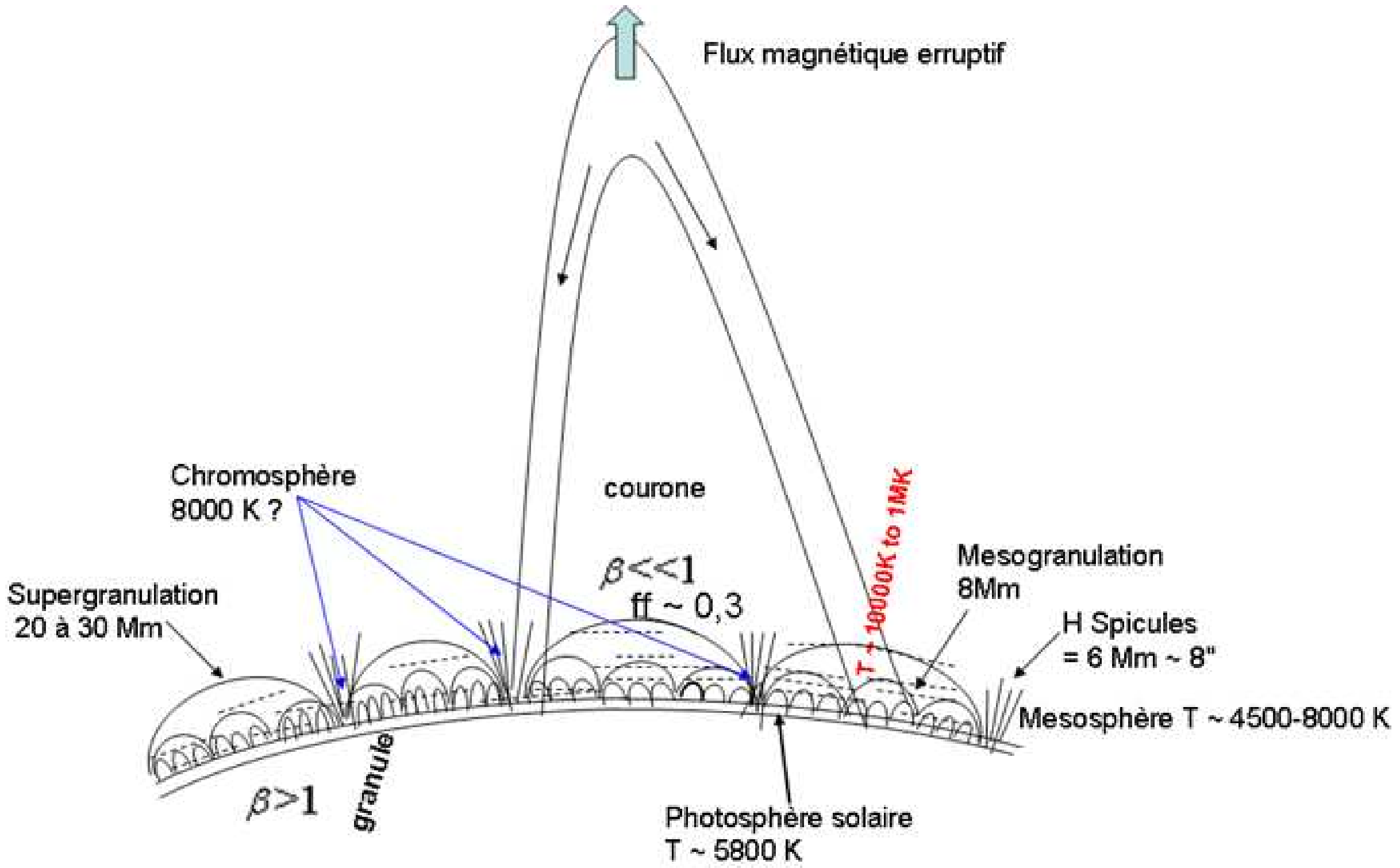

**Figure V-3-4:** *schéma possible de la structuration des couches atmosphériques solaire, vue sur le limbe, déduit d'après les observations des spectres éclair d'éclipses, en l'absence de lumière parasite diffusée.*

Le limbe solaire est composé de cellules d'échelles variables, allant de la granulation de plusieurs Mm, à la mésogranulation de 10 Mm jusqu'à la supergranulation allant jusqu'à 30 Mm. Le terme de mésosphère utilisé dans cette thèse est associé aux cellules de mésogranulation, et dans lesquelles seraient concentrées les raies « low FIP ». Le schéma figure V-3-4 permet de représenter une structuration possible des couches de l'atmosphère solaire en observant sur le limbe, par analogie avec les croissants des raies des spectres éclairs. Dans le cadre de cette thèse, nous n'utilisons pas les observations sur le disque, mais sur le limbe. Les valeurs du béta du plasma inférieures à 1 indiquent que le champ magnétique est dominant dans ces régions, il sert à « guider » le plasma vers la couronne, et peut contribuer à alimenter la couronne en éléments low FIP. Le brusque saut de température, $10^4$ à $10^6$ K, de la région de transition chromosphère-couronne est indiqué, pour présenter la problématique du chauffage de la couronne solaire.

## V-4) Discussion sur les mécanismes d'excitation et ionisation des raies He I 4713 Å et He II 4686 Å et leur structuration sous forme d'enveloppes

L'observation des enveloppes d'hélium grâce aux spectres éclair CCD a été une nouveauté, surtout pour la raie de l'hélium ionisé 4686Å dans les raies optiquement minces, où l'on ne s'attend pas à l'observer aussi comme une enveloppe. Des analyses des spicules et leur étendue avec le SOT de Hinode dans la raie H du Ca II ont été effectuées pour comparer les extensions des enveloppes d'hélium, et essayer de rechercher des corrélations entre les raies « low FIP » et « high FIP », et discuter sur les altitudes de formation des raies. La figure V-4-1 montre une comparaison entre les mesures d'extensions des spicules et les extensions des enveloppes d'hélium aux éclipses.

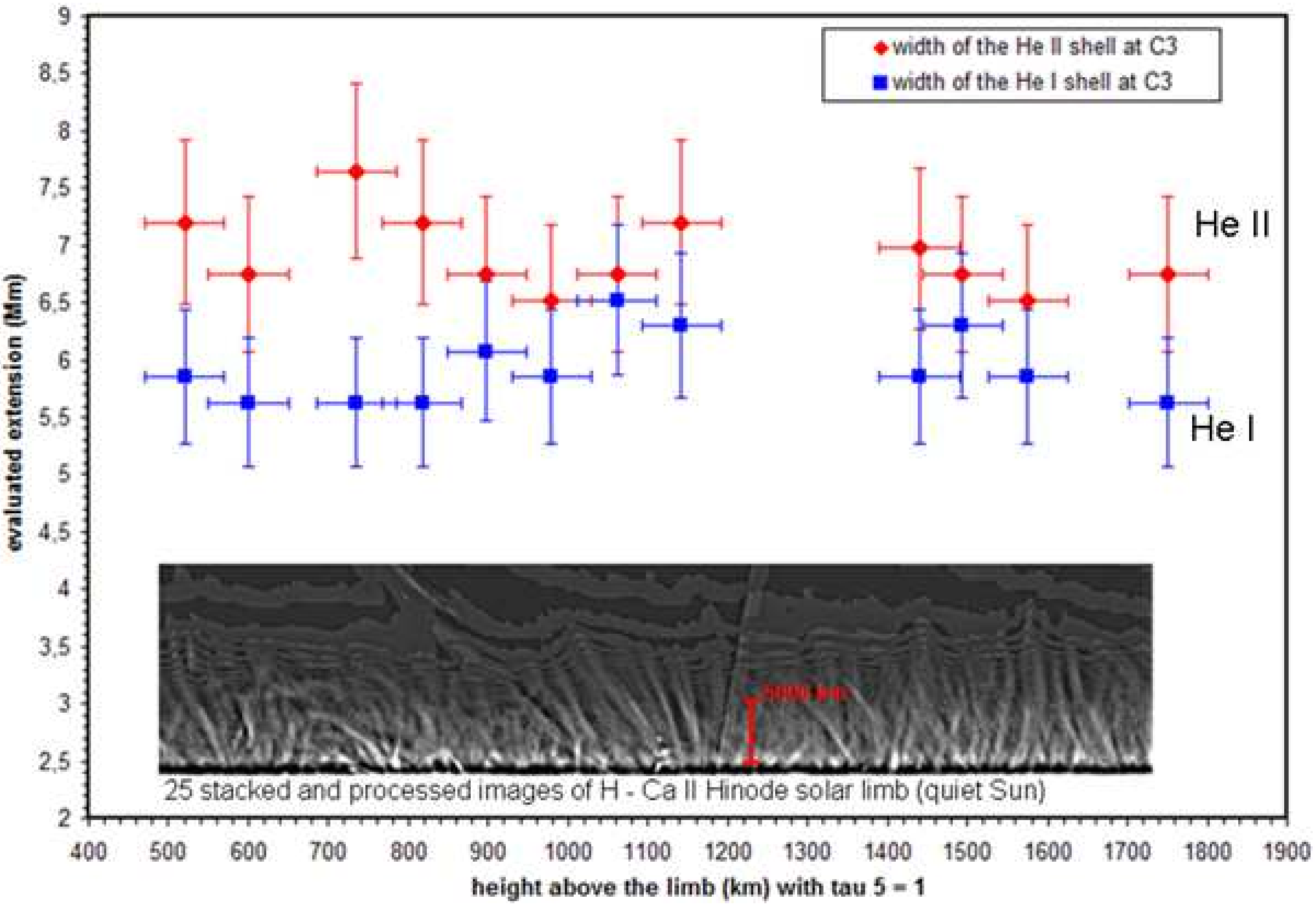

**Figure V-4-1:** *comparaison des étendues des enveloppes de spicules dans la raie H – Ca II et des étendues des enveloppes d'hélium He II 4686 Å et He I 4713 Å dans les raies optiquement minces.*

Ce graphique V-4-1 avec une image traitée du SOT de Hinode en H – Ca II sert à montrer grâce au renforcement des images des spicules, une corrélation entre l'étendue des spicules H – Ca II qui est une raie optiquement plus épaisse et l'éxtension des enveloppes d'hélium optiquement minces. Une corrélation est constatée pour l'extension de l'hélium neutre He I 4713Å évaluée à 5.5 Mm et l'extension des spicules autour de 5 Mm. Une interprétation peut être faite en faisant l'hypothèse que les spicules plus froids pourraient être associés à l'hélium neutre, et notament pour discuter sur le milieu interspiculaire, dont la nature du plasma, l'environnement sont encore mal compris à ce jour. L'hélium pourrait être aussi photo-ionisé par la couronne ambiante et chaude ( T > 1 MK) qui interpénètre les régions du plasma

« froid» composée de spicules ( T < 30000 K), qui se formeraient au dessus du minimum de température vers 700 km et se terminent vers 5500 km. L’extension des raies des spectres éclairs et les courbes de lumière éffectuées pour quelques ions de masses atomiques différentes, peuvent donner une éventuelle indication sur le milieu interspiculaire, car les altitudes sont semblables, mais l’interprétation reste difficile.
Les macrospicules visibles aux altitudes plus élevées jusqu’à 8000 km, pourraient aussi être associés au extensions d’enveloppes d’hélium neutre produit par excitation et une fois ionisé où se mélangent des courants chauds (couronne à 1 MK) et courants « froids » liées aux spicules et macrospicules.
Cette analyse contribue paralèllement à traiter du problème du milieu entre les spicules, où l’hélium penètre des les couches au dessus du minimum de température, où la température commence à remonter. Le milieu entre les spicules, et depuis leurs pieds, apparaît vers 600 à 700 km au dessus du limbe dans le minimum de température. Ce milieu interspiculaire pourrait aussi commencer à apparaître à partir des altitudes du minimum de température qui correspond aussi à la mésosphère où sont formées une myriade de petites raies d’émission low FIP et ayant des potentiels d’excitation très faibles, inférieurs à 10 eV. Les spectres éclair ont permis de sonder simultanément ces régions en séparant les différentes composantes. Le champ magnétique est concentré dans ces régions profondes, et ce champ émergeant expliquerait la structuration de ces couches et surtout les phénomènes dynamiques que sont les spicules et macrospicules, mais aussi les jets, et petites boucles.
Les observations des spicules montrent clairement que les modèles proposés par A. Gabriel 1976, qui dépendent de l’altitude voir figure V-4-2 ne correspondent pas à ces récentes observations, car aucun effet d’étranglement ou de « constriction » n’a été observé.

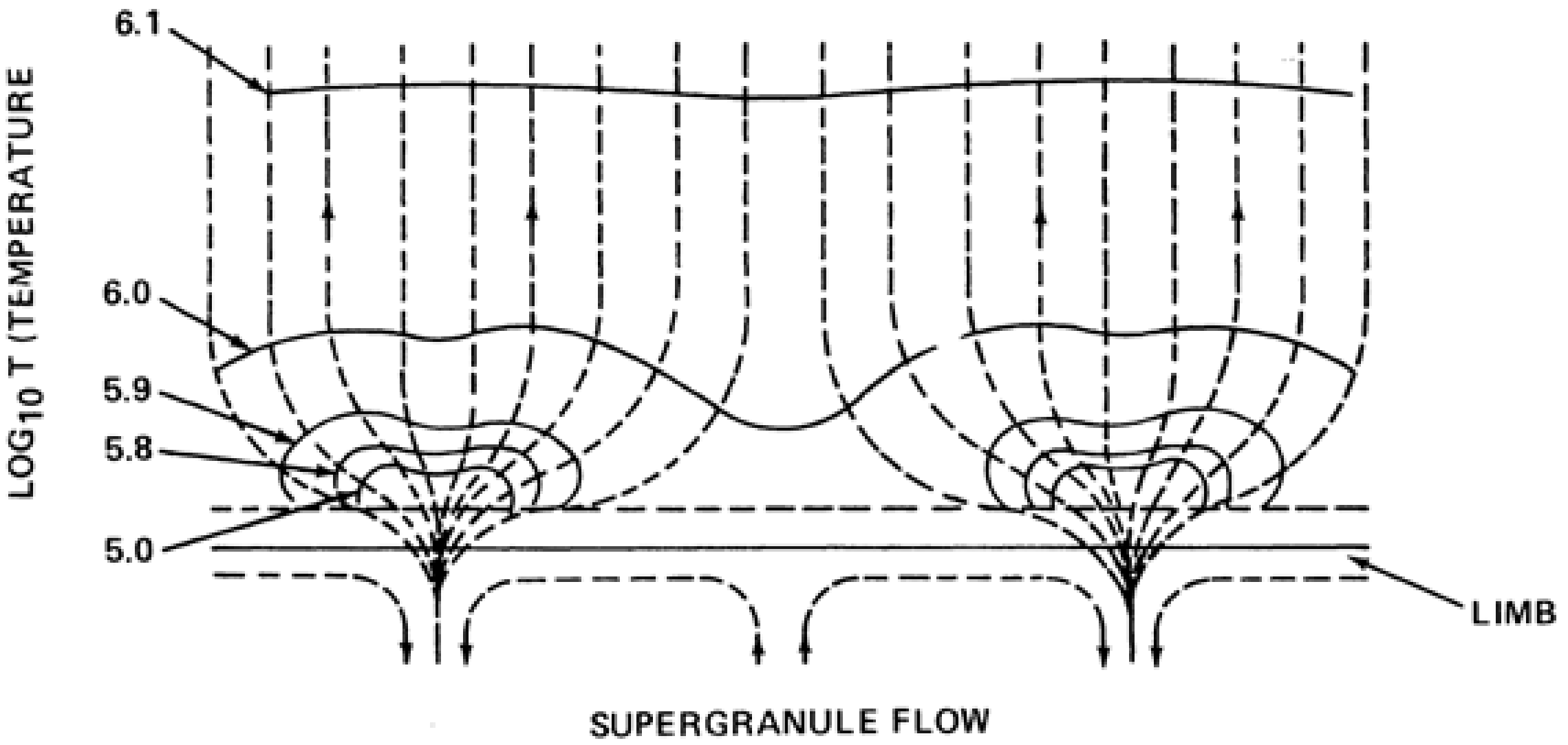

**Figure V-4-2:** *Modèle 2D de l’interface surface-couronne calme proposé par A. Gabriel (1976), avec constriction du champ magnétique pour réduire la conduction de la chaleur depuis la couronne vers la surface. Les contours de températures entre Log(T) = 5.4 et Log(T) = 6.1 sont indiqués. D’après A. H. Gabriel 1976. Les lignes de champ émergent des limites du réseau où elles sont concentrées par le flot de supergranulation et d’où elles divergent rapidement avec l’altitude jusqu’à ce qu’elles deviennent uniformes et verticales dans la couronne. Tout le flux magnétique qui émerge de la surface peut atteindre la couronne.*

Les ions positifs et les électrons, qui sont des particules chargées s'enroulent le long des lignes de force et contribuent à la conduction du chauffage provenant de la couronne. La conduction devient faible lorsque la température est très inférieiure à 1 MK.
Le modèle hydrostatique 1D, homogène et stratifié VAL (Vernazza, Avrett, Loeser 1981) n'est pas adapté car il ne tient pas compte des spicules, et du milieu interspiculaire. Athay et Menzel 1956 ont aussi modélisé ces couches mais leur modèle de structuration des couches ne correspond pas aux phénomènes dynamiques observés avec Skylab.
Ces modèles 1D ne correspondent pas non plus aux observations récentes réalisées avec AIA/SDO, Hinode, car les lignes de champ magnétique pourraient correspondre aussi à des boucles, avec des dimensions variables, comme le montre le schéma de la figure V-4-3 extraite de l'article de Dowdy et al 1986:

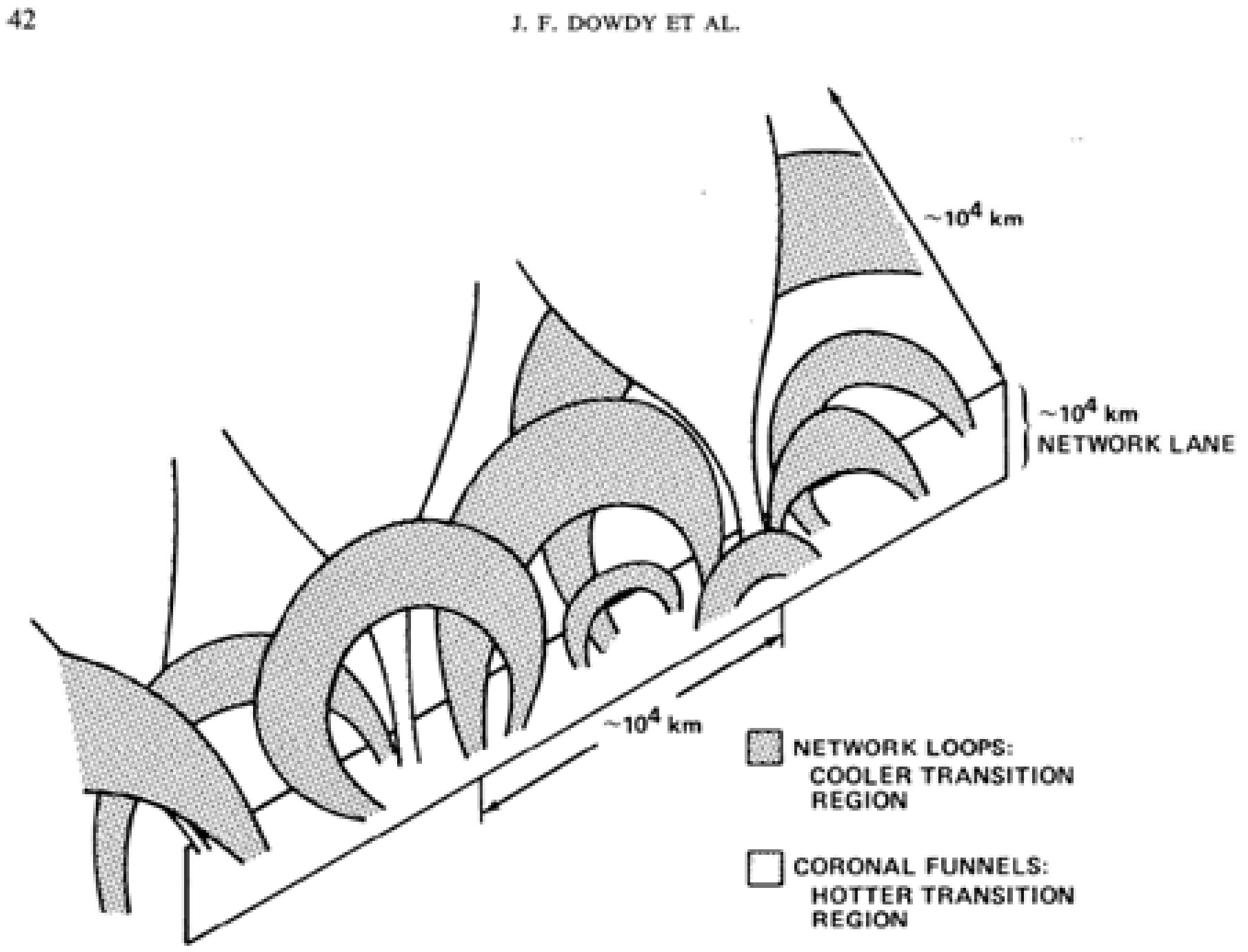

**Figure V-4-3 :** *Image en 3 dimensions de la structuration du plasma par le champ magnétique dans une région de transition "calme". 2 populations de structures magnétiques sont présentes : les boucles dans le réseau chromosphérique, des cheminées coronales comprenant des lignes de champ ouvertes rejoignant la couronne. D'après James F. Dowdy, Douglas Rabin, Ronald R. Moore 1986. La majorité de la région de transition « froide » est située dans ces réseaux de boucles. La partie supérieure des boucles vers $10^4$ km ressemble aux structures de la figure standard de (A. H. Gabriel, 1976) et contiennent une région de transition conductrice, mais ces conduits sont resserrés « constriction » à cause de la multitude de boucles dans le réseau. Cela étant, bien que du plasma froid de la région de transition soit contenu aux pieds de ces conduits, la surface dans la partie inférieure des tubes de flux est trop réduite pour donner une contribution majeure à l'émission.*
*D'autre part, le plasma de la région de transition plus froide dans le réseau de boucles serait chauffé magnétiquement de « l'intérieur » plutôt que par transfert thermique depuis la couronne, par des processus encore mal connus.*

Ce modèle de Dowdy est plus récent et mieux adapté aux observations des spectres éclairs. Le travail de cette thèse a été essentiellement consacré à l'étude des régions et altitudes situées aux pieds des boucles représentées sur cette figure V-4-3. Nous nous sommes intéressés à ces structures sur le limbe, aux altitudes inférieures à 10 Mm et c'est pourquoi au chapitre V-5 nous avons effectué des déconvolutions sur des images sommées et redressées en EUV pour les analyser.

## V– 5) Déconvolutions des limbes SDO/AIA 131 Å, 171 Å, 193 Å et 304 Å réalisés simultanément pendant l'éclipse totale du 11 Juillet 2010

L'objectif de ce chapitre est de comparer des profils des limbes en imagerie EUV du Soleil, réalisés à des altitudes comparables à celles des spectres éclairs, au second et troisième contact. La structuration inhomogène et dynamique du bord solaire, où se mélangent des régions froides en absorption très près du bord et des régions chaudes en émission, plus étendues est discutée.
La méthode employée pour réaliser ces analyses est de renforcer la visibilité des structures qui produisent indirectement l'embrillancement du limbe, en déterminant le nombre de pixels optimal, pour effectuer la meilleure deconvolution possible, et la plus adaptée pour améliorer et optimiser la visibilité des structures fines lors de ce processus de traitement des images.
Ce procédé de traitement d'images consiste à utiliser des images de référence du disque solaire observé par SDO/AIA qui ont été partiellement éclipsées par la Lune:

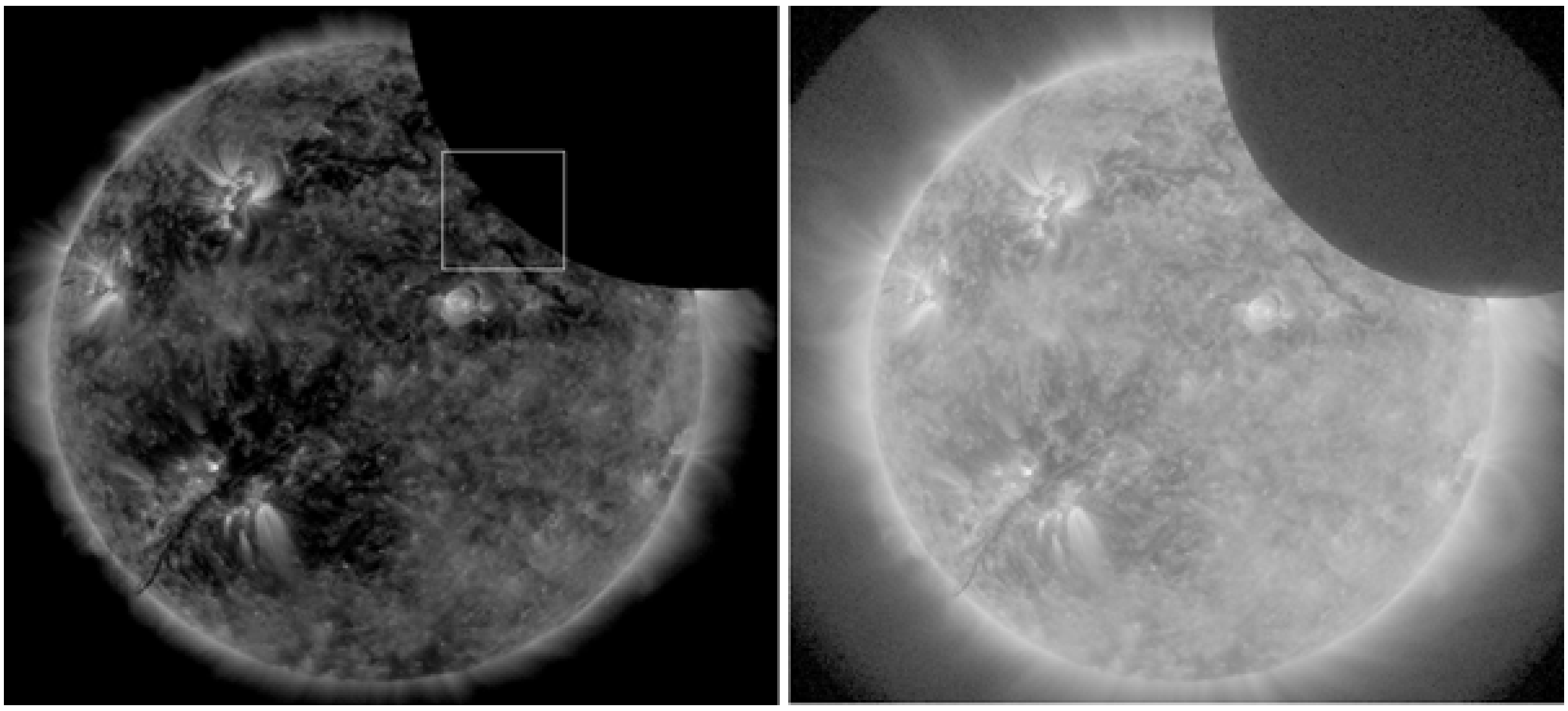

**Figure V-5-1:** Image SDO/AIA 171A, (3 secondes d'exposition) du 6 Décembre 2010 à 13h15 TU partiellement éclipsée, à gauche, le réctangle blanc indique la région examinée, et à droite, niveaux d'intensité convertis en échelle logarithmique ajustés au minimum pour montrer la lumière diffusée sur le disque lunaire.

Des mesures d'intensité ont été effectuées dans la partie de la Lune et aux limites du champ, et ont donné des valeurs comparables proches de 8 ADU, ce qui confirme l'effet de lumière diffusée.
Comme le montrent les extraits des figures V-5-2 à V-5-5, le bord net de la lune sert de référence pour ajuster les paramètres de déconvolution et obtenir la meilleure visibilité. Nous avons pris des images du 6 Décembre 2010 entre 3h12 et 3h13 TU où la Lune a partiellement éclipsé les images en EUV de SDO. Les déconvolutions ont été réalisées à partir de la fonction « masque flou » sous le logiciel Iris sur les images 131 Å, 171 Å,  193 Å, 304 Å. La valeur de 1 pixel a été un compromis et optimum entre d'une part des effets d'artefacts altérant le limbe, où des pixels négatifs apparaissent, si une valeur supérieure à 1 pixel est prise, et d’autre part, si un fenétrage inférieur à 1 pixel est choisi, la déconvolution conduisait à une image encore floue, et insuffisamment contrastée. Pour l’image à 131 Å, la valeur a été de 2 pixels pour ce compromis. Les images suivantes montrent la déconvolution effectuée comme référence sur le bord de la Lune:

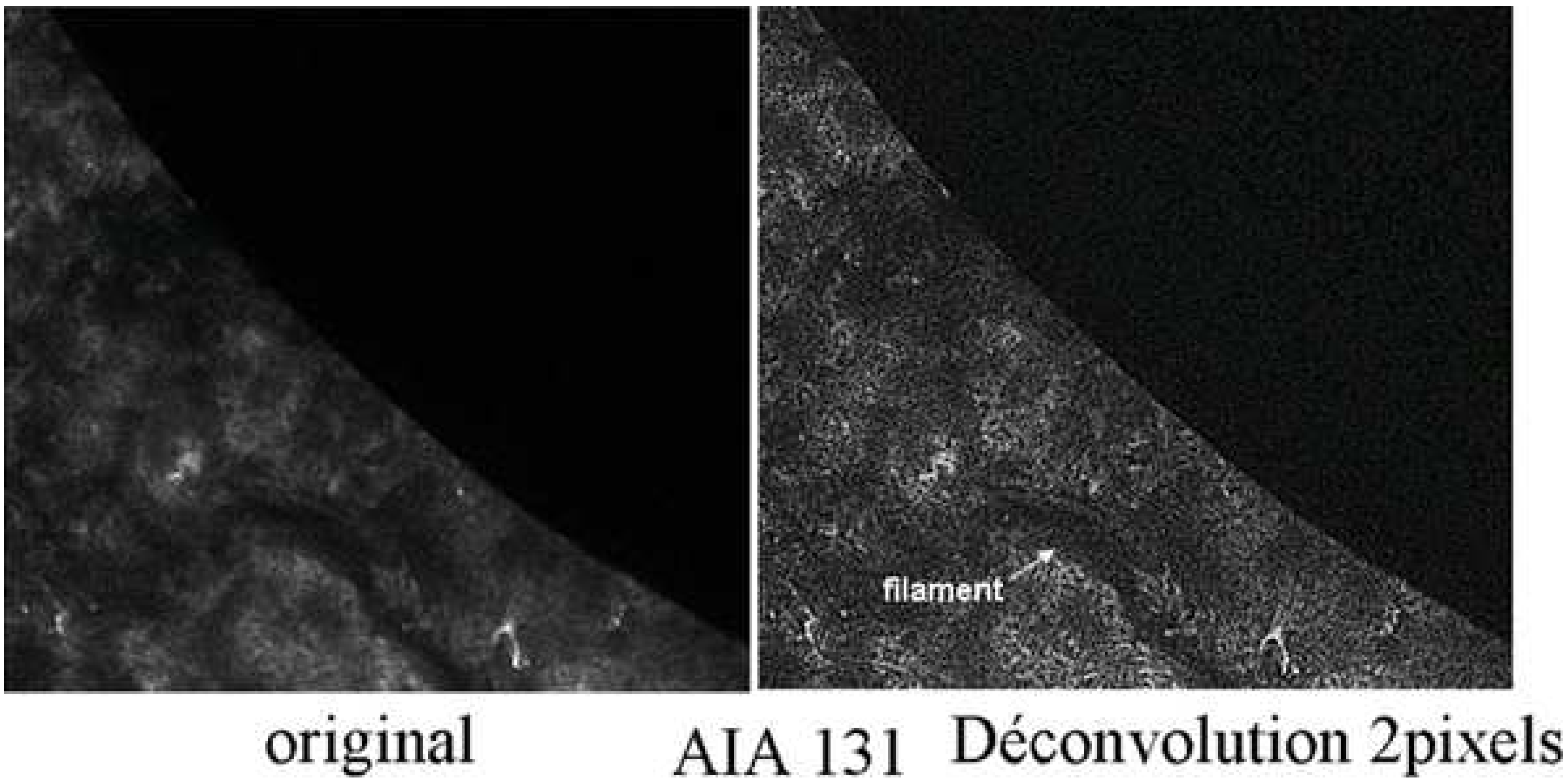

**Figure V-5-2**: *comparaison de l’image originale à 131 A du 6 Décembre 2010 à 3h12TU, sans déconvolution et après déconvolution, avec une partie du disque lunaire en noir.*

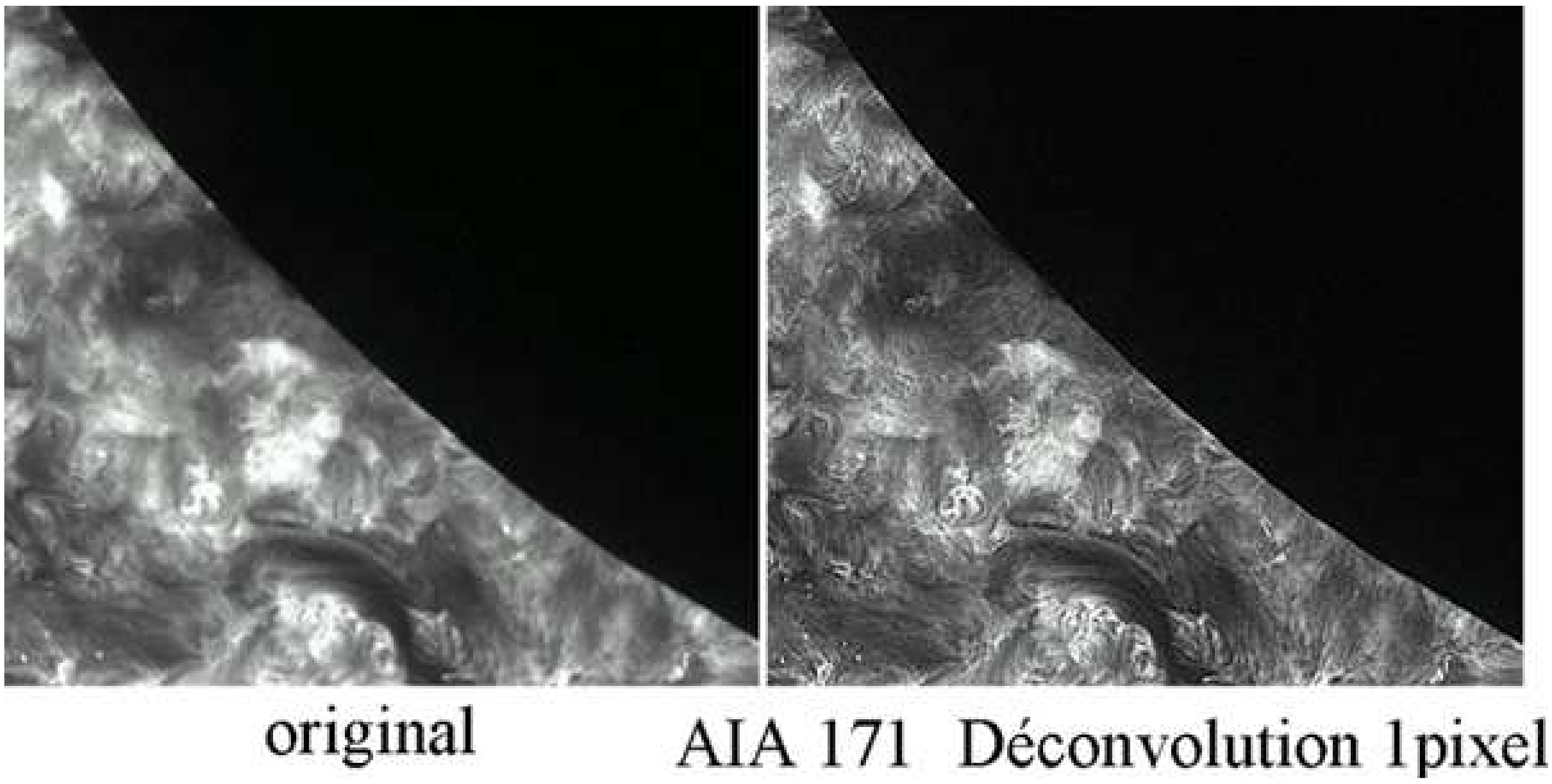

**Figure V-5-3:** *comparaison de l’image originale à 171 A du 6 Décembre 2010 à 3h12TU, sans déconvolution et après déconvolution avec une partie du disque lunaire en noir*

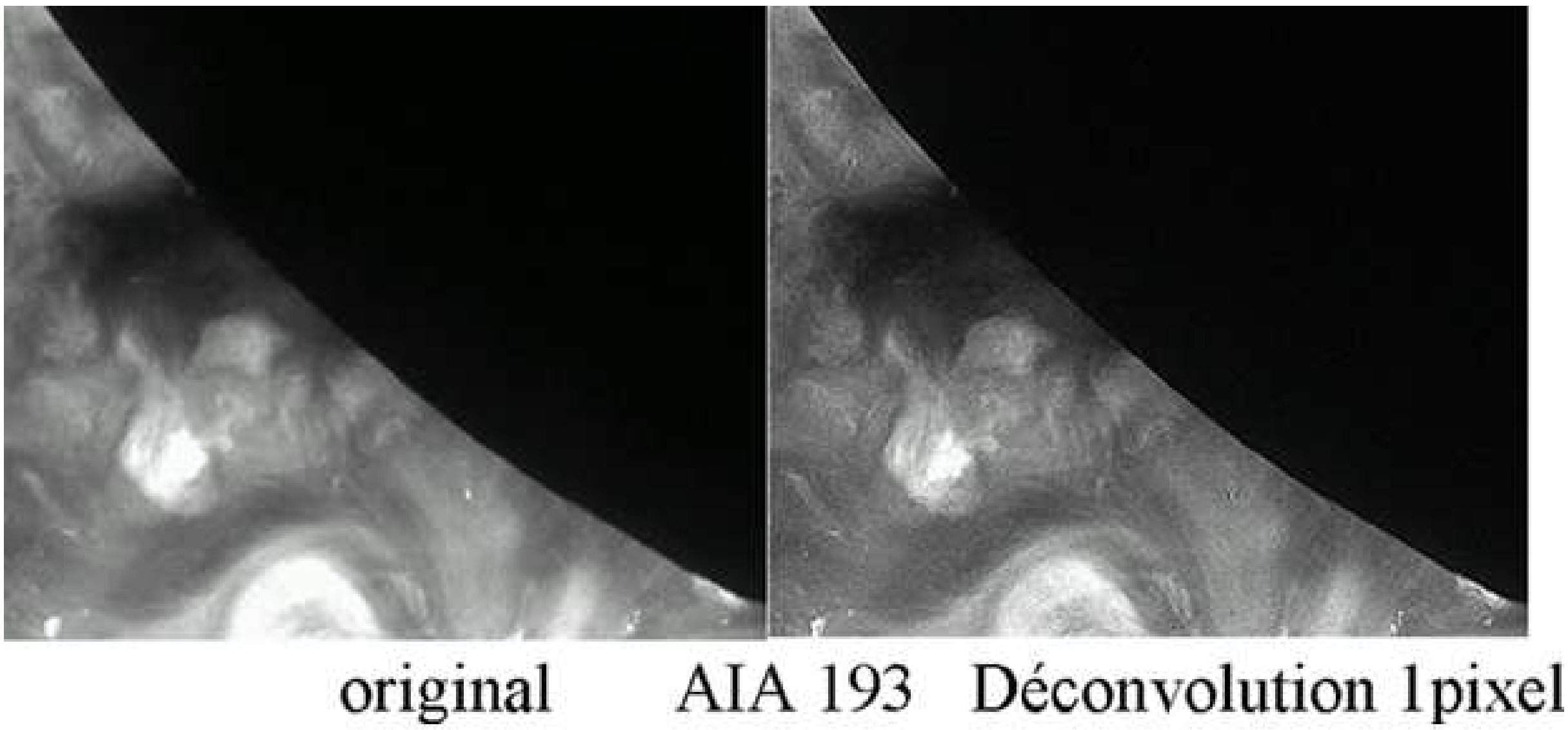

**Figure V-5-4:** *comparaison de l’image originale à 193 Å (Fe XII) du 6 Décembre 2010 à 3h12TU, sans déconvolution et après déconvolution avec une partie du disque lunaire en noir afin de montrer l’amélioration et renforcement des fins détails.*

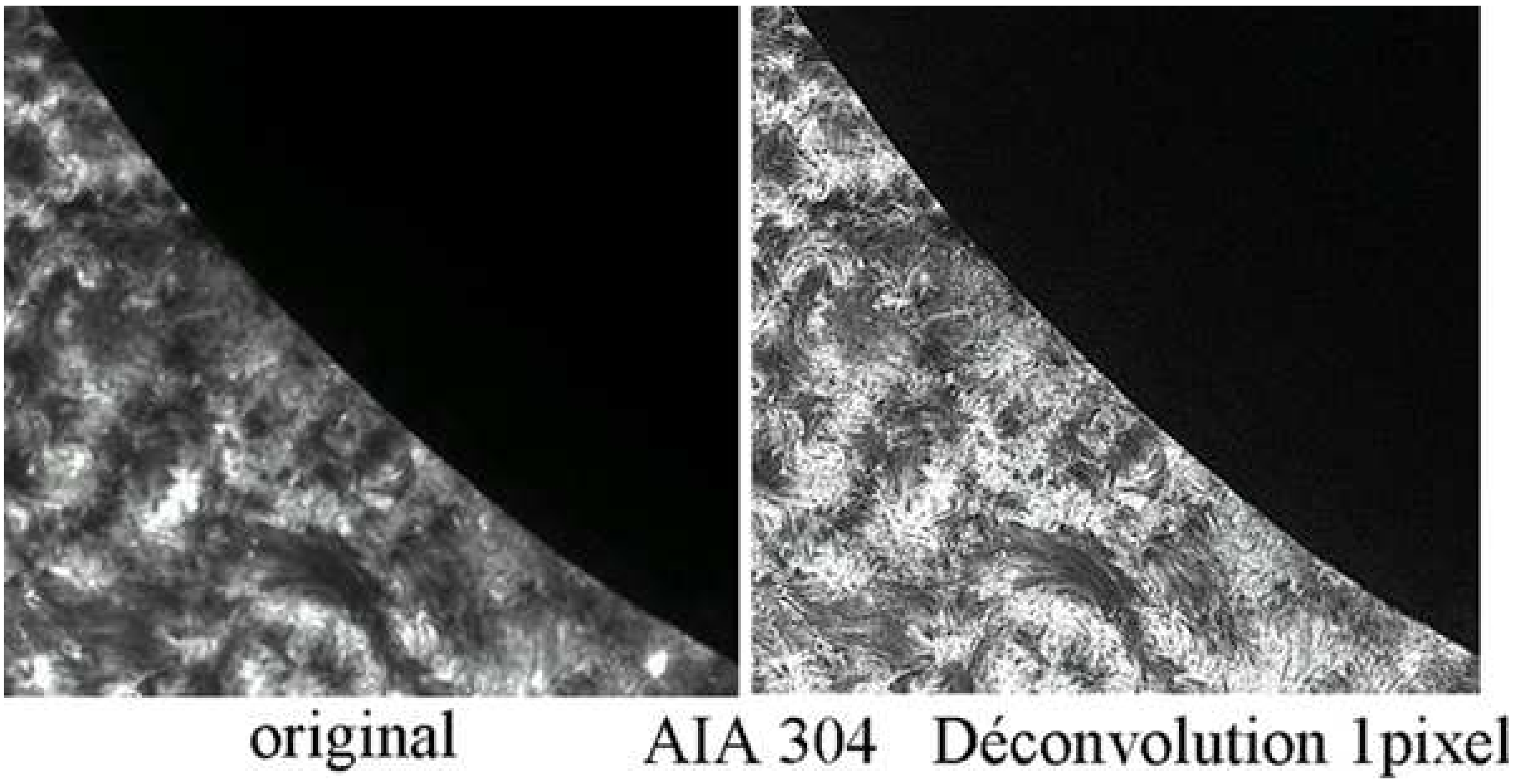

**Figure V-5-5:** *comparaison de l’image originale à 304Å du 6 Décembre 2010 à 3h12 TU, sans déconvolution et après déconvolution avec une partie du disque lunaire en noir afin de montrer l’amélioration et renforcement des détails fins.*

Les images brutes des disques solaires partiellement éclipsés ont été étudiées en vue d’évaluer la lumière diffusée, et sa répartition dans le champ.
Le bord lunaire a été redressé, puis des profils d’intensité ont été effectués dans un rectangle ayant permis de moyenner sur 36 pixels et améliorer le rapport signal sur bruit d’un facteur 6.
La figure V-5-6 présente les profils obtenus à partir de chacune des images de SDO/AIA, et les temps d’exposition ont été indiqués.

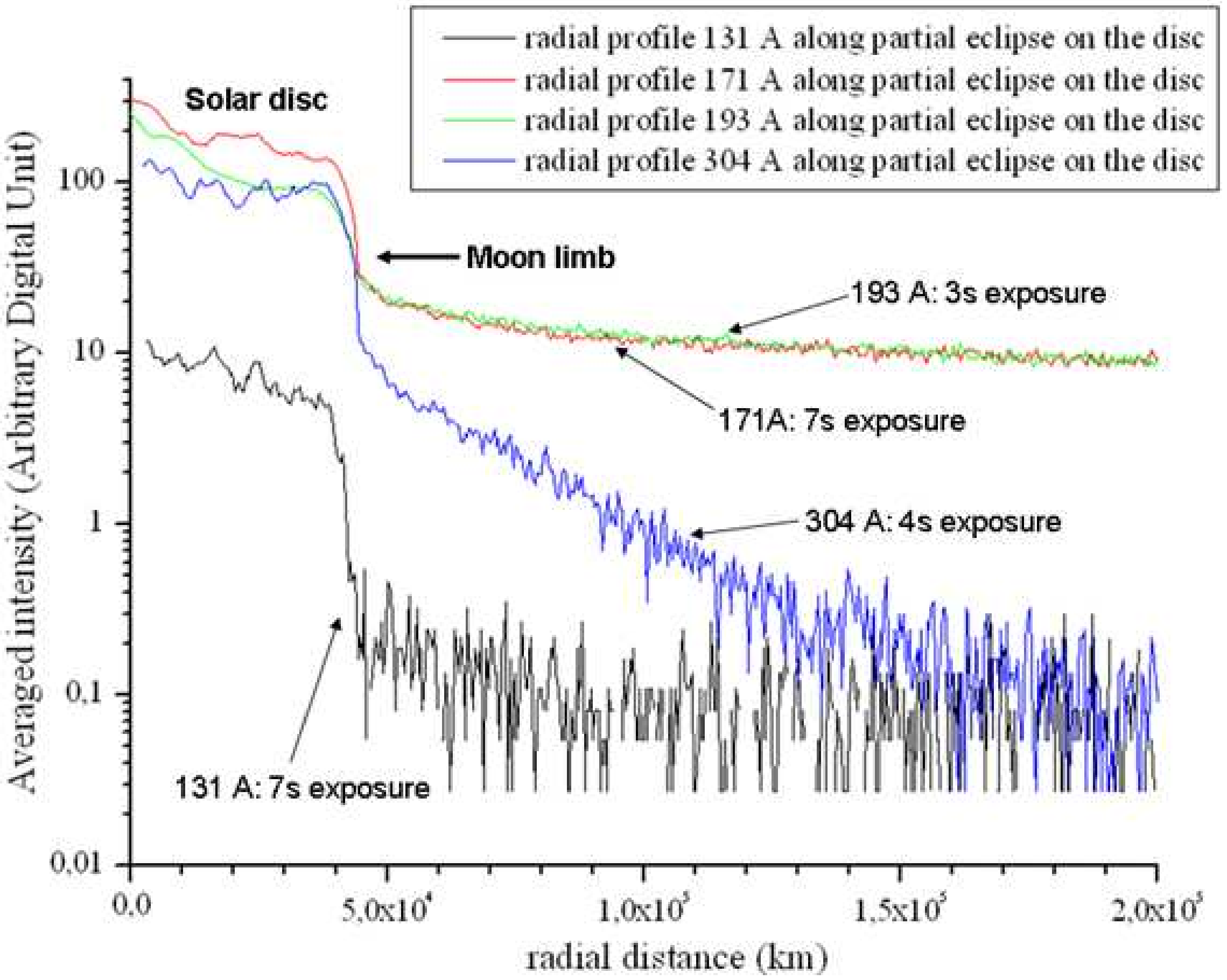

**Figure V-5-6:** *profils d'intensité radiales effectués sur le disque solaire partiellement eclipsé. Du 6 Décembre 2010 à 13h 15 TU.Le bord de la Lune est indiqué, afin de comparer le taux de lumière diffusée.*

Les profils sont liés aux réponses des filtres AIA indiqués à la figure IV-6-2-3. La lumière diffusée peut être due à un mélange de raies d'émission EUV à des gammes de températures étendues $10^5$ à $10^6$ K. cela a pour effet de mélanger des basses températures et des plus hautes températures, ce qui produit une émission uniforme, et donc une lumière diffusée plus intense. Bien que cette lumière diffusée soit présente, des analyses et traitements sur les images SDO/AIA ont été possibles. Des sommations de 16 images ont été effectuées pour améliorer le rapport signal sur bruit, améliorer la visibilité des structures de plus faible intensité, connaissant les niveaux de lumière diffusée.

Une fois les paramètres de déconvolution définis grâce aux figures V-5-2 à V-5-5, les images ont été redressées, pour effectuer les prises des profils sans déconvolution, et ensuite les mêmes images redressées ont subi le procédé de déconvolution par masque flou, avec 1 pixel pour les images en 171 Å, 193 Å et 304 Å et 2 pixels pour l'image en 131 Å. Les figures suivantes montrent après sommation d'un certain nombre d'images l'aspect des limbes redressés à partir desquels les profils d'intensité sont ensuite réalisés. Les images des limbes linéarisés ont permis d'effectuer plus facilement une intégration en sommant et moyennant les limbes dans le sens radial inscrits dans les boîtes rectangulaires de 350 pixels d'étendue. Les sommations et intégrations ont permis d'améliorer le rapport signal sur bruit d'un facteur au moins 40.

Plusieurs régions du limbe solaire ont été sommées, linéarisés, puis un masque flou a été appliqué, voir les figures V-5-7 à V-5-10. Dans les légendes, « Qr », signifie quiet region,

« Ch » signifie coronal hole. Cela permet de comparer ces différentes régions, et le limbe après traitements, et de comparer les effets de ces régions sur les profils d'intensité des limbes. Ces images redréssées et les profils des limbes en figures V-5-11 démontrent que le bord du Soleil est composé de structures froides au plus à 30000 K comme les macrospicules, vus en absorption dans la raie du Fe XII à 2 MK, et où la couronne chaude interpénètre. Cela provoque une photoionisation qui serait responsable de l'embrillancement des limbes observés, sur une échelle de hauteur entre 10 et 30 Mm. Ces résultats révélant les inhomogénéités, jets, spicules et macrospicules en absorption ne peuvent pas être décrits avec un modèle hydrostatique stratifié.

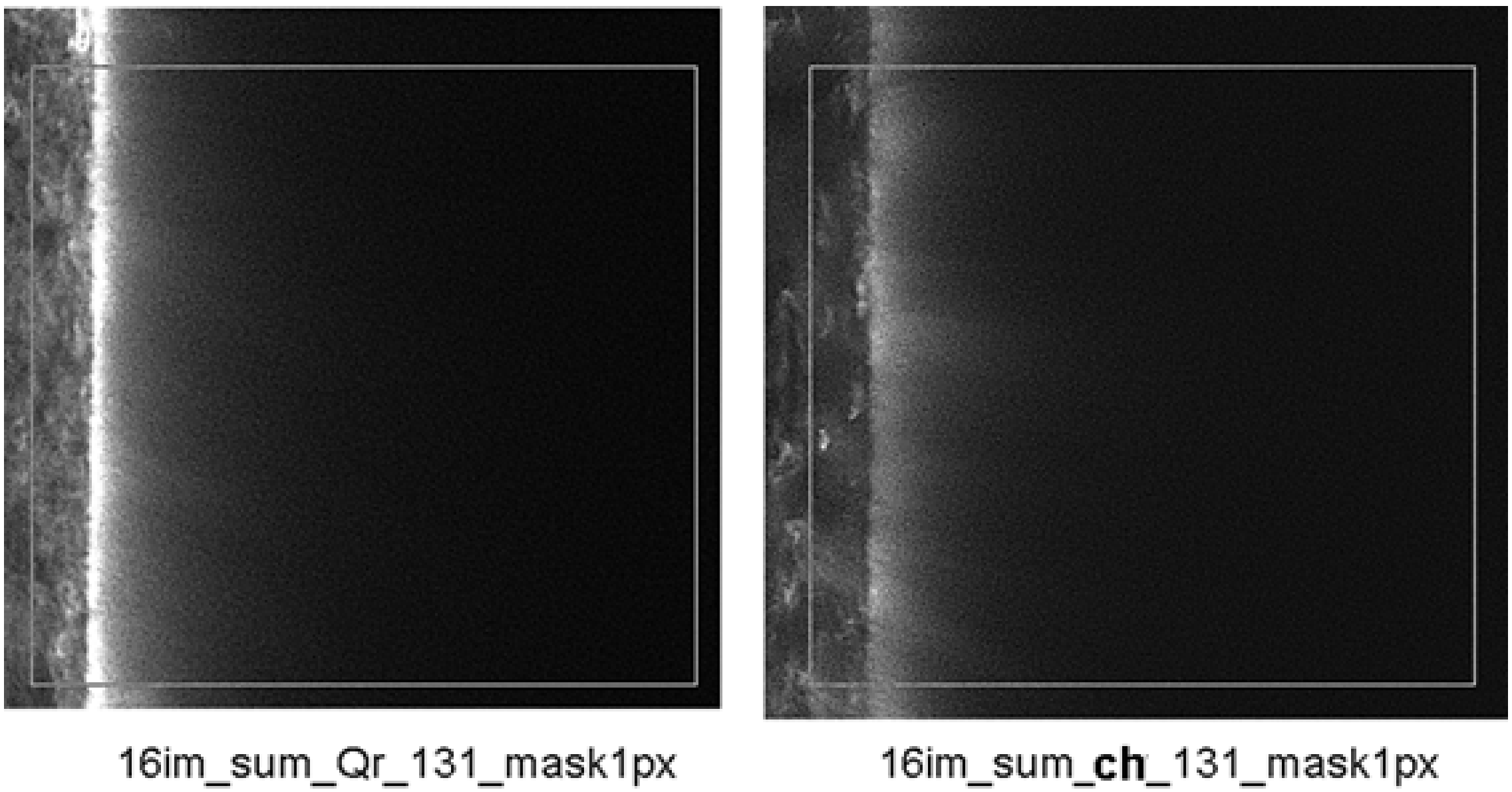

**Figure V-5-7:** *limbes solaires redréssés à 131Å après sommation de 12 images pour améliorer le rapport signal sur bruit, selon une région calme ou un trou coronal.*

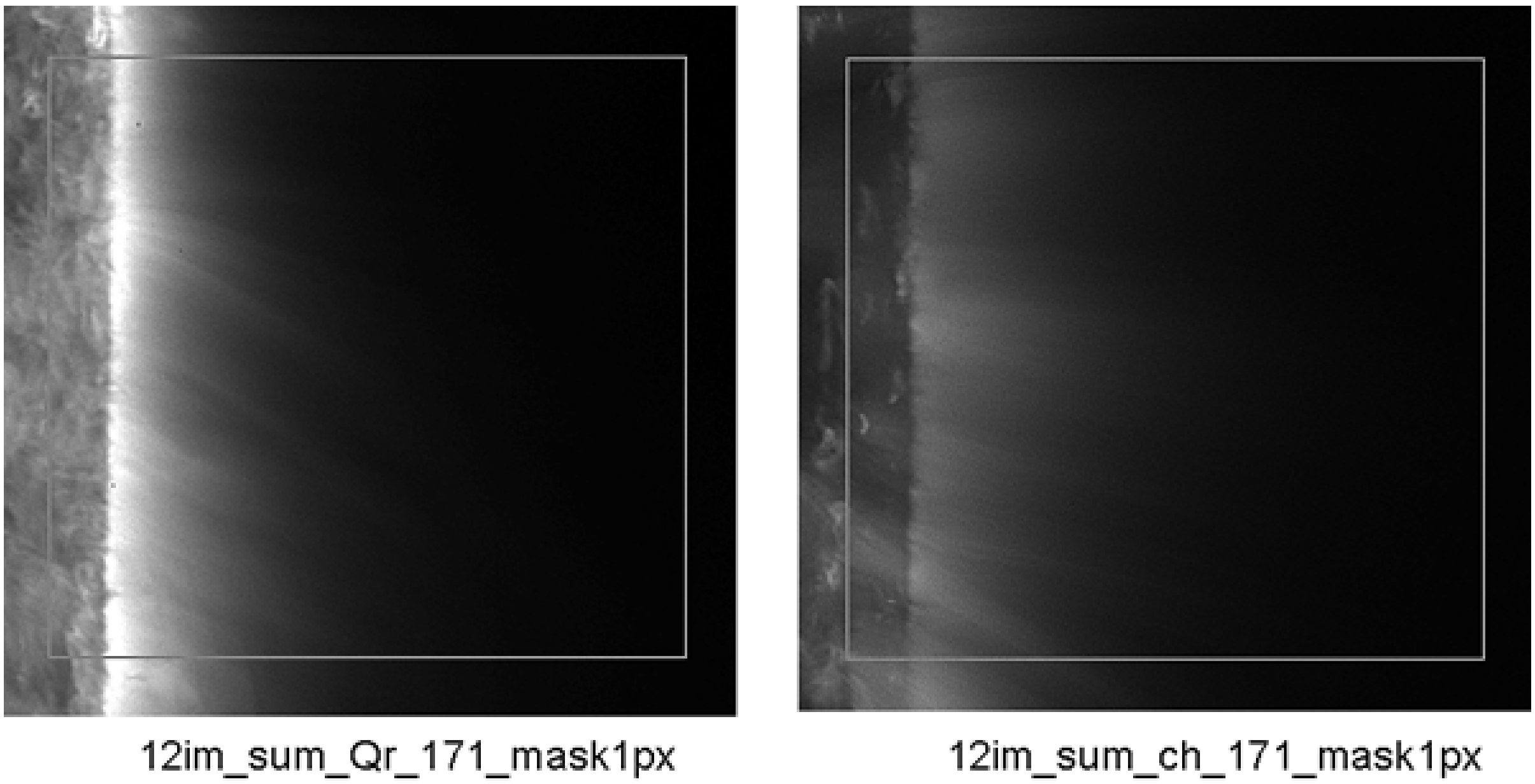

**Figure V-5-8:** *limbes solaires redréssés à 171Å après sommation de 12 images pour améliorer le rapport signal sur bruit, selon une région calme ou un trou coronal.*

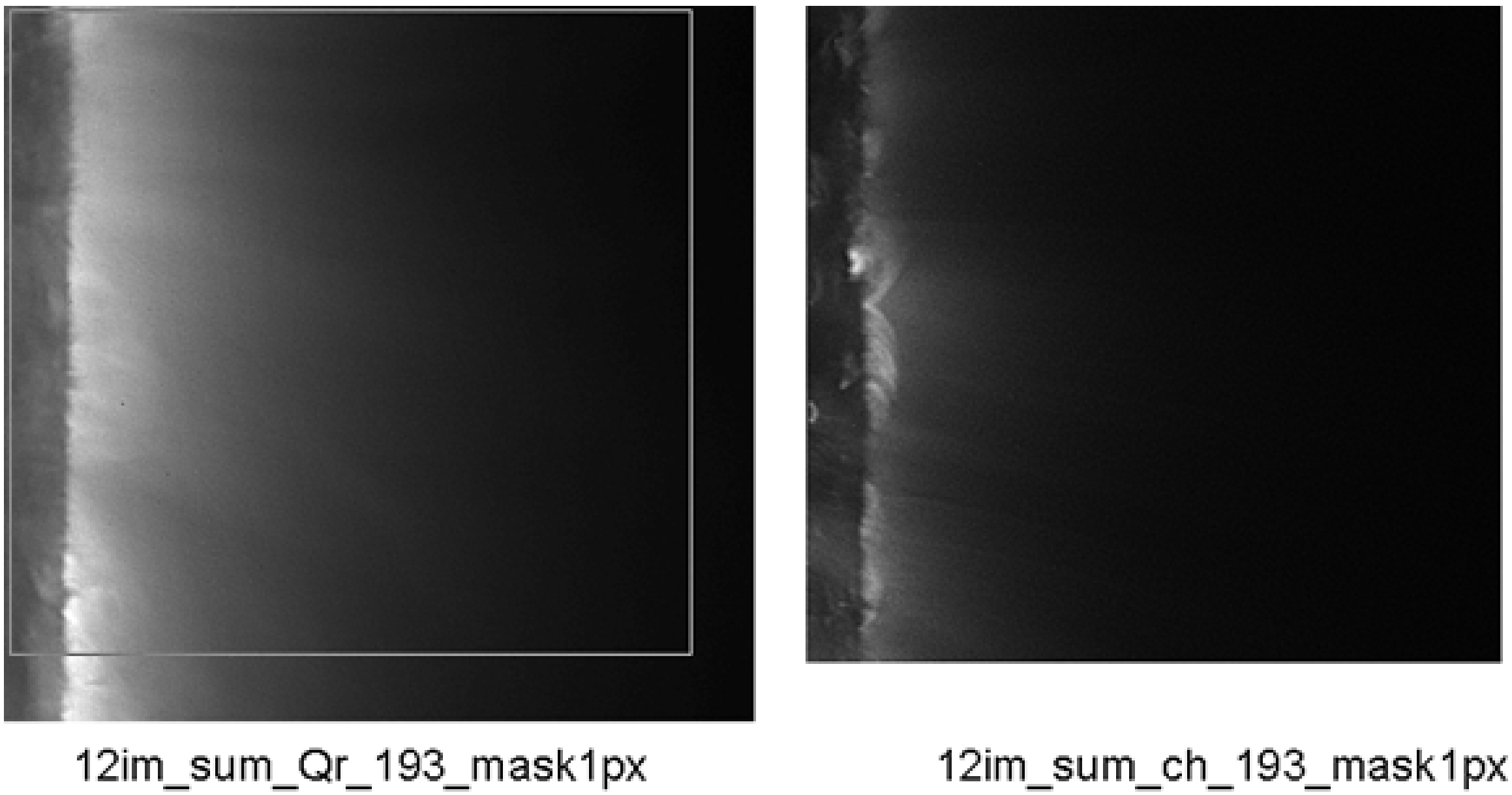

**Figure V-5-9:** *limbes solaires redréssés à 193 Å après sommation de 12 images pour améliorer le rapport signal sur bruit, selon une région calme ou un trou coronal.*

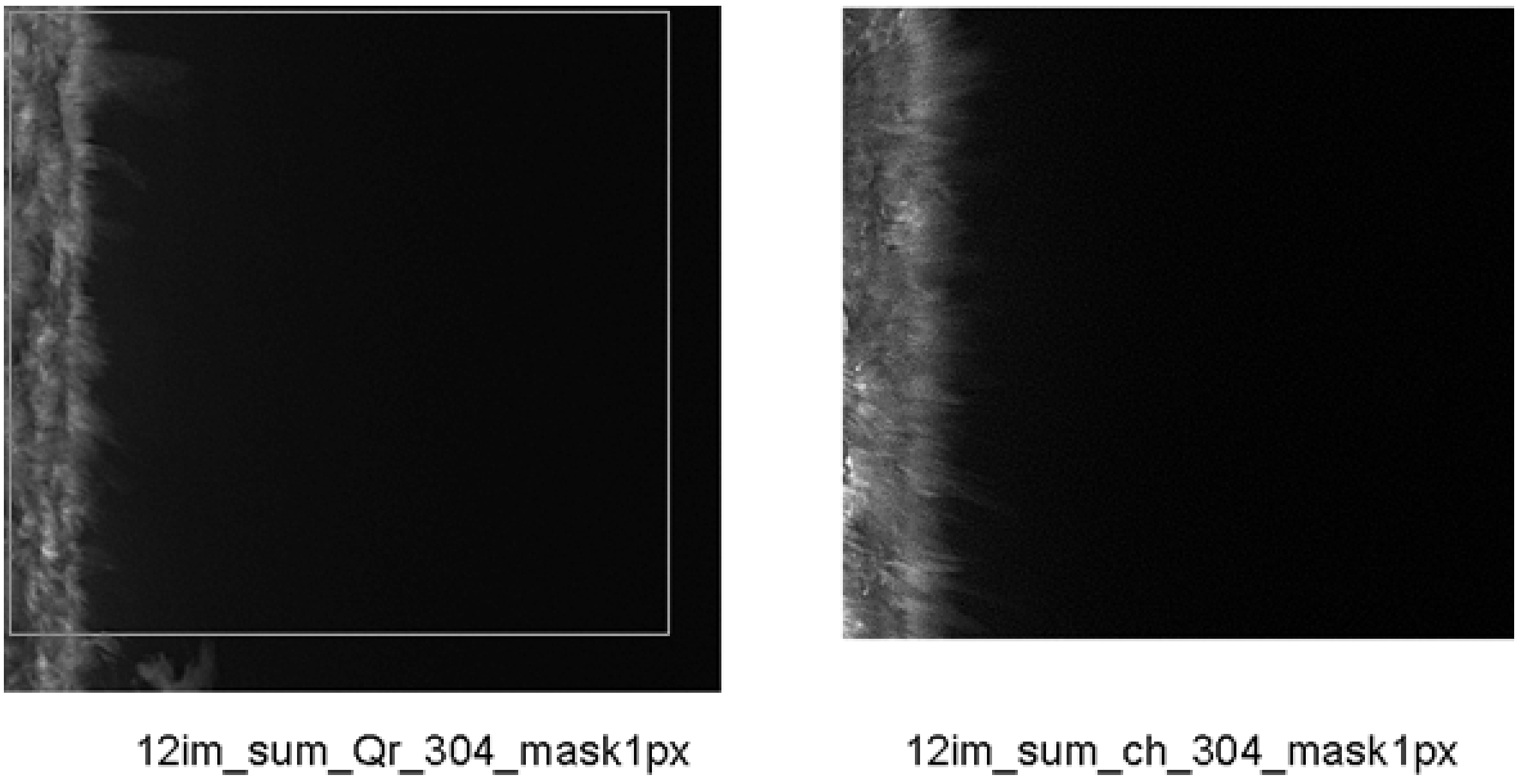

**Figure V-5-10:** *limbes solaires redréssés à 304Å après sommation de 12 images pour améliorer le rapport signal sur bruit, selon une région calme ou un trou coronal.*

A partir des limbes solaires redressés, la comparaison des images après déconvolution permet de différentier les régions calmes et des trous coronaux. Le limbe observé dans les trous coronaux rend compte des structures filamenteuses de plasma « froid » observé en absorption dans les images en émission des raies EUV 193 Å et 171 Å. Des boucles de plasma sont observées dans les trous coronaux dans le Fe XII à 193 Å. Les profils d'intensité moyennés sont ensuite réalisés sur chacun des limbes, selon les régions calmes et trous coronaux, pour différentes raies EUV, et comparer les profils avant et après déconvolution.

La figure V-5-11 montre les profils des limbes de AIA originaux, c'est à dire n'ayant pas subi de déconvolution comme référence, pour les comparer ensuite après déconvolution, dans le but de montrer l'étendue de l'absorption au niveau du limbe proche de $h$ = 0:

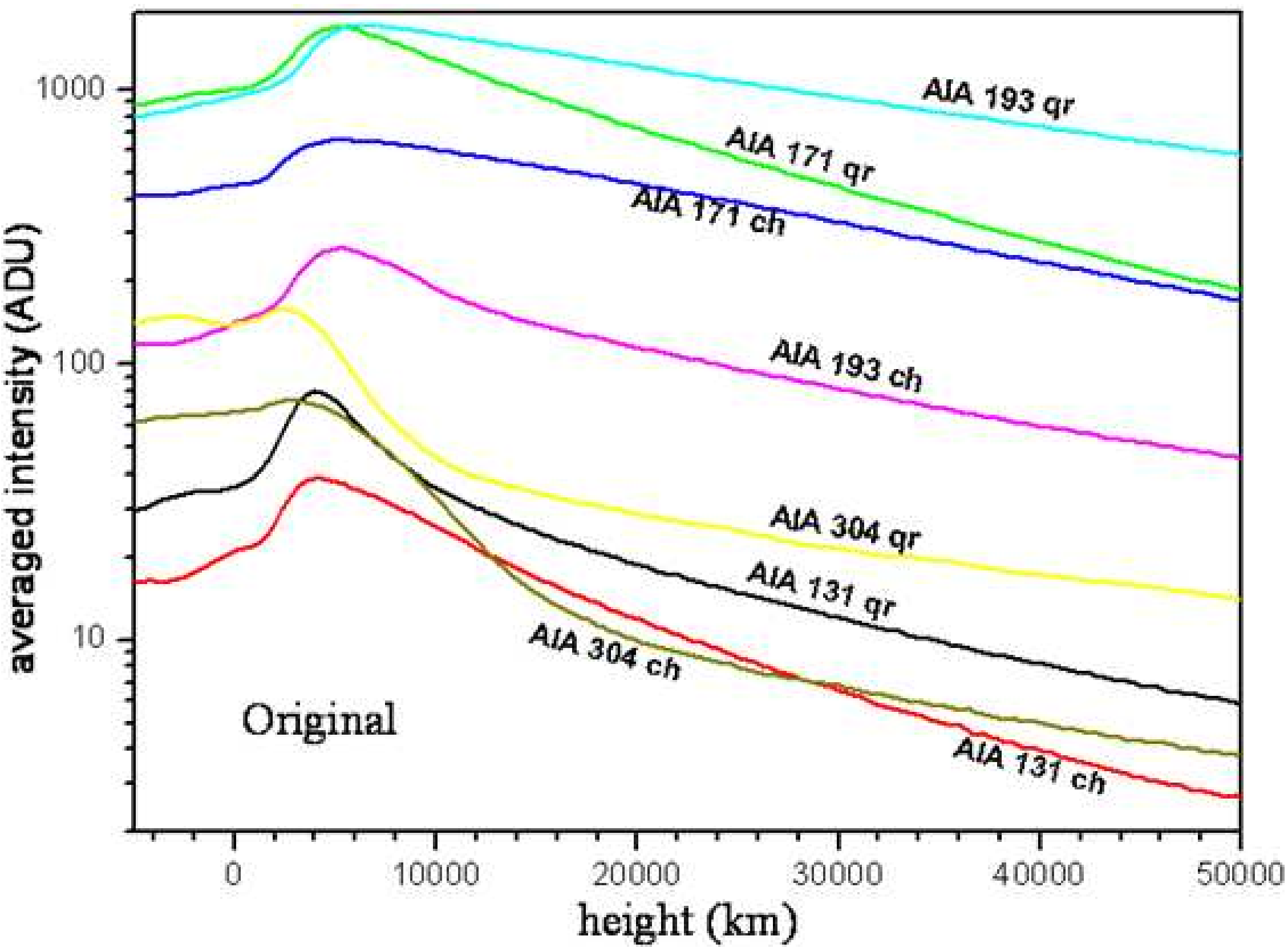

**Figure V-5-11:** *profils moyennés originaux des limbes de SDO/AIA, après intégration sur 350 lignes de pixels, dans un trou coronal – CH et région calme- QR pour les raies 131Å, 171Å, 193Å et 304 Å (avant déconvolution) du 11 Juillet 2010.*

Ces profils en Figure V-5-11 sont effectués à partir des limbes redréssés avant d'effectuer le traitement de  type masque flou (qui se rapprochant d'une déconvolution, où un filtrage des hautes fréquences spatiales est effectué). Ils présentent un embrillancement du limbe pour les températures coronales. Mais pour l'hélium ionisé He II à 304 Å optiquement épais, cet embrillancement est difficile à déceler, car il est mélangé au réseau chromosphérique dont la température est de l'ordre de 30000 K.
Les profils de la figure V-5-12 ont été tracés à partir des images ayant subi une déconvolution. Ils montrent une meilleure visibilité de l'étendue de l'absorption au dessus du limbe où sont formés les macrospicules froids.

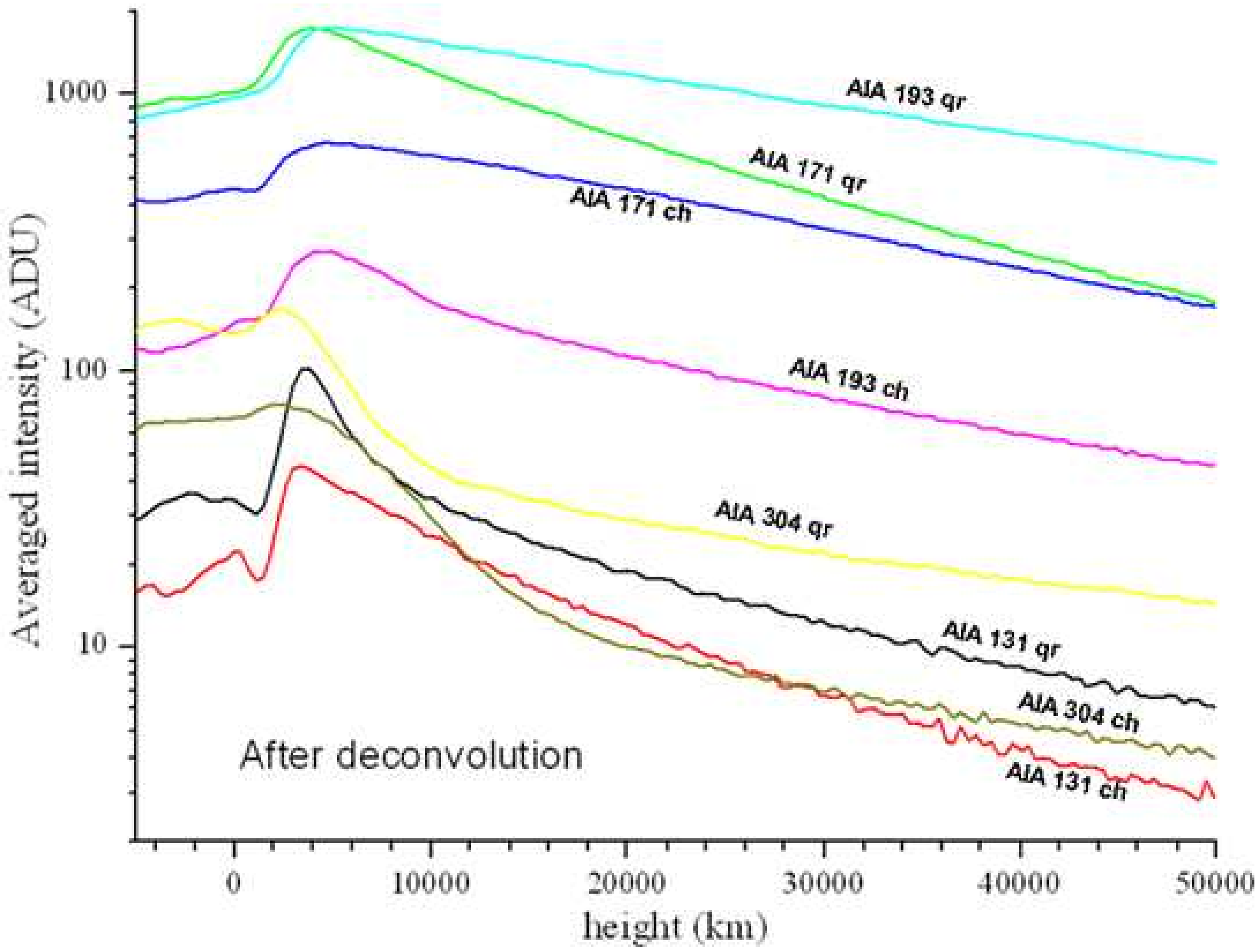

**Figure V-5-12:** *profils moyennés après déconvolution des limbes de SDO/AIA, après intégration sur 350 lignes de pixels, d'après les figures précédentes des limbes redréssés dans un trou coronal – CH et région calme- QR pour les raies 131 Å, 171 Å, 193 Å et 304 Å.*

Les profils ayant subi la déconvolution, montrent non seulement l'embrillancement du limbe, mais aussi une absorption, due aux spicules, macrospicules à 30000 K, lorsqu'on observe dans les raies coronales 193 Å, 171 Å. En 131Å, la température de la raie du Si XII est inférieure à 1 MK, et l'absorption est davantage visible et réhaussée. Les absorptions des limbes sont visibles directement sur l'extrait du bord solaire figure V-5-14, ayant subi un masque flou de 2 pixels, ayant permis de renforcer le limbe et converti en échelle logarithmique, pour montrer les niveaux plus faibles.
A partir des profils V-5-11 et V-5-12, des analyses des contrastes et largeurs à mi-hauteur sont effectués pour déterminer et quantifier sur quelles images, et quelles régions, l'embrillancement du limbe a été renforcé, et analyser les effets d'étalement sur les images de AIA/SDO. La figure V-5-13 résume les mesures effectuées sur les profils, afin de comparer les différentes régions, longueurs d'onde étudiée, et selon avant et après déconvolution.

| Longueur d'onde (Å) | Région | Contraste | FWHM (km) | Type |
|---|---|---|---|---|
| 131 | CH | 0.27 | 5750 | Avant déconvolution |
| 131 | QR | 0.38 | 3600 | Avant déconvolution |
| 171 | CH | 0.18 | 9160 | Avant déconvolution |
| 171 | QR | 0.25 | 4920 | Avant déconvolution |
| 193 | CH | 0.25 | 5260 | Avant déconvolution |
| 193 | QR | 0.23 | 11500 | Avant déconvolution |
| 304 | CH | 0.04 | 2760 | Avant déconvolution |
| 304 | QR | 0.06 | 2340 | Avant déconvolution |
| | | | | |
| 131 | CH | 0.39 | 4260 | Après déconvolution |
| 131 | QR | 0.47 | 5140 | Après déconvolution |
| 171 | CH | 0.16 | 8440 | Après déconvolution |
| 171 | QR | 0.23 | 5460 | Après déconvolution |
| 193 | CH | 0.24 | 5220 | Après déconvolution |
| 193 | QR | 0.22 | 10520 | Après déconvolution |
| 304 | CH | 0.04 | 3130 | Après déconvolution |
| 304 | QR | 0.11 | 2000 | Après déconvolution |

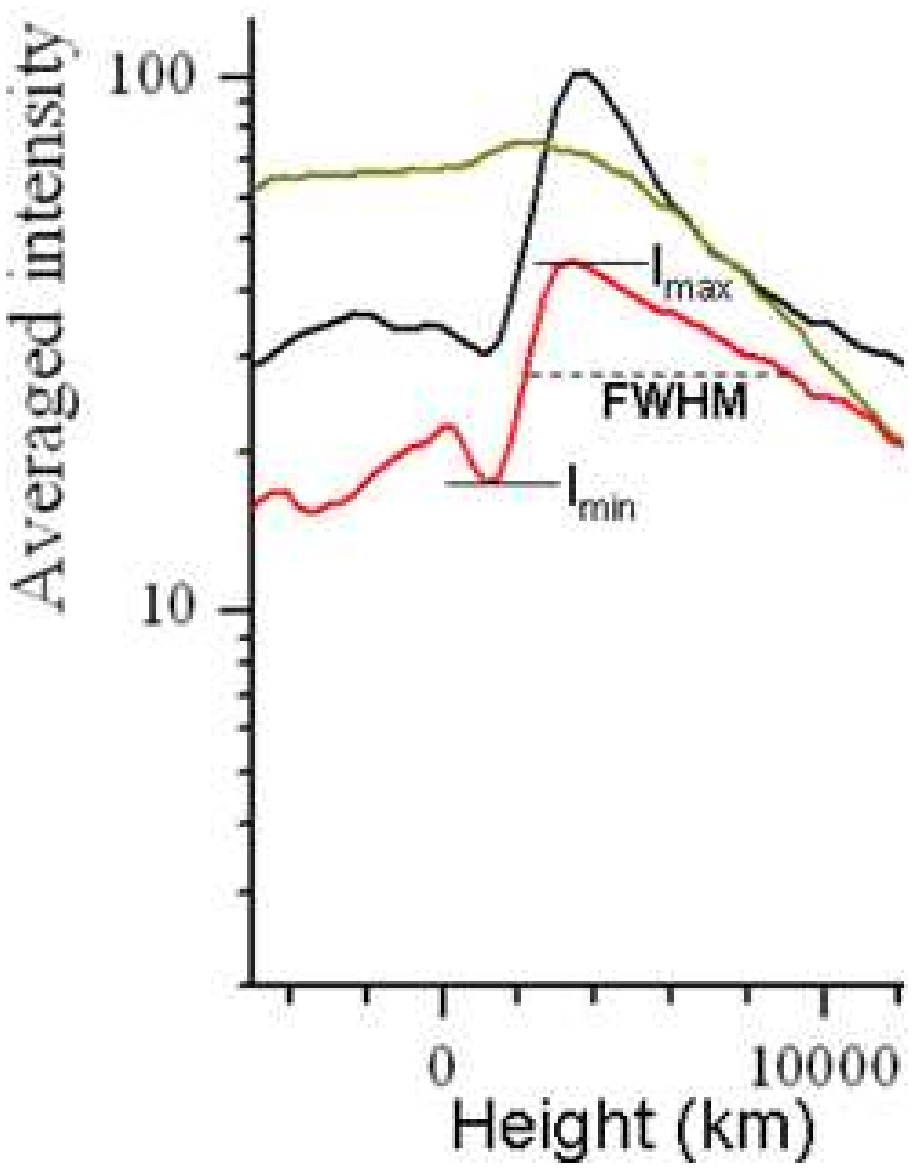

**Figure V-5-13:** *à gauche, tableau des contrastes, FWHM, avant et après déconvolutions selon le type de régions examinées sur le limbe, et à droite extrait des graphiques figure V-5-12 pour montrer où sont pris les intensités min et max et FWHM sur les profils des limbes embrillancés.*

Les contrastes C sont calculés en considérant le calcul suivant à partir de smesures des intensités $I_{min}$ et $I_{max}$ :

$$C = \left( \frac{I_{max} - I_{min}}{I_{max} + I_{min}} \right),$$

et la FWHM est prise à $(I_{max} + I_{min})/2$.

Ces résultats montrent que le contraste est faiblement amélioré (facteur 1.3) uniquement pour 131Å. Pour les autres longueurs d'onde, aucune accentuation notoire n'est constatée. Les FWHM relevées sur les parties des profils correspondant aux embrillancements sont réduites d'environ 1000 km après la déconvolution. L'élargissement des profils est essentiellement dus aux effets d'étalement de la lumière diffusée (nombreuses raies EUV et à des températures étendues). Le processus de déconvolution par une méthode de masque flou a permis d'améliorer la visibilité des limbes solaires, et de montrer une étendue des régions d'embrillancements, dont l'étendue moyenne de 5000 km est comparable à celles des spicules (voir figures II-4-10). La FWHM des embrillancements des limbes aux différentes températures (effets d'opacité sur la ligne de visée) sont corrigées des effets d'étalement.
Ces résultats montrent qu'indépendament du traitement de déconvolution, la largeur du pic d'embrillancement augmente avec la température pour les régions calmes « QR ». En effet, la FWHM est de 2340 km pour He II 304 à 0.03 MK, puis devient 0.5 MK dans le Si XII à 131 Å, 4920 km dans le Fe IX/Fe X à 171 Å, et 11500 km dans le Fe XII (193 Å) à 2MK. Par contre on ne peut pas conclure pour les trous coronaux « CH », car dans le Fe XII 193 Å, ce devrait être une FWHM supérieure à 9160 km, et la valeur est de 5220 km.
Le tableau figure V-5-13 montre aussi, indépendament du traitement de déconvolution que la largeur des pics d'embrillancement dans les trous coronaux est de l'ordre de 1.5 à 1.8 fois plus large dans les régions calmes « QR », sauf pour les profils à 193 Å. Cet effet est probablement du à des effets d'opacité intégrés sur la ligne de visée, mais aussi est du au plasma qui émet du rayonnement en avant plan et arrière plan, dans les régions émissives.
Cet embrillancement du limbe correspond à une enveloppe « shell » en EUV. Ce phénomène peut être dû aux pieds des boucles coronales (éléments non résolus) Aschwanden 2001 et Ascwanden 2008. La figure V-4-3 représente leur structuration. L’embrillancement peut être associé à des petites boucles coronales où la distance entre les pieds est du même ordre de grandeur que leur hauteur. Des fils « threads » quasi horizontaux comparables au champ magnétique horizontal peuvent aussi contribuer à cet embrillancement.
La figure V-5-14 dans une région calme montre le limbe solaire après la sommation de 12 images, où la région correspondant aux altitudes d’embrillancements est visible, avec des structures sombres, (dont certaines sont non résolues), traduisant du plasma plus « froid » qui absorbe du rayonnement EUV coronal. Cette figure montre la présence de filaments associés à la protubérance vue en absorption, la cavité coronale avec déficit de plasma autour, limitée par les cornes de plasma. Cette figure montre la nature inhomogène du limbe solaire, où les profils d’embrillancement du limbe ont été effectués.

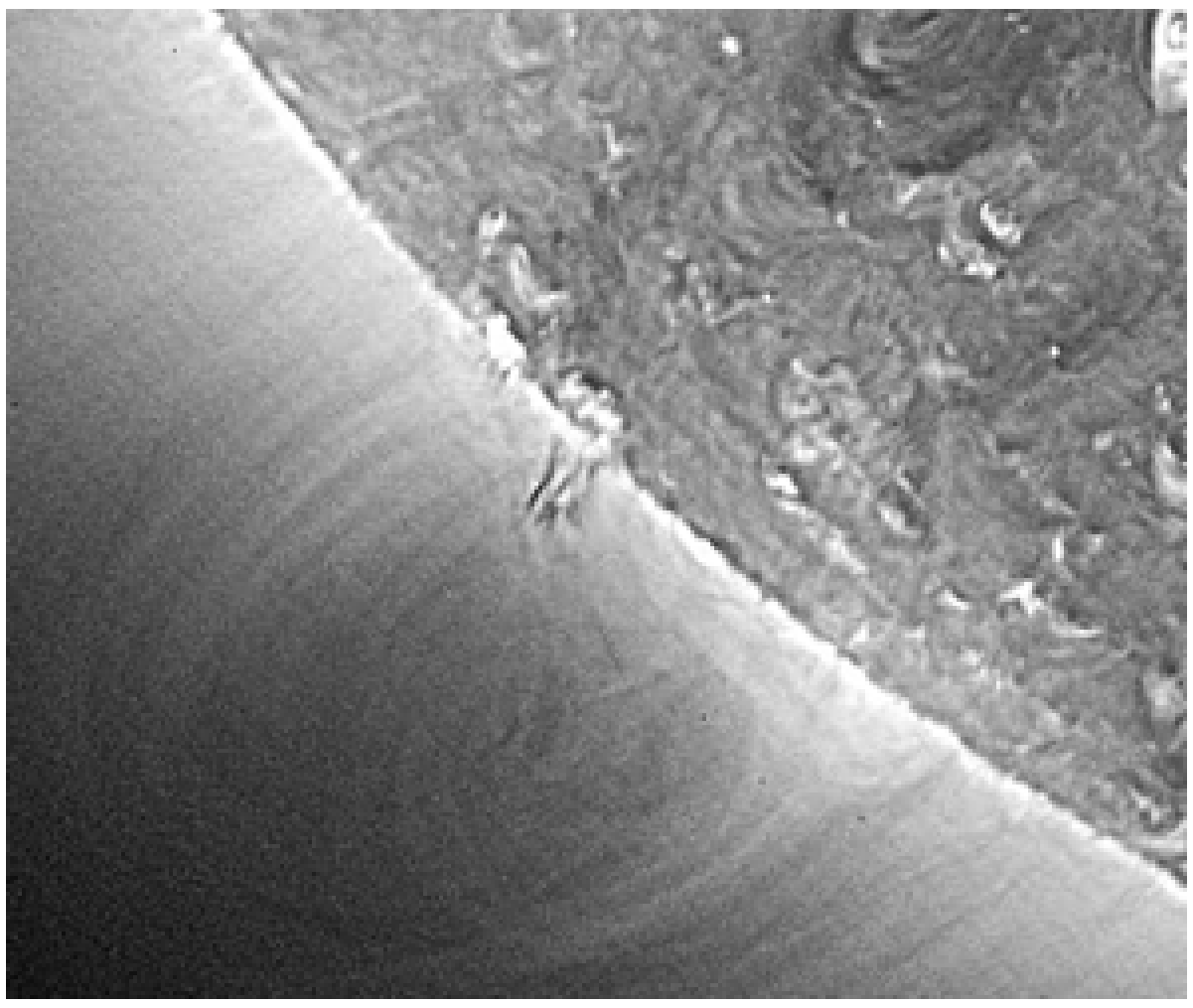

**Figure V-5-14:** *région calme en 193 Å à latitude 45°, le 11 Juillet 2010 à 18h45 TU. 12 images sommées, masque flou 2 pixels, renforcé en échelle logarithmique. L’embrillancement du limbe est précédé d’une fine zone étroite en absorption. Au dessus se situe la cavité coronale.*

L’embrillancement du limbe EUV dans la raie du Fe XII 193 Å (2 MK) peut être une cause de la photo-ionisation des enveloppes d’hélium neutre et ionisé, au dessus du minimum de température.
Cette frange sombre vue en absorption correspond aux spicules froids Nishikawa, Kanno 1979. Et au dessus se trouve l'embrillancement du limbe dont les profils précédents ont été tracés. Ces phénomènes dynamiques plus froids sont bien observés dans les régions polaires, dans les trous coronaux où le déficit de plasma sur la ligne de visée permet de les rendre davantage visibles, comme le montre la figure V-5-15. Ce type d’image une fois sommée, puis ayant subi un filtrage réduisant des hautes fréquences spatiales, permet de mieux explorer les structures dynamiques, les petites boucles sur le limbe.

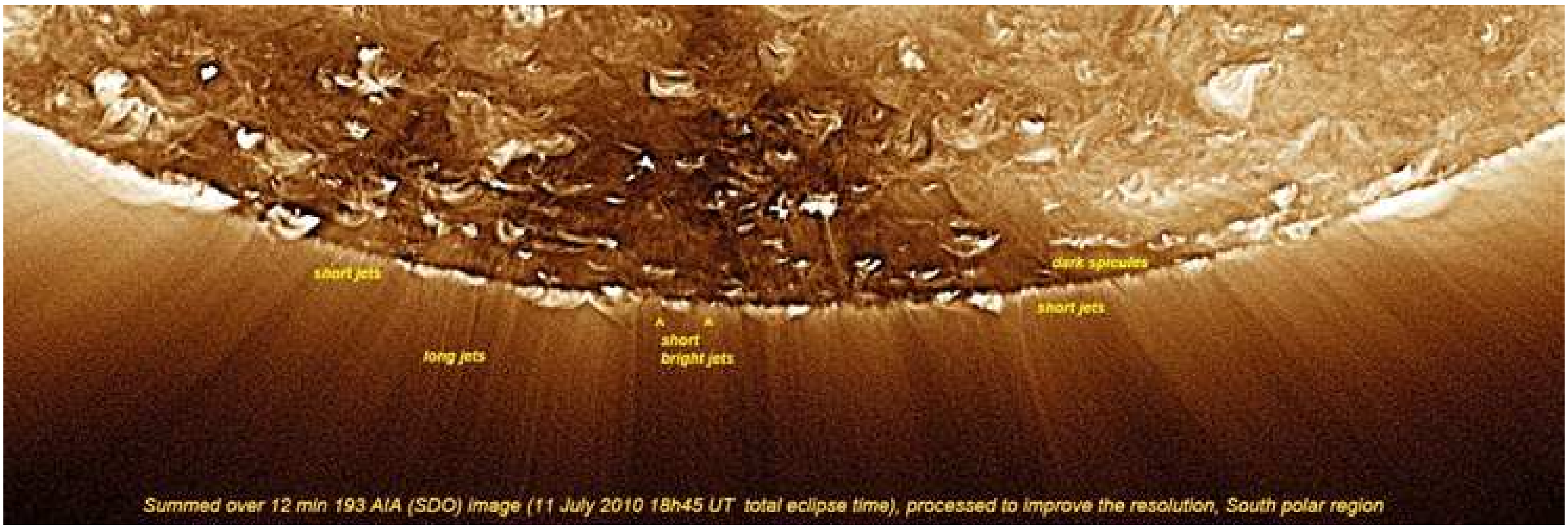

**Figure V-5-15:** *région du trou coronal Sud, 11 Juillet 2010 à 18h45 TU, 12 images sommées, masque flou 2 pixels, renforcé en échelle logarithmique.*

Cette région montre la complexité des structures avec les spicules et les macrospicules vus en absorption (situés en dessous de l'indication en jaune « dark spicules » de la figure V-5-15), de jets et petites boucles en émission, qui s'étendent au-delà de l'interface photosphère-couronne. La forte absorption constatée peut être due à He II – He III à 228 Å.
Des petits jets sont observés en absorption. Ces petits jets pourraient être constitués de plasma froid comme décrit dans l'article de Veselovsky, Triskova et Koutchmy 1994.
La très basse couronne est fortement structuée, mais il est impossible de proposer un modèle statique. Tout est dynamique. Aux éclipses, seule une séquence instantannée était réalisée, ce qui limite les observations durant la courte totalité.
Les échelles de hauteurs correspondant aux températures de l'ordre de 1 à 2 MK se situent entre 50 et 100 Mm, dans le cas d'un modème hydrostatique et homogène. Les largeurs des zones embrillancée, sont de 5 à 10 Mm, soit dix fois moins étendues, ce qui montre que l'embrillancement traduit une nature complexe et inhomogène du limbe solaire, où des phénomènes dynamiques, ont lieu.

# VI) Conclusion

Les expériences de spectres éclair ont contribué à définir le bord du Soleil, sans lumière parasite, grâce aux éclipses totales, de diagnostiquer les gradients de densité, les structurations dans les différentes couches montrant une interface complexe. Les émissions des raies « low FIP » observées dans les protubérances montrent une similitude avec l'interface photosphère-couronne. Les analyses des spectres éclairs ont montré la surabondance des éléments « low FIP » qui proviennent des basses couches. Nous avons sans doute pour la première fois, grâce aux nouvelles caméras CCD à cadence rapide, réussi à obtenir des courbes de lumière de qualité photométrique, qui ont permis de révéler l'existance d'enveloppes, comme le Ba II, et notament celle de l'enveloppe d'hélium ionisée He II 4686 Å (produite par photo-ionisation les rayonnements EUV coronaux et/ou par effets des électrons rapides en provenance de la couronne) avec une résolution suffisante. Ces nouvelles observations ont aussi permis de définir une nouvelle couche de l'atmosphère solaire, la mésosphère où la couronne ne domine pas encore, pour mieux définir la couche renversante, et qui autrefois était mal résolue et définie. Grâce à la résolution spatiale, spectrale, et cadence rapide des caméras CCD ces couches ont pu être examinées en détails et avec une résolutionn extrême.

De plus, les mesures du continu entre les protubérances, et le continu coronal ambiant sur les spectres éclair sommés, ont permis de mieux quantifier le déficit de plasma dans le visible d'un facteur 10 autour des protubérances par rapport aux cavités, et d'un facteur inférieur à 2 sur des profils effectués sur des images AIA à 193 Å le long et en dehors de la cavité.

Ces travaux ont contribué à mieux définir les phénomènes de cavités coronales associées à l'environnement des protubérances, et un renforcement du chauffage d'un facteur 2 au centre de la cavité a été constaté.

Les courbes de lumière $I = f(h)$ sur chaque ion (Fe II, Ti II, Ba II), se situent à des altitudes relativement distinctes dans l'atmosphère solaire.

Les raies plus chaudes d'hélium pour ces couches optiquement minces ont été aussi observées et analysées dans ces basses altitudes à partir de 800 km, ce qui n'avait pas été réalisé de façon aussi précise à cause de la résolution temporelle limitée (plaques photographiques).

Concernant les ions Fe II et Ti II, leur rôle est très important, car leur abondance a été mise en évidence dans les basses couches de l'interface photosphère-couronne, grâce aux spectres éclairs, et pourraient contribuer à alimenter la couronne et les protubérances en « low FIP ».

Les variations des taux d'ionisation de l'hélium ont permis d'étudier comment la couronne peut interpénétrer dans les couches plus basses de l'interface photosphère-couronne aux altitudes inférieures à 2000 km où la gravité domine et dans l'interface protubérance-couronne où le champ magnétique domine. Ces variations du taux d'ionisation de l'hélium He II ont contribué à comparer les couches profondes avec les interfaces protubérance (de plasma froid) et couronne ambiante à 2 MK. La photo-ionisation des raies optiquement minces d'hélium par les raies coronales EUV a permis de montrer une structuration en forme d'enveloppes, et celles-ci sont situées au dessus de la myriade de raies low FIP observées dans les spectres éclair. L'embrillancement du limbe solaire est associé à ces enveloppes et des phénomènes inhomogènes, dynamiques de petites échelles (boucles, spicules) ont lieu dans ces régions. Les spicules, et les macrospicules dans les raies optiquement minces et froides du spectre visible sont associés aux structures observées en absorption dans les raies EUV et sont des structures dynamiques. Les éclipses totales ont une limitation car elles ne permettent d'obtenir que des spectres instantannés.

Le bord du Soleil est défini comme le spectre continu, relevé entre les raies d'émission low FIP lors des contacts d'éclipses, sans lumière parasite. Les tranches monochromatiques extraites de chaque spectre éclair prises aux contacts de l'éclipse de 2010 montrent la structuration complexe du bord du soleil avec le profil du bord lunaire, en l'absence de lumière parasite.
Enfin les expériences des spectres éclairs à cadence CCD rapide, a permis d’apporter une contribution sur l’analyse des couches des interfaces photosphère- chromosphère et couronne, dans des conditions uniques d’éclipses.

# VII) Perspectives

Des expériences de spectres éclair polarisé pourraient être tentées dans les éclipses solaires totales  à venir, dans le but de mesurer les effets du champ magnétique sur la lumière polarisée provenant des raies low FIP de l’interface photosphère- chromosphère et protubérance - couronne. Cependant une des difficultés majeures de ce type d’expérience de spectre éclair polarisé, est le taux de lumière parasite polarisée induite par le réseau par transmission et d’origine instrumentale, dont les valeurs peuvent être supérieures aux taux de polarisation ou dépolarisation d’origine solaire.
Une autre expérience qui pourrait être tentée aux prochaines éclipses totales concerne des acquisitions à cadence très élevée pour améliorer encore la résolution entre 2 spectres consécutifs limités par le mouvement de la Lune dans l’atmosphère solaire. En effet, ce type d’expérience avec une cadence de 60 images/ seconde (sans binning) à 155 images/ seconde (avec binning) pourrait permettre de définir le bord du Soleil avec un pas de 6 km dans l’atmosphère du Soleil, afin de mieux mesurer des infimes variations du flux intégré sur la ligne de visée. Cette résolution temporelle pourrait être améliorée, en effectuant un binning de 4 pixels par 4 pixels, où  on perdrait en résolution spatiale.
Un diviseur optique pourrait être installé à la sortie du réseau objectif où ces 2 caméras réaliseraient les acquisitions simultanément sur les mêmes spectres éclair, sur 2 ordinateurs séparés. La caméra Skynyx enregistrerait à 15 images/s comme guidage sur les spectres et la caméra CCD Lumenera 75 U ayant un binning 4x4 pixels, réaliserait des acquisitions avec une cadence de 155 images par seconde. Cela permettrait de réaliser des courbes de lumière du bord solaire avec une résolution spatiale/temporelle encore jamais atteinte, mais dont la résolution spatiale dans les profils des grains de Baily serait moins bonne. 1 pixel rebinné 4x4 de cette caméra correspond à un champ d’environ 6x6 Mm sur la surface du limbe solaire indépendamment du mouvement de la Lune qui recouvre le disque solaire.

Une nouvelle mission spatiale vient d’être lancée, du nom de IRIS – Interface Region Imaging Spectrograph, et où plus de détails sont donnés en Annexe 35. Cette mission va effectuer des spectres sur le bord du Soleil dans les raies du Mg II vers 2796 et 2804 Å, du C IV 1550 Å, Si IV 1400 Å, dans les régions d’interface chromosphère –couronne que j’ai étudié aux éclipses avec les spectres éclair. J’envisage à l’avenir de nouvelles collaborations avec les investigateurs de ce programme,  pour accéder aux données, comparer les spectres UV obtenus avec IRIS avec les spectres éclair dans le visible obtenus aux prochaines éclipses totales de 2013, 2015, 2016 et 2017, et réaliser des publications dans des revues internationales.

Une autre application d’utilisation de la Lune comme procédé d’occultation naturel pourrait concerner des recherches d’occultation d’étoiles pour tenter d’améliorer la détection des planètes extrasolaires. Les éclipses de Lune (ou en utilisant le bord lunaire du côté de la lumière cendrée lors de la phase en croissant), présentent un avantage de faible lumière réfléchie.
Des spectres pourraient être tentés, en ayant le sens de dispersion radial, dans le sens du déplacement du limbe lunaire (comme pour les spectres éclair) au moyen de grands télescopes (diamètres supérieurs à 1 m) munis au foyer de caméras CCD rapides (100 images/seconde en 16 bit minimum) et de sensibilité élevées. Les courbes de lumière de haute résolution temporelle pourraient atteindre un pas d'échantillonnage de quelques milli-seconde d'arc et moins. Ces courbes de lumière d'étoiles occultées seraient obtenues dans le continu des spectres et dans les vallées lunaires, (le bord lunaire s'apparente d'une certaine manière à un occulteur à bord accidenté), pourraient être analysées en vue de rechercher soit d'éventuelles signatures de planètes extrasolaire après inversion des courbes, où leur atmosphère réfléchirait 1/100000 à 1/10000000 du flux de l'étoile hôte et en comparant aussi avec nos spectres réalisés lors de la sortie de Vénus (egress) au dernier transit de Vénus du 6 Juin 2012.
Ou encore, grâce à ces courbes de lumière, les chromosphères d'autres étoiles de type solaire pourraient être étudiées, pour comparer d'autres interfaces de transition photosphère-chromosphère stellaires.

Ces résultats de spectres éclair pourraient contribuer a améliorer les images de la mission Picard pour définir le bord solaire en vue de mesurer son diamètre, où grâce aux courbes de lumière d'éclipse sans lumière diffusée à 4700 Å et 4500 Å, celles-ci pourraient être introduites dans un processus de traitement des images de Picard pour éliminer la lumière parasite, et permettre enfin de mieux apprécier le bord solaire des images de Picard pour mesurer le « vrai » diamètre solaire.
En effet, l'expérience Picard réalisant des images du Soleil hors conditions d'éclipses totales est hélas affectée par la lumière parasite du disque solaire, malgré les méthodes utilisées, et il n'est pas possible de de définir le « vrai » bord solaire à partir des données que cette mission a fourni. Le « vrai » diamètre solaire n'a ainsi pas pu être mesuré, car les conditions de « vrai » bord solaire sans lumière parasite n'ont pas été atteintes comme aux éclipses totales. Par ailleurs d'autres problèmes de dérive de transmission des filtres, contraintes mécaniques ont été constatés sur cette mission.

Enfin, les expériences des spectres éclairs réalisées le 3 Novembre 2013 lors de la récente éclipse totale de Soleil en Ouganda ont bien réussies. Cette éclipse avait une durée de totalité de 20 secondes, et a permis d'observer la myriade de raies low FIP sous forme de croissants beaucoup plus étendus que lors des éclipses plus longues. En effet, les arcs faisaient un secteur de 110 à 140°, ce qui va permettre l'analyse des profils des raies dans des régions plus proches des pôles. Ces nouvelles données ont montré les enveloppes d'hélium He I 4713 Å et He II 4686 Å sous forme d'anneaux entiers durant la totalité. Les images ont montré le limbe Est dans ces raies d'hélium avec la superposition du limbe Ouest de la raie H béta, décalée de 150 Å, et qui est dans le même champ. Ces nouvelles données seront analysées et feront l'objet d'articles et contributions, suite à mon travail de thèse, lors des prochaines perspectives du Programme Nationnal Soleil Terre – PNST, dans la thématique « couplages entre les différentes enveloppes de plasmas ».